\documentclass[onecolumn,fleqn]{revtex4}

\usepackage{ifthen}
\usepackage{ifpdf}

\usepackage{latexsym}
\usepackage{amsmath} 
\usepackage{amssymb} 
\usepackage{bm}
\usepackage{wasysym}

\ifpdf
\usepackage{graphicx}
\usepackage{epstopdf}
\newcommand{\rang}{90}
\graphicspath{{./Figs/}{./}}
\else
\usepackage{graphicx}
\usepackage{epsfig}
\newcommand{\rang}{0}
\renewcommand{\includegraphics}[2][0]{FIGURE}
\fi

\usepackage[colorlinks,linkcolor=blue,urlcolor=blue,bookmarks=true]{hyperref}

\def\dbar{{\mathchar'26\mkern-12mu d}}
\newcommand{\vecb}[1]{\overset{{}_{\leftarrow}}{#1}}
\newcommand{\vecm}[1]{\overset{{}_{\leftrightarrow}}{#1}}
\newcommand{\sinc}{\mathrm{sinc}}
\newcommand{\const}{\mathrm{const}}
\newcommand{\trc}{\mathsf{trace}}

\newcommand{\im}{\mathrm{Im}}
\newcommand{\re}{\mathrm{Re}}
\newcommand{\eexp}{\mathrm{e}^}
\newcommand{\varphiJ}{\bm{\varphi}}
\newcommand{\thetaJ}{\bm{\theta}}

\newcommand{\mass}{\mathsf{m}} 
\newcommand{\gdos}{\varrho}

\newcommand{\tbox}[1]{\text{#1}}
 
\newcommand{\amatrix}[1]{\begin{matrix} #1 \end{matrix}}

\newcommand{\fracd}[2]{\frac{\partial #1}{\partial #2}}
\newcommand{\bra}[1]{\left\langle #1 \right|}
\newcommand{\ket}[1]{\left| #1 \right\rangle}

\newcommand{\Braket}[2]{\left\langle #1 \middle| #2 \right\rangle}
\newcommand{\BraKet}[3]{\left\langle #1 \middle| #2 \middle| #3 \right\rangle}

\newcommand{\ora}{\protect\overrightarrow}

\newcommand{\bitem}{$\bullet$ \ \ \ }
\newcommand{\drawline}{\begin{picture}(500,1)\line(1,0){500}\end{picture}}

\newcommand{\Cn}[1]{\begin{center} #1 \end{center}}

\newcommand{\mpgt}[2][1.0\hsize]{\begin{minipage}[t]{#1}{#2}\end{minipage}}

\newcommand{\putgraph}[2][0.30\hsize]{\includegraphics[width=#1]{#2}} 
\newcommand{\putgraphv}[2][0.30\hsize]{\includegraphics[height=#1]{#2}}
\newcommand{\putgraphr}[2][0.30\hsize]{\includegraphics[angle=\rang,width=#1]{#2}}

\newcommand{\beq}{\begin{eqnarray}}
\newcommand{\eeq}{\end{eqnarray}}

\usepackage{tikz}

\newcommand{\ExternalLink}{    \tikz[x=1.2ex, y=1.2ex, baseline=-0.05ex]{        \begin{scope}[x=1ex, y=1ex]
            \clip (-0.1,-0.1) 
                --++ (-0, 1.2) 
                --++ (0.6, 0) 
                --++ (0, -0.6) 
                --++ (0.6, 0) 
                --++ (0, -1);
            \path[draw, 
                line width = 0.5, 
                rounded corners=0.5] 
                (0,0) rectangle (1,1);
        \end{scope}
        \path[draw, line width = 0.5] (0.5, 0.5) 
            -- (1, 1);
        \path[draw, line width = 0.5] (0.6, 1) 
            -- (1, 1) -- (1, 0.6);
        }
    }

\renewcommand{\thesection}{\arabic{section}}
\renewcommand{\thesubsection}{\arabic{subsection}}
\renewcommand{\thesubsubsection}{\arabic{subsubsection}}
\newcommand{\sheadA}[1]
{
\addtocounter{section}{1}
\newpage 
\begin{center} 
\pdfbookmark[1]{#1}{shdA\thesection}
{\bf {\LARGE #1}} 
\end{center} 
\addtocontents{toc}{\protect\contentsline{section}{{\Large #1}}{}{}}
\phantomsection
}

\newcommand{\sheadB}[1]
{
\addtocounter{subsection}{1}
\setcounter{subsubsection}{0} 
\setcounter{equation}{0}
\vspace{5mm}
\pdfbookmark[2]{#1}{shdB\thesubsection}
{\bf\LARGE[\thesubsection] \ #1} 
\nopagebreak
\addtocontents{toc}{\protect\contentsline{subsection}{\thesubsection \ \ #1}{\thepage}{shdB\thesubsection.2}}
\phantomsection
}

\newcommand{\sheadC}[1]
{
\addtocounter{subsubsection}{1}
\vspace{5mm}
{\Large\bf $=\!=\!=\!=\!=\!=\;$ [\thesubsection.\thesubsubsection] \ #1}  
\nopagebreak
\phantomsection
}

\begin{document} 
 
\pdfbookmark[1]{Abstract and Topics}{abstrct}

\title{Lecture Notes in Quantum Mechanics} 
 
\author{Doron Cohen} 
 
\affiliation{Department of Physics, Ben-Gurion University, Beer-Sheva 84105, Israel} 
 
\makeatletter
\def\Dated@name{\,\!}
\makeatother
\date{\href{http://arxiv.org/abs/quant-ph/0605180}{\texttt{arXiv:quant-ph/0605180}}}

\begin{abstract} 
These are the \href{http://physics.bgu.ac.il/~dcohen/ARCHIVE/qmc.pdf}{lecture notes \ExternalLink} 
of quantum mechanics \href{http://physweb.bgu.ac.il/COURSES/QuantumMechCohen}{courses \ExternalLink}
that are given by DC at Ben-Gurion University. 
They cover textbook topics that are listed below, 
and also additional advanced topics (marked by~*) 
at the same level of presentation. 
\end{abstract}

\maketitle 

\drawline

\mpgt[0.48\hsize]{

\hyperlink{shdA1.1}{\bf Fundamentals I} \\

\bitem The classical description of a particle  \\
\bitem Hilbert space formalism  \\
\bitem A particle in an $N$ site system  \\
\bitem The continuum limit ($N=\infty$) \\
\bitem Translations and rotations \\

\hyperlink{shdA2.1}{\bf Fundamentals II} \\

\bitem Quantum states / EPR / Bell \\
\bitem The 4 postulates of the theory   \\
\bitem The evolution operator \\
\bitem The rate of change formula \\
\bitem Finding the Hamiltonian for a physical system \\
\bitem The non-relativistic Hamiltonian \\
\bitem The "classical" equation of motion \\
\bitem Symmetries and constants of motion \\

\hyperlink{shdA3.1}{\bf Fundamentals III} \\

\bitem Group theory, Lie algebra \\
\bitem Representations of the rotation group \\
\bitem Spin 1/2, spin 1 and $Y^{\ell,m}$ \\
\bitem Multiplying representations \\
\bitem Addition of angular momentum (*) \\
\bitem The Galilei group (*) \\
\bitem Transformations and invariance (*) \\

\hyperlink{shdA4.1}{\bf Dynamics and driven systems} \\

\bitem Systems with driving \\ 
\bitem The interaction picture \\
\bitem The transition probability formula \\ 
\bitem Fermi golden rule \\
\bitem Markovian master equations \\
\bitem Cross section / Born \\

\bitem The adiabatic equation \\
\bitem The Berry phase \\
\bitem Theory of adiabatic transport (*) \\
\bitem Linear response theory and Kubo (*) \\
\bitem The Born-Oppenheimer picture (*) 

}
\mpgt[0.48\hsize]{

\hyperlink{shdA5.1}{\bf The Green function approach (*)} \\

\bitem The evolution operator  \\ 
\bitem Feynman path integral  \\
\bitem The resolvent and the Green function  \\ 

\bitem Perturbation theory for the resolvent  \\
\bitem Perturbation theory for the propagator  \\
\bitem Complex poles from perturbation theory  \\

\hyperlink{shdA6.1}{\bf Scattering theory (*)} \\

\bitem Scattering: $T$ matrix formalism  \\
\bitem Scattering: $S$ matrix formalism  \\
\bitem Scattering: $R$ matrix formalism  \\
\bitem Cavity with leads `mesoscopic' geometry \\ 
\bitem Spherical geometry, phase shifts  \\ 
\bitem Cross section, optical theorem, resonances \\

\hyperlink{shdA7.1}{\bf Quantum mechanics in practice} \\

\bitem The dynamics of a two level system \\
\bitem Fermions and Bosons in a few site system (*) \\
\bitem Quasi 1D network systems (*) \\

\bitem Approximation methods for $H$ diagonalization \\ 
\bitem Perturbation theory for $H=H_0+V$ \\
\bitem Wigner decay, LDOS, scattering resonances \\

\bitem The Aharonov-Bohm effect \\
\bitem Magnetic field (Landau levels, Hall effect) \\
\bitem Motion in a central potential, Zeeman \\
\bitem The Hamiltonian of spin 1/2 particle, implications \\

\hyperlink{shdA10.1}{\bf Special Topics (*) } \\

\bitem Quantization of the EM field \\   
\bitem Fock space formalism \\

\bitem The Wigner Weyl formalism \\
\bitem Theory of quantum measurements \\ 
\bitem Theory of quantum computation \\
\bitem The foundations of Statistical Mechanics  

}

\drawline

\newpage

\pdfbookmark[1]{Opening Remarks}{opnrmrk}

\Cn{\LARGE Opening remarks}

These lecture notes are based on 3 courses in 
non-relativistic quantum mechanics that are 
given at BGU: "Quantum~2" (undergraduates), 
"Quantum~3" (graduates), and 
"Selected topics in Quantum and Statistical Mechanics" (graduates). 
The lecture notes are self contained, 
and give the {\em road map} to quantum mechanics. 
However, they do not intend to come instead 
of the standard textbooks. In particular I recommend:

[1] L.E.Ballentine, Quantum Mechanics (library code: QC 174.12.B35).

[2] J.J. Sakurai, Modern Quantum mechanics (library code: QC 174.12.S25).

[3] Feynman Lectures Volume III.

[4] A. Messiah, Quantum Mechanics. [for the graduates]

The major attempt in this set of lectures was 
to give a self contained presentation 
of quantum mechanics, {\em which is not based 
on the historical "quantization" approach}. 
The main inspiration comes from Ref.[3] and Ref.[1].
The challenge was to find a compromise 
between the over-heuristic approach of Ref.[3] 
and the too formal approach of Ref.[1]. 

Another challenge was to give a presentation 
of scattering theory that goes well beyond the common 
undergraduate level, but still not as intimidating 
as in Ref.[4]. A major issue was to {\em avoid 
the over emphasis on spherical geometry}. 
The language that I use is much more suitable 
for research with ``mesoscopic" orientation.

Some {\em highlights} for those who look for original 
or advanced pedagogical pieces: 
The EPR paradox, Bell's inequality, and the notion of quantum state;    
The 4~postulates of quantum mechanics; 
Berry phase and adiabatic processes; 
Linear response theory and the Kubo formula; 
Wigner-Weyl formalism; 
Quantum measurements; 
Quantum computation; 
The foundations of Statistical mechanics. 
Note also the following example problems: 
Analysis of systems with~2 or~3 or more sites;
Analysis of the Landau-Zener transition; 
The Bose-Hubbard Hamiltonian;  
Quasi 1D networks; 
Aharonov-Bohm rings; 
Various problems in scattering theory.

Additional topics are covered by:

[5]  D. Cohen, {\em Lecture Notes in Statistical Mechanics and Mesoscopic}, \href{http://arxiv.org/abs/1107.0568}{arXiv:1107.0568}

\Cn{\LARGE Credits}

The first drafts of these lecture notes were prepared 
and submitted by students on a weekly basis during~2005. 
Undergraduate students were requested 
to use HTML with ITEX formulas.  
Typically the text was written in Hebrew. 
Graduates were requested to use Latex. 
The drafts were corrected, integrated, 
and in many cases completely re-written   
by the lecturer. The English translation  
of the undergraduate sections has been    
prepared by my former student {\em Gilad Rosenberg}.  
He has also prepared most of the illustrations.
The current version includes further contributions 
by my PhD students {\em Maya Chuchem} and {\em Itamar Sela}. 
I also thank my colleague Prof. {\em Yehuda Band} 
for some comments on the text.
The arXiv versions are quite remote from 
the original (submitted) drafts, 
but still I find it appropriate to list 
the names of the students who have participated:
Natalia Antin,
Roy Azulai,
Dotan Babai,
Shlomi Batsri,
Ynon Ben-Haim,
Avi Ben Simon,
Asaf Bibi,
Lior Blockstein,
Lior Boker,
Shay Cohen,
Liora Damari,
Anat Daniel,
Ziv Danon,
Barukh Dolgin,
Anat Dolman,
Lior Eligal,
Yoav Etzioni,
Zeev Freidin,
Eyal Gal,
Ilya Gurwich, 
David Hirshfeld,
Daniel Hurowitz,
Eyal Hush,
Liran Israel,
Avi Lamzy,
Roi Levi,
Danny Levy,
Asaf Kidron,
Ilana Kogen,
Roy Liraz,
Arik Maman,
Rottem Manor,
Nitzan Mayorkas,
Vadim Milavsky,
Igor Mishkin,
Dudi Morbachik,
Ariel Naos,
Yonatan Natan,
Idan Oren,
David Papish,
Smadar Reick Goldschmidt,
Alex Rozenberg,
Chen Sarig,
Adi Shay,
Dan Shenkar,
Idan Shilon,
Asaf Shimoni,
Raya Shindmas,
Ramy Shneiderman,
Elad Shtilerman,
Eli S. Shutorov,
Ziv Sobol,
Jenny Sokolevsky,
Alon Soloshenski,
Tomer Tal,
Oren Tal,
Amir Tzvieli,
Dima Vingurt,
Tal Yard,
Uzi Zecharia,
Dany Zemsky,
Stanislav Zlatopolsky.

\newpage

{\footnotesize \tableofcontents}

\sheadA{Fundamentals (part I)}

\sheadB{Introduction}

\sheadC{The building blocks of the universe}

The universe consists of a variety of particles which 
are described by the "standard model". 
The known particles are divided into two groups:

\bitem Quarks: constituents of the proton and the neutron, 
which form the $\sim100$ nuclei known to us. \\
\bitem Leptons: include the electrons, muons, taus, and the neutrinos.

The interaction between the particles is via fields 
(direct interaction between particles is contrary to the 
principles of the special theory of relativity). 
These interactions are responsible for the way material is "organized". 
We shall consider in this course the electromagnetic interaction. 
The electromagnetic field is described by the Maxwell equations.
Within the framework of the "standard model" there are 
additional gauge fields that can be treated on equal footing.
In contrast the gravity field has yet to be incorporated 
into quantum theory. 

\sheadC{A particle in an electromagnetic field} 

Within the framework of classical electromagnetism, the electromagnetic 
field is described by the scalar potential ${V(x)}$ and the 
vector potential ${\vec{A}(x)}$. In addition one defines:
\beq
\mathcal{B} 
&=& \nabla \times \vec{A} 
\\ \nonumber
\mathcal{E} 
&=& - \frac{1}{c} \frac{ \partial \vec{A}}{\partial {t}} - \nabla{V}
\eeq
We will not be working with natural units in this 
course, but from now on we are going to absorb 
the constants $c$ and $e$ in the definition 
of the scalar and vector potentials: 
\beq
\frac{e}{c} A  \ \rightarrow \ A,
&\hspace*{2cm}&
eV \ \rightarrow \ V 
\\ \nonumber
\frac{e}{c} \mathcal{B}  \ \rightarrow \ \mathcal{B}, 
&\hspace*{2cm}&
e\mathcal{E} \ \rightarrow \ \mathcal{E}
\eeq
In classical mechanics, the effect of the electromagnetic 
field is described by Newton's second law with the Lorentz
force. Using the above units convention we write: 
\beq
\ddot{x} = \frac{1}{\mass} \left(\mathcal{E} 
- \mathcal{B} \times v \right)
\eeq
The Lorentz force dependents on the velocity of 
the particle. This seems arbitrary and counter intuitive, 
but we shall see in the future how it can be derived 
from general and fairly simple considerations.   

In analytical mechanics it is customary to derive 
the above equation from a Lagrangian. Alternatively, 
one can use a Legendre transform and derive the equations 
of motion from a Hamiltonian:
\beq
\dot{x} &=& \frac{\partial{\mathcal{H}}}{\partial{p}} 
\\ \nonumber
\dot{p} &=& -\frac{\partial{\mathcal{H}}}{\partial{x}} 
\eeq
where the Hamiltonian is: 
\beq
\mathcal{H}(x,p) = \frac{1}{2\mass}(p-A(x))^2 + V(x) 
\eeq

\sheadC{Canonical quantization} 

The historical method of deriving the quantum description 
of a system is canonical quantization. In this method 
we assume that the particle is described by a "wave function" 
that obeys the equation:  
\beq
\frac{\partial\Psi(x) }{\partial{t}} 
= -\frac{i}{\hbar} {\mathcal{H}}\left({x,-i \hbar \frac{\partial}{\partial {x}}}\right)\Psi(x) 
\eeq
This seems arbitrary and counter-intuitive. 
In this course we shall abandon the historical approach. 
Instead we shall construct quantum mechanics using 
simple heuristic considerations. Later we shall see 
that classical mechanics can be obtained as a special 
limit of the quantum theory.

\sheadC{Second quantization} 

The method for quantizing the electromagnetic 
field is to write the Hamiltonian as a sum of 
harmonic oscillators (normal modes) and then to 
quantize the oscillators. It is exactly the same as 
finding the normal modes of spheres connected with springs. 
Every normal mode has a characteristic frequency. 
The ground state of the field (all the oscillators 
are in the ground state) is called the "vacuum state". 
If a specific oscillator is excited to level $n$, 
we say that there are $n$ photons with frequency $\omega$ in the system. 

A similar formalism is used to describe a many particle 
system. A vacuum state and occupation states are defined. 
This formalism is called "second quantization". 
A better name would be "formalism of quantum field theory".
One important ingredient of this formulation 
is the distinction between fermions and bosons.

In the first part of this course we regard the electromagnetic field 
as a classical entity, where ${V(x), A(x)}$ are given as an input.
The distinction between fermions and bosons will be obtained 
using the somewhat unnatural language of "first quantization". 

\ \\

\sheadC{Definition of mass} 

The "gravitational mass" is defined using a weighting apparatus. 
Since gravitational theory is not includes in this course, 
we shall not use that definition. 
Another possibility is to define "inertial mass". 
This type of mass is determined by considering the 
collision of two bodies: 
\beq
{\mass_1 v_1 + \mass_2 v_2 = \mass_1 u_1 + \mass_2 u_2} 
\eeq
Accordingly one can extract the mass ratio of the two bodies:
\beq
\frac{\mass_1}{\mass_2} = -\frac{u_2-v_2}{u_1-v_1} 
\eeq
In order to give information on the inertial mass of an object, 
we have to agree on some reference mass, say the "kg", to set the units.

Within the framework of quantum mechanics the above Newtonian 
definition of inertial mass will not be used. 
Rather we define mass in an absolute way, 
that does not require to fix a reference mass.
We shall define mass as a parameter in the "dispersion relation".

\sheadC{The dispersion relation} 

It is possible to prepare a "monochromatic" beam of (for example) electrons 
that all have the same velocity, and the same De-Broglie wavelength. 
The velocity of the particles can be measured by using a pair of 
rotating circular plates (discs). The wavelength of the beam can 
be measured using a diffraction grating. We define the momentum of the 
moving particles ("wave number") as:
\beq
p = 2\pi / \mbox{wavelength} 
\eeq
It is possible to find (say by an experiment) the relation 
between the velocity of the particle and its momentum. 
This relation is called the "dispersion relation". 
Here is a plot of what we expect to observe:

\putgraph{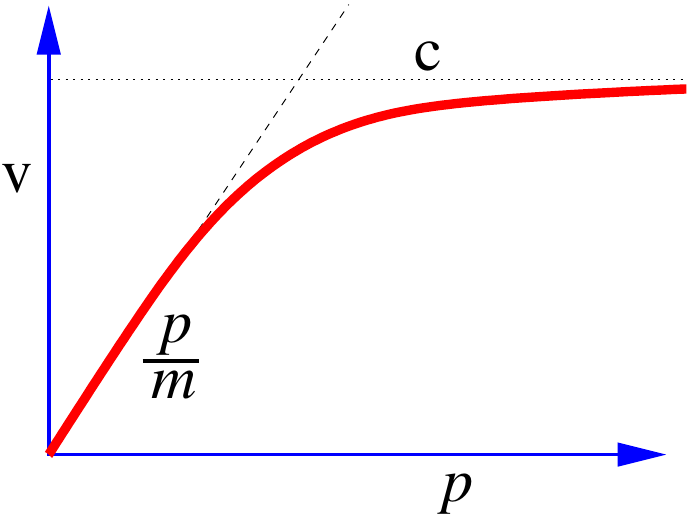}

For low (non relativistic) velocities the relation is approximately linear:
\beq
v \ \ = \ \  \frac{cp}{\sqrt{(mc^2)^2+(cp)^2}} \, c   \ \  \approx  \ \ \frac{1}{\mass} p 
\eeq
This relation defines the "mass" parameter. 
The implied units of mass are  
\beq
{[\mass]}=\frac{T}{L^2} 
\eeq
If we use arbitrary units for measuring mass, say "kg", 
then the conversion prescription is:
\beq
\mass[\mbox{kg}] \ \ = \ \ \hbar \, \mass \left[\frac{\mbox{second}}{\mbox{meter}^{2}}\right],
\ \ \ \ \ \ \ \ \ \ \hbar = \frac{h}{2\pi}
\eeq
where $\hbar$ is known as the Planck constant.

\sheadC{Spin} 

Apart from the degrees of freedom of being in space, the particles also 
have an inner degree of freedom called "spin". We say that a particle 
has spin $s$ if its inner degree of freedom is described by 
a representation of the rotations group of dimension~${2s{+}1}$. 
For example, "spin ${\frac{1}{2}}$" can be described by 
a representation of dimension~2, and "spin~1" can be described 
by a representation of dimension~3. In order to make this abstract 
statement clearer we will look at several examples. 

\bitem Electrons have spin ${\frac{1}{2}}$, 
hence $180^o$ difference in polarization  
("up" and "down") means orthogonality.

\bitem Photons have spin 1, hence $90^o$ difference 
in linear polarizations means orthogonality.

If we position two polarizers one after the other in the 
angles that were noted above, no particles will pass through. 
We see that an abstract mathematical consideration 
(representations of the rotational group) has very realistic consequences.

\newpage

\sheadB{Digression: The classical description of nature}

\sheadC{The electromagnetic field} 

The electric field $\mathcal{E}$ and the magnetic 
field $\mathcal{B}$ can be derived from the vector potential $A$ 
and the electric potential $V$:
\beq
\mathcal{E} 
&=& -\nabla V 
- \frac{1}{c} \frac{\partial \vec{A}}{\partial t} 
\\ \nonumber
\mathcal{B} 
&=&\nabla\times \vec{A} 
\eeq
The electric potential and the 
vector potential are not uniquely determined, 
since the electric and the magnetic fields 
are not affected by the following changes:
\beq
V \,\,\mapsto\,\, \tilde{V} &=& V - \frac{1}{c}\frac{\partial \Lambda}{\partial t} 
\\ \nonumber
A\,\,\mapsto\,\, \tilde{A} &=& A+ \nabla\Lambda 
\eeq
where $\Lambda(x,t)$ is an arbitrary scalar function. 
Such a transformation of the potentials is called "gauge". 
A special case of "gauge" is changing the 
potential $V$ by an addition of a constant.  

Gauge transformations do not affect the 
classical motion of the particle since the 
equations of motion contain only the derived 
fields ${\mathcal{E},\mathcal{B}}$.
\beq
\frac{d^2x}{dt^2}
= \frac{1}{\mass}\left[ e\mathcal{E} 
- \frac{e}{c}\mathcal{B} \times \dot{x} \right] 
\eeq
This equation of motion can be derived from the Lagrangian:
\beq
\mathcal{L}(x,\dot{x})
= \frac{1}{2}\mass \dot{x}^2 + \frac{e}{c}\dot{x}A(x,t) - eV(x,t) 
\eeq
Or, alternatively, from the Hamiltonian:
\beq
\mathcal{H}(x,p) = \frac {1}{2\mass}(p-\frac{e}{c}A)^2 + eV 
\eeq

\sheadC{The Lorentz Transformation} 

The Lorentz transformation takes 
us from one reference frame to the other. 
A Lorentz boost can be written in matrix form as:
\beq
S=\left( \amatrix{ \gamma & -\gamma\beta & 0 & 0 \cr 
-\gamma \beta & \gamma & 0 & 0 \cr 
0 & 0 & 1 & 0 \cr 0 & 0 & 0 & 1 }\right) 
\eeq
where $\beta$ is the velocity of our reference 
frame relative to the reference frame of the lab, and  
\beq
\gamma=\frac{1}{\sqrt{1- \beta ^2}} 
\eeq
We use units such that the speed of light is ${c=1}$. 
The position of the particle in space is:
\beq
\mathbf{x}=\left( \amatrix{ t \cr x \cr y \cr z }\right) 
\eeq
and we write the transformations as:
\beq
\mathbf{x}'=S\mathbf{x} 
\eeq
We shall see that it is convenient 
to write the electromagnetic field as: 
\beq
F=\left( \amatrix{ 0 & \mathcal{E}_1 & \mathcal{E}_2 & \mathcal{E}_3 \cr 
\mathcal{E}_1 & 0 & \mathcal{B}_3 & -\mathcal{B}_2 \cr \mathcal{E}_2 & -\mathcal{B}_3 & 0 & \mathcal{B}_1 \cr 
\mathcal{E}_3 & \mathcal{B}_2 & -\mathcal{B}_1 & 0 }\right) 
\eeq
We shall argue that this transforms as:
\beq
F'=SFS^{-1} 
\eeq
or in terms of components:
\beq
\begin{array}{ll}
\mathcal{E}_1'
=\mathcal{E}_1 \ & \mathcal{B}_1'
=\mathcal{B}_1 
\\ \nonumber
\mathcal{E}_2'
= \gamma(\mathcal{E}_2-\beta \mathcal{B}_3) \ & \mathcal{B}_2'
=\gamma(\mathcal{B}_2+\beta \mathcal{E}_3) 
\\ \nonumber
\mathcal{E}_3'
= \gamma(\mathcal{E}_3+\beta \mathcal{B}_2) \ & \mathcal{B}_3'
=\gamma(\mathcal{B}_3-\beta \mathcal{E}_2) 
\end{array}
\eeq

\sheadC{Momentum and energy of a particle} 

Let us write the displacement of the particle as:
\beq
d\mathbf{x}=\left( \amatrix{ dt \cr dx \cr dy \cr dz }\right) 
\eeq
We also define the proper time 
(as measured in the particle frame) as:
\beq
d\tau^2=dt^2-dx^2-dy^2-dz^2 = (1-{v_x}^2-{v_y}^2-{v_z}^2) dt^2 
\eeq
or:
\beq
d\tau = \sqrt{1-v^2}dt 
\eeq
The relativistic velocity vector is: 
\beq
\mathbf{u} = \frac{d\mathbf{x}}{d\tau}, 
\hspace{2cm}  [u_t^2-u_x^2-u_y^2-u_z^2 = 1]
\eeq
It is customary to define the {\em non-canonical} momentum as 
\beq
\mathsf{p} = \mass \mathbf{u}
= \left( \amatrix{ \epsilon \cr \mathsf{p}_x \cr \mathsf{p}_y \cr \mathsf{p}_z }\right) 
\eeq
According to the above equations we have:
\beq
\epsilon^2- \mathsf{p}_x^2-\mathsf{p}_y^2-\mathsf{p}_z^2 = \mass^2 
\eeq
and write the dispersion relation:
\beq
\epsilon&=&\sqrt{\mass^2 + \mathsf{p}^2} 
\\ \nonumber
v&=& \frac{p}{\sqrt{\mass^2+\mathsf{p}^2}}
\eeq
We note that for non-relativistic 
velocities ${\mathsf{p}_i \approx \mass v_i}$ for  ${i=1,2,3}$ while:
\beq
\epsilon \ \ = \ \ \mass \frac{dt}{d\tau}
\ \ = \ \ \frac{\mass}{\sqrt{1-v^2}} 
\ \ \approx \ \ \mass+\frac{1}{2} \mass v^2+\dots 
\eeq

\sheadC{Equations of motion for a particle}

The non-relativistic equations 
of motion for a particle in 
an electromagnetic field are:
\beq
\mass\frac{d\vec{v}}{dt} 
\ \ = \ \ e\mathcal{E}-e\mathcal{B}\times\vec{v} 
\eeq
The right hand side is the so-called Lorentz force~$\vec{f}$.
It gives the rate of change of the {\em non-canonical} momentum.
The rate of change of the associated non-canonical energy ${E}$ is 
\beq
\frac{d\epsilon}{dt} \ \ = \ \ \vec{f} \cdot \vec{v}
\ \ = \ \ e\mathcal{E}\cdot\vec{v} 
\eeq
The electromagnetic field has equations 
of motion of its own: the Maxwell equations. 
We shall see shortly that Maxwell equations 
are Lorentz invariant. But what Newton's 
second law as written above is not Lorentz invariant. 
In order for the Newtonian equations of motion to 
be Lorentz invariant we have to adjust them. 
It is not difficult to see that the obvious 
required revision is:
\beq
\mass\frac{d\mathbf{u}}{d\tau}
\ \ = \ \ eF\mathbf{u} 
\eeq
To prove the invariance 
under the Lorentz transformation we write:
\beq
\frac{du'}{d\tau} 
\ = \ \frac{d}{d\tau}(Su) 
\ = \ S\frac{d}{d\tau}u 
\ = \ S \frac{e}{\mass}F\mathbf{u} 
\ = \ \frac{e}{\mass}SFS^{-1} (S\mathbf{u}) 
\ = \ \frac{e}{\mass} F' \mathbf{u}' 
\eeq
Hence we have deduced the transformation ${F'=SFS^{-1} }$ 
of the electromagnetic field.

\sheadC{Equations of motion of the field} 

Back to the Maxwell equations. A simple way of writing them is 
\beq
\partial^{\dag} F = 4 \pi J ^{\dag} 
\hspace{3cm}
\eeq
where the derivative operator ${\partial}$, 
and the four-current ${J}$, are defined as:
\beq
\partial=\left( \amatrix{\frac{\partial}{\partial t} \cr 
-\frac{\partial}{\partial x} \cr 
-\frac{\partial}{\partial y} \cr 
-\frac{\partial}{\partial z} }\right) 
\hspace{3cm}
\partial^{\dag} 
= \left(\frac{\partial}{\partial t}, 
\frac{\partial}{\partial x}, 
\frac{\partial}{\partial y}, 
\frac{\partial}{\partial z}\right)
\eeq
and:
\beq
\mathbf{J}=\left( \amatrix{ \rho \cr J_x \cr J_y \cr J_z }\right) 
\hspace{3.3cm}
\mathbf{J}^{\dag} =
(\rho, -J_x, -J_y, -J_z )
\eeq
The Maxwell equations are invariant 
because ${\mathbf{J}}$ and ${\partial}$ transform 
as vectors. For more details see Jackson. An important 
note about notations: in this section we have 
used what is called a "contravariant" representation 
for the column vectors. 
For example ${u=\mbox{column}(u_t,u_x,u_y,u_z)}$. 
For the "adjoint" we use the "covariant" 
representation  ${u = \mbox{row}(u_t,-u_x,-u_y,-u_z)}$. 
Note that $u^{\dag}u = (u_t)^2-(u_x)^2-(u_y)^2-(u_z)^2$ 
is a Lorentz scalar.

\sheadC{The full Hamiltonian} 

The Hamiltonian that describes a system of charged particles 
including the electromagnetic field will be discussed in a 
dedicated lecture (see ``special topics"). Here we just cite 
the bottom line expression:
\beq
\mathcal{H}(\mathbf{r},\mathbf{p},A,\mathcal{E}) \ \ = \ \ 
\sum_i \frac{1}{2\mass_i} (\mathbf{p}_i-e_iA(\mathbf{r}_i))^{2} 
+ \frac{1}{8\pi} \int (\mathcal{E}^{2}+ c^2 (\nabla \times A)^{2}) d^3x
\eeq
The canonical coordinates of the particles are ${(\mathbf{r},\mathbf{p})}$, 
and the canonical coordinates of the field are ${(A,\mathcal{E}_{\perp})}$. 
Note that the radiation field satisfies ${\mathcal{E}_{\perp}=-\dot{A}}$, 
which is conjugate to the magnetic field ${\mathcal{B}=\nabla \times A}$. 
The units of $\mathcal{E}$ as well as the prefactor $1/(8\pi)$ 
are determined via Coulomb law as in the Gaussian convention. 
The units of $\mathcal{B}$ are determined via the Lorentz force formula 
as in the SI convention. Note that for the purpose if conceptual clarity 
we do not make the replacement ${A \mapsto (1/c)A}$, 
hence $\mathcal{B}$ and $\mathcal{E}$ do not have the same units.  

In the absence of particles the second term of the Hamiltonian describes 
waves that have a dispersion relation ${\omega=c|k|}$. 
The strength of the interaction is determined by the coupling 
constants $e_i$. Assuming that all the particles have elementary 
charge ${e_i = \pm e}$, it follows that in the quantum treatment 
the above Hamiltonian is characterized by a single dimensionless 
coupling constant $e^2/c$, which is knows as the ``fine-structure constant".

\newpage

\sheadB{Hilbert space}

\sheadC{Linear algebra} 

In Euclidean geometry, three dimensional 
vectors can be written as: 
\beq
\vec{u} \ \ = \ \ u_1\vec{e}_1+u_2\vec{e}_2+u_3\vec{e}_3
\eeq
Using Dirac notation we can write the same as: 
\beq
|u\rangle \ \ = \ \ u_1|e_1\rangle+u_2|e_2\rangle+u_3|e_3\rangle
\eeq
We say that the vector has the representation: 
\beq
|u\rangle \ \ \mapsto \ \ u_i 
\ = \ \left( \amatrix{ u_1 \cr u_2 \cr u_3 } \right) 
\eeq
The operation of a linear operator ${A}$ is 
written as ${| v \rangle = A| u \rangle}$ 
which is represented by:
\beq
\left( \amatrix{ v_1 \cr v_2 \cr v_3 } \right) 
\ = \ \left( \amatrix{ 
A_{11} & A_{12} & A_{13} \cr 
A_{21} & A_{22} & A_{23} \cr 
A_{31} & A_{32} & A_{33} 
} \right) 
\left( \amatrix{ u_1\cr u_2\cr u_3 } \right) 
\eeq
or shortly as ${v_i = A_{ij} u_j}$. 
Thus a linear operator is represented by a matrix:
\beq
A \ \ \mapsto \ \ A_{ij} 
\ = \ \left( \amatrix{ 
A_{11} & A_{12} & A_{13} \cr 
A_{21} & A_{22} & A_{23} \cr 
A_{31} & A_{32} & A_{33} 
} \right) 
\eeq

\sheadC{Orthonormal basis} 

We assume that an inner 
product ${\langle u | v \rangle}$ has been defined. 
From now on we assume 
that the basis has been 
chosen to be orthonormal: 
\beq
\langle e_i|e_j \rangle \ \ = \ \ \delta_{ij} 
\eeq
In such a basis the inner product (by linearity) 
can be calculated as follows: 
\beq
\langle u|v\rangle \ \ = \ \ u_1^{*}v_1+u_2^{*}v_2+u_3^{*}v_3
\eeq
It can also be easily proved that the 
elements of the representation vector 
can be calculated as follows: 
\beq
u_j \ \ = \ \ \langle e_j | u \rangle 
\eeq
And for the matrix elements we can prove: 
\beq
A_{ij} \ \ = \ \ \langle e_i | A | e_j \rangle 
\eeq

\sheadC{Completeness of the basis} 

In Dirac notation the expansion 
of a vector is written as: 
\beq
|u\rangle 
\ \ = \ \ 
|e_1\rangle \langle e_1|u\rangle 
+|e_2\rangle \langle e_2|u\rangle 
+|e_3\rangle \langle e_3|u\rangle 
\eeq
which implies
\beq
\mathbf{1} \ \ = \ \ |e_1\rangle \langle e_1| +|e_2\rangle \langle e_2| +|e_3\rangle \langle e_3| 
\eeq
Above ${\mathbf{1} \mapsto \delta_{ij}}$ 
stands for the identity operator, 
and  ${P^j = |e_j\rangle \langle e_j|}$ 
are called "projector operators", 
\beq
\mathbf{1} \mapsto \left( \amatrix{ 1 & 0 & 0  \cr 0 & 1 & 0  \cr 0 & 0 & 1 } \right), 
\ \ \ \ P^1 \mapsto \left( \amatrix{1 & 0 & 0 \cr 0 & 0 & 0 \cr 0 & 0 & 0} \right),
\ \ \ \ P^2 \mapsto \left( \amatrix{0 & 0 & 0 \cr 0 & 1 & 0 \cr 0 & 0 & 0} \right),
\ \ \ \ P^3 \mapsto \left( \amatrix{0 & 0 & 0 \cr 0 & 0 & 0 \cr 0 & 0 & 1} \right),
\eeq
Now we can define the "completeness of the basis" as the requirement  
\beq
\sum_j P^j \ \ = \ \ \sum_j |e_j \rangle \langle e_j| \ \ = \ \ \mathbf{1}
\eeq
From the completeness of the basis it follows e.g. 
that for any operator 
\beq
A \ \ = \ \ \left[\sum_i P^i\right] \, A \, \left[\sum_j P^j\right]
\ \ = \ \ \sum_{i,j} |e_i \rangle \langle e_i|A|e_j \rangle \langle e_j| 
\ \ = \ \ \sum_{i,j} |e_i \rangle A_{ij} \langle e_j| 
\eeq

\sheadC{Operators} 

In what follows we are interested in "normal" 
operators that are diagonal in some orthonormal basis.  
Say that we have an operator ${A}$. 
By definition, if it is normal, there exists 
an orthonormal basis ${\{ |a\rangle \}}$ such that $A$ 
is diagonal. Hence we write 
\beq
A \ \ = \ \ \sum_{a}|a\rangle a\langle a| \ \ = \ \ \sum_{a}aP^{a} 
\eeq
In matrix representation it means:
\beq
\left( \amatrix{ a_1 & 0 & 0 \cr 0 & a_2 & 0 \cr 0 & 0 & a_3 } \right) 
\ \ = \ \ 
a_1 \left( \amatrix{ 1 & 0 & 0 \cr 0 & 0 & 0 \cr 0 & 0 & 0 } \right) 
+a_2 \left( \amatrix{ 0 & 0 & 0 \cr 0 & 1 & 0 \cr 0 & 0 & 0 } \right) 
+ a_3 \left( \amatrix{ 0 & 0 & 0 \cr 0 & 0 & 0 \cr 0 & 0 & 1} \right) 
\eeq
It is useful to define what is meant 
by ${\hat{B}=f(\hat{A})}$ where ${f()}$ is an 
arbitrary function. Assuming 
that ${\hat{A}=\sum |a\rangle a\langle a|}$, 
it follows by definition 
that ${\hat{B}=\sum |a\rangle f(a )\langle a|}$. 
Another useful rule to remember is that 
if ${A |k\rangle = B |k\rangle}$ for some 
complete basis ${k}$, then it follows by 
linearity that ${A |\psi\rangle = B |\psi\rangle}$ 
for any vector, and therefore ${A=B}$.

With any operator $A$, we can associate an ``adjoint operator" $A^{\dagger}$. 
By definition it is an operator that satisfies the following relation: 
\beq
\langle u|A v \rangle = \langle A^{\dagger} u|v\rangle 
\eeq
If we substitute the basis vectors in the above 
relation we get the equivalent matrix-style definition 
\beq
(A^{\dagger})_{ij} \ \ = \ \ A_{ji}^{*}
\eeq
If $A$ is normal then it is diagonal in some 
orthonormal basis, and then also $A^{\dagger}$
is diagonal in the same basis.
It follows that a normal operator has 
to satisfy the {\em necessary} condition 
${A^{\dagger}A=AA^{\dagger}}$.
As we show below this is also a {\em sufficient} condition 
for "normality".

We first consider Hermitian operators, 
and show that they are "normal". 
By definition they satisfy ${A^{\dagger}=A}$.
If we write this relation in the eigenstate basis 
we deduce after one line of algebra 
that~${(a^*-b)\langle a|b\rangle=0}$, 
where $a$ and $b$ are any two eigenvalues.
If follows (considering ${a=b}$) that the eigenvalues 
are {\em real}, and furthermore (considering ${a\ne b}$)
that eigenvectors that are associate 
with different eigenvalues are orthogonal.  
This is called the {\em spectral theorem}:
one can find an orthonormal basis in which $A$ is diagonal.

We now consider a general operator~$Q$.
Always we can write it as 
\beq
Q=A+iB, \ \ \ \ \  
\mbox{with} \ A=\frac{1}{2}(Q+Q^{\dagger}), \ \ \ \ \ 
\mbox{and} \ B=\frac{1}{2i}(Q-Q^{\dagger})
\eeq
One observes that $A$ and $B$ are Hermitian operators. 
It is easily verified that ${Q^{\dagger}Q=QQ^{\dagger}}$
iff $AB=BA$. It follows that there is 
an orthonormal basis in which both $A$ and $B$ are diagonal, 
and therefore $Q$ is a normal operator.

We see that an operator is normal iff it satisfies the 
commutation ${Q^{\dagger}Q=QQ^{\dagger}}$ and iff it 
can be written as a function $f(H)$ of an Hermitian operator~$H$.     
We can regard any $H$ with non-degenerate 
spectrum as providing a specification of a basis, 
and hence any other operator that is diagonal in that basis 
can be expressed as a function of this~$H$.

Of particular interest are unitary operators. 
By definition they satisfy ${U^{\dagger}U=\mathbf{1}}$, 
and hence they are "normal" and can be diagonalized in an orthonormal basis. 
Hence their eigenvalues satisfy ${\lambda_{r}^{*}\lambda _{r}=1}$, 
which means that they can be written as:
\beq
U \ \ = \ \ \sum_{r} |r \rangle \eexp{ i \varphi_{r}} \langle r| \ \ = \ \ \eexp{iH}
\eeq
where $H$ is Hermitian. This is an example for the general 
statement that any normal operator can be written as a function 
of some Hermitian operator~$H$.

\sheadC{Conventions regarding notations} 

In Mathematica there is a clear distinction between   
dummy indexes and fixed values. For example $f( x\_ )=8$ 
means that $f(x)=8$ for any $x$, hence $x$ is a dummy index.   
But if $x=4$ then $f(x)=8$ means that only one element 
of the vector $f(x)$ is specified.  Unfortunately  in the 
printed mathematical literature there are no clear conventions. 
However the tradition is to use notations such 
as $f(x)$ and $f(x')$ where $x$ and $x'$ are dummy indexes,  
while $f(x_0)$ and $f(x_1)$ where $x_0$ and $x_1$ are fixed values.   
Thus 
\beq
A_{ij} &=& \left(\amatrix{2 & 3 \cr 5 & 7}\right) 
\\ \nonumber
A_{i_0j_0} &=& 5  \ \ \ \ \ \ \ \mbox{for $i_0=2$ and $j_0=1$}
\eeq
Another typical example is 
\beq
T_{x,k} \ &=& \  \langle x | k \rangle \ \ = \ \ \text{matrix}
\\
\Psi(x) \ &=& \ \langle x | k_0 \rangle \ \ = \ \ \text{column}
\eeq
In the first equality we regard $\langle x | k \rangle$ 
as a matrix: it is the transformation matrix form the 
position to the momentum basis. In the second equality 
we regard the same object (with fixed $k_0$) as a column, 
or as a "wave-function". 

We shall keep the following extra convention: 
The "bra" indexes would appear as subscripts (used for representation), 
while the "ket" indexes would appear as superscripts (reserved for the specification of the state). 
For example: 
\beq
Y^{\ell m}(\theta,\varphi) \ &=& \  \langle \theta,\varphi | \ell m \rangle 
\ \ \ = \ \mbox{spherical harmonics} 
\\
\varphi^n(x) \ &=& \ \langle x | n \rangle
\ \ \ = \ \mbox{harmonic oscillator eigenfunctions}
\\
\psi_n \ &=& \ \langle n | \psi \rangle
\ \ \ = \ \mbox{representation of wavefunction in the $n$ basis}  
\eeq
Sometime it is convenient to use the Einstein summation 
convention, where summation over repeated dummy indexes is implicit. 
For example:
\beq
f(\theta,\varphi) 
\ \ = \ \ 
\sum_{\ell m}       
\langle \theta,\varphi | \ell m \rangle 
\langle \ell m | f \rangle 
\ \ = \ \ 
f_{\ell m} Y^{\ell m}(\theta,\varphi) 
\eeq
In any case of ambiguity it is best 
to translate everything into Dirac notations.

\sheadC{Change of basis}

\ \\
{\bf Definition of $T$:}

Assume we have an "old" basis and a "new" basis for 
a given vector space. In Dirac notation: 
\beq
\mbox{old basis}\,\,\, &=& \,\,\, \{ \,\, |a=1\rangle, \,\, |a=2\rangle, \,\, |a=3\rangle, \,\, \dots  \} 
\\ \nonumber
\mbox{new basis}\,\,\, &=& \,\,\, \{ \,\, |\alpha=1\rangle, \,\, |\alpha=2\rangle, \,\, |\alpha=3\rangle, \,\, \dots  \} 
\eeq
The matrix ${T_{a,\alpha}}$ whose columns 
represent the vectors of the new basis in 
the old basis is called 
the "transformation matrix from the old basis 
to the new basis". In Dirac notation this may be written as:
\beq
| \alpha \rangle \,\, = \,\, \sum_a T_{a,\alpha} \, | a \rangle 
\eeq
In general, the bases do not have to be orthonormal. 
However, if they are orthonormal then ${T}$ must be 
unitary and we have
\beq
T_{a,\alpha} \,\, = \,\, \langle a | \alpha \rangle 
\eeq
In this section we will discuss the general case, 
not assuming orthonormal basis, but in the future 
we will always work with orthonormal bases.

\ \\
{\bf Definition of $S$:} 

If we have a vector-state then we can 
represent it in the old basis or in the new basis:
\beq
| \psi \rangle \,\, 
&=& \,\, \sum_a \psi_a \, | a \rangle 
\\ \nonumber
| \psi \rangle \,\, 
&=& \,\, \sum_{\alpha} \tilde{\psi}_{\alpha} \, | \alpha \rangle 
\eeq
So, the change of representation 
can be written as: 
\beq
\tilde{\psi}_{\alpha} \,\, 
= \,\, \sum_a S_{\alpha,a} \psi_a 
\eeq
Or, written abstractly:
\beq
\tilde{\psi} \,\, 
= \,\, S \psi 
\eeq
The transformation matrix from the old representation 
to the new representation is: ${S=T^{-1}}$.

\ \\
{\bf Similarity Transformation:} 

A unitary operation can be represented 
in either the new basis or the old basis:
\beq
\varphi_{a} \,\, 
&=& \,\, \sum_a A_{a,b} \psi_b 
\\ \nonumber
\tilde{\varphi}_{\alpha} \,\, 
&=& \,\, \sum_{\alpha} \tilde{A}_{\alpha,\beta} \tilde{\psi}_{\beta} 
\eeq
The implied transformation between 
the representations is:
\beq
\tilde{A} \,\, = \,\, S A S^{-1} \,\, = \,\, T^{-1} A T 
\eeq
This is called a similarity transformation.

\sheadC{Generalized spectral decompositions}

Not any operator is normal: that means that not any matrix 
can be diagonalized by a unitary transformation. In particular 
we have sometime to deal with non-Hermitian Hamiltonian that 
appear in the reduced description of open systems. 
For this reason and others it is important to know how 
the spectral decomposition can be generalized. The generalization 
has a price: either we have to work with non-orthonormal basis 
or else we have to work with two unrelated orthonormal sets.
The latter procedure is known as singular value decomposition (SVD).

Given a matrix $A$ we can find its eigenvalues $\lambda_r$, 
which we assume below to be non degenerate. without making any 
other assumption we can always define a set $|r\rangle$ 
of right eigenstates that satisfy ${A|r\rangle = \lambda_r |r\rangle}$.
We can also define a set $|\tilde{r}\rangle$ of left eigenstates  
that satisfy ${A^{\dag}|\tilde{r}\rangle = \lambda_r^* |\tilde{r}\rangle}$.
Unless $A$ is normal, the $r$ basis is not orthogonal, 
and therefore ${\langle r|A| s\rangle}$ 
is not diagonal. But by considering ${\langle \tilde{r}|A| s\rangle}$ 
we can prove that ${\langle \tilde{r}|s\rangle =0}$ if ${r \ne s}$.  
Hence we have dual basis sets, and without loss of generality 
we adopt a normalization convention such that 
\beq
\langle \tilde{r} | s\rangle = \delta_{r,s}
\eeq
so as to have the generalized spectral decomposition:
\beq
A \ \ = \ \ \sum_r  | r \rangle \lambda_r \langle \tilde{r} | 
\ \ = \ \ T \,\left[\mbox{diag}\{\lambda_r\}\right]\,T^{-1}
\eeq
where $T$ is the transformation matrix 
whose columns are the right eigenvectors, 
while the rows of $T^{-1}$ are the left eigenvectors. 
In the standard decomposition method $A$ 
is regarded as describing stretching/squeezing 
in some principal directions, where~$T$ 
is the transformation matrix.   
The SVD procedure provides a different 
type of decompositions. Within the SVD framework $A$ is 
regarded as a sequence of 3~operations: 
a generalized "rotation" followed by stretching/squeezing, 
and another generalized "rotation". Namely:   
\beq
A \ \ = \ \ \sum_r  |U_r\rangle \sqrt{p_r} \langle V_r| 
\ \ = \ \ U\,\sqrt{\mbox{diag}\{p_r\}}\,V^{\dagger}
\eeq
Here the positive numbers $p_r$ are called singular values, 
and $U_r$ and $V_r$ are not dual bases 
but unrelated orthonormal sets. The corresponding 
unitary transformation matrices are $U$ and $V$.

\sheadC{The separation of variables theorem} 

Assume that the operator ${\mathcal{H}}$ commutes 
with an Hermitian operator ${A}$. 
It follows that if ${| a, \nu \rangle }$ is a basis 
in which $A$ is diagonalized, then the operator $\mathcal{H}$ is block 
diagonal in that basis:
\beq
\langle a,\nu | A | a', \nu' \rangle 
& \ \ = \ \ & a\delta_{aa'} \delta_{\nu\nu'} 
\\
\langle a,\nu | \mathcal{H} | a', \nu' \rangle 
& \ \ = \ \ & \delta_{aa'} \mathcal{H}_{\nu\nu'}^{(a)} 
\eeq
where the top index indicates which is the 
block that belongs to the eigenvalue $a$. \\
To make the notations clear consider the following example:
\beq
A=\left(\amatrix{
2 & 0 & 0 & 0 & 0 \cr     
0 & 2 & 0 & 0 & 0 \cr
0 & 0 & 9 & 0 & 0 \cr
0 & 0 & 0 & 9 & 0 \cr
0 & 0 & 0 & 0 & 9  
}\right) 
\ \ \ \ \ \ \ \ \ \ 
H=\left(\amatrix{
5 & 3 & 0 & 0 & 0 \cr     
3 & 6 & 0 & 0 & 0 \cr
0 & 0 & 4 & 2 & 8 \cr
0 & 0 & 2 & 5 & 9 \cr
0 & 0 & 8 & 9 & 7 
}\right)
\ \ \ \ \ \ \ \ \ \ 
H^{(2)} 
=\left(\amatrix{
5 & 3  \cr     
3 & 6  \cr
}\right)
\ \ \ \ \ \ \ \ \  \ 
H^{(9)} 
=\left(\amatrix{
4 & 2 & 8 \cr
2 & 5 & 9 \cr
8 & 9 & 7 
}\right)
\eeq

\beq
\mbox{{\bf Proof:}} 
\ \ \ \ \ \ \ \ \ \ \ 
&&[\mathcal{H},A] = 0 
\\ \nonumber
&&\langle a,\nu|\mathcal{H}A-A\mathcal{H}|a', \nu'\rangle= 0 
\\ \nonumber
&&a'\langle a,\nu|\mathcal{H}|a', \nu'\rangle -a\langle a,\nu|\mathcal{H}|a', \nu'\rangle = 0
\\ \nonumber
&&(a-a')\mathcal{H}_{ a \nu, a' \nu'} = 0 
\\ \nonumber
&&a\neq a' \,\,\, \Rightarrow \,\,\, \mathcal{H}_{ a \nu, a' \nu'}= 0 
\\ \nonumber
&&\langle a,\nu|\mathcal{H}|a', \nu'\rangle \ = \ \delta_{aa'} \, \mathcal{H}^{(a)}_{\nu\nu'}
\eeq
It follows that there is a basis in which 
both $A$ and $\mathcal{H}$ are diagonalized. 
This is because we can diagonalize the matrix~$H$  
block by block (the diagonalizing of 
a specific block does not affect the rest 
of the matrix).

\sheadC{Separation of variables - examples}

The best know examples for ``separation of variables" are  
for the Hamiltonian of a particle in a centrally symmetric
field in 2D and in 3D. In the first case $L_z$ is constant 
of motion while in the second case both $L^2$ and $L_z$ are 
constants of motion. The separation of the Hamiltonian 
into blocks is as follows:

{\bf  Central symmetry in 2D:}   
\beq
\mbox{standard basis} \ \ &=& \ \ |x,y\rangle  \ \ = \ \ |r, \varphi\rangle \\
\mbox{constant of motion} \ \ &=& \ \ L_z  \\
\mbox{basis for separation} \ \ &=& \ \ |m,r\rangle \\
\langle m,r|\mathcal{H}|m', r'\rangle \ \ &=& \ \ \delta_{m,m'} \ \mathcal{H}^{(m)}_{r,r'}  
\eeq
The original Hamiltonian and its blocks:
\beq
\mathcal{H} 
\ \ &=& \ \ \frac{1}{2}\bm{p}^2 + V(r) 
\ \ = \ \ \frac{1}{2}\left(p_r^2 + \frac{1}{r^2}L_z^2\right) + V(r)
\\
\mathcal{H}^{(m)} 
\ \ &=& \ \  \frac{1}{2}p_r^2 +  \frac{m^2}{2 r^2} + V(r)
\ \ \ \ \ \ \ \ \ \ \ \ \ \ \ \
\mbox{where} \ p_r^2 \mapsto -\frac{1}{r}\frac{\partial}{\partial r}\left(r \frac{\partial}{\partial r}\right)
\eeq

{\bf  Central symmetry in 3D:}
\beq
\mbox{standard basis} \ \ &=& \ \ |x,y,z\rangle  \ \ = \ \ |r, \theta, \varphi\rangle \\
\mbox{constants of motion} \ \ &=& \ \ L^2, \ L_z  \\
\mbox{basis for separation} \ \ &=& \ \ |\ell m, r\rangle \\
\langle \ell m,r|\mathcal{H}|\ell' m', r'\rangle \ \ &=& \ \ \delta_{\ell,\ell'} \ \delta_{m,m'} \ \mathcal{H}^{(\ell m)}_{r,r'} 
\eeq
The original Hamiltonian and its blocks:
\beq
\mathcal{H} 
\ \ &=& \ \ \frac{1}{2}\bm{p}^2 + V(r) 
\ \ = \ \ \frac{1}{2}\left(p_r^2 + \frac{1}{r^2}L^2\right) + V(r)
\\
\mathcal{H}^{(\ell m)} 
\ \ &=& \ \  \frac{1}{2}p_r^2 +  \frac{\ell(\ell+1)}{2 r^2} + V(r)
\ \ \ \ \ \ \ \ \ \ \ \ \ \ \ \
\mbox{where} \ p_r^2 \mapsto -\frac{1}{r}\frac{\partial^2}{\partial r^2}r 
\eeq

\newpage

\sheadB{A particle in an $N$ site system}

\sheadC{N site system} 

A site is a location where a particle can be positioned. 
If we have ${N=5}$ sites it means that we have 
a 5-dimensional Hilbert space of quantum states. 
Later we shall assume that the particle can "jump" between sites. 
For mathematical reasons it is conveneint to assume torus topology. 
This means that the next site after ${x=5}$ is ${x=1}$. 
This is also called periodic boundary conditions. 

The standard basis is the position basis. 
For example: ${|x\rangle}$ with ${x = 1,2,3,4,5}$. 
So we can define the position operator as follows: 
\beq
\hat{x} |x\rangle = x |x\rangle 
\eeq
In this example we get:
\beq
\hat{x}\mapsto \left( \amatrix{ 1 & 0 & 0 & 0 & 0 \cr 
0 & 2 & 0 & 0 & 0 \cr 0 & 0 & 3 & 0 & 0 \cr 
0 & 0 & 0 & 4 & 0 \cr 
0 & 0 & 0 & 0 & 5 } \right) 
\eeq
The operation of this operator on a state vector is for example:
\beq
&&|\psi\rangle 
= 7|3\rangle + 5|2\rangle 
\\ \nonumber
&&\hat{x}|\psi\rangle 
=21 |3\rangle + 10 |2\rangle
\eeq

\sheadC{Translation operators} 

The one-step translation operator 
is defined as follows: 
\beq
\hat{D} | x\rangle = |x+1\rangle 
\eeq
For example: 
\beq
D \mapsto \left( \amatrix{ 
0 & 0 & 0 & 0 & 1 \cr 
1 & 0 & 0 & 0 & 0 \cr 
0 & 1 & 0 & 0 & 0 \cr 
0 & 0 & 1 & 0 & 0 \cr 
0 & 0 & 0 & 1 & 0 } \right) 
\eeq
and hence ${D|1\rangle =|2\rangle}$ 
and ${ D|2\rangle =|3\rangle }$ 
and ${ D|5\rangle = |1\rangle}$. 
Let us consider the superposition:
\beq
|\psi\rangle =\frac{1}{\sqrt{5}}[|1\rangle +|2\rangle +|3\rangle +|4\rangle +|5\rangle ]
\eeq
It is clear that ${D|\psi\rangle = |\psi \rangle}$. 
This means that ${\psi}$ is an eigenstate of the translation 
operator (with eigenvalue ${\eexp{i0}}$). 
The translation operator has other eigenstates 
that we will discuss in the next section.

\sheadC{Momentum states} 

The momentum states are defined as follows: 
\beq
&&|k\rangle \rightarrow \frac{1}{\sqrt{N}}\eexp{ikx} 
\\ \nonumber
&&k = \frac{2\pi}{N} n , \,\,\,\, n = \mbox{integer $\mod(N)$} 
\eeq
In the previous section we have encountered the ${k=0}$ momentum state. 
In Dirac notation this is written as:
\beq
|k\rangle =\sum_x \frac{1}{\sqrt{N}}\eexp{ikx}|x\rangle 
\eeq
or equivalently as:
\beq
\langle x|k\rangle =\frac{1}{\sqrt{N}}\eexp{ikx} 
\eeq
while in old fashioned notation it is written as: 
\beq
\psi^k_x = \langle x|k\rangle 
\eeq
where the upper index $k$ identifies the state, 
and the lower index $x$ is the representation index. 
Note that if $x$ were continuous then it would 
be written as ${\psi^k(x)}$.

The ${k}$ states are eigenstates of the 
translation operator. This can be proved 
as follows: 
\beq
D|k\rangle = \sum_x D|x\rangle \langle x|k\rangle 
= \sum_x |x+1\rangle \frac{1}{\sqrt{N}}\eexp{ikx}
= \sum_{x'}| x' \rangle \frac{1}{\sqrt{N}}\eexp{ik{(x'-1})}
= \eexp{-ik}\sum_{x'}|x'\rangle \frac{1}{\sqrt{N}}\eexp{ikx'}
= \eexp{-ik}|k\rangle 
\eeq
Hence we get the result:
\beq
D|k\rangle = \eexp{-ik}|k\rangle 
\eeq
and conclude that ${|k\rangle}$ is an eigenstate of ${\hat{D}}$ 
with an eigenvalue ${\eexp{-ik}}$. Note that the number of independent 
eigenstates is $N$. For exmaple for a 5-site system we have ${\eexp{i k_6}=\eexp{i k_1}}$.

\sheadC{Momentum operator} 

The momentum operator is defined as follows:
\beq
\hat{p}|k\rangle  \ \ \equiv \ \ k|k\rangle
\eeq
From the relation ${\hat{D}|k\rangle = \eexp{-ik} |k\rangle}$ 
it follows that  ${\hat{D}|k\rangle = \eexp{-i\hat{p}} |k\rangle}$. 
Therefore we deduce the operator identity:
\beq
\hat{D} \ \ = \ \ \eexp{-i\hat{p}} 
\eeq
We can also define $2$-step, $3$-step, 
and $a$-step translation operators as follows: 
\beq
\hat{D}(2) \ \ = \ \ (\hat{D})^2 \ \ = \ \ \eexp{-i2\hat{p}}
\\ \nonumber
\hat{D}(3) \ \ = \ \  (\hat{D})^3 \ \ = \ \ \eexp{-i3\hat{p}}
\\ \nonumber
\hat{D}(a) \ \ = \ \ (\hat{D})^a \ \ = \ \ \eexp{-ia\hat{p}}
\eeq

\newpage

\sheadB{The continuum limit}

\sheadC{Definition of the Wave Function} 

Below we will consider a site system 
in the continuum limit. 
${\epsilon \rightarrow 0}$ is the distance between 
the sites, and $L$ is the length of the system.
So, the number of sites is $N = {L}/{\epsilon} \rightarrow \infty$.
The eigenvalues of the position operator 
are ${x_i = \epsilon \times \mbox{integer}}$.
We use the following recipe for changing 
a sum into an integral:
\beq
\sum_i \mapsto \int \frac{dx}{\epsilon} 
\eeq

\begin{center}
\putgraph[0.5\hsize]{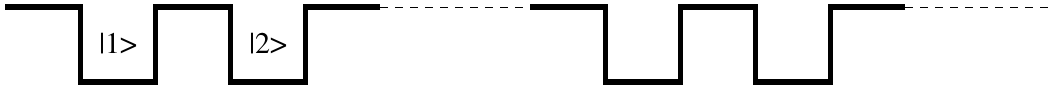}
\end{center}

The definition of the position operator is:
\beq
\hat{x} | x_i \rangle = x_i | x_i \rangle 
\eeq
The completeness of the basis can be written as follows
\beq
\bm{1} \ \ = \ \ \sum  |x_i\rangle \langle x_i | \ \ = \ \ \int   |x\rangle  dx \langle x |
\eeq
In order to get rid of the $\epsilon$ in the integration measure 
we have re-defined the normalization of the basis states as follows:
\beq
|x\rangle \ = \ \frac{1}{\sqrt{\epsilon}} | x_i \rangle \ \ \ \ \ \ \ \mbox{[infinite norm!]}
\eeq
Accordingly the orthonormality relation takes the following form,  
\beq
\langle x|x'\rangle \ \ = \ \ \delta(x-x')
\eeq
where the Dirac delta function is defined as $\delta(0)=1/\epsilon$ 
and zero otherwise.
Consequently the representation of a quantum state is:
\beq
| \psi \rangle  \ \ = \ \ \sum_i \psi_i |x_i\rangle  \ \ = \ \ \int dx \, \psi(x) |x\rangle
\eeq
where
\beq
\psi(x) \ \ \equiv \ \  \langle x | \psi \rangle  \ \ = \ \ \frac{1}{\sqrt{\epsilon}} \psi_x 
\eeq
Note the normalization of the "wave function" is:
\beq
\langle \psi | \psi \rangle \ \ = \ \ \sum_x | \psi_x | ^2 
\ \ = \ \ \int \frac{dx}{\epsilon} | \psi_x | ^2 \ \ = \ \ \int dx | \psi(x) | ^2 \ \ = \ \  1 
\eeq

\sheadC{Momentum States} 

The definition of the momentum states 
using this normalization convention is:
\beq
\psi^k(x) = \frac{1}{\sqrt{L}} \eexp{ikx } 
\eeq
where the eigenvalues are:
\beq
k = \frac{2\pi}{L} \, \times \, \mbox{integer} 
\eeq
We use the following recipe for changing a sum into an integral:
\beq
\sum_k \mapsto \int \frac{dk}{2\pi/L} 
\eeq
We can verify the orthogonality of the momentum states:
\beq
\langle k_2 | k_1 \rangle = \sum_x \langle k_2 | x \rangle \langle x | k_1 \rangle 
= \sum_x  {\psi^{k_2}_x}^*  {\psi^{k_1}_x} 
= \int dx {\psi^{k_2}(x)}^* {\psi ^{k_1}(x)}
= \frac{1}{L} \int dx \mbox{e}^{i(k_1-k_2)x} 
= \delta_{k_2,k_1} 
\eeq
The transformation from the position basis to the momentum basis is:
\beq
\Psi_k= \langle k | \psi \rangle 
= \sum_x \langle k | x \rangle \langle x | \psi \rangle 
= \int {\psi^k(x)}^* \psi(x) dx 
= \frac{1}{\sqrt{L}} \int \psi(x) \mbox{e}^{-ikx} dx 
\eeq
For convenience we will define:
\beq
\Psi(k) = \sqrt{L} \Psi_k 
\eeq
Now we can write the above relation as a Fourier transform:
\beq
\Psi(k) = \int \psi(x) \mbox{e}^{-ikx} dx 
\eeq
Or, in the reverse direction:
\beq
\psi(x) = \int \frac{dk}{2\pi} \Psi(k) \mbox{e}^{ikx}
\eeq

\sheadC{Translations} 

We define the translation operator:
\beq
D(a) | x \rangle = | x+a \rangle 
\eeq
We now proof the following:
\beq
\mbox{Given that:} \ \ \ \ & |\psi\rangle \mapsto \psi(x)  &\\
\mbox{It follows that:} \ \ \ \ & D(a)|\psi\rangle \mapsto \psi(x-a) &
\eeq
In Dirac notation we may write:
\beq
\langle x | D(a) | \psi \rangle = \langle x-a | \psi \rangle 
\eeq
This can obviously be proved easily by operating ${D^{\dagger}}$ on the "bra". 
However, for pedagogical reasons we will also present a longer proof: Given 
\beq
| \psi \rangle = \sum_x \psi(x) |x \rangle 
\eeq
Then 
\beq
D(a) | \psi \rangle = \sum_x \psi(x) |x+a \rangle = \sum_{x'} \psi(x'-a) | x' \rangle = \sum_{x} \psi(x-a) | x \rangle 
\eeq

\sheadC{The Momentum Operator} 

The momentum states are eigenstates of the translation operators:
\beq
D(a) | k \rangle = \mbox{e}^{-iak} |k\rangle 
\eeq
The momentum operator is defined the same as in the discrete case:
\beq
\hat{p}|k\rangle = k| k\rangle 
\eeq
Therefore the following operator identity emerges:
\beq
\hat{D}(a) = \mbox{e}^{-ia\hat{p}} 
\eeq
For an infinitesimal translation:
\beq
D( \delta a) = 1 - i \delta a \hat{p} 
\eeq
We see that the momentum operator is the generator of the translations.

\sheadC{The differential representation}

The matrix elements of the translation operator are:
\beq
\langle x|D(a)|x' \rangle \ \ = \ \ \delta((x-x')-a)
\eeq
For an infinitesimal translation we write:
\beq
\langle x|(\hat{1}-i\delta a \hat{p}) |x' \rangle 
\ \ = \ \ \delta(x-x') - \delta a \delta'(x-x')
\eeq
Hence we deduce:
\beq
\langle x|\hat{p}|x' \rangle \ \ = \ \ -i\delta '(x-x')
\eeq
We notice that the delta function is symmetric, 
so its derivative is anti-symmetric. 
In analogy to multiplying a matrix with 
a column vector we write: ${\hat{A}|\Psi \rangle \mapsto \sum_j A_{ij}\Psi_j }$. 
Let us examine how the momentum opertor 
operates on a "wavefunction":
\beq
\hat{p}|\Psi \rangle \,\, &\mapsto& \,\, \sum_{x'}\hat{p}_{xx'}\Psi_{x'} 
= \int \langle x|\hat{p}|x' \rangle\Psi(x')dx'= 
\\ \nonumber
&=& -i\int\delta '(x-x') \Psi(x')dx' = i \int\delta '(x'-x) \Psi(x')dx'  
\\ \nonumber
&=& -i\int\delta(x'-x) \frac{\partial}{{\partial}x'}\Psi(x')dx' 
= -i\frac{\partial} {\partial x} \Psi(x) 
\eeq
Therefore:
\beq
\hat{p}|\Psi \rangle  \ \ \mapsto \ \  -i\frac{\partial}{\partial x} \Psi(x) 
\eeq

We see that in the continuum limit the operation of $p$ can be realized 
by a differential operator. Let us perform a consistency check.  
We have already proved in a previous section that:
\beq
D(a)|\psi\rangle \ \ \mapsto \ \ \psi(x-a)
\eeq
For an infinitesimal translation we have:
\beq
\Big(1 - i \delta a \hat{p} \Big)| \psi \rangle \ \ \mapsto \ \ \psi(x) - \delta a \frac{d}{dx}\psi(x) 
\eeq
From here it follows that 
\beq
\langle x | p | \psi \rangle \ \ = \ \  -i \frac{d}{dx} \psi(x)
\eeq
This means: the operation of $p$ on a wavefunction is realized 
by the differential operator ${-i(d/dx)}$.

\sheadC{Algebraic characterization of translations} 

If ${|x \rangle}$ is an eigenstate of ${\hat{x}}$ with 
eigenvalue ${x}$, then ${D|x \rangle}$ is an eigenstate 
of ${\hat{x}}$ with eigenvalue ${x+a}$.
In Dirac notations:
\beq
\hat{x} \Big[ D|x \rangle \Big] \ \ = \ \ (x+a) \Big[ D|x \rangle \Big] 
\ \ \ \ \ \ \ \ \mbox{for any $x$} 
\eeq
We have $(x+a) D = D (x+a)$, and $x|x\rangle = \hat{x}|x\rangle$.
Therefore the above equality can be re-written as  
\beq
\hat{x}  D|x \rangle \ \ = \ \ D \, (\hat{x}+a) |x \rangle  
\ \ \ \ \ \ \ \ \mbox{for any $x$}
\eeq
Therefore the following operator identity is implied:
\beq
\hat{x} \ D \ \ = \ \ D \ (\hat{x}+a) 
\eeq
Which can also be written as 
\beq
[\hat{x},D] \ \ = \ \ aD  
\eeq
The opposite is correct too: if an operator $D$ fulfills 
the above relation with another operator $x$, 
then the former is a translation operator with respect to the latter, 
where~$a$ is the translation distance. 

The above characterization applies to any type 
of translation operators, include "raising/lowering" 
operators which are not necessarily unitary.   
A nicer variation of the algebraic relation that 
characterizes a translation operator is obtained  
if $D$ is unitary: 
\beq
D^{-1} \hat{x} D \ \ = \ \ \hat{x}+a
\eeq
If we write the infinitesimal version of this operator relation, 
by substituting ${D(\delta a) = 1 - i \delta a \hat{p}}$ 
and expanding to the first order,  
then we get the following commutation relation:
\beq
[\hat{x},\hat{p}] \ = \ i 
\eeq
The commutation relations allow us to understand 
the operation of operators without having to actually 
use them on wave functions.

\sheadC{Particle in a 3D space} 

Up to now we have discussed the representation of a 
a particle which is confined to move in a one dimensional geometry. 
The generalization to a system with three geometrical dimensions 
is straightforward: 
\beq
| {x,y,z} \rangle &=& | x \rangle \otimes |y \rangle \otimes |z \rangle 
\\ \nonumber
\hat{x} | {x,y,z} \rangle &=& x | {x,y,z} \rangle 
\\ \nonumber
\hat{y} | {x,y,z} \rangle &=& y | {x,y,z} \rangle 
\\ \nonumber
\hat{z} | {x,y,z} \rangle &=& z | {x,y,z} \rangle  
\eeq
We define a "vector operator" 
which is actually a "package" of three operators:
\beq
\hat{\mathbf{r}} = (\hat{x},\hat{y},\hat{z})
\eeq
And similarly:
\beq
\hat{\mathbf{p} } &=& (\hat{p}_x, \hat{p}_y ,\hat{p}_z ) 
\\ \nonumber
\hat{\mathbf{v} } &=& (\hat{v}_x, \hat{v}_y ,\hat{v}_z ) 
\\ \nonumber
\hat{\mathbf{A} } &=& (\hat{A}_x, \hat{A}_y ,\hat{A}_z ) 
\eeq
Sometimes an operator is defined as a function of other operators:
\beq
\hat{\mathbf{A}}= \mathbf{A}( \hat {\mathbf{r}} ) 
= (A_x( \hat{x},\hat{y},\hat{z}) , A_y ( \hat{x},\hat{y},\hat{z}) ,A_z ( \hat{x},\hat{y},\hat{z}) ) 
\eeq
For example $\hat{\mathbf{A}} = \hat{\mathbf{r}} / |\hat{\mathbf{r}}|^3$. 
We also note that the following notation is commonly used:
\beq
\hat{\mathbf{p}}^2 = \hat{\mathbf{p }} \cdot \hat{\mathbf{p}}= \hat{p}_x^2+\hat{p}_y^2+\hat{p}_z^2 
\eeq

\sheadC{Translations in 3D space} 

The translation operator in 3-D is defined as:
\beq
\hat{D} (\mathbf{a}) | \mathbf{r} \rangle \, = \, |\mathbf{r}+ \mathbf{a} \rangle
\eeq
An infinitesimal translation can be written as:
\beq
\hat{D} (\mathbf{\delta a}) 
&=& 
\eexp{-i{\delta}a_x\hat{p}_x} 
\eexp{-i{\delta}a_y\hat{p}_y} 
\eexp{-i{\delta}a_z\hat{p}_z} 
\\ \nonumber
&=& \hat{1} - i \delta a_x \hat{p}_x - i\delta a_y \hat {p}_y - i \delta a_z \hat{p}_z 
= \hat{1} - i \mathbf{\delta a} \cdot \hat{\mathbf{p}} 
\eeq
The matrix elements of the translation operator are:
\beq
\langle \mathbf{r}|D( \mathbf{a}) | \mathbf{r}' \rangle 
= \delta^3 ( \mathbf{r}-( \mathbf{r}'+ \mathbf{a})) 
\eeq
Consequently, the differential representation of the momentum operator is:
\beq
\hat{\mathbf{p}}|\Psi\rangle &\mapsto& \left( - i\frac{\partial}{{\partial}x}\Psi , 
- i\frac{\partial}{{\partial}y}\Psi , 
- i\frac{\partial}{{\partial}z}\Psi \right)
\eeq
or in simpler notation  ${\hat{\mathbf{p}} |\Psi \rangle \mapsto - i \nabla \Psi}$.
We also notice that ${\mathbf{p}^2 |\Psi \rangle \mapsto -{ \nabla }^2\Psi}$.

\newpage

\sheadB{Rotations}

\sheadC{The Euclidean Rotation Matrix}

The Euclidean Rotation Matrix ${R^E(\vec{\Phi})}$ 
is a ${3\times3}$ matrix that rotates the vector ${\mathbf{r}}$.
\beq
\left( \amatrix{x' \cr y' \cr z'} \right) 
= \left( \amatrix{ROTATION \cr MATRIX} \right) \left( \amatrix{x \cr y \cr z} \right) 
\eeq
The Euclidean matrices constitute  
a representation of dimension~$3$
of the rotation group.
The parametrization of a rotation 
is requires three numbers that are kept in a vector~${\vec{\Phi}}$. 
These are the rotation axis orientation $(\theta, \varphi)$, 
and the rotation angle~ ${ \Phi }$. Namely, 
\beq
\vec{\Phi} \ \ =  \ \ \Phi\vec{n} \ \ = \ \ \Phi(\sin\theta\cos\phi,\ \sin\theta\sin\phi,\ \cos\theta) 
\eeq
An infinitesimal rotation $\delta \vec{\Phi}$ can be written as:
\beq
R^E \mathbf{r} \ \ = \ \ \mathbf{r}+ \delta \vec{\Phi} \times \mathbf{r} 
\eeq
Recalling the definition of a cross product we write 
this formula using matrix notations:
\beq
\sum_j R^E_{ij} \mathbf{r}_j \ \ = \ \ 
\sum_j \left[ 
\delta_{ij} + \sum_k \delta\Phi_k \, \epsilon_{kji}
\right] \mathbf{r}_j 
\eeq
Hence we deduce that the matrix that represents 
an arbitrary {\em infinitesimal} rotations is   
\beq
R^E_{ij} \ \ = \ \ \delta_{ij} + \sum_k \delta\Phi_k \, \epsilon_{kji}
\eeq
To find the matrix representation for a {\em finite} rotation 
is more complicated. In the future we shall learn a simple 
recipe how to construct a matrix that represents 
an arbitrary large rotation around an arbitrary axis. 
For now we shall be satisfied in writing the matrix that represents 
an arbitrary large rotation around the Z axis:
\beq
R(\Phi\vec{e}_z)
\ \ = \ \ \left(\amatrix{ 
\cos(\Phi) & -\sin(\Phi) & 0 \cr 
\sin(\Phi) & \cos(\Phi) & 0 \cr 
0 & 0 & 1 }  
\right) 
\ \ \equiv \ \ R^z(\Phi) 
\eeq
Similar expressions hold for X axis and Y axis rotations.
We note that $\vec{\Phi}=R^z(\varphi) R^y(\theta)\vec{e}_z$, 
hence by similarity transformation it follows that    
\beq
R(\vec{\Phi}) \ \ = \ \ R^z(\varphi) R^y(\theta) R^z(\Phi) R^y(-\theta) R^z(-\varphi)
\eeq
This shows that it is enough to know the rotations matrices 
around Y and Z to construct any other rotation matrix.  
However, this is not an efficient way to construct rotation matrices. 
Optionally a rotation matrice can be parameterized by its 
so-called "Euler angles" 
\beq
R(\vec{\Phi}) \ \ = \ \ R^z(\alpha) \ R^x(\beta) \ R^z(\gamma)
\eeq
This reflects the same idea (here we use the common ZXZ convention). 
To find the Euler angles can be complicated, and the advantage is not clear.

\sheadC{The Rotation Operator Over the Hilbert Space}

The rotation operator over the Hilbert space 
is defined (in analogy to the translation operator) as:
\beq
\hat{R}(\vec{\Phi}) | \mathbf{r} \rangle \ \ \equiv \ \ | R^E (\vec{\Phi}) \mathbf{r} \rangle 
\eeq
This operator operates over an infinite 
dimension Hilbert space (the standard basis is 
an infinite number of "sites" in the three-dimensional 
physical space). Therefore, it is represented 
by an infinite dimension matrix:
\beq
R_{r'r} \,= \, \langle r' | \hat{R} | r \rangle \,= \, \langle r' | R^E r \rangle \,= \, \delta ( r'- R^E r ) 
\eeq
That is in direct analogy to the translation operator which is represented by the matrix:
\beq
D_{r'r} \,= \, \langle r' | \hat{D} | r \rangle \,= \, \langle r' | r+a \rangle \,= \, \delta ( r' - (r+a) ) 
\eeq
Both operators $\hat{R}$ and $\hat{D}$ can be regarded as "permutation operators". 
When they act on some superposition (represented by a "wavefunction") their 
effect is to shift it somewhere else.  
As discussed in a previous section if a wavefunction $\psi(\mathbf{r})$ 
is translated by $D(a)$ then it becomes $\psi(\mathbf{r}-a)$. 
In complete analogy, 
\beq
\mbox{Given that:} \ \ \ \ & |\psi\rangle \ \ \mapsto \ \ \psi(r)  &\\
\mbox{It follows that:} \ \ \ \ & \hat{R}(\Phi)|\psi\rangle \ \ \mapsto \ \ \psi(R^E(-\Phi)r) &
\eeq

\sheadC{Which Operator is the Generator of Rotations?} 

The generator of rotations (the "angular momentum operator") 
is defined in analogy to the definition of the generator 
of translations (the "linear momentum operator"). 
In order to define the generator of rotations around 
the axis ${n}$ we will look at an infinitesimal rotation 
of an angle ${\delta \Phi \vec{n}}$. 
An infinitesimal rotation is written as:
\beq
R( \delta \Phi \vec{n} ) \, = \, \mathbf{1} - i\delta \Phi L_n 
\eeq
Below we will prove that the generator of rotations around the axis $n$ is:
\beq
L_n = \vec{n} \cdot ( \mathbf{r} \times \mathbf{p} ) 
\eeq
where:
\beq
\hat{\mathbf{r}} &=& (\hat{x}, \hat{y},\hat{z}) 
\\ \nonumber
\hat{\mathbf{p}} &=& (\hat{p}_x, \hat{p}_y,\hat{p}_z ) 
\eeq

{\bf Proof:} 
We shall show that both sides of the equation
give the same result if they operate on any 
basis state $| \mathbf{r} \rangle$. This means that we 
have an operator identity. 
\beq
R( \delta \vec{\Phi} ) | \mathbf{r} \rangle 
&=& | R^E (\vec{\delta\Phi}) \mathbf{r} \rangle
= | \mathbf{r} + \delta \vec{\Phi} \times \mathbf{r} \rangle 
= D( \delta \vec{\Phi} \times \mathbf{r} ) | \mathbf{r} \rangle 
\\ \nonumber
&=& [ \hat{1}- i ( \delta \vec{\Phi} \times \mathbf{r} ) \cdot \hat{\mathbf{p}} ] | \mathbf{r} \rangle 
= [ \hat{1}- i \hat{\mathbf{p}}\cdot \delta \vec{\Phi} \times \mathbf{r} ] | \mathbf{r} \rangle 
= [ \hat{1} - i \hat{\mathbf{p}} \cdot \delta\vec{\Phi} \times \hat{\mathbf{r}} ] | \mathbf{r} \rangle 
\eeq
So we get the following operator identity:
\beq
R( \delta \vec{\Phi}) 
= \hat{1} - i \hat{\mathbf{p}} \cdot \delta \vec{\Phi} \times \hat{\mathbf{r}} 
\eeq
Which can also be 
written (by exploiting the cyclic property 
of the triple vectorial multiplication):
\beq
R( \delta \vec{\Phi}) 
= \hat{1} - i \delta \vec{\Phi} \cdot (\hat{\mathbf{r}} \times \hat{\mathbf{p}} ) 
\eeq
From here we get the desired result.
Note: The more common procedure to derive this identity  
is based on expanding the rotated 
wavefunction ${\psi(R^E(-\delta\Phi)\mathbf{r})=\psi(r-\delta\Phi \times \mathbf{r})}$, 
and exploiting the association ${p \mapsto -i\nabla}$.

\sheadC{Algebraic characterization of rotations}

A unitary operator $\hat{D}$ realizes a translation~$a$ 
in the basis which is determined by an observable $\hat{x}$  
if we have the equality ${\hat{D}^{-1}\hat{x}\hat{D} = \hat{x}+a}$.
Let us prove the analogous statement for rotations:
A unitary operator $\hat{R}$ realizes rotation $\Phi$
in the basis which is determined by an observable $\hat{r}$ 
if we have the equality 
\beq
\hat{R}^{-1}  \hat{\bm{r}}_i \hat{R} \ \ = \ \ \sum_j R^E_{ij} \hat{\bm{r}}_j 
\eeq
where $R^E$ is the Euclidean rotation matrix.
This relation constitutes an algebraic characterization
of the rotation operator. As a particular example we write 
the characterization of an operator that induce $90^o$ rotation 
around the Z~axis:
\beq
\hat{R}^{-1}  \hat{x} \hat{R}  = -\hat{y}, \ \ \ \ \ \ \ \ \ \ \ 
\hat{R}^{-1}  \hat{y} \hat{R}  = \hat{x}, \ \ \ \ \ \ \ \ \ \ \ 
\hat{R}^{-1}  \hat{z} \hat{R}  = \hat{z}
\eeq
This should be contrasted, say, with the characterization
of translation in the $X$ direction:
\beq
\hat{D}^{-1}  \hat{x} \hat{D}  = \hat{x}+a, \ \ \ \ \ \ \ \ \ \ \ 
\hat{D}^{-1}  \hat{y} \hat{D}  = \hat{y}, \ \ \ \ \ \ \ \ \ \ \ 
\hat{D}^{-1}  \hat{z} \hat{D}  = \hat{z}
\eeq

{\bf Proof:}
The proof of the general statement with regard 
to the algebraic characterization of the rotation 
operator is totally analogous to that in the case of translations. 
We first argue that $\hat{R}$ is a rotation operator iff
\beq
\hat{R}|\bm{r}\rangle \ \ =  \ \ | R^E\bm{r}\rangle  
\ \ \ \ \ \mbox{for any $\bm{r}$} 
\eeq
This implies that  
\beq
\hat{r}_i \Big[ \hat{R}|r \rangle \Big] \ \ = \ \ \left(\sum_j R^E_{ij}r_j\right) \, \Big[ \hat{R}|r \rangle \Big] 
\ \ \ \ \ \ \ \ \mbox{for any $r$} 
\eeq
By the same manipulation as in the case of translations 
we deduce that 
\beq
\hat{r}_i  \, \hat{R} \, |r \rangle \ \ = \ \ \sum_j R^E_{ij} \, \hat{R} \, \hat{r}_j \, |r\rangle  
\ \ \ \ \ \ \ \ \mbox{for any $r$}
\eeq
From here, operating on both sides with $\hat{R}^{-1}$, we get the identity that we wanted to prove.

\sheadC{The algebra of the generators of rotations}

Going on in complete analogy with the case 
of translations we write the above algebraic characterization 
for an {\em infinitesimal} rotation: 
\beq
\left[1+i\delta\Phi_j L_j \right] \, \hat{r}_i \, \left[1-i\delta\Phi_j L_j \right] 
\ \ = \ \ \hat{r}_i + \epsilon_{ijk} \, \delta \Phi_j \, \hat{r}_k 
\eeq
where we used the Einstein summation convention. 
We deduce that 
\beq
[L_j,r_i ] \ \ = \ \ -i \ \epsilon_{ijk} \, \hat{r}_k 
\eeq
Thus we deduce that in order to know if a set of operators ${(J_x,J_y,J_z)}$ 
generate rotations of eigenstates of a 3-component observable~$A$, 
we have to check whether the following algebraic relation is satisfied:
\beq
[\hat{J}_i,\hat{A}_j ] \ \ = \ \  i \, \epsilon_{ijk} \, \hat{A}_k 
\eeq
Note that for a stylistic convenience we have interchanged the order 
of the indexes. In particular we deduce that the algebra that
characterized the generators of rotations is  
\beq
[\hat{J}_i,\hat{J}_j ] \ \ = \ \  i \, \epsilon_{ijk} \, \hat{J}_k 
\eeq
This is going to be the starting point for constructing 
other representations of the rotation group.

\sheadC{Scalars, Vectors, and Tensor Operators}

We can classify operators according to the way that 
they transform under rotations. The simplest possibility 
is a scalar operator $C$. It has the defining property
\beq
\hat{R}^{-1}\hat{C} \hat{R} \ \ = \ \ \hat{C}, 
\ \ \ \ \ \ \ \ \ \mbox{for any rotation} 
\eeq
which means that
\beq
[J_i,C] \ \ = \ \ 0 
\eeq
Similarly the defining property of a vector is
\beq
\hat{R}^{-1} \hat{A}_i \hat{R} \ \  = \ \  R^E_{ij} \hat{A}_j
\ \ \ \ \ \ \ \ \ \mbox{for any rotation} 
\eeq
or equivalently 
\beq
[\hat{J}_i,\hat{A}_j ] \ \ = \ \ i\epsilon_{ijk}\hat{A}_k 
\eeq

The generalization of this idea leads to the 
notion of a tensor. A multi-component observer 
is a tensor of rank $\ell$, if it transforms 
according to the $R^{\ell}_{ij}$ representation  
of rotations. Hence a tensor of rank $\ell$ should
have $2\ell+1$ components.
In the special case of a 3-component "vector", 
as discussed above, the transformation is done using 
the Euclidean matrices $R^E_{ij}$. 

It is easy to prove that if $A$ and $B$ are 
vector operators, then $C=A\cdot B$ is a 
scalar operator. We can prove it either directly, 
or by using the commutation relations.   
The generalization of this idea to tensors 
leads to the notion of "contraction of indices".

\sheadC{Wigner-Eckart Theorem}

If we know the transformation properties of an operator, 
it has implications on its matrix elements.
This section assumes that the student is already familiar 
with the representations of the rotation group.

Let us assume that the representation of the rotations 
over our Hilbert (sub)space is irreducible of dimension dim=$2j{+}1$.
The basis states are ${|m\rangle}$ with ${m=-j...+j}$. 
Let us see what are the implications 
with regard to scalar and vector operators.  

The representation of a scalar operator $C$ should be trivial, 
i.e. proportional to the identity, i.e. a "constant":   
\beq
C_{m'm} \,\,=\,\, c \, \delta_{m'm}  
\ \ \ \ \ \ \mbox{within a given $j$ irreducible subspace} 
\eeq
else it would follow from the ``separation of variables theorem" 
that all the generators ($J_i$) are block-diagonal in the same basis. 
Note that within the pre-specified subspace 
we can write $c = \langle C \rangle$, 
where the expectation value can be taken with any state.

A similar theorem applies to a vector operator $A$. Namely,  
\beq
[A_k]_{m'm} \,\,=\,\, g \times [J_k]_{m'm}  
\ \ \ \ \ \ \mbox{within a given $j$ irreducible subspace} 
\eeq
How can we determine the coefficient $g$? 
We simply observe that from the last equation 
it follows that 
\beq
[A \cdot J]_{m'm} 
\,\,=\,\, g \, [J^2]_{m'm} 
\,\,=\,\,   g \, j(j+1) \, \delta_{m'm}
\eeq
in agreement with what we had claimed regarding 
scalars in general. Therefore we get the formula 
\beq
g \,\,=\,\, \frac {\langle J \cdot A\rangle}{j(j+1)} 
\eeq
where the expectation value of the scalar 
can be calculated with any state. 

The direct proof of the Wigner-Eckart theorem, 
as e.g. in Cohen-Tannoudji, is extremely lengthy. 
Here we propose a very short proof that can be 
regarded as a variation on what we call 
the "separation of variable theorem".

{\bf Proof step (1):} 
From ${[A_x,J_x]=0}$ we deduce that $A_x$ is diagonal 
in the $J_x$ basis, so we can write this relation  
as ${A_x=f(J_x)}$. The rotational invariance implies 
that the same function ${f()}$ related $A_y$ to $J_y$ 
and $A_z$ to $J_z$. This invariance is implied 
by a similarity transformation and using the defining 
algebraic property of vector operators. 

{\bf Proof step (2):} 
Next we realize that for a vector 
operator ${[J_z,A_{+}]=A_{+}}$ 
where ${A_{+}=A_x+iA_y}$. It follows that $A_{+}$
is a raising operator in the $J_z$ basis, 
and therefore must be expressible 
as ${A_{+}=g(J_z)[J_x+iJ_y]}$, where $g()$ is some 
function. 

{\bf Proof step (3):}
It is clear that the only way to satisfy   
the equality ${f(J_x)+if(J_y)=g(J_z)[J_x+iJ_y]}$, 
is to have $f(X)=gX$ and $g(X)=g$, where $g$ is a 
constant. Hence the Wigner-Eckart theorem is proved.

\sheadA{Fundamentals (part II)}

\sheadB{Quantum states / EPR / Bell / postulates}

\sheadC{The two slit experiment}

If we have a beam of electrons, that have been prepared 
with a well defined velocity, and we direct it to a screen 
through two slits, then we get an interference pattern 
from which we can determine the "de-Broglie wavelength" 
of the electrons. I will assume that the student is familiar 
with the discussion of this experiment from introductory courses.  
The bottom line is that the {\em individual} electrons 
behave {\em like~wave} and can be characterized by a 
wavefunction $\psi(x)$. This by itself does not mean that our
world in not classical. We still can speculate that $\psi(x)$ 
has a classical interpenetration. Maybe our modeling of the system 
is not detailed enough. Maybe the two slits, if they are both open,
deform the space in a special way that makes the electrons 
likely to move only in specific directions? Maybe, if we had 
better experimental control, we could {\em predict~with~certainty} 
where each electron will hit the screen. 

The modern interpenetration of the two slit experiment is not classical. 
The so called "quantum picture" is that the electron can be at the same time at two different places:
it goes via both slits and interferes with itself. This sounds strange. 

Whether the quantum interpenetration is correct we cannot establish:
maybe in the future we will have a different theory.
What we {\em can} establish is that a classical interpretation
of reality is {\em not} possible. This statement is based on 
a different type of an experiment that we discuss below.

\sheadC{Is the world classical? (EPR, Bell)}

We would like to examine whether the world we live 
in is ``classical" or not. The notion of 
classical world includes mainly two ingredients: 
(i) realism (ii) determinism.  
By realism we means that any quantity that can 
be measured is well defined even if we do not 
measure it in practice. By determinism we mean 
that the result of a measurement is determined 
in a definite way by the state of the system 
and by the measurement setup. We shall see later 
that quantum mechanics is not classical in both 
respects: In the case of spin~$1/2$ we cannot 
associate a definite value of $\hat{\sigma}_y$ 
for a spin which has been polarized in the $\hat{\sigma}_x$  
direction. Moreover, if we measure the $\hat{\sigma}_y$
of a $\hat{\sigma}_x$ polarized spin, 
we get with equal probability $\pm1$ as the result.
 
In this section we would like to assume that our 
world is "classical". Also we would like to assume    
that interactions cannot travel faster than light.  
In some textbooks the latter is called "locality 
of the interactions" or "causality". It has been found 
by Bell that the two assumptions lead to an inequality 
that can be tested experimentally. It turns out 
from actual experiments that Bell's inequality are violated. 
This means that our world is either non-classical or else we have 
to assume that interactions can travel faster than light.

If the world is classical it follows  
that for any set of initial conditions 
a given measurement would yield a definite result. 
Whether or not we know how to predict or calculate 
the outcome of a possible measurement is not assumed.
To be specific let us consider a particle of zero spin, 
which disintegrates into two particles going
in opposite directions, each with spin~$1/2$. 
Let us assume that each spin is described 
by a set of state variables. 
\beq
\mbox{state of particle A} &=&  x^A_1, x^A_2, ... 
\\ \nonumber
\mbox{state of particle B} &=&  x^B_1, x^B_2, ...
\eeq
The number of state variables might be very big,
but it is assumed to be a finite set. Possibly we 
are not aware or not able to measure some of these ``hidden'' variables.

Since we possibly do not have total control over the
disintegration, the emerging state of the two particles 
is described  by a joint probability function 
$\rho\left(x^A_1,...,x^B_1,...\right)$.
We assume that the particles do not affect each other  
after the disintegration (``causality" assumption).
We measure the spin of each of the particles using
a Stern-Gerlach apparatus. The measurement can yield
either $1$ or $-1$. For the first particle the measurement 
outcome will be denoted as $a$, and for the second particle 
it will be denoted as $b$.  It is assumed that 
the outcomes $a$ and $b$ are determined in a deterministic 
fashion. Namely, given the state variables of the particle 
and the orientation $\theta$ of the apparatus we have 
\beq
a &=&  f(\theta_A, x^A_1, x^A_2, ... ) = \pm1 
\\ \nonumber
b &=&  f(\theta_B, x^B_1, x^B_2, ... ) = \pm1
\eeq
where the function $f()$ is possibly very complicated. 
If we put the Stern-Gerlach machine in a different orientation 
then we will get different results:
\beq
a' &=&   f\left(\theta_A', x^A_1, x^A_2, ...  \right)=\pm1 
\\ \nonumber
b' &=&   f\left(\theta_B', x^B_1, x^B_2, ...  \right)=\pm1
\eeq
We have the following innocent identity:
\beq
ab+ab'+a'b-a'b' = \pm 2
\eeq
The proof is as follows: if $b=b'$ the sum is $\pm 2a$, 
while if $b=-b'$ the sum is $\pm 2a'$. 
Though this identity looks innocent, it is completely 
non trivial. It assumes both "reality" and "causality".
The realism is reflected by the assumption 
that both $a$ and $a'$ have definite values, 
as implied by the function $f()$, 
even if we do not measure them. 
In the classical context it is not an issue 
whether there is a practical possibility 
to measure both $a$ and $a'$ at a single run 
of the experiment.
As for the causality: it is reflected by assuming 
that $a$ depends on $\theta_A$ but not 
on the distant setup parameter $\theta_B$.

Let us assume that we have conducted this experiment many times. 
Since we have a joint probability distribution $\rho$, 
we can calculate average values, for instance:
\beq
\langle ab\rangle = \int{
\rho\left(x^A_1,...,x^B_1,...\right)
f\left(\theta_A, x^A_1,...\right)
f\left(\theta_B, x^B_1,...\right)}
\eeq
Thus we get that the following inequality should hold: 
\beq
\left|
\langle ab\rangle +\langle ab'\rangle +\langle a'b\rangle -\langle a'b'\rangle 
\right|
\leq2\eeq
This is called Bell's inequality (in fact it is a variation of the original version).  
Let us see whether it is consistent with quantum mechanics. 
We assume that all the pairs 
are generated in a singlet (zero angular momentum) state.  
It is not difficult to calculate the expectation values. 
The result is   
\beq
\langle ab \rangle 
\ \ = \ \ -\cos(\theta_A-\theta_B) 
\ \ \equiv \ \ C(\theta_A-\theta_B)
\eeq
we have for example
\beq
C(0^o) = -1, \hspace{15mm}
C(45^o) = -\frac{1}{\sqrt{2}}, \hspace{15mm}
C(90^o) = 0, \hspace{15mm}
C(180^o) = +1.
\eeq
If the world were classical the Bell's inequality would imply  
\beq
|C(\theta_A-\theta_B)+C(\theta_A-\theta_B')
+C(\theta_A'-\theta_B)-C(\theta_A'-\theta_B')| \le 2
\eeq
Let us  take $\theta_A=0^o$ 
and $\theta_B=45^o$ and  $\theta_A'=90^o$ and $\theta_B'=-45^o$. 
Assuming that quantum mechanics holds we get  
\beq
\left|
 \left(-\frac{1}{\sqrt{2}}\right)
+\left(-\frac{1}{\sqrt{2}}\right)
+\left(-\frac{1}{\sqrt{2}}\right)
-\left(+\frac{1}{\sqrt{2}}\right)
\right|
\ = \ 2\sqrt{2} \ > \ 2
\eeq
It turns out, on the basis of celebrated experiments 
that Nature has chosen to violate Bell's inequality. 
Furthermore it seems that the results of the experiments 
are consistent with the predictions of quantum mechanics.
Assuming that we do not want to admit that interactions 
can travel faster than light it follows that our world   
is not classical.

\sheadC{Optional tests of realism}

Mermin and Greenberger-Horne-Zeilinger have proposed optional tests for realism.
The idea is to show that the {\em feasibility} of preparing some quantum states 
cannot be explained within the framework of a {\em classical} theory.
We provide below two simple examples. The spin~1/2 mathematics that is required 
to understand these examples will be discussed in later lecture. What we need 
below is merely the following identities that express polarizations 
in the X and Y directions as a superposition of polarizations in the Z direction:
\beq
| x \rangle \ &=& \ \frac{1}{\sqrt{2}}\left(| z \rangle + | \bar{z}\rangle\right) \\
| \bar{x} \rangle \ &=& \ \frac{1}{\sqrt{2}}\left(| z \rangle - | \bar{z}\rangle\right) \\
| y \rangle \ &=& \ \frac{1}{\sqrt{2}}\left(| z \rangle +i | \bar{z}\rangle\right) \\
| \bar{y} \rangle \ &=& \ \frac{1}{\sqrt{2}}\left(| z \rangle -i | \bar{z}\rangle\right)
\eeq
We use the notations $| z \rangle$ and  $| \bar{z} \rangle$ for denoting "spin up" 
and "spin down" in Z~polarization measurement, and similar convection 
for polarization measurement in the other optional directions X and Y.

{\bf Three spin example.-- } 
Consider 3 spins that are prepared in the following superposition state: 
\beq
|\psi\rangle \ \ = \ \ 
\frac{1}{\sqrt{2}}\left(|\uparrow\uparrow\uparrow\rangle - |\downarrow\downarrow\downarrow\rangle\right)
\ \ \equiv \ \ 
\frac{1}{\sqrt{2}}\left(| zzz \rangle - | \bar{z}\bar{z}\bar{z}\rangle\right)
\eeq
If we measure the polarization of 3 spins 
we get ${a=\pm1}$ and  ${b=\pm1}$ and ${c=\pm1}$, 
and the product would be ${C=abc=\pm1}$. 
If the the measurement is in the ZZZ basis 
the result might be either ${C_{ZZZ}=+1}$ or ${C_{ZZZ}=-1}$ 
with equal probabilities. 
But optionally we can perform an XXX measurement or XYY, 
or YXY, or YYX measurement. If for example we perform XYY measurement 
it is useful to write the state in the XYY basis:
\beq
|\psi\rangle \ \ = \ \ \frac{1}{2}\left(| \bar{x} y \bar{y} \rangle + | \bar{x} \bar{y} y \rangle + | x \bar{y} \bar{y} \rangle + | x y y \rangle\right)
\eeq 
We see that the product of polarization is always $C_{XYY}=+1$. 
Similarly one can show that $C_{YXY}=+1$ and $C_{YYX}=+1$.
If the world were classical we could predict the result 
of an XXX measurement:
\beq
C_{XXX} \ \ = \ \ a_xb_xc_x \ \ = \ \  a_x \ b_x \ c_x \ a_y^2 \ b_y^2 \ c_y^2 \ \ = \ \ C_{XYY}C_{YXY} C_{YYX} \ \ = \ \ 1
\eeq
But quantum theory predicts a contradicting result. 
To see what is the expected result we write the state in the XXX basis:  
\beq
|\psi\rangle \ \ = \ \ \frac{1}{2}\left(| \bar{x} x x \rangle + | x \bar{x} x \rangle + | x x \bar{x} \rangle + | \bar{x} \bar{x} \bar{x} \rangle\right)
\eeq 
We see that the product of polarization is always $C_{XXX}=-1$. 
Thus, the experimental feasibility of preparing such quantum state 
contradicts classical realism.

{\bf Two spin example.-- }
Consider 2 spins that are prepared in the following superposition state: 
\beq
|\psi\rangle 
\ \ &=& \ \ \frac{1}{\sqrt{3}}\left(| z\bar{z} \rangle + | \bar{z}z \rangle - | \bar{z}\bar{z} \rangle  \right) \\
\ \ &=& \ \ \frac{1}{\sqrt{6}}\left(| zx \rangle - | z\bar{x} \rangle + 2 | \bar{z}\bar{x} \rangle  \right) \\
\ \ &=& \ \ \frac{1}{\sqrt{6}}\left(| xz \rangle - | \bar{x}z \rangle + 2 | \bar{x}\bar{z} \rangle  \right) \\
\ \ &=& \ \ \frac{1}{\sqrt{12}}\left(| xx \rangle + | x\bar{x} \rangle +  | \bar{x}x \rangle - 3 | \bar{x}\bar{x} \rangle \right)
\eeq
Above we wrote the state in the optional bases ZZ and ZX and XZ and XX.
By inspection we see the following: \\
{\bf \hspace*{5mm} (1)} The ZZ measurement result $| zz \rangle$ is impossible. \\
{\bf \hspace*{5mm} (2)} The ZX measurement result $| \bar{z}x \rangle$ is impossible. \\
{\bf \hspace*{5mm} (3)} The XZ measurement result $| x\bar{z} \rangle$ is impossible. \\
{\bf \hspace*{5mm} (4)} All XX measurement results are possible with finite probability. \\
We now realize that in a classical reality observation (4) is in contradiction 
with observations (1-3). The argument is as follow: in each run of the experiment 
the state ${\vec{a}=(a_x,a_z)}$ of the first particle is determined by some set of hidden variables. 
The same applies with regard to the ${\vec{b}=(b_x,b_z)}$ of the second particle. 
We can define a {\em joint} probability function $f\left(\vec{a},\vec{b}\right)$ that gives the 
probabilities to have any of the $4\times4$ possibilities (irrespective 
of what we measure in practice). 
It is useful to draw a $4\times4$ truth table and to indicate all 
the possibilities that are not compatible with (1-3). Then it turns out 
that the remaining possibilities are all characterized by 
having ${a_x=-1}$ or  ${b_x=-1}$. This means that in a classical
reality the probability to measure $| xx \rangle$ is zero.
This contradicts the quantum prediction~(4). Thus, the experimental feasibility 
of preparing such quantum state contradicts classical realism.

\sheadC{The notion of quantum state}

A-priory we can classify the possible "statistical states" 
of a prepared system as follows: 
\begin{itemize}
\item Classical state: any measurement gives a definite value.
\item Pure state: there is a complete set of measurements that 
give definite value, while any other measurement gives an uncertain value.
\item Mixture: it is not possible to find a complete set of 
measurements that give a definite value. 
\end{itemize}
When we go to Nature we find that classical states do not exist. 
The best we can get are "pure states". For example:

{\bf (1)} The best we can have with the spin of an electron 
is 100\% polarization (say) in the $X$ direction, 
but then any measurement in any different direction 
gives an uncertain result, except the $-X$ direction which we 
call the "orthogonal" direction. Consequently we are inclined 
to postulate that polarization (say) in the non-orthogonal $Z$ direction 
is a superposition of the orthogonal $X$ and $-X$ states.

{\bf (2)} With photons we are inclined to postulate that 
linear polarization in the $45^o$ direction is a superposition 
of the orthogonal $X$~polarization and $Y$~polarization states.
Note however that contrary to the electronic spin, 
here the superposition of linear polarized states can optionally 
give {\em different~type} of polarization (circular / elliptic).   
  
{\bf (3)} With the same reasoning, and on the basis of the ``two slit experiment" 
phenomenology, we postulate that a particle can be in a superposition 
state of two different locations.  The subtlety here is that superposition 
of different locations is not another location but rather (say) a momentum state, 
while superposition of different polarizations states is still another 
polarization state. 

Having postulated that all possible pure states can be regarded  
as forming an Hilbert space, it still does not help us to define 
the notion of quantum state in the statistical sense.
We need a second postulate that would imply the following: 
If a full set of measurements is performed (in the statistical sense), 
then one should be able to predict (in the statistical sense) the result 
of any other measurement. 

{\bf Example:} In the case of spins 1/2,  say that one measures 
the average polarization $M_i$ in the $i=X,Y,Z$ directions.  
Can one predict the result for $M_n$, where $\bm{n}$ is 
a unit vector pointing in an arbitrary direction?    
According to the second postulate of quantum mechanics (see next section) 
the answer is positive. Indeed experiments reveal that $M_n = \bm{n} \cdot M$.
Taking together the above two postulates, our objective would be 
to derive and predict such linear relations from 
our conception of Hilbert space. In the spin~$1/2$ example 
we would like to view $M_n = \bm{n} \cdot M$ as arising from 
the $\mbox{dim}{=}2$ representation of the rotation group. 
Furthermore, we would like to derive more complicated relations that 
would apply to other representations (higher spins).

\sheadC{The four Postulates of Quantum Mechanics}

The 18th century version of classical mechanics can be derived 
from three postulates: The three laws of Newton.
The better formulated 19th century version 
of classical mechanics can be derived from three postulates: 
(1) The state of classical particles is determined 
by the specification of their positions and its velocities; 
(2) The trajectories are determined by an ``action principle", 
hence derived from a Lagrangian. 
(3) The form of the Lagrangian of the theory is determined by symmetry considerations,  
namely Galilei invariance in the non-relativistic case.
See the Mechanics book of Landau and Lifshitz for details.

Quantum mechanics requires four postulates: 
two postulates define the notion of quantum state, 
while the other two postulates, in analogy with 
classical mechanics, are about the laws that 
govern the evolution of quantum mechanical systems.
The four postulates are: 

{\bf (1)} 
The collection of "pure" states is a linear space (Hilbert).

{\bf (2)} 
The expectation values of observables obey linearity:   
${\langle \alpha \hat{X} + \beta \hat{Y} \rangle = \alpha \langle \hat{X}\rangle + \beta \langle \hat{Y} \rangle}$ 

{\bf (3)} 
The evolution in time obey the superposition principle:
${\alpha | \Psi^0 \rangle + \beta | \Phi^0 \rangle 
\, \rightarrow\, \alpha | \Psi^t \rangle + \beta | \Phi^t \rangle}$

{\bf (4)} 
The dynamics of a system is invariant 
under specific transformations ("gauge", "Galilei"). \\

The first postulate refers to "pure states". These 
are states that have been filtered. 
The filtering is called "preparation". 
For example: we take a beam of electrons. 
Without "filtering" the beam is not polarized. 
If we measure the spin we will 
find (in any orientation of the measurement apparatus) 
that the polarization is zero. On the other hand, 
if we "filter" the beam (e.g. in the left direction) 
then there is a direction for which we will get 
a definite result (in the above example, in the right/left direction). 
In that case we say that there is full polarization - a pure state. 
The "uncertainty principle" tells us that if in a specific 
measurement we get a definite result 
(in the above example, in the right/left direction), 
then there are different measurements 
(in the above example, in the up/down direction) 
for which the result is uncertain. 
The uncertainty principle is implied by the first postulate.

The second postulate use the notion of "expectation value" 
that refers to "quantum measurement". In contrast with classical 
mechanics, the measurement has meaning only in a statistical sense.   
We measure "states" in the following way: 
we prepare a collection of systems that 
were all prepared in the same way. We make the measurement 
on all the "copies". The outcome of the measurement is an 
event ${\hat{x} = x}$ that can be characterized by a distribution 
function. The single event can show that a particular outcome 
has a non-zero probability, but cannot provide full information 
on the state of the system.
For example, if we measured the spin of a single electron and 
get ${\hat{\sigma_z} = 1}$, it does not mean that the state is polarized "up". 
In order to know if the electron is polarized we must measure 
a large number of electrons that were prepared in an identical way. 
If only $50\%$ of the events give ${\hat{\sigma_z} = 1}$ we should  
conclude that there is no definite polarization in the direction we measured!

\sheadC{Observables as random variables} 

Observable is a random variable that can have upon measurement 
a real numerical value. In other words ${\hat{x} = {x}}$
is an {\em event}. Let us assume, for example, that 
we have a particle that can be in one of five 
sites: ${x = 1 , 2 , 3 , 4 , 5}$. An experimentalist 
could measure ${\mbox{Prob}( \hat{x} = 3 )}$ 
or ${\mbox{Prob}( \hat{p} = 3(2\pi/5) )}$. 
Another example is a measurement of the 
probability ${\mbox{Prob}( \hat{\sigma_z} = 1)}$ 
that the particle will have spin up. 

The collection of values of ${x}$ is called the spectrum 
of values of the observable. We make the distinction 
between random variables with a discrete spectrum, 
and random variables with a continuous spectrum. 
The probability function for a random variable 
with a discrete spectrum is defined as:
\beq
f (x) \,=\, \mbox{Prob}( \hat{x} =x ) 
\eeq
The probability density function for a random 
variable with a continuous spectrum is defined as:
\beq
f(x)dx \, = \, \mbox{Prob}( x < \hat{x}< x+dx ) 
\eeq
The expectation value of a variable is defined as:
\beq
\langle \hat{x} \rangle \,\,\,=\,\,\, 
\sum_x f(x) x  
\eeq
where the sum should be understood 
as an integral $\int dx$ in the case 
the $x$ has a continuous spectrum. 
Of particular importance is the random variable  
\beq
\hat{P}^x \ \ = \ \ \delta_{\hat{x},x}
\eeq
This random variable equals~$1$ if ${\hat{x}=x}$ 
and zero otherwise. Its expectation value is the 
probability to get~$1$, namely 
\beq
f(x) \ \ = \ \ \langle \hat{P}^x \rangle 
\eeq
Note that $\hat{x}$ can be expressed as 
the linear combination $\sum_x x \hat{P}^x$.

\sheadC{Quantum Versus Statistical Mechanics}

Quantum mechanics stands opposite classical statistical mechanics. 
A particle is described in classical statistical mechanics by a probability function:
\beq
\rho(x,p)dxdp \ \ = \ \ \mbox{Prob}( x < \hat{x}<x+dx, p < \hat{p} < p + dp) 
\eeq
Optionally this definition can be expressed as the expectation value 
of a phase space projector 
\beq
\rho(x,p) \ \ =  \ \ \langle \,\delta(\hat{x}-x)\,\delta(\hat{p}-p) \,\rangle
\eeq
The expectation value of a random 
variable ${\hat{A} = A(\hat{x}, \hat{p})}$ 
is implied:
\beq
\langle \hat{A} \rangle \ \ = \ \ \int A(x, p) \rho(x, p) dxdp 
\eeq
From this follows the linear relation:
\beq
\langle \alpha \hat{A} + \beta \hat{B} \rangle 
\ \ = \ \ \alpha \langle \hat{A}\rangle + \beta \langle \hat{B} \rangle
\eeq
We see that the linear relation of the expectation 
values is a trivial result of classical probability 
theory. It assumes that a {\em joint probability function}  
can be defined. But in quantum mechanics we 
cannot define a "quantum state" using a joint 
probability function, as implied by the observation  
that our world is not ``classical". For example 
we cannot have both the location and the momentum 
well defined simultaneously: a momentum state, 
by definition, is spread all over space.
For this reason, we have to use a more sophisticated 
definition of ${\rho}$. The more sophisticated definition 
regards $\rho$ as set of expectation values,
from which all other expectation values can be deduced,    
taking the linearity of the expectation value as a postulate.

\sheadC{Observables as operators}

In the quantum mechanical treatment 
we regard an observable~$\hat{x}$ as an operator.
Namely we {\em define} its operation on 
the basis states as $\hat{x}|x\rangle=x|x\rangle$, 
and by linearity its operation is defined on any other state.
We can associate with the basis states projectors $\hat{P}^x$.
For example  
\beq
\hat{x} \mapsto \left( \amatrix{1&0&0 \cr 0&2&0 \cr 0&0&3}\right);  
\ \ \ \ \ \ \ \
\hat{P}^1 \mapsto \left( \amatrix{1&0&0 \cr 0&0&0 \cr 0&0&0}\right);
\ \ \ \ \ \ \ \
\hat{P}^2 \mapsto \left( \amatrix{0&0&0 \cr 0&1&0 \cr 0&0&0}\right);
\ \ \ \ \ \ \ \
\hat{P}^3 \mapsto \left( \amatrix{0&0&0 \cr 0&0&0 \cr 0&0&1}\right);
\eeq
In order to further discuss the implications of the first two postulates 
of quantum mechanics it is useful to consider the simplest example, which is spin~$1/2$.
Motivated by the experimental context, we make the following associations between random variables and operators:
\beq
\hat{\sigma}_n \ \ \mapsto \ \ \left( \amatrix{1&0 \cr 0&-1}\right)_{\mbox{in the $n$ basis}} 
\ \ \ \ \ \ \ \ \ \mbox{$n=x,y,z$ or any other direction}
\eeq
Optionally we can define the projectors 
\beq
\hat{P}^n \ \ \mapsto \ \ \left( \amatrix{1&0 \cr 0&0}\right)_{\mbox{in the $n$ basis}} 
\ \ \ \ \ \ \ \ \ \mbox{$n=x,y,z$ or any other direction}
\eeq
Note that 
\beq
\hat{\sigma}_n \ \ = \ \ 2\hat{P}^n -\hat{1}
\eeq
It follows from the first postulate that the polarization state $|n\rangle$ 
can be expressed as a linear combination of, say, "up" and "down" polarizations. 
We shall see that the mathematical theory of the rotation group representation, 
implies that in the standard (up/down) basis 
the operators $\hat{\sigma}_x$, $\hat{\sigma}_y$, and $\hat{\sigma}_z$
are represented by the Pauli matrices $\sigma_x$, $\sigma_y$, and $\sigma_z$. 
We use the notations: 
\beq
\ \ \sigma_x = \left(\amatrix{0&1\cr 1&0}\right), 
\ \ \sigma_y = \left(\amatrix{0&-i\cr i&0}\right), 
\ \ \sigma_z = \left(\amatrix{1&0\cr 0&-1}\right)
\eeq
Furthermore, the mathematical theory of the rotation group representation,    
allows us to write any $\sigma_n$ as a linear combination of ${(\sigma_x, \sigma_y, \sigma_z)}$.

Taking a more abstract viewpoint we point out that any spin operator 
is represented by a ${2\times 2}$ matrix, 
that can be written as a linear combination of standard basis matrices as follows:
\beq
\left( \amatrix{a&b\cr c&d}\right)
\ \ = \ \ a  \left(\amatrix{1&0\cr 0&0}\right) 
+ b  \left(\amatrix{0&1\cr 0&0}\right) 
+ c \left(\amatrix{0&0\cr 1&0}\right) 
+ d  \left(\amatrix{0&0\cr 0&1} \right) 
\ \ = \ \ 
\sum_{nm} A_{nm} \mathbb{P}^{mn} 
\eeq
In the example above we denote the basis matrices  
as ${ \{P^{z},S^{+},S^{-},P^{-z}\} }$.
In general we use the notation $\mathbb{P}^{mn} = |n\rangle\langle m|$. 
Note that ${P^n = \mathbb{P}^{nn}}$ are projectors, 
while the $n\ne m$ operators are not even hermitian, 
and therefore cannot be interpreted as representing observables.   

Instead of using the standard basis ${ \{P^{z},S^{+},S^{-},P^{-z}\} }$  
it is possibly more physically illuminating  
to take ${\{\bm{1},\hat{\sigma}_x, \hat{\sigma}_y, \hat{\sigma}_z\}}$ as the basis set. 
Optionally one can take ${\{\bm{1},\hat{P}^x, \hat{P}^y, \hat{P}^z\}}$ 
or ${ \{ \hat{P}^z, \hat{P}^{-z} ,\hat{P}^x, \hat{P}^y \}}$ as the basis set. 

The bottom line is that the operators that act on an $N$ dimensional Hilbert space  
form an $N^2$ dimensional space. We can span this space in the standard basis, 
but physically it is more illuminating, and always possible, to pick a basis set of $N^2$ hermitian operators. 
Optionally we can pick a complete set of $N^2$ linearly independent projectors. 
The linear relations between sets of states (as implied by the first postulate of quantum mechanics)   
translate into linear relations between sets of operators.
One should be careful not to abuse the latter statement: 
in the above example the projectors ${ \{ P^x, P^y, P^z \} }$ are linearly independent, 
while the associated states ${ \{ |x\rangle, |y\rangle, |z\rangle \} }$ are not.

\sheadC{The tomography of quantum states}

The first postulate of Quantum Mechanics implies that 
with any {\em observable} we can associate an Hermitian 
operator that belongs to the $N^2$-dimensional space of operators.
We can span the whole space of observables 
by any set of $N^2$~independent operators $\hat{P}^r$. 
The standard basis is not physically illuminating,  
because the matrices $\mathbb{P}^{mn}$ are not hermitian, 
and therefore cannot be associated with random variables. 
But without loss of generality we can 
always assume an optional basis of hermitian matrices, 
possibly $N^2$~independent projectors. 

From the second postulate of Quantum mechanics 
it follows that if ${\hat{A}=\sum_r a_r \hat{P}^r}$ then 
\beq
\langle\hat{A}\rangle \ \ = \ \ \sum_r a_r \rho_r  
\ \ \ \ \ \ \ \ \mbox{where} \ \rho_r \equiv \langle\hat{P}^r\rangle
\eeq
The set of $N^2$ expectation values fully characterizes the quantum state. 
Hence we can say the $\rho$ represents the quantum state, 
in the same sense that a probability function represent 
a statistical state in classical mechanics.  
The determination of the state $\rho$ on the basis of a set 
of measurements is called "quantum tomography".
It should be clear that if we know $\rho$ we can predict the 
expectation value of any other measurement. 

In the dim$=2$ spin case any operator can be written 
as a linear combination of the Pauli matrices:
\beq
\hat{A} \ \ = \ \ a_0 \hat{1} + a_x \sigma_x + a_y \sigma_y + a_z \sigma_z \ \ = \ \ a_0 + \bm{a} \cdot \bm{\sigma}
\eeq
It is implied by the second postulate of quantum mechanics that 
\beq
\langle\hat{A}\rangle \ \ = \ \ a_0  + a_x \langle \sigma_x \rangle + a_y \langle \sigma_y \rangle + a_z \langle \sigma_z \rangle
\ \ = \ \  a_0+\bm{a}\cdot\bm{M}
\eeq
where the polarization vector is defined as follows:
\beq
\bm{M} \ \ = \ \ \Big(\langle\hat{\sigma}_x\rangle, \langle\hat{\sigma}_y\rangle, \langle\hat{\sigma}_z\rangle\Big)
\eeq
In the general case, to be discussed below, 
we define a package ${\rho=\{\rho_r\}}$ of expectation values 
that we call {\em probability matrix}. 
The polarization vector $\bm{M}$ can be regarded as the simplest 
example for such matrix. The term ``matrix" is used because 
in general the label~$r$ that distinguishes the $N^2$ basis operators 
is composed of two indexes. We note that a measurement of 
a non-degenerate {\em observable} provides $N{-}1$ independent  
expectation values (the probabilities to get any of the possible outcomes).
Accordingly quantum tomography requires the measurement of $N{+}1$ non-commuting observables.

\sheadC{Definition of the probability matrix} 

The definition of ${\rho}$ in quantum mechanics 
is based on the trivial observation that 
any observable $A$ can be written as a linear 
combination of ${N^2{-}1}$ independent projectors. 
If we know the associated ${N^2{-}1}$ independent probabilities,  
or any other set of ${N^2{-}1}$ independent expectation values,   
then we can predict the result of any other measurement (in the statistical sense). 
The possibility to make a prediction is based 
on taking the linearity of the expectation value as a postulate.
The above statement is explained below, but the best 
is to consider the $N=2$ example that comes later.  

Any Hermitian operator can be written 
as a combination of ${N^2}$ operators 
as follows:
\beq
\hat{A} \ \ = \ \ \sum_{n,m} |n \rangle \langle n|A|m \rangle \langle m| 
\ \ = \ \ \sum_{n,m}A_{nm} \, \hat{\mathbb{P}}^{mn} 
\eeq
where ${\hat{\mathbb{P}}^{mn} = |n \rangle \langle m|}$. 
These $N^2$ operators are not all hermitians, 
and therefore, strictly speaking, they do not represent a set of observables.  
However we can easily express them using a set of $N^2$ hermitian 
observables as follows:
\beq
\mathbb{P}^{nn} & \ = \ &  P^{n}   \\
\mathbb{P}^{mn} & \ = \ &  \frac{1}{2}(X^{r} + iY^{r})  \ \ \ \ \ \ \ \  \mbox{for $m{>}n$} \\
\mathbb{P}^{mn} & \ = \ &  \frac{1}{2}(X^{r} - iY^{r})  \ \ \ \  \ \ \ \ \mbox{for $m{<}n$} 
\eeq
where $r=(nm)=(mn)$ is a composite index that runs over 
the $[N(N{-}1)/2]$ possible combinations of $n$~and~$m$. 
The definitions of $X^{r}$ and $Y^{r}$ are 
implied by the above expressions:
in the dim$=2$ case $X$ and $Y$ are called Pauli matrices, 
while here we shall call them generalized Pauli matrices. 
It is useful to notice that the generalized Pauli matrices
can be expressed as combinations of elementary Projectors. 
Define $Q^{r}$ as a projector on the state  ${|n\rangle+|m\rangle}$
and $R^{r}$ as a projector on the state ${|n\rangle+i|m\rangle}$.
Then we have:
\beq
X^{(nm)} \ \ &=& \ \ 2Q^{(nm)}-P^{n}-P^{m} \\
Y^{(nm)} \ \ &=& \ \ 2R^{(nm)}-P^{n}-P^{m}
\eeq

Accordingly the set of generalized projectors ${\mathbb{P}}^{mn}$, 
is in fact equivalent to a set of $N^2$ 
proper projectors ${(P^{n},Q^{r},R^{r})}$.   
If one can extract from measurements the associated probabilities, 
it becomes possible to predict the result of any other expectation value  
according to the equation:
\beq
\langle A \rangle \ \ =  \ \ \sum_{n,m}A_{nm} \, \rho_{mn} \ \ = \ \ \trc(A\rho) 
\eeq
where ${\rho}$ is named {\em the probability matrix}. Each entry in the probability 
matrix is a linear combination of expectation values of projectors.  
Note that the expectation value of a projector $P=|\psi\rangle\langle\psi|$ 
is the probability to find the systems in the state $|\psi\rangle$, 
while the expectation values of a generalized projector $\mathbb{P}^{r}$ 
is defined as $(\langle X^{r} \rangle \pm i \langle Y^{r}\rangle )/2$, 
as implied by the definitions above.

\sheadC{Example: probability matrix for spin}

Let us relate the two common ways in which one can specify 
the quantum state of a spin~$1/2$ entity.
On possibility is to specify the polarization vector, 
as discussed in previous section. 
The other way is to define a $2\times2$ 
probability matrix as follows. 
The generalized projectors are 
\beq
\mathbb{P}^{ \uparrow \uparrow} 
&=& |\uparrow \rangle\langle \uparrow| 
\ \ = \ \ \left(\amatrix{1&0\cr 0&0}\right)
\ \ = \ \ \frac{1}{2}(1 + \sigma_z) 
\ \ = \ \ P^{z}
\\ \nonumber
\mathbb{P}^{ \downarrow \downarrow } 
&=&  |\downarrow \rangle\langle \downarrow| 
\ \ = \ \ \left(\amatrix{0&0\cr 0&1}\right) 
\ \ = \ \ \frac{1}{2}(1 - \sigma_z) 
\ \ = \ \ 1-P^{z}
\\ \nonumber
\mathbb{P}^{\downarrow \uparrow} 
&=& |\uparrow \rangle\langle \downarrow| 
\ \ = \ \ \left(\amatrix{0&1\cr 0&0}\right)
\ \ = \ \ \frac{1}{2}(\sigma_x + i \sigma_y) 
\ \ = \ \ \frac{1}{2}(2P^x-1)+\frac{i}{2}(2P^y-1)
\\ \nonumber
\mathbb{P}^{ \uparrow \downarrow} 
&=& |\downarrow \rangle\langle \uparrow| 
\ \ = \ \ \left(\amatrix{0&0\cr 1&0}\right)
\ \ = \ \ \frac{1}{2}(\sigma_x -i \sigma_y) 
\ \ = \ \ \frac{1}{2}(2P^x-1)-\frac{i}{2}(2P^y-1)
\eeq
The elements of the probability matrix 
are the expectation values of the above 
generalized projectors. 
Consequently we deduce the following relation 
between the probability matrix 
and the polarization vector:
\beq
\rho 
\ \ = \ \ 
\langle P^{ji} \rangle 
\ \ = \ \ 
\left( \amatrix{ \frac{1}{2}(1+M_3)& \frac{1}{2}(M_1-iM_2)\cr \frac{1}{2}(M_1+iM_2)& \frac{1}{2}(1-M_3)} \right) 
\ \ = \ \ 
\frac{1}{2}(\hat{1}+\vec{M} \cdot \vec{\sigma}) 
\eeq

\sheadC{Pure states as opposed to mixed states}

After diagonalization, the probability matrix can be written as:
\beq
\rho \ \ \rightarrow \ \ \left(\amatrix{p_1&0&0&.\cr 0&p_2&0&.\cr 0&0&p_3&.\cr .&.&.&.\cr &&&&.\cr &&&&&.}\right) 
\eeq
The convention is to order the diagonal elements in descending order.
Using the common jargon we say that the state represented 
by $\rho$ is a mixture of ${|1 \rangle , |2 \rangle , |3 \rangle, \dots }$
with weights ${p_1, p_2, p_3, \dots }$.
The most well known mixed state is the canonical state:
\beq
p_r = \frac{1}{Z} \eexp{- \beta E_r}, 
\ \ \ \ \ \ \ \ \ \ \beta\equiv1/T, 
\ \ \ Z\equiv \sum_r \eexp{- \beta E_r}
\eeq
A "pure state" is the special case where the probability matrix 
after diagonalization is of the form:
\beq
\rho \ \ \rightarrow \ \ \left(\amatrix{1&0&0&.\cr 0&0&0&.\cr 0&0&0&.\cr .&.&.&.\cr &&&&.\cr &&&&&.} \right) 
\eeq
This may be written in a more compact 
way as ${\rho = |1 \rangle \langle 1| =| \psi \rangle \langle \psi| = P^{\psi} }$. 
Note that $\langle P^{\psi} \rangle = 1$. This means a definite  
outcome for a measurement that is aimed in checking whether 
the particle is in state~"1". That is why we say that the state is pure.

\sheadC{Various versions of the expectation value formula}

[1] The standard version of the expectation value formula: 
\beq
\langle A \rangle \ \ = \ \  tr(A \rho)
\eeq
[2] The "mixture" formula: 
\beq
\langle A \rangle \ \ = \ \  \sum_r p_r \langle r|A|r \rangle
\eeq
[3] The "sandwich" formula: 
\beq
\langle A \rangle_{ \psi} \ \ = \ \  \langle \psi|A| \psi \rangle 
\eeq
[4] The "projection" formula:
\beq
\mbox{Prob}( \phi| \psi) \ \ = \ \  |\langle \phi| \psi \rangle|^2  
\eeq
The equivalence of statements 1-4 can be proved. In particular 
let us see how we go from the fourth statement to the third:
\beq
\langle A \rangle_{ \psi} 
\ \ = \ \  \sum_a \mbox{Prob}( a | \psi) a  
\ \ = \ \  \sum_a |\langle a | \psi \rangle|^2 a 
\ \ = \ \  \langle \psi|A| \psi \rangle
\eeq

\newpage

\sheadB{The evolution of quantum mechanical states}

\sheadC{The Evolution Operator and the Hamiltonian}

Consider a particle in an $N$~site system. If we redefine 
the first basis state as ${|\tilde{1} \rangle = -8| 1 \rangle}$
we get a non-orthonormal basis, and therefore not convenient 
for practical mathematical calculations. 
{\em Explanation:} if we represent the state ${| \psi \rangle}$ as 
a linear combination of normalized basis 
vectors ${|\psi\rangle=\sum_j\psi_j|j\rangle}$, 
then we can find the coefficients of the combination by 
using the formula ${ \psi_i=\langle i|\psi\rangle }$.

Even if we decide to work with a set of 
orthonormal states, still there is some freedom left 
which is called "gauge freedom" or "phase freedom": 
multiplying a state with a phase factor does not 
imply a different state, $\rho$~stays the same, and 
hence all physical expectation values remain the same too.
{\em Example:} The spin states ${|\uparrow\rangle }$ 
and ${\mbox{e}^ { \frac{\pi}{8}i } |\uparrow\rangle}$  
are represented by the same polarization vector,  
or optionally by the same~${\rho}$. 

From the superposition principle, and from the above 
remarks regarding the normalization, it follows that 
the evolution in quantum mechanics is described by a unitary operator: 
\beq
&&|\psi^{t=0} \rangle \ \ \rightarrow \ \ |\psi^t \rangle 
\\ \nonumber
&&|\psi^t\rangle \ \ = \ \ U|\psi^{t=0}\rangle 
\eeq
It is assumed here that the system is {\em isolated}. 
In order to simplify the discussion below we further assume 
that the external fields are constant in time. 
In such a case, the evolution operator must fulfill 
the following "group property":
\beq
U(t_2+t_1) \ \ = \ \ U(t_2) \ U(t_1) 
\eeq
It follows that the evolution operator can be written as 
\beq
U(t) \ \ = \ \ \eexp{-it \mathcal{H}}
\eeq
where $\mathcal{H}$ is called the Hamiltonian or "generator" 
of the evolution.

{\bf Proof:} 
The constructive way of proving the last formula is as follows:
In order to know the evolution of a system from ${t_1}$ to ${t_2}$ 
we divide the time into many small intervals 
of equal size ${dt = (t_2-t_1)/N}$. This means that:
\beq
U(t_2,t_1) \ \ = \ \ U(t_2,t_2{-}dt) \cdot\cdot\cdot  U(t_1{+}2dt,t_1+dt) \ U(t_1{+}dt,t_1)
\eeq
The evolution during an infinitesimal time interval can be written as:
\beq
U(dt) \ \ \equiv \ \ \hat{1} - i dt \mathcal{H} 
\eeq
where the Hamiltonian can be regarded as the "evolution per unit of time". 
Or we may say that $\mathcal{H}$ is the log derivative of $U$ with respect 
to time. By multiplying many infinitesimal time steps we get:
\beq
\hat{U} \ \ =  \ \ (1-i dt\mathcal{H})\cdot \cdot \cdot (1-i dt\mathcal{H}) (1-idt\mathcal{H})
\ \ = \ \ 
\left(1-i \frac{t}{\#steps}\mathcal{H}\right)^{\#steps} 
\ \ = \ \ \eexp{-it\mathcal{H}} 
\eeq
where we have assumed that the Hamiltonian does not 
change in time, so that the multiplication of exponents 
can be changed into a single exponent with a sum of powers. 
We have used above the definition of the 
exponential function in mathematics: ${\exp(t) = \lim (1+t/N)^N}$, 
that follows from the assumed 
multiplicative property ${\exp(t_1)\exp(t_2)=\exp(t_1+t_2)}$.

\sheadC{The Schr\"{o}dinger Equation}

Consider the evolution of a pure state:
\beq
\psi^{t} \ \ = \ \ U \psi^{t{=}0} 
\eeq
From the rerlation ${\psi^{t+dt} =(1-idt\mathcal{H})\psi^{t}}$ it follows that  
\beq 
\frac{d\psi}{dt} \ \ = \ \ -i\mathcal{H}\psi
\eeq
This is the Schr\"{o}dinger equation.
We have allowed ourselves above to use 
sloppy notations (the "ket" has been omitted). 
One should realize that the Schr\"{o}dinger equation
reflects the definition of the Hamiltonian 
as the generator of the evolution.
 
We now consider the evolution of a general mixture:
\beq
\rho \ \ = \ \ \sum_r |r\rangle p_r \langle r|
\eeq
we have $|r\rangle \rightarrow U|r\rangle$ and $\langle r| \rightarrow \langle r|U^\dagger$, 
therefore the evolution of $\rho$ in time is:
\beq
\rho^t \ \ = \ \ U \rho^{t{=}0}U^{\dagger} 
\\ 
\frac{d\rho}{dt} \ \ = \ \ -i[\mathcal{H}, \rho]
\eeq
This is the Liouville von-Neumann equation. 
One of its advantages is that the correspondence 
between the formalism of statistical mechanics  
and quantum mechanics becomes explicit. 
The difference is that in quantum mechanics we deal 
with a probability matrix whereas in statistical mechanics 
we deal with a probability function.

\sheadC{Stationary States (the "Energy Basis")}

We can find the eigenstates ${|n \rangle}$ and 
the eigenvalues $E_n$ of a Hamiltonian, 
which is called diagonalization: 
\beq
&& \mathcal{H} |n \rangle \ \ = \ \ E_n |n \rangle 
\\ 
&& U |n \rangle \ \ = \ \ \eexp{-iE_n t} |n \rangle 
\\ 
&& U \ \ \rightarrow \ \ \delta_{n,m} \eexp{-iE_n t}
\eeq
Using Dirac notations:
\beq
\langle n| U | m \rangle \ \ = \ \ \delta_{nm} \eexp{-iE_n t}
\eeq
If we prepare a superposition of basis states:
\beq
| \psi^{t=0} \rangle \ \ = \ \ \sum_n \psi_n \ |n \rangle
\eeq
we get after time $t$ 
\beq
|\psi(t) \rangle \ \ = \ \ \sum_n \eexp{-iE_n t} \psi_n \ | n \rangle
\eeq

\sheadC{How do we construct the Hamiltonian?} 

We construct the Hamiltonian from "symmetry" considerations. 
In the next lecture our object will be to show that 
the Hamiltonian of a non-relativistic particle is of the form:
\beq
\mathcal{H} \ \ = \ \ \frac{1}{2\mass}(p-A(x))^2+V(x) 
\eeq
In this lecture we discuss a simpler case: 
the Hamiltonian of a particle in a two-site system. 
We make the following assumptions about 
the two-site dynamics:

\bitem The system is symmetric with respect to reflection. \\ 
\bitem The particle can move from site to site. \\

\begin{center}
\putgraph{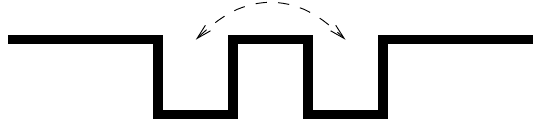}
\end{center}

These two assumptions determine 
the form of the Hamiltonian. 
In addition, we will see how "gauge" considerations 
can make the Hamiltonian simpler, 
without loss of generality. 

In advance we note that because 
of gauge considerations, the Hamiltonian 
can only be determined up to a constant. 
\beq
\mathcal{H} \ \ \rightarrow \ \ \mathcal{H} + \epsilon_0 \hat{1} 
\eeq
Namely, if we add a constant to a Hamiltonian, 
then the evolution operator 
only changes by a global phase factor:
\beq
U (t) \ \ \rightarrow \ \ \eexp{- i t ( \mathcal{H} + \epsilon_0 \hat{1})} 
\ \ = \ \ \eexp{ -i \epsilon_0t } \ \eexp{-i t \mathcal{H}} 
\eeq
This global phase factor can be gauged away by  
means of time dependent gauge transformation. 
We shall discuss gauge transformations in the next sections.

\sheadC{The Hamiltonian of a two-site system} 

It would seem that the most general Hamiltonian 
for a particle in a two-site system includes 4 parameters:
\beq
\mathcal{H} \ \ = \ \ \left(\amatrix{ \epsilon_1 & c \eexp{-i\phi} \cr c \eexp{i\phi} & \epsilon_2 } \right) 
\eeq
The first observation is that thanks 
to gauge freedom we can define a 
new basis such that ${\phi=0}$. The new basis is 
\beq
| \tilde {1} \rangle & \ = \ & |1 \rangle 
\\ \nonumber
| \tilde{2} \rangle & \ = \ & \eexp{i\phi} | 2 \rangle 
\eeq
and we see that:
\beq
\langle \tilde{2} | \mathcal{H} | \tilde{1} \rangle 
\ \ = \ \ \eexp{- i\phi }\langle 2| \mathcal{H} | 1 \rangle 
\ \ = \ \ \eexp{-i\phi} c \eexp{i\phi} \ \ = \ \ c 
\eeq
Next we can make change the Hamiltonian by a constant $\epsilon_1$. 
This can be regarded as gauge transformation in time. 
This means that the basis in time~$t$ is identified 
as $|\tilde{1}\rangle=\exp(-i\epsilon_1 t)|1\rangle$ 
and $|\tilde{2}\rangle=\exp(-i\epsilon_1 t)|2\rangle$.
In this time dependent basis the diagonal matrix elements 
of the Hamiltonian becomes $0$ instead of $\epsilon_1$, 
and  ${\epsilon=\epsilon_2-\epsilon_1}$ instead of $\epsilon_2$.
It should be emphasized that on physical grounds   
one cannot say whether the old or new basis 
is "really" time dependent. All we can say is that 
the new basis is time dependent relative to the old basis. 
This is just another example of the relativity principle.

The bottom line is that without loss of generality 
we can set ${\phi=0}$ and ${\epsilon_1=0}$, hence 
the most general Hamiltonian of a two site systems 
included 2~physical parameters:
\beq
\mathcal{H} \ \ = \ \ 
\left(\amatrix{ 0 & c \cr c & \epsilon} \right) 
\ \ = \ \ c \bm{\sigma}_1 \ - \  (\epsilon/2)\bm{\sigma}_3  \ + \ \const
\eeq
where $\bm{\sigma}_1$ and $\bm{\sigma}_3$ 
are the Pauli matrices. 
If we further assume reflection symmetry, 
then ${\epsilon=0}$. In the next subsection 
we discuss the dynamics that is generated.

\sheadC{The evolution of a two-site system} 

The eigenstates of the mirror symmetric (${\epsilon=0}$) Hamiltonian 
are the states that are symmetric or anti-symmetric with 
respect to reflection:
\beq
|+ \rangle \ \ &=& \ \  \frac{1}{\sqrt{2}} (|1 \rangle + |2 \rangle ) 
\\ 
|- \rangle \ \ &=& \ \  \frac{1}{\sqrt{2}} (|1 \rangle - |2 \rangle ) 
\eeq
The Hamiltonian in the new basis is:
\beq
\mathcal{H} 
\ \ = \ \ \left( \amatrix{ c & 0 \cr 0 & -c} \right) 
\ \ \equiv \ \  \left( \amatrix{ E_{+} & 0 \cr 0 & E_{-}} \right)
\eeq
Let us assume that we have prepared 
the particle in the first site:
\beq
| \psi^{t=0} \rangle \ \ =  \ \ |1 \rangle 
\ \ = \ \ \frac {1}{\sqrt{2}} (| + \rangle + | - \rangle ) 
\eeq
The state of the particle, after time $t$ will be:
\beq
|\psi^t \rangle 
\ \ = \ \ \frac {1}{ \sqrt{2}} ( \eexp{ -ic t } | + \rangle + \eexp{ - i(-c) t }| - \rangle ) 
\ \ = \ \ \cos( c t) | 1 \rangle - i\sin( c t) | 2 \rangle 
\eeq
We see that a particle in a two-site system makes coherent 
oscillations between the two sites. In particular the probability
to find it in the initial site is:
\beq
P(t) \ \ = \ \ |\cos( c t)|^2 \ \ = \ \ \frac{1}{2}\left[1+\cos(\Omega t)\right]
\eeq
where the frequency of oscillations is 
\beq 
\Omega \ \ = \ \ E_{+}-E_{-} \ \ = \ \ 2c 
\eeq
This result should be contrasted with classical stochastic evolution 
where the probability to be in each site (if we wait long enough time) 
would become equal. 
In the future we will see that the tendency to make a transition from 
site to site is characterized by a parameter that is called "inertial mass".

\newpage

\sheadB{The non-relativistic Hamiltonian}

\sheadC{$N$ Site system in the continuum Limit}

In the last lecture we deduced the Hamiltonian ${\hat{\mathcal{H}}}$ 
that describes a two-site system using gauge and symmetry considerations. 
Now we are going to generalize this deduction for an $N$-site system. 
We start with 1D modeling, where the distance between two adjacent 
sites is~$a$. The assumption is that the particle is able  
to move from site to site. We want to write an expression for 
the generator $\hat{\mathcal{H}}$ of the particle movement, such that 
\beq
U_{ij}(dt) \ \ = \ \ \delta_{ij} - idt\mathcal{H}_{ij} 
\eeq

\begin{center}
\putgraph[0.5\hsize]{NSiteSystem}
\end{center}

The Hamiltonian should reflect the possibility that 
the particle can either stay in its place or
move one step right or one step left. Say that $N=4$. 
Taking into account that $\hat{\mathcal{H}}$ should be Hermitian 
it has to be of the form
\beq
\mathcal{H}_{ij} \ \ = \ \ 
\left(\amatrix{
v   & c^* & 0   & c \\ 
c   & v   & c^* & 0 \\
0   & c   & v   & c^* \\
c^* & 0   & c   & v  
}\right)
\ \ \equiv \ \ K\mbox{(so called kinetic part)} + V\mbox{(so called potential part)}
\eeq
For a moment we assume that all the diagonal elements (``on sites energies") are the same, 
and that also all the hopping amplitudes are the same. Thus for general $N$ we can write      
\beq
\hat{\mathcal{H}} 
\ \ = \ \ cD + c^*D^{-1} + \mbox{Const} 
\ \ = \ \ c\mbox{e}^{-ia\hat{p}} + c^* \mbox{e}^{ia\hat{p}} + \mbox{Const} 
\eeq
We define ${c=c_0 \eexp{i\phi}}$, where $c_0$ is real, and get:
\beq
\hat{\mathcal{H}} \ \ = \ \ c_0 \mbox{e}^{-i(a\hat{p}-\phi)} + c_0 \mbox{e}^{i(a\hat{p}-\phi)} + \mbox{Const} 
\eeq
We define ${A= \phi / a}$ (phase per unit distance) and get:
\beq
\hat{\mathcal{H}} \ \ = \ \ c_0 \mbox{e}^{-ia(\hat{p}-A)} + c_0 \mbox{e}^{ia(\hat{p}-A)} + \mbox{Const} 
\eeq
By using the identity ${\mbox{e}^{ix} \approx 1 + ix - (1/2)x^2}$ we get:
\beq
\hat{\mathcal{H}} \ = \ \frac{1}{2\mass} (\hat{p}-A)^2 + V, 
\ \ \ \ \ \ \ \ \ \ \ \
\frac{1}{2\mass}\equiv-c_0a^2, 
\ \ \ \ \
V \equiv \mbox{Const}+2c_0
\eeq
The above expression for $\mathcal{H}$ has three constants: ${\mass, A, V}$. 
If we assume that the space is homogeneous then the constants 
are the same all over space. But, in general, it does not 
have to be so, therefore:
\beq
\hat{\mathcal{H}} \ \ = \ \ \frac{1}{2\mass} (\hat{p}-A(x))^2 + V(x) 
\eeq
If $A$ and $V$ depend on~$x$ we say that there is a field in space. 
In fact also $\mass$ can be a function of~$x$, but then one should be careful 
to keep $\hat{\mathcal{H}}$ hermitian, caring for appropriate ``symmetrization".  
A mass that changes from place to place could perhaps describe an electron 
in a non-uniform metal. Here we discuss a particle whose mass~$\mass$ 
is the same all over space, for a reason that will be explained in the next section.

\sheadC{The Hamiltonian of a Particle in 3-D Space} 

In 3D we can adopt the same steps in order to deduce 
the Hamiltonian. We write:
\beq
\mathcal{H} & \ \ = \ \ & cD_x + c^*D_x^{-1} + cD_y + c^* D_y^{-1}+ cD_z+c^*D_z^{-1} 
\\ \nonumber
& \ \ = \ \ & c \eexp{-ia\hat{p}_x}
+ c^* \eexp{ia\hat{p}_x}
+ c \eexp{-ia\hat{p}_y}
+ c^* \eexp{ia\hat{p}_y}
+ c \eexp{-ia\hat{p}_z}
+ c^* \eexp{ia\hat{p}_z} 
\eeq
After expanding to second order and allowing space dependence we get:
\beq \nonumber
\mathcal{H} 
& \ \ = \ \ & 
\frac{1}{2\mass}( \hat{p}_x-A_x(\hat{x}, \hat{y}, \hat{z}))^2 
+ \frac{1}{2\mass}( \hat{p}_y-A_y(\hat{x}, \hat{y}, \hat{z}))^2 
+ \frac{1}{2\mass}( \hat{p}_z-A_z(\hat{x}, \hat{y}, \hat{z}))^2 
+ V(\hat{x},\hat{y},\hat{z})
\\
& \ \ = \ \ &
\frac{1}{2\mass}( \hat{\mathbf{p}}-\mathbf{A}( \hat{\mathbf{r}}))^2+ V (\hat{\mathbf{r}}) 
\ \ = \ \
\frac{1}{2\mass}( \hat{\mathbf{p}}-\hat{\mathbf{A}} )^2+ \hat{V} 
\eeq
This is the most general Hamiltonian that is invariant 
under Galilein transformations. Note that having a mass
that is both isotropic and position-independent is 
required by this invariance requirement.  
The Galilein group includes translations, rotations and boosts. 
The relativistic version of the Galilein group is the Lorentz group 
(not included in the syllabus of this course). 
In addition, we expect the Hamiltonian to be invariant under 
gauge transformations, which is indeed the case as further 
discussed below.

\sheadC{Invariance of the Hamiltonian} 

The definition of "invariance" is as follows:  
Given that ${ \mathcal{H} = h(x,p; V, A) }$ 
is the Hamiltonian of a system in the laboratory reference frame, 
there exist $\tilde{V}$ and $\tilde{A}$ such that the Hamiltonian 
in the "new" reference frame is ${\tilde{\mathcal{H}}=h(x,p; \tilde{V}, \tilde{A}) }$.
The most general Hamiltonian that is invariant 
under translations, rotations and boosts is:
\beq
\hat{\mathcal{H}} \ \ = \ \ h(\hat{x},\hat{p}; V, A) 
\ \ = \ \ \frac{1}{2\mass} ( \hat{p} - A(\hat{x}) )^2 + V(\hat{x}) 
\eeq
Let us demonstrate the invariance of the Hamiltonian under translations: 
in the original basis ${|x\rangle}$ we have the fields $V(x)$ and $A(x)$. 
In the translated reference frame the new basis is ${|\tilde{x}\rangle \equiv |x+a\rangle}$, 
hence ${\langle \tilde{x} |\mathcal{H} | \tilde{x}' \rangle = \langle x |\tilde{\mathcal{H}} | x' \rangle}$ 
with ${\tilde{\mathcal{H}}=D^{\dag}\mathcal{H}D}$. Hence we deduce that 
the Hamiltonian is "invariant" (keeps its form) with   
\beq
\tilde{V}(x) \ \ &=& \ \ V(x+a) \\
\tilde{A}(x) \ \ &=& \ \ A(x+a)
\eeq
In order to verify that we are not confused with the signs, 
let us consider the potential ${V(x)=\delta (x)}$. 
If we make a translation with ${a=7}$, 
then the basis in the new reference frame will 
be ${|\tilde{x} \rangle = |x+7 \rangle }$, 
and we get ${\tilde{V}(x)=\delta {(x+7)}}$ 
which means a delta at~${x=-7}$.

For completeness we cite how the Hamiltonian transform
under Galilean transformation $T$ to a moving frame whose 
velocity is $v_0$. This transformation is composed of 
a translation ${\eexp{-iv_0t \hat{p}}}$ and a boost ${\eexp{i\mass v_0 \hat{x}}}$.   
The following result is derived in a dedicated lecture: 
\beq
\tilde{\mathcal{H}} 
\ \ = \ \ T^{\dag} \, \mathcal{H} \, T \ - \ v_0 \cdot \hat{\mathbf{p}}
\ \ = \ \ h(\hat{\mathbf{x}},\hat{\mathbf{p}}; \tilde{V}, \tilde{A}) 
\eeq   
where 
\beq
\tilde{V}(x) \ &=& \ V(x+v_0t) \ - \  v_0\cdot A(x+v_0t) \\
\tilde{A}(x) \ &=& \ A(x+v_0t) 
\eeq
From here follows the well know no-relativistic 
transformation ${\tilde{\mathcal{E}}=\mathcal{E}+v_0\times \mathcal{B}}$, 
and  ${\tilde{\mathcal{B}}=\mathcal{B}}$.

\newpage
\sheadC{The dynamical phase, electric field}

Consider the case where there is no hopping between sites ($c_0=0$). 
Accordingly the Hamiltonian ${\mathcal{H} = V(x)}$ 
does not include a kinetic part, 
and the evolution operator is ${\hat{U}(t)=\exp[-itV(\hat{x})]}$.   
A particle that is located in a given point in space 
will accumulate so called a dynamical phase:  
\beq
\hat{U}(t) | x_0 \rangle \ \  = \ \ \mbox{e}^{-itV(x_0)} | x_0 \rangle 
\eeq
The potential $V$ is the "dynamical phase" that 
the particle accumulates per unit time. 
The $V$ in a specific site is called "binding energy" 
or "on site energy" or "potential energy"  
depending on the physical context.
A $V(x)$ that changes from site to site reflects   
the non-homogeneity of the space, or the presence 
of an "external field". 
To further clarify the significance of~$V$ 
let us consider a simple prototype example.
Let us assume that $V(x)=-\mathcal{E}x$. 
If the particle is initially prepared  
in a momentum eigenstate we get after some time 
\beq
\hat{U}(t) | p_0 \rangle \ \  = \ \ \mbox{e}^{-it V} | p_0 \rangle  
\ \  = \ \ \mbox{e}^{i \mathcal{E}t \hat{x}} | p_0 \rangle  
\ \  = \ \ | p_0+\mathcal{E}t\rangle
\eeq
This means that the momentum changes with time. 
The rate of momentum increase equals $\mathcal{E}$.
We shall see later that this can be interpreted 
as the "second law of Newton".

\sheadC{The geometric phase, magnetic field}

Once we assume that the particle can move 
from site to site we have a hopping amplitude 
which we write as ${c=c_0 \eexp{i\phi}}$. 
It includes both a {\em geometric} phase ${\phi}$ 
and an {\em inertial} parameter ${c_0}$. 
The latter tells us what is the tendency of the 
particle to "leak" from site to site.
In order to have a Galileo invariant Hamiltonian 
we assume that ${c_0}$ is isotropic and the 
same all over space, hence we can characterize 
the particle by its inertial mass $\mass$. 

Still, in general the hopping amplitudes 
might be complex numbers ${c_{i \rightarrow j} \propto \eexp{i \phi_{i \rightarrow j} }}$.
These phases do not threaten Galileo invariance.
Accordingly, as the particle moves from site to site, 
it accumulates in each jump an additional phase $\phi_{i \rightarrow j}$ 
that can vary along its path. This is called "geometric phase".
By definition the vector potential $A$ is the "geometric phase" 
that the particle accumulates per unit distance. 
It is defined via the following formula:
\beq 
\phi_{i \rightarrow j} \ \ \equiv \ \ \vec{A}\cdot (\mathbf{r}_j-\mathbf{r}_i) 
\ \ = \ \ A_x dx + A_y dy + A_z dz
\eeq
The three components of $\vec{A}$ give the accumulated phase per distance 
for motion in the X, Y, and Z directions respectively.
An infinitesimal jump in a diagonal direction is regarded as a sum 
of infinitesimal jumps in the X, Y, and Z directions, hence the scalar product.

In the following section we shall clarify that using gauge freedom 
we can describe the same system by a different $\vec{A}$. 
However, the circulation of $\vec{A}$, which gives the total phase 
that is accumulated along a closed trajectory, is gauge invariant. 
Consequently we can define the circulation per unit area:
\beq 
\mathcal{B} \ \ = \ \ \nabla \times A
\eeq
We shall see later that this can be interpreted as "magnetic field".
This field characterizes $\vec{A}$ in a gauge invariant way.

\sheadC{What determines $V(x)$ and $A(x)$}

We have defined $V(x)$ and $A(x)$ as the phase that is accumulated 
by a particle per time or per distance, respectively. In the next section 
we shall see that the presence of $V(x)$ and $A(x)$ in the Hamiltonian 
are required in order to ensure {\em gauge invariance}: the Hamiltonian 
should have the same form irrespective of the way we gauge the "phases" 
of the position basis states. Next comes the {\em physical} question: 
what determines $V(x)$ and $A(x)$ in reality. A-priori the particle 
can accumulate phase either due to a non-trivial {\em geometry} of 
space-time, or due to the presence of {\em fields}. The former effect is 
called {\em Gravitation}, while the latter are implied by the {\em standard model}. 
For our purpose it is enough to consider the electromagnetic (EM) field.
It is reasonable to postulate that different particles have different 
couplings to the EM field. Hence we characterize a particle by its charge~($e$)
and write in CGS units:
\beq 
V(x) \ \ &=& \ \  V(x) \ + \ eV^{\tbox{EM}}(x) \\
A(x) \ \ &=& \ \  A(x) \ + \ (e/c)A^{\tbox{EM}}(x)
\eeq
With regard to the gravitation we note that on the surface of Earth  
we have ${V=\mass gh}$, where~$h$ is the vertical height of the particle, 
and ${\nabla \times A = 2\mass\Omega}$ which is responsible for the Coriolis force.   
Note that by the {\em equivalence principle}, so-called fictitious 
forces are merely a simple example of a gravitational effect, 
i.e. they reflect a non-trivial metric tensor that describes 
the space-time geometry in a given coordinate system.

From now on, unless stated otherwise, $V(x)$ and $A(x)$ refer to the electromagnetic field. 
We note that in the Feynman path-integral formalism, which we describe 
in a different lecture, the probability amplitude of a particle 
to get form point~1 to point~2, namely ${\langle x_2|U(t_2,t_1)|x_1\rangle}$,  
is expressed as a sum over amplitudes $\exp[i\phi[x(t)]]$. The sum extends 
over all possible paths. The contribution of the electromagnetic field to the 
accumulated phase along a give path is
\beq 
\phi[x(t)] \ \ = \ \ \phi_0[x(t)] \ + \ \int_1^2 A\cdot dx \ - \ \int_1^2 V\cdot dt    
\eeq
One realizes that this is the so called "action" in classical mechanics.

\sheadC{Invariance under Gauge Transformation}

Let us define a new basis:
\beq
| \tilde{x}_1 \rangle \ \ = \ \ \mbox{e}^{-i \Lambda_1} | x_1 \rangle 
\\ \nonumber
| \tilde{x}_2 \rangle \ \ = \ \ \mbox{e}^{-i \Lambda_2} | x_2 \rangle 
\eeq
and in general:
\beq
| \tilde{x} \rangle \ \ = \ \ \mbox{e}^{-i \Lambda(x)} | x \rangle 
\eeq
The hopping amplitudes in the new basis are:
\beq
\tilde{c}_{1 \rightarrow 2} 
\ \ = \ \ \langle \tilde{x}_2 | \hat{\mathcal{H}} | \tilde{x}_1 \rangle 
\ \ = \ \ \mbox{e}^{i( \Lambda_2 - \Lambda_1)} \langle x_2 | \hat{\mathcal{H}} | x_1 \rangle 
\ \ = \ \ \mbox{e}^{i( \Lambda_2 - \Lambda_1)} c_{1 \rightarrow 2} 
\eeq
We can rewrite this as:
\beq
\tilde{\phi}_{1\rightarrow 2} \ \ = \ \ \phi_{1\rightarrow 2} + ( \Lambda_2 - \Lambda_1 ) 
\eeq
Dividing by the size of the step and taking 
the continuum limit we get:
\beq
\tilde{A}(x) \ \ = \ \ A(x) + \frac{d}{dx} \Lambda (x) 
\eeq
Or, in three dimensions:
\beq
\tilde{A}(x) \ \ = \ \ A(x) + \nabla \Lambda (x) 
\eeq
We see that the Hamiltonian is invariant (keeps its form) under gauge. 
As we have said, there is also invariance for all the Galilei 
transformations (notably boosts). This means that it is possible 
to find transformation laws that connect the fields in the "new" 
reference frame with the fields in the "laboratory"  reference frame.

\sheadC{Is it possible to simplify the Hamiltonian further?}

Is it possible to find a gauge transformation of the basis 
so that $A$ will disappear?
We have seen that for a two-site system the answer is yes: 
by choosing ${\Lambda(x)}$ correctly, we can eliminate ${A}$ 
and simplify the Hamiltonian. On the other hand, if there is 
more than one route that connects two points, the answer becomes no.
The reason is that in any gauge we may choose, 
the following expression will always be gauge invariant:
\beq
\oint{\tilde{A} \cdot dl} \ \ = \ \ \oint{A \cdot dl} \ \ = \ \ \mbox{gauge invariant} 
\eeq
In other words: it is possible to change each of the 
phases separately, but the sum of phases along a closed loop 
will always stay the same. We shall demonstrate this 
with a three-site system:

\begin{center}
\putgraph[0.2\hsize]{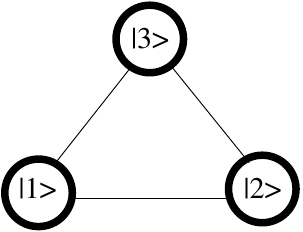}
\end{center}

\beq
&& | \tilde{1} \rangle \ \ = \ \ \mbox{e}^{-i \Lambda_1} | 1 \rangle 
\\ \nonumber
&& | \tilde{2} \rangle \ \ = \ \ \mbox{e}^{-i \Lambda_2} | 2 \rangle 
\\ \nonumber
&& | \tilde{3} \rangle \ \ = \ \ \mbox{e}^{-i \Lambda_3} | 3 \rangle 
\\ \nonumber
&& \tilde{\phi}_{1\rightarrow 2} \ \ = \ \ \phi_{1\rightarrow 2} + ( \Lambda_2 - \Lambda_1 ) 
\\ \nonumber
&& \tilde{\phi}_{2\rightarrow 3} \ \ = \ \ \phi_{2\rightarrow 3} + ( \Lambda_3 - \Lambda_2 ) 
\\ \nonumber
&& \tilde{\phi}_{3\rightarrow 1} \ \ = \ \ \phi_{3\rightarrow 1} + ( \Lambda_1 - \Lambda_3 ) 
\\ \nonumber
&& \tilde{\phi}_{1\rightarrow 2} + \tilde{\phi}_{2\rightarrow 3} + \tilde{\phi}_{3\rightarrow 1} 
\ \ = \ \ \phi_{1\rightarrow 2} + \phi_{2\rightarrow 3} + \phi_{3\rightarrow 1} 
\eeq
If the system had three sites but with an open topology, 
then we could have gotten rid of $A$ like in the two-site system. 
That is also generally true of all the one dimensional problems, 
if the boundary conditions are "zero" at infinity. 
Once the one-dimensional topology is closed ("ring" boundary conditions) 
such a gauge transformation cannot be made. Furthermore, 
when the motion is in two or three dimensional space, there is 
always more than one route that connects any two points, without 
regard to the boundary conditions. Consequently 
we can define the gauge invariant field $\mathcal{B}$ that 
characterizes~$A$. It follows that in general one cannot eliminate~$A$ 
from the Hamiltonian.

\sheadC{Gauging away the $V(x)$ potential}

We have concluded that in general it is impossible to gauge away the $A(x)$. 
But we can ask the opposite question, whether it is possible to gauge away the $V(x)$. 
The answer here is trivially positive, but the "price" is getting time dependent 
hopping amplitudes. Namely, the gauge transformation that does the job is 
\beq
| \tilde{x} \rangle = \mbox{e}^{-i \Lambda(x,t)} | x \rangle 
\ \ \ \ \ \mbox{with} \ \ \ \ \Lambda(x,t)=V(x) \, t
\eeq  
This is a temporal gauge of the basis. It is analogous to a transforming into a moving frame 
(in both cases the new basis is time dependent relative to the lab frame). 
In the new frame we have 
\beq
\tilde{V}(x;t) \ &=& \ 0 \\
\tilde{A}(x;t) \ &=& \ A(x) + t\nabla V(x)   
\eeq 
Hence we get the same magnetic and electric fields, 
but now both derived from a time dependent $A(x;t)$.

\newpage

\sheadB{Getting the equations of motion}

\sheadC{Rate of change of the expectation value}

Given an Hamiltonian, with any operator ${\hat{A}}$ 
we can associate an operator ${\hat{B}}$,  
\beq
\hat{B} \ \ =  \ \ i[\hat{\mathcal{H}},\hat{A}] + \frac{\partial {A}}{\partial{t}} 
\eeq
such that
\beq
\frac{d \langle \hat{A} \rangle}{dt} \ \ = \ \ \langle \hat{B} \rangle 
\eeq
{\bf proof:} From the expectation value formula:
\beq
\langle \hat{A} \rangle_t = \trc(\hat{A}\rho (t)) 
\eeq
Using the Liouville equation and the cyclic property of the trace, we get:
\beq
\frac{d}{dt}\langle \hat{A} \rangle_t 
&=& \trc\left(\frac{\partial {A}}{\partial {t}} \rho(t)\right) + \trc\left(\hat{A} \frac{d \rho(t) }{dt}\right) 
\\ \nonumber
&=& \trc\left(\frac{\partial {A}}{\partial {t}} \rho(t)\right) - i \trc\left(A[\mathcal{H},\rho(t)]\right)
\\ \nonumber
&=& \trc\left(\frac{\partial {A}}{\partial {t}} \rho(t)\right) + i \trc\left([\mathcal{H},A]\rho(t)\right) 
\\ \nonumber
&=& \left\langle\frac{\partial {A}}{\partial {t}}\right\rangle+ i\left\langle [\mathcal{H},A] \right\rangle 
\eeq
Optionally, if the state is pure, we can write:
\beq
{\langle \hat{A} \rangle}_t \ \ = \ \ \langle \psi{(t)} | \hat{A} | \psi{(t)} \rangle 
\eeq
Using the Schr\"{o}dinger equation, we get
\beq
\frac{d}{dt} \Big\langle \hat{A} \Big\rangle 
& \ \ = \ \ & \Big\langle \frac{d}{dt} \psi \Big| \hat{A} \Big| \psi \Big\rangle 
+ \Big\langle \psi \Big| A \Big| \frac{d}{dt} \psi \Big\rangle 
+ \Big\langle \psi \Big| \frac{\partial A }{\partial t} \Big| \psi \Big\rangle 
\\ \nonumber
& \ \ = \ \ & i \Big\langle \psi \Big| \mathcal{H}A \Big| \psi \Big\rangle 
- i \Big\langle \psi \Big| A\mathcal{H} \Big| \psi \Big\rangle 
+ \Big\langle \psi \Big| \frac{ \partial {A} }{\partial {t}} \Big| \psi \Big\rangle \ \ = \ \ ...
\eeq
We would like to highlight the distinction between a full derivative 
and a partial derivative. Let us assume that there is an operator 
that perhaps represents a field that depends on the time~$t$:
\beq
\hat{A} = \hat{x}^2 + t \hat{x}^8 
\eeq
Then the partial derivative with respect to $t$ is:
\beq
\frac{ \partial {A} }{\partial {t}} = \hat{x}^8 
\eeq
While the total derivative of ${\langle \hat{A} \rangle}$ 
takes into account the change in the quantum state too.

\sheadC{The classical equations of motion}

If ${\hat{x}}$ is the location of a particle, 
then its rate of change is called velocity. 
By the rate of change formula we identify
the velocity operator $v$ as follows: 
\beq
\hat{v} \ \ = \ \ i [\hat{\mathcal{H}},\hat{x}] 
\ \ = \ \ i \left[\frac{1}{2\mass} ( \hat{p} - A(\hat{x}) )^2,\hat{x}\right] 
\ \ = \ \ \frac{1}{\mass} (\hat{p} - A(x) ) 
\eeq
and we have:
\beq
\frac{d \langle \hat{x} \rangle}{dt} = \langle \hat{v} \rangle,  
\hspace{2cm} \mbox{3D:} \ \ \  \hat{x}:=(\hat{x},\hat{y},\hat{z}), \ \ \ \hat{v}:=(\hat{v}_x,\hat{v}_y,\hat{v}_z)
\eeq
It is useful to relaize that the kinetic part 
of the Hamiltonian can be written as $(1/2)\mass \hat{v}^2$. 
The commutaion relations of the velocity components are  
\beq
[\mass \hat{v}_i, \mass \hat{v}_j] \ \ &=& \ \ i(\partial_i A_j- \partial_j A_i) \ \ = \ \ i\epsilon_{ijk}B_k
\\ 
\ [V(\hat{x}), \mass \hat{v}_j] \ \ &=& \ \ i\, \partial_j V(\hat{x}) 
\eeq
The rate of change of the velocity ${\hat{v}}$ is 
called acceleration. By the rate of change formula
\beq
\frac{d \langle \hat{v} \rangle}{dt} 
\ \ = \ \ 
\left\langle 
i \left[\frac{1}{2}\mass \hat{v}^2+V(\hat{x}), \ \hat{v}\right] + \frac{\partial {\hat{v} }}{ \partial{t} }
\right\rangle 
\ \ = \ \ 
\frac{1}{\mass} 
\left\langle 
\frac{1}{2} ( v \times \mathcal{B} 
- \mathcal{B} \times v) 
+ \mathcal{E} 
\right\rangle 
\eeq
Note that this is dense vector-style writing  of 3~equations, for the rate of change 
of $\langle \hat{v}_x \rangle$  and $\langle \hat{v}_y \rangle$  and $\langle \hat{v}_z \rangle$.
Above we have used the follwing notations:
\beq
\mathcal{B} &=& \nabla \times A 
\\ \nonumber
\mathcal{E} &=& - \frac{ \partial{A} }{ \partial{t} } - \nabla{V} 
\eeq
We would like to emphasize that the Hamiltonian is 
the "generator" of the evolution of the system, 
and therefore all the equations of motion can be derived from it. 
From the above it follows that in case of a "minimal" wavepacket 
the expectation values of  $\hat{x}$ and $\hat{v}$ 
obey the classical equations approximately. 
The classical Lorentz force equation becomes exact if 
the $\mathcal{B}$ and $\mathcal{E}$ are {\em constants}
in the region where the motion takes place:
\beq
\frac{d \langle \hat{v} \rangle}{dt} 
\ \ = \ \ 
\frac{1}{\mass} 
\Big[
- \mathcal{B}_0 \times \langle v \rangle
+ \mathcal{E}_0 
\Big]
\eeq

\sheadC{Heuristic interpretation}

In the expression for the acceleration we have two terms: 
the ``electric" force and the ``magnetic" (Lorentz) force.
These forces bend the trajectory of the particle.   
It is important to realize that the bending of trajectories 
has to do with {\em interference} and has a very intuitive 
heuristic explanation. This heuristic explanation is due to 
Huygens: We should regard each front of the propagating 
beam as a point-like source of waves. The next front (after time $dt$) 
is determined by interference of waves that come from all 
the points of the previous front. For presentation purpose 
it is easier to consider first the interference of $N=2$ points, 
then to generalize to $N$ points, and then to take 
the continuum limit of plane-wave front. 
The case $N=2$ is formally equivalent to a two slit experiment. 
The main peak of constructive interference is in the 
forward direction. We want to explain why the presence of 
an electric or a magnetic field can shift the main peak. 
A straightforward generalization of the argument explains 
why a trajectory of a plane-wave is bent.

{\bf The Huygens deflection formula:}
Consider a beam with wavenumber $k$ that propagates in the $x$ direction.
It goes through two slits that are located on the $y$ axis. 
The transverse distance between the slits is ${d}$. 
A detector is positioned at some angle $\theta$ very far away from the slits.
The phase difference between the oscillations 
that arrive to the detector from the two slits is  
\beq
\phi_2-\phi_1 \ \ = \ \ \Delta \phi \ - \ k \cdot d \cdot \theta 
\eeq
In this formula it is assumed that after the slits there 
is a "scattering region" of length $\Delta x$ where fields 
may be applied. We define $\Delta \phi$ as the phase difference 
after this "scattering region", while  ${\phi_2-\phi_1}$ is the 
phase difference at the very distant detector.   
The ray propagation direction is defined by the requirement ${\phi_2-\phi_1=0}$, 
leading to the optical deflection formula $\theta=\Delta \phi/(kd)$. 
Changing notations ${k\mapsto p_x}$ and ${d\mapsto \Delta y}$ 
we write the deflection formula as follows:   
\beq
\theta \ \ &=& \ \ \frac{\Delta p_y}{p_x} \ \ = \ \ \mbox{Deflection angle}
\\
\Delta p_y \ \ &\equiv& \ \ \frac{\Delta \phi}{\Delta y} \ \ = \ \ \mbox{Optical Impulse} 
\eeq
From the definition of the dynamical and the geometrical phases 
it follows that if there are electric and magnetic fields in the 
scattering region then   
\beq
\Delta \phi \ \ = \ \ (V_2-V_1) \Delta t \ + \ \Phi_B
\eeq
where $\Phi_B = \mathcal{B}\Delta x\Delta y $ is the magnetic flux
that is enclosed by the interfering rays, 
and $V_2-V_1=\mathcal{E} \Delta y$ is the potential difference between the two slits, 
and $\Delta t = \Delta x / v$ is the travel time via the scattering region. 
From here follows that the optical impulse is 
\beq
\Delta p_y 
\ \ = \ \ \frac{\Delta \phi}{\Delta y}
\ \ = \ \ \mathcal{E} \Delta t \ + \ \mathcal{B}  \Delta x 
\eeq

{\bf The Newtonian deflection formula:}
In the Newtonian perspective the deflection of 
a beam of particles is 
\beq
\theta \ \ &=& \ \ \frac{\Delta p_y}{p_x} \ \ = \ \ \mbox{Deflection angle}
\\
\Delta p_y \ \ &\equiv& \ \ F \Delta t \ \ = \ \ \mbox{Newtonian Impulse} 
\eeq
where $F$ is called the "Newtonian force". 
By comparison with the Huygens analysis 
we deduce that the "force" on the particles is  
\beq
F 
\ \ = \ \ \frac{\Delta p_y}{\Delta t}
\ \ = \ \ \mathcal{E} + \mathcal{B} v \ \ = \ \ \mbox{Lorentz force} 
\eeq
It is important to realize that the deflection is due to an interference effect:  
The trajectory is bending either due to a gradient in $V(x)$, 
or due to the presence of an enclosed magnetic flux. 
Unlike the classical point of view, it is not $\mathcal{B}(x)$ that matters 
but rather $A(x)$, which describes the geometric 
accumulation of the phase along the interfering rays.  
We further discuss this interference under the headline ``The Aharonov Bohm effect":

\sheadC{The continuity Equation (conservation of probability)}

The Schrodinger equation is traditionally written as follows: 
\beq
&&\hat{\mathcal{H}} 
\ \ = \ \ \mathcal{H}(\hat{x},\hat{p})
\\ \nonumber
&&\frac{\partial|{\Psi\rangle}}{{\partial}t} 
\ \ = \ \ -i\hat{\mathcal{H}}|\Psi\rangle
\\ \nonumber
&&\frac{\partial\Psi}{{\partial}t} 
\ \ = \ \ -i\mathcal{H}\left(x,-i \frac{\partial}{{\partial}x }\right)\Psi
\\ \nonumber
&&\frac{\partial\Psi}{\partial t} 
\ \ = \ \ -i \left[ \frac{1}{2\mass} (-i \nabla - A(x))^2 + V(x) \right] \Psi(x) 
\eeq
Using the "rate of change formula" for 
the probability density we 
can obtain a continuity equation:
\beq
\frac{ \partial\rho (x)}{{\partial}t} \ \ = \ \ - \nabla\cdot J(x)
\eeq
In this formula the probability density and probability current are 
defined as the expectation values of the following operators:
\beq
\hat{\rho}(x) &=& \delta(\hat{x}-x) 
\\ 
\hat{J}(x) &=& \frac{1}{2}(\hat{v}\delta (\hat{x}-x) + \delta (\hat{x}-x) \hat{v}))
\eeq
leading to 
\beq
\rho(x) &=& 
\langle\Psi|\hat{\rho} (x) |\Psi\rangle 
\ \ = \ \ |\Psi (x)|^2 
\\
J(x) &=& 
\langle\Psi|\hat{\mathbf{J}} (x) |\Psi\rangle
\ \ = \ \ \re\left[{\Psi}^*(x)\frac{1}{\mass}(-i \nabla - A(x))\Psi(x)\right] 
\eeq
The procedure to get this result can be optionally 
applied to a particle in an $N$-site system 
(see appropriate "QM in practice" section).

\sheadC{Definition of generalized forces}

We would like to know how the system energy changes 
when we change a parameters~$X$ on which the Hamiltonian $\mathcal{H}$ depends. 
We define the generalized force $\mathcal{F}$ as
\beq
\mathcal{F} \ \ = \ \ - \frac{\partial \mathcal{H}}{\partial X}
\eeq
We recall that the rate of change formula for an operator $A$ is:
\beq
\frac{d\langle\hat{A}\rangle }{dt}
\ \ = \ \ \left\langle i[\hat{\mathcal{H}},\hat{A}]+\frac{\partial\hat{ A}}{\partial t}\right\rangle 
\eeq
In particular, the rate of change of the energy is:
\beq
\frac{dE}{dt} \ \ = \ \ \frac{d\langle \hat{\mathcal{H}}\rangle}{dt}
\ \ = \ \ \left\langle i [\hat{\mathcal{H}},\hat{\mathcal{H}}]
+\frac{\partial \mathcal{H}}{\partial t}\right\rangle
\ \ = \ \ \left\langle\frac{\partial \mathcal{H}}{\partial t}\right\rangle
\ \ = \ \ \dot{X} \left\langle\frac {\partial \mathcal{H}}{\partial X }\right\rangle
\ \ = \ \ -\dot{X} \left\langle\mathcal{F}\right\rangle
\eeq
If ${E(0)}$ is the energy at time ${t=0}$ 
we can calculate the energy ${E(t)}$ 
at a later time. Using standard phrasing 
we say that an external work  $E(t)-E(0)$ 
is involved in the process. Hence it 
is customary to define the work ${W}$ 
which is done by the system as:
\beq
W \ \ = \ \  -(E(t) - E(0)) 
\ \ = \ \ \int \left\langle\mathcal{F}\right\rangle\dot{X}dt
\ \ = \ \ \int \left\langle\mathcal{F}\right\rangle dX 
\eeq
A "Newtonian force" is associated with the displacement of a piston.
A generalized force called "pressure" is associated with the change of the volume of a box.
A generalized force called "polarization" is associated with the change in an electric field.
A generalized force called "magnetization" is associated with the change in a magnetic field.

\sheadC{Definition of currents}

There are two ways to define "current" operators. 
The "probability current" is defined via the rate of change 
of the occupation operator (see discussion of the "continuity equation"). 
The "electrical current" is defined as the generalized force 
associated with the change in a magnetic flux, as explained below.

Let us assume that at a moment ${t}$ the flux is ${\Phi}$, 
and that at the moment ${t+dt}$ the flux is ${\Phi +d\Phi}$. 
The electromotive force (measured in volts) is according to Faraday's law:
\beq
\mbox{EMF} \,\,=\,\, -\frac{d\Phi}{dt}
\eeq
If the electrical current is ${I}$ then the amount of charge that has been displaced is:
\beq
dQ\,\,=\,\, I dt 
\eeq
The ("external") work which is done by the electric 
field on the displaced charge, is   
\beq
-dW \ \ = \ \ \mbox{EMF} \times dQ \ \ = \ \  -Id\Phi 
\eeq
This formula implies that the generalized force 
which is associated with the change of magnetic flux 
is in fact the electrical current. Note the analogy  
between flux and magnetic field, and hence between 
current and magnetization. In fact one can regard 
the current in the ring as the "magnetization" 
of a spinning charge.

\newpage
\sheadC{The Concept of Symmetry} 

{\em Pedagogical remark:} In order to motivate and to clarify 
the abstract discussion in this section it is recommended 
to consider the problem of finding the Landau levels in Hall geometry, 
where the system is invariant to $x$ translations 
and hence $p_x$ is a constant of motion. Later the ideas 
are extended to discuss motion in centrally symmetrical potentials.  

We emphasize that symmetry and invariance are two different concepts. 
Invariance means that the laws of physics and hence the form 
of the Hamiltonian do not change. But the fields in the Hamiltonian 
may change. In contrast to that in case of a symmetry we 
requite ${ \tilde{\mathcal{H}} = \mathcal{H} }$, meaning 
that the fields look the literally same. As an example 
consider a particle that moves in the periodic potential  
$V(x;R)=\cos(2\pi(x-R)/L)$. The Hamiltonian is invariant 
under translations: If we make translation $a$ then the 
new Hamiltonian will be the same but with $R=R-a$. 
But in the special case that $R/L$ is an integer we have 
symmetry, because then $V(x;R)$ stays the same.

\sheadC{What is the meaning of commutativity?}

Let us assume for example that ${[\mathcal{H},p_x]=0}$. 
We say in such case that the Hamiltonian commutes with 
the generator of translations. What are the implication 
of this statement? The answer is that in such case: 

\bitem The Hamiltonian is symmetric under translations \\
\bitem The Hamiltonian is block diagonal in the momentum basis \\
\bitem The momentum is a constant of motion \\
\bitem There might be systematic degeneracies in the spectrum  

The second statement follows from the ``separation of variables" theorem. 
The third statement follows from the expectation value rate of change formula:
\beq
\frac {d \langle p_x \rangle}{dt} \,\,=\,\, \langle i [\mathcal{H},p_x] \rangle \,\,= \,\, 0 
\eeq
For time independent Hamiltonians $E=\langle \mathcal{H} \rangle$ 
is a constant of the motion because ${[\mathcal{H},\mathcal{H}]=0}$. 
Thus $\langle \mathcal{H} \rangle = \const$ is associated with 
symmetry with respect to ``translations" in time, 
while $\langle p \rangle = \const$ 
is associated with symmetry with respect to translations in space, 
and $\langle L \rangle = \const$ is associated with symmetry 
with respect to rotations.
In the follwing two subsection we further dwell on 
the first and last statements in the above list.

\sheadC{Symmetry under translations and rotations} 

We would like to clarify the algebraic characterization of symmetry. 
For simplicity of presentation we consider translations. 
We claim that the Hamiltonian is symmetric under translations iff  
\beq
[\mathcal{H},p_i]=0 \,\,\,\, \mbox{for} \,\, i=x,y,z
\eeq
This implies that   
\beq
[\mathcal{H},D(a)]=0, \,\,\,\, \mbox{for any $a$} 
\eeq
which can be written as $\mathcal{H}D-D\mathcal{H} = 0$,
or optionally as 
\beq
D^{-1}\mathcal{H}D \ \  = \ \  \mathcal{H} 
\eeq
If we change to a translated frame of reference, 
then we have a new basis which is defined as follows:
\beq
|\tilde{x}\rangle \ \ = \ \ |x+a\rangle \ \ = \ \ D|x\rangle 
\eeq
The transformation matrix is 
\beq
T_{x_1,x_2} 
\ \ \equiv \ \  \langle x_1 | \tilde{x}_2  \rangle 
\ \ = \ \ \Big[ D(a) \Big]_{x_1,x_2} 
\eeq
and the Hamiltonian matrix in the new basis is 
\beq
\tilde{\mathcal{H}}_{x_1,x_2}
\ \ \equiv \ \  \langle \tilde{x}_1 | \mathcal{H} | \tilde{x}_2  \rangle   
\ \ = \ \ \Big[ T^{-1} \mathcal{H} T \Big]_{x_1,x_2} 
\eeq
The commutation of $\mathcal{H}$ with $D$ implies 
that the transformed Hamiltonian $\tilde{\mathcal{H}}$ 
is the same in the translated frame of reference. 

An analogous statement holds for rotations.
The algebraic characterization of symmetry in this case is
\beq
[\mathcal{H},L_i]=0 \,\,\,\, \mbox{for} \,\, i=x,y,z
\eeq
which implies 
\beq
[\mathcal{H},R(\vec{\Phi})]=0, \,\,\,\, \mbox{for any} \,\, \vec{\Phi} 
\eeq
leading as before to the conclusion that 
the Hamiltonian remains the same if we transform
it to a rotated frame of reference.

\sheadC{Symmetry implied degeneracies} 

Let us assume that $\mathcal{H}$ is symmetric under 
translations~$D$. Then if $|\psi\rangle$ is 
an eigenstate of $\mathcal{H}$ then also ${|\varphi\rangle = D|\psi\rangle}$
is an eigenstate with the same eigenvalue. 
This is because 
\beq
\mathcal{H}|\varphi\rangle 
= \mathcal{H}D|\psi\rangle 
= D\mathcal{H}|\psi\rangle 
= E |\varphi\rangle 
\eeq
Now there are two possibilities. One possibility is 
that $|\psi\rangle$ is an eigenstate of $D$, 
and hence  $|\varphi\rangle$ is the same state 
as $|\psi\rangle$.  In such case we say that 
the symmetry of $|\psi\rangle$ is the same as 
of $\mathcal{H}$, and a degeneracy is not implied.
The other possibility is that $|\psi\rangle$ 
has lower symmetry compared with $\mathcal{H}$.   
Then it is implied that $|\psi\rangle$ and $|\varphi\rangle$ 
span a subspace of degenerate states.
If we have two symmetry operations $A$ and $B$, then 
we might suspect that some eigenstates would have 
both symmetries: that means 
both ${A|\psi\rangle \propto |\psi\rangle}$
and ${B|\psi\rangle \propto |\psi\rangle}$. 
If both symmetries hold for all the eigenstates, then it follows 
that ${[A,B]=0}$, because both are diagonal in the same basis.

In order to argue a symmetry implied degeneracy  
the Hamiltonian should commute with a 
non-commutative group of operators. 
It is simplest to explain this statement 
by considering an example. 
Let us consider particle on a clean ring. 
The Hamiltonian has symmetry under translations (generated by $\hat{p}$) 
and also under reflections ($R$). 
We can take the $k_n$ states as a basis. 
They are eigenstates of the Hamiltonian, 
and they are also eigenstates of $\hat{p}$.      
The ground state $n=0$ has the same symmetries 
as that of the Hamiltonian and therefore there 
is no implied degeneracy. But $|k_n\rangle$ 
with $n\ne0$ has lower symmetry compared 
with $\mathcal{H}$, and therefore there 
is an implied degeneracy with its mirror 
image $|k_{-n}\rangle$. These degeneracies are 
unavoidable. If all the states were non-degenerated 
it would imply that both $\hat{p}$ and $R$ 
are diagonal in the same basis. This cannot be the 
case because the group of translations together 
with reflection is non-commutative.

The dimension of a degenerate subspace must be equal 
to the dimension of a representation of the symmetry group. 
This is implied by the following argument: 
One can regard $\mathcal{H}$ as a mutual 
constant of motion for all the group operators;  
therefore, by the ``separation of variables theorem",  
it induces a block-decomposition of the group representation.
The dimensions of the blocks are the dimensions 
of the degenerate subspaces, and at the same time  
they must be compatible with the dimensions of the 
irreducible representations of the group.

Above we were discussing only the systematic 
degeneracies that are implied by the 
symmetry group of the Hamiltonian. 
In principle we can have also ``accidental" 
degeneracies which are not implied by symmetries. 
The way to "cook" such symmetry is as follows:  
pick two neighboring levels, and change 
some parameters in the Hamiltonian so as to make 
them degenerate. It can be easily argued 
that in general we have to adjust 3~parameters 
in order to cook a degeneracy. If the system 
has time reversal symmetry, then the Hamiltonian 
can be represented by a real matrix. In such 
case it is enough to adjust 2~parameters   
in order to cook a degeneracy.

\sheadA{Fundamentals (part III)}

\sheadB{Group representation theory}

\sheadC{Groups} 

A group is a set of elements with a binary operation: 

\bitem The operation is defined by a multiplication table for ${\tau* \tau'}$. \\
\bitem There is a unique identity element ${\mathbf{1}}$. \\
\bitem Every element has an inverse element such that ${\tau \tau^{-1}=\mathbf{1}}$ \\
\bitem Associativity: ${\tau *(\tau'*\tau'') = (\tau*\tau')*\tau''}$

Commutativity does not have to be obeyed:  
this means that in general ${\tau *\tau'\neq\tau'*\tau}$. 

The Galilein group is our main interst. It includes translations, rotations, and boosts. 
A translation is specified uniquely by three parameters ${\mathbf{a}=(a_1,a_2,a_3)}$. 
Rotations are specified by ${(\theta,\varphi,\Phi)}$, 
from which ${\mathbf{\Phi}}$ is constructed.
A boost is parametrized by the relative velocity ${\mathbf{u}=(u_1,u_2,u_3)}$.
A general element is any translation, rotation, boost, or any combination of them. 
Such a group, that has a general element that can be defined using 
a set of parameters is called a Lie group. 
The Galilein group is a Lie group with $9$~parameters. 
The rotation group is a Lie group with $3$~parameters.

\sheadC{Realization of a Group} 

If there are ${\aleph}$ elements in a group, 
then the number of rows in the full multiplication 
table will be ${(\aleph^9)^2=\aleph^{18}}$. 
The multiplication table is too big for us 
to construct and use. Instead we use the following 
strategy: each element of the group 
is regarded as a {\em transformation} over some space:
\beq
\mbox{element of the group} \ \ &=& \ \ \bm{\tau} = (\tau_1,\tau_2, ... ,\tau_d) \\
\mbox{corresponding transformation} \ \ &=& \ \ U(\bm{\tau}) \\
\mbox{the "group property"} \ \ &=& \ \ U(\bm{\tau}) U(\bm{\tau}') =  U(\bm{\tau}*\bm{\tau}') 
\eeq
The association $\bm{\tau}\mapsto U(\bm{\tau})$ is called a {\em realization}. 
For convenience we assume that the parameterization is constructed 
such that ${\tau=(0,0,...,0)}$ is the identity,  
and $\tau*\tau*...*\tau \equiv n\tau$, 
where $n$ is the number of repetitions. 
Hence ${U(0)=\bm{1}}$, and ${U(n\tau)=U(\tau)U(\tau)...U(\tau) = U(\tau)^n}$. 
For example $(\Phi_x,\Phi_y,\Phi_z)$ is good parameterization of 
a rotation, that has this property, as opposed to $(\theta,\varphi,\Phi)$.

It is best to consider an example. 
The realization that defines the Galilein group is over 
the six dimensional phase space ${(\mathbf{x},\mathbf{v})}$. 
The realization of a translation is 
\beq
\tau_{\mathbf{a}} : \,\,\, \left\{\amatrix{\tilde{\mathbf{x}}&=&\mathbf{x}&+&\mathbf{a}\cr 
\tilde{\mathbf{v}}&=&\mathbf{v}& & }\right. 
\eeq
The realization of a boost is 
\beq
\tau_{\mathbf{u}}: \,\,\, \left\{\amatrix{\tilde{\mathbf{x}}&=&\mathbf{x}& &\cr
\tilde{\mathbf{v}}&=&\mathbf{v}&+&\mathbf{u} }\right. 
\eeq
and the realization of a rotation is 
\beq
\tau_{\mathbf{\Phi}}:\,\,\, \left\{\amatrix{\tilde{\mathbf{x}}
=R^E(\Phi)\mathbf{x}\cr \tilde{\mathbf{v}}
= R^E(\Phi) \mathbf{v} }\right. 
\eeq
A translation by ${\mathbf{b}}$, 
and afterward a translation by ${\mathbf{a}}$, 
gives a translation by ${\mathbf{b}+\mathbf{a}}$. 
This composition rule is simple. 
More generally the "multiplication"  
of group elements ${\tau = \tau'*\tau''}$  
implies a very complicated function  
\beq
(\mathbf{a},\mathbf{\Phi},\mathbf{u}) 
= f(\mathbf{a}',\mathbf{\Phi}',\mathbf{u}' ; \mathbf{a}'',\mathbf{\Phi}'',\mathbf{u}'') 
\eeq
We notice that this function requires 18~input parameters, 
and outputs 9~parameters. It is not practical to construct this 
function explicitly. Rather, in order to "multiply" elements 
we compose transformations.   

\sheadC{Realization using linear transformations}

As mentioned above, a realization means that we regard  
each element of the group as an operation over a space. 
We treat the elements as transformations. Below 
we will discuss the possibility of finding a realization 
which consists of {\em linear} transformations. 

First we will discuss the concept of linear transformation, 
in order to clarify it. As an example, we will check 
whether ${f(x) = x + 5}$ is a linear function. 
A linear function must fulfill the condition:
\beq
f(\alpha X_1+\beta X_2)=\alpha f(X_1)+\beta f(X_2)
\eeq
Checking ${f(x)}$:
\beq
&& f(3) = 8 , f(5) = 10, f(8)=13
\\ \nonumber
&& f(3+5) \neq f(3) + f(5)
\eeq
Hence we realize that ${f(x)}$ is not linear. 

In the defining realization of the Galilein group 
over phase space, rotations are linear transformations, 
but translations and boosts are not. If we want to realize 
the Galilein group using linear transformations, the most natural 
way would be to define a realization over the function space. 
For example, the translation of a function is defined as:
\beq
\tau_{\mathbf{a}} : \,\,\, \tilde{\Psi}(x) = \Psi(x-a) 
\eeq
The translation of a function is a linear operation. 
In other words, if we translate the function ${\alpha\Psi^a(x)+\beta\Psi^b(x)}$ 
a spatial distance~$d$, then we get the appropriate linear combination 
of the translated functions ${\alpha\Psi^a(x-d)+\beta\Psi^b(x-d)}$.

Linear transformations are represented by matrices. 
This leads us to the concept of a "representation", 
which we discuss in the next section.

\sheadC{Representation of a group using matrices} 

A representation is a realization of the elements of a group 
using matrices. With each element ${\bm{\tau}}$ of the group, 
we associate a matrix ${U_{ij}(\bm{\tau})}$. 
The requirement is to have the "multiplication table" 
for the matrices would be be in one-to-one correspondence  
to the multiplication table of the elements of the group. 
Below we "soften" this requirement, being satisfied with having 
a "multiplication table" that is the same "up to a phase factor": 
\beq
U(\tau'*\tau'') \,\,=\,\, \eexp{i\mbox{(phase)}} U(\tau') U(\tau'') 
\eeq
It is natural to realize the group elements using orthogonal 
transformations (over a real space) or unitary 
transformations (over a complex space). Any realization 
using a linear transformation is automatically a "representation". 
The reason for this is that linear transformations are always 
represented by matrices. For example, we may consider the realization 
of translations over the function space. Any function 
can be written as a combination of delta functions:
\beq
\Psi(x) = \int{\Psi(x')\delta(x-x')dx}
\eeq
In Dirac notation this can be written as:
\beq
|\Psi\rangle = \sum_x \Psi_x |x\rangle 
\eeq
In this basis, each translation is represented by a matrix:
\beq
D_{x,x'} \,\, =\,\, \langle x | D(a) |x'\rangle \,\, = \,\, \delta(x-(x'+a)) 
\eeq
Finding a "representation" for a group is very useful: 
the operative meaning of "multiplying group elements" 
becomes "multiplying matrices". This means that we can deal 
with groups using linear algebra tools.

\sheadC{Generators}

Every element in a Lie group is specified by a set of parameters. 
Below we assume that we have a "unitary representation" 
of the group. That means that there is a mapping
\beq
\bm{\tau} \ \ \mapsto \ \ U( \tau_1, \tau_2,...) 
\eeq
We also use the convention:
\beq
\bm{1} \ \mapsto \ U(0, 0,...) \ \ = \ \ \hat{1} \ = \ \mbox{identity matrix}
\eeq
We define a set of generators ${\hat{G}_{\mu}}$ in the following way:
\beq
U(\delta\tau_1,0, 0,...) 
\ \ = \ \ \hat{1} - i \delta\tau_{1} \hat{G}_{1} 
\ \ = \ \  \eexp{-i\delta\tau_{1}\hat{G}_{1}}, 
\ \ \ \ \ \ \ \ \ \mbox{[etc.]} 
\eeq
The number of independent generators is the same 
as the number of parameters that specify the 
elements of the group, e.g. 3 generators for the rotation group. 
In the case of the Galieli we have 9~generators, 
but since we allow arbitrary phase factor in the multiplaication table, 
we have in fact~10~generators: 
\beq
\hat{P}_x, \hat{P}_y, \hat{P}_z, 
\hat{J}_x, \hat{J}_y, \hat{J}_z, 
\hat{Q}_x, \hat{Q}_y, \hat{Q}_z,  
\ \ \mbox{and} \ \ \hat{\bm{1}}.
\eeq
Consider the standard representation over wavefunctions.
It follows from the commutation relation ${[\hat{x},\hat{p}]=i}$ 
that ${-\hat{x}}$ is the generator of translations in ${\hat{p}}$.
Hence it follows that he generators of the boosts are ${\hat{Q}_x = -\mass\hat{x}}$,  
where ${\mass}$ is the mass.

\sheadC{Generating a representation}

In general a transformation that is generated by a generator $A$ would 
not commute with a transformation that is generated by a different generator $B$,  
\beq
\eexp{\hat{A}}\eexp{\hat{B}} \ \ \neq \ \ \eexp{\hat{B}}\eexp{\hat{A}} \ \ \neq \ \ \eexp{\hat{A}+\hat{B}}
\eeq
But if the generated transformations are infinitesimal then:
\beq
\eexp{\epsilon(\hat{A}+\hat{B})}
\ \ = \ \ \eexp{\epsilon \hat{A}}\eexp{\epsilon \hat{B}} 
\ \ = \ \ \eexp{\epsilon \hat{B}}\eexp{\epsilon \hat{A}}
\ \ = \ \ \hat{1}+\epsilon \hat{A} + \epsilon \hat {B} + O(\epsilon^2) 
\eeq
If $\epsilon$ is not small we can still do the exponentiation in "small steps" 
using the so called "Trotter formula", namely, 
\beq
\eexp{\hat{A}+\hat{B}} \ \ = \ \ \lim_{N\rightarrow\infty} \left[\eexp{(1/N)\hat{A}} \ \eexp{(1/N)\hat{B}}\right]^N
\eeq 
We can use this observation in order to show that 
any transformation ${U(\mathbf{\tau})}$ 
can be generated using the complete set of 
generators that has been defined in the previous section. 
This means that it is enough to know what 
are the generators in order to calculate all the 
matrices of a given representation. Defining ${\delta \tau = \tau/N}$. 
where $N$ is arbitrarily large, the calculation goes as follows: 
\beq
U( \mathbf{\tau} ) 
&=& [ U(\delta \mathbf{\tau} ) ]^N 
\\ \nonumber
&=& [ U(\delta\tau_1) U(\delta\tau_2) ... ]^N = 
\\ \nonumber
&=& [ \eexp{-i\delta\tau_1 \hat{G}_1} \eexp{-i\delta\tau_2 \hat{G}_2} ... ]^N = 
\\ \nonumber
&=& [ \eexp{-i\delta\tau_1 \hat{G}_1 - i\delta\tau_2 \hat{G}_2 ... } ]^N = 
\\ \nonumber
&=& [ \eexp{-i\delta\mathbf{\tau} \cdot \hat{G} } ]^N \ = \ \eexp{-i \mathbf{\tau} \cdot \hat{G} } 
\eeq
In the above manipulation infinitesimal transformations are involved, 
hence it was allowed to perform $\mathcal{O}(N)$ transpositions
in the ordering of their multiplication.   
Obviously the number of transpositions should not reach $\mathcal{O}(N^2)$, 
else the accumulated error would invalidate the procedure.
The next issue is how to multiply transformations. 
For this we have to learn about the algebra of the generators.

\sheadC{Combining generators} 

It should be clear that if ${A}$ and ${B}$ generate (say) rotations, 
it does not imply that (say) the hermitian operator ${AB+BA}$  
is a generator of a rotation.  On the other hand we have the following 
important statement: if ${A}$ and ${B}$ are generators of group elements, 
then also ${G = \alpha A + \beta B}$ 
and ${G= i[A,B]}$ are generators of group elements. 

{\bf Proof:} By definition ${G}$ is a generator 
if ${\eexp{-i \epsilon G}}$ is a matrix that 
represents an element in the group. We will 
prove the statement by showing that the 
infinitesimal transformation ${\eexp{-i \epsilon G}}$ 
can be written as a multiplication of matrices that 
represent elements in the group. In the first case:
\beq
\mbox{e}^{-i \epsilon( \alpha A + \beta B )} 
\ \ = \ \ \mbox{e}^{ - i \epsilon\alpha A } \, \mbox{e}^{ - i \epsilon\beta B } 
\eeq
In the second case we use the identity:
\beq
\mbox{e}^{ \epsilon [A,B] } 
\ \  =  \ \
\mbox{e}^{ - i \sqrt{\epsilon} B }
\,\mbox{ e}^{ - i \sqrt{\epsilon} A }
\,\mbox{e}^{i \sqrt{\epsilon} B }
\,\mbox{e}^{i \sqrt{\epsilon} A } 
\ + \ \mathcal{O}(\epsilon^2) 
\eeq
This identity can be proved as follows:
\beq
1+\epsilon(AB-BA) \ \  = \ \
(1 -i\sqrt{\epsilon} B - \frac{1}{2} \epsilon B^2) 
(1 -i\sqrt{\epsilon} A - \frac{1}{2} \epsilon A^2) 
(1 +i\sqrt{\epsilon} B - \frac{1}{2} \epsilon B^2) 
(1 +i\sqrt{\epsilon} A - \frac{1}{2} \epsilon A^2) 
\eeq

\sheadC{Structure constants}

Any element in the group can be written 
using the set of basic generators:
\beq
U(\mathbf{\tau})=\eexp{-i\mathbf{\tau} \cdot \hat{G}} 
\eeq
From the previous section it follows that ${i[\hat{G}_{\mu},\hat{G}_{\nu}]}$ is a generator. 
Therefore, it must be a linear combination of the basic generators. 
In other words, there exit constants ${c_{\mu \nu}^{\lambda }}$  
such that the following closure relation is satisfied:
\beq
[G_{\mu}, G_{\nu}] \ \ = \ \ i \sum_{\lambda} c_{\mu \nu}^{\lambda} G_{\lambda} 
\eeq
The constants ${c_{\mu \nu}^\lambda}$ are called 
the "structure constants" of the group. 
Every "Lie group" has its own structure coefficients. 
If we know the structure constants, we can 
reconstruct the group "multiplication table".
In the next lecture we shall deduce the 
structure constants of the rotation group
from its defining representation, 
and then we shall learn how to build up 
all its other representations, just  
from our knowledge of the structure constants.

\sheadC{The structure constants and the multiplication table} 

In order to find the group multiplication 
table based on our knowledge of the generators, 
we can use, in principle, the following formula:
\beq
\eexp{A} \eexp{B} \,\,=\,\, \mbox{e}^{ A+B+C} 
\eeq
Here ${C}$ is an expression that includes only commutators. 
There is no simple expression for ${C}$. 
However, it is possible to find an explicit 
expression up to any desired accuracy. 
This is called the {\em Baker-Campbell-Hausdorff formula}.
By Taylor expansion  up to the third order we deduce:
\beq
C \ \ = \ \ \log( \eexp{A} \eexp{B})-A-B 
\ \ = \ \ \frac{1}{2}[A,B] + \frac{1}{12} [(A-B),[A,B]] + ... 
\eeq
From this we conclude that:
\beq
\eexp{-i\alpha\cdot G -i\beta\cdot G} 
\,\,=\,\, \mbox{e}^{-i\gamma \cdot G} 
\eeq
where
\beq
\gamma_{\lambda} \ \  = \ \ \alpha_{\lambda}+\beta_{\lambda} 
+ \frac{1}{2}c^{\lambda}_{\mu\nu}\alpha_{\mu}\beta_{\nu}
- \frac{1}{12} c^{\lambda}_{\kappa\sigma} c^{\kappa}_{\mu\nu} (\alpha-\beta)_{\sigma}\alpha_{\mu}\beta_{\nu} +... 
\eeq
For more details see Wikipedia, or a paper by Wilcox (1967) that is available in the course site.

\sheadC{The group of rotations} 

In the next lecture we are going to concentrate of the group of rotations.
More precisely we consider on equal footing the groups SO(3) and SU(2). 
Both a characterized by the same Lie algebra, and consequently share 
the same multiplication table up to phase factors.

\sheadC{The Heisenberg group} 

In the present section we relate to the group of translations and boosts. 
Without loss of generality we assume a one-dimensional geometrical space. 
One option is to define this group is via a realization 
over an ${(x,v)}$ phase space. If we adopt this 
definition we get a Abelian group whose elements commute with each other.
Let us call it the "Galilein version" of the group. 

Optionally we can consider a realization of translations and boosts
over complex wavefunctions $\Psi(x)$. 
In the standard representation the space of wavefunctions  
is spanned by the eigenstates of the $\hat{x}$ operator, 
and the translations are generated by ${P=\hat{p}}$.
It follows from ${[\hat{x},\hat{p}]=i}$ that the generator 
of boosts is ${Q=-\hat{x}}$.  Hence the group is non-Abelian  
and characterized by the Lie algebra ${[P,Q]=i}$. 
It is known as the "Heisenberg group". 
It has the same multiplication table as the Abelian "Galilein version", 
up to phase factors.  
  
If we believe (following the two slit experiment) that the state 
of particle is described by a wavefunction, it follows, 
as argued above, that boost and translations do not commute.
Let us illuminate this point using an intuitive physics language. 
Say we have a particle in the laboratory reference frame 
that is described by a wavefunction ${\Psi(x)}$
that is an eigenstate of the translation operators. 
In other words, we are talking about a momentum eigenstate 
that has a well defined wavenumber~${k}$. 
Let us transform to a moving reference frame. 
Assuming that boosts were commuting with translations, 
if follows that boost is a symmetry operation, 
and hence the transformed state ${\tilde{\Psi}(x)}$ 
is still a momentum eigenstate with the same ${k}$. 
From this we would come to the absurd conclusion that 
the particle has the same momentum in all reference frames.  

It is interesting to look for a {\em realization} of the Heisenberg 
group over a finite-dimensional "phase-space". For this purpose 
we have to assume that phase space has a third coordinate: 
instead of ${(x,v)}$ we need ${(a,b,c)}$. 
The way to come to this conclusion is based on the 
observation that any element of the Heisenberg group
can be written in the standard representation as 
\beq
U(a,b,c) \ \ = \ \ \exp\left(iaQ+ibP+ic\right)
\eeq
Hence the multiplication ${U(a_1,b_1,c_1) \, U(a_2,b_2,c_c)}$ is realized 
by the composition law
\beq
\Big(a_1,b_1,c_1\Big) \star \Big(a_2,b_2,c_2\Big) \ \ = \ \ \left(a_1+a_2, \ b_1+b_2, \ c_1+c_2,+\frac{1}{2}(a_1b_2-b_1a_2) \right) 
\eeq
Here $a$ is position displacement, and $b$ is momentum displacement, 
and $c$ is an additional coordinate. If the order of a translation and a boost 
is reversed the new result is different by a phase factor $\exp(iC)$, 
where $C$ equals the symplectic area of the encircled phase-space cell.
This non-commutativity is reflected in the $c$ coordinate.

\newpage

\sheadB{The group of rotations}

\sheadC{The rotation group SO(3)} 

The rotation group ${SO(3)}$ is a non-commutative group. 
That means that the order of rotations is important. 
Despite this, it is important to remember that infinitesimal 
rotations commute. We have already proved this statement in general, 
but we shall prove it once again for the specific case of rotations:
\beq
R(\delta\mathbf{\Phi})\mathbf{r} \ \ = \ \ \mathbf{r}+\delta\mathbf{\Phi}\times\mathbf{r}
\eeq
Therefore
\beq
R(\delta\mathbf{\Phi}^2)R(\delta\mathbf{\Phi}^1)\mathbf{r} 
& \ \ = \ \ &
(\mathbf{r}+\delta\mathbf{\Phi}^1\times\mathbf{r}) 
+ \delta\mathbf{\Phi}^2 \times (\mathbf{r}+\delta\mathbf{\Phi}^1\times\mathbf{r}) =
\\ \nonumber
&\ \ = \ \ & \mathbf{r} + (\delta\mathbf{\Phi}^1 + \delta\mathbf{\Phi}^2) \times \mathbf{r} =
\\ \nonumber
& \ \ = \ \ & R(\delta\mathbf{\Phi}^1)R(\delta\mathbf{\Phi}^2)\mathbf{r} 
\eeq
Obviously, this is not correct for finite non-infinitesimal rotations:
\beq
R(\vec{\Phi}^1)R(\vec{\Phi}^2) \,\,\neq\,\, 
R(\vec{\Phi}^1 + \vec{\Phi}^2) \,\,\neq\,\, 
R(\vec{\Phi}^2)R(\vec{\Phi}^1) 
\eeq
We can construct any infinitesimal rotation 
from small rotations around the major axes:
\beq
R(\delta\vec{\Phi})
\ \ = \ \ R(\delta\Phi_x\vec{e}_x + \delta\Phi_y\vec{e}_y + \delta\Phi_z\vec{e}_z ) 
\ \ = \ \ R(\delta \Phi_x \vec{e}_x ) R(\delta \Phi_y \vec{e}_y ) R(\delta \Phi_z \vec{e}_z ) 
\eeq
Using the vector notation ${\vec{M} = (M_x,M_y,M_z)}$, 
we conclude that a finite rotation around any axis can be written as:
\beq
R(\Phi \vec{n}) 
& \ \ = \ \ & 
R(\vec{\Phi}) = R(\delta \vec{\Phi})^N 
\ \ = \ \ ( R(\delta \Phi_x) R(\delta \Phi_y) R(\delta \Phi_z) )^N 
\\ \nonumber
& \ \ = \ \ & 
( \eexp{-i\delta\vec{\Phi} \cdot \vec{M}} )^N 
\ \ = \ \ \eexp{-i\vec{\Phi} \cdot \vec{M}}= \eexp{-i\Phi M_n} 
\eeq
Hence we conclude that ${M_n = \vec{n} \cdot \vec{M}}$ 
is the generator of the rotations around the axis ${\vec{n}}$.

\sheadC{Structure constants of the rotation group}

We would like to find the structure constants 
of the rotation group ${SO(3)}$, 
using its defining representation. 
The ${SO(3)}$ matrices induce rotations without 
performing reflections and all their elements are real. 
The matrix representation of a rotation around the ${z}$ axis is:
\beq
R(\Phi\vec{e}_z)
\ \ = \ \ \left(\amatrix{ 
\cos(\Phi) & -\sin(\Phi) & 0 \cr 
\sin(\Phi) & \cos(\Phi) & 0 \cr 
0 & 0 & 1 } 
\right) 
\eeq
For a small rotation:
\beq
R(\delta\Phi\vec{e}_z) 
\ \ = \ \ \left( \amatrix{ 
1 & -\delta\Phi & 0 \cr 
\delta\Phi & 1 & 0 \cr 
0 & 0 & 1} \right) 
\ \ = \ \ \hat{1}+\delta\Phi \left( \amatrix{ 
0 & -1 & 0 \cr 
1 & 0 & 0 \cr 
0 & 0 & 0} 
\right) 
\ \ = \ \ \hat{1} -i\delta\Phi M_z 
\eeq
where:
\beq
M_z \ \ = \ \ \left( \amatrix{ 0 & -i & 0 \cr i & 0 & 0 \cr 0 & 0 & 0 } \right) 
\eeq
We can find the other generators in the same way:
\beq
M_x = \left( \amatrix{ 0 & 0 & 0 \cr 0 & 0 & -i \cr 0 & i & 0 } \right),  
\hspace{2cm}
M_y = \left( \amatrix{ 0 & 0& i \cr 0 & 0 & 0 \cr -i & 0 & 0 } \right ) 
\eeq
In compact notation the 3 generators are 
\beq
\Big[ M_k \Big]_{ij} \ \ = \ \ -i\epsilon_{ijk}
\eeq
Now we can calculate the structure constants. 
For example ${[M _ x, M_y ] = iM_z}$, and in general: 
\beq
[M_i,M_j] \ \ = \ \ i\epsilon_{ijk}M_k 
\eeq
We could of course use a different representation of the rotation group 
in order to deduce this Lie algebra. In particular we could use the 
differential representation of the $L_i$ over the infinite-dimensional 
Hilbert space of wavefunctions.

\sheadC{Motivation for finding dim$=$2 representation} 

We have defined the rotation group by the Euclidean realization 
over 3D space. Obviously, this representation can be used 
to make calculations ("to multiply rotations"). The advantage 
is that it is intuitive, and there is no need for complex numbers. 
The disadvantage is that they are $3\times 3$ matrices 
with inconvenient algebraic properties, so a calculation 
could take hours. It would be convenient if we could "multiply rotations" 
with simple $2 \times 2$ matrices. In other words, we are interested 
in a dim$=$2 representation of the rotation group. The mission is to find 
three simple $2 \times 2$ matrices that fulfill:
\beq
[ J_x,J_y] \ = \ iJ_z \,\,\,\,\,\,\,\, \mbox{etc.} 
\eeq
In the next lecture we will learn a systematic approach to building 
all the representations of the rotation group. In the present lecture, 
we will simply find the requested representation by guessing. 
It is easy to verify that the matrices
\beq
S_x = \frac{1}{2}\sigma_x , 
\ \ \ S_y = \frac{1}{2}\sigma_y , 
\ \ \ S_z = \frac{1}{2}\sigma_z 
\eeq
fulfill the above commutation relations. So, we can use them 
to create a dim$=$2 representation of the rotation group. 
We construct the rotation matrices using the formula:
\beq
R \ \ = \ \ \eexp{-i \vec{\Phi} \cdot \vec{S}} 
\eeq
The matrices that are generated necessarily satisfy  
the group multiplication table.

We should remember the distinction between a realization 
and a representation: in a realization it matters what we 
are rotating. In a representation it only matters to us that 
the correct multiplication table is fulfilled. Is it possible 
to regard any representation as a realization? 
Is it possible to say what the rotation matrices rotate? 
When there is a dim$=$3 Euclidean rotation matrix
we can use it on real vectors that represent points in space. 
If the matrix operates on complex vectors, then we must look for another
interpretation for the vectors. This will lead us to the definition of 
the concept of spin (spin~1). When we are talking about a dim$=$2 representation 
it is possible to give the vectors an interpretation. The interpretation 
will be another type of spin (spin~1/2).

\sheadC{How to calculate a general rotation matrix} 

The calculation of a ${2 \times 2}$ rotation 
matrix is extremely simple. 
All the even powers of a given 
Pauli matrix are equal to the identity matrix, while  
all the odd powers are equal to the original matrix. 
From this (using Taylor expansion and separating 
into two partial sums), we get the result:
\beq
R(\Phi) \ \ = \ \ R(\Phi \vec{n}) = \eexp{-i\Phi S_n} 
\ \ = \ \ \cos(\Phi/2) \hat{1} -i\sin (\Phi/2) \sigma_n
\eeq
where ${\sigma_n = \vec{n} \cdot \vec{\sigma}}$, 
and ${S_n=(1/2)\sigma_n}$ is the generator of 
a rotation around the ${\vec{n}}$ axis.

The analogous formula for constructing 
a ${3 \times 3}$ rotation matrix is:
\beq
R(\vec{\Phi}) \ \ = \ \ R(\Phi \vec{n}) = \eexp{-i\Phi M_n} 
\ \ = \ \  1-(1-\cos(\Phi))M_n^2-i\sin(\Phi)M_n 
\eeq
where ${M_n = \vec{n} \cdot \vec{M}}$ 
is the generator of a rotation 
around the ${\vec{n}}$ axis. 
The proof is based on a Taylor expansion. 
We notice that ${M_z^3 = M_z}$, from this it follows 
that for all the odd powers ${M_z^k = M_z}$, while 
for all the even powers ${M_z^k = M_z^2}$ where ${k>0}$. 
The same properties apply to any~${M_n}$, 
because all the rotations are "similar" one 
to the other (moving to another reference 
frame is done by means of a similarity 
transformation that represents a change of basis).

\sheadC{An example for multiplication of rotations}

Let us make a $90^o$ rotation $R(90^{0}e_z)$ around the $Z$ axis, 
followed by a $90^o$ rotation $R(90^{0}e_y)$  around the $Y$ axis. 
We would like to know what this sequence gives.  
Using the Euclidean representation
\beq
R=1-i\sin {\Phi}M_n-(1-\cos {\Phi})M_n^{2} 
\eeq
we get
\beq
R (90^{0}e_z) =1- iM_z- M_z^{2} 
\\ \nonumber
R (90^{0}e_y) =1- iM_y- M_y^{2} 
\eeq
We do not wish to open the parentheses, and add up $9$~terms 
which include multiplications of ${3 \times 3}$ matrices. 
Therefore we abandon the Euclidean representation and 
try and do the same thing with a dim$=$2 representation, 
working with the ${2 \times 2}$ Pauli matrices. 
\beq
&& R(\Phi) = \cos(\Phi/2) \hat{1} -i\sin (\Phi/2) \sigma_n 
\\ \nonumber
&& R (90^{0}e_z) =\frac {1} {\sqrt {2}}(\hat{1}-i \sigma_z) 
\\ \nonumber
&& R (90^{0}e_y) =\frac {1} {\sqrt {2}}(\hat{1}-i \sigma_y) 
\eeq
Hence
\beq
R = R (90^{0}e_y) R (90^{0}e_z) 
= \frac{1}{2}(1-i\sigma_x-i\sigma_y-i\sigma_z) 
\eeq
where we have used the fact that ${\sigma_y\sigma_z = i \sigma_x}$.
We can write this result as:
\beq
R= \cos \frac {120^o}{2} - i \sin \frac {120^o}{2} \vec{n} \cdot \vec{\sigma} 
\eeq
where ${n=\frac {1} {\sqrt{3}}(1,1,1)}$. This defines 
the equivalent rotation which is obtained by combining 
the two $90^o$ rotations.

\newpage

\sheadB{Building the representations of rotations}

\sheadC{Irreducible representations} 

A reducible representation is a representation for which a basis 
can be found such that each matrix in the group decomposes into 
the same block structure. In this basis the set of single-block 
sub-matrices obeys the same multiplication table as that of the 
full matrices, hence we say that the representation is the 
sum of smaller representations.
For a commutative group a basis can be found in which all 
the matrices are diagonal. In the latter case we can say that the 
representation decomposes into one-dimensional representations. 

The rotation group is not a commutative group. We are interested 
in finding all its irreducible representations. 
Let us assume that someone hands us an irreducible representation.
We can find its generators, written as matrices 
in some "standard basis", and we can verify that they 
satisfy the desired Lie algebra. 
We shall see that it is enough to know the {\em dimension} of the 
irreducible representation in order to figure out 
what are the matrices that we have in hand (up to a choice of basis). 
In this way we establish that there is one, and only one, 
irreducible representation for each dimension.

\sheadC{First Stage - determination of basis}

If we have a representation of the rotation group,  
we can look at infinitesimal rotations and define 
generators. For a small rotation around the ${X}$ axis, we can write:
\beq
R(\delta\Phi\vec{e}_x) \ \ = \ \ \hat{1} - i \delta\Phi \hat{J}_x
\eeq
In the same way we can write rotations round 
the $Z$ and $Y$ axes. So, we can find the 
matrices ${\hat{J}_x, \hat{J}_y, \hat{J}_z}$. 
How can we check that the representation that 
we have is indeed a representation of the 
rotation group? All we have to do is check that 
the following equation is obeyed:
\beq
[\hat{J}_i,\hat{J}_j] \ \ = \ \ i\epsilon_{ijk}\hat{J}_k 
\eeq
We define:
\beq
\hat{J}_{\pm} &=& \hat{J}_x \pm i\hat{J}_y 
\\ \nonumber
\hat{J}^2 &=& \hat{J}_x^2+\hat{J}_y^2+\hat{J}_z^2 
= \frac{1}{2}(\hat{J}_{+}\hat{J}_{-} + \hat{J}_{-}\hat{J}_{+})+\hat{J}_z^2 
\eeq
We notice that the operator ${\hat{J}^2}$ commutes 
with all the generators, and therefore also with 
all the rotation matrices. 
From the ``separation of variable" theorems it follows  
that if ${\hat{J}^2}$ has (say) two different eigenvalues, 
then it induces a decomposition of all the rotation 
matrices into two blocks. So in such case the 
representation is reducible. Without loss of generality 
our interest is focused on irreducible    
representations for which we necessarily have 
${\hat{J}^2 = \lambda  \bm{1}}$, 
where $\lambda$ is a constant. Later we shall argue 
that $\lambda$ is uniquely determined by the {\em dimension} 
of the irreducible representation.

If we have a representation, we still have the freedom 
to decide in which basis to write it. Without loss of generality, 
we can decide on a basis that is determined by the operator ${\hat{J}_z}$: 
\beq
\hat{J}_z | m \rangle &=& m| m\rangle
\eeq
Obviously, the other generators, or a general rotation 
matrix will not be diagonal in this basis,  so we have  
\beq
\langle  m | \hat{J}^2 | m' \rangle 
&=& \lambda \delta_{mm'} 
\\ \nonumber
\langle  m | \hat{J}_z |  m' \rangle 
&=& m \delta_{mm'} 
\\ \nonumber
\langle  m | R | m' \rangle 
&=&  R^{\lambda}_{mm'} 
\eeq

\sheadC{Reminder: Ladder Operators}

Given an operator ${\hat{D}}$, which does not have 
to be unitary or Hermitian, and an observable ${\hat{x}}$ 
that obeys the commutation relation 
\beq
[\hat{x},\hat{D}] \ \ = \ \ a\hat{D} 
\eeq
we prove that ${\hat{D}}$ is a ``ladder" operator 
that shifts between eigenstates of ${\hat{x}}$. 
\beq
\hat {x}\hat{D}-\hat{D}\hat{x} \ &=& \ a\hat{D}
\\ \nonumber
\hat {x}\hat{D} \ &=& \ \hat{D}(\hat{x}+a)
\\ \nonumber
\hat{x}\hat{D}| x\rangle \ &=& \ \hat{D}(\hat{x}+a)|x\rangle 
\\ \nonumber
\hat{x}\hat{D}| x\rangle \ &=& \ \hat{D}(x+a)|x\rangle
\\ \nonumber
\hat{x} [\hat{D}| x\rangle] \ &=& \ (x+a) [\hat{D}|x\rangle] 
\eeq
So the state ${|\Psi\rangle = \hat{D}|x\rangle}$ 
is an eigenstate of ${\hat{x}}$ with eigenvalue ${(x+a)}$.
The normalization of  $|\Psi\rangle$ is determined by:
\beq
|| \Psi || \ \ = \ \ \sqrt{\langle\Psi | \Psi\rangle} 
\ \ = \ \ \sqrt{\langle x| \hat{D}^{\dagger}\hat{D} | x \rangle}  
\eeq

\sheadC{Second stage: identification of ladder operators} 

It follows from the commutation relations of the generators that:
\beq
[\hat{J}_z,\hat{J}_{\pm}] \ \ = \ \ \pm\hat{J}_{\pm} 
\eeq
So ${\hat{J}_{\pm}}$ are ladder operators in the 
basis that we have chosen. Using them we 
can shift from a given eigenstate ${|m\rangle}$ to 
other eigenstates: ${..., |m-2\rangle, |m-1\rangle , |m+1\rangle, |m+2\rangle, |m+3\rangle, ... }$.

From the commutation relations of the generators 
\beq
(\hat{J}_+\hat{J}_-) - (\hat{J}_-\hat{J}_+) \ \ = \ \ [\hat{J}_+,\hat{J}_-] = 2\hat{J}_z
\eeq
From the definition of ${\hat{J}^2}$ 
\beq
(\hat{J}_+\hat{J}_-) + (\hat{J}_-\hat{J}_+) \ \ = \ \ 2(\hat{J}^2- (\hat{J}_z)^2)
\eeq
By adding and subtracting these two identities we get respectively:
\beq
\hat{J}_-\hat{J}_+ \ \ = \ \ \hat{J}^2- \hat{J}_z ( \hat{J}_z + 1 )
\\ \nonumber
\hat{J}_+\hat{J}_-  \ \ = \ \ \hat{J}^2- \hat{J}_z ( \hat{J}_z - 1 )
\eeq
Now we can find the normalization of the states 
that are found by using the ladder operators:
\beq
||\hat{J}_{+}|m \rangle||^2
\ \ = \ \ \langle m|\hat{J}_{-} \hat{J}_{+} |m\rangle 
\ \ = \ \ \langle m|\hat{J}^2|m\rangle - \langle m|\hat{J}_z(\hat{J}_z +1)|m\rangle
\ \ = \ \ \lambda-m(m + 1)
\\
||\hat{J}_{-}|m \rangle||^2
\ \ = \ \ \langle m|\hat{J}_{+} \hat{J}_{-} |m\rangle 
\ \ = \ \ \langle m|\hat{J}^2|m\rangle - \langle m|\hat{J}_z(\hat{J}_z - 1)|m\rangle
\ \ = \ \ \lambda-m(m - 1)
\eeq
It will be convenient from now on 
to write the eigenvalue of ${\hat{J}^2}$ as ${\lambda=j(j+1)}$. 
Therefore:
\beq
\hat{J}_{+}|m\rangle \ \ = \ \ \sqrt{j(j+1)-m(m+1)} \ |m+1\rangle
\\ 
\hat{J}_{-}|m\rangle \ \ = \ \ \sqrt{j(j+1)-m(m-1)} \ |m-1\rangle
\eeq

\sheadC{Third stage - deducing the representation}

Since the representation is of a finite dimension, the shift 
process of raising or lowering cannot go on forever. By looking 
at the results of the last section we may conclude that there 
is only one way that the raising process could stop: 
at some stage we should get ${m=+j}$. 
Similarly, there is only one way that the lowering process  
could stop: at some stage we should get ${m=-j}$. 
Hence from the raising/lowering  process we get a ladder 
that includes ${\text{dim} = 2j+1}$ basis states. This number 
must be an integer number. Therefore ${j}$ must be either 
an integer or half integer number. 

For a given ${j}$ the matrix representation of the generators 
is determined uniquely. This is based on the formulas 
of the previous section, from which we conclude:
\beq
[\hat{J}_{+}]_{m'm} \ &=& \ \sqrt{j(j+1)-m(m+1)} \ \delta_{m',m+1} 
\\ \nonumber
[\hat{J}_{-}]_{m'm} \ &=& \ \sqrt{j(j+1)-m(m-1)} \ \delta_{m',m-1} 
\eeq
And all that is left to do is to write:
\beq
[\hat{J}_{x}]_{m'm} \ &=& \ \frac{1}{2} \left[ ( \hat{J}_{+})_{m'm} + (\hat{J}_{-})_{m'm} \right] 
\\ \nonumber
[\hat{J}_{y}]_{m'm} \ &=& \ \frac{1}{2i} \left[ ( \hat{J}_{+})_{m'm} - (\hat{J}_{-})_{m'm} \right] 
\\ \nonumber
[\hat{J}_{z}]_{m'm} \ &=& \ m \ \delta_{m'm} 
\eeq
And then we get every rotation matrix in the representation by:
\beq
R_{m'm} \ \ = \ \ \eexp{-i \vec{\Phi} \cdot \vec{J}} 
\eeq

{\bf A technical note:} 
In the raising/lowering process described above we get a "multiplet" of ${m}$ states. 
Can we get several independent multiplets?  Without loss of generality we had assumed 
that we are dealing with an irreducible representation, and therefore there is only one multiplet.

\newpage

\sheadB{Rotations of spins and of wavefunctions}

\sheadC{Building the dim$=$2 representation (spin $1/2$)}

Let us find the $j=1/2$ representation. 
This representation can be interpreted 
as a realization of spin~$1/2$.
We therefore we use from now on 
the notation~$S$ instead on~$J$.  
\beq
&& S^2 | m \rangle \ \ = \ \ \frac {1}{2} \left(\frac {1}{2}+1\right) | m \rangle 
\\ 
&& S_z \ \ = \ \ \left( \amatrix{ \frac{1}{2} &0 \cr 0& -\frac{1}{2}} \right) 
\eeq
Using formulas of the previous section we find ${S_{+}}$ and ${S_{-}}$ 
and hence $S_x$ and $S_y$
\beq
S_{+}&=& \left( \amatrix{0&1\cr 0&0} \right) 
\\ 
S_{-}&=& \left( \amatrix{0&0\cr 1&0} \right) 
\\ 
S_x & \ \ = \ \ & \frac {1}{2} \left( \left(\amatrix{ 0 & 1 \cr 0 & 0}\right) 
+ \left( \amatrix{ 0 & 0 \cr 1 & 0 }\right) \right) 
\ \ = \ \ \frac {1}{2} \sigma_x 
\\ 
S_y & \ \ = \ \ & \frac {1}{2i} \left( \left(\amatrix{ 0 & 1 \cr 0 & 0}\right) 
- \left( \amatrix{ 0 & 0 \cr 1 & 0 }\right) \right)
\ \ = \ \ \frac{1}{2} \sigma_y 
\eeq

We recall that  
\beq
R(\Phi) \ \ = \ \ R(\Phi \vec{n}) \ \ = \ \ \eexp{-i\Phi S_n} 
\ \ = \ \ \cos(\Phi/2) \hat{1} -i\sin (\Phi/2) \sigma_n
\eeq
where
\beq
&& \vec{n} \ \ = \ \ (\sin {\theta}\cos {\varphi}, \ \sin {\theta}\sin{\varphi}, \ \cos {\theta})
\\ 
&& \sigma_n \ \ = \ \ \vec{n} \cdot \vec{\sigma}
\ \ = \ \ \left(
\amatrix{
\cos{\theta} & \eexp{-i\varphi} \sin{\theta} \cr 
\eexp{i{\varphi}} \sin{\theta} & -\cos{\theta} } 
\right) 
\eeq
Hence
\beq
R(\vec{\Phi}) 
\ \ = \ \ \left( 
\amatrix{ 
\cos(\Phi/2)-i\cos(\theta)\sin(\Phi/2) & -i\eexp{-i\varphi}\sin(\theta)\sin(\Phi/2) \cr 
-i\eexp{i\varphi}\sin(\theta)\sin(\Phi/2) & \cos(\Phi/2)+i\cos(\theta)\sin(\Phi/2) } 
\right ) 
\eeq
In particular a rotation around the $Z$ axis is given by:
\beq
R \ \ = \ \ \eexp{-i{\Phi}S_z} 
\ \ = \ \ \left( \amatrix { \eexp{-i{\Phi}/{2}} & 0 \cr 0 & \eexp{i{\Phi}/{2}} } \right) 
\eeq
And a rotation round the $Y$ axis is given by:
\beq
R \ \ = \ \ \eexp{-i{\Phi}S_y} 
\ \ = \ \ \left( \amatrix{ \cos(\Phi/2)& -\sin(\Phi/2) \cr \sin(\Phi/2) & \cos(\Phi/2) } \right ) 
\eeq

\sheadC{Polarization states of Spin $1/2$} 

We now discuss the physical interpretation 
of the "states" that the $s=1/2$ matrices rotate. 
Any state of "spin 1/2" is represented 
by a vector with two complex number.  
That means we have 4 parameters.
After gauge and normalization, we are 
left with 2 physical parameters 
which can be associated with the polarization 
direction ${(\theta,\varphi)}$.
Thus it makes sense to represent the state 
of spin~1/2 by an arrow that points to some direction 
in space.

The eigenstates of ${S_z}$ 
do not change when we rotate them around 
the $Z$ axis (aside from a phase factor). 
Therefore the following interpretation comes to mind: 
\beq
\left|m=+\frac{1}{2}\right\rangle
\ \ = \ \ 
\ \ \ |\vec{z} \rangle 
& \ = \ & 
| \uparrow \rangle  
\ \ \mapsto \ \ 
\left( \amatrix{1 \cr 0} \right)
\\ 
\left|m=-\frac{1}{2}\right\rangle 
\ \ = \ \ 
| \vecb{z} \rangle 
& \ = \ & 
| \downarrow \rangle 
\ \ \mapsto \ \  
\left( \amatrix{0 \cr 1} \right)
\eeq
This interpretation is confirmed by rotating 
the "up" state by 180 degrees, and getting the "down" state. 
\beq
R \ \ = \ \ \eexp{-i{\pi}S_y} \ \  = \ \ \left(\amatrix{0 & -1 \cr 1 & 0 } \right) 
\eeq
We see that:
\beq
\left( \amatrix{ 1 \cr 0} \right) 
\ \ \leadsto  \ \ 180^{0} 
\ \ \leadsto  \ \  { \left(\amatrix{0\cr 1}\right) } 
\ \ \leadsto  \ \  180^{0} 
\ \ \leadsto  \ \  - \left(\amatrix{1\cr 0}\right) 
\eeq
With two rotations of $180^o$  we get back the "up" state, with a minus sign.
Optionally one observes that 
\beq
\eexp{-i 2\pi S_z} \ \ = \ \ \eexp{-i{\pi}\sigma_z} \ \ = \ \ -1 
\eeq
and hence by similarity this holds for any $2\pi$ rotation.
We see that the representation that we found is not 
a one-to-one representation of the rotation group. 
It does not obey the multiplication table in 
a one-to-one fashion! In fact, we have found a representation 
of $SU(2)$ and not $SO(3)$. The minus sign has a physical significance. 
In a two slit experiment it is possible to turn destructive 
interference into constructive interference by placing 
a magnetic field in one of the paths. The magnetic field 
rotates the spin of the electrons. 
If we induce $360^o$ rotation, then the relative 
phase of the interference change sign, 
and hence constructive interference becomes 
destructive and vice versa. The relative phase is important! 
Therefore, we must not ignore the minus sign.

It is important to emphasize that the physical degree of freedom 
that is called "spin~$1/2$" cannot be visualized  
as a arising from the spinning of small rigid body 
around some axis like a top. 
If it were possible, then we could say that the spin 
can be described by a wave function. In this case, 
if we would rotate it by ${360^o}$ we would get 
the same state, with the same sign. But in the representation 
we are discussing we get minus the same state. 
That is in contradiction with the definition of a (wave) function 
as a single valued object.

We can get from the "up" state all the other 
possible states merely by using the appropriate 
rotation matrix.  In particular we can get any spin 
polarization state 
by combining a rotation round the $Y$ axis and a rotation 
round the $Z$ axis. The result is:
\beq
|\vec{n}_{\theta,\varphi}\rangle
\ \ = \ \ R(\varphi)R(\theta) |\uparrow\rangle 
\ \ = \ \ \eexp{-i\varphi S_z} \eexp{-i\theta S_y} |\uparrow\rangle 
\ \ \mapsto \ \ \left( \amatrix{ \eexp{-i\varphi/2}\cos(\theta/2) \cr \eexp{i\varphi/2}\sin(\theta/2) } \right ) 
\eeq

\newpage \sheadC{Building the dim$=$3 representation (spin 1)} 

Let us find the $j=1$ representation. 
This representation can be interpreted 
as a realization of spin~$1$, and hence 
we use the notation $S$ instead of $J$ 
as in the previous section.  
\beq
S^2|m\rangle & \ \ = \ \ & 1(1+1)|m\rangle
\\
S_z & \ \ = \ \ & \left( \amatrix{1&0&0\cr 0&0&0\cr 0&0&-1} \right)
\\
S_+ & \ \ = \ \ & \left( \amatrix{0&\sqrt{2}&0\cr 0&0&\sqrt{2}\cr 0&0&0} \right) 
\eeq
The standard representation of the generators is:
\beq
S \ \ \rightarrow \ \ \left[ \frac{1}{\sqrt{2}} 
\left( \amatrix{0 & 1& 0 \cr 1 & 0 & 1 \cr 0& 1 & 0} \right), 
\frac{1}{\sqrt{2}} 
\left( \amatrix{0 & -i & 0 \cr i & 0 & -i \cr 0 & i & 0} \right), 
\left( \amatrix{1 & 0 & 0 \cr 0& 0 & 0 \cr 0& 0 & -1} \right) \right] 
\eeq
We remember that the Euclidean representation is:
\beq
M \ \ \rightarrow \ \ \left[ 
\left( \amatrix { 0& 0& 0 \cr 0 & 0 & -i \cr 0& i & 0 } \right), 
\left( \amatrix { 0 &0 &i \cr 0& 0 &0\cr -i &0& 0 } \right), 
\left( \amatrix {0 &-i& 0 \cr i& 0 & 0 \cr 0& 0 & 0 } \right) \right] 
\eeq
Now we have two different dim$=$3 representations that 
represent the rotation group. They are actually the
same representation in a different basis. 
By changing bases (diagonalizing ${M_z}$) it is 
possible to move from the Euclidean representation 
to the standard representation. It is obvious that 
diagonalizing ${M_z}$ is only possible over the 
complex field. In the defining realization, the 
matrices of the Euclidean representation rotate 
points in the real space. But it is possible also to 
use them on complex vectors. In the latter case it is 
a realization for spin~1.

For future use we list some useful matrices:
\beq
S^2_x = \frac{1}{2} 
\left( \amatrix{ 1 & 0 & 1 \cr 0 & 2 & 0 \cr 1 & 0 & 1 } \right), 
\ \ \ \ \ 
S^2_y = \frac{1}{2} 
\left( \amatrix{ 1 & 0 & -1 \cr 0 & 2 & 0 \cr -1 & 0 & 1 } \right), 
\ \ \ \ \ 
S^2_z = 
\left(\amatrix{ 1 & 0 & 0 \cr 0 & 0 & 0 \cr 0 & 0 & 1 } \right) 
\eeq
As expected for the $s=1$ representation we have  
\beq
S^2 \ \  = \ \ S^2_x+S^2_y+S^2_z \ \ = \ \  
\left( \amatrix{ 2 & 0 & 0 \cr 0 & 2 & 0 \cr 0 & 0 & 2 } \right) 
\ \ = \ \ s(s+1) \delta _{m,m'}
\eeq
We note that one can define a projector ${P^z=1-S_z^2}$
on the $m=0$ state. Similarly we can define projectors ${P^x=1-S_x^2}$ and ${P^y=1-S_y^2}$.
We see that this set is complete ${P^x+P^y+P^z=\bf{1}}$, 
and one can verify that these projectors are orthogonal. 
In the next section we shall see that they define the Euclidean basis 
that consist of so-called linearly polarized states. 
In this basis the generators are represented by the matrices ${M_x,M_y,M_z}$.

Having found the generators we can construct 
any rotation of spin~1. We notice the following equation:
\beq
S^3_i \ \ = \ \ S^2_i S_i \ \ = \ \ S_i \ \ \ \ \  \mbox{for} \ i=x,y,z 
\eeq
From this equation we conclude that all 
the odd powers ${(1,3,5,...)}$ 
are the same and are equal to ${S_i}$, 
and all the even powers ${(2,4,6,...)}$ 
are the same and equal to ${S^2_i}$. 
It follows (by way of a Taylor expansion) that:
\beq
U(\vec{\Phi}) \ \ = \ \ \eexp{-i\vec{\Phi} \cdot \vec{S}} 
\ \ = \ \ \hat{1}-i\sin (\Phi) S_n - (1-\cos(\Phi))S_n^2 
\eeq
where:
\beq
S_n \ \ = \ \ \vec{n}\cdot \vec{S}
\eeq
Any rotation can be given by a combination of a rotation round the ${z}$ 
axis and a rotation round the ${y}$ axis. We mark the rotation 
angle around the ${y}$ axis by ${\theta}$, and the rotation angle 
around the ${z}$ axis by ${\varphi}$, and get:
\beq
&& U(\varphi \vec{n}_z) \ \ = \ \ \eexp{-i \varphi S_z}
\ \ = \ \ \left( \amatrix{ \eexp{-i \varphi} & 0 & 0 \cr 0 & 1 & 0 \cr 0 & 0 & \eexp{i \varphi} } \right) 
\\ 
&& U(\theta \vec{n}_y) \ \ = \ \ \eexp{-i \theta S_y}
\ \ =  \ \ \left(\amatrix{ \frac{1}{2}(1+\cos \theta) 
& -\frac{1}{\sqrt{2}}\sin \theta 
& \frac{1}{2}(1-\cos \theta) \cr \frac{1}{\sqrt{2}}\sin \theta 
& \cos \theta 
& -\frac{1}{\sqrt{2}}\sin \theta \cr \frac{1}{2}(1-\cos \theta) 
& \frac{1}{\sqrt{2}}\sin \theta 
& \frac{1}{2}(1+\cos \theta) } \right) 
\eeq

\sheadC{Polarization states of a spin 1}

The states of "Spin 1" cannot be represented 
by simple arrows. This should be obvious  
in advance because it is represented by 
a vector that has three complex components. 
That means we have 6 parameters. 
After gauge and normalization, we still have 4 physical parameters.  
Hence it is not possible to find all the possible 
states of spin~1 by using only rotations.
Below we further discuss the physical 
interpretation of spin~1 states. This discussion 
suggest to use the following notations  
for the basis states of the standard representation:   
\beq
|m=1 \rangle 
\ \ = \ \  
\ \ \ | \vec{z} \rangle 
& \ = \ &   | \Uparrow \rangle 
\ \  \mapsto \ \  
\left(\amatrix{1 \cr 0 \cr 0 }\right) 
\\ 
|m=0 \rangle 
\ \ = \ \
\ \ \ | \vecm{z} \rangle 
& \ = \ & | \Updownarrow \rangle 
\ \  \mapsto \ \ 
\left(\amatrix{0 \cr 1 \cr 0 }\right) 
\\ 
|m=-1 \rangle 
\ \ = \ \
|\vecb{z} \rangle 
& \ = \ & | \Downarrow \rangle 
\ \  \mapsto \ \ 
\left(\amatrix{0 \cr 0 \cr 1 }\right) 
\eeq
The first and the last states represent circular polarizations. 
By rotating the first state by ${180^o}$ around the Y~axis 
we get the third state. This means that we have $180^o$ degree 
orthogonality. However, the middle state is different:  
it describes linear polarization. Rotating the middle 
state by $180^o$ degrees around the Y~axis
gives the same state again! This explains the reason 
for marking this state with a double headed arrow.  
  
If we rotate the linear polarization 
state ${|\Updownarrow \rangle}$ by ${90^o}$, 
once around the Y~axis and once around the X~axis,  
we get an orthogonal set of states:
\beq
| \vecm{x} \rangle 
& \ = \ & \frac{1}{\sqrt{2}}(-| \Uparrow \rangle + | \Downarrow \rangle ) 
\ \  \mapsto \ \  
\frac{1}{\sqrt{2}} \left(\amatrix{-1 \cr 0 \cr 1 }\right)
\\ 
| \vecm{y} \rangle 
& \ = \ & \frac{i}{\sqrt{2}}(| \Uparrow \rangle + | \Downarrow \rangle ) 
\ \  \mapsto \ \  
\frac{1}{\sqrt{2}} \left(\amatrix{i \cr 0 \cr i }\right)
\\ 
| \vecm{z} \rangle 
& \ = \ & | \Updownarrow \rangle 
\ \  \mapsto \ \  
\left(\amatrix{0 \cr 1 \cr 0 }\right) 
\eeq
This basis is called the linear basis. States 
of "spin~1" can be written either in the standard basis
or in the basis of linear polarizations. The latter   
option, where we have $90^o$ orthogonality of 
the basis vectors, corresponds to the Euclidean representation.

We can rotate the state ${| \Uparrow \rangle }$ 
in order to get other circularly polarized states:
\beq
|\vec{n}_{\theta,\varphi}\rangle 
\ \ = \ \ U( \varphi \vec{n}_z) U(\theta \vec{n}_y) | \Uparrow \rangle 
\ \ = \ \ \left( 
\amatrix{ 
\frac{1}{2}(1+\cos \theta) \eexp{-i \varphi} \cr 
\frac{1}{\sqrt{2}}\sin \theta \cr 
\frac{1}{2}(1-\cos \theta) \eexp{i \varphi} } 
\right) 
\eeq
Similarly, we can rotate the state ${| \Updownarrow \rangle }$ 
in order to get other linearly polarized states:
\beq
|\vecm{n}_{\theta,\varphi}\rangle
\ \ = \ \ U( \varphi \vec{n}_z) U(\theta \vec{n}_y) | \Updownarrow \rangle 
\ \ = \ \ \left(\amatrix{ 
-\frac{1}{\sqrt{2}}\sin \theta \eexp{-i \varphi} \cr 
\cos \theta \cr 
\frac{1}{\sqrt{2}}\sin \theta \eexp{i \varphi} } 
\right) 
\eeq
In particular we note that linearly polarized states in the XY plane 
can be written as follows:
\beq
|\vecm{n}_{\varphi}\rangle
\ \ = \ \ \frac{1}{\sqrt{2}}(- \eexp{-i \varphi} | \Uparrow \rangle + \eexp{i \varphi} | \Downarrow \rangle ) 
\ \ = \ \ \cos(\varphi) | \vecm{x} \rangle + \sin(\varphi) | \vecm{y} \rangle 
\eeq

As defined above the circularly polarized states 
are obtained by rotating the ${| \Uparrow \rangle}$  
state, while the linearly polarized states are obtained 
by rotating the ${| \Updownarrow \rangle}$ state.
But a general polarization state will not necessarily 
be circularly polarized, neither linearly polarized.  
We shall argue below that any polarization state 
of spin~1 can be obtained by rotation of so-called
elliptically polarized state 
\beq
|\text{elliptic}\rangle \ \ \equiv \ \ \frac{1}{\sqrt{2}} \left(\sqrt{1+q}| \Uparrow \rangle - \sqrt{1-q}| \Downarrow \rangle \right)
\eeq
where $0<q<1$.
This is an interpolation between ${q=1}$ circular polarization in the Z direction, 
and ${q=0}$ linear polarization in the X direction.  
The states that are obtained by rotation of the $q=1$ state are the circularly polarized states, 
while the states that are obtained by rotation of the $q=0$ state are the linearly polarized states. 
From this follows that a general spin~1 state is 
characterized by 4~parameters: $q$ and the 3~angles
that define rotation.  This is consistent with the 
discussion in the opening of this section.

In order to establish that the most general polarization
state is elliptic we define the following procedure.  
Given an arbitrary state vector 
${(\psi_{+},\psi_0,\psi_{-})}$ we can re-orient 
the $z$ axis in a $(\theta,\varphi)$ direction 
such that $\psi_0=0$. 
We say that this $(\theta,\varphi)$ direction 
defines a ``polarization plane".
Using $\phi$ rotation around the new Z~axis, 
and taking gauge freedom into account, 
we can arrange that the amplitudes  ${(\psi_{+},\psi_{-})}$  
would be real with opposite sign (representing X polarization).
Hence we have establish that after suitable  
rotation we get what we called elliptic polarization
that can be characterized by a number ${0<q<1}$. 
Thus we see that indeed an arbitrary state 
is characterized by the four parameters ${(\theta,\varphi,\phi,q)}$. 
We can represent the state by an ellipse as follows: 
The angles $(\theta,\varphi)$ define the plane of the ellipse, 
while $\phi$~describes the angle of major axes in this plane, 
and $q$ describes the ratio of the major radii.  
Note that ${180^o}$ rotation in the polarization plane 
leads to the same state (with minus sign). 
The special case ${q=0}$ is called circular polarization 
because any rotation in the polarization plane leads to the 
same state (up to a phase). The special case ${q=0}$ 
is called linear polarization: the ellipse becomes a double headed arrow.

It follows from the above procedure that is possible 
to find one-to-one relation between polarization states 
of spin~1 and ellipses. 
The direction of the polarization is the orientation 
of the ellipse. When the ellipse is a circle, the spin  
is circularly polarized, and when the ellipse shrinks down 
to a line, the spin is linearly polarized.  
In the latter case the orientation of the polarization plane is ill defined.

\newpage
\sheadC{Translations and rotations of wavefunctions}

We first consider in this section 
the space of functions that live on a torus (bagel).  
We shall see that the representation of {\em translations} 
over this space decomposes 
into one-dimensional irreducible representations,  
as expected in the case of a commutative group.
Then we consider the space of functions that live 
on the surface of a sphere. We shall see that 
the representation of {\em rotations} 
over this space decomposes as ${1 \oplus 3 \oplus 5 \oplus  \dots}$.
The basis in which this decomposition becomes 
apparent consists of the spherical harmonics.

Consider the space of functions that live on a torus (bagel).
This functions can represent the motion  
of a particle in a 2-D box of size ${L_x \times L_y}$  
with periodic boundary conditions. 
Without loss of generality we assume that 
the dimensions of the surface are ${L_x=L_y=2\pi}$, 
and use ${\bm{x}=(\theta,\varphi)}$ as the coordinates.  
The representation of the state of a particle 
in the standard basis is:
\beq
|\Psi \rangle \ \ = \ \ \sum_{\theta,\varphi} \psi(\theta,\varphi) |\theta,\varphi\rangle 
\eeq
The momentum states are labeled as $\bm{k}=(n,m)$. 
The representation of a wavefunction in this basis is:
\beq
|\Psi \rangle \ \ = \ \ \sum_{n,m} \Psi_{n,m} |n,m\rangle 
\eeq
where the transformation matrix is:
\beq
\langle \theta,\varphi |n,m\rangle \ \ = \ \ \frac{1}{2\pi}\eexp{i (n\theta  + m \varphi)} 
\eeq
The displacement operators in the standard basis are not diagonal:
\beq
D_{\bm{x},\bm{x}'} \ \ = \ \ 
\delta (\theta-(\theta'+a))\delta(\varphi-(\varphi'+b)) 
\eeq
However, in the momentum basis we get diagonal matrices:
\beq
D_{\bm{k},\bm{k}'} \ \ = \ \  
\delta_{n,n'} \delta_{m,m'} \, \eexp{-i(a n + b m)} 
\eeq
In other words, we have decomposed the translations 
group into 1-D representations. This is possible 
because the group is commutative. If a group is 
not commutative it is not possible to find a basis 
in which all the matrices of the group are diagonal 
simultaneously.

Now we consider the space of functions that live on 
the surface of a sphere.
This functions can represent the motion  
of a particle in a 2-D spherical shell. 
Without loss of generality we assume that 
the radius of the sphere is unity. 
In full analogy with the case of a torus, 
the standard representation of the states of a particle 
that moves on the surface of a sphere is:
\beq
| \Psi \rangle \ \ = \ \  \sum_{ \theta ,\varphi}\psi ( \theta ,\varphi) |\theta , \varphi \rangle 
\eeq
Alternatively, we can work with a different basis:
\beq
| \Psi \rangle \ \ = \ \  \sum_{\ell m} \Psi_{\ell m} |\ell, m \rangle 
\eeq
where the transformation matrix is:
\beq
\langle \theta, \varphi | \ell, m \rangle \ \ = \ \  Y^{\ell m}(\theta , \varphi) 
\eeq
The "displacement" matrices are actually "rotation" matrices. 
They are not diagonal in the standard basis:
\beq
R_{\mathbf{x},\mathbf{x}'} \ \ = \ \ 
\delta(\theta -f(\theta',\varphi')) \delta(\varphi -g(\theta', \varphi')) 
\eeq
where $f()$ and $g()$ are complicated functions.
But if we take $Y^{\ell m}(\theta,\varphi)$ 
to be the spherical harmonics then 
in the new basis the representation 
of rotations becomes simpler:
\beq
R_{\ell m, \ell' m'} \ \ = \ \   
\left(
\amatrix{
1\times1 & 0 & 0 & 0 \cr 
0 & 3\times3 & 0 & 0 \cr 
0 & 0 & 5\times5 &0 \cr 
0 & 0 & 0 &  \dots } 
\right) 
\ \ = \ \   \mbox{block diagonal} 
\eeq
When we rotate a function, each block stays "within itself". 
The rotation does not mix states that have different~${\ell}$. 
In other words: in the basis ${|\ell,m\rangle}$ 
the representation of rotations decomposes 
into a sum of irreducible representations 
of finite dimension:
\beq
1 \oplus 3 \oplus 5 \oplus  \dots  
\eeq
In the next section we show how the general procedure 
that we have learned for decomposing representations, 
does indeed help us to find the ${Y^{\ell m}(\theta, \varphi)}$ functions.

\sheadC{The spherical harmonics}

We have already found the  
representation of the generators 
of rotations over the 3D space 
of wavefunctions. 
Namely we have proved that $\vec{L}=\vec{r}\times \vec{p}$. 
If we write the differential 
representation of $L$ is spherical 
coordinates we find as expected that 
the radial coordinate~$r$ is not involved: 
\beq
 L_z & \ \rightarrow \ & - i \frac{\partial}{\partial \varphi }
\\ 
 L_\pm  & \ \rightarrow \ &  \eexp{\pm i \varphi} \left(\pm \frac {\partial} { \partial \theta} 
+ i \cot (\theta)\frac{\partial}{\partial \varphi}\right)
\\ 
L^2 & \ \rightarrow \ & - \left[\frac{1}{\sin (\theta)} 
\frac{\partial}{\partial \theta} (\sin (\theta) \frac{\partial}{\partial \theta}) 
+ \frac {1}{\sin^2 \theta} \frac{\partial^2}{\partial \varphi ^2}\right] 
\eeq
Thus the representation trivially 
decompose with respect to~$r$, 
and without loss of generality 
we can focus on the subspace of 
wavefunctions ${\Psi(\theta,\varphi)}$  
that live on a spherical shell of a given radius. 
We would like to find the basis 
in which the representation decomposes,    
as defined by 
\beq
&& L^2\Psi = \ell(\ell+1)\Psi
\\ 
&& L_z\Psi = m\Psi
\eeq
The solution is:
\beq
Y^{\ell m}(\theta , \phi) \,\ \ = \ \ \, 
\left[ \frac{2\ell+1}{4\pi} \frac{(\ell-m)!}{(\ell+m)!} \right]^{1/2} \, 
\left[ (-1)^m P_{\ell m}(\cos(\theta))\right] \, \eexp{im\varphi } 
\eeq
It is customary in quantum textbooks to absorb 
the factor ${(-1)^m}$ in the definition of the Legendre polynomials. 
We note that it is convenient to start with 
\beq
\, Y^{\ell \ell}(\theta , \phi) 
\,\, \propto \,\, 
(\sin(\theta))^{\ell} \, \eexp{i \ell \varphi } 
\eeq
and then to find the rest of the functions using the lowering operator:
\beq
| \ell,m \rangle 
\,\, \propto \,\, 
L_{-}^{(\ell-m)} 
\, |\ell,\ell\rangle 
\eeq

Let us give some examples for Spherical Functions. 
The simplest function is spread uniformly 
over the surface of the sphere, 
while a linear polarization state along the Z~axis 
is concentrated mostly at the poles:
\beq
Y^{0,0} = \frac{1}{\sqrt{4\pi}}, 
\hspace*{2cm}
Y^{1,0} = \sqrt{\frac{3}{4\pi}} \cos(\theta) 
\eeq
If we rotate the polar wave function by 90 degrees we get:
\beq
Y^{1,x} = \sqrt{\frac{3}{4\pi}} \sin(\theta)\cos(\varphi), 
\hspace*{2cm}
Y^{1,y} = \sqrt{\frac{3}{4\pi}} \sin(\theta)\sin(\varphi) 
\eeq
While according to the standard "recipe" the circular polarizations are:
\beq
Y^{1,1} = -\sqrt{\frac{3}{8\pi}} \sin(\theta) \eexp{i \varphi},
\hspace*{2cm}
Y^{1,-1} = \sqrt{\frac{3}{8\pi}} \sin(\theta) \eexp{-i \varphi} 
\eeq

The $Y^{lm}$ can be visualized as a free-wave with $m$ periods 
in the azimuthal $\varphi$ direction, and $\ell{-}m$ nodal circles in 
the $\theta$ direction.  Here is an illustration that is taken from Wikipedia:

\includegraphics[width=8cm]{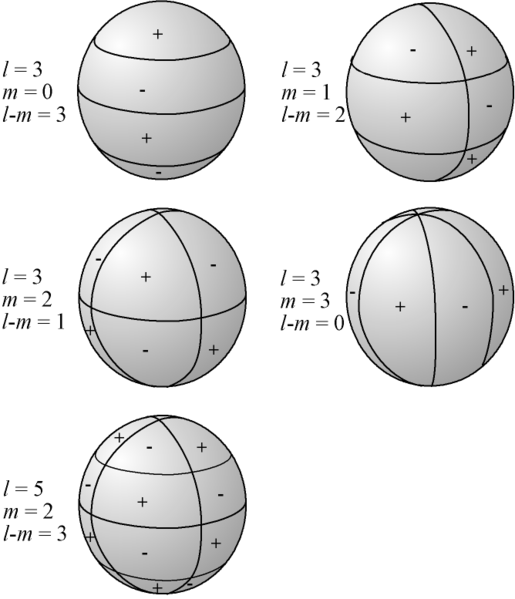}

\newpage

\sheadB{Multiplying representations}

\sheadC{Product space} 

Let us assume that we have two Hilbert spaces. 
One is spanned by the basis ${| i\rangle}$ and the 
other is spanned by the basis ${| \alpha\rangle}$. 
We can multiply the two spaces "externally" 
and get a space with a basis defined by:
\beq
| i,\alpha\rangle \ \ = \ \ | i\rangle \otimes | \alpha\rangle 
\eeq
The dimension of the Hilbert space that 
we obtain is the multiplication of the dimensions.
For example, we can multiply the "position" space ${x}$ 
by the spin space ${m}$. We assume that the space 
contains three sites ${x=1,2,3}$, and that the particle 
has spin $1/2$ with ${m=\pm 1/2}$. 
The dimension of the space that we get from the external 
multiplication is ${2 \times 3 = 6}$. The basis states are 
\beq
| x,m \rangle \ \ = \ \ | x \rangle \otimes | m \rangle
\eeq
A general state is represented by a column vector:
\beq
| \Psi \rangle \ \ \rightarrow \ \  
\left( 
\amatrix{ 
\Psi_{1 \uparrow} \cr 
\Psi_{1 \downarrow} \cr 
\Psi_{2 \uparrow} \cr 
\Psi_{2 \downarrow} \cr 
\Psi_{3 \uparrow} \cr 
\Psi_{3 \downarrow} }
\right) 
\eeq
Or, in Dirac notation:
\beq
| \Psi \rangle \ \ = \ \ \sum_{x,m} \Psi _{x,m} | x,m\rangle 
\eeq
If $x$ has a continuous spectrum then 
the common notational style is  
\beq
|\Psi\rangle  \ \ = \ \ \sum_{x,m} \Psi_m(x) |x,m\rangle 
\ \ \ \mapsto \ \ \ 
\Psi_m(x) 
= \left( \amatrix{ 
\Psi_{\uparrow}(x) \cr 
\Psi_{\downarrow}(x)} 
\right) 
\eeq
If we prepare separately the position 
wavefunction as $\psi_x$ and the momentum 
polarization as $\chi_m$, then the state
of the particle is:
\beq
|\Psi\rangle  
\ \ = \ \   
|\psi\rangle \otimes |\chi\rangle
\ \ \longmapsto \ \  
\Psi_{x,m}
\ = \ 
\psi_x\chi_m
\ \ = \ \ 
\left( 
\amatrix{ 
\psi_{1} \chi_{\uparrow} \cr 
\psi_{1} \chi_{\downarrow} \cr 
\psi_{2} \chi_{\uparrow} \cr 
\psi_{2} \chi_{\downarrow} \cr 
\psi_{3} \chi_{\uparrow} \cr 
\psi_{3} \chi_{\downarrow} }
\right) 
\eeq
It should be clear that in general an 
arbitrary $|\Psi\rangle$ of a particle 
cannot be written as a product 
of some $|\psi\rangle$ with some $|\chi\rangle$. 
In other words: the space and spin degrees of freedom 
of the particle might be {\em entangled}.
Similarly one observed that different subsystems 
might be {\em entangled}: this is unusually the 
case after the subsystems interact with each other.

\sheadC{External multiplication of operators} 

Let us assume that in the Hilbert space 
that is spanned by the basis ${|\alpha\rangle}$, 
an operator is defined, with the representation ${\hat{A} \rightarrow A_{\alpha,\beta}}$. 
This definition has a natural extension over the 
product space ${|i,\alpha\rangle}$, 
namely ${\hat{A} \rightarrow \delta_{i,j} A_{\alpha,\beta}}$.
Similarly for an operator that acts over 
the second Hilbert space ${\hat{B} \rightarrow B_{i,j} \delta_{\alpha,\beta}}$.
Formally if we have operators $A$ and $B$ that are defined 
over the respective Hilbert spaces, 
we can use the notations $\hat{1} \otimes \hat{A}$ 
and  $\hat{B} \otimes \hat{1}$ for their extension over the 
product space, and define their external product as 
\beq
\hat{C} \ \ \equiv \ \ \hat{B} \otimes \hat{A} 
\ \ \equiv \ \ (\hat{B} \otimes \hat{1}) \, (\hat{1} \otimes \hat{A}),
\hspace{2cm}
C_{i\alpha,j\beta} \ \ \rightarrow \ \ B_{i,j} \ A_{\alpha,\beta} 
\eeq
In Dirac notation:
\beq
\langle i\alpha | \hat{C} | j \beta\rangle \ \ = \ \ \langle i | \hat{B} |j \rangle \, \langle \alpha | \hat{A} | \beta \rangle 
\eeq
For example, let us assume that we have 
a particle in a three-site system:
\beq
\hat{x} \ \ = \ \ \left(\amatrix{1&0&0\cr 0&2&0\cr 0&0&3}\right) 
\eeq

\begin{center}
\putgraph{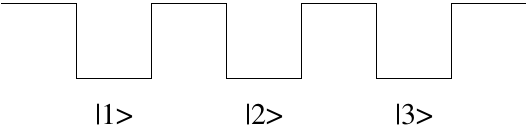}
\end{center}

If the particle has spin $1/2$ we must define the position operator as:
\beq
\hat{x} \ \ = \ \ \hat{x}\otimes \hat{ 1} 
\eeq
That means that:
\beq
\hat{x} |x,m \rangle \ \ = \ \  x | x, m \rangle 
\eeq
And the matrix representation is:
\beq
\hat{x} \ \ = \ \ \left(\amatrix{1&0&0\cr 0&2&0\cr 0&0&3}\right) \otimes \left( \amatrix{1&0\cr 0&1} \right) 
\ \ = \ \ \left(\amatrix{1&0&0&0&0&0\cr 0&1&0&0&0&0\cr 0&0&2&0&0&0\cr 0&0&0&2&0&0\cr 0&0&0&0&3&0 \cr 0&0&0&0&0&3 }\right) 
\eeq
The system has a 6 dimensional basis. We notice 
that in physics textbooks there is no distinction 
between the notation of the operator in the original 
space and the operator in the space that includes 
the spin. We must understand the "dimension" of the 
operator representation by the context. A less trivial 
example of an external multiplication of operators:
\beq
\left(\amatrix{1&0&4\cr 0&2&0\cr 4&0&3}\right) \otimes \left(\amatrix{2&1\cr 1&2}\right) 
\ \ = \ \ \left(\amatrix{2&1&0&0&8&4\cr 1&2&0&0&4&8\cr 0&0&4&2&0&0\cr 0&0&2&4&0&0\cr 8&4&0&0&6&3 \cr 4&8&0&0&3&6} \right) 
\eeq
Specifically, we see that if the operator 
that we are multiplying externally is diagonal, 
then we get a block diagonal matrix.

\sheadC{External multiplication of spin spaces}

Let us consider two operators 
in different Hilbert spaces ${\hat{L}}$ and ${\hat{S}}$. 
The bases of the spaces are ${|m_{\ell} \rangle , |m_s \rangle}$. 
The eigenvalues are ${m_{\ell} = 0,\pm1}$ and ${m_s = \pm 1/2}$.
We label the new states as follows:
${ |\Uparrow \uparrow\rangle }$, \
${ |\Uparrow \downarrow\rangle }$, \ 
${ |\Updownarrow \uparrow\rangle }$, \
${ |\Updownarrow \downarrow\rangle }$, \
${ |\Downarrow \uparrow\rangle }$, \
${ |\Downarrow \downarrow\rangle }$
This basis consists of 6 states. Therefore, 
each operator is represented by a $6\times 6$ matrix. 
For example: 
\beq
\hat{J}_x \ \ = \ \ \hat{S}_x + \hat{L}_x 
\eeq
A mathematician would write it as follows:
\beq
\hat{J}_x \ \ = \ \ \hat{1}\otimes \hat{S}_x + \hat{L}_x \otimes \hat{1} 
\eeq
Let us consider for example the case $\ell=1$ and $s=1/2$. 
In the natural basis we have     
\beq  
S_z &\rightarrow 
\left( 
\amatrix{ 
1 &  &  \cr  & 1 &  \cr  &  & 1 } 
\right) 
\otimes 
\frac{1}{2} 
\left( \amatrix{ 1 & 0 \cr 0 & -1 } \right) 
&\ \ = \ \ \frac{1}{2} 
\left( 
\amatrix{ 
1 & 0 &   &   &   &   \cr 
0 & -1 &   &   &   &   \cr 
  &   & 1 & 0 &   &   \cr 
  &   & 0 & -1 &   &   \cr 
  &   &   &   & 1 & 0 \cr 
  &   &   &   & 0 & -1 } 
\right) 
\\
S_x &\rightarrow 
\left( 
\amatrix{ 
1 &  &  \cr  & 1 &  \cr  &  & 1 }  
\right) 
\otimes 
\frac{1}{2} 
\left( \amatrix{ 0 & 1 \cr 1 & 0 } \right) 
&\ \  = \ \ \frac{1}{2} 
\left( 
\amatrix{ 
0 & 1 &   &   &   &   \cr 
1 & 0 &   &   &   &   \cr 
  &   & 0 & 1 &   &   \cr 
  &   & 1 & 0 &   &   \cr 
  &   &   &   & 0 & 1 \cr 
  &   &   &   & 1 & 0 } 
\right) 
\\
L_z &\rightarrow 
\left( \amatrix{ 1 & 0 & 0 \cr 0 & 0 & 0 \cr 0 & 0 & -1 } \right) 
\otimes \left( \amatrix{ 1 &  \cr  & 1 } \right) 
&\ \ = \ \ \left( 
\amatrix{ 
1 &   & 0 &   & 0 &   \cr 
  & 1 &   & 0 &   & 0 \cr 
0 &   & 0 &   & 0 &   \cr 
  & 0 &   & 0 &   & 0 \cr 
0 &   & 0 &   & -1 &   \cr 
  & 0 &   & 0 &   & -1 } 
\right) 
\\
L_x &\rightarrow 
\left( \amatrix{ 0 & 1 & 0 \cr 1 & 0 & 1 \cr 0 & 1 & 0 } \right) 
\otimes \left( \amatrix{ 1 &  \cr  & 1 } \right) 
&\ \ = \ \ \left( 
\amatrix{ 
0 &   & 1 &   & 0 &   \cr 
  & 0 &   & 1 &   & 0 \cr 
1 &   & 0 &   & 1 &   \cr 
  & 1 &   & 0 &   & 1 \cr 
0 &   & 1 &   & 0 &   \cr 
  & 0 &   & 1 &   & 0 } 
\right) 
\eeq
etc. Empty entries means a zero value that is implied by the fact 
that the operator act only on the "other" degree of freedom. 
This way of writing highlights the block structure of the matrices.

\sheadC{Rotations of a Composite system}

We assume that we have a spin $\ell$ entity whose 
states are represented by the basis 
\beq
|m_{\ell}=-\ell...+\ell \rangle
\eeq
and a spin $s$ entity whose states are represented 
by the basis 
\beq
|m_{s}=-s...+s \rangle
\eeq
The natural $(2\ell+1) \times (2s+1)$ basis 
for the representation of the composite system 
is defined as 
\beq
|m_{\ell},m_s\rangle \ \ = \ \ |m_{\ell}\rangle \otimes |m_s\rangle 
\eeq
A rotation of the $\ell$ entity
is represented by the matrix
\beq
R \ \ = \ \ R^{\ell} \otimes \hat{1} 
\ \ = \ \  \eexp{-i\Phi L_n} \otimes \hat{1}  
\ \ = \ \ \eexp{-i\Phi L_n \otimes \hat{1}}
\eeq
We have used the identity 
${f(A)\otimes\hat{1} = f(A\otimes\hat{1}) }$ 
which is easily established by considering 
the operation of both sides on a basis ${|a,b\rangle}$
that diagonalizes~$A$.
More generally we would like to 
rotate both the $\ell$ entity 
and the $s$ entity. This two operations 
commute since they act on different degrees 
of freedom (unlike two successive rotation 
of the same entity). Thus we get 
\beq
R \ \ = \ \  
\eexp{-i\Phi \hat{1} \otimes S_n}
\ 
\eexp{-i\Phi L_n \otimes \hat{1}}
\ \ = \ \  
\eexp{-i \vec{\Phi} \cdot J}
\eeq
where  
\beq
J \,\,=\,\, L \otimes\hat{1} + \hat{1} \otimes S \,\,=\,\, L + S 
\eeq
From now on we use the conventional sloppy notations of physicists 
as in the last equality: the space over which the operator 
operates and the associated dimension of its matrix representation 
are implied by the context.  
Note that in full index notations the 
above can be summarized as follows: 
\beq
\langle m_{\ell}m_s|R|m_{\ell}'m_s'\rangle 
\ \ &=& \ \ 
R^{\ell}_{m_{\ell},m_{\ell}'} R^s_{m_s,m_s'} 
\\
\langle m_{\ell}m_s|J_i|m_{\ell}'m_s'\rangle 
\ \ &=& \ \ 
[L_i]_{m_{\ell},m_{\ell}'} \delta_{m_s,m_s'} 
+ \delta_{m_{\ell},m_{\ell}'} [S_i]_{m_s,m_s'}  
\eeq

\sheadC{Addition of angular momentum}

It is important to realize that the basis states $|m_{\ell},m_s\rangle$ 
are eigenstates of $J_z$, but not of~$J^2$. 
\beq
J_{z}\big|m_{\ell},m_{s}\big\rangle
\,\,=\,\,(m_{\ell}+m_{s}) \big|m_{\ell},m_{s}\big\rangle 
\,\, \equiv \,\, m_{j} \big|m_{\ell},m_{s}\big\rangle
\\
J^2\big|m_{\ell},m_{s}\big\rangle
\,\,=\,\, \mbox{superposition}\Big[ \big|m_{\ell}[\pm1],m_{s}[\mp1]\big\rangle \Big]
\eeq
The second expression is based on the observation that 
\beq
&& J^{2} = J_{z}^{2} + \frac{1}{2} (J_{+}J_{-}+J_{-}J_{+}) 
\\
&& J_{\pm} = L_{\pm} + S_{\pm} 
\eeq
This means that the representation is reducible, 
and can be written as a sum of irreducible representations. 
Using a conventional procedure we shall show later in this 
lecture that 
\beq
(2\ell+1)\otimes(2s+1) \,\,=\,\, 
(2|\ell+s|+1) \oplus \dots \oplus (2|l-s|+1)
\eeq
It is called the "addition of angular momentum" theorem.  
The output of the "addition of angular momentum" procedure 
is a new basis $|j, m_j\rangle$ that satisfies 
\beq
J^2|j, m_j\rangle &=& j(j+1) |j, m_j\rangle \\ 
J_{z}|j, m_j\rangle &=& m_j |j, m_j\rangle
\eeq

In the next sections we learn how to make the following decompositions:
\beq
&& 2\otimes 2 = 3\oplus 1
\\ \nonumber
&& 2\otimes 3 = 4\oplus 2
\eeq
The first decomposition is useful in connection 
with the problem of two particles with spin $1/2$, 
where we can define a basis that includes 
three ${j=1}$ states (the "triplet") 
and one ${j=0}$ state (the "singlet"). 
The second example is useful in analyzing 
the Zeeman splitting of atomic levels.

\sheadC{The Clebsch-Gordan-Racah coefficients}

We shall see how to efficiently find
the transformation matrix between 
the $|m_{\ell},m_s\rangle$ and the $|j,m_j\rangle$ bases. 
The entries of the transformation matrix are called 
the Clebsch-Gordan-Racah Coefficients, 
and are commonly expressed using the Wigner 3j-symbol:
\beq
T_{m_{\ell}m_s,jm_j} 
\ \ = \ \ 
\langle m_{\ell}, m_s | j, m_j \rangle
\ \ = \ \ 
(-1)^{\ell-s+m} \sqrt{2j+1}
\left( \begin{array}{ccc} 
\ell &  s &  j \\  m_{\ell} & m_s &  -m 
\end{array}\right)
\eeq
Note that the Wigner 3j-symbol is non-zero 
only if its entries can be regarded 
as the sides of a triangle which is formed 
by 3 vectors of length $\ell$ and $s$ and $j$ 
whose projections $m_{\ell}$ and $m_s$ and $-m$  
sum to zero. This geometrical picture 
is implied by the addition of angular momentum theorem.

With the Clebsch-Gordan-Racah transformation matrix 
we can transform states and the operators 
between the two optional representations. 
In particular, note that $J^2$ is diagonal 
in the "new" basis, while in the "old basis" 
it can be calculated as follows: 
\beq
[J^2]_{\tbox{old basis}} \ \ \mapsto \ \   
\langle m_{\ell}',m_{s}'|J^{2}|m_{\ell},m_{s}\rangle
\ \ = \ \ \sum \langle m_{l}',m_{s}'|j',m_{j}'\rangle
\langle j',m_{j}'|J^{2}|j,m_{j}\rangle
\langle j,m_{j}|m_{\ell},m_{s}\rangle 
\eeq
or in short
\beq
[J^2]_{\tbox{old basis}} \ \ = \ \ T [J^2]_{\tbox{diagonal}} T^{\dag}
\eeq
We shall see that in practical applications 
each representation has its own advantages.

\sheadC{The inefficient decomposition method}

Let us discuss as an example the case $\ell=1$ and $s=1/2$. 
In order to find $J^2$ we apparently have to do 
the following calculation: 
\beq
J^2 
\ \ = \ \  
J_x^2+J_y^2+J_z^2
\ \ = \ \ 
\frac{1}{2}\left(J_{+}J_{-} + J_{-}J_{+} \right) + J_z^2
\eeq
The simplest term in this expression 
is the square of the diagonal matrix 
\beq
J_z \ \ =  \ \ L_z + S_z \ \ \rightarrow \ \ \left[ 6 \times 6 \ \ \mbox{matrix} \right] 
\eeq
We have additional terms in the $J^2$ expression 
that contain non-diagonal ${ 6\times6 }$ matrices. 
To find them in a straightforward fashion can be time consuming.  
Doing the calculation of the matrix elements 
in the $|m_{\ell},m_{s}\rangle$ basis, 
one realizes that most of the off-diagonal 
elements are zero: it is advised to use 
the $(m_{\ell},m_{s})$ diagram of the next section   
in order to identify which basis states are coupled.
The result is 
\beq
J^2 \rightarrow 
\left(
\begin{array}{cccccc}
{15}/{4} & 0 & 0 & 0 & 0 & 0 \\
0 & {7}/{4} & \sqrt{2} & 0 & 0 & 0 \\
0 & \sqrt{2} & 11/4 & 0 & 0 & 0 \\
0 & 0 & 0 & 11/4 & \sqrt{2} & 0 \\
0 & 0 & 0 & \sqrt{2} & {7}/{4} & 0 \\
0 & 0 & 0 & 0 & 0 & {15}/{4} 
\end{array}
\right)
\eeq
Next we have to diagonalize $J^2$ in order to get 
the "new" basis in which the representations decomposes 
into its irreducible components.  

In the next section we shall introduce a more efficient procedure 
to find the matrix elements of $J^2$, and the "new" basis, 
using a "ladder operator picture". Furthermore, it is implied by the 
"addition of angular momentum" theorem that $ 3\otimes 2 = 4 \oplus 2$, 
meaning that we have a $j=3/2$ subspace and a $j=1/2$ subspace.  
Therefore it is a-priori clear that after diagonalization we should get 
\beq
J^2 \rightarrow 
\left( 
\amatrix{ 
(15/4) & 0 & 0 & 0 & 0 & 0 \cr 
0 & (15/4) & 0 & 0 & 0 & 0 \cr 
0 & 0 & (15/4) & 0 & 0 & 0 \cr 
0 & 0 & 0 & (15/4) & 0 & 0 \cr 
0 & 0 & 0 & 0 & (3/4) & 0 \cr 
0 & 0 & 0 & 0 & 0 & (3/4) } 
\right) 
\eeq
This by itself is valuable information. Furthermore, if we know the transformation 
matrix $T$ we can switch back to the old basis by using a similarity transformation.

\sheadC{The efficient decomposition method}

In order to explain the procedure to build the new basis 
we consider, as an example, the addition of $\ell=2$ and 
$s=\frac{3}{2}$. The figure below serves to clarify 
this example. Each point in the left panle represents a basis
state in the $|m_{\ell},m_s\rangle$ basis. 
The diagonal lines connect states that span the same $J_z$ subspace, 
namely ${m_{\ell}+m_s=\const \equiv m_j}$. 
Let us call each such subspace a "floor". The upper floor 
$m_j=\ell+s$ contains only one state. The lower floor 
also contains only one state.

\begin{center}
\putgraph[0.45\hsize]{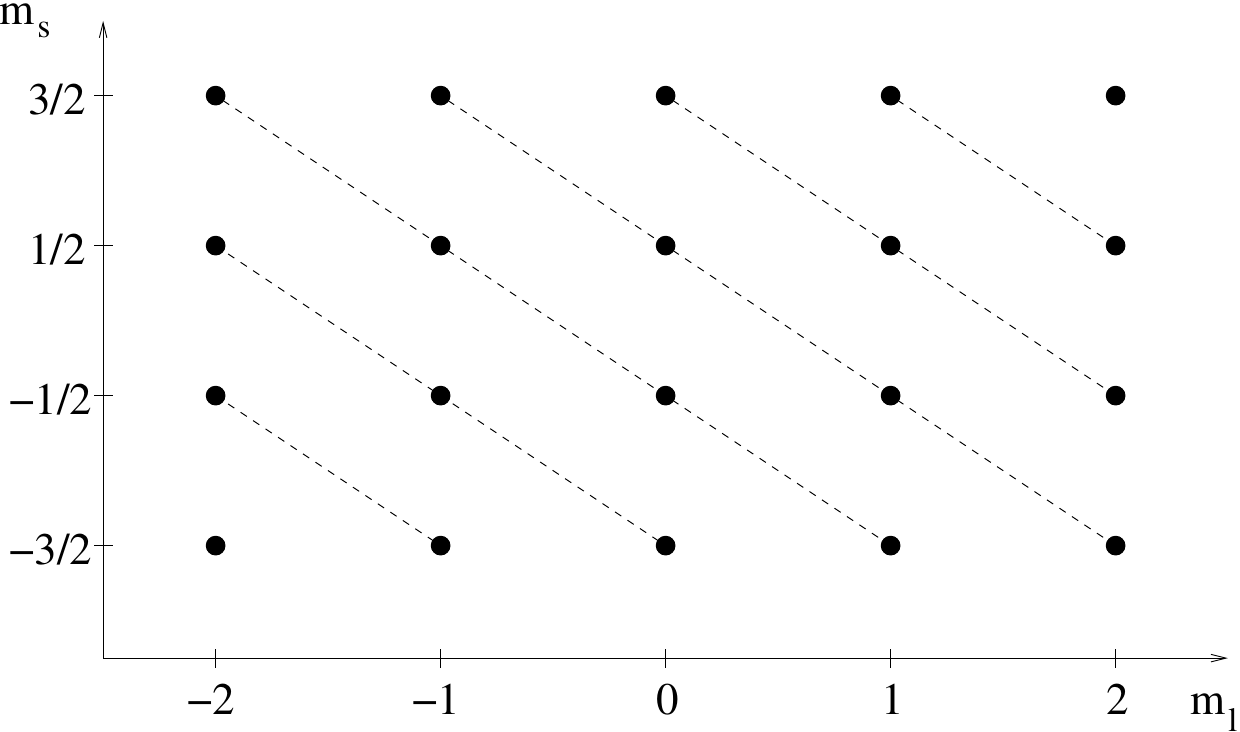}
\hspace*{0.08\hsize}
\putgraph[0.35\hsize]{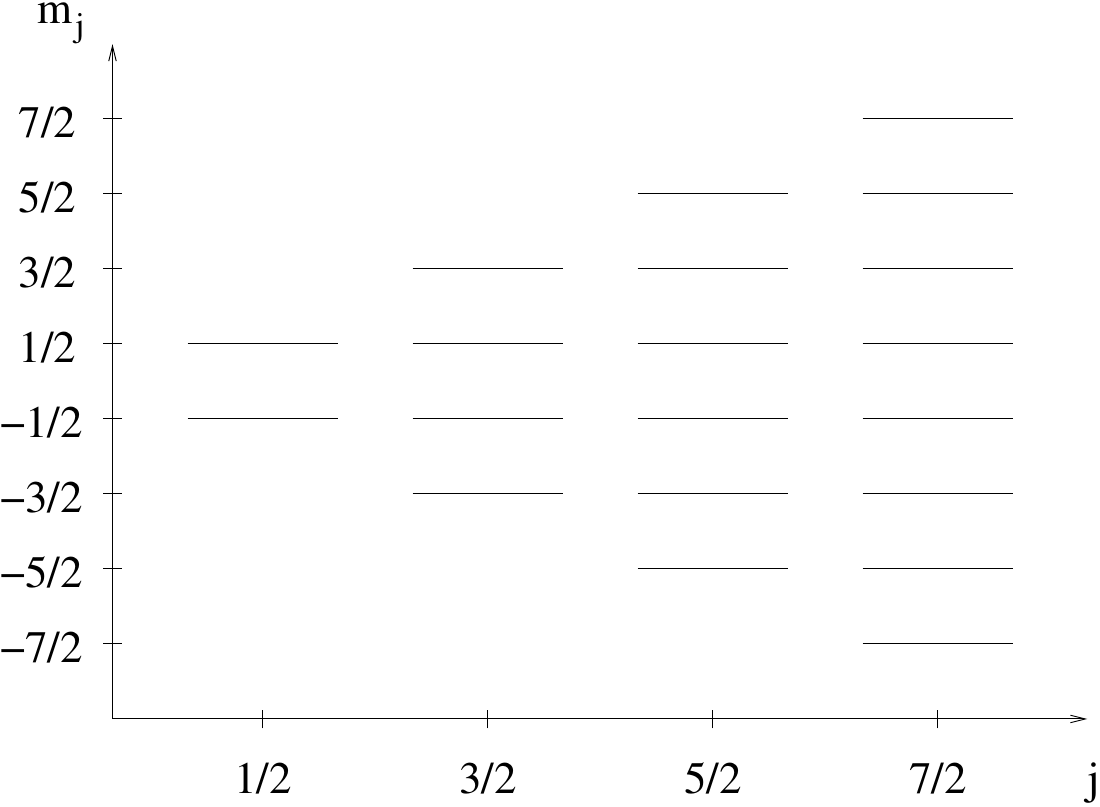}
\end{center}

We recall that 
\beq
&& J_z|m_{\ell},m_s\rangle = (m_{\ell}+m_s) |m_{\ell},m_s\rangle \\
&& S_{-}|m_{\ell},m_s\rangle = \sqrt{s(s+1)-m_s(m_s-1)} |m_{\ell},m_s-1\rangle \\
&& L_{-}|m_{\ell},m_s\rangle = \sqrt{\ell(\ell+1)-m_{\ell}(m_{\ell}-1)} |m_{\ell}-1,m_s\rangle \\ 
&& J_{-} = S_{-} + L_{-} \\
&& J^{2} = J_{z}^{2} + \frac{1}{2} (J_{+}J_{-}+J_{-}J_{+}) 
\eeq
Applying  $J_{-}$ or $J_{+}$ on a state takes us either 
one floor down or one floor up.
By inspection we see that if $J^2$ operates on the state 
in the upper or in the lower floor, then we stay "there". 
This means that these states are eigenstates of $J^2$ 
corresponding to the eigenvalue $j=\ell+s$. Note that 
they could not belong to an eigenvalue $j>\ell+s$ 
because this would imply having larger (or smaller) $m_j$ values.

Now we can use $J_{-}$ in order to obtain the multiplet 
of $j=\ell+s$ states from the $m_{j}=\ell+s$ state. 
Next we look at the second 
floor from above and notice that we know 
the $|j=\ell+s,m_j=\ell+s-1\rangle$ state, 
so by orthogonalization we can find the 
$|j=\ell+s-1,m_j=\ell+s-1\rangle$ state. 
Once again we can get the whole multiplet 
by applying $J_{-}$. Going on with this procedure 
gives us a set of states as arranged in the 
right graph. 

By suggesting the above procedure we have in fact proven  
the "addition of angular momentum" statement.
In the displayed illustration we end up with 4 multiplets 
($j=\frac{7}2,\frac{5}2,\frac{3}2,\frac{1}2$) 
so we have $5\otimes4=8\oplus6\oplus4\oplus2$. 
In the following sections we review some basic examples in detail.

\sheadC{The case of $2\otimes 2 = 3 \oplus 1$}

Consider the addition of $\ell=\frac{1}{2}$ and $s=\frac{1}{2}$ 
(for example, two electrons). In this case the "old" basis is 
\beq
|m_{\ell},m_{s}\rangle =  
|\uparrow\uparrow\rangle, \ 
|\uparrow\downarrow\rangle, \ 
|\downarrow\uparrow\rangle, \ 
|\downarrow\downarrow\rangle
\eeq
The "new" basis we want to find is 
\beq
|j,m_{j}\rangle = 
\underbrace{|1,1\rangle, |1,0\rangle, |1,-1\rangle}, 
\underbrace{|0,0\rangle}
\eeq
These states are called triplet and singlet states.

\begin{center}
\putgraph[0.3\hsize]{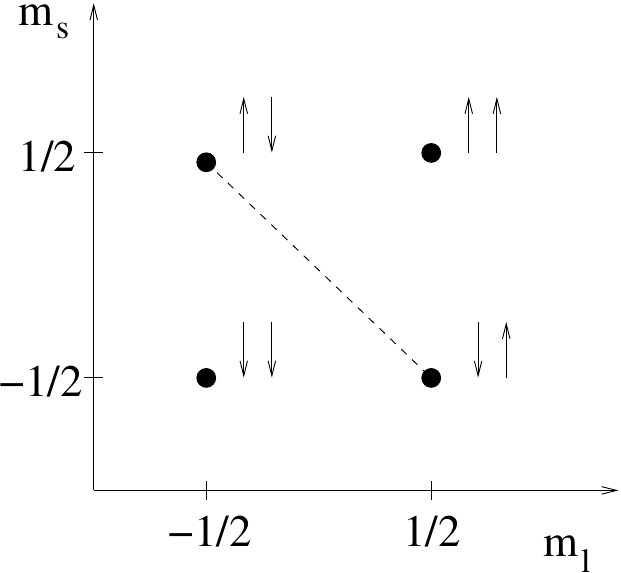}
\hspace*{0.08\hsize}
\putgraph[0.32\hsize]{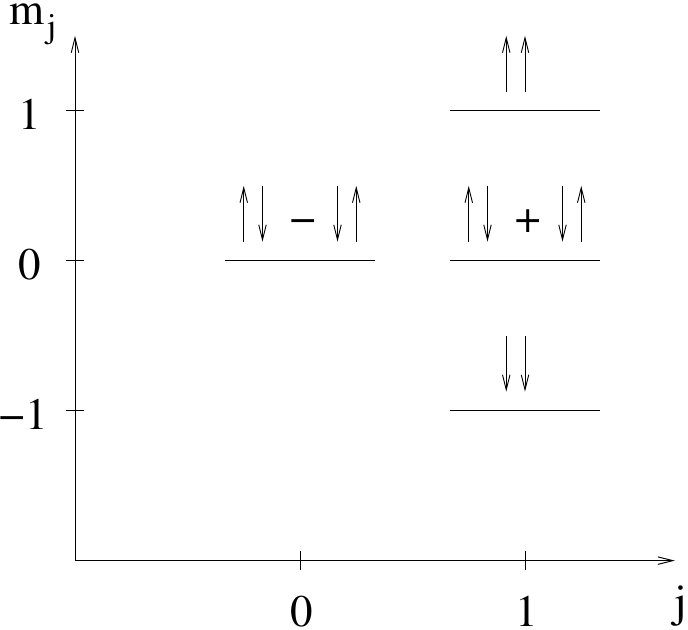}
\end{center}

The decomposition procedure gives:
\beq
&& |1,1 \rangle \,\, = \,\,
|\uparrow\uparrow\rangle \\ 
&& |1,0 \rangle \,\,\propto\,\,
J_{-} |\uparrow\uparrow\rangle 
= |\uparrow\downarrow\rangle + |\downarrow\uparrow\rangle  \\
&& |1,-1 \rangle \,\,\propto\,\,
J_{-} J_{-} |\uparrow\uparrow\rangle 
= 2 |\downarrow\downarrow\rangle 
\eeq
By orthogonaliztion we get the singlet state, which after normalization is 
\beq
|0,0\rangle  \,\, = \,\,  \frac{1}{\sqrt{2}}
\left( |\uparrow\downarrow\rangle - |\downarrow\uparrow\rangle \right)
\eeq
Hence the transformation matrix from the old to the new basis is 
\beq
T_{m_{\ell}, m_{s} | j, m_{j}}=
\left(
\begin{array}{cccc}
1 & 0 & 0 & 0 \\
0 & \frac{1}{\sqrt{2}} & 0 & \frac{1}{\sqrt{2}} \\
0 & \frac{1}{\sqrt{2}} & 0 & -\frac{1}{\sqrt{2}} \\
0 & 0 & 1 & 0 \\
\end{array}
\right)
\eeq
The operator $J^{2}$ in the $|m_{\ell},m_{s}\rangle$ basis is
\beq
\langle m_{\ell}',m_{s}'|J^{2}|m_{\ell},m_{s}\rangle
\,\,=\,\, 
T
\left(
\begin{array}{cccc}
2 & 0 & 0 & 0 \\
0 & 2 & 0 & 0 \\
0 & 0 & 2 & 0 \\
0 & 0 & 0 & 0 \\
\end{array}
\right)
T^{\dag}
=\left(
\begin{array}{cccc}
2 & 0 & 0 & 0 \\
0 & 1 & 1 & 0 \\
0 & 1 & 1 & 0 \\
0 & 0 & 0 & 2 \\
\end{array}
\right)
\eeq

\sheadC{The case of $3\otimes 2 = 4 \oplus 2$}

Consider the composite system of $\ell=1$ and $s=\frac{1}{2}$.  \\
In this case the "old" basis is 
\beq
|m_{\ell},m_{s}\rangle= 
|\Uparrow\uparrow\rangle, 
|\Uparrow\downarrow\rangle, 
|\Updownarrow\uparrow\rangle, 
|\Updownarrow\downarrow\rangle, 
|\Downarrow\uparrow\rangle, 
|\Downarrow\downarrow\rangle
\eeq
The "new" basis we want to find is 
\beq
|j,m_{j}\rangle = 
\underbrace{ 
\left|\frac{3}{2},\frac{3}{2}\right\rangle,
\left|\frac{3}{2},\frac{1}{2}\right\rangle,
\left|\frac{3}{2},-\frac{1}{2}\right\rangle,
\left|\frac{3}{2},-\frac{3}{2}\right\rangle},
\underbrace{ 
\left|\frac{1}{2},\frac{1}{2}\right\rangle,
\left|\frac{1}{2},-\frac{1}{2}\right\rangle}
\eeq

\begin{center}
\putgraph[0.35\hsize]{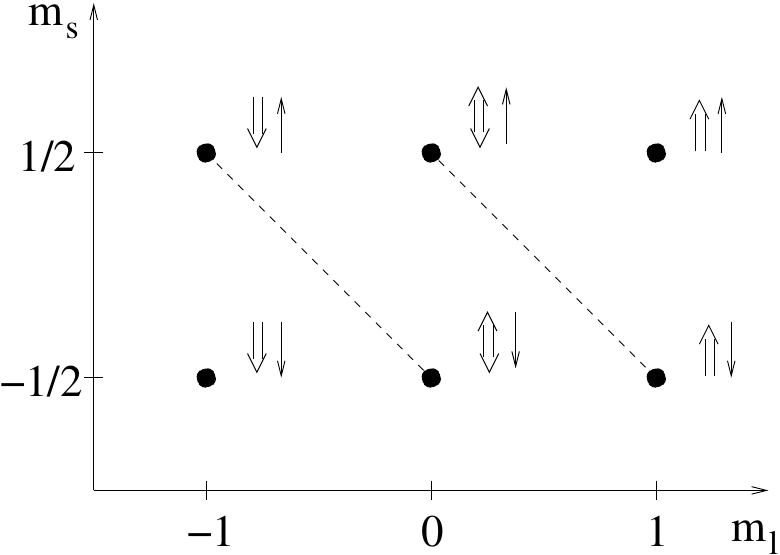}
\hspace*{0.08\hsize}
\putgraph[0.25\hsize]{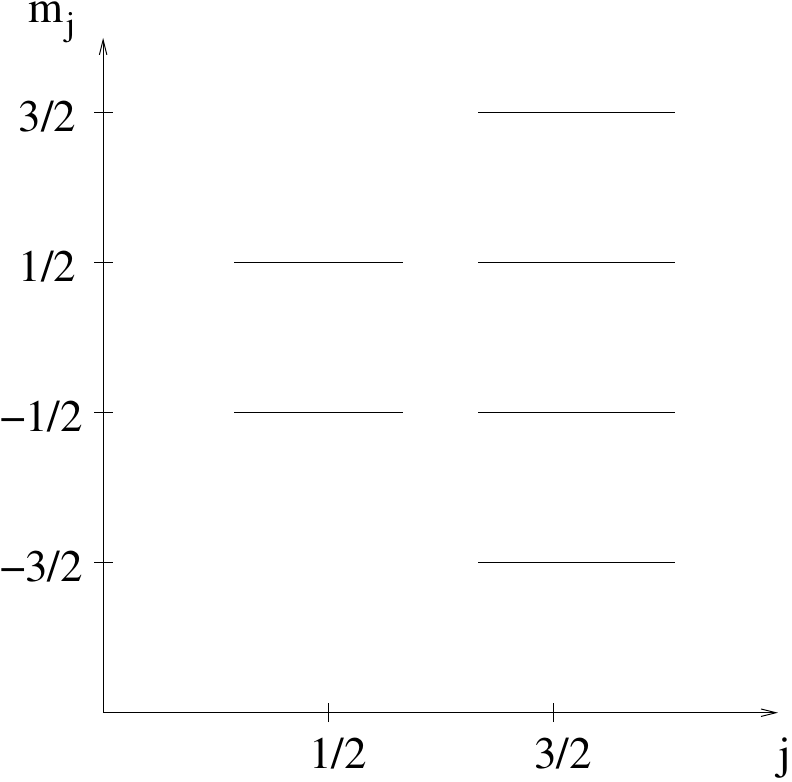}
\end{center}

The decomposition procedure is applied as in the previous section. 
Note that here the lowering operator $L_{-}$ 
is associated with a $\sqrt{2}$ prefactor: 
\beq
&& \left|\frac{3}{2},\frac{3}{2}\right\rangle
 \,\, = \,\, |\Uparrow\uparrow\rangle 
\\ 
&& \left|\frac{3}{2},\frac{1}{2}\right\rangle
\,\,\propto\,\, 
J_{-} |\Uparrow\uparrow\rangle 
=|\Uparrow\downarrow\rangle + \sqrt{2} |\Updownarrow\uparrow\rangle
\\
&& \left|\frac{3}{2},-\frac{1}{2}\right\rangle
\,\,\propto\,\,
J_{-}J_{-} |\Uparrow\uparrow\rangle 
= 2\sqrt{2}|\Updownarrow\downarrow\rangle + 2|\Downarrow\uparrow\rangle
\\
&& \left|\frac{3}{2},-\frac{3}{2}\right\rangle
\,\,\propto\,\,
J_{-}J_{-}J_{-}| \Uparrow\uparrow\rangle 
= 6|\Downarrow\downarrow\rangle
\eeq
By orthogonalization we get the starting point of the next multiplet, 
and then we use the lowering operator again:
\beq
&& \left|\frac{1}{2},\frac{1}{2}\right\rangle
\,\, \propto \,\,
-\sqrt{2}|\Uparrow\downarrow\rangle + |\Updownarrow\uparrow\rangle
\\ 
&& \left|\frac{1}{2},-\frac{1}{2}\right\rangle
\,\,\propto\,\,
-\sqrt{2} |\Updownarrow\downarrow\rangle+|\Downarrow\uparrow\rangle
\eeq
Hence the transformation matrix from the 
old to the new basis is 
\beq
T_{m_{\ell},m_{s} | j,m_{j}}=\left(
\begin{array}{cccccc}
1 & 0 & 0 & 0 & 0 & 0 \\
0 & \sqrt{\frac 13} & 0 & 0 & -\sqrt{\frac 23} & 0 \\
0 & \sqrt{\frac 23} & 0 & 0 & \sqrt{\frac 13} & 0 \\
0 & 0 & \sqrt{\frac 23} & 0 & 0 & -\sqrt{\frac 13} \\
0 & 0 & \sqrt{\frac 13} & 0 & 0 & \sqrt{\frac 23} \\
0 & 0 & 0 & 1 & 0 & 0 \\
\end{array}
\right)
\eeq
and the operator $J^{2}$ in the $|m_{\ell},m_{s}\rangle$ basis is
\beq
\langle m'_{\ell},m'_{s}|J^{2}|m_{\ell},m_{s}\rangle=
T\left(
\begin{array}{cccccc}
\frac {15}{4} & 0 & 0 & 0 & 0 & 0 \\
0 & \frac {15}{4} & 0 & 0 & 0 & 0 \\
0 & 0 & \frac {15}{4} & 0 & 0 & 0 \\
0 & 0 & 0 & \frac {15}{4} & 0 & 0 \\
0 & 0 & 0 & 0 & \frac {3}{4} & 0 \\
0 & 0 & 0 & 0 & 0 & \frac {3}{4} \\
\end{array}
\right)T^{\dag}=\left(
\begin{array}{cccccc}
\frac {15}{4} & 0 & 0 & 0 & 0 & 0 \\
0 & \frac {7}{4} & \sqrt{2} & 0 & 0 & 0 \\
0 & \sqrt{2} & \frac {11}{4} & 0 & 0 & 0 \\
0 & 0 & 0 & \frac {11}{4} & \sqrt{2} & 0 \\
0 & 0 & 0 & \sqrt{2} & \frac {7}{4} & 0 \\
0 & 0 & 0 & 0 & 0 & \frac {15}{4} \\
\end{array}
\right)
\eeq
This calculation is done in the Mathematica file \mbox{\em zeeman.nb}.

\sheadC{The case of $(2\ell+1) \otimes 2 = (2\ell+2) \oplus (2\ell) $}

The last example was a special case of a more general result 
which is extremely useful in studying the Zeeman Effect in atomic physics.  
We consider the addition of integer $\ell$ (angular momentum) 
and $s=\frac{1}{2}$ (spin). The procedure is exactly as in the 
previous example, leading to two multiplets: The $j=\ell+\frac{1}{2}$ 
multiplet and the  $j=\ell-\frac{1}{2}$ multiplet. 
The final expression for the new basis states is:
\beq
\left| j=\ell + \frac{1}{2} ,m \right\rangle
\,\, &=& \,\, 
+\beta \left|m+\frac{1}{2},\downarrow \right\rangle
+\alpha \left|m-\frac{1}{2},\uparrow \right\rangle
\\
\left| j=\ell - \frac{1}{2} ,m \right\rangle
\,\, &=& \,\, 
-\alpha \left|m+\frac{1}{2},\downarrow \right\rangle
+\beta \left|m-\frac{1}{2},\uparrow \right\rangle
\eeq
where
\beq
\alpha = \sqrt{ \frac {\ell + (1/2)  + m } {2\ell+1} } , 
\hspace*{3cm}       
\beta = \sqrt{ \frac {\ell + (1/2) - m } {2\ell+1} }
\eeq

\newpage 

\sheadB{Galilei group and the non-relativistic Hamiltonian}

\sheadC{The Representation of the Galilei Group} 

The defining realization of the Galilei group 
is over phase space. Accordingly, that natural 
representation is with functions that "live" in  
phase space. Thus the $a$-translated $\rho(x,v)$  
is $\rho(x-a,v)$ while the $u$-boosted $\rho(x,v)$ 
is $\rho(x,v-u)$ etc. 

The generators of the displacements 
are denoted $\mathbf{P}_{x}, \mathbf{P}_{y}, \mathbf{P}_{z}$, 
the generators of the boosts are 
denoted $\mathbf{Q}_{x}, \mathbf{Q}_{y}, \mathbf{Q}_{z}$, 
and the generators of the rotations are 
denoted $\mathbf{J}_{x}, \mathbf{J}_{y}, \mathbf{J}_{z}$. 
Thus we have 9 generators. It is clear 
that translations and boosts commute, so the only non-trivial 
structure constants of the Lie algebra have to do 
with the rotations:
\beq
&& \left[ \mathbf{P}_{i}, \mathbf{P}_{j}\right] = \mathbf{0} \\ 
&& \left[ \mathbf{Q}_{i}, \mathbf{Q}_{j}\right] = \mathbf{0} \\ 
&& \left[ \mathbf{P}_{i}, \mathbf{Q}_{j}\right] = \mathbf{0} 
\ \ \ \ \ \ \ \ \ \ \ \ \ \ \ \mbox{(to be discussed)} \\ 
&& \left[ \mathbf{J}_{i}, \mathbf{A}_{j}\right] = i \mathbf{\epsilon}_{ijk} \mathbf{A}_{k} 
\ \ \ \ \ \ \ \ \ \ \ \ \ \ \ \mbox{for} \ \mathbf{A}=P,Q,J 
\eeq

Now we ask the following question: is it 
possible to find a faithful representation  
of the Galilei group that "lives" 
in configuration space. We already know 
that the answer is "almost" positive: 
We can represent pure quantum states 
using "wavefunctions" $\psi(x)$.  
These wavefunctions can be translated and rotated.  
On a physical basis it is also clear that 
we can talk about "boosted" states: 
this means to give the particle a different velocity. 
So we can also boost wavefunctions. 
On physical grounds it is clear that 
the boost should not change $|\psi(x)|^2$. 
In fact it is not difficult to figure out 
that the boost is realized 
by a multiplication of $\psi(x)$ 
by $\eexp{i(\mass u)x}$.  Hence we get the 
identifications $P_x \mapsto -i(d/dx)$ 
and $Q_x \mapsto -\mass x$ for the generators.    
Still the wise reader should realize that 
in this "new" representation boosts and translations 
do not commute, while in case of the strict 
phase space realization they do commute!

On the mathematical side it would be nice 
to convince ourselves that the price of not having 
commutation between translations and boosts 
is inevitable, and that there is a unique 
representation (up to a gauge)  
of the Galilei group using "wavefunctions". 
This mathematical discussion should clarify 
that the "compromise" for having such a representation 
is: (1) The wavefunctions have to be complex; 
(2) The boosts commute with the translations 
only up to a phase factor. We shall see that 
the price that we have to pay is 
to add $\hat{\mathbf{1}}$ as a tenth generator 
to the Lie algebra. This is similar to 
the discussion of the relation 
between SO(3) and SU(2). The elements of 
the latter can be regarded as "rotations" provided  
we ignore an extra "sign factor". 
Here rather than ignoring a "sign factor" 
we have to ignore a complex "phase factor".

Finally, we shall see that the most general 
form of the non-relativistic Hamiltonian  
of a spinless particle, and in particular its mass, 
are implied by the structure of the 
quantum Lie algebra.

\sheadC{The Mathematical Concept of Mass} 

An element $\tau$ of the Galilei group  
is parametrized by 9 parameters. To find 
a strict (unitary) representation means 
to associate with each element a linear 
operator $U(\tau)$ such that $\tau ^{1}\otimes\tau ^{2}=\tau ^{3}$ 
implies 
\beq
U(\tau^{1})U(\tau^{2})=U(\tau^{3}) 
\eeq
Let us see why this strict requirement 
cannot be realized if we want a representation 
with "wavefunctions". Suppose that we have 
an eigenstate of $\hat{P}$ such that 
$\hat{P}|k\rangle=k|k\rangle$. 
since we would like to assume that 
boosts commute with translations 
it follows that also $U_{boost}|k\rangle$ 
is an eigenstate of $\hat{P}$ 
with the same eigenvalue. This is absurd, 
because it is like saying 
that a particle has the same momentum 
in all reference frames. 
So we have to replace the strict requirement by  
\beq
U(\tau^{1})U(\tau^{2})= \eexp{i\times\mbox{phase}} U(\tau^{3}) 
\eeq
This means that now we have an extended 
group that "covers" the Galilei group, 
where we have an additional parameter (a phase), 
and correspondingly an additional generator ($\hat{\mathbf{1}}$). 
The Lie algebra of the ten generators is 
characterized by 
\beq
\left[ G_{\mu },G_{\nu }\right] 
= i\sum_{\lambda} c_{\mu \nu}^{\lambda} G_{\lambda}
\eeq
where $G_0=\hat{\mathbf{1}}$ and the other nine 
generators are $P_i,Q_i,J_i$ with $i=x,y,z$.   
It is not difficult to convince ourselves that 
without loss of generality this introduces 
one "free" parameter into the algebra 
(the other additional structure constants 
can be set to zero via appropriate re-definition 
of the generators). The "free" non-trivial 
structure constant $\mass$ appears in the commutation  
\beq
\left[ \mathbf{P}_{i},\mathbf{Q}_{j}\right] =i \mass\delta_{ij}
\eeq
which implies that boosts do not commute with translations.

\sheadC{Finding the Most General Hamiltonian}

Assume that we have a spinless particle for which 
the standard basis for representation is $|x\rangle$. 
With appropriate gauge of the $x$ basis 
the generator of the translations is $P \mapsto -i(d/dx)$. 
From the commutation relation $[P,Q]=i\mass$ we 
deduce that $Q = -\mass\hat{x} + g(\hat{p})$, where $g()$ 
is an arbitrary function. With appropraite gauge 
of the momentum basis we can assume $Q = -\mass\hat{x}$. 

The next step is to observe that the effect 
of a boost on the velocity operator should be  
\beq
U_{boost}(u)^{-1} \hat{v} U_{boost}(u) = \hat{v} + u
\eeq
which implies that $[Q,\hat{v}]=-i$. The simplest possibility 
is $\hat{v}={\hat{p}}/{\mass}$. But the most general 
possibility is 
\beq
\hat{v}=\frac{1}{\mass}(\hat{p}-A(\hat{x}))
\eeq
where $A$ is an arbitrary function. 
This time we cannot gauge away $A$.

The final step is to recall the rate of change 
formula which implies the relation $\hat{v}=i[\mathcal{H},\hat{x}]$. 
The simplest operator that will give the 
desired result for $v$ is $\mathcal{H}=\frac{1}{2\mass}(p-A(x))^2$. 
But the most general possibility involves 
a second undetermined function:
\beq
\mathcal{H}=\frac{1}{2\mass}(\hat{p}-A(\hat{x}))^2 + V(x)
\eeq
Thus we have determined the most general Hamiltonian 
that agrees with the Lie algebra of the Galilei group.  
In the next sections we shall see that this 
Hamiltonian is indeed invariant under Galilei transformations.

\newpage 

\sheadB{Transformations and invariance}

\sheadC{Transformation of the Hamiltonian} 

First we would like to make an important 
distinction between passive ["Heisenberg"] 
and active ["Schr\"{o}dinger"] points of view 
regarding transformations. The failure 
to appreciate this distinction is an endless 
source of confusion. 

In classical mechanics 
we are used to the passive point of view. 
Namely, to go to another reference frame 
(say a displaced frame) is like a change 
of basis. Namely, we relate the new  
coordinates to the old ones (say ${\tilde{x}=x-a}$),  
and in complete analogy we relate 
the new basis $|\tilde{x}\rangle$ to the old 
basis $|x\rangle$ by a transformation 
matrix $T=\eexp{-ia\hat{p}}$ such that 
${|\tilde{x}\rangle=T|x\rangle=|x+a\rangle}$.

However we can also use an active point of view.
Rather than saying that we "change the basis"  
we can say that we "transform the wavefunction". 
It is like saying that "the tree is moving backwards"  
instead of saying that "the car is moving forward". 
In this active approach the transformation 
of the wavefunction is induced by $S=T^{-1}$, 
while the observables stay the same. 
So it is meaningless to make a distinction between 
old ($x$) and new ($\tilde{x}$) coordinates!

From now on we use the more convenient 
active point of view. It is more convenient 
because it is in the spirit of the Schr\"{o}dinger 
(rather than Heisenberg) picture. In this 
active point of view observables do not transform. 
Only the wavefunction transforms ("backwards").     
Below we discuss the associated 
transformation of the evolution operator 
and the Hamiltonian. 

Assume that the transformation of the state 
as we go from the "old frame" to the "new frame" is 
$\tilde{\psi}=S\psi$. The evolution operator that 
propagates the state of the system from $t_{0}$ to $t$ 
in the new frame is: 
\beq
\tilde{U}(t,t_{0})=S(t)U(t,t_{0})S^{-1}(t_{0})
\eeq
The idea is that we have to transform the state 
to the old frame (laboratory) by $S^{-1}$, 
then calculate the evolution there, 
and finally go back to our new frame.
We recall that the Hamiltonian is defined as the 
generator of the evolution. By definition 
\beq
\tilde{U}(t+\delta t,t_{0})=(1-i\delta t \tilde{\mathcal{H}}(t) ) \tilde{U}(t,t_{0})
\eeq
Hence 
\beq
\tilde{\mathcal{H}} \ = \ i\dfrac{\partial \tilde{U}}{\partial t}\tilde{U}^{-1} 
\ = \ 
i\left[ 
\dfrac{\partial S\left( t\right) }{\partial t}US\left( t_{0}\right) ^{-1}
+S(t) \dfrac{\partial U}{\partial t} S(t_{0})^{-1} \right] 
S(t_{0}) U^{-1}S(t)^{-1}
\eeq
and we get the result
\beq
\tilde{\mathcal{H}} \ = \ S \mathcal{H} S^{-1} + i \dfrac{\partial S}{\partial t} S^{-1}
\eeq

In practice we assume a Hamiltonian of the form $\mathcal{H}=h(x,p;V,A)$.  
Hence we get that the Hamiltonian in the new frame is 
\beq
\tilde{\mathcal{H}} \ = \  h(SxS^{-1},SpS^{-1};V,A)  + i \dfrac{\partial S}{\partial t} S^{-1}
\eeq
Recall that "invariance" means that the Hamiltonian 
keeps its form, but the fields in the 
Hamiltonian may have changed. So the question is 
whether we can write the new Hamiltonian as   
\beq
\tilde{\mathcal{H}} = h(x,p;\tilde{A},\tilde{V})
\eeq
To have "symmetry" rather than merely "invariance" 
means that the Hamiltonian remains the same 
with $\tilde{A}=A$ and $\tilde{V}=V$.    
We are going to show that the following Hamiltonian 
is invariant under translations, rotations, boosts and gauge transformations:
\beq
\mathcal{H} = \frac{1}{2\mass} \left( \hat{p}-\vec{A}(x) \right)^{2}+ V(x)
\eeq
We shall argue that this is the most general 
non-relativistic Hamiltonian for a spinless particle.
We shall also discuss the issue of time reversal 
(anti-unitary) transformations.

\sheadC{Invariance Under Translations} 

\beq
T \ &=& \ D(a) \ = \ \eexp{-i a \hat{p}} 
\\ \nonumber 
S \ &=& \ T^{-1} \ = \ \eexp{i a \hat{p}} 
\eeq
The coordinates (basis) transform with $T$,  
while the wavefunctions are transformed with $S$. 
\beq
&& S\hat{x}S^{-1} = \hat{x+a} 
\\ \nonumber 
&& S\hat{p}S^{-1} = \hat{p} 
\\ \nonumber 
&& Sf(\hat{x},\hat{p})S^{-1} = f(S\hat{x}S^{-1},S\hat{p}S^{-1}) = f(\hat{x}+a,\hat{p}) \\
\eeq
Therefore the Hamiltonian is invariant with 
\beq
\tilde{V}(x) &=& V(x+a) 
\\ \nonumber 
\tilde{A}(x) &=& A(x+a) 
\eeq

\sheadC{Invariance Under Gauge} 

\beq 
&& T \ = \ \eexp{-i \Lambda(x) } 
\\ \nonumber 
&& S \ = \ \eexp{i \Lambda(x) } 
\\ \nonumber
&& S\hat{x}S^{-1} = \hat{x} 
\\ \nonumber
&& S\hat{p}S^{-1} = \hat{p} - \nabla\Lambda(x) 
\\ \nonumber
&& Sf(\hat{x},\hat{p})S^{-1} 
=  f(S\hat{x}S^{-1},S\hat{p}S^{-1}) 
= f(\hat{x},\hat{p}-\nabla\Lambda(x) )
\eeq
Therefore the Hamiltonian is invariant with 
\beq
\tilde{V}(x) &=& V(x) 
\\ \nonumber
\tilde{A}(x) &=& A(x) + \nabla\Lambda(x)
\eeq
Note that the electric and the magnetic 
fields are not affected by this transformation.

More generally we can consider 
time dependent gauge transformations 
with $\Lambda(x,t)$. 
Then we get in the "new" Hamiltonian 
an additional term, leading to 
\beq
\tilde{V}(x) &=& V(x) - (d/dt)\Lambda(x,t)  
\\ \nonumber
\tilde{A}(x) &=& A(x) + \nabla\Lambda(x,t) 
\eeq
In particular we can use the very simple 
gauge $\Lambda=ct$ in order to change the 
Hamiltonian by a constant ($\tilde{\mathcal{H}}=\mathcal{H}-c$).

\sheadC{Boosts and Transformations to a Moving System} 

From an algebraic point of view a boost 
can be regarded as a special case of gauge:
\beq
&& T \ = \ \eexp{i (\mass u) x } 
\\ \nonumber
&& S \ = \ \eexp{-i (\mass u) x } 
\\ \nonumber
&& S\hat{x}S^{-1} = \hat{x} 
\\ \nonumber
&& S\hat{p}S^{-1} = \hat{p} + \mass u 
\eeq
Hence $\tilde{V}(x)=V(x)$ and $\tilde{A}(x)=A(x)-\mass u$. 
But a transformation to a moving frame is not quite 
the same thing. The latter combines a boost and 
a time dependent displacement. The order of these 
operations is not important because we get   
the same result up to a constant phase factor 
that can be gauged away:
\beq
&& S \ = \eexp{i \mbox{phase}(u)} \  \eexp{-i (\mass u) x } \  \eexp{i (ut) p } 
\\ \nonumber
&& S\hat{x}S^{-1} = \hat{x} + ut 
\\ \nonumber 
&& S\hat{p}S^{-1} = \hat{p} + \mass u 
\eeq
The new Hamiltonian is
\beq 
\tilde{\mathcal{H}}=S\mathcal{H}S^{-1}+i\frac{\partial S}{\partial t}S^{-1}
=S\mathcal{H}S^{-1}-u\hat{p} 
= \frac{1}{2\mass} ( \hat{p}-\tilde{A}(x))^{2} + \tilde{V}(x) + \mbox{const}(u)
\eeq
where
\beq
\tilde{V}(x,t) &=& V(x+ut,t)-u \cdot A(x+ut,t) 
\\ \nonumber 
\tilde{A}(x,t) &=& A(x+ut,t)
\eeq
Thus in the new frame the magnetic field is the same 
(up to the displacement) while the electric field is:
\beq
\tilde{\mathcal{E}} 
= -\frac{\partial \tilde{A}}{\partial t}-\nabla\tilde{V}  
=  \mathcal{E} + u \times \mathcal{B}
\eeq
In the derivation of the latter we used the identity
\beq
\nabla ( u \cdot A ) - (u \cdot \nabla) A  
= u \times (\nabla \times A)
\eeq
Finally we note that if we do not include 
the boost in $S$, then we get essentially 
the same results up to a gauge. By including 
the boost we keep the same dispersion relation: 
If in the lab frame $A=0$ and we have $v=p/\mass$,  
then in the new frame we also have $\tilde{A}=0$ 
and therefore $v=p/\mass$ still holds.

\sheadC{Transformations to a rotating frame} 

Let us assume that we have a spinless 
particle held by a a potential $V(x)$.   
Assume that we transform to a rotating frame.     
We shall see that the transformed 
Hamiltonian will have in it a Coriolis force 
and a centrifugal force. 

The transformation that we consider is 
\beq 
S \ \ = \ \ \eexp{i(\vec{\Omega} t) \cdot \hat{L}}
\eeq
The new Hamiltonian is 
\beq 
\tilde{\mathcal{H}}
\ \ = \ \ S\mathcal{H}S^{-1}+i\frac{\partial S}{\partial t}S^{-1}
\ \ = \ \ \frac{1}{2\mass}p^2 + V(x) - \Omega \cdot \hat{L} 
\eeq
It is implicit that the new $x$ coordinate 
is relative to the rotating frame of reference. 
Without loss of generality we assume $\vec{\Omega}=(0,0,\Omega)$.
Thus we got Hamiltonian that looks very similar to that 
of a particle in a uniform magnetic field (see appropriate lecture): 
\beq
\mathcal{H} 
\ \ = \ \ 
\frac{1}{2\mass}(p-A(x))^{2} + V(x) 
\ \ = \ \ 
\frac{p^2}{2\mass} - \frac{\mathcal{B}}{2\mass} L_{z}
+\frac{\mathcal{B}^2}{8\mass}(x^2+y^2) + V(x)
\eeq
The Coriolis force is the 
"magnetic field" $\mathcal{B} = 2\mass\Omega$. 
By adding and subtracting a quadratic term we can 
write the Hamiltonian $\tilde{\mathcal{H}}$ 
in the standard way with 
\beq
\tilde{V} &=& V - \frac{1}{2}\mass \Omega^2 (x^2+y^2) 
\\ \nonumber
\tilde{A} &=& A + \mass \vec{\Omega} \times r  
\eeq
The extra $-(1/2)\mass\Omega^2(x^2+y^2)$ term 
is called the centrifugal potential.

\sheadC{Time Reversal transformations}

Assume for simplicity that the Hamiltonian is time independent. 
The evolution operator is $U=\eexp{-i\mathcal{H}t}$. 
If we make a {\em unitary} transformation $T$ we get  
\beq
\tilde{U} 
\ \ = \ \ T^{-1} \eexp{-i\mathcal{H}t} T 
\ \  = \ \ \eexp{-i (T^{-1} \mathcal{H} T) t} 
\ \ = \ \ \eexp{-i\tilde{\mathcal{H}}t}
\eeq
where $\tilde{\mathcal{H}} = T^{-1} \mathcal{H} T$. 
Suppose we want to reverse the evolution in our laboratory. 
Apparently we have to engineer $T$ such that $T^{-1} \mathcal{H} T = -\mathcal{H}$.
If this can be done the propagator $\tilde{U}$ will take 
the system backwards in time. We can name such $T$ operation 
a "Maxwell demon" for historical reasons. Indeed for systems 
with spins such transformations have been realized using NMR techniques. 
But for the "standard" Hamiltonian $\mathcal{H}=\hat{p}^{2}/(2\mass)$ 
it is impossible to find a {\em unitary} transformation that do the trick 
for a reason that we explain below.

At first sight it seems that in classical mechanics there is a transformation 
that reverses the dynamics. All we have to do is to invert 
the sign of the velocity. Namely $p \mapsto -p$ while $x\mapsto x$. 
So why not to realize this transformation in the laboratory? 
This was Loschmidt's claim against Boltzman. 
Boltzman's answer was "go and do it". 
Why is it "difficult" to do? 
Most people will probably say that to reverse the sign 
of an Avogadro number of particles is tough. 
But in fact there is a better answer. 
In a sense it is impossible to reverse the sign 
even of one particle! If we believe that the dynamics 
of the system are realized by a Hamiltonian, 
then {\em the only physical transformations} 
are proper canonical transformations.
Such transformations preserves the Poisson brackets, 
whereas ${\{ p \mapsto -p, \ x\mapsto x\}}$ inverts sign.
So it cannot be physically realized.  
  
In quantum mechanical language we say that any physical 
realizable evolution process is described 
by a unitary operator.
We claim that the transformation 
$p \mapsto -p$ while $x \mapsto x$ cannot be realized 
by any physical Hamiltonian. 
Assume that we have a unitary transformation $T$ 
such that $T\hat{p}T^{-1} = -\hat{p}$ while  $T\hat{x}T^{-1}=\hat{x}$. 
This would imply $T [\hat{x},\hat{p}] T^{-1}  = - [\hat{x},\hat{p}]$. 
So we get $i=-i$. This means that such a transformation 
does not exist.  

But there is a way out. Wigner 
has proved that there are two types of transformations 
that map states in Hilbert space such that the overlap 
between states remains the same. These are 
either unitary transformations or anti-unitary transformations.
The time reversal transformations that we are going to 
discuss are anti-unitary.  
They cannot be realized in an actual laboratory experiment. 
This leads to the distinction between "microreversibility" 
and actual "reversibility": It is one thing to say 
that a Hamiltonian has time reversal symmetry. 
It is a different story to actually reverse the evolution.  

We shall explain in the next section that 
the "velocity reversal" transformation 
that has been mentioned above can be 
realized by an antiunitary transformation. 
We also explain that in the case of an 
antiunitary transformation we get
\beq
\tilde{U} \ \ = \ \ T^{-1} \eexp{-i\mathcal{H}t} T \ \ = \ \  \eexp{+i (T^{-1} \mathcal{H} T) t} \ \ = \ \ \eexp{-i\tilde{\mathcal{H}}t}
\eeq
where $\tilde{\mathcal{H}} = -T^{-1} \mathcal{H} T$. Thus in order 
to reverse the evolution we have to engineer $T$ 
such that $T^{-1} \mathcal{H} T = \mathcal{H}$, or equivalently 
$[\mathcal{H},T]=0$. If such a $T$ exists then we say that $\mathcal{H}$ 
has time reversal symmetry. In particular we shall explain 
that in the absence of a magnetic field the 
non-relativistic Hamiltonian has a time reversal symmetry.

There is a subtle distinction between the physical notion 
of time reversal invariance, as opposed to invariance under unitary operation.
In the latter case, say "rotation", the given transformation $T$ is well defined 
irrespective of the dynamics. Then we can check whether the "physical law" 
of the dynamics is "invariant"  
\beq
\mbox{Given $T$}, \ \ \forall A, \ \ \exists \tilde{A}, \ \ T^{-1}U[A]T = U[\tilde{A}] 
\eeq
In contrast to that time reversal invariance means:
\beq
\exists T, \ \ \forall A, \ \ \exists \tilde{A}, \ \  U[A]^{-1} =  T U[\tilde{A}] T^{-1}  
\eeq
For example, if $A$ is the vector potential, time reversal invariance 
implies the transformation ${\tilde{A}=-A}$, which means that 
the time reversed dynamics can be realized by inverting the magnetic field.   
Thus, the definition of time reversal transformation is implied by the dynamics, 
and cannot be introduced out of context.

\sheadC{Anti-unitary Operators} 
 
An anti-unitary operator has an anti-linear 
rather than linear property. Namely,  
\beq
T\left( \alpha \left\vert \phi \right\rangle 
+\beta \left\vert \psi \right\rangle \right) 
= \alpha^* T\left\vert \phi \right\rangle 
+ \beta^* T\left\vert \psi \right\rangle  
\eeq
An anti-unitary operator can be represented 
by a matrix $T_{ij}$ whose columns are the images 
of the basis vectors. 
Accordingly ${|\varphi\rangle = T |\psi\rangle}$ 
implies ${ \varphi_i = T_{ij}\psi_j^* }$.
So its operation is complex conjugation followed 
by linear transformation. 
It is useful to note that ${ \tilde{H} = T^{-1}HT }$ 
implies ${ \tilde{H}_{\mu\nu} = T^{*}_{i\mu} H_{ij}^{*} T_{j\nu} }$.
This is proved by pointing out that the effect 
of double complex conjugation when operating on 
a vector is canceled as far as its elements are concerned.

The simplest procedure to construct 
an anti-unitary operator is as follows: 
We pick an arbitrary basis ${|r\rangle}$ 
and define a diagonal anti-unitary operator $K$ 
that is represented by the unity matrix. 
Such operator maps $\psi_r$ to $\psi_r^{*}$,  
and has the property $K^2=1$.
Note also that for such operator ${ \tilde{H}_{ij} = H_{ij}^{*}}$.    
In a sense there is only one anti-unitary 
operator per choice of a basis. Namely, assume 
that $T$ is represented by the diagonal 
matrix ${\{ \eexp{i \phi_{r}} \}}$. That means  
\beq
T\left\vert r\right\rangle = \eexp{i \phi_{r}} \left\vert r\right\rangle 
\eeq
Without loss of generality we can assume 
that $\phi_{r}=0$. This is because we can 
gauge the basis. Namely, we can define 
a new basis $|\tilde{r}\rangle = \eexp{i\lambda_r} |r\rangle$ 
for which  
\beq
T\left\vert \tilde{r} \right\rangle 
= \eexp{ i (\phi_{r} - 2\lambda_r) } \left\vert \tilde{r} \right\rangle 
\eeq
By setting $\lambda_r = \phi_r/2$ we can make 
all the eigenvalues equal to one, and hence $T=K$.
Any other antiunitary operator can be 
written trivially as $T=(TK)K$ where $TK$ is unitary.
So in practice any $T$ is represented 
by complex conjugation followed by a unitary 
transformation.  Disregarding the option of having 
the "extra" unitary operation, time reversal 
symmetry $T^{-1} \mathcal{H} T = \mathcal{H}$
means that in the particular basis 
where $T$ is diagonal the Hamiltonian matrix 
is real ($\mathcal{H}_{r,s}^* = \mathcal{H}_{r,s}$), 
rather than complex.

Coming back to the "velocity reversal" transformation 
it is clear that $T$ should be diagonal in the position 
basis ($x$ should remain the same). Indeed we can verify 
that such a $T$ automatically reverses the sign of the momentum:
\beq 
&& \left\vert k\right\rangle = \sum_{x}\eexp{ikx}\left\vert x\right\rangle 
\\ \nonumber  
&& T\left\vert k\right\rangle = \sum_{x}T\eexp{ikx}\left\vert x\right\rangle 
=\sum_{x}\eexp{-ikx}\left\vert x\right\rangle =\left\vert -k\right\rangle 
\eeq
In the absence of a magnetic field the kinetic term $p^2$ 
in the Hamiltonian has symmetry with respect to this $T$. 
Therefore we say that in the absence of a magnetic field 
we have time reversal symmetry. In which case the Hamiltonian 
is real in the position representation.

What happens if we have a magnetic field? Does it mean 
that there is no time reversal symmetry?  
Obviously in particular cases the Hamiltonian may 
have a different anti-unitary symmetry:  
if ${V(-x)=V(x)}$ then the Hamiltonian is 
symmetric with respect to the transformation 
${x\mapsto -x}$ while ${p\mapsto p}$. 
The anti-unitary $T$ in this case is diagonal 
in the $p$ representation. It can be regarded 
as a product of "velocity reversal" 
and "inversion" (${x\mapsto -x}$ and ${p\mapsto -p}$).            
The former is anti-unitary while the latter 
is a unitary operation.

If the particle has a spin we can define $K$ 
with respect to the standard basis. The 
standard basis is determined by $\hat{x}$ and $\sigma_3$. 
However, $T=K$ is not the standard time reversal 
symmetry: It reverse the polarization if it 
is in the Y~direction, but leave it unchanged 
if it is in the~Z or in the~X direction.   
We would like to have $T^{-1}\sigma T=-\sigma$.
This implies that 
\beq
T \ \ = \ \ \eexp{-i\pi S_y} K \ \ = \ \ -i\sigma_y K
\eeq
Note that $T^2 = (-1)^N$ where $N$ is the number 
of spin~$1/2$ particles in the system.
This implies Kramers degeneracy for odd~$N$. 
The argument goes as follows: If $\psi$ 
is an eigenstate of the Hamiltonian, 
then symmetry with respect to $T$ implies 
that also $T\psi$ is an eigenstate. 
Thus we must have a degeneracy unless 
${T\psi = \lambda\psi}$, where $\lambda$ 
is a phase factor. But this would 
imply that $T^2\psi = \lambda^2\psi$ 
while for odd $N$ we have ${T^2=-1}$.
The issue of time reversal for particles 
with spin is further discussed in {\bf [Messiah p.669]}.

\sheadA{Dynamics and Driven Systems}

\sheadB{Transition probabilities}

\sheadC{Time dependent Hamiltonians} 

To find the evolution which is generated by 
a time independent Hamiltonian is relatively easy. 
Such a Hamiltonian has eigenstates ${|n\rangle}$ 
which are the "stationary" states of the system. 
The evolution in time of an arbitrary state is:
\beq
|\psi(t)\rangle \ \ = \ \ \sum_{n} \eexp{-iE_n t} \psi_n |n\rangle 
\eeq
But in general the Hamiltonian can be 
time-dependent ${[\mathcal{H}(t_1),\mathcal{H}(t_2)]\neq0}$. 
In such case the strategy that was described above for 
finding the evolution in time loses its significance. 
In this case, there is no simple expression for 
the evolution operator: 
\beq
\hat{U}(t,t_0) \ \ = \ \  
\eexp{-idt_N \mathcal{H}(t_N)} 
\cdots 
\eexp{-idt_2 \mathcal{H}(t_2)}
\eexp{-idt_1 \mathcal{H}(t_1)} 
\ \ \equiv \ \ 
\text{Texp}\left[-i\int_{t_0}^{t} \mathcal{H}(t') dt'\right] 
\eeq
where Texp denotes time-ordered exponential, 
that can be replaced by the ordinary $exp$ 
if $\mathcal{H}$ is constant, but not in general.

Of special interest is the case where the Hamiltonian can be written 
as the sum of a time independent  part $\mathcal{H}_0$ 
and a time dependent perturbation $V(t)$. In such case 
we can make the substitution 
\beq
\eexp{-idt_n \mathcal{H}(t_n)} \ \ = \ \  \eexp{-idt_n \mathcal{H}_0} \left( 1 - i dt_n V(t_n) \right) 
\eeq
Next we can expand and rearrange the Texp sum as follows:
\beq
\hat{U}(t) \ \ = && U_0(t,t_0) \ \ + \ \ (-i)\int_{t_0<t_1<t} dt_1 \ U_0(t,t_1)V(t_1)U_0(t_1,t_0) \\ 
&+& (-i)^2\iint_{t_0<t_1<t_2<t} dt_2 dt_1 \ U_0(t,t_2)V(t_2)U_0(t_2,t_1)V(t_1)U_0(t_1,t_0) \ \ + \ \  \cdots    
\eeq
There are few cases where the calculation can be carried out analytically 
to infinite order (a nice example is the Landau-Zener problem).
In many case, if the perturbation is weak enough, one is satisfied 
with the leading order approximation.   Below we assume that 
the Hamiltonian can be written as a sum of a time independent 
part $\mathcal{H}_0$ and a time dependent perturbation. Namely, 
\beq
\mathcal{H} \ \ =\ \  \mathcal{H}_0 + V \ \ =\ \  \mathcal{H}_0 + f(t)W 
\eeq
Using first order perturbation theory, we shall introduce a formula 
for calculating the probability of transition between unperturbed eigenstates. 
The formula can be obtained directly in "one line" derivation 
from the above expansion. But for pedagogical reason we shall repeat 
its derivation using a traditional iterative scheme.

\sheadC{The interaction picture} 

We would like to work in a basis 
such that ${\mathcal{H}_0}$ is diagonal:
\beq
&& \mathcal{H}_0 |n \rangle \ = \  E_n |n \rangle
\\ \nonumber
&& |\Psi(t)\rangle \ = \ \sum_{n}\Psi_n(t) \, |n\rangle 
\eeq
The evolution is determined by the Schr\"{o}dinger's equation:
\beq
i\frac{d\psi_n}{dt} \ \ = \ \ E_n\psi_n \ + \ \sum_{n'} V_{nn'} \Psi_{n' }
\eeq
which can be written in a matrix style as follows:
\beq
i\frac{d}{dt} \left(\amatrix{ \Psi_1& \cr \Psi_2 \cr \vdots } \right)
\ \ = \ \ 
\left(\amatrix{ E_1\Psi_1& \cr E_2 \Psi_2 \cr \vdots } \right)
\ + \  
\left(
\amatrix{
V_{11} & V_{12} & \cdots \cr 
V_{21} & V_{22} & \cdots \cr 
\vdots & \vdots &\ddots } 
\right) 
\left(\amatrix{ \Psi_1& \cr \Psi_2 \cr \vdots } \right) 
\eeq
Without the perturbation we would 
get ${\psi_n(t)= c_n \eexp{-iE_n t}}$, 
where ${c_n}$ are constants.  
It is therefore natural to use 
the variation of parameters method, 
and to write 
\beq
\Psi_n(t) \ \ = \ \ c_n(t) \ \eexp{-iE_n t} 
\eeq
In other words, we represent the "wave function" 
by the amplitudes ${ c_n(t) = \psi_n(t) \eexp{iE_n t}}$ 
rather than by the amplitudes $\Psi_n(t)$. 
The Schr\"{o}dinger's equation in the new 
representation takes the form  
\beq
i\frac{d c_n}{dt} \ \ = \ \ \sum_{n'} \eexp{i(E_{n}-E_{n'})t} \ V_{nn'} \ c_{n'}(t) 
\eeq
This is called the Schr\"{o}dinger's equation 
in the "interaction picture". It is a convenient 
equation because the term on the right 
is assumed to be "small". Therefore, the amplitudes ${c_n(t)}$ change 
slowly. This equation can be solved using 
an iterative scheme which leads naturally 
to a perturbative expansion. The iteration are done 
with  the integral version of the above equation:
\beq
c_n(t)\ \ = \ \  c_n(0) \ - i \sum_{n'} \int_0^{t} 
\eexp{i E_{nn'} t'} \ V_{n,n'} \ c_{n'}(t') \ dt' 
\eeq
where ${ E_{nn'} = E_n-E_{n'} }$. In each iteration we get 
the next order. Let us assume that the system has 
been prepared in level $n_0$. This means that the zero 
order solution is 
\beq
c_n^{[0]}(t) \ \  = \ \  c_{n}(0) \ \ =\ \  \delta_{n,n_0 } 
\eeq
We iterate once and get the first-order solution:
\beq
c_n(t) \ \ = \ \ \delta_{n,n_0} \ - i\int_0^t \eexp{iE_{nn_0} t'} V_{n,n_0} dt'
\eeq

It is instructive to consider constant perturbation ($V$~does not depend on time). 
Doing the integral and going back to the standard representation 
we get the following result for the first-order transition amplitude:
 \beq 
\psi_n(t) \ \ = \ \ V_{n,n_0}\frac{\eexp{-iE_{n}t}-\eexp{-iE_{n_0}t}}{E_{n}-E_{n_0}}, 
\ \ \ \ \ \ \ \ \ \mbox{[constant perturbation, $n \neq n_0$]}
\eeq
We notice that for very short times we get ${\psi_n(t) \approx -iV_{n,n_0} t}$, 
which reflects the definition of the matrix elements of the Hamiltonian as 
the "hopping" amplitude per unit of time. As long as the 
energy difference ${(E_n-E_{n_0})}$ is not resolved, 
this expression is merely modulated  ${\psi_n(t) \approx [-iV_{n,n_0} t] \eexp{-iE_nt}}$,  
reflecting the choice of the energy reference.  
But for longer times the accumulation of the amplitude 
is suppressed due to the temporal oscillations of the integrand. 
If the perturbation is constant, the oscillation frequency 
of the integrand is ${\omega= (E_n-E_{n'})}$. 
If the perturbation is characterized by a frequency $\Omega$, 
this oscillation frequency becomes ${\nu = (E_n-E_{n'}-\Omega)}$.

In order to illustrate the effect of the oscillating integrand, consider 
a problem in which the unperturbed Hamiltonian is "to be in some site". 
We can regard the possibility to move from site to site as a perturbation. 
The energy differences in the site problem are the gradients 
of the potential energy, which we call ``electric field". 
The hopping amplitude is the same in each "hop" even if the hop 
into a potential wall. However, due to the potential difference 
the probability amplitude does not "accumulate" and therefore 
the particle is likely to be reflected. The reflection is not due 
to the lack of coupling between site, but due to the potential 
difference that induces ``fields" that suppresses the transition probability.

\sheadC{The transition probability formula}

The expression we found for the wavefunction amplitudes  
using first-order perturbation theory can be written as:
\beq
c_n(t) \ \approx \ 1  \hspace*{6cm}  &&  \mbox{for} \ n=n_0
\\ 
c_n(t) \  = \ -i W_{n,n_0} \int_0^t f(t') \eexp{i E_{nn_0} t'} dt' \hspace{2cm} && \mbox{otherwise} 
\eeq
The latter expression for the transition amplitude 
can be written optionally as a Fourier transform~(FT):  
\beq
c_n(t) \ \ \approx \ \  -iW_{n,n_0} \ \mbox{FT} \left[ f(t) \right] 
\eeq
We use here the convention that ${f(t)=0}$ before and after the pulse. 
For example, if we turn on a constant perturbation for a finite duration, 
then ${f(t)}$ is a rectangle function.
It is implicit that the FT is calculated at the 
frequency of the transition ${\omega=E_n-E_{n_0}}$, namely, 
\beq
\mbox{FT} \left[f(t)\right] \ \ = \ \ \int_{-\infty}^{\infty}f(t') \ \eexp{iE_{n,n_0} t'} \ dt' 
\eeq
The associated expression for the transition probability is:
\beq
P_t(n|n_0) \ \  \approx \ \  |W_{n,n_0}|^2 \ \times \ \Big| \mbox{FT} \left[ f(t) \right] \Big|^2 
\eeq

\sheadC{The effect of a constant or harmonic perturbation} 

We consider the following scenario: A particle is prepared 
in the state ${n_0}$, and then a constant perturbation is 
turned on for a time ${t}$. We want to know what is the 
probability of finding the particle at some later time in the state ${n}$.
Using the transition probability formula we get
\beq
P_t(n|n_0) 
\ \ = \ \  
|c_n(t)|^2 
\ \ = \ \  
|W_{nn_0}|^2 
\left| \frac{1-\eexp{i(E_n-E_{n_0})t}} {E_n-E_{n_0}} \right|^2 
\eeq
We notice that the transition amplitude is larger  
to closer levels and smaller for distant levels.

The above expression can be regarded as a special case 
of a more general result that concerns harmonic perturbation: 
\beq
f(t') \ \ =\ \ \eexp{-i\Omega t'} \ \ \ \ \ \ \ \ \  \mbox{for} \ \ \ t' \in [0,t] 
\eeq
Recall that the Hamiltonian should be hermitian.  
Therefore this perturbation has a physical meaning 
only if it appears together with a conjugate 
term ${\eexp{+i\Omega t'}}$. In other words, 
the driving is done by a real field ${2\cos(\Omega t')}$ 
that changes periodically. Below we treat only "half" 
of the perturbation: the effect of the second half 
is obtained by making the replacement ${\Omega \mapsto -\Omega}$.

Using the transition probability formula with finite $\Omega$ 
we get the same integral as for constant perturbation 
with  ${E_n-E_{n_0}}$ replaced by ${E_n-E_{n_0}-\Omega}$. 
Accordingly the result is  
\beq
P_t(n|n_0) 
\ \ &=& \ \  
|W_{nn_0}|^2 \left| 
\frac{1-\eexp{i(E_n-E_{n_0}-\Omega)t}} {E_n-E_{n_0}-\Omega} 
\right|^2 
\ \ = \ \ 
|W_{n,n_0}|^2 \frac{2[1-\cos(E_n-E_{n_0}-\Omega)t)]}{(E_n-E_{n_0}-\Omega)^2} 
\\
\ \ &=& \ \   
|W_{n,n_0}|^2 \frac {4 \sin^2( (E_n-E_{n_0})-\Omega)t/2) }{ (E_n-E_{n_0}-\Omega)^2 } 
\ \ = \ \ 
|W_{n,n_0}|^2 \, t^2 \, \sinc^2((E_n-E_{n_0}-\Omega)t/2) 
\eeq
Schematically we can write this result as follows:
\beq
P_t(n|n_ 0 )
\ \ =\ \  
2\pi \ |W_{n,n_0} |^2 \ \delta_{2\pi/t} (E_n-E_{n_0}-\Omega) \ \times \ t
\eeq
The schematic way of writing emphasizes that if~$t$ is large enough 
the $\sinc^2$ function becomes a narrow function of width $2\pi/t$
that resembles a delta function. Namely,    
\beq
&& \sinc(\nu) \ \ \equiv \ \ \frac{\sin(\nu)}{\nu}
\\ 
&& \int_{-\infty}^{\infty} 
\frac{d\nu}{2\pi}
\mbox{sinc}^2 \left( \frac{\nu}{2} \right)  
\ \ = \ \ 1 
\eeq

\sheadC{The Fermi golden rule (FGR)} 

The main transitions which are induced by a purely harmonic driving  
is to energy levels that obey the "resonance condition":
\beq
(E_n-E_{n_0}) \ \ \sim \ \ \Omega 
\eeq
From the expression we found, we see that 
the probability of transition to levels that obey 
${|E_n-(E_{n_0}+\Omega)| < 2\pi/t}$ 
is proportional to ${t^2}$. That is what 
we would expect to get by the definition 
of the Hamiltonian as the probability amplitude 
for transitions per unit time. But the "width" 
of the area that includes these levels is proportional 
to ${2\pi/t}$. From this we conclude that the 
probability of transition to other levels grows linearly. 
We call the rate of transition to other levels ${\Gamma}$.  
\beq
\Gamma \ \ = \ \  
\frac{2\pi}{\Delta} \, |W_{n,n_0}|^2 
\ \ = \ \  
2\pi \gdos(E) \ |W_{n,n_0}|^2 
\eeq
The formula can be proved by calculating 
the probability to stay in level ${n_0}$:
\beq
P(t) \ \ = \ \ 1-\sum_{n (\neq n_0)} P_t(n|n_0) 
\ \ = \ \ 1-\int \frac {dE}{\Delta} P_t(E|E_0) 
\ \ = \ \ 1-\frac{2\pi t}{\Delta}|W_{n,n_0}|^2 
\ \ = \ \ 1-\Gamma t 
\eeq
It is implicit in the above derivation that we assume 
a dense spectrum with well defined density of states. 
We also assume that the relevant matrix elements are 
all of the same order of magnitude.

Let us discuss the conditions for the validity 
of the Fermi golden rule picture. 
First-order perturbation theory is valid 
as long as  ${P(t) \approx 1}$, 
or equivalently ${\Gamma t \ll 1}$. 
Hence a relevant time scale is the Wigner time   
\beq
t_{\Gamma} \ \ = \ \ \frac{1}{\Gamma}
\eeq
Another important time scale that gets into 
the game is the Heisenberg time which is defined as: 
\beq
t_H =\frac {2\pi}{\Delta} 
\eeq
We distinguish below between the  
case of weak perturbation (${|W| \ll \Delta}$),
for which $t_H \ll t_{\Gamma}$  
from the case of strong perturbation (${|W| > \Delta}$), 
for which $t_H \gg t_{\Gamma}$.

In the case  ${|W| \ll \Delta}$ first-order perturbation theory 
is still valid when ${t=t_H}$. 
If perturbation theory is valid up to this time 
then it is valid at any time, since after 
the Heisenberg time the (small) probability that 
has moved from the initial state to other energy 
levels oscillates, and does not grow further. 
This argument is based on the assumption  
that the difference ${\nu = E_n-(E_{n_0} + \Omega)}$ 
is of the order ${\Delta}$, even for levels 
in the middle of the resonance. If there is 
an "exact" resonance, it is possible to show 
that the probability will oscillate between 
the two energy levels, as in the "two-site" 
problem, and there is no "cumulative" leakage 
of the probability to other levels.

In the case ${|W| > \Delta}$ first-order perturbation theory 
breaks down before the Heisenberg time. 
Then we must go to higher orders of perturbation theory. 
With some limitation we find the result:
\beq
P(t) \ \ =\ \  \eexp{-\Gamma t} 
\eeq
This means that there is a decay. In another  
lecture we analyze a simple model where we can 
get this result exactly. In the general case, 
this is an approximation that has, at best, limited validity. 
For an isolated system it can be justified on 
the basis of rezolvent theory (see separate lecture) 
assuming that a single pole is dominating. 
For a non-isolated system it can be justified on   
the basis of a Markovian picture.

Finally it should be clear that the FGR
is irrelevant if the time is very short. 
Let us assume that $\Delta_b$ is the width 
of the energy-band whose levels~$E_n$ are 
coupled by $W_{n,n_0}$ to the initial level $E_{n_0}$.  
Or optionally let us assume that the density 
of states has variation on some energy scale $\Delta_b$. 
We define an associate time scale 
\beq
t_c \ \ = \ \ \frac{2\pi}{\Delta_b}
\eeq
During the time ${t<t_c}$ the total transition 
probability is possibly growing like~$t^2$.
The FGR holds only after the bandwidth $\Delta_b$ 
is "resolved", which can be written as $t\gg t_c$. 
If the driving source is noisy (the "other" version of 
Fermi golden rule) one identifies $t_c=\tau_c$ 
as the correlation time of the noise. 
The bottom line is that the Fermi golden rule 
in any of its variations is likely to be 
applicable during a finite time interval
\beq
t_c \ \ \ll \ \  t \ \  \ll \ \ t_H
\eeq

\sheadB{Transition rates} 

In the previous subsection we have derived an expression 
for the transition probability if the driving is 
purely harmonic. In this section we assume 
a noisy driving. The ``noise"  is characterized by  
a correlation function 
\beq
\langle f(t)f(t') \rangle \ \ = \ \  F(t-t')  
\eeq
where the average is taken over realizations of $f(t)$.  
The Fourier transform of of the correlation function 
is the power spectrum  of the noise $\tilde{F}(\omega)$.
From the  transition probability formula we get 
\beq
P_t(n|m) \ \ = \ \ 
|W_{nm}|^2 \int_0^t\int_0^t  \langle f(t')f(t'')\rangle \ \eexp{iE_{nm}(t'-t'')} \ dt'dt''
\ \ \approx \ \ 
|W_{nm}|^2  \, \tilde{F}(\omega{=}E_{nm}) \times t
\eeq
where the approximation assumes that the duration of the noise 
is much larger compared with its correlation time~$\tau_c$.  
In practice we can distinguish 3 cases of interest: 
{\bf (a)}~White noise; 
{\bf (b)}~Low frequency noise; 
{\bf (c)}~Noisy periodic driving. 
In the first case $\tilde{F}(\omega)$ is flat and wide, 
so it is enough to characterize it by its spectral 
intensity ${\nu=\tilde{F}(\omega)}$, and to write the above 
result as ${P_t(n|m) = w_{nm}t}$, where the transition rates  
are ${w_{nm}=\nu|W_{nm}|^2}$. 
In the second case $\tilde{F}(\omega)$ is narrow, 
with some spectral width $2\pi/\tau_c$, that is determined 
by the correlation time $\tau_c$. 
We write ${F(0)= \langle f^2 \rangle \equiv \varepsilon^2}$
where $\varepsilon$ is the RMS intensity of the noise. 
Then it follows that its peak spectral intensity
is ${\nu=\tilde{F}(0)=\varepsilon^2 \tau_c}$. 
The third case refers to noisy harmonic driving 
that has a correlation time ${\tau_c \gg 2\pi/\Omega}$. 
In the latter case $\tilde{F}(\omega)$ is concentrated 
around the frequency~$\Omega$,  
and it is suggestive to re-write the expression for 
the transition rate schematically as follows:  
\beq
w_{nm}
\ \  = \ \ 
\varepsilon^2 \, |W_{n,m}|^2  \ \ 2\pi \delta_{2\pi/\tau_c}((E_{n}-E_{m})-\Omega) 
\eeq
Note that $\varepsilon$ can be absorbed into the definition of $W$. 
The above result for constant rate transition between levels  
is commonly regarded as a special version of the Fermi golden rule 
that we discussed in the previous lecture.

\sheadC{Master equations} 

The above version of the FGR is commonly used in order to determine 
the rates constants in Master equations that 
provide a reduced description for the dynamics   
of a system that is coupled to a bath or to a noise source. 
Here we demonstrates how a Master equation is derived 
for the simplest case: system that is subjected  
to the influence of a white noise source. 

Consider the Hamiltonian ${\mathcal{H}=\mathcal{H}_0+f(t)W}$, 
were  $f(t)$ represents white noise: that means that upon 
ensemble average ${\langle f(t)\rangle=0}$, 
while ${\langle f(t)f(t') \rangle = \nu\delta(t-t')}$. 
The Liouville von-Neumann equation for the time evolution 
of~$\rho$ can be solved iteratively in order 
to determine $\rho(t+dt)$ given $\rho(t)\equiv\rho$, where $dt$ 
is small time interval:
\beq
\rho(t+dt) \ \ = \ \ \rho -i\int_t^{t+dt} dt' \, [\mathcal{H}, \rho] 
-\int_t^{t+dt} \int_t^{t'}  dt'dt''  \, [\mathcal{H}, [\mathcal{H}, \rho]] + ... 
\eeq
Averaging over realizations of $f(t)$ all the odd orders in this expansion 
vanish, while the leading $dt$ contribution comes only from the 
zero order term that involves $\mathcal{H}_0$ 
and from the second order term that involves~$W$.
Consequently we get the following Master equation:
\beq 
\frac{d\rho}{dt} = -i[\mathcal{H}_0,\rho] - \frac{1}{2}\nu[W,[W,\rho]]
\eeq
In the most standard example $W=\hat{x}$, 
and the second term corresponds to the 
diffusion term ${(\nu/2)\partial^2\rho/\partial p^2}$ 
in the classical Fokker-Plank equation.

In the weak noise limit the rate of the noise induced 
transitions becomes much smaller compared with the Bloch 
frequency of the coherent transitions. 
Then we can ignore all the highly oscillating terms 
that involve the off-diagonal terms, because they average to zero. 
Consequently we get the so-called Pauli mater equation 
for the probabilities $p_n$  
\beq 
\frac{d\bm{p}}{dt} \ = \ \mathcal{W} \ \bm{p}
\ \ \ \ \ \ \ \ \ \ \ \ 
\mathcal{W} = \left( 
\amatrix{
-\Gamma_1 & w_{12} & ... \cr
w_{21} & -\Gamma_2 & ... \cr
... & ... & ...
} 
\right)
\eeq
In this equation the $\mathcal{W}$ matrix is in agreement 
with the FGR picture. Namely, the rate constants 
are $w_{nm} = \nu |W_{nm}|^2$,  
and the decay constants are $\Gamma_n=\sum_m w_{nm}$. 
We note that if we had interaction with a bath, 
then the ratio $w_{12}/w_{21}$ would not be unity, 
but would favor downwards transitions.

The noise induced ``diffusion" term in the master equation 
that we have derived is ${W\rho W-(1/2)(W W \rho + \rho W W)}$. 
We can regard it as a special case of the so-called Lindblad equation.
The evolution of a non-isolated system is further 
discussed under {\em Special topics} in the lecture 
regarding {\em Quantum states, operations and measurements}. 
The presentation of the Master Equation formalism for 
a system that is coupled to a bath has been differed to    
{\em Lecture Notes in Statistical Mechanics and Mesoscopic}, 
\href{http://arxiv.org/abs/1107.0568}{arXiv:1107.0568}

\newpage

\sheadB{The cross section in the Born approximation}

\sheadC{Cross Section} 

In both classical mechanics and quantum mechanics there 
are two types of problems: closed systems and open systems. 
We will discuss an open system. The dynamical problem 
that we will analyze is called a "scattering problem". 
For example, a wave that is scattered on a sphere. 
In a problem of this type the energy is given. We assume 
that there is an "incident particle flux" and we ask what 
is the "scattered flux". 

\begin{center}
\putgraph{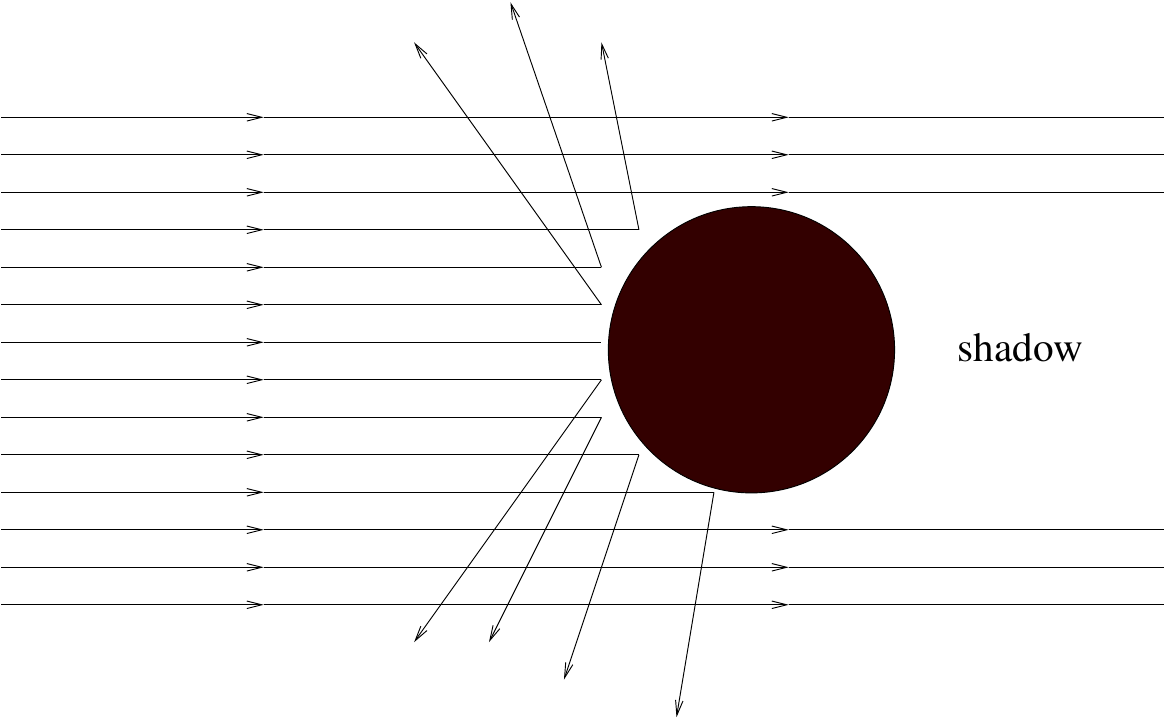} 
\end{center}

We notice that the sphere "hides" a certain area of the beam. The total 
hidden area is called the "total cross section"~${\sigma_{\text{total}}}$. 
Let us assume that we have a beam of particles with energy ${E}$ 
and velocity ${v_E}$, so that the current density is:
\beq
J \text{[particles/time/area]}  \ \ = \ \  \rho_0 v_E 
\eeq
where ${\rho_0}$ is the particle density. 
We write the scattered current as:
\beq
I_{\text{scattered}}  \ \ = \ \  [\sigma_{\text{total}}] \times J 
\eeq
where the cross section ${\sigma_{\text{total}}}$ is defined 
as the ratio of the scattered current ${I_{\text{scattered}}}$ 
to the incident particle flux density ${J}$. We notice that 
each area element of the sphere scatters to a different direction. 
Therefore, it is more interesting to talk about the differential 
cross section ${\sigma(\Omega) }$. In full 
analogy ${\sigma(\Omega)d\Omega}$ is defined by the formula:
\beq
I(k \in d\Omega | k_0)  \ \ = \ \  [\sigma(\Omega) d\Omega] \times J 
\eeq
where $k_0$ denotes the initial momentum, while $k$ is the final momentum.
Here only the scattering into the angular element ${d\Omega }$ is detected.

\sheadC{Cross section and rate of transition} 

For the theoretical discussion that will follow, 
it is convenient to think of the space as if it has a finite 
volume ${ L^3 = L_x L_y L_z}$ with periodic boundary 
conditions. In addition we assume that the "incident" 
beam takes up the whole volume. If we normalize the particle 
density according to the volume then ${\rho_0=1/L^3}$. 
With this normalization, the flux $J$ (particles per unit time) 
is actually the "probability current" (probability per unit time), 
and the current $I_{\text{scattered}}$ is in fact 
the scattering rate. Therefore an equivalent definition 
of the cross section is:
\beq
\Gamma( k \in d\Omega | k_0 )  \ \ = \ \  [\sigma(\Omega)d\Omega] \times \frac{1}{L^3}v_{\tbox{E}} 
\eeq
Given the scattering potential $U(r)$ we can calculate 
its Fourier transform ${\tilde{U}(q)}$
\beq
\tilde{U}(q)  \ \ = \ \  \text{FT} [U(r)]  
 \ \ = \ \  \int\int\int U(r) \eexp{-i\vec{q} \cdot \vec{r}} d^3r 
\eeq
Then we get from the Fermi golden rule (see derivation below) 
a formula for the differential cross section 
which is called the "Born approximation":
\beq
\sigma(\Omega)
\ \ = \ \ 
\frac{1}{ (2\pi)^2 } \, \left(\frac{k_E}{v_E}\right)^2 \, |\tilde{U}(\vec{k}_{\Omega}-\vec{k}_0)|^2 
\ \ = \ \ 
\left(\frac{\mass}{2\pi}\right)^2 \Big|\tilde{U}(\vec{k}_{\Omega}-\vec{k}_{0})\Big|^2 
\eeq
The second expression assumes the non-relativistic 
dispersion relation ${v_E=k_E/\mass}$.   
The Born approximation is a first-order 
perturbation theory approximation. 
It can be derived with higher order corrections  
within the framework of scattering theory.

\sheadC{The DOS for a free particle} 

In order to use the Fermi golden rule we need an expression 
for the density of states of a free particle. In the past 
we defined ${\gdos(E)dE}$ as the number of states 
with energy ${E<E_k<E+dE}$. But in order to calculate 
the differential cross section we need a refined definition:
\beq
\gdos(\Omega,E) d\Omega dE
 \ \  = \ \  
\text{Number of states} \ \vec{k} \in d\Omega  \ \mbox{with energy} \ E< E_k<E+dE  
\eeq
If we have a three-dimensional space with 
volume ${ L^3 = L_x L_y L_z }$ and periodic  
boundary conditions, then the momentum states are:
\beq
k_{n_x,n_y,n_z}
 \ \ = \ \  
\left( \frac{2\pi}{L_x}n_x, \frac{2\pi}{L_y}n_y, \frac{2\pi}{L_z}n_z\right) 
\eeq

\begin{center}
\putgraph[0.4\hsize]{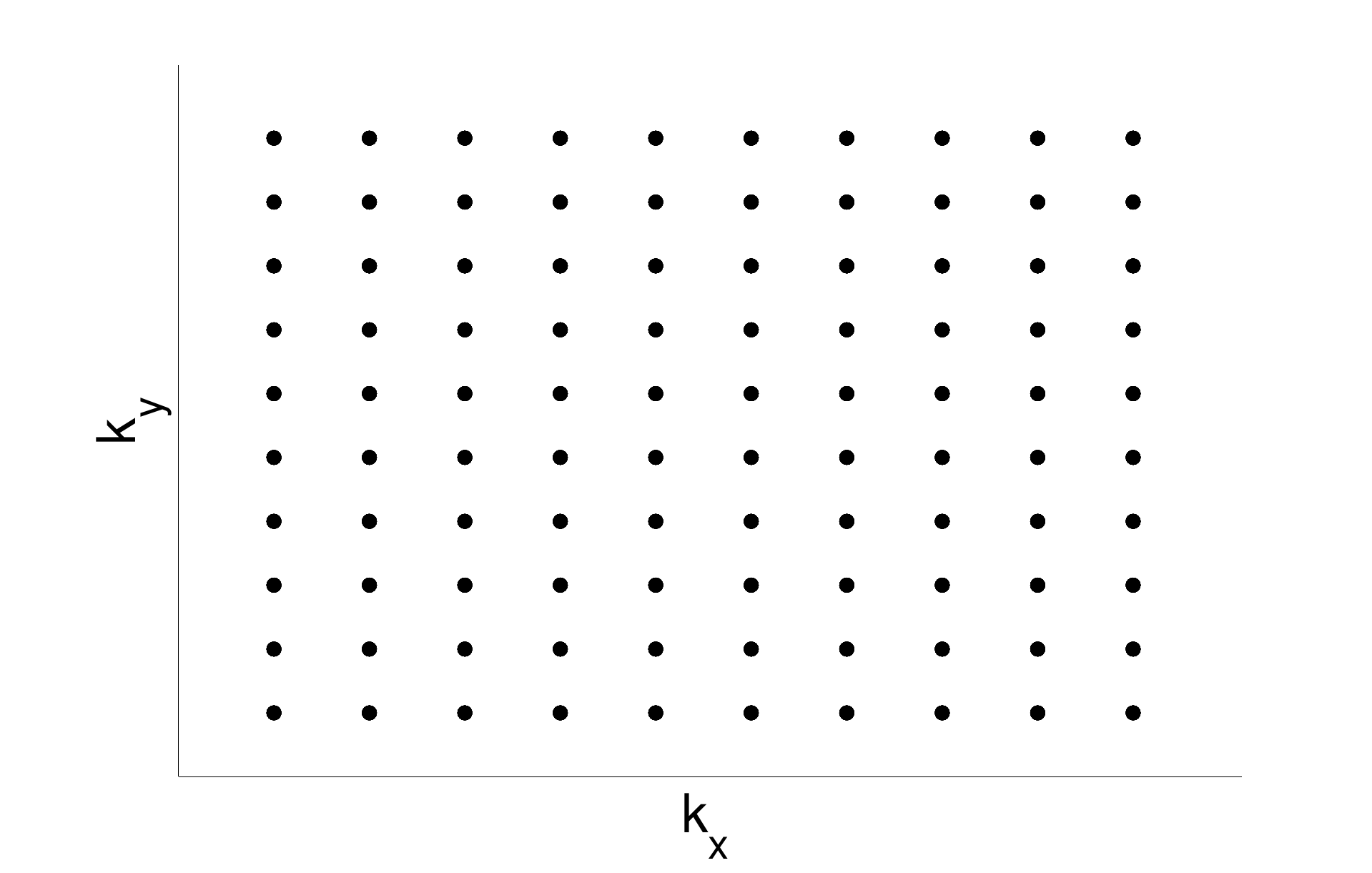}
\end{center}

The number of states with a momentum 
that in a specified region of $k$~space is:
\beq
\frac{ dk_x \, dk_y \, dk_z } 
{\frac{2\pi}{L_x} \, \frac{2\pi}{L_y} \, \frac{2\pi}{L_z}} 
 \ \ = \ \  
\frac{L^3}{(2\pi)^3} \, d^3k 
 \ \  =  \ \  
\frac{L^3}{(2\pi)^3} \, k^2 d\Omega \, dk
 \ \  =  \ \  
\frac{L^3}{(2\pi)^3} \, k_E^2 d\Omega \, \frac{dE}{v_E}
\eeq
where we have moved to spherical coordinates 
and used the relation ${dE = v_E dk}$. 
Therefore, we find the result:
\beq
\gdos(\Omega,E) d\Omega dE 
 \ \  = \ \  
\frac{L^3}{(2\pi)^3} \frac{k_E^2}{v_E} d\Omega dE 
\eeq

\newpage \sheadC{Derivation of the Born formula} 

Let us assume that we have a flux of particles that are moving 
in a box with periodic boundary conditions in the $z$~direction.  
As a result of the presence of the scatterer there are transitions  
to other momentum states (i.e. to other directions of motion). 
According to the Fermi golden rule the transition rate is:
\beq
\Gamma( k \in d\Omega | k_0 )  \ \ = \ \  2\pi \, [ \gdos(\Omega,E)d\Omega ] \, |U_{k_{\Omega},k_0}|^2 
\eeq
By comparing with the definition of a cross section we get the formula:
\beq
\sigma(\Omega)  \ \ = \ \  \frac{2\pi}{v_E} \, L^3 \, \gdos(\Omega,E) \, |U_{k_{\Omega},k_0}|^2 
\eeq
We notice that the matrix elements of the scattering potential are:
\beq
\langle\vec{k}|U(r)|\vec{k_0}\rangle 
 \ \ = \ \  
\int \frac{d^3x}{L^3} 
\, \eexp{-ik \cdot r}U(r) 
\, \eexp{ik_0 \cdot r} 
 \ \ = \ \  
\frac {1}{L^3} \int U(r) 
\, \eexp{-i(k-k_0) \cdot r} d^3r 
=\frac{1}{L^3}\tilde{U}(k-k_0) 
\eeq
By substituting this expression and using the result 
for the density of states we get the Born formula.

\sheadC{Scattering by a spherically symmetric potential} 

In order to use the Born formula in practice we define 
our system of coordinates as follows: the incident
wave propagates in the ${z}$ direction, and the 
scattering direction is ${\Omega = (\theta_{\Omega} , \varphi_{\Omega})}$. 
The difference between the ${k}$ of the scattered wave 
and the ${k_0}$ of the incident wave is ${\vec{q}=\vec{k}-\vec{k}_0}$. 
Next we have to calculate $\tilde{U}(q)$ which is the 
Fourier transform of ${U(r)}$.  
If the potential is spherically symmetric 
we can use a rotated coordinate system 
for the calculation of the Fourier transform integral. 
Namely, we can use spherical coordinates 
such that ${\theta=0}$ is the direction 
of ${\vec{q}}$. Consequently 
\beq
\tilde{U}(q)  \ \ = \ \  
\int\int\int U(r) \eexp{-iqr \cos(\theta)} d\varphi d\cos(\theta) r^2 dr 
 \ \ = \ \  4\pi \int_{0}^{\infty} U(r) \, \text{sinc}(qr) \, r^2 dr 
\eeq
where the angular integration has been done using 
\beq
\int_{-1}^{1} \eexp{-i \lambda s}ds 
= \left[ \frac{\eexp{-i \lambda s}}{-i\lambda}\right]_{-1}^{1} 
= \frac{\eexp{i\lambda}-\eexp{-i\lambda}}{i\lambda} 
= \frac{2\sin(\lambda)}{\lambda} 
= 2\text{sinc}(\lambda) 
\eeq
As an example consider $U(r)=1/r$, for which we get $\tilde{U}(q)=4\pi/q^2$ (Rutherford formula). \\
Next, we can go on calculating the total cross section: 
\beq
\sigma_{\text{total}} 
 \ \ = \ \  
\int\int\sigma(\Omega)d\Omega 
 \ \ = \ \  
\frac{1}{(2\pi)^2} 
\left(\frac{k_E}{v_E} \right)^2 
\int_{0}^{\pi} |\tilde{U}(q)|^2 \, 2\pi \sin{\theta_{\Omega}}  d\theta_{\Omega}
\eeq
We note that by simple trigonometry:
\beq
q = 2 k_E \, \sin\left(\frac{\theta_{\Omega}}{2}\right) 
\ \ \ \ \ \ \ \ \leadsto \ \ \ \ \ \ \ \ 
dq = k_E \, \cos\left(\frac{\theta_{\Omega}}{2} \right) d\theta_{\Omega} 
\ \ \ \ \ \ \ \ \leadsto \ \ \ \ \ \ \ \ 
\sin{\theta_{\Omega}}  d\theta_{\Omega} = \frac{qdq}{k_E^2}
\eeq
Hence we can write the integral of the cross section as:
\beq
\sigma_{\text{total}} 
 \ \ = \ \  
\frac{1}{2\pi v_{E}^2} \int_{0}^{2k_E} |\tilde U(q)|^2qdq 
\eeq

\newpage

\sheadB{Dynamics in the adiabatic picture}

\sheadC{The notion of adiabaticity} 

Consider a particle in a one dimensional box 
with infinite walls. We now move the wall. 
What happens to the particle? 
Let us assume that the particle has been prepared 
in a certain level. It turns out that if the wall 
is displaced slowly, the particle stays in the same level. 
This is called the "adiabatic approximation".
We notice that staying in the same energy level means 
that the state of the particle changes! 
If the wall is moved very fast 
then the state of the particle does 
not have time to change. This is called 
the "sudden approximation". In the latter case 
the final state (after the displacement of the wall) 
is not an eigenstate of the (new) Hamiltonian. 
After a sudden displacement of the wall, 
the particle has to "ergodize" its state inside the box. 

The fact that the energy of the particle decreases 
when we move the wall outwards, means that the particle 
is doing work. If the wall is displaced adiabatically, 
and then displaced back to its original location, 
then there is no net work done. In such case 
we say that the process is reversible. 
But if the displacement of the wall is not adiabatically slow, 
the particle makes transitions to other energy levels. 
This scattering to other energy levels is in general irreversible. 

In the problem that we have considered above, 
the parameter $X$ that we change is the length $L$ 
of the box. Therefore ${V=\dot{X}}$ 
is the velocity at which the wall (or the "piston") 
is displaced. In other problems ${X}$ could 
be any field. An important example is 
a particle in a ring where ${X=\Phi}$ is the 
magnetic flux through the ring, 
and ${\mbox{EMF}=-\dot{X}}$ is the electro motive force 
(by Faraday law). In problems of this type, the 
change in the parameter ${X}$ can be very large, 
so we cannot use standard perturbation theory to analyze 
the evolution in time. Therefore, we would like to 
find another way to write Schr\"{o}dinger's equation, 
so that ${\dot{X}}$ is the small parameter.

\sheadC{The Schr\"{o}dinger equation in the adiabatic basis} 

We assume that we have Hamiltonian $\mathcal{H}(\hat{Q},\hat{P};X)$ 
that depends on a parameter~$X$. The adiabatic states 
are the eigenstates of the instantaneous Hamiltonian:
\beq
\mathcal{H}(X) \ \left|n(X) \right\rangle \ \ = \ \ E_n(X) \ \left|n(X) \right\rangle 
\eeq
It is natural in such problems to work with the adiabatic 
basis and not with a fixed basis. 
We write the state of the system as:
\beq
\left|\Psi \right\rangle \ \ = \ \ \sum_n a_n(t) \left|n(X(t))\right\rangle, 
\hspace{2cm} a_n(t) \equiv \left\langle n(X(t))|\Psi(t)\right\rangle
\eeq
If we prepare the particle in the energy level~$n_0$ 
and change~$X$ in adiabatically, 
then we shall see that ${|a_n(t)|^2 \approx \delta_{n,n_0}}$ at later times. 
We shall find a slowness condition for the validity of this approximation. 
Our starting point is the Schr\"{o}dinger's equation  
\beq
\frac{d\Psi}{dt} \ \ = \ \ -i\mathcal{H}(x,p;X(t)) \ \Psi
\eeq
from which we get:
\beq
\frac{d a_n}{dt} 
\ \ &=& \ \ 
\left\langle n \Big| \frac{d}{dt} \Psi\right\rangle 
+ \left\langle\frac{d}{dt}n \Big| \Psi\right\rangle 
\ \ = \ \ 
-i\left\langle n\Big|\mathcal{H} \psi \right\rangle 
+ \sum_m\left\langle\frac{d}{dt}n\Big|m\right\rangle \left\langle m \Big| \Psi \right\rangle 
\\
\ \ &=& \ \ 
-i E_n a_n + \dot{X}\sum_m\left\langle \frac{\partial}{\partial X}n\Big|m\right\rangle a_m 
\ \ \equiv \ \ 
-i E_n a_n + i \dot{X}\sum_m A_{nm} a_m  
\eeq
We conclude that in the adiabatic basis the effective Hamiltonian 
acquires an additional term, 
namely ${\mathcal{H} \mapsto \mathcal{H} - \dot{X} A}$. 
As an example consider the case where $X$ is the position 
of a box that confine a particle, 
then we get in the moving frame ${\tilde{H} = \mathcal{H} - \dot{X} \hat{p}}$.  
See additional remark at the conclusion of the next sub-section.

\sheadC{The calculation of ${A_{nm}}$}

Before we make further progress we would like 
to dwell on the calculation of the perturbation 
matrix~${A_{nm}}$. First of all we notice that for any $X$ 
\beq
\left\langle n\Big|m\right\rangle=\delta_{nm}
\ \ \ \ \ \leadsto \ \ \ \ \ 
\frac{\partial}{\partial X}\left\langle n\Big|m\right\rangle=0
\ \ \ \ \ \leadsto \ \ \ \ \ 
\left\langle\frac{\partial}{\partial X}n\Big|m\right\rangle+\left\langle n\Big|\frac{\partial}{\partial X}m\right\rangle=0 
\eeq
Which means that ${\left\langle \partial n|m \right\rangle}$ is anti-Hermitian, 
and therefore ${-i\left\langle \partial n |m \right\rangle}$ is Hermitian. 
If follows that 
\beq
A_{nm} \ \ = \ \ -i \Braket{\frac{\partial}{\partial X}n}{m}  
\ \ = \ \ i \Braket{n}{\frac{\partial}{\partial X}m} 
\ \ = \ \ i \BraKet{n}{T^{\dag} \frac{\partial T}{\partial X}}{m}
\eeq
where $T(X)$ is the transformation to the moving frame, 
namely, ${\ket{n(X)} = T(X)\ket{n(0)}}$.  
For the diagonal elements we use the notation
\beq
A_n(X) \ \ = \ \ A_{nn} \ \ = \ \  i \left\langle n\Big|\frac{\partial}{\partial X}n \right\rangle
\eeq
With regard  to the off-diagonal elements one observes that 
\beq
A_{nm} \ =  \ \frac{-i V_{nm}}{E_n-E_m}, 
\hspace{2cm} 
V_{nm} \ \equiv \ \left(\frac{\partial \mathcal{H}}{\partial X} \right)_{nm}
\eeq
This is a very practical formula. Its proof is based on the following:
\beq
&& \left\langle n\Big|\mathcal{H}\Big|m \right\rangle =0 
\ \ \ \ \mbox{for $n\ne m$, for any $X$}
\\ \nonumber
\leadsto && 
\frac{\partial}{\partial X} \left\langle n\Big|\mathcal{H}\Big|m \right\rangle =0
\\ \nonumber
\leadsto && 
\left\langle \frac{\partial}{\partial X}n\Big|\mathcal{H}\Big|m \right\rangle 
+ \left\langle n\Big|\frac{\partial}{\partial X}\mathcal{H}\Big|m \right\rangle 
+ \left\langle n\Big|\mathcal{H}\Big|\frac{\partial}{\partial X}m \right\rangle 
=0 
\\ \nonumber
\leadsto && 
E_m \left\langle \frac{\partial}{\partial X}n\Big|m \right\rangle 
+V_{nm}
+E_n \left\langle n\Big|\frac{\partial}{\partial X}m \right\rangle
=0 
\eeq

It should be noticed that the Hamiltonian in the adibatic basis
can be regarded as the transformed Hamiltonian in a moving frame. 
Namely we have in fact re-derived the following result:
\beq
\tilde{\mathcal{H}} \ \ = \ \ T^{\dag}\mathcal{H} T \ - i T^{\dag}\frac{\partial T}{\partial t}
\eeq  
See lecture ``Transformations and invariance" for further discussion.  
The second term (our "$A$") is simply the generator of the transformation 
to the moving frame.

\sheadC{The adiabatic condition}

We found that in the adiabatic basis 
the Schr\"{o}dinger's equation takes the form
\beq
\frac{d a_n}{dt} 
\ \ = \ \ -i E_n a_n \ + \ i \dot{X}\sum_m A_{nm} a_m 
 \ \ \equiv \ \ -i(E_n-\dot{X}A_n)a_n \ -i\sum_m W_{nm} a_m 
\eeq
For sake of analysis we separated above  
the diagonal part of the perturbation.
We use the simplified notation $A_n=A_{nn}$, 
and packed the off-diagonal elements 
into the matrix $W_{nm}=-\dot{X}A_{nm}$,
whose diagonal elements are zero. 
It should be noticed that the strength of the perturbation in this 
representation is determined by the rate $\dot{X}$ and not by 
the amplitude of the driving. 

Let us write again the Schrodinger equation 
in the adiabatic basis, with the substitution 
of the explicit expression that we have derived 
for the perturbation $W_{nm}$,  
\beq
\frac{d a_n}{dt} \ \ = \ \ 
-i(E_n-\dot{X}A_n)a_n 
\ + \ 
\dot{X} \sum_{m (\ne n)} \left(\frac{V_{nm}}{E_n-E_m}\right) \ a_m
\eeq
If $\dot{X}$ is small enough,  
the perturbation matrix $W$ is not able to induce 
transitions between levels, 
and we get the adiabatic 
approximation $|a_n(t)|^2 \approx \const$.
This means that the probability distribution 
does not change with time. In particular, 
if the particle is prepared in level $n$, 
then is stays in this level all the time.

From the discussion of first-order 
perturbation theory we know that we can neglect the coupling 
between two different energy levels if the absolute value of 
the matrix element is smaller compared with the 
energy difference between the levels. 
Assuming that all the matrix elements are 
comparable the main danger to the adiabaticity 
are transitions to neighboring levels.  
Therefore the adiabatic condition is ${|W| \ll \Delta}$ or 
\beq
\dot{X} \ll \frac{\Delta^2}{\hbar \sigma} 
\eeq
where $\sigma$ is the estimate for the 
matrix element $V_{nm}$ that couples 
neighbouring levels.

An example is in order. Consider a particle in a box of length ${L}$. 
The wall is displaced at a velocity ${\dot{X}}$. 
Given that the energy of the particle is~$E$ we recall that 
the energy level spacing is ${\Delta = (\pi/L)v_E }$,   
while the coupling of neighbouring levels, based on a formula 
that we derive in the \hyperref[sWalls]{Hard Walls} subsection (p.\pageref{sWalls}), is 
\beq
\sigma \ \ = \ \ \frac{1}{\mass L} k_n^2 
\ \ = \ \ \frac{1}{\mass L} (\mass v_E)^2
\ \ = \ \ \frac{1}{L} \mass v_E^2
\eeq
It follows that the adiabatic condition is 
\beq
\dot{X} \ \ \ll \ \ \frac{\hbar}{\mass L} 
\eeq
Note that the result does not depend on~$E$. 
This is not the typical case. In typical cases 
the density of states increases with energy, 
and consequently it becomes more difficult 
to satisfy the adiabatic condition.  

It is worth pointing out that in the classical framework 
the adiabatic condition for a moving wall is ${\dot{X} \ll v_E}$.
If the classical condition is violated the sudden approximation 
applies. If the classical condition is obeyed, but not the 
quantum adiabatic condition, the dynamics is semi-classical.
Namely, in each collision with the moving wall 
the change of energy is $\Delta_{cl}=2\mass v_E \dot{X}$.
In the  semi-classical regime the energy changes 
in steps $\Delta_{cl}$ that are larger than~$\Delta$.
This gives just the opposite condition to quantum adibaticity.
For more details see [\href{http://arxiv.org/abs/cond-mat/0605591}{arXiv:cond-mat/0605591}].

\sheadC{The zero order adiabatic approximation}

Assuming we can ignore the coupling between 
different levels, the adiabatic equation becomes
\beq
\frac{d a_n}{dt} \ \ = \ \ -i( E_n -\dot{X} A_n)a_n 
\eeq
And its solution is:
\beq
a_n(t) \ \ = \ \  \eexp{i\Phi_n(t)} \ a_n(0) 
\eeq
where the accumulated phase is 
\beq
\Phi_n(t) 
\ \  = \ \  
\int_0^t \left( -E_n + \dot{X}A_n \right) dt' 
\ \ = \ \  
-\int_0^t E_n dt' \ + \ \int_{X(0)}^{X(t)} A_n(X')dX' 
\eeq
As already observed the probability ${|a_n(t)|^2 }$ to be in a specific 
energy level does not change in time. This is the adiabatic 
approximation. But it is interesting to look at the phase 
that the particle accumulates. Apart from the dynamical phase, 
the particle also accumulates a geometrical phase. 
An interesting case is when we several parameters in a cyclic manner. 
In this case, just as in the Aharonov-Bohm effect, 
the ``geometrical phase" is regarded as ``topological phase" 
or ``Berry phase" 
\beq
\Phi_{\text{Berry}}  \ \ \equiv \ \ \oint A_n(X) \, dX 
\eeq
In fact, the Aharonov-Bohm effect can be viewed as 
a special case of the topological effect that 
has been explained above. 
In order to discuss further topological effects 
we have to generalize the derivation of the adiabatic equation.  
This will be done in the next lecture.

\sheadC{Beyond the adiabatic approximation}

Here we assume that we deal with one-parameter driving, 
accordingly the $A_n$ can be gauged away, 
and the~$\Phi_n$ is simply the integral over~$E_n$.   
We can handle the adiabatic Hamiltonian within the 
framework of an ``interaction picture", meaning that 
we substitute 
\beq
a_{n}(t) \ \ = \ \ c_{n}(t) \ \eexp{i\Phi_n(t)}     
\eeq
hence getting the equation
\beq
\frac{d c_n}{dt} \ \ = \ \ 
\dot{X} \sum_{m (\ne n)}  \frac{V_{nm}}{E_n-E_m}\eexp{-i(\Phi_n-\Phi_m)}  \ c_m 
\eeq
Note the implicit time dependence of the energies and the matrix elements, 
due to their parametric dependence on ${X=X(t)}$.
The leading order evaluation of the 
non-adiabatic transition probability is 
\beq
P_t(n|m) \ \ = \ \ \left|\int_0^{t} 
\frac{V_{nm}}{E_n-E_m}\eexp{-i(\Phi_n-\Phi_m)}   
\ \dot{X} dt\right|^2
\eeq
It is illuminating to consider 
a constant rate process ${\dot{X}=\const \ ({>}0)}$ 
and use $dX$ rather than $dt$ integration.
Then the above expression takes the following 
schematic form:
\beq
P_t \ \ = \ \ \left|\int_{X_1}^{X_2} 
\frac{dX}{g(X)} \ \exp\left[-\frac{i}{\dot{X}} \phi(X) \right] \right|^2
\ \ \sim \ \ \eexp{-\const/\dot{X}}
\eeq
The approximation is based on the assumption that the integral 
is dominated by a single complex pole where ${g(X)=0}$.
For the simplest example of such calculation 
see the \hyperref[sZener]{Landau-Zener dynamics} subsection, 
where we present the analysis of the non-adiabatic two-level crossing.

\newpage

\sheadB{The Berry phase and adiabatic transport}

In the following lectures we discuss the adiabatic formalism 
and the linear-response (Kubo) theory adopting the presentation 
and the notations as in [\href{http://arxiv.org/abs/cond-mat/0307619}{cond-mat/0307619}]. 
Note that unlike most textbooks we write the Schr\"{o}dinger equation
in the adiabatic representation without switching to the ``interaction picture".

\sheadC{Geometric phase}
 
Consider a parametric cycle in Hilbert space ${C(t)=\ket{\psi(t)}\!\bra{\psi(t)}}$, where $t$ is parameter, 
and it is assumed that the final "state" is the same as the initial "state". The question how the cycle is 
realized is irrelevant, hence an Hamiltonian is not specified. We can associate with the cycle 
so-called geometric phase as follows:
\beq
\gamma[C] \ \ = \ \ i\oint \Braket{\psi(t)}{\partial_t \psi(t)} dt \ \ \equiv \ \ \oint A_{\psi} dt  
\eeq
This is the "phase" that is accumulated by $\psi(t)$ during the cycle.     
The definition is non-ambiguous because it is invariant under gauge. 
Namely, if $\psi(t)$ is replaced by $\eexp{-i\Lambda(t)}\psi(t)$ 
we get the same result. In is also independent of the $t$-parametrization.  
If $\psi(t)$ is an eigenstate of some parametric Hamiltonian $\mathcal{H}(t)$, 
then $\gamma$ is called "Berry phase", and $A_{\psi}$ is known as the "Berry connection".

\sheadC{Parallel transport perspective}
 
Let us assume that we want to perform "parallel transport" of an arrow (or of a spin) upon Earth. The coordinates of the manifold are ${r=(\theta,\varphi)}$, and the trajectory along which we go is ${r(t)}$. We can define a simple-minded parallel transport, such that the arrow keeps the same orientation with respect to a fixed reference frame. But we are interested in a different notion of "parallel transport" where an "up" arrow with respect to the surface of Earth remains "up", while a "down" arrow remains "down". We shall see that the two notions of "parallel transport" formally correspond to the "sudden limit" and to the "adiabatic limit" of quantum-mechanical processes. Most importantly, we shall see that the "Berry phase" naturally appears in the analysis of "parallel transport" irrespective of the quantum-mechanical context: it is merely a geometrical phase that is implied by the topology of the manifold.   

There are several equivalent methods to visualize "parallel transport" on a sphere. Some methods (a,b) assume that the 2D manifold is embedded in a 3D Euclidean space, while the other methods (c,d,e) relay on the intrinsic topology of the surface. In options (a-c) we consider an SO(3) arrow. The transport of its perpendicular component is trivial, hence without loss of generality we consider arrows that are tangent to the surface. One find out that for a closed-cycle the rotation angle of a tangent vector equals the solid angle $\Omega$ of the surface-region that is encircled. 
In method (e) we consider an SU(2) spin, whose orientation is a superposition of "up" and "down" states. In the latter case the same rotation angle ($\Omega$) is deduced from the "Berry phases" of the "up" and "down" polarizations. Below we provide details on methods (a-e).

\hspace*{1cm}
\includegraphics[height=4cm]{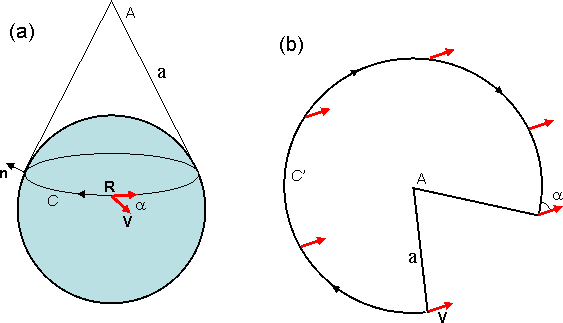}
\hspace*{2cm}
\includegraphics[height=4cm]{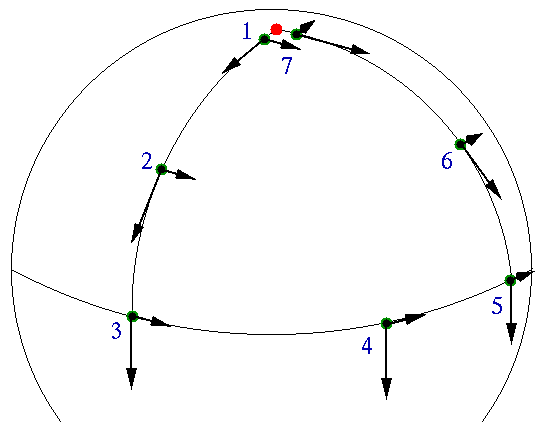}

{\bf (a) Projection method.-- } 
We divide the trajectory $r(t)$ into small infinitesimal steps. In each step the arrow is parallel transported in the simple-minded Euclidean sense, but then it is projected back to the tangent direction. Note that the shortening of the length is second order in the step size, hence the arrows keeps its length in the limit of a continuous process. 

{\bf (b) Tangent plane method.-- }  At each point along the trajectory ${r(t)}$ we attach a flat piece of tangent plane. For example, if we move along a line of constant latitude the union of all these pieces is a conical surface. This surface can be regarded as flat, and we can perform on it an Euclidean "parallel transport". See left panels of the figure [\href{http://www.princeton.edu/~npo/SurveyTopics/Berry_examples_files/Berryphase.html}{N.P.Ong website}].

{\bf (c) Geodesic walk method.-- } 
We can always approximated $r(t)$ as a sum of infinitesimal geodesic steps. In the case of a sphere "geodesic" means to go along a big circle. The law of "parallel transport" is simple: the arrow should keep fixed angle with the geodesic. See right panel of the figure [\href{http://www.mi.infm.it/manini/berryphase.html}{N.Manini website}].

{\bf (d) Metric connection method.-- }  
To each point ${\bm{r}=(x_1,x_2)}$ on the surface we attach a set of "unit vectors" with which a metric is associated:
\beq
\ora{e}_n = \partial_n \bm{r} = \fracd{\bm{r}}{x_n}, 
\hspace*{3cm}
g_{nm} = \Braket{\ora{e}_n}{\ora{e}_m} =  \Braket{\partial_n \bm{r}}{\partial_m \bm{r}}  
\eeq 
Then we can define so called {\em "connection"}, that can be derived from the metric:  
\beq
\Gamma_{n,km} \ = \  \Braket{\ora{e}_n}{\partial_k \ora{e}_m} 
\ = \ \Braket{\partial_n \bm{r}}{ \partial_k \partial_m\bm{r}}
\ = \ \frac{1}{2}\left[\partial_k g_{nm} + \partial_m g_{kn} - \partial_n g_{km} \right]
\eeq
The metric connection allows us to calculate full derivatives: 
\beq
\ket{v} = \sum_n v_n \ket{\ora{e}_n}, 
\hspace*{3cm}
D_k\ket{v} = \sum_n  \left[ \partial_k v_n + \sum_{m} \Gamma_{n,km}v_m \right] \ \ket{\ora{e}_n} 
\eeq   
The construct ${D=\partial + \Gamma}$ is known as the "covariant derivative". 
It takes into account the explicit variation of the vector-components 
as well as the implicit parametric variation of the basis-vectors. 
We say that the vector $v$ is parallel transported if ${Dv=0}$.  
Accordingly $\Gamma$ can be regarded as the generator of an orthogonal transformation
that preservers the scalar product. 

{\bf (e) Hermitian connection method.-- } 
We can repeat the formal derivation of the parallel transport law, considering transport of  SU(2) spins instead of SO(3) arrows.
To each point ${\bm{r}=(x_1,x_2)}$ on the surface we attach a set basis vectors $\ket{n(\bm{r})}$. 
For example:    
\beq
\hspace*{2cm}
\ket{\uparrow(\theta,\varphi)} = \left(\amatrix{\cos(\theta/2) \\ \eexp{i\varphi}\sin(\theta/2)}\right),
\hspace*{2cm}
\ket{\downarrow(\theta,\varphi)} = \left(\amatrix{\sin(\theta/2) \\ -\eexp{i\varphi}\cos(\theta/2)}\right)
\eeq
Then we can define an Hermitian connection, and an associate ”covariant derivative” as follows: 
\beq
D_k\text{[simple minded parallel transport]} 
\ \ = \ \ \partial_k + \Braket{n(\bm{r})}{\partial_k m(\bm{r})} 
\ \ \equiv \ \ \partial_k -iA_{nm}^k  
\eeq
The requirement $D\psi=0$ defines a simple-minded parallel transport, 
corresponding to the sudden-approximation in the quantum-dynamics context. 
We would like to define a different notion of "parallel transport" 
that respects "up" and "down" relative to the surface. 
Therefore the "connection" in the definition of the covariant derivative
should be modified. In a fixed-basis frame we can write ${D=\partial-i\Gamma}$, 
while in the $n$-basis frame it should transforms into a diagonal matrix 
\beq
D_k\text{[parametric parallel transport]} 
\ \ = \ \ \partial_k -i \left[\Gamma_{nm}^{k}+A^k_{nm}\right] 
\ \ \equiv \ \  \partial_k - i\tilde{A}_{nm}^k
\eeq
where $\tilde{A}=\text{diag}\{A_n\}$ is a diagonal matrix in the $n$-representation.
In the concluding paragraph of this section we argue 
that $A_n$ should be identified as the Berry connection, 
hence ${\tilde{A}_{nm}^k=\delta_{n,m}A^k_{nm}}$.   
Once this is established it follows automatically that 
for a cyclic parallel transport process ${C_n(t)=\ket{n(t)}\!\bra{n(t)}}$, 
the transported $\ket{n}$ accumulates a phase that equals ${\gamma_n=\gamma[C_n]}$.  
Consequently, if we start at ${r_0}$ and go along a trajectory $r(t)$ back to ${r_0}$, 
then a tangent spin will rotate as follows:
\beq
\ket{\uparrow(r_0)}+\ket{\downarrow(r_0)} 
\ \ \  \leadsto \ \ \ 
\eexp{i\gamma_{\uparrow}[r]}\ket{\uparrow(r_0)}+\eexp{i\gamma_{\downarrow}[r]}\ket{\downarrow(r_0)}
\hspace*{1cm}
\text{RotationAngle} = - (\gamma_{\uparrow}-\gamma_{\downarrow}) 
\eeq
Specifically if we go along the constant $\theta$ latitude line, 
the connection is ${A_{\uparrow}^{\varphi}=-[\sin(\theta/2)]^2}$, 
and one obtains ${\gamma_{\uparrow}=-\Omega/2}$, where ${\Omega=2\pi[1{-}\cos(\theta)]}$.   
Similarly for a "down" spin  one obtains ${\gamma_{\downarrow}=+\Omega/2}$. 
Hence the expected result for the rotation angle is recovered.

{\bf Natural hermitian connection.-- } 
To deduce the connection for an SO(3) arrow given the metric of the manifold is easy.
To deduce the connection for an SU(2) spin requires further reasoning. 
What we want is to have a  parallel transport law that preserves $\ket{n}\!\bra{n}$. 
This means that "up" remains "up" and "down" remains "down". 
Namely ${\ket{\psi(t)} = \eexp{i\gamma_n(t)} \ket{n(t)}}$ should be qualified 
as a parallel-transport process, where $\gamma_n(t)$ still has to be determined. 
From such transport law it follows that $A_n=\partial_t\gamma_n$. 
We would like to argue that the proper determination of $\gamma_n(t)$ implies 
that $A_n$ is the Berry connection.  

At first sight it seems that the problem is ill-defined, and that $\gamma_n(t)$ 
can be chosen in an arbitrary way. The naive prescription would be ${\gamma_n(t)=0}$. 
Clearly such prescription is gauge dependent: 
if we define a different basis ${ \ket{\tilde{n}(r)}=\eexp{-i\Lambda_n(r)} \ket{n(r)} }$ 
we get a different rotation angle for a tangent vector.    
Even if we allow such gauge-dependent law of transport, still there is a problem. 
For a manifold with non-trivial topology (here it is a sphere) 
there is no way to have a well-defined gauge globally.
With our choice of gauge the $n$~states at the SouthPole remains ambiguous.  
Any gauge requires the exclusion of at least one point from the sphere, 
which is like saying that we are not dealing with a sphere...

The remedy for all this gauge-related complications is to define 
a parallel-transport law that does not depend on the gauge.
Saying it differently, we claim that there is a natural way to define a gauge-invariant connection, 
such that "up" remains "up" and "down" remains "down".  
We first note that a change ${\ket{\delta\psi} = \ket{\psi(t{+}dt)}- \ket{\psi(t)}}$
can be decomposed into "parallel" and "transverse" components as follows:
\beq
\ket{\delta\psi}_{\parallel}  \ \ &=& \ \ \ket{\psi}\Braket{\psi}{\delta \psi} \ \ = \ \ -iA_{\psi} \ket{\psi} \\  
\ket{\delta \psi}_{\perp} \ \ &=& \ \ \ket{\delta \psi} -  \ket{\psi}\Braket{\psi}{\delta \psi} \ \ = \ \ \ket{\delta \psi} + iA_{\psi} \ket{\psi}
\eeq  
Inspired by the ``projection method" for constructing a parallel-transport process,   
the requirement for parallel transport becomes ${\ket{\delta\psi}_{\parallel}=0}$.
Note that in the SO(3) treatment the projection was onto the tangent direction, 
while here we are keeping the spin in the perpendicular ("parallel") direction.   
We realize that $\ket{n(t)}$ is not a proper parallel transport process. 
The remedy is to consider ${\ket{\psi(t)} = \eexp{i\gamma_n(t)} \ket{n(t)}}$
with ${\partial_t\gamma_n = A_n(t)}$, which is a gauge-invariant prescription.   
This implies that $\tilde{A}=\text{diag}\{A_n\}$ is the hermitian connection.

\sheadC{The Berry curvature}
 
We consider a set of basis states $\bra{n(x)}$ that depend 
on a set of parameters ${x=(x_1,x_2,x_3,...)}$.    
In the previous section we have motivated the definition
of the hermitian connection matrix 
\beq
A^{j}_{nm}(x) \ \ = \ \ 
i\hbar \left\langle n(x) \Big|
\frac{\partial}{\partial x_j} m(x)\right\rangle
\eeq
The hermiticity  of the matrix can be deduced via differentiation by parts 
of $\partial_j \langle n(x)|m(x) \rangle  = 0$.
Furthermore, from differentiation by parts of $\partial_j \langle n(x)|{\mathcal{H}}|m(x) \rangle  = 0$,
we get for $n\ne m$ the following expressions:
\beq
A^{j}_{nm}(x) \ \ = \ \ \frac{i\hbar}{E_m - E_n}
\left\langle n \left|\frac{\partial {\mathcal{H}}}{\partial x_j}\right|m\right\rangle
\ \ \equiv \ \ -\frac{i\hbar \mathcal{F}^j_{nm}}{E_m - E_n}
\eeq
It should be realized that $A^{j}_{nm}$ is not gauge invariant.  
The effect of gauge transformation is
\beq
|n(x)\rangle \ & \ \mapsto \ & \ \eexp{-i\frac{\Lambda_n(x)}{\hbar}} \ |n(x)\rangle 
\\ \nonumber
A^{j}_{nm} \ & \ \mapsto \ & \
\eexp{i\frac{\Lambda_n{-}\Lambda_m}{\hbar}} A^{j}_{nm}
+ (\partial_j \Lambda_n)\delta_{nm} \nonumber
\eeq
We now focus on the the diagonal elements $A^{j}_{n} \equiv A^{j}_{nn}$. 
These are real, and transform as $A^{j}_{n} \mapsto A^{j}_{n} + \partial_j \Lambda_n$.
Associated with $A_n(x)$ is the gauge invariant 2-form 
which is called the "Berry curvature": 
\beq
B^{kj}_{n} 
\ = \ 
\partial_k A^{j}_{n} - \partial_j A^{k}_{n} 
\ = \ -2\hbar \im\langle\partial_k n|\partial_j n\rangle 
\ = \ -\frac{2}{\hbar}\im \sum_{m}
A^{k}_{nm} A^{j}_{mn}
\eeq
This can be written in abstract notation as $B=\partial \wedge A$, 
and if not singular (see below) it has zero divergence 
(${\partial \wedge B=  \partial \wedge \partial \wedge A = 0}$).  
The derivation of the following identity is straightforward: 
\beq
B^{kj}_{n} \ \ = \ \ 2\hbar \sum_{m(\ne n)}
\frac{\im\left[
\mathcal{F}^k_{nm}{\mathcal{F}}^j_{mn}\right]}
{(E_m-E_n)^2}
\eeq
From this expression it is evident that the sources of the 
divergenceless field $B_n$ are located at point where it 
encounters a degeneracy with a neighbouring level.  
Also it is useful to notice that ${\sum B_n =0}$, 
because the double summation results in a real expression inside the $\im[]$.   
The Berry phase can be regarded as the "flux" of $B$, namely, 
\beq
\gamma_n \ \ = \ \ \oint A_n \cdot dr  \ \ = \ \ \iint B_n \cdot ds
\eeq
Using the same argumentation as in the discussion of Dirac monopoles (see \hyperref[sMonopoles]{[here]})
we deduce that
\beq
\frac{1}{2\pi}\oiint B_n \cdot ds \ \ = \ \ \text{integer} \ \ \equiv \ \ \text{Chern number} 
\eeq
The mathematical digression below clarifies the analogy between~$B$ 
and the notion of magnetic field.

\sheadC{Digression - Geometrical Forms}

Geometrical forms are the ``vector analysis"  
generalization of the length, area and volume concepts 
to any dimension. 
In Euclidean geometry with three dimensions 
the basis for the usual vector space 
is $\hat{e}_1,\hat{e}_2,\hat{e}_3$. 
These are called 1-forms. We can also define 
a basis for surface elements (2-forms) and volume (3-form).
\beq
\hat{e}_{12} = \hat{e}_1\wedge\hat{e}_2 
\hspace{2cm}
\hat{e}_{23} = \hat{e}_2\wedge\hat{e}_3 
\hspace{2cm}
\hat{e}_{31} = \hat{e}_3\wedge\hat{e}_1
\hspace{2cm}
\hat{e}_1\wedge\hat{e}_2\wedge\hat{e}_3
\eeq
By definition $\hat{e}_{21} = -\hat{e}_{12}$, and $\hat{e}_1\wedge\hat{e}_2\wedge\hat{e}_3=-\hat{e}_2\wedge\hat{e}_1\wedge\hat{e}_3$ etc. 
There is a natural duality between 2-forms and 1-forms, 
namely $\hat{e}_{12} \mapsto \hat{e}_3$ 
and  $\hat{e}_{23} \mapsto \hat{e}_1$ 
and $\hat{e}_{31} \mapsto \hat{e}_2$. 
This duality between surface elements and 1-forms does not hold in Euclidean geometry of higher dimension. 
For example in 4 dimensions the surface elements (2-forms) constitute $C^2_4=6$ dimensional space. 
In the latter case we have duality between the hyper-surface elements (3-forms) and the 1-forms, which are both 4 dimensional spaces.   
There is of course the simplest example of Euclidean geometry in 2 diemnsional space 
where 2-forms are regarded as either area or as volume and not as 1-form vectors.  
In general for $N$ dimensional Euclidean geometry the $k$ forms constitute a $C^k_N$ 
dimensional vector space, and they are dual to the $(N-k)$ forms. 
We can multiply 1-forms as follows:
\beq
\left[\sum A_i \hat{e_i} \right] \wedge \left[ \sum B_i\hat{e_j} \right]
\ \ = \ \ \sum_{i,j} A_i A_j \hat{e_i}\wedge\hat{e_j}\
\ \ = \ \ \sum_{i<j} (A_i A_j-A_j A_i)\hat{e_i}\wedge\hat{e_j}
\eeq
Note that $\hat{e_i}\wedge\hat{e_i}=\mbox{Null}$. 
This leads to the practical formula for a wedge product  
\beq
(A\wedge B)_{ij} \ \ = \ \ A_i B_j-A_j B_i
\eeq
Similarly we use the notation $(\partial \wedge A)_{ij}=\partial_i A_j - \partial_j A_i$. 
Note that in 3 dimensional Euclidean geometry we have the following association 
with conventional "vector analysis" notations:  
\beq
\partial \wedge A \ \ \mapsto & \nabla \times A  \ \ \ \ & \mbox{if $A$ is a 1-forms}
\\ \nonumber
\partial \wedge B \ \ \mapsto & \nabla \cdot B \ \ \ \ \ & \mbox{if $B$ is a 2-forms}
\eeq
The above identifications are implied by the following:
\beq
\partial &=& \partial_1\hat{e}_1+\partial_1\hat{e}_2+\partial_1\hat{e}_3
\\ \nonumber
A &=& A_1\hat{e}_1+A_2\hat{e}_2+A_3\hat{e}_3 
\\ \nonumber
B &=& B_{12} \hat{e}_{12}+B_{23} \hat{e}_{23}+B_{31} \hat{e}_{31}
\eeq
We now introduce a more flexible notations. We regard any scalar function $f(x)$ as a 0-form, 
and associate differential quantities with 1-forms and 2-forms as follows:
\beq 
w_f = f(x)
\hspace{15mm}
w_A = \left[\sum_j A_j dx_j\right] \equiv A \cdot dr
\hspace{15mm}     
w_B = \left[\sum_{i<j} B_{ij} dx_i \wedge dx_j\right] \equiv B \cdot ds
\eeq
we also use the notation $d= \sum_j \partial_j dx_j$ for full differential.
With this notation a relation like ${B=\partial \wedge A}$ 
can be written as  ${w_B=d \wedge w_A}$.
The Stokes integral theorem is conventionally written as 
\beq
\oint A \cdot dr \ \ = \ \ \iint\partial\wedge A\cdot ds
\eeq
This concerns 1-forms. 
The Divergence integral theorem concerns 2-forms.  
The generalized Stokes theorem relates the closed 
boundary integral over a $k$-form to an integral 
over a $(k{+}1)$-form within the interior. 
With the above notations it can be written as 
\beq
\oint w_A \ \ = \ \ \iint d \wedge w_A
\eeq
where $A$ is any $k$-form.

\sheadC{Adiabatic evolution}

The adiabatic equation is obtained from the
Schr\"{o}dinger equation by expanding the wavefunction
in the $x$-dependent adiabatic basis:
\beq
\frac{d}{dt}|\psi\rangle \ \ &=&
\  -\frac{i}{\hbar} {\mathcal{H}}(x(t)) \ |\psi\rangle
\\ \nonumber
|\psi\rangle \ \ &=& \ \ \sum_n a_n(t) \ |n(x(t))\rangle
\\ \nonumber
\frac{da_n}{dt} \ \ &=& \ \ -\frac{i}{\hbar} E_n a_n
+\frac{i}{\hbar} \sum_m \sum_j
\dot{x}_j A^{j}_{nm} a_m 
\eeq
We define the perturbation matrix as
\beq
W_{nm} \ \ = \ \ -\sum_j\dot{x}_j A^j_{nm}
\ \ \ \ \ \ \ \ \mbox{for $n\ne m$}
\eeq
and $W^{j}_{nm}=0$ for $n=m$.
Then the adiabatic equation can be re-written
as follows:
\beq
\frac{da_n}{dt} = -\frac{i}{\hbar} (E_n - \dot{x} A_n) a_n
-\frac{i}{\hbar} \sum_m
W_{nm} a_m
\eeq
If we neglect the perturbation $W$,
then we get the strict adiabatic solution:
\beq
| \psi(t) \rangle \ \ = \ \ 
\eexp{
-\frac{i}{\hbar} \left(
\int_0^t E_n(x(t'))dt'
-\int_{x(0)}^{x(t)} A_n(x)\cdot dx
\right)
}
\ |n(x(t))\rangle
\eeq
The time dependence of this solution is 
exclusively via the $x$ dependence of the  
basis states. On top, due to $A_n(x)$ we have 
the so called geometric phase.
This can be gauged away unless we consider a closed
cycle. For a closed cycle, the gauge invariant
phase $\frac{1}{\hbar} \oint \vec{A}\cdot \vec{dx}$ 
is called the Berry phase.

With the above zero-order solution
we can obtain the following result:
\beq
\langle \mathcal{F}^k \rangle \ \ = \ \
\left\langle  \psi(t)   
\left|-\frac{\partial \mathcal{H}}{\partial x_k}  
\right| \psi(t)  \right\rangle
\ = \ 
-\frac{\partial}{\partial x_k}
\left\langle n(x)|\ \mathcal{H}(x) \ |n(x)\right\rangle
\eeq
In the standard examples this corresponds to a conservative force 
or to a persistent current. From now on we ignore this
trivial contribution to $\langle \mathcal{F}^k \rangle$,
and look for the a first order contribution.

\sheadC{Adiabatic Transport}

For linear driving (unlike the case of a cycle) the
$A_n(x)$ field can be gauged away.
Assuming further that the adiabatic equation
can be treated as parameter independent
(that means disregarding the dependence
of $E_n$ and ${W}$ on $x$) one realizes
that the Schr\"{o}dinger equation in the adiabatic basis 
possesses stationary solutions.
To first order these are:
\beq
| \psi(t) \rangle \ = \
|n\rangle +
\sum_{m(\ne n)}
\frac{W_{mn}}
{E_n-E_m} |m\rangle
\eeq
Note that in a {\em fixed-basis representation}
the above stationary solution is in fact time-dependent.
Hence the explicit notations $|n(x(t))\rangle$ 
and $|m(x(t))\rangle$ are possibly more appropriate.

With the above solution
we can write $\langle \mathcal{F}^k \rangle$ as a sum of
zero order and first order contributions.
From now on we ignore the
zero order contribution,
but keep the first order contribution:
\beq
\langle \mathcal{F}^k \rangle 
\ = \
-\sum_{m(\ne n)}
\frac{W_{mn}}{E_n-E_m}
\left\langle n \Big|
\frac{\partial \mathcal{H}}{\partial x_k}
\Big| m \right\rangle + \mbox{CC}
\ = \ 
\sum_j \left(i\sum_{m}
A^k_{nm} A^{j}_{mn} + \mbox{CC}\right) \ \dot{x}_j
\ = \ 
-\sum_j B_n^{kj} \ \dot{x}_j
\eeq
For a general {\em stationary} preparation,
either pure or mixed, one obtains 
\beq
\langle \mathcal{F}^k \rangle = -\sum_j G^{kj} \ \dot{x}_j
\eeq
with
\beq
G^{kj} \ = \ \sum_n f(E_n) \ B_n^{kj}
\eeq
where $f(E_n)$ are weighting factors,
with the normalization $\sum_n f(E_n)=1$.
For a pure state preparation $f(E_n)$ distinguishes
only one state $n$, while for a canonical
preparation $f(E_n)\propto\eexp{-E_n/T}$,
where $T$ is the temperature.
For a many-body system of non-interacting particles 
$f(E_n)$ is re-interpreted as
the occupation function,
so that $\sum_n f(E_n)=N$ is the total
number of particles.

Thus we see that the assumption of a stationary
first-order solution leads to a non-dissipative (antisymmetric)
conductance matrix. This is known as either
"adiabatic transport" or "geometric magnetism".
In the next lecture we are going to see 
that "adiabatic transport" is in fact a special limit 
of the Kubo formula.

\newpage

\sheadB{Linear response theory and the Kubo formula}

\sheadC{Linear Response Theory}

We assume that the Hamiltonian depends on 
several parameters, say three parameters: 
\beq
\mathcal{H} \ = \ \mathcal{H}(\vec{r},\vec{p} ; \ x_1(t),x_2(t),x_3(t)) 
\eeq
and we define generalized forces 
\beq
\mathcal{F}^k \ = \ -\frac{\partial \mathcal{H}}{\partial x_k}
\eeq
Linear response means that 
\beq
\langle \mathcal{F}^k \rangle_t \ = \ 
\sum_j \int_{-\infty}^{\infty} \alpha^{kj}(t-t') \ \delta x_j(t')dt'
\eeq
where $\alpha^{kj}(\tau)=0$ for $\tau<0$.
The expression for the response Kernel is 
known as the Kubo formula:  
\beq
\alpha^{kj}(\tau) = \Theta(\tau) \, 
\frac{i}{\hbar} \langle [\mathcal{F}^k(\tau),\mathcal{F}^j(0)]\rangle_0
\eeq
where the average is taken with 
the assumed zero order stationary solution. 
Before we present the standard derivation 
of this result we would like to illuminate 
the {\em DC limit} of this formula, and to 
further explain the {\em adiabatic limit} 
that was discussed in the previous section.

\sheadC{Susceptibility and DC Conductance}

The Fourier transform of  $\alpha^{kj}(\tau)$
is the generalized susceptibility $\chi^{kj}(\omega)$.
Hence 
\beq
[\langle \mathcal{F}^k \rangle]_{\omega} \ = \ 
\sum_j \chi_0^{kj}(\omega) [x_j]_{\omega}
- \sum_j \mu^{kj}(\omega) [\dot{x}_j]_{\omega}
\eeq
where the dissipation coefficient is defined as
\beq
\mu^{kj}(\omega)  =
\frac{\im[\chi^{kj}(\omega)]}{\omega} =
\int_0^{\infty} \alpha^{kj}(\tau) \frac{\sin(\omega\tau)}{\omega}d\tau
\eeq
In the "DC limit" ($\omega\rightarrow 0$) 
it is natural to define the generalized conductance matrix:
\beq
G^{kj} \ = \ 
\mu^{kj}(\omega\sim0) \ = \
\lim_{\omega\rightarrow 0}
\frac{\im[\chi^{kj}(\omega)]}{\omega}
\ = \ \int_0^{\infty} \alpha^{kj}(\tau) \tau d\tau
\eeq
Consequently the non-conservative part of the 
response can be written as a generalized Ohm law.
\beq
\langle \mathcal{F}^k \rangle =
-\sum_{j} G^{kj} \ \dot{x}_j
\eeq
It is convenient to write the conductance matrix as  
\beq
G^{kj} \equiv \ \eta^{kj} + B^{kj}
\eeq

where $\eta^{kj}=\eta^{jk}$ is the symmetric part
of the conductance matrix,
while $B^{kj}=-B^{jk}$ is the antisymmetric part.
In our case there are three parameters so we can arrange the
elements of the antisymmetric part
as a vector ${\vec{B}=(B^{23},B^{31},B^{12})}$.
Consequently the generalized Ohm law 
can be written in abstract notation as
\beq
\langle \mathcal{F} \rangle \ = \ 
-\eta \cdot \dot{x} \ 
- \ B \wedge \dot{x}
\eeq
where the dot product should be interpreted as matrix-vector
multiplication, which involves summation over the index~$j$.
The wedge-product can also be regarded as a matrix-vector
multiplication. It reduces to the more familiar cross-product
in the case we have been considering - $3$ parameters.
The dissipation, which is defined as the rate at which energy
is absorbed into the system, is given by
\beq
\dot{W}  \ = \
-\langle \mathcal{F} \rangle \cdot \dot{x} \ = \
\sum_{kj} \eta^{kj} \ \dot{x}_k\dot{x}_j
\eeq
which is a generalization of Joule's law.
Only the symmetric part contributes
to the dissipation. The contribution
of the antisymmetric part is identically zero.

The conductance matrix is essentially a synonym 
for the term "dissipation coefficient".
However, "conductance" is a better 
(less misleading) terminology:
it does not have the (wrong) connotation
of being  specifically  associated with dissipation,
and consequently it is less confusing to say that
it contains a non-dissipative component.
We summarize the various definitions by the following diagram: \\

\begin{center}
\setlength{\unitlength}{2000sp}
\begin{picture}(4725,5767)(751,-7112)
\put(1501,-1861){\vector( 0,-1){375}}
\put(1501,-3061){\vector(-2,-3){242.308}}
\put(1801,-3061){\vector( 4,-1){1641.177}}
\put(4801,-4261){\vector( 0,-1){525}}
\put(4951,-5611){\vector( 1,-1){375}}
\put(4651,-5611){\vector(-1,-1){375}}
\put(1126,-1561){$\alpha^{kj}(t-t')$}
\put(1201,-2761){$\chi^{kj}(\omega)$}
\put(751,-3886){$\mbox{Re}[\chi^{kj}(\omega)]$}
\put(3301,-3886){$(1/\omega) \mbox{Im}[\chi^{kj}(\omega)]$}
\put(3676,-6511){$\eta^{kj}$}
\put(5326,-6511){$B^{kj}$}
\put(5476,-7036){(non-dissipative)}
\put(4576,-5386){$G^{kj}$}
\put(3001,-7036){(dissipative)}
\end{picture}
\end{center}

\sheadC{Derivation of the Kubo formula}

If the driving is not strictly adiabatic 
the validity of the stationary adiabatic solution
that has been presented in the previous lecture becomes questionable. 
In general we have to take non-adiabatic transitions between levels 
into account. This leads to the Kubo formula 
for the response which we discuss below.  
The Kubo formula has many type of derivations. 
One possibility is to use the same procedure 
as in the previous lecture starting with
\beq
| \psi(t) \rangle \ = \
\eexp{-iE_nt}|n\rangle 
&+& \sum_{m(\ne n)}
\left[-i\bm{W}_{mn}\int_0^t\eexp{i(E_n{-}E_m)t'}dt'\right]
\eexp{-iE_mt} |m\rangle
\eeq
We shall not expand further on this way of 
derivation, which becomes quite subtle once 
we go beyond the stationary adiabatic approximation. 
A one line derivation of the Kubo formula 
is based on the interaction picture, and is 
presented in a \hyperref[sKubo]{dedicted section} 
There are various different looking derivations 
of the Kubo formula that highlight 
the quantum-to-classical correspondence  
and/or the limitations of this formula.    
The advantage of the derivation below is that 
it also allows some extensions within the 
framework of a master equation approach 
that takes the environment into account.

For notational simplicity we write the Hamiltonian as 
\beq
\mathcal{H} \ = \ \mathcal{H}_0 \ - f(t)V
\eeq
We assume that the system, in the absence
of driving, is prepared in a stationary state $\rho_0$.
In the presence of driving we look for a first order solution
$\rho(t) = \rho_0 +\tilde{\rho}(t)$.
The equation for $\tilde{\rho}(t)$ is:
\beq
\frac{\partial \tilde{\rho}(t)}{\partial t} \ \approx \
-i[\mathcal{H}_0, \tilde{\rho}(t)] + if(t)[V,\rho_0]
\eeq
Next we use the substitution 
$\tilde{\rho}(t) = U_0(t) \tilde{\tilde{\rho}}(t) U_0(t)^{-1}$, 
where $U_0(t)$ is the evolution operator
which is generated by $\mathcal{H}_0$.
Thus we eliminate from the equation the zero order term: 
\beq
\frac{\partial \tilde{\tilde{\rho}}(t) }{\partial t}
\ \approx \
if(t)[  U_0(t)^{-1}V U_0(t),\rho_0]
\eeq
The solution of the latter equation is straightforward and leads to
\beq
{\rho}(t)
\ \approx \ {\rho}_0 +
\int^{t}   i \ [V(-(t{-}t')), \rho_0]  \ f(t')dt'
\eeq
where we use the common notation 
$V(\tau)=U_0(\tau)^{-1}V U_0(\tau)$.

Consider now the time dependence of the expectation value
\mbox{$\langle \mathcal{F} \rangle_t = \trc(\mathcal{F}\rho(t))$}
of an observable. Disregarding the zero order contribution,
the first order expression is
\beq \nonumber
\langle \mathcal{F} \rangle_t
\ \approx \
\int^{t}  i \ \trc\left(\mathcal{F} [V(-(t{-}t')), \rho_0]\right)  \ f(t')dt'
\ \ = \ \ 
\int^{t}   \alpha(t-t')  \ f(t')dt'
\eeq
where the response kernel $\alpha(\tau)$ is
defined for $\tau>0$ as
\beq
\alpha(\tau) = i\ \trc\left(\mathcal{F} [V(-\tau), \rho_0]\right)
\ = i\ \trc\left([\mathcal{F}, V(-\tau)] \rho_0\right)
\ = i \langle [\mathcal{F}, V(-\tau)] \rangle
\ = i \langle [\mathcal{F}(\tau), V] \rangle
\eeq
where we have used the cyclic property of the trace operation;
the stationarity $U_0\rho_0 U_0^{-1}=\rho_0$ of
the unperturbed state; and the notation 
$\mathcal{F}(\tau)=U_0(\tau)^{-1}\mathcal{F}U_0(\tau)$.

\newpage

\sheadB{The Born-Oppenheimer picture}

We now consider a more complicated problem,
where $x$ becomes a dynamical variable.
The standard basis for the representation of the
composite system is $ |x,Q\rangle = |x\rangle \otimes |Q\rangle$.
We assume a total Hamiltonian of the form
\beq
\mathcal{H}_{\tbox{total}} \ = \ 
\frac{1}{2\mass}\sum_j p_j^2 + \mathcal{H}(Q,P;x) - f(t)V(Q)
\eeq
Rather than using the standard basis,  we can use
the Born-Oppenheimer basis
$|x,n(x)\rangle = |x\rangle \otimes |n(x)\rangle$.
Accordingly the state of the combined system
is represented by the wavefunction $\Psi_n(x)$, namely
\beq
|\Psi\rangle \ = \ 
\sum_{n,x} \Psi_n(x) \ |x,n(x)\rangle
\eeq
The matrix elements of ${\mathcal{H}}$ are
\beq
\langle x,n(x)|\mathcal{H}|x_0,m(x_0) \rangle \ = \ 
\delta(x - x_0)  \times  \delta_{nm}E_n(x)
\eeq
The matrix elements of $V(Q)$ are
\beq
\langle x,n(x)|V(Q)|x_0,m(x_0) \rangle \ = \ 
\delta(x - x_0)  \times  V_{nm}(x)
\eeq
The matrix elements of $p$ are
\beq
\langle x,n(x)|p_j|x_0,m(x_0)  \rangle \ = \ 
(-i\partial_j \delta(x{-}x_0)) \times  \langle n(x) | m(x_0)  \rangle
\nonumber
\eeq
The latter can be manipulated "by parts" leading to
\beq
\langle x,n(x)|p_j|x_0,m(x_0)  \rangle \ = \ 
-i \partial_j\delta(x - x_0)\delta_{nm}
- \delta(x - x_0)  A^j_{nm}(x)
\eeq
This can be summarized by saying that
the operation of $p_j$ on a wavefunction is
like the differential operator
\mbox{$-i\partial_j-A^j_{nm}(x)$}.
Thus in the Born-Oppenheimer basis
the total Hamiltonian takes the form
\beq
\mathcal{H}_{\tbox{total}} \ = \ 
\frac{1}{2\mass}\sum_j(p_j-A^j_{nm}(x))^2
+ \delta_{nm}E_n(x) - f(t)V_{nm}(x)
\eeq
Assuming that the system is prepared in
energy level $n$, and disregarding the effect
of $A$ and $V$, the adiabatic
motion of $x$ is determined by the effective
potential $E_n(x)$. This is the standard
approximation in studies of diatomic molecules,
where $x$ is the distance between the nuclei.
If we treat the slow motion as classical,
then the interaction with $A$ can be written as
\beq
\mathcal{H}^{(1)}_{\tbox{interaction}}
\ = \ -\sum_j \dot{x}_j A^j_{nm}(x)
\eeq
This brings us back to the theory
of driven systems as discussed
in previous sections.
The other interaction that can induce
transitions between levels is
\beq
\mathcal{H}^{(2)}_{\tbox{interaction}}
\ = \ -f(t) V_{nm}(x)
\eeq
The analysis of molecular "wavepacket dynamics"
is based on this picture.

\sheadA{The Green function approach}

\sheadB{The propagator and Feynman path integral}

\sheadC{The propgator}

The evolution of a quantum mechanical system 
is described by a unitary operator 
\beq
|\Psi(t)\rangle \ \ = \ \ U(t,t_0)  \ |\Psi(t_0)\rangle
\eeq
The Hamiltonian is defined by 
writing the infinitesimal evolution as  
\beq
U(t+dt,t) = 1 - i dt \mathcal{H}(t)
\eeq
This expression has an imaginary $i$ in order to 
make $\mathcal{H}$ a Hermitian matrix.  
If we want to describe continuous evolution 
we can "break" the time interval 
into $N$ infinitesimal steps:
\beq
U(t,t_0) \ \ = \ \ 
(1-idt_N\mathcal{H}) \dots (1-idt_2\mathcal{H})(1-idt_1\mathcal{H}) 
\ \ \equiv \ \ 
\mathcal{T}\eexp{-i \int^t_{t_0} \mathcal{H}(t') dt'} 
\eeq
For a time independent Hamiltonian we get simply $U(t)=\eexp{-it \mathcal{H}}$ 
because of the identity $\eexp{A} \eexp{B}=\eexp{A+B}$ if $[A,B]=0$.

If we consider a particle and use the standard 
position representation then the unitary operator is 
represented by a matrix $U(x|x_0)$. The time interval $[t_0,t]$ 
is implicit. We alway assume that $t>t_0$. 
Later it would be convenient to define the propagator 
as $U(x|x_0)$ for $t>t_0$ and as zero otherwise. 
The reason for this convention is related to the formalism 
that we are going to introduce later on.

\sheadC{The Propagator for a free particle} 

Consider a free particle in one dimension. 
Let us find the propagator using a direct calculation. The hamiltonian is:
\beq
\mathcal{H}=\frac{p^2}{2\mass}
\eeq
As long as the Hamiltonian has a quadratic form, 
the answer will be a Gaussian kernel: 
\beq
U(x|x_0) = \left\langle x \left| \eexp{-i\frac{t}{2\mass} p^2 } \right |x_0 \right\rangle =
\left(\frac{\mass}{2\pi i t}\right)^\frac{1}{2}\eexp{i \frac{\mass}{2t}(x-x_0)^2}
\eeq
We note that in the case of a harmonic oscillator 
\beq
\mathcal{H}=\frac{p^2}{2\mass}+\frac{1}{2}\mass\Omega^2 x^2
\eeq
The propagator is  
\beq
U(x|x_0)=\left(\frac{\mass\Omega }{2\pi i \sin{\Omega t}}\right)^\frac{1}{2}
\eexp{i \frac{\mass\Omega }{2\sin{\Omega t}}[\cos{\Omega t}(x^2+x_0^2) - 2 x x_0]}
\eeq
If we take $ t \rightarrow 0$ then $U \rightarrow \hat{1}$, 
and therefore $U(x|x_0) \rightarrow \delta (x-x_0)$.

The derivation of the expression for the propagator 
in the case of a free particle goes as follows. 
For clarity we use units such that ${\mass=1}$:
\beq
\langle x|\eexp{-\imath \frac{1}{2}t p^2}|x_0 \rangle
=
\sum_k \langle x|k\rangle \eexp{-\imath \frac{1}{2}t k^2} \langle k|x_0 \rangle
=
\int \frac{dk}{2\pi} \eexp{-\imath \frac{1}{2}t k^2 + ik(x-x_0)} 
\eeq
This is formally an inverse FT of a Gaussian. 
One obtains the desired result, which is formally  
a Gaussian that has an imaginary variance $\sigma^2=it$.  
We note that we can write the result in the form
\beq
\langle x|\eexp{-\imath \frac{1}{2}t p^2}|x_0\rangle 
=\frac{1}{\sqrt{2\pi \imath t }} \eexp{\frac{\imath (x-x_0)^2}{2t }}
=\frac{1}{\sqrt{2\pi \imath t }}
\left[\cos{\frac{1}{2t }(x-x_0)^2}+
\imath \sin{\frac{1}{2t }(x-x_0)^2}\right]
\eeq
If $t \rightarrow 0$ we should get 
a delta function. This is implied by the FT, 
but it would be nice to verify this statement 
directly. Namely we have to show that 
in this limit we get a narrow 
function whose "area" is unity. 
For this purpose we use identity  
\beq
\int \cos{r^2}  \dots  dr \ = \ \int \frac{\cos{u}}{2\sqrt{u}}  \dots  du
\eeq
and a similar expression in case of $\sin$ function. 
Then we recall the elementary integrals
\beq
\int_0 ^\infty \frac{\sin{u}}{\sqrt{u}}du
=\int_0 ^\infty \frac{\cos{u}}{\sqrt{u}}du
=\sqrt{\frac{\pi}{2}}
\eeq
Thus the "area" of the two terms in the square brackets 
is proportional to $(1+i)/\sqrt{2}$ which cancels 
the $\sqrt{i}$ of the prefactor.

\sheadC{Feynman Path Integrals} 

How can we find the propagator $U(x|x_0)$ for the general 
Hamiltonian $\mathcal{H}=\frac{p^2}{2\mass}+V(x)$? 
The idea is to write $\langle x|\eexp{-\imath t\mathcal{H}}|x_0 \rangle$ 
as a convolution of small time steps:
\beq
\langle x|\eexp{-\imath t\mathcal{H}}|x_0 \rangle = 
\sum_{x_1,x_2, \dots x_{N-1}}
\langle x |\eexp{-\imath \delta t_N \mathcal{H}} |x_{N-1} \rangle \dots 
\langle x_2|\eexp{-\imath \delta t_2\mathcal{H}}|x_1 \rangle
\langle x_1|\eexp{-\imath \delta t_1\mathcal{H}}|x_0 \rangle
\eeq
Now we have to find the propagator for each infinitesimal step. 
At first sight it looks as if we just complicated 
the calculation. But then we recall that 
for infinitesimal operations we have:   
\beq
\eexp{\varepsilon A+\varepsilon B} 
\approx \eexp{\varepsilon A}\eexp{\varepsilon B} 
\approx \eexp{\varepsilon B}\eexp{\varepsilon A} 
\ \ \ \ \ \ \ \mbox{for any $A$ and $B$}
\eeq
This is because the higher order correction can be made 
as small as we want. So we write
\beq
\langle x_j|\eexp{-\imath \delta t(\frac{p^2}{2\mass}+V(x))}|x_{j-1} \rangle 
\approx  
\langle x_j|
\eexp{-\imath \delta tV(x))}
\eexp{-\imath \delta t\frac{p^2}{2\mass}}
|x_{j-1} \rangle 
\approx 
(\frac{\mass}{2\pi \imath dt_j})^{\frac{1}2}
\eexp{\imath \frac{\mass}{2dt_j}(x_j-x_{j-1})^2 - dt_jV(x_j)}
\eeq
and get: 
\beq
U(x|x_0)=
\int dx_1 dx_2  \dots  dx_{N-1} 
(\frac{\mass}{2\pi \imath dt})^{\frac{N}2}
\eexp{i\mathcal{A}[x]}
\ \equiv \
\int d[x] \ \eexp{i\mathcal{A}[x]}
\eeq
where $\mathcal{A}[x]$ is called the \textbf{action}.
\beq
\mathcal{A}[x]=\sum _{j=1}^{N-1} 
\left[ \frac{\mass}{2dt}(x_j-x_{j-1})^2-dtV(x) \right] 
= \int (\frac{1}{2}\mass \dot{x}^2-V(x))dt
= \int \mathcal{L}(x,\dot{x}) dt
\eeq
More generally, if we include the vector potential 
in the Hamiltonian, then we get the Lagrangian
\beq
\mathcal{L}(x,\dot{x}) \ \ = \ \ \frac{1}{2}\mass \dot{x}^2-V(x)+A(x)\dot{x}
\eeq
leading to:
\beq
\mathcal{A}[x]= \int (\frac{1}{2}\mass \dot{x}^2-V(x))dt + \int A(x) \cdot dx
\eeq

\sheadC{Stationary Point Approximation}

The Stationary Point Approximation is an estimate for an integral 
of the form $\int \eexp{\imath \mathcal{A}(x)}dx$. 
The main contribution to the integral comes from the point $x=\bar{x}$, 
called a {\textbf stationary point}, where $\mathcal{A}'(x)=0$. 
We expand the function $\mathcal{A}(x)$ near the stationary point:
\beq
\mathcal{A}(x) \ \ = \ \ \mathcal{A}(\bar{x})
\ + \ \frac{1}{2}\mathcal{A}''(\bar{x})(x-\bar{x})^2+ \dots 
\eeq
Leading to 
\beq
\int \eexp{\imath \mathcal{A}(x)}dx \ \ \approx \ \ 
\eexp{\imath \mathcal{A}(\bar{x})} 
\int \eexp{\imath \frac{1}{2}\mathcal{A}^{"}(\bar{x})(x-\bar{x})^2}dx
\ \ = \ \ 
\sqrt{\frac{i2\pi}{\mathcal{A}^{"}(\bar{x})}} \eexp{\imath \mathcal{A}(\bar{x})} 
\eeq
where the exponential term is a leading order term, 
and the prefactor is an "algebraic decoration".

The generalization of this method to multi-dimensional integration 
over $d[x]$ is immediate. The stationary point is in fact the trajectory 
for which the first order variation is zero ($\delta \mathcal{A} = 0$). 
This leads to Lagrange equation, which implies that the 
"stationary point" is in fact the classical trajectory.
Consequently we get the so called semiclassical (Van-Vleck) approximation:     
\beq
U(x|x_0) \ \ = \ \ 
\int d[x] \ \eexp{i\mathcal{A}[x]}
\ \ \approx  \ \ 
\sum_{cl}\left(\frac{1}{i 2\pi\hbar }\right)^{{d}/{2}}
\left|\det\left( -\frac {\partial^{2}\mathcal{A}_{cl}} {\partial x\partial x_0} \right)\right|^{1/2} 
\ \eexp{i \mathcal{A}_{cl}(x,x_0) - i(\pi/2)\nu_{cl}}
\eeq
where $d$ is the number of degrees of freedom, and 
\beq
\mathcal{A}_{cl}(x,x_{0})\equiv \mathcal{A}[x_{cl}] 
\eeq
is the action of the classical trajectory as a function 
of the two end points. In $d$ dimensions the associated 
determinant is $d \times d$. 
The Morse-Maslov index $\nu_{cl}$ counts the number of 
conjugate points along the classical trajectory.
If the time is short it equals zero and then both this 
index and the absolute value that enclosed the determinant
can be omitted. Otherwise the recipe for the determination 
of the Morse index is as follows:
The linearized equation of motion 
for $x(t) = x_{cl}(t) + \delta x(t)$ is
\beq
\mass \delta \ddot{x} + V''(x_{cl}) \delta x = 0
\eeq
A conjugate point (in time) is defined as the time when the 
linearized equation has a non-trivial solution.
With this rule we expect that in case of a reflection from 
a wall, the Morse index is {\bf +1} for each collision. 
This is correct for ``soft" wall. In case of ``hard" wall 
the standard semiclassical result for $\nu_{cl}$ breaks down, 
and the correct result turns out to be {\bf +2} for each collision. 
The latter rule is implied by the {\bf Dirichlet} boundary conditions. 
Note, that it would be {\bf +0} in the case of {\bf Neumann} boundary conditions.
The Van-Vleck semiclassical approximation is exact for 
quadratic Hamiltonians because the "stationary phase integral" is exact 
if there are no higher order terms in the Taylor expansion.

Let us compute again $U(x|x_0)$ for a free particle, 
this time using the Van-Vleck expression: 
The action $\mathcal{A}[x]$ for the free particle is  
\beq
\mathcal{A}[x]=\int_{0}^{t}\frac{1}{2}\mass \dot{x}^{2}dt'.
\eeq
Given the end points we find the classical path  
\beq
x_{cl}=x_{0}+\frac{x-x_{0}}{t}t'
\eeq
and hence 
\beq
\mathcal{A}_{cl}(x,x_0)=\int_{0}^{t}\frac{1}{2}
\mass(\frac{x-x_{0}}{t})^{2}dt' =
\frac{\mass}{2t}(x-x_0)^2.
\eeq
We also observe that 
\beq
-\frac{\partial^{2}\mathcal{A}_{cl}}{\partial x\partial x_0} 
\ \ = \ \ \frac{\mass}{t}
\eeq
which leads to the exact result.
The big advantage of this procedure 
emerges clearly once we try to 
derive the more general expression 
in case of a harmonic oscillator.

\newpage

\sheadB{The resolvent and the Green function}

\sheadC{The resolvent}

The resolvent is defined in the complex plane as 
\beq
G(z)=\frac{1}{z-\mathcal{H}}
\eeq
In case of a bounded system it has poles at 
the eigenvalues. We postpone for later 
the discussion of unbounded systems. 
It is possibly more illuminating 
to look on the matrix elements of the resolvent 
\beq
G(x|x_0) 
\ = \ 
\langle x |G(z)|x_0\rangle 
\ = \
\sum_n \frac{ {\psi^n(x)} {\psi^n(x_0)}^{*} }{z-E_n}
\ \equiv \ 
\sum_n \frac{q_n}{z-E_n}
\eeq
where $\psi^n(x)=\langle x|n \rangle$ are the eigenstates of 
the Hamiltonian. If we fix $x$ and $x_0$ and regard this expression 
as a function of $z$ this is formally the complex representation 
of an electric field in a two dimensional electrostatic problem.

We can look on $G(x|x_0)$, with fixed $z=E$ and $x_0$,  
as a wavefunction in the variable $x$.
We see that $G(x|x_0)$ is a superposition of eigenstates. 
If we operate on it with $(E-\mathcal{H})$ the coefficients 
of this superposition are multiplied by $(E-E_k)$, 
and because of the completeness of the basis 
we get $\delta(x-x_0)$. This means that $G(x|x_0)$ 
satisfies the Schrodinger equation 
with the complex energy~$E$ and with 
an added source at $x=x_0$. Namely,    
\beq
(E-\mathcal{H}) G(x|x_0) = \delta(x-x_0)
\eeq
This is simply the standard representation 
of the equation ${(z-\mathcal{H})G=\hat{1}}$ 
which defines the matrix inversion ${ G=1/(z-\mathcal{H}) }$. 
The wavefunction $G(x|x_0)$ should satisfy 
that appropriate boundary conditions. 
If we deal with a particle in a box this means Dirichlet 
boundary conditions.  In the case of an unbounded system 
the issue of boundary conditions deserves 
further discussion (see later).

The importance of the Green functions 
comes from its Fourier transform relation 
to the propagator. Namely,  
\beq
FT\Big[ \ \Theta(t)\eexp{-\gamma t} \  U(t) \Big] 
\ \ = \ \ iG(\omega+i\gamma)
\eeq
where $\Theta(t)$ is the step function,  $\Theta(t)U(t)$ 
is the "propagator", and $\eexp{-\eta t}$ is an envelope 
function that guarantees convergence of the FT integral.
Later we discuss the limit $\eta\rightarrow0$.

We note that we can extract from the resolvent 
useful information. For example, we can get 
the energy eigenfunction by calculation the residues 
of $G(z)$. The Green functions, which we discuss 
in the next section are obtained (defined) as follows: 
\beq
G^{\pm}(\omega) \ = \ G(z=\omega \pm i0) \ = \ \frac{1}{\omega - \mathcal{H} \pm i0}
\eeq
From this definition follows that 
\beq
\im[G^{+}]  \equiv  -\frac{i}{2} (G^{+}-G^{-}) = -\pi \delta(E-\mathcal{H})
\eeq
From here we get expression for 
the density of states $\gdos(E)$ and 
for the local density of states $\rho(E)$,    
where the latter is with respect 
to an arbitrary reference state~$\Psi$ 
\beq
\gdos(E) &=&  -\frac{1}{\pi} \trc\Big( \ \im[G^{+}(E)] \ \Big)
\\ \nonumber
\rho(E) &=& -\frac{1}{\pi} \Big\langle \Psi \Big| \ \im[G^{+}(E)] \ \Big| \Psi \Big\rangle
\eeq
Further applications of the Green functions 
will be discussed later on.

Concluding this section we note that from the 
above it should become clear that 
there are three methods of calculating 
the matrix elements of the resolvent: 

\bitem Summing as expansion in the energy basis  \\
\bitem Solving an equation (Helmholtz for a free particle) \\
\bitem Finding the Fourier transform of the propagator 

Possibly the second method is the simplest, 
while the third one is useful in semiclassical schemes.

\sheadC{Mathematical digression}
 
Let us summarize some useful mathematical facts 
that are related to the theory of the resolvent. 
First of all we have the following identity for 
a decomposition into "principal" and "singular" parts: 
\beq
\frac{1}{\omega+i0} \ \ = \ \ \frac{\omega}{\omega^2+0^2}-i\frac{0}{\omega^2+0^2} 
\ \ = \ \  \frac{1}{\omega}-i\pi\delta(\omega)
\eeq
where $0$ is an infinitesimal. An optional proof of this 
identity in the context of contour integration is based 
on deforming the contour into a semicircle in the upper 
complex half plane, and then displacing the pole to be on  
the real axis. Then the first term is contributed 
by the principal part, while the second term is contributed by the semicircle.

The following Fourier transform relation is extremely useful, 
and we would like to establish it both in the forward and in 
the backward directions:
\beq
\mbox{FT}\left[ \Theta(t) \right] \ \ = \ \ \frac{i}{\omega+i0} 
\eeq
An equivalent relation is  
\beq
\mbox{FT}\left[ \Theta(t)-\frac{1}{2} \right] \ \ = \ \ \frac{i}{\omega} 
\ \ \ \ \ \ \ \ \ \ \ \ \
\mbox{[the FT that gives the principal part]} 
\eeq
In the forward direction the proof is based on 
taking the $\gamma\rightarrow0$ limit of 
\beq
\mbox{FT}\left[ \Theta(t)\eexp{-\gamma t} \right]
= \int_{0}^{\infty}\eexp{-(\gamma-i\omega) t} dt 
= \frac{1}{\gamma-i\omega} 
\eeq
Note that this is not the same as 
\beq
\mbox{FT}\left[ \eexp{-\gamma |t|} \right]
= \frac{2\gamma}{\omega^2+\gamma^2} 
\ \ \ \ \ \ \ \ \ \ \ \ \
\mbox{[the FT that gives the singular part]} 
\eeq
In the backward direction the simplest is to use contour integration:
\beq
\int_{-\infty}^{\infty}\frac{d\omega}{2\pi} \left[\frac{i \eexp{-i\omega t}}{\omega+i0}\right] 
\ = \ 2\pi i \times \mbox{Residue}
\ \ \ \ \ \ \ \ \ \ \ \ \
\mbox{[for $t>0$ the contour is closed in the lower half plane]} 
\eeq
The other possibility is to stay with the original contour 
along the real axis, but to split the integrand into principal part 
and singular part. The integral over the principal part is in-fact 
the inverse Fourier transform of $i/\omega$. Its calculation is 
based on the elementary integral  
\beq
\int_{-\infty}^{+\infty} \frac{\sin(ax)}{x}dx \ = \ \mbox{Sign}(a) \, \pi  \ = \ \pm\pi
\eeq

\sheadC{The Green function of a bounded particle}

In order to get insight into the mathematics 
of $G^{+}(z)$ we first consider how $G(z)$ looks 
like for a particle in a very large box.
To be more specific we can consider a particle 
in a potential well or on a ring. In the latter case 
it means periodic boundary conditions rather 
than Dirichlet boundary conditions.  
Later we would like to take the length $L$ 
of the box to be infinite so as to have a "free particle". 
Expanding ${\psi(x)=\langle x|G(z)|x_0 \rangle}$ 
in the energy basis we get the 
following expressions:  
\beq
\langle x|G_{\tbox{well}}(z)|x_0\rangle &=& 
\frac{2}{L}\sum_n \frac{\sin(k_n x)\sin(k_n x_0)}{z-E_n} 
\\ \nonumber
\langle x|G_{\tbox{ring}}(z)|x_0\rangle &=& 
\frac{1}{L}\sum_n \frac{\eexp{ik_n(x-x_0)}}{z-E_n}
\eeq
where the real $k_n$ correspond 
to a box of length $L$. 
As discussed in the previous lecture, 
this sum can be visualized as  
the field which is created 
by a string of charges along the real axis. 
If we are far enough from the real axis 
we get a field which is the same as 
that of a smooth distribution of "charge". 
Let us call it the "far field" region.
As we take the volume of the box to infinity 
the "near field" region, whose width 
is determined by the level spacing, 
shrinks and disappears. Then we are left  
with the "far field" which is the resolvent of a free particle.
The result should not depend on whether 
we consider Dirichlet of periodic boundary conditions.

The summation of the above sums 
is technically too difficult.  
In order to get an explicit 
expression for the resolvent 
we recall that ${\psi(x)=\langle x|G(z)|x_0 \rangle}$ 
is the solution of a Schrodinger 
equation with complex energy~$z$ 
and a source at ${x=x_0}$. 
The solution of this equation is 
\beq
\langle x|G(z)|x_0\rangle = -i\frac{\mass}{k}\eexp{ik|x-x_0|} + A \eexp{ikx} + B \eexp{-ikx} 
\eeq
where $k=(2\mass z)^{1/2}$ corresponds to the complex 
energy~$z$. The first term satisfies the matching 
condition at the source, while 
the other two terms are "free waves" 
that solve the associated homogeneous equation. 
The coefficients $A$ and $B$ should be adjusted 
such that the boundary conditions are satisfied. 
For the "well" we should ensure 
the Dirichlet boundary conditions $\psi(x)=0$ for $x=0,L$,  
while for the "ring" we should ensure 
the periodic boundary conditions $\psi(0)=\psi(L)$ 
and $\psi'(0)=\psi'(L)$.

Let us try to gain some insight for the solution.
If $z$ is in the upper half plane then we can write 
$k = k_E + i\alpha$ where both $k_E$ and $\alpha$ are 
positive(!) real numbers.  
This means that a propagating wave (either right going 
or left going)  exponentially decays to zero 
in the propagation direction, 
and exponentially explodes in the opposite direction. 
It is not difficult to conclude that in the limit 
of a very large $L$ the coefficients $A$ and $B$ 
become exponentially small. In the strict 
$L\rightarrow \infty$ limit we may say that $\psi(x)$ 
should satisfy "outgoing boundary conditions".  
If we want to make analytical continuation of $G(z)$ 
to the lower half plane, we should stick to these  
"outgoing boundary conditions". The implication is that 
$\psi(x)$ in the lower half plane exponentially 
explodes at infinity.

An optional argument that establishes the application 
of the outgoing boundary conditions is based 
on the observation that the FT of the retarded $G^{+}(\omega)$
gives the propagator. The propagator is identically 
zero for negative times. If we use the propagator 
to propagate a wavepacket, we should get a non-zero result 
for positive times and a zero result for negative times. 
In the case of an unbounded particle only outgoing waves are 
consistent with this description.

\sheadC{Analytic continuation}

The resolvent is well defined for any $z$ away 
from the real axis. We define $G^{+}(z) = G(z)$  
in the upper half of the complex plane. 
As long as we discuss bounded systems this "definition" 
looks like a duplication. The mathematics becomes 
more interesting once we consider unbounded systems 
with a continuous energy spectrum.
In the latter case there are circumstances that allow 
analytic continuation of $G^{+}(z)$ into the lower half 
of the complex plane.
This analytical continuation, if exists, would not  
coincide with $G^{-}(z)$.

In order to make the discussion of analytical continuation  
transparent let us assume, without loss of generality,    
that we are interested in the following object: 
\beq
f(z) \ = \  \langle \Psi |G(z)| \Psi \rangle 
\ = \ \sum_n \frac{q_n}{z-E_n} 
\eeq
The function $f(z)$ with $z=x+iy$ can be regarded 
as describing the electric field in a two dimensional  
electrostatic problem. The field is created by charges 
that are placed along the real axis.
As the system grows larger and larger the charges become 
more and more dense, and therefore in the "far field" 
the discrete sum $\sum_n$ can be replaced 
by an integral $\int \gdos(E)dE$ where $\gdos(E)$ 
is the smoothed density of states.  
By "far field" we mean that $\im[z]$ is much larger  
compared with the mean level spacing and therefore 
we cannot resolve the finite distance between the charges. 
In the limit of an infinite system this becomes exact for 
any finite (non-zero) distance from the real axis.

In order to motivate the discussion of analytical 
continuation let us consider a typical problem.
Consider a system built of two weakly coupled 1D regions. 
One is a small "box" and the other is a very large "surrounding". 
The barrier between the two regions is a large delta function.  
According to perturbation theory the zero order states 
of the "surrounding" are mixed with the zero order bound 
states of the "box". The mixing is strong if the energy 
difference of the zero order states is small. 
Thus we have mixing mainly in the vicinity 
of the energies $E_r$ where we formerly 
had bound states of the isolated "box". 
Let us assume that $\Psi$ describes the initial 
preparation of the particle inside the "box". 
Consequently we have large $q_n$ only for 
states with $E_n \approx E_r$.  
This means that we have an increased "charge density"  
in the vicinity of energies $E_r$. It is the LDOS rather 
than the DOS which is responsible for this increased 
charge density. Now we want to calculate $f(z)$. 
What would be the implication of the increased 
charge density on the calculation?

In order to understand the implication of the increased 
charge density on the calculation we recall a familiar 
problem from electrostatics. Assume that we have a conducting 
metal plate and a positive electric charge. Obviously there will 
be an induced negative charge distribution on the metal plate. 
We can follow the electric field lines through the plate, from above
the plate to the other side. We realize that we can replace all the
charge distribution on the plate by a single negative electric charge (this
is the so called "image charge").

\begin{center}
\putgraph[0.8\hsize]{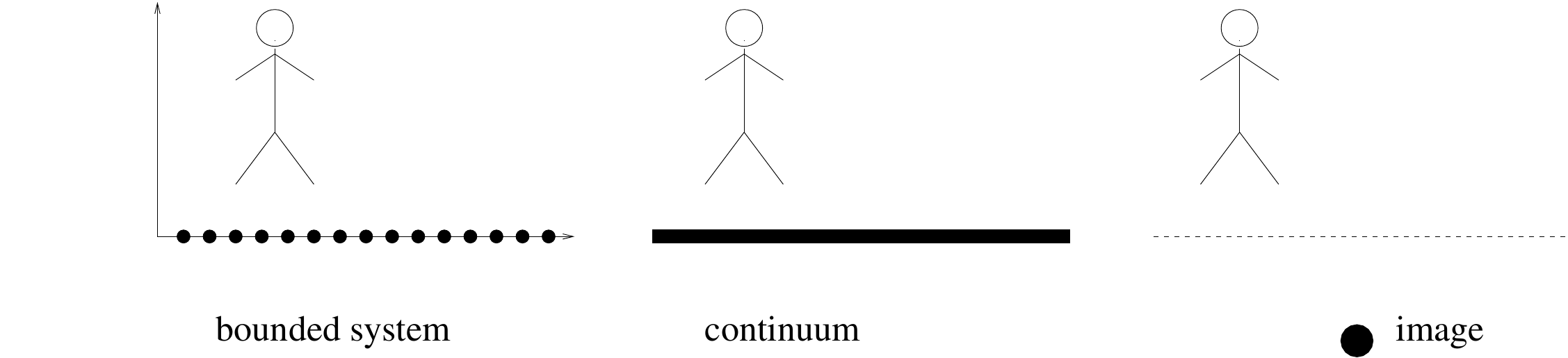} 
\end{center}

Returning to the resolvent, we realize that we can 
represent the effect of the increased "charge density"  
using an "image charge" which we call a "resonance pole". 
The location of the "resonance pole" is written as $E_r-i(\Gamma_r/2)$.
Formally we say that the resonance poles are 
obtained by the analytic continuation of $G(z)$ from the 
upper half plane into the lower half plane. 
In practice we can find these poles by looking for 
complex energies for which the Schrodinger equation 
has solutions with "outgoing" boundary conditions.    
In another section we give an explicit solution 
for the above problem. Assume that this way or another 
we find an approximation for $f(z)$ using such "image charges":
\beq
f(E) \ = \ \langle \Psi | G^{+}(E) | \Psi \rangle 
\ = \ \sum_n \frac{q_n}{E-(E_n-i0)} 
\ = \  \sum_r \frac{Q_r}{E-(E_r-i(\Gamma_r/2))}
+ \mbox{smooth background} 
\eeq
We observe that the sum over $n$ is in fact an integral, 
while the sum over $r$ is a discrete sum. 
So the analytic continuation provides a simple 
expression for $G(z)$. Now we can use this 
expression in order to deduce physical information. 
We immediately find the LDOS can 
be written as as sum over Lorentzians. 
The Fourier transform of the LDOS is the 
survival amplitude, which comes out a sum over 
exponentials. In fact we can get the result 
for the  survival amplitude directly by 
recalling that $\Theta(t)U(t)$ is the FT of $iG^{+}(\omega)$.  
Hence 
\beq
\langle x| U(t) | x_0 \rangle \ = \ 
\sum_r Q_r \eexp{ -iE_rt - (\Gamma_r/2)t }
+ \mbox{short time corrections} 
\eeq
If $\Psi$ involves contribution from only  
one resonance, then the probability to stay 
inside the "box"  decreases exponentially with time. 
This is the type of result that we would expect 
form either Wigner theory or from the Fermi Golden rule. 
Indeed we are going later to develop  
a perturbation theory for the resolvent, 
and to show that the expression for $\Gamma$ 
in leading order is as expected.

\sheadC{The Green function via direct diagonalization}

For a free particle the eigenstates are known and we can calculate 
the expression by inserting a complete set and integrating.
From now on we set $\mass=1$ in calculations, but we restore it in 
the final result.
\beq
G^+(x|x_0)&=&
\sum_k\langle x|k\rangle \frac{1}{E-\frac 12 k^2 +i0}\langle k|x_0 \rangle 
\\ \nonumber
&=&\int\frac{d\mathbf{k}}{(2\pi)^d}\eexp{i\mathbf{k}\cdot\mathbf{r}}
\frac{1}{E-\frac 12 k^2 +i0}
\eeq
where $d$ is the dimension of the space. In order to compute this expression 
we define $\vec{r}=\vec{x}-\vec{x_0}$
and choose our coordinate system in such a way 
that the $\hat{z}$ direction will coincide with the direction of $\mathbf{r}$.
\beq
G^+(x|x_0)=\int \frac{2\eexp{ikr\cos\theta}}{{k_E}^2-k^2+i0}\frac{d\Omega k^{d-1}dk}{(2\pi)^d}
\eeq
where $k_E=\sqrt{2 \mass E}$ is the wavenumber for a particle with energy $E$. The integral is a
$d$-dimensional spherical integral. The solutions of $|\mathbf{k}|=k$ in 1D give two $k$'s, 
while in 2D and 3D the $k$'s lie on a circle and on a sphere respectively. 
We recall that $\Omega_d$ is $2,2\pi,4\pi$ in 1D, 2D and 3D respectively, and define   
averaging over all directions for a function $f(\theta)$ as follows:
\beq
\langle f(\theta)\rangle_d=\frac{1}{\Omega_d}\int f(\theta) d\Omega
\eeq
With this definition we get
\beq
\langle \eexp{ikr\cos\theta}\rangle_d
=\left\{
\begin{array}{ll}
\cos(kr), & \hbox{d=1;} \\
\mathrm{J_0}(kr), & \hbox{d=2;} \\
\sinc(kr), & \hbox{d=3.} \\
\end{array}
\right.
\eeq
where $\mathrm{J_0}(x)$ is the zero order \textit{Bessel function of the first kind}. 
Substituting these results and using the notation $z=kr$ we get in the 3D case:
\beq
G^+(\mathbf{r})
&=&
\frac{1}{\pi^2 r} 
\frac 12 {\int_{-\infty}^\infty} 
\frac{z\sin z}{{z_E}^2-z^2+i0}dz
=
\frac{1}{\pi^2 r} 
\frac {1}{4i} 
{\int_{-\infty}^\infty} 
\frac{z(\eexp{iz}-\eexp{-iz})}{{z_E}^2-z^2+i0}dz 
\\ \nonumber
&=&
\frac{1}{\pi^2 r} 
\frac {1}{4i}                     
\left[-\int \frac{z \eexp{iz}}{(z-(z_E+i0))(z+(z_E+i0))}dz                          
+\int \frac{z \eexp{-iz}}{(z-(z_E+i0))(z+(z_E+i0))}dz\right] 
\\ \nonumber
&=&
\frac{1}{2\pi r}
\sum_{\tbox{poles}}\mathrm{Res}[f(z)]
=
\frac{1}{2\pi r} 
\left[-\frac 12 \eexp{iz_E}-\frac 12 \eexp{-i(-z_E)}\right]
=
-\frac{\mass}{2\pi} \frac{\eexp{ik_E r}}{r}
\eeq

\begin{center}
\putgraph[0.4\hsize]{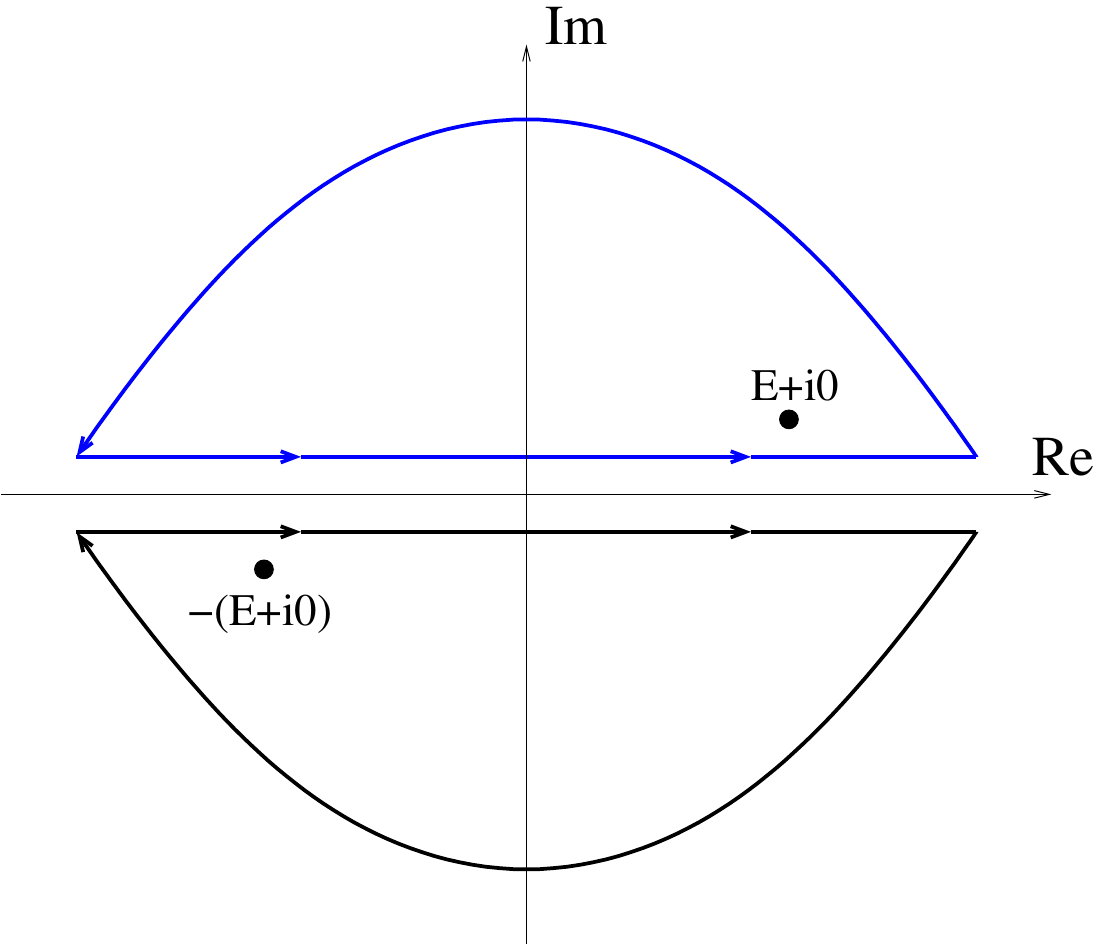}
\end{center}

The contour of integration (see figure) has been closed 
in the upper half of the plane for the term containing $\eexp{iz}$,  
and in the lower half for the term with $\eexp{-iz}$, 
so the arc at infinity add zero contribution. 
Then the contour has been deformed, so the contribution for 
each term comes from one of the poles at $\pm(z_E+i0)$. 
We see that the solution is a variation on the familiar Coulomb law.

\sheadC{Solving the Helmholtz equation in 1D/2D/3D}

The simplest method to find the Green function 
is by solving the Schrodinger equation with 
a source and appropriate boundary conditions. 
In the case of a free particle we get 
the Helmholtz equation which is a generalization 
of the Poisson equation of electrostatic problems:  
\beq
(\nabla^2 +{k_E}^2)G(\mathbf{r}|\mathbf{r}_0)=
-q\delta(\mathbf{r}-\mathbf{r}_0)
\eeq
where the "charge" in our case is $q=-2\mass/\hbar^2$. 
For $k_E=0$ this is the Poisson equation 
and the solution is the Coulomb law. 
For $k_E\neq0$ the solution is a variation on Coulomb law.
We shall explore below the results in case of 
a particle in 3D, and then also for 1D and 2D.

\ \\
{\bf The 3D case:}

In the 3D case the "Coulomb law" is:
\beq
G(\mathbf{r}|\mathbf{r}_0)=
\frac{q}{4\pi |\mathbf{r}-\mathbf{r}_0|}\cos(k_E|\mathbf{r}-\mathbf{r}_0|)
\eeq
This solution still has a gauge freedom, just as in electrostatics where 
we have a "free constant". We can add to this solution any "constant", which in our case means 
an arbitrary (so called "free wave") solution of the homogeneous equation. 
Note that any "free wave" can be constructed from a superpostion of planar waves.   
In particular the "spherical" free wave is obtained 
by averaging $\eexp{i\mathbf{k} \cdot \mathbf{r}}$ 
over all directions. If we want to satisfy 
the "outgoing wave" boundary conditions we get:  
\beq
G(r)=\frac{q}{4\pi r}\cos(k_E r)+i\frac{q}{4\pi r}\sin(k_E r)=
\frac{q}{4\pi r}\eexp{ik_Er} = -\frac{\mass}{2\pi r}\eexp{ik_Er}
\eeq
The solutions for 1D and 2D can be derived in the same way.

\ \\
{\bf The 1D case:}

In the one dimensional case the equation that determines the Green function is   
\beq
\left(\frac{\partial^2}{\partial x^2}+k_E^2\right)G(x)=-q \delta(x)
\eeq
where for simplicity we set $x_0=0$.
The delta function source requires a jump of  
the derivative ${G'(+0)-G'(-0)=-q}$.  
Using different phrasing: in order for the second derivative 
to be a delta function the first derivative 
must have a step function. 
Thus, when $k_E=0$ we get 
the 1D Coulomb law $G(x)=-(q/2)|x|$, 
but for $k_E\neq0$ we 
have a variation on Coulomb law
\beq
G(x)=-\frac{q}{2k_E}\sin(k_E|x|)
\eeq
(see figure). To this we can add any 1D free wave.  
In order to satisfy the "outgoing waves" boundary conditions 
we add $\cos(kx)=\cos(k|x|)$ to this expression, 
hence we get the retarded Green's function in 1D
\beq
G(x)=i\frac{q}{2k_E}\eexp{ik_E|x|} = -i\frac{\mass}{k_E}\eexp{ik_E|x|} 
\eeq

\begin{center}
\putgraph[0.4\hsize]{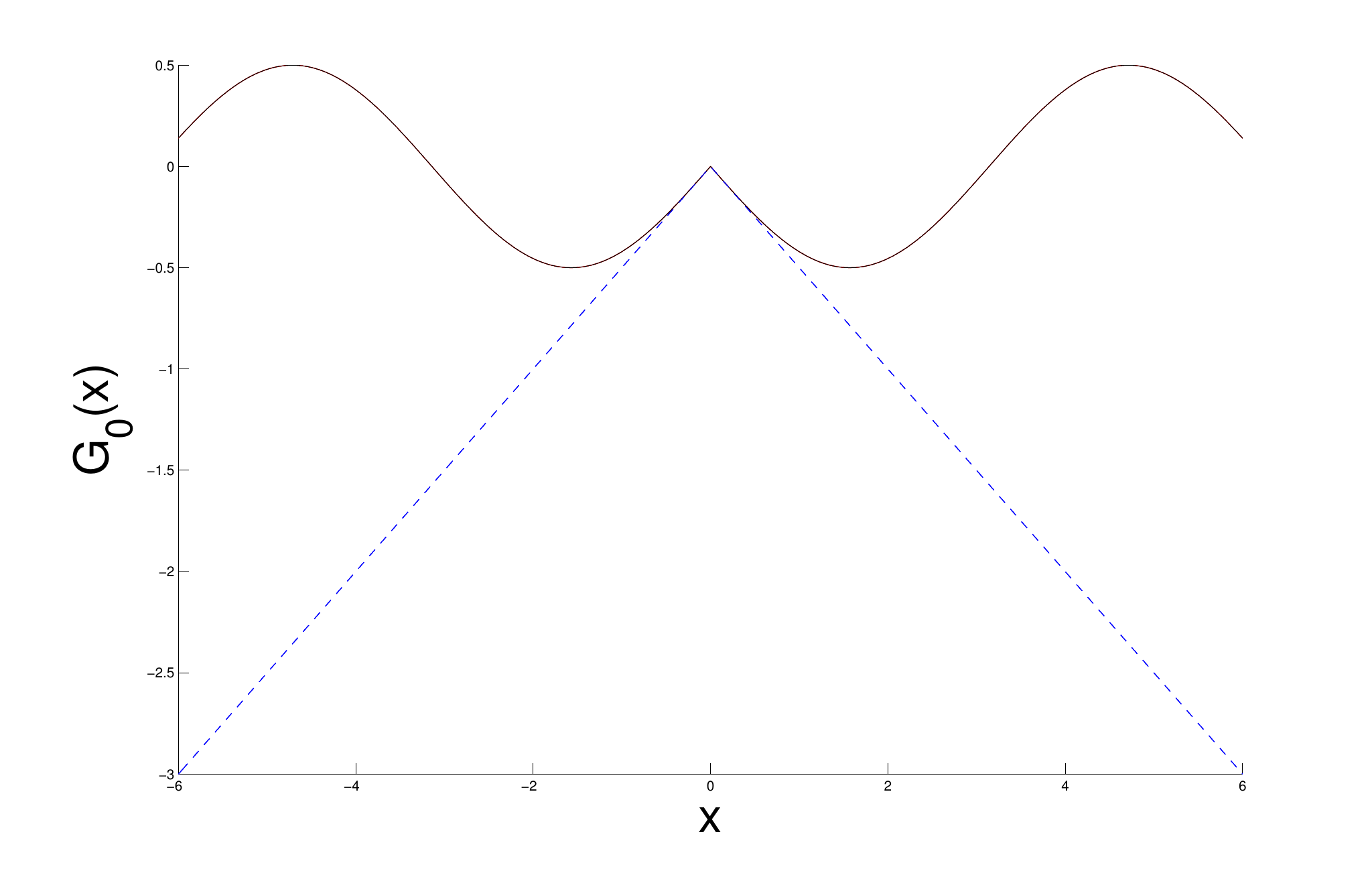}
\end{center}

\ \\
{\bf The 2D case:}

In two dimensions for $k_E=0$ 
we use Gauss' law to calculate the 
electrostatic field that goes 
like $1/r$, and hence the 
electrostatic potential is 
$G(r)=-(1/(2\pi))\ln r$.
for $k_E\neq0$ we get the modulated result
\beq
G(r) = -\frac{q}{4}\mathrm{Y_0}(k_Er)  
\eeq
where $\mathrm{Y_0}(x)$ 
is the Bessel function of the second kind
\beq
\mathrm{Y_0}'(x) &=& -\mathrm{Y}_1(x) 
\\ \nonumber
\mathrm{Y_0}(x) &\sim& \sqrt{\frac{2}{\pi x}} \sin(x-\frac \pi 4)
\;\;\;,\;\; \textrm{for large $x$}
\eeq
The second independent solution of the associated 
homogeneous equation is the 
Bessel function of the first kind $\mathrm{J_0}(x)$. 
Adding this "free wave" solution leads to 
a solution that satisfies the "outgoing wave" boundary 
conditions. Hence we get the retarded Green's function in 2D
\beq
G(r)=i\frac{q}{4}\mathrm{H_0}(k_Er) = -i\frac{\mass}{2}\mathrm{H_0}(k_Er)
\eeq
where $\mathrm{H_0}(x)=\mathrm{J_0}(x)+i\mathrm{Y_0}(x)$ 
is the Hankel function.

\sheadC{Green function for a quasi 1D network}

There is a very simple expression for the Green function of a particle 
that has energy~$E$ in a quasi 1D network (see "QM is practice" section):
\beq
G(x|x_0) = -i\frac{1}{v_E} \sum_p A_p(E) \, \eexp{ik_E L_p}
\eeq
where $p$ labels trajectories that start at~$x_0$ and end at~$x$.  
The length of a given trajectory is~$L_p$, while $A_p(E)$ 
is the product of transmission and reflection amplitudes 
due to scattering at the nodes of the network.
The proof of this formula is a straightforward generalization 
of the standard 1D case: The function $\Psi(x)=G(x|x_0)$ satisfies the Schrodinger 
equation for any $x\ne x_0$, and in particular the matching 
conditions at the nodes are satisfied by construction.
At $x_0$ the function $\Psi(x)$ has the same discontinuity 
as in the standard 1D case, because for the direct trajectory $L_p=|x-x_0|$ 
has a jump in its derivative. 
This discontinuity reflects having a source at $x=x_0$ as required.

{\bf Delta scatterer.-- } 
Possibly the simplest quasi 1D networks, is a line with one node. 
We can describe the node as a delta function potential:
\beq
V(x) \ \ = \ \ u\delta(x) \ \ = \ \   |0 \rangle u \langle 0|
\eeq
In the last equality we point out that $V(x)$ is essentially 
a projector $|\psi\rangle\langle \psi|$. For this type of potential   
there is a simple inversion formula that we discuss below,  
that allows a direct evaluation of the resolvent.     
Optionally one can calculate the reflection and the transmission 
coefficients for a delta function potential, and use the 
general formula of the previous paragraph, leading to 
\beq
G(x|x_0) \ \ = \ \ -i\frac{1}{v_E} \left[ \eexp{ik_E (x-x_0)} + r \eexp{ik_E(|x|+|x_0|)}\right] 
\eeq
where $r$ is the reflection coefficient of a delta scatterer.

{\bf Inversion formula.-- }
Given a state $\psi$  we have the identity 
\beq
\frac{1}{1-|\psi\rangle\lambda\langle \psi|} 
\ \ = \ \ 
\bf{1}+\frac{1}{1-\lambda}|\psi\rangle\lambda\langle \psi|
\eeq
This is proved by selecting an orthonormal basis that contains $\psi$ 
as one of its states, such that the matrix in the denominator 
is ${\mbox{diag}\{1{-}\lambda,1,1,1,...\}}$.  
A generalization of this formula for any invertible $A$, 
and $B=|\psi\rangle\lambda\langle\psi|$ is
\beq
\frac{1}{A-B} 
\ \ = \ \ 
\frac{1}{\sqrt{A}} \left[1- \frac{1}{\sqrt{A}}B\frac{1}{\sqrt{A}}\right]^{-1} \frac{1}{\sqrt{A}}
\ \ = \ \ 
\frac{1}{A}+  \left(\frac{1}{1-\trc(A^{-1}B)}\right) \frac{1}{A}B\frac{1}{A}
\eeq
We can use this formula in order to find the resolvent 
in the case of a delta function:
\beq
G \ \ = \ \ 
\frac{1}{(E-H_0)-V} 
\ \ = \ \ 
G_0 +  \left(\frac{1}{1-u\mathcal{G}(E)}\right) G_0 V G_0 
\ \ \equiv \ \ 
G_0 +  G_0 T G_0  
\eeq
where $\mathcal{G}(E)=G_0(0|0)$, and $T$ is like an effective interaction   
\beq
T \ \ = \ \ \left(\frac{1}{1-u\mathcal{G}(E)}\right) V \ \ = \ \  \frac{u}{1+i(u/v_E)}\delta(x) 
\eeq
By inspection one can verify that this result is 
consistent with the result that we have cited at the 
beginning of this section.   
We shall further discuss the significance of this expression 
and the details of the $\mathcal{G}(E)$ calculation in 
the lecture regarding the $T$~matrix formalism.

\sheadC{Semiclassical evaluation of the Green function}

The semiclassical procedure, based on the Van-Vleck formula, 
provides a common approximation for all the special cases 
that have been discussed above, and also allows further 
generalization. The Green function is the FT of the propagator:
\beq
G(x|x_0)
\ \ \sim \ \ \int_0^{\infty} \eexp{i\mathcal{A}_{\tbox{cl}}(x,x_0)+iEt}dt 
\ \ \sim \ \ \eexp{i\mathcal{S}_{\tbox{cl}}(x,x_0)}
\eeq
In order to derive the last expression one applies the stationary 
phase approximation for the evaluation of the $dt$ integral. 
The stationary point is determined by 
the equation ${E_{\tbox{cl}}=E}$ 
where ${E_{\tbox{cl}} \equiv -d\mathcal{A}_{\tbox{cl}} / dt}$.
The reduced action is defined as 
\beq
\mathcal{S}_{\tbox{cl}}(x,x_0) 
= \mathcal{A}_{\tbox{cl}} + E 
= \int_{0}^{t}[\mathcal{L}_{\tbox{cl}} + E]dt 
= \int_{x(0)}^{x(t)} p_{\tbox{cl}} \cdot dx
\eeq
It is implicit in this definition that the 
reduced action is evaluated for a classical trajectory 
that has energy~$E$ and connects the points $x(0)$ and $x(t)$,  
and likewise the classical momentum  $p_{\tbox{cl}}$ 
is evaluated along this trajectory.

\sheadC{The boundary integral method}

The Schr\"{o}dinger equation $\mathcal{H}\psi=E\psi_0$ can be 
written as $\mathcal{H}_E\psi=0$ where 
\beq
\mathcal{H}_E=-\nabla^2 + U_E(r) 
\eeq
with $U_E(r)=U(r)-E$. Green's function solves 
the equation $\mathcal{H}_E G(r|r_0)=-q\delta(r-r_0)$ 
with $q=-2\mass/\hbar^2$. In this section we follow the convention 
of electrostatics and set $q=1$. From this point on we use 
the following (generalized) terminology:
\beq
\mbox{Laplace equation [no source]:} 
\hspace*{2cm} &&
\mathcal{H}_E \psi(r) = 0
\\
\mbox{Poisson equation:} 
\hspace*{2cm} &&
\mathcal{H}_E \psi(r) = \rho(r)
\\
\mbox{Definition of the Coulomb kernel:}
\hspace*{2cm} &&
\mathcal{H}_E G(r|r_0)=\delta(r-r_0) 
\eeq
The solution of the Poisson equation is unique 
up to an arbitrary solution of the associated Laplace equation.  
If the charge density $\rho(r)$ is real, 
then the imaginary part of $\psi(r)$ is a solution 
of the Laplace equation. Therefore without loss of generality we 
can assume that $\psi(r)$ and $G(r|r_0)$ are real. 
From the definition of the Coulomb kernel it follows 
that the solution of the Poisson equation is 
\beq
\mbox{Coulomb law:}
\hspace*{2cm}
\psi(r) \ \ = \ \ \int G(r|r') \rho(r') dr'
\eeq
In particular we write the solution which is obtained 
if we have a charged closed boundary ${r=r(s)}$. 
Assuming that the charge density on the boundary   
is~$\sigma(s)$ and that the dipole density 
(see discussion below) is~$d(s)$ we get:
\beq
\psi(r) \ \ = \ \ 
\oint \Big( [G(r|s)] \sigma(s) + [\partial_s G(r|s)] d(s) \Big) ds
\eeq
We use here the obvious notation ${G(r|s)=G(r|r(s))}$.  
The normal derivative ${\partial=\vec{n}\cdot\nabla}$ 
is taken with respect to the source coordinate,  
where $\vec{n}$ is a unit vector that points outwards.

It should be obvious that the obtained $\psi(r)$ solves 
the Laplace equation in the {\em interior} region, 
as well as in the {\em exterior} region, 
while across the boundary it satisfies the matching conditions    
\beq
\mbox{Gauss law:} 
\hspace*{2cm} 
\partial\psi(s^{+})-\partial\psi(s^{-}) &=& -\sigma(s)
\\
\psi(s^{+}) - \psi(s^{-}) &=& d(s)
\eeq
These matching conditions can be regarded 
as variations of Gauss law. They are obtained 
by integrating the Poisson equation 
over an infinitesimal range across the boundary.  
The charge density  $\sigma(s)$ implies 
a jump in the electric field $-\partial\psi(s)$.
The dipole density $d(s)$ is formally like a very 
thin parallel plates capacitor, and it implies 
a jump in the potential $\psi(s)$.

Let us ask the inverse question: 
Given a solution of Laplace equation 
in the interior region, can we find $\sigma(s)$ 
and $d(s)$ that generate it? The answer is {\em yes}. 
In fact, as implied from the discussion below, 
there are infinitely many possible choices. 
But in particular there is one unique choice that 
gives $\psi(r)$ inside and {\em zero} outside.
Namely, ${\sigma(s)=\partial \psi(s)}$ and ${d(s)=-\psi(s)}$, 
where $\psi(s) = \psi(r(s))$. Thus we get: 
\beq
\mbox{The bounday integral formula:} 
\hspace*{2cm} 
\psi(r) \ \ = \ \ 
\oint \Big( [G(r|s)] \partial\psi(s) - [\partial_s G(r|s)] \psi(s) \Big) ds
\eeq
The bounday integral formula allows to express the $\psi(r)$ 
at an arbitrary point inside the domain 
using a boundary integral over $\psi(s)$ 
and its normal derivative $\partial \psi(s)$.

The standard derivation of the boundary integral formula
is based on formal algebraic manipulations with Green's theorem. 
We prefer below a simpler physics-oriented argumentation. 
If $\psi(r)$ satisfies Laplace equation in the {\em interior},  
and it is defined to be {\em zero} in the exterior, 
then it satisfies (trivially) the Laplace equation also 
in the {\em exterior}. On top it satisfies the Gauss matching 
conditions with ${\sigma(s)=\partial \psi(s^{-})}$ and ${d(s)=-\psi(s^{-})}$. 
Accordingly it is a solution of the Poisson equation
with $\sigma(s)$ and $d(s)$ as sources. 
But for the same $\sigma(s)$ and $d(s)$ we can optionally 
obtain another solutions of the Poisson equation
from the bounday integral formula. 
The two solutions can differ by 
a solution of the associated Laplace equation.  
If we supplement the problem with zero boundary conditions 
at infinity, the two solutions have to coincide.

For the case where the wave function vanishes 
on the boundary $\psi(s)=0$
the expression becomes very simple
\beq
\psi(r) = \oint G(r|s') \varphi(s') ds'
\eeq
where $\varphi(s)=\partial\psi(s)$, 
and in particular as we approach 
the boundary we should get: 
\beq
\int G(s|s')\varphi(s')ds' = 0
\eeq
An obvious application of this formula 
leads to a powerful numerical method 
for finding eigenfunctions. This is 
the so called \textit{boundary integral method}.
Let us consider the problem of a
particle in a billiard potential. 
Our Green's function is (up to a constant) 
\beq
G(s|s')=\mathrm{Y_0}(k_E|r(s)-r(s')|)
\eeq
If we divide the boundary line into $N$ segments 
then for any point on the boundary the 
equality $\int G(s|s')\varphi(s')ds' = 0$ should hold, so:
\beq
\sum_j A_{ij} \varphi_j = 0
\ \ \ \ \ \ \ \ \ \ \mbox{with} \ \ 
A_{ij} = \mathrm{Y_0}(k_E|r(s_i)-r(s_j)|) 
\eeq
Every time the determinant $\det(A)$
vanishes we get a non trivial solution 
to the equation, and hence we can construct 
an eigenfunction. So all we have to do is 
plot the determinant $\det(A)$ as a function of $k_E$. 
The points where $\det(A)$ equals zero are the values of $k_E$
for which the energy $E$ is an eigenvalue of the Hamiltonian $\mathcal{H}$.

\newpage

\sheadB{Perturbation theory}

\sheadC{Perturbation theory for the resolvent}

We first generalize the "inversion formula" that has been 
introduced in the previous lecture. There it has been 
applied to the exact calculation of the Green function for 
a delta function. Here we consider general matrices $A$ and $B$, 
and point out the formal identity  
\beq
\frac{1}{(1-B)}
\ \ &=& \ \ (1-B)^{-1}
\ \ = \ \ \sum_{n=0}^{\infty}{B^n}
\\ \nonumber
\frac{1}{A-B}
\ \ &=& \ \ (A(1-A^{-1}B))^{-1}
\ \ = \ \ (1-A^{-1}B)^{-1}A^{-1}
\ \ = \ \ \sum_n (A^{-1}B)^n A^{-1}
\\ 
\ \ &=& \ \ \frac{1}{A}+\frac{1}{A}B\frac{1}{A}
+ \frac{1}{A}B\frac{1}{A}B\frac{1}{A} + \dots 
\eeq
Applying the above formal expansion for the calculation 
of the resolvent one obtains: 
\beq
G(z) \ \ &=& \ \ \frac{1}{z-\mathcal{H}}  \ \ =  \ \  \frac{1}{z-(\mathcal{H}_0+V)} \ \ = \ \ \frac{1}{(z-\mathcal{H}_0)-V} 
\\ 
\ \ &=& \ \ G_{0}(z)+G_{0}(z)VG_{0}(z)+G_{0}(z)VG_{0}(z)VG_{0}(z)+ \dots 
\eeq
Or, in matrix representation
\beq
G(x|x_0) 
=  G_0(x|x_0)
+ \int  G_0(x|x_2) dx_2 \langle {x_{2}}|V|x_{1} \rangle dx_1 G_0(x_1|x_0) +  \dots 
\eeq
Note that for the scalar potential $\hat{V}=u(\hat{x})$ we get 
\beq
G(x|x_0) 
=  G_0(x|x_0)
+ \int dx' G_0(x|x') u(x') G_0(x'|x_0) +  \dots 
\eeq

\sheadC{Perturbation Theory for the Propagator}

For the Green function we get
\beq
G^+(\omega)=G^+_{0}(\omega)
+G^+_{0}(\omega)VG^+_{0}(\omega)
+G^+_{0}(\omega)VG^+_{0}(\omega)VG^+_{0}(\omega)+ \dots 
\eeq
Recall that 
\beq
G^{+}(\omega)\rightarrow FT\rightarrow -i\Theta(\tau)U(\tau)
\eeq
Then from the convolution theorem it follows that 
\beq
(-i)[\Theta(t)U(t)]
= (-i)[\Theta U_{0}(t)] 
+(-i)^2\int dt' [\Theta(t-t^{'}) U(t-t^{'})]  
\, [V] \, [\Theta(t^{'}) U_0(t^{'})] + \dots
\eeq
which leads to 
\beq
U(t)=U_{0}(t)
+\sum_{n=1}^{\infty}(-i)^{n}
\int_{0<t_{1}<t_{2}<\dots<t_{n}<t}
dt_{n} \dots dt_{2}dt_{1} \,
U_{0}(t-t_{n}) V \dots U_{0}(t_{2}-t_{1})V
U_{0}(t_{1})
\eeq
for $t>0$ and zero otherwise. 
This can be illustrated diagrammatically 
using Feynman diagrams.

Let us see how we use this expression in order 
to get the transition probability formula. 
The first order expression for the evolution operator is 
\beq
U(t)= U_{0}-i\int dt^{'} U_{0}(t-t^{'})V U_{0}(t^{'})
\eeq
Assume that the system is prepared in an eigenstate~$m$ 
of the unperturbed Hamiltonian, we get the amplitude 
to find it after time~$t$ in another eigenstate~$n$ is 
\beq
\langle n|U(t)|m \rangle
= \eexp{-iE_{n}t}\delta_{nm}
-i\int dt^{'} 
\eexp{-iE_{n}(t-t^{'})} 
\langle n|V|m \rangle
\eexp{-i E_{m}t^{'}}
\eeq
If $n \neq m$ it follows that:
\beq
P_t(n|m) = 
|\langle n|U(t)|m \rangle|^2
= \left| 
\int_{0}^{t} dt' V_{nm}
\eexp{ i (E_n-E_m) t'}
\right|^2
\eeq

\sheadC{Perturbation theory for the evolution}

In this section we review the elementary approach 
to solve the evolution problem via an iterative 
scheme with the Schrodinger equation. Then we 
make the bridge to a more powerful procedure.  
Consider 
\beq
|\Psi \rangle =\sum_{n} \Psi_{n}(t)|n \rangle
\eeq
It follows that:
\beq
i\frac {\partial \Psi_{n}}{\partial
t}=E_{n}\Psi_{n}+\sum_{n^{'}}V_{nn^{'}}\Psi_{n'}
\eeq
We want to solve in the method which 
is called "variation of parameters", so we set:
\beq
\Psi_{n}(t)=c_{n}(t)\eexp{-iE_{n}t}
\eeq
Hence:
\beq
\frac{dc_{n}}{dt} \ = \ -i \sum_{n'} \eexp{i(E_{n}-E_{n'})t} V_{nn'}c_{n'}(t)
\eeq
From the zero order solution $c_{n}(t)=\delta_{nm}$ 
we get after one iteration:
\beq
c_{n}(t)=\delta_{nm}-i\int_{0}^{t}dt' \eexp{i(E_{n}-E_{m})t'}V_{nm}
\eeq
In order to make the connection with the formal approach 
of the previous and of the next section we write  
\beq
c_{n} = \eexp{iE_{n}t} \Psi_n 
= \BraKet{ n }{ U_{0}^{-1} U(t) }{ m } 
\equiv \BraKet{ n }{ U_I(t) }{ m } 
\eeq
and note that 
\beq
\eexp{ i(E_{n}-E_{m})t }  V_{nm}   
\,\, = \,\, 
\BraKet{ n }{ U_0(t)^{-1} V U_0(t) }{ m }
\,\, \equiv \,\,
\BraKet{ n }{ V_I(t) }{ m }
\eeq
Hence the above first order result can be written as 
\beq
\BraKet{ n }{ U_I(t) }{ m }
= \delta_{nm} - i\int_0^t  \BraKet{ n }{ V_I(t') }{ m }  dt'
\eeq
In the next sections we generalize this result 
to all orders.

\sheadC{The Interaction Picture}

First we would like to recall 
the definition of time ordered exponentiation
\beq
U(t,t_0)=(1-idt_N \mathcal{H}(t_N)) \dots (1-idt_2\mathcal{H}(t_2))(1-idt_1\mathcal{H}(t_1)) 
\equiv \mathcal{T}\eexp{-i \int^t_{t_0} \mathcal{H}(t') dt'} 
\eeq
Previously we have assumed that the Hamiltonian is not time dependent. 
But in general this is not the case, so we have to keep the time order. 
The parenthesis in the above definition can be "opened" and then we 
can assemble the terms of order of $dt$. Then we get the expansion
\beq
U(t,t_0) &=& 
1-i(dt_{N}\mathcal{H}(t_{N}))\dots-i(dt_{1} \mathcal{H} (t_{1}))
+(-i)^2 (dt_{N}\mathcal{H}(t_{N}))(dt_{N-1}\mathcal{H}(t_{N-1}))+\dots
\\ \nonumber
&=& 1-i\int_{t_0<t'<t} \mathcal{H}(t')dt' 
+ (-i)^{2}\int_{t_0<t'<t''<t} \mathcal{H}(t'')\mathcal{H}(t')dt''dt'+\dots 
\\ \nonumber
&=& \sum_{n=0}^{\infty}(-i)^{n} \int_{t_0<t_{1}<t_{2}\dots <t_{n}<t}
dt_{n}\dots dt_{1} \mathcal{H}(t_{n})\dots \mathcal{H}(t_{1})
\eeq
Note that if $\mathcal{H}(t')=\mathcal{H}$ is not time dependent then we simply get 
the usual Taylor expansion of the exponential function where 
the $1/n!$ prefactors would come from the time ordering limitation.

The above expansion is not very useful because the sum is likely 
to be divergent. What we would like to consider is the case  
\beq
\mathcal{H} \ \ = \ \ \mathcal{H}_0+V 
\eeq
where $V$ is a small perturbation. The perturbation $V$ can be either 
time dependent or time independent. It is convenient 
to regard the evolution as a sequence of "free" evolution during ${[t_0,0]}$, 
an interaction at ${t=0}$, and a later "free" evolution during ${[0,t]}$. 
Consequently we define a unitary operator $U_I$ such that
\beq
U(t,t_0) \ \ = \ \ U_0(t,t_0) \, U_{I} \, U_0(0,t_0)
\eeq 
Accordingly $U_I=\bm{1}$ if $V=0$. Otherwise $U_I$ gives the 
effect of the interaction as if it happened at the reference time~${t=0}$.
For simplicity we adopt from now on the convention ${t_0=0}$, 
and use the notation $U(t)$ instead of $U(t,t_0)$. Thus 
\beq
U_{I} \ \ \equiv \ \ U_{0}(t)^{-1} \ U(t) 
\eeq 
We also define the following notation
\beq
V_{I}(t) \ \ = \ \  U_{0}(t)^{-1} \ V \ U_{0}(t)  
\eeq
By definition of $\mathcal{H}$ as the generator of the evolution:
\beq
\frac{d}{dt}U(t) \ \ = \ \ -i(\mathcal{H}_0 + V) \ U(t)
\eeq
Consequently:
\beq
\frac{d}{dt}U_I \ \ =  \ \ -iV_I(t) \ U_I
\eeq
The solution is by time ordered exponentiation 
\beq
U_{I} \ \ = \ \ 
\mathcal{T}\exp\left(-i\int_{0}^{t} V_{I}(t')dt'\right) 
\ \ = \ \ 
\sum_{n=0}^{\infty}(-i)^{n} \int_{0<t_{1}<t_{2}\dots <t_{n}<t}
dt_{n}\dots dt_{1} \, V_I(t_{n})\dots V_I(t_{1})
\eeq
Which can be written formally as 
\beq
U_{I} \ \ = \ \ 
\sum_{n=0}^{\infty}\frac{(-i)^{n}}{N!} 
\int_{0}^{t} dt_{n}\dots dt_{1} \, \mathcal{T} V_I(t_{n})\dots V_I(t_{1})
\eeq
Optionally we can switch back to the Schrodinger picture: 
\beq
U(t) \ \ = \ \ \sum_{n=0}^{\infty}(-i)^{n}
\int_{0<t_{1}<t_{2}<\dots <t_{n}<t}
dt_{n}\dots dt_2 dt_{1} \,
U_{0}(t-t_{n}) V \dots  U_{0}(t_{2}-t_{1})V
U_{0}(t_{1})
\eeq
The latter expression is more general than the one 
which we had obtained via FT of the resolvent expansion, 
because here $V$ is allowed to be time dependent.

\sheadC{The Kubo formula}
\label{sKubo}

Consider the special case of having 
a time dependent perturbation $V=-f(t)\hat{B}$. 
The first order expression for the 
evolution operator in the interaction picture is  
\beq
U_I(t) \ \ = \ \ 1 + i\int_{0}^{t} A_I(t') f(t')dt' 
\eeq
where $A_I(t)$ is $A$ in the interaction picture. 
If our interest is in the evolution of an expectation value 
of another observable $\hat{A}$, then we have the identity
\beq
\langle A \rangle_t  
\ \ = \ \ 
\langle\psi(t)|A|\psi(t)\rangle 
\ \ = \ \ 
\langle\psi|A_H(t)|\psi\rangle
\ \ = \ \ 
\langle\psi|U_I(t)^{-1} A_I(t) U_I(t)|\psi\rangle
\eeq
To leading order we find 
\beq
\langle A \rangle_t  
\ \ = \ \ 
\langle A_I(t) \rangle \ + \ \int_0^t \alpha(t,t') f(t') dt'
\eeq
where the linear response kernel 
is given by the Kubo formula: 
\beq
\alpha(t,t') \ \ = \ \ 
i\langle[A_I(t),B_I(t')]\rangle
\eeq
In the above formulas expectation values 
without subscript are taken with the state $\psi$. 
It should be clear that the Kubo formula is 
merely the interaction picture version 
of the ``rate of change" formula: 
The rate of change of the expectation value of~$A$ 
is determined by the expectation value of 
the commutator $[\mathcal{H},A]$, hence in 
the interaction picture it is determined 
by the expectation value of $[V_I,A_I]$.

\sheadC{The $S$ operator}

The formula that we have found for the evolution 
operator in the interaction picture can 
be re-written for an evolution that starts 
at ${t=-\infty}$ (instead of $t=0$) and ends at ${t=\infty}$. 
For such scenario we use the notation $\hat{S}$ 
instead of $U_I$ and write: 
\beq
\hat{S} \ \ = \ \ 
\sum_{n=0}^{\infty}\frac{(-i)^{n}}{N!} 
\int_{-\infty}^{+\infty} dt_{n}\dots dt_{1} \, \mathcal{T} V_I(t_{n})\dots V_I(t_{1})
\eeq
It is convenient to regard ${t=-\infty}$ and ${t=\infty}$ 
as two well defined points in the far past and in the 
far future respectively. The limit 
${t\rightarrow-\infty}$ and  ${t\rightarrow\infty}$ 
is ill defined unless $\hat{S}$ is sandwiched 
between states in the same energy shell, 
as in the context of time independent scattering.

The $S$ operator formulation is useful 
for the purpose of obtaining an expression for 
the temporal cross-correlation of (say) two observables~$A$ and~$B$.
The type of system that we have in mind is (say) a Fermi sea 
of electrons. The interest is in the the ground state $\psi$   
of the system which we can regard as the ``vacuum state". 
The object that we want to calculate, written in the 
Heisenberg / Schrodinger / Interaction pictures is   
\beq
C_{BA}(t_2,t_1) 
& \ \ = \ \ & 
\langle\psi|B_H(t_2)A_H(t_1)|\psi\rangle 
\\
& \ \ = \ \ &
\langle\psi| \, U(t_2)^{-1} \, B \, U(t_2,t_1) \, A \, U(t_1) \, |\psi\rangle 
\\
& \ \ = \ \ &
\langle\psi| \, U_I(t_2)^{-1} \, B_I(t_2) \, U_I(t_2,t_1) \, A_I(t_1) \, U_I(t_1) \, |\psi\rangle 
\eeq
where it is implicitly assumed that ${t_2 > t_1 > 0}$, 
and we use the notation ${U_I(t_2,t_1)=U_I(t_2)U_I(t_1)^{-1}}$, 
which is the evolution operator  in the interaction picture
referenced to the time $t_1$ rather than to $t{=}0$. 
Note that ${\hat{S}=U_I(\infty,-\infty)}$.
Assuming that $\psi$ is obtained from the non-interacting 
ground state $\phi$ via an adiabatic switching 
of the perturbation during the time ${-\infty < t < 0}$ 
we can rewrite this expression as follows:   
\beq
C_{BA}(t_2,t_1) 
& \ \ = \ \ &
\langle\phi| \, U_I(\infty,-\infty)^{-1} \, U_I(\infty,t_2)   \, B_I(t_2) \, U_I(t_2,t_1) \, A_I(t_1) \, U_I(t_1,-\infty) \, |\phi\rangle
\\
& \ \ = \ \ &
\langle\phi| \, \hat{S}^{\dag} \, \mathcal{T} \, \hat{S} \, B_I(t_2) \, A_I(t_1) \, |\phi\rangle 
\eeq
In the last line it is implicit that 
one should first substitute the expansion for $\hat{S}$,  
and then (prior to integration) perform 
the time ordering of each term.
The Gell-Mann-Low theorem rewrites the 
above expression as follows:  
\beq
\langle\psi|\mathcal{T}B_H(t_2)A_H(t_1)|\psi\rangle 
\ \ = \ \ 
\frac{\langle\phi|\mathcal{T} \hat{S} B_I(t_2) A_I(t_1) |\phi\rangle} 
{\langle\phi| \hat{S} |\phi\rangle} 
\ \ = \ \ 
\langle\phi|\mathcal{T} \hat{S} B_I(t_2) A_I(t_1) |\phi\rangle_{\tbox{connected}}
\eeq
The first equality follows from the observation 
that due to the assumed adiabaticity 
the operation of $\hat{S}$ on $\phi$ 
is merely a multiplication by a phase factor. 
The second equality is explained using a diagrammatic   
language. Each term in the perturbative expansion 
is illustrated by a Feynman diagram. 
It is argued that the implicit unrestricted summation 
of the diagrams in the numerator equals to 
the restricted sum over all the connected diagrams, 
multiplied by the unrestricted summation of 
vacuum-to-vacuum diagrams (as in the numerator). 
The actual diagrammatic calculation is carried out 
by applying Wick's theorem. The details of the 
diagrammatic formalism are beyond the scope of our presentation.

\newpage

\sheadB{Complex poles from perturbation theory}

\sheadC{Models of interest}

In this section we solve the particle decay problem using 
perturbation theory for the resolvent. We are going to show that 
for this Hamiltonian the analytical continuation of the resolvent 
has a pole in the lower part of the complex $z$-plane. 
The imaginary part ("decay rate") of the pole that we find 
is the same as we found by either the Fermi golden rule 
or by the exact solution of the Schr\"{o}dinger equation.

We imagine that we have an "unperturbed problem" with one energy
level $|0\rangle$ of energy $E_0$ and a continuum of levels $|k\rangle$ with
energies $E_k$. A model of this kind may be used for describing 
tunneling from a metastable state. Another application is the 
decay of an excited atomic state due to emission of photons.
In the latter case the initial state would be the excited
atomic state without photons, while the continuum are states 
such that the atom is in its ground state and there is a
photon. Schematically we have
\beq
\mathcal{H}=\mathcal{H}_0+V
\eeq
where we assume
\beq
&& \langle k|V|0 \rangle = \sigma_k  \ \ \ \ \ \mbox{(coupling to the continuum)} 
\\ \nonumber
&& \langle k'|V|k \rangle = 0 \ \ \ \ \ \mbox{(no transitions within the continuum)}
\eeq
Due to gauge freedom we can assume that the coupling coefficients are 
real numbers without loss of generality. 
The Hamiltonian matrix can be illustrated as follows:
\beq
\mathcal{H}=\left( \begin{array}{ccccccccc} 
E_0&\sigma_1 &\sigma_2&\sigma_3&.&.&.&\sigma_k &. \\
\sigma_1&E_1&0&0&.&.&.&0 &.\\
\sigma_2&0&E_2&0&.&.&.&0 &.\\
\sigma_3&0&0&E_3&.&.&.&0 &.\\
.&.&.&.&.&.&.&0 &.\\
.&.&.&.&.&.&.&0 &.\\
.&.&.&.&.&.&.&0 &.\\
\sigma_k&0&0&0&0&0&0&E_k &.\\
.&.&.&.&.&.&.&.&. 
\end{array} \right)
\eeq
We later assume that the system under study is prepared in state $|0\rangle$ 
at an initial time ($t=0$), and we want to obtain the probability amplitude 
to stay in the initial state at a later times.

It is important to notice the following (a) we take one state and
neglect other states in the well. (b) we assume that $V$ allows 
transitions from this particular state in the well (zero state $|0\rangle $) 
into the $|k\rangle$ states in the continuum, while we do not have transitions 
in the continuum itself. These assumptions allow us to make exact 
calculation using perturbation theory to infinite order.
The advantage of the perturbation theory formalism 
is that it allows the treatment of more complicated 
problems for which exact solutions cannot be found.

\sheadC{Heuristic approach}

The Hamiltonian of the previous section is of the form 
\beq 
\mathcal{H} = \mathcal{H}_0 + V = 
\left( 
\begin{array}{cc} 
H_0^P & 0 \\
0 & H_0^Q 
\end{array} 
\right) 
+
\left( 
\begin{array}{cc} 
0 & V^{PQ} \\
V^{QP} & 0 
\end{array} 
\right)
\eeq
and the wavefunction can be written as 
\beq
\Psi = \left(\amatrix{\psi \cr \chi}\right) 
\eeq 
In the decay problem of the previous section $H_0^P$ is a $1\times 1$ matrix, 
and $H_0^Q$ is an $\infty\times\infty$ matrix. 
The perturbation allows transitions between $P$ and $Q$ states.
In the literature it is customary to define 
projectors $\mathcal{P}$ and $\mathcal{Q}$ on the respective sub-spaces.
\beq 
\mathcal{P}+ \mathcal{Q} \ = \ \hat{1} 
\eeq
Our interest is focused only in the $\psi$ piece 
of the wavefunction, which describes the 
state of the particle in the $P$~space.
On the other hand we assume that~$\mathcal{H}$    
has a continuous spectrum, and therefore any~$E$ 
above the ground state defines 
a stationary state via the equation
\beq
\left(\amatrix{H_0^{P} & V^{PQ} \cr V^{QP} & H_0^{Q}}\right)   
\left(\amatrix{\psi \cr \chi}\right) 
= E \left(\amatrix{\psi \cr \chi}\right) 
\eeq 
From the second raw of the above equation 
it follows that $\chi=G^{Q} V^{QP} \psi$, 
where $G_0^{Q}=1/(E-H_0^{Q})$. 
Consequently the reduced equation for $\psi$ is 
obtained from the first raw of the matrix equation 
\beq
\Big[ H_0^{P} + V^{PQ} G^{Q} V^{QP} \Big] \psi = E \psi
\eeq 
The second term in the square brackets $\Sigma(E)=V^{PQ} G_0^{Q} V^{QP}$
is the so called "self energy". Using the vague 
prescription ${E\mapsto E+i0}$  we can split it into 
an effective potential due to so called "polarization cloud" 
(also known as Lamb shift) and an imaginary part due to FGR decay:
\beq
H^{P} &=& H_0^{P} + \Delta -i \Gamma/2 \\
\Delta_{ij} &=& \sum_{k \in Q} \frac{V_{ik} V_{kj}}{E-E_k} \\
\Gamma_{ij} &=& 2\pi \sum_{k \in Q} V_{ik} V_{kj} \delta(E-E_k) 
\eeq 
We argue that $H^{P}$ generates the time evolution of $\psi$ 
within~$P$ space. This effective Hamiltonian is non-hermitian 
because probability can escape to the~$Q$ space.  
It can be diagonalized so as to get the decay modes of the 
system. Each mode is characterized by its 
complex energy ${E_r-i(\Gamma_r/2)}$. Note that the 
effective Hamiltonian is non-hermitian and therefore 
the eigen-modes are in general non-orthogonal  [see discussion of generalized 
spectral decomposition in the Fundementals~I section].

The advantage of the above treatment is its technical simplicity.
However, the mathematical significance of the ${E\mapsto E+i0}$ 
prescription is vague. We therefore turn to a more formal derivation 
of the same result, which illuminates how the non-hermiticity 
emerges once the $Q$~continuum is eliminated.

\sheadC{The $P+Q$ formalism}

We want to calculate the resolvent.
But we are interested only in the single matrix element $(0,0)$ 
because we want to know the probability to stay 
in the $|0\rangle$ state: 
\beq 
\mbox{survival probability} \ = \  
\Big| \mbox{FT} \Big[ \ \langle 0  | G(\omega)  |0  \rangle \ \Big] \Big|^2
\eeq
Here $G(\omega)$ is the retarded Green function. 
More generally we may have several states in the well. 
In such a case instead of $P=|0 \rangle \langle 0|$ 
we have 
\beq 
P=\sum_{n\in \mbox{well}}  |n\rangle \langle n| 
\eeq
If we prepare the particle in an arbitrary state $\Psi$ 
inside the well, then the probability to survive 
in the same state is 
\beq 
\mbox{survival probability} \ = \  
\Big| \mbox{FT} \Big[ \ \langle \Psi  | G(\omega)  | \Psi  \rangle \ \Big] \Big|^2
\ = \ 
\Big| \mbox{FT} \Big[ \ \langle \Psi  | G^P(\omega)  | \Psi  \rangle \ \Big] \Big|^2
\eeq
Namely, we are interested only in one block of the resolvent which we call 
\beq
G^P(z) = {\cal P} G(z) {\cal P}
\eeq
Using the usual expansion we can write 
\beq 
G^P &=& PGP 
= P\frac{1}{z-(H_0+V)}P
= P(G_0 + G_0 V G_0 + \dots )P 
\\ \nonumber
&=& PG_0P + PG_0(P+Q)V(P+Q)G_0 P + PG_0(P+Q)V(P+Q)G_0(P+Q)V(P+Q)G_0P+\dots  
\\ \nonumber
&=& G_0^P + G_0^P \Sigma^P G_0^P+G_0^P \Sigma^P G_0^P \Sigma^P G_0^P +\dots  
\\ \nonumber
&=& \frac{1}{z-(H_0^P+\Sigma^P)}
\eeq
where the "self energy" term 
\beq
\Sigma^P = V^{PQ} G_0^Q V^{QP} 
\eeq
represents the possibility of making a round trip out of the well.
Note that only even order terms contribute to this perturbative expansion 
because we made the simplifying assumption  
that the perturbation does not have "diagonal blocks". 
In our problem $\Sigma^P$ is a $1\times 1$ matrix that 
we can calculate as follows: 
\beq 
{\Sigma}^P &=& \sum_k 
\langle 0 | V |k \rangle
\langle k | G_0 | k \rangle 
\langle k | V  |0 \rangle =
\sum_k \frac{|V_k|^2}{E-E_k+i0} 
\\ \nonumber
&=& 
\sum_k
\frac{|V_k|^2}{E-E_k} 
- i \pi \sum_k|V_k|^2 \delta (E-E_k) 
\,\,\equiv\,\,
\Delta_0 - i (\Gamma_0/2)
\eeq
In the last step we have implicitly assumed that our interest is in the retarded Green function. 
We have identified as before the Fermi golden rule rate 
\beq 
\Gamma_0 = 
2\pi \sum_k|V_k|^2 \delta (E-E_k) 
\,\,\equiv\,\,
2\pi \gdos(E) \, |V|^2 
\eeq
Thus, the resolvent is the $1\times 1$ matrix.
\beq 
G^P(z)  = \frac{1}{z-(H_0^P+{\Sigma}^P)}=
\frac{1}{z-(\varepsilon_0+\Delta_0)+i(\Gamma_0/2)} 
\eeq
We see that due to the truncation of the $Q$ states we get 
a complex Hamiltonian, and hence the resolvent 
has a pole in the lower plane. The Fourier transform 
of this expression is the survival amplitude. After squaring 
it gives a simple exponential decay $\eexp{-\Gamma t}$.

\sheadA{Scattering Theory}

\sheadB{The plane wave basis}

There are several different conventions for 
the normalization of plane waves: \\
\bitem  Box normalized plane waves $|n\rangle$ \\
\bitem  Density normalized plane waves $|k\rangle$ \\
\bitem  Energy shell normalized plane waves $|E,\Omega \rangle$

We are going to be very strict in our notations, 
else errors are likely. We clarify the three 
conventions first in the 1D case, 
and later in the 3D case, and then to remark on 
energy shell bases in general.

\sheadC{Plane waves in 1D}

The most intuitive basis set 
originates from  quantization in a box with periodic 
boundary conditions (a torus):
\beq
|n\rangle \longrightarrow\frac{1}{\sqrt{L}} \eexp{ik_nx}
\eeq
where
\beq
k_n=\frac{2\pi}{L}n
\eeq
Orthonormality:
\beq
\langle n|m\rangle =\delta_{nm}
\eeq
Completeness:
\beq
\sum_n|n \rangle \langle n|=\hat{1}
\eeq

The second convention is to have   
the density normalized to unity: 
\beq
|k\rangle \longrightarrow \eexp{ikx}
\eeq
Orthonormality:
\beq
\langle k|k'\rangle =2\pi\delta(k-k')
\eeq
Completeness:
\beq
\int |k\rangle \frac{dk}{2\pi}\langle k|=\hat{1}
\eeq

Yet there is a third convention which 
assumes that the states are labeled 
by their energy, and by another 
index that indicate the direction. 
\beq
|E,\Omega\rangle 
=  \frac{1}{\sqrt{v_E}} | k_{\Omega} \rangle
\longrightarrow 
\frac{1}{\sqrt{v_E}} 
\eexp{i k_{\Omega} x}
\eeq
where $0 < E < \infty$ and  $\Omega$ is the direction 
of propagation with ${n_{\Omega} = \pm 1}$, hence 
\beq
k_{\Omega} = n_{\Omega} k_E = \pm \sqrt{2mE}
\eeq
Orthonormality:
\beq
\langle E, \Omega|E',\Omega'\rangle  
= 2\pi \delta(E-E') \delta_{\Omega\Omega'}
\eeq
Completeness:   
\beq
\int  \frac{dE}{2\pi}  \sum_\Omega 
|E,\Omega\rangle\langle E,\Omega| \,\,=\,\, \hat{1}
\eeq
In order to prove the orthonormality 
we note that $dE = v_E dk$  and therefore 
\beq
\delta(E-E')=\frac{1}{v_E}\delta(k-k')
\eeq
The energy shell normalization of plane waves 
in 1D is very convenient also for another reason.
We see that the probability flux of the plane waves 
is normalized to unity. We note that this does not hold in 
more than 1D. Still also in more than 1D, the $S$ matrix 
formalism reduces the scattering problem to 1D channels, 
and therefore this property is very important in general.

\sheadC{Plane waves in 3D}

The generalization of the box normalization convention 
to the 3D case is immediate. The same applied to 
the density normalized plane waves: 
\beq
|\vec{k}\rangle \longrightarrow \eexp{i\vec{k}\vec{x}}
\eeq
Orthonormality:
\beq
\langle \vec{k}|\vec{k}'\rangle = (2\pi)^3 \delta^3(\vec{k}-\vec{k}')
\eeq
Completeness:
\beq
\int |\vec{k} \rangle \frac{d^3k}{(2\pi)^3} \langle \vec{k}| = \hat{1}
\eeq

The generalization of the energy shell  
normalization convention is less trivial:
\beq
|E,\Omega \rangle 
= \frac{1}{2\pi}\frac{k_E}{\sqrt{v_E}}
|\vec{k}_{\Omega} \rangle
\longrightarrow
\frac{1}{2\pi}\frac{k_E}{\sqrt{v_E}}
\eexp{i\vec{k}_\Omega \cdot \vec{x}}
\eeq
where we define the direction by $\Omega=(\theta,\varphi)$, 
with an associated unit vector $\vec{n}_{\Omega}$ and a wavenumber 
\beq
\vec{k}_{\Omega} = k_E \vec{n}_{\Omega}
\eeq
Orthonormality:
\beq
\langle E,\Omega|E',\Omega'\rangle =2\pi\delta(E-E')\delta^2(\Omega-\Omega')
\eeq
Completeness:
\beq
\int \frac{dE}{2\pi} \int     
|E,\Omega\rangle d\Omega  \langle E,\Omega|=\hat{1}
\eeq
To prove the identities above we note that  
\beq
d^3k = k^2 dk d\Omega 
\ = \ k_E^2 \frac{dE}{v_E} d\Omega 
\ = \ k_E^2 \frac{dE}{v_E} d\varphi d\cos\theta
\eeq
and 
\beq
\delta^3(\vec{k}-\vec{k}') 
\ = \ \frac{v_E}{k_E^2}\delta(E-E') \delta^2(\Omega-\Omega')
\ = \ \frac{v_E}{k_E^2}\delta(E-E') \delta(\varphi-\varphi') \delta(\cos\theta-\cos\theta')
\eeq
In general we have to remember that any change 
of the "measure" is associated with a compensating 
change in the normalization of the delta functions.

\sheadC{Optional energy shell bases}

Instead of the standard energy shell basis $|E,\Omega\rangle$ one 
can use some other basis $|E,a\rangle$, where $a$ is a quantum number 
that labels the different basis states. In particular 
in 3D problem it is convenient to use the $|E,\ell m\rangle$ basis:
\beq
\langle E',\Omega | E,\ell m\rangle \ \ = \ \ 2\pi\delta(E'-E) \ Y^{\ell m}(\Omega) 
\eeq
Note that in position representation we have  
\beq 
\langle r,\Omega | E,\ell m\rangle  \ \ = \ \  2\frac{k_E}{\sqrt{v_E}} \ j_{\ell}(k_Er)  \ Y^{\ell m}(\Omega) 
\eeq
This wavefunction is properly normalized as an energy-shell basis state. 
The asymptotic behaviour of the spherical Bessel function 
is ${j_{\ell}(kr) \sim  \sin(kr-\mbox{phase})/(kr)}$,  
and accordingly, as in the 1D case, this wavefunction 
is also "flux normalized": the flux of both the ingoing 
and the outgoing wave components is unity.

\newpage

\sheadB{Scattering in the $T$-matrix formalism}

\sheadC{The Scattering States}

Our purpose is to solve the Schr\"{o}dinger's 
equation for a given energy $E$.  
\beq
(\mathcal{H}_0+V)\Psi=E\Psi
\eeq
If we rearrange the terms, we get:
\beq
(E-\mathcal{H}_0-V)\Psi=0
\eeq
In fact we want to find scattering solutions.  
These are determined uniquely if we define 
what is the "incident wave" and require outgoing 
boundary conditions for the scattered component. 
Thus we write $\Psi$ as a superposition 
of a free wave and a scattered wave,
\beq
\Psi=\phi + \Psi^{scatt}
\eeq
The scattered wave $\Psi^{scatt}$ is required to 
satisfy outgoing boundary conditions.  
The free wave $\phi$ is any solution of:
\beq
\mathcal{H}_0\phi=E\phi
\eeq
Substituting, we obtain:
\beq
(E-\mathcal{H}_0-V)\Psi^{scatt}=V\phi
\eeq
with the solution $\Psi^{scatt}=G^{+} V\phi$, leading to:
\beq
\Psi \ \ = \ \ (1+G^{+}V)\phi
\eeq

\sheadC{The Lippman Schwinger equation}

The explicit solution for $\Psi$ that was found 
in the previous section is in typically useless, 
because it is difficult to get~$G$. 
A more powerful approach is to write 
an integral equation for $\Psi$. For this 
purpose we re-arrange the differential equation as 
\beq
(E-\mathcal{H}_0)\Psi=V\Psi
\eeq
Using exactly the same procedure 
as in the previous section we get 
\beq
\Psi = \phi + G_0^{+}V\Psi
\eeq
This Lippman Schwinger equation 
can be solved for $\Psi$ using 
standard techniques (see example in the next section).  
More generally closed analytical solutions cannot be obtained. 
Still if $V$ is small we can try to find a solution iteratively, 
starting with the free wave as a zero order solution.  
This leads to a perturbative expansion for $\Psi$ which we are going 
to derive in a later section using a simpler approach.

\sheadC{Example: scattering by a regularized delta function}

Let us demonstrate the procedure of solving Lippman Schwinger equation  
for scattering on a delta potential $u\delta(x)$ in one~dimension.  
The relation $\Psi=\phi+G_0^{+}V\Psi$ is written as 
\beq
\Psi(x) \ \ = \ \ \eexp{ikx} -i\frac{u}{v_E} \Psi(0) \eexp{ik|x|}
\eeq
Setting $x=0$ we obtain a closed equation for the unknown 
wave amplitude $\Psi(0)$, whose solution leads to  
\beq
\Psi(x) \ \ = \ \ \eexp{ikx} + r \eexp{ik|x|}, 
\ \ \ \ \ \ \ \ \ \ \ \ \ \  
r=\frac{-i(u/v_E)}{1+i(u/v_E)}
\eeq
where $r$ is identified as the reflection coefficient, 
while the transmission coefficient is ${t=1+r}$.

It is instructive to consider the more general case of 
scattering  by a regularized delta function in any dimension, 
and with any dispersion relation. 
This generalization is also known as $s$-scattering.
In particular it illuminates how a divergent 
series for the $T$ matrix (see next section) can give a finite result, 
and why in the absence of regularization 
a delta function does not scatter in more 
than one dimension. By a regularized delta function 
we mean a potential $V(x)$ whose non-zero matrix elements are 
\beq
V_{k,k'} \ \ = \ \ u,
\hspace{2cm} \text{for} \ |k|,|k'| < \Lambda 
\eeq
where $\Lambda$ is a large momentum cutoff.  
The Lippman Schwinger equation reads 
\beq
\Psi(x) \ \ = \ \ \phi(x) +  \sum_{k,k'}   \langle x|G_0^{+}|k \rangle  V_{k,k'} \langle k'|\Psi\rangle 
\eeq
The sum over $k$ and $k'$ factorizes, hence the second term 
in the right hand side can be approximated as $uG_0^{+}(x|0)\Psi(0)$.
Setting $x=0$ we obtain a closed equation for the unknown 
wave amplitude $\Psi(0)$, whose solution leads to  
\beq
\Psi(x) \ \ = \ \ \phi(x) + G_0^{+}(x|0) \ u_{\tbox{eff}} \phi(0), 
\ \ \ \ \ \ \ \ \ \ \ \ \ \  
u_{\tbox{eff}} \ \ = \ \ \frac{u}{1-u\mathcal{G}(E)}
\eeq
where $\mathcal{G}(E) = G_0^{+}(0|0)$.  
The above expression for $\Psi(x)$ can be written as ${\Psi=(1+G_0^{+}T)\phi}$ where 
\beq
T \  = \ u_{\tbox{eff}} \ \delta(x)
\eeq
In the next section we further discuss the significance of~$T$, 
which can be regarded as a renormalized version of the potential ${V=u\delta(x)}$.

We now turn to discuss the practical calculation 
of ${\mathcal{G}(E)}$. From the definition it follows that 
\beq
\mathcal{G}(E) \ \ = \ \  G_0^{+}(0|0)  \ \ = \ \ \sum_k \frac{1}{E-E_k+i0} 
\eeq
The $k$ summation should be treated with the appropriate integration measure.
It equals~$-i/v_E$ for a non-regularized Dirac delta function 
in one dimension.  
In higher dimensions $\mathcal{G}(E)$
has a real part that diverges, 
which implies that the scattering goes 
to zero.  The regularization makes $\mathcal{G}(E)$ 
finite. In three dimensions we 
get $\mathcal{G}(E)=-(\mass/\pi^2)\Lambda_E$ where 
\beq
\Lambda_E = 
-2\pi^2 
\int_0^{\Lambda} 
\frac{d^3k}{(2\pi)^3} 
\frac{1}{k_E^2-k^2+i0} 
= 
\Lambda
-\frac{1}{2}k_E\log\left(\frac{\Lambda+k_E}{\Lambda-k_E}\right)
+i\frac{\pi}{2}k_E 
\eeq
Another interesting case to consider is having  
(say) a constant density of states with a threshold energy. 
The divergence of $\mathcal{G}(E)$ near a threshold 
is logarithmic.  It is quite amusing that 
the second order as well as all the higher 
terms in perturbation theory are divergent, 
while their sum goes to zero...

\sheadC{Perturbation Theory for the Scattering State}

Going back to the formal solution for $\Psi$ 
we can substitute there the perturbative expansion of $G$
\beq
G=G_0+G_0VG_0+G_0VG_0VG_0+\dots 
\eeq
leading to  
\beq
\Psi=(1+G^{+}V)\phi=\phi+G_0V\phi+G_0VG_0V\phi+ \dots 
\eeq
As an example consider the typical case 
of scattering by a potential $V(x)$. 
In this case the above expansion to leading 
order in space representation is:
\beq
\Psi(x)=\phi(x)+\int G_0(x,x')V(x')\phi(x')\,dx'
\eeq

\sheadC{The $T$ matrix}

It is customary to define the $T$ matrix as follows
\beq
T \ \ = \ \ V+VG_0V+VG_0VG_0V + \dots \ \ = \ \ V+VGV 
\eeq
The $T$ matrix can be regarded as a "corrected"  
version of the potential $V$, so as to make the 
following first order look-alike expression exact:  
\beq
G \ \ = \ \ G_0+G_0TG_0
\eeq
Or the equivalent expression for the wavefunction:
\beq
\Psi=\phi + G_0^{+} T \phi
\eeq
Later it is convenient to take matrix elements in the unperturbed basis of free waves:  
\beq
V_{\alpha\beta} &=& \left\langle \phi^{\alpha} \left| V \right |\phi^{\beta}\right\rangle
\\ \nonumber
T_{\alpha\beta}(E) &=& \left\langle \phi^{\alpha} \left| T(E) \right |\phi^{\beta} \right\rangle
\eeq
In principle we can take the matrix elements 
between any states. But in practice 
our interest is in states that have the same energy,  
namely $E_{\alpha}=E_{\beta}=E$. Therefore it is 
convenient to use two indexes $(E,a)$, where  
the index $a$ distinguishes different free waves 
that have the same energy. In particular $a$ may 
stand for the "direction" ($\Omega$) of the plane wave.    
Thus in practice we are interested only 
in matrix elements "on the energy shell":
\beq
T_{ab}(E) =
\left\langle \phi^{E,a} \left| T(E) \right |\phi^{E,b} \right\rangle
\eeq
One should be very careful to handle correctly 
the different measures that are associated  
with different type of indexes. In particular note that in 3D:
\beq
T_{\Omega,\Omega_0} 
= \frac{1}{v_E}\left(\frac{k_E}{2\pi}\right)^2  
T_{k_{\Omega},k_{\Omega_0}}
\eeq

\sheadC{Scattering states in 1D problems} 

In this section we look for a scattering 
solution that originates from the free wave $|k_0\rangle$.  
Using the result of a previous section 
we write ${\Psi=\phi^{k_0}+G_0^+T\phi^{k_0}}$ with 
\beq
\phi^{k_0}(r) \ \ = \ \ \eexp{ i k_0 x }
\ \ \ \ \ \ \ \ \mbox{[density normalized]}
\eeq
In Dirac notations:
\beq
|\Psi\rangle \ \ = \ \ |\phi^{k_0}\rangle + G_0^{+} T|\phi^{k_0}\rangle
\eeq
In space representation: 
\beq
\langle x|\Psi\rangle \ \ = \ \ 
\langle x|\phi^{k_0}\rangle
+\langle x|G_0^+T|\phi^{k_0}\rangle
\eeq
or in "old style" notation:   
\beq
\Psi(x) \ \ = \ \ 
\phi^{k_0}(x)+ \int G_0^{+}(x|x_0) dx_0  \langle x_0 | T | k_0 \rangle 
\eeq
In 1D problems the Green function is 
\beq
G_0^{+}(x|x_0) \ \ = \ \ \langle x|G_0^{+}|x_0\rangle 
\ \ = \ \  - \frac{i}{v_E} \, \eexp{ik_E|x-x_0|}
\eeq
Thus we get:
\beq
\psi(x) \ \ = \ \ \eexp{ i \vec {k_0} x }
- \frac{i}{v_E} \, \int \eexp{ik_E|x-x_0|} dx_0 \langle x_0 | T | k_0 \rangle 
\eeq
By observation of the  left asymptotic region we deduce that the reflection coefficient is 
\beq
r \ \ =  \ \ - \frac{i}{v_E} \, \langle -k_0 | T | k_0 \rangle 
\eeq
By observation of the right asymptotic region we deduce that the transmission coefficient is 
\beq
t \ \ = \ \  1 - \frac{i}{v_E} \, \langle k_0 | T | k_0 \rangle  \ \ = \ \ 1+r
\eeq
We note that conservation of probability implies $|r|^2 +|t|^2=1$.
With the substitution of the above relation (${t=1+r}$) it follows that 
\beq
|r|^2 \ \ = \ \ -\re[r]
\eeq
This is a special example for an "optical theorem" that we shall discuss later on: 
it relates the total cross section (here $2|r|^2$) to the "forward" 
scattering amplitude. In order to understand this terminology note that $r$ 
is the amplitude of the scattered wave both in the "backward" and "forward" 
directions while $t$ is the sum of the forward scattering with the incident wave.  
The above relation implies  
\beq
\text{scattering phase shift}  \ \ = \ \ \arccos(r) 
\eeq

\sheadC{Scattering states in 3D problems}

The derivation is the 3D case follows the same procedure as in the 1D case.
The incident wave is 
\beq
\phi^{k_0}(r) \ \ = \ \ \eexp{ i \vec {k_0} \cdot \vec{r} }
\ \ \ \ \ \ \ \ \mbox{[density normalized]}
\eeq
and the implied scattering state is  
\beq
\Psi(r) \ \ =  \ \ \phi^{k_0}(r)+ \int G_0^{+}(r|r_0) dr_0  \langle r_0 | T | k_0 \rangle 
\eeq
where
\beq
G_0^{+}(r|r_0) = \langle r|G_0^{+}|r_0\rangle = 
-\frac{\mass}{2\pi}\,\frac{\eexp{ik_E|r-r_0|}}{|r-r_0|}
\eeq
Thus we get
\beq
\Psi(r)=\phi^{k_0}(r)
-  \frac{\mass}{2\pi} \int 
\frac{\eexp{ik_E |r-r_0|}}{|r-r_0|} 
\langle r_0|T|k_0 \rangle \, dr_0 
\eeq
So far everything is exact.
Now we want to get a simpler expression for 
the asymptotic form of the wavefunction. 
Note that from the experimental point of view 
only the "far field" region (far away from the target) 
is of interest. 
The major observation is that the $dr_0$ integration 
is effectively bounded to the scattering region $|r|<r_0$ 
where the matrix elements of $V$ and hence of $T$ are 
non-zero. Therefore for $|r|\gg |r_0|$ we can use the approximation
\beq
|\vec{r}-\vec{r}_0| = \sqrt{(\vec{r}-\vec{r}_0)^2} 
= \sqrt{|r|^2 - 2\vec{r} \cdot \vec{r}_0 + \mathcal{O}(|r_0|^2)} 
\approx |r| \left[ 1 - \vec{n}_{\Omega} \cdot \frac{\vec{r_0}}{|r|} \right]
= |r| - \vec{n}_{\Omega} \cdot \vec{r}_0
\eeq
Here and below we use the following notations:
\beq
\vec{r}  &\equiv&  |r| \vec{n}_{\Omega}
\\ \nonumber
\Omega &=& (\theta, \varphi) = \mbox{spherical coordinates}
\\ \nonumber
\vec{n}_{\Omega} &=& (\sin\theta\cos\phi, \sin\theta\sin\phi, \cos\theta)
\\ \nonumber
\vec{k}_{\Omega} &=& k_E \vec{n}_{\Omega}
\eeq

\begin{center}
\putgraph[0.6\hsize]{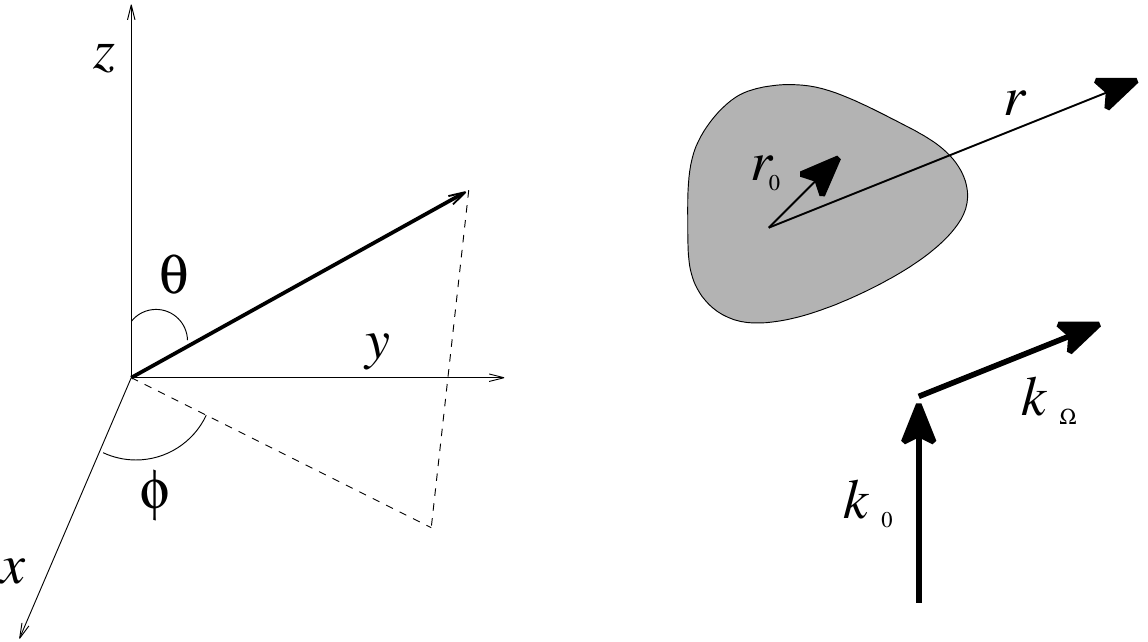}
\end{center}

With the approximation above we get:
\beq
\Psi(r) \ \ \approx \ \  
\eexp{i \vec{k}_0\cdot\vec r}
-\frac{\mass}{2\pi}
\frac{ \eexp{i k_E |r|} }{r}
\int \eexp{-i\vec{k}_{\Omega} \cdot \vec{r}_0} 
\langle r_0|T|k_0\rangle \, dr_0
\ \ \equiv \ \ 
\eexp{i \vec{k}_0\cdot \vec{r}}
+ f(\Omega)\frac{ \eexp{i k_E |r|} }{|r|}
\eeq
where
\beq
f(\Omega) = -\frac{\mass}{2\pi} 
\langle \vec{k}_{\Omega} | T(E)| \vec{k}_0 \rangle
\eeq
It follows that the differential cross section is 
\beq
\frac {d\sigma}{d\Omega}\,\,=\,\, 
|f(\Omega)|^2 \,\,=\,\, 
\left(\frac{\mass}{2\pi\hbar}\right)^2
\, \left| \frac{1}{\hbar} T_{k_{\Omega},k_0}\right|^2
\eeq
This formula assumes density normalized plane waves.
It relates the scattering which is described by $f(\Omega)$ 
to the $T$ matrix. It can be regarded as a special case 
of a more general relation between the $S$ matrix 
and the $T$ matrix, as discussed in later sections 
(and see in particular the discussion of the "optical theorem" below).

\sheadC{Born approximation and beyond}

For potential scattering the first order approximation $T\approx V$ 
leads to the Born approximation:
\beq
f(\Omega) = -\frac{\mass}{2\pi} 
\langle \vec{k}_{\Omega} | T(E)| \vec{k}_0 \rangle
\approx -\frac{\mass}{2\pi} \tilde{V}(q)
\eeq
where $\vec{q}=\vec{k}_{\Omega}-\vec{k}_0$ and $\tilde{V}(\vec{q})$ 
is the Fourier transform of $V(\vec{r})$. The corresponding formula 
for the cross section is consistent with the Fermi golden rule.

It is customary in high energy physics 
to take into account higher orders. 
The various terms in the expansion are 
illustrated using Feynman diagrams. 
However, there are circumstance where 
we can gain better insight by considering 
the analytical properties of the Green function.  
In particular we can ask what happens if $G$ 
has a pole at some complex energy ${z=E_r-i(\Gamma_r/2)}$. 
Assuming that the scattering is dominated 
by that resonance we get that the cross section 
has a Lorentzian line shape. More generally, 
If there is an interference between the resonance 
and the non-resonant terms then we get a Fano line shape. 
We shall discuss further resonances later on 
within the framework of the $S$ matrix formalism.

\sheadC{The Optical Theorem}

From the energy-shell matrix elements 
of the $T$ operator we can form an object 
that later we identify as the $S$ matrix:
\beq
S_{ab}=\delta_{ab} - i \langle E,a|T|E,b\rangle 
\eeq
Using a more sloppy notation the equation above 
can be written as $S=1-iT$. We should remember 
that $S$ is not an operator, and that $S_{ab}$ 
are not the matrix elements of an operator. 
In contrast to that $T_{ab}$ are the on-shell matrix elements 
of an operator. This relation assumes that 
the energy shell basis states are {\em properly} normalized.

The $S$ matrix, as we define it later, is unitary. 
So we have $S^{\dagger}S=1$, and therefore we conclude 
that the $T$ matrix should satisfy the following equality:
\beq
(T^{\dagger}-T) = iT^{\dagger}T
\eeq
In particular we can write:
\beq
\langle a_0|T-T^{\dagger}|a_0\rangle 
=-i\sum_a\langle a_0|T|a\rangle \langle a|T^{\dagger}|a_0\rangle 
\eeq
and we get:
\beq
\sum_a |T_{aa_0}|^2 = -2\im[T_{a_0a_0}]
\eeq
This so called "optical theorem" establishes 
a connection between the "cross section" and 
the forward scattering amplitude $T_{a_0a_0}$. 
It should be clear that if this relation holds in 
one particular energy-shell basis, then it holds 
also in any other (properly normalized) energy 
shell basis. In 3D scattering problems we can write 
one of the following expressions:
\beq
\sum_{\ell,m}|\langle \phi^{E,\ell,m} |T| \phi^{E,\ell_0,m_0} \rangle |^2 
&=&-2 \im [ \langle \phi^{E,l_0,m_0}|T|\phi^{E,l_0,m_0}\rangle ]
\\ \nonumber
\sum_{\Omega}|\langle \phi^{E,\Omega} |T| \phi^{E,\Omega_0} \rangle |^2 
&=&-2 \im [ \langle \phi^{E,\Omega_0}|T|\phi^{E,\Omega_0}\rangle ]
\eeq
Using the relations 
\beq
&& |E,\Omega\rangle = \frac{1}{\sqrt{v_E}} \frac{k_E}{2\pi} |k_{\Omega}\rangle 
\\ \nonumber
&& f(\Omega) = -\frac{\mass}{2\pi} 
\langle \vec{k}_{\Omega} | T(E)| \vec{k}_0 \rangle
\eeq
we get 
\beq 
\int|\langle \vec{k}_{\Omega}|T|\vec{k}_0\rangle |^2d\Omega
\ \ = \ \  -2v_E\left(\frac{2\pi}{k_E}\right)^2
\im[\langle \vec{k}_0|T|\vec{k}_0\rangle ]
\eeq
Or the more familiar version: 
\beq
\sigma_{\tbox{total}}
\ \ = \ \  \int|f(\Omega)|^2 d\Omega
\ \ = \ \ \frac{4\pi}{k_E} \im [f(0)]
\eeq

\newpage

\sheadB{Scattering in the $S$-matrix formalism}

\sheadC{Channel Representation}

Before we define the $S$ matrix, Let us review 
some of the underlying assumptions of the $S$ matrix 
formalism. Define 
\beq
\rho(x) = {| \Psi (x)| }^{2} 
\eeq
The continuity equation is
\beq
\frac{\partial\rho}{\partial{t}} = -{\nabla\cdot{J}} 
\eeq
We are working with a time-independent Hamiltonian 
and looking for stationary solutions, hence: 
\beq
\nabla\cdot{J}=0 
\eeq
The standard basis for representation is ${|x\rangle}$. 
We assume that the wave function is separable outside 
of the scattering region. Accordingly we arrange the basis as follows:
\beq
|x \in \mbox{\small inside} \rangle = \mbox{the particle is located inside the scattering region}
\\ \nonumber
|a,r\rangle = \mbox{the particle is located along one of the outside channels}
\eeq
and write the quantum state in this representation as 
\beq
|\Psi\rangle \ \ = \ \ 
\sum_{x \in \mbox{\small inside}} \varphi(x) \ |x\rangle
\ + \ \sum_{a,r} R_a(r) \ |a,r\rangle
\eeq
The simplest example for a system that has (naturally) this type 
of structure is a set of 1D wires connected together to some "dot".  
In such geometry the index $a$ distinguishes the different wires. 
Another, less trivial example, is a lead connected to a "dot". 
Assuming for simplicity 2D geometry, the wavefunction 
in the lead can be expanded as 
\beq
\Psi(x) \,\,=\,\, 
\Psi(r,s) \,\,=\,\, \sum_a R_a(r) \chi^a(s) 
\eeq
where the channel functions (waveguide modes) are:
\beq
\chi^a(s) \,\,=\,\,
\sqrt{\frac{2}{\ell}}\sin\left(\left(\frac{\pi}{\ell}a\right)s\right) 
\ \ \ \ \ \ \mbox{with $a=1,2,3 \dots $ and $0<s<\ell$} 
\eeq
In short, we can say that the wavefunction outside 
of the scattering region is represented by a set of radial functions:
\beq
\Psi(x) \,\, \mapsto \,\, R_a(r) 
\ \ \ \ \ \ \mbox{where $a=1,2,3 \dots $ and $0<r<\infty$}
\eeq

The following figures illustrate several examples 
for scattering problems (from left to right): three connected wires, 
dot-waveguide system, scattering in spherical 
geometry, and inelastic scattering. 
The last two systems will be discussed 
below and in the exercises.

\begin{center}
\putgraph[0.1\hsize]{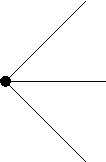} 
\hspace{0.06\hsize}
\putgraph[0.2\hsize]{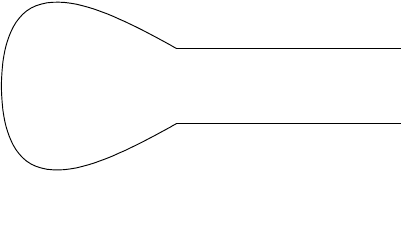} 
\hspace{0.06\hsize}
\putgraph[0.2\hsize]{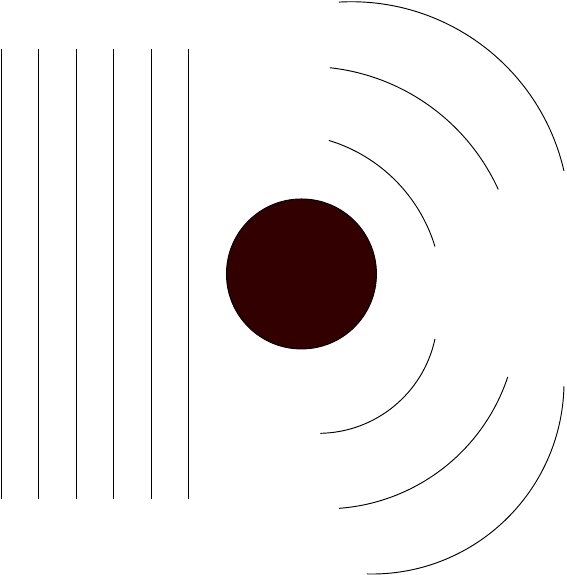} 
\hspace{0.06\hsize}
\putgraph[0.25\hsize]{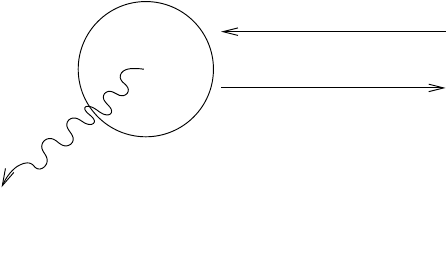} 
\end{center}

\sheadC{The Definition of the $S$ Matrix}

Our Hamiltonian is time-independent, so the energy $E$ 
is a good quantum number, and therefore 
the Hamiltonian $\mathcal{H}$ is block diagonal 
if we take $E$ as one index of the representation. 
For a given energy $E$ the Hamiltonian has 
an infinite number of eigenstates which form 
the so called "energy shell". 
For example in 3D for a given 
energy $E$ we have all the plane waves 
with momentum $|\vec{k}|=k_E$, and all the 
possible superpositions of these waves.    
Once we are on the energy shell, 
it is clear that the radial functions 
should be of the form
\beq
R_a(r) \ \ = \ \ A_a R^{E,a,-}(r) - B_a R^{E,a,+}(r)
\eeq
For example, in case of a waveguide
\beq
R^{E,a,\pm}(r) \ \ = \ \ \frac{1}{\sqrt{v_a}} \ \eexp{\pm i k_a r}
\ \ \ \ \ \ \ \ \mbox{[flux normalized]}
\eeq
where the radial momentum in channel $a$ 
corresponds to the assumed energy $E$,  
\beq
k_a \ \ = \ \ \sqrt{2\mass\left(E-\frac{1}{2\mass}\left(\frac{\pi}{\ell}a\right)^2\right)}
\eeq
and the velocity $v_a=k_a/\mass$  in channel~$a$ 
is determined by the dispersion relation. 
Thus on the energy shell the wavefunctions 
can be represented by a set of ingoing and outgoing amplitudes:
\beq
\Psi(x) \ \ \longmapsto \ \ (A_a,B_a)
\ \ \ \ \ \ \ \ \ \mbox{with $a=1,2,3 \dots $}
\eeq
But we should remember that not all sets of amplitudes 
define a stationary energy state. In order to have 
a valid energy eigenstate we have to match 
the ingoing and outgoing amplitudes on the 
boundary of the scattering region. The matching 
condition is summarized by the $S$ matrix.
\beq
B_b \ \ = \ \ \sum_a S_{ba} A_a
\eeq
By convention the basis radial functions are "flux normalized".
Consequently  the current in channel $a$ is  $I_a = |B_a|^2-|A_a|^2$ 
and from the continuity equation it follows that 
\beq
\sum_a |B_a|^2 \ \ = \ \ \sum_a |A_a|^2
\eeq
From here follows that the $S$ matrix is unitary. 
The unitarity of the~$S$ matrix sometimes implies 
surprising consequences. For example: in 1D the 
transmission from-left-to-right must be equal 
to the transmission from-right-to-left. 
More generally: if we have a system with two 
different leads $\bm{1}$ and $\bm{2}$ 
then ${\sum_{a\in\bm{1},b\in\bm{2}} |S_{ba}|^2 =\sum_{a\in\bm{2},b\in\bm{1}} |S_{ba}|^2}$. 
The latter observation is essential 
for the theory of the two-terminal Landauer conductance.

In order to practice the definition of the $S$ matrix 
consider a system with a single 2D lead. 
Let us assume that the lead has 3 open channels. 
That means that $k_a$ is a real number 
for $a=1,2,3$, and becomes imaginary for $a>3$. 
The $a>3$ channels are called "closed channels"  
or "evanescent modes". They should not 
be included in the $S$ matrix because 
if we go far enough they contribute nothing 
to the wavefunction (their contribution decays exponentially). 
Thus we have a system with 3 open channels, 
and we can write 
\beq
R_1(r) &=& \frac{1}{\sqrt{v_1}} (A_1 \eexp{-ik_1r} -B_1\eexp{+k_1r}) 
\\ \nonumber
R_2(r) &=& \frac{1}{\sqrt{v_2}} (A_2 \eexp{-ik_2r} -B_2\eexp{+k_2r}) 
\\ \nonumber
R_3(r) &=& \frac{1}{\sqrt{v_3}} (A_3 \eexp{-ik_3r} -B_3\eexp{+k_3r})
\eeq
and
\beq
\left(\amatrix{B_1\cr B_2\cr B_3}\right) 
\ \ = \ \ \mathbf{S} 
\left(\amatrix{A_1\cr A_2\cr A_3}\right)
\eeq
Simple examples for $S$ matrices are provided 
in the ``delta junction" sections 
of the ``Boxes and Networks" lecture.

\sheadC{Scattering states}

Let us define the unperturbed Hamiltonian $\mathcal{H}_0$ as that 
for which the particle cannot make transitions between channels.
Furthermore without loss of generality the phases of $R^{a\pm}(r)$  
is chosen such that  $S_{ab}=\delta_{ab}$, or equivalently $B_a=A_a$, 
should give the "free wave" solutions.  
We label the "free" energy states that correspond 
to the Hamiltonian $\mathcal{H}_0$ as $|\phi\rangle$.
In particular we define a complete set $|\phi^{\alpha}\rangle$, 
that are indexed by $\alpha=(E,a)$, Namely, we define  
\beq
|\phi^{\alpha}\rangle 
\ \ = \ \ 
|\phi^{E_{\alpha} a_{\alpha}}\rangle 
\ \ \longmapsto \ \  
\delta_{a,a_{\alpha}} (R^{a-}(r) - R^{a+}(r)) 
\eeq
The following figure illustrates how 
the "free wave" $|\phi^{E,2}\rangle$ l
of a three wire system looks like.

\begin{center}
\putgraph[0.2\hsize]{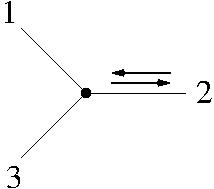}
\end{center}

It is clear that the states $|\phi^{E,a}\rangle$ form a complete basis. 
Now we take the actual Hamiltonian $\mathcal{H}$ that permits 
transitions between channels. The general solution is written as
\beq
|\Psi^{\alpha}\rangle  
\ \ \longmapsto \ \  
A_a R^{a-}(r) - B_a R^{a+}(r) 
\eeq
where $B_a = S_{ab}A_b$ or the equivalent relation $A_a=(S^{-1})_{ab}B_b$. 
In particular we can define the following sets of solutions: 
\beq
|\Psi^{\alpha+} \rangle 
& \ \ = \ \ & 
|\Psi^{E_{\alpha}a_{\alpha}+}\rangle 
\ \ \longmapsto \ \ 
\delta_{a,a_{\alpha}} R^{a-}(r) - S_{a,a_{\alpha}} R^{a+}(r) 
\\ \nonumber
|\Psi^{\alpha-} \rangle 
& \ \ = \ \ & 
|\Psi^{E_{\alpha}a_{\alpha}-}\rangle 
\ \ \longmapsto \ \ 
(S^{-1})_{a,a_{\alpha}} R^{a-}(r) - \delta_{a,a_{\alpha}} R^{a+}(r) 
\eeq
The set of $(+)$ states describes an incident wave 
in the $a_{\alpha}$ channel ($A_a=\delta_{a,a_{\alpha}}$)  
and a scattered wave that satisfies "outgoing" 
boundary conditions. The sign convention is such 
that ${|\phi^{\alpha}\rangle}$ is obtained for ${\bm{S}=1}$.
The set of $(-)$ states is similarly defined.
We illustrate some of these states 
in the case of a three wire system.

\begin{center}
$|\Psi^{E,1,+}\rangle$ 
\hspace{0.22\hsize} 
$|\Psi^{E,2,+}\rangle$ 
\hspace{0.25\hsize} 
$|\Psi^{E,1,-}\rangle$ \\
\putgraph[0.2\hsize]{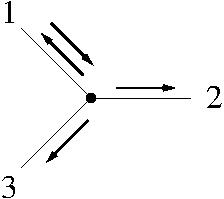}
\hspace{0.1\hsize} 
\putgraph[0.2\hsize]{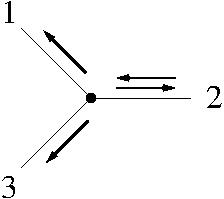}
\hspace{0.1\hsize} 
\putgraph[0.2\hsize]{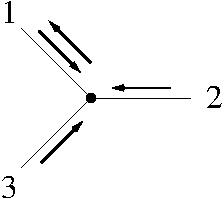}
\end{center}

\sheadC{Time reversal in scattering theory}

It is tempting to identify the $(-)$ scattering states  
as the time reversed version of the $(+)$
scattering states.
This is indeed correct in the absence 
of a magnetic field, when we have time reversal 
symmetry. Otherwise it is wrong.  
We shall clarify this point below.

Assuming that we have the solution 
\beq
|\Psi^{E,a_0,+}\rangle 
\ \ \longmapsto \ \ 
\delta_{a,a_0}\eexp{-ikx} - S_{a,a_0}\eexp{+ikx}
\eeq
The time reversed state is obtained via complex conjugation: 
\beq
\mathcal{T}|\Psi^{E,a_0,+}\rangle 
\ \ \longmapsto \ \ 
\delta_{a,a_0}\eexp{ikx} - (S^*)_{a,a_0}\eexp{-ikx}
\eeq
This should be contrasted with 
\beq
|\Psi^{E,a_0,-}\rangle 
\ \ \longmapsto \ \ 
(S^{-1})_{a,a_0} \eexp{-ikx} - \delta_{a,a_0}\eexp{ikx} 
\eeq
We see that the two coincide (disregarding a global minus) only if 
\beq
S^* \ \ = \ \ S^{-1}
\eeq
which means that the $S$ matrix should be symmetric ($S^{\tbox{T}}=S$). 
This is the condition for having time reversal symmetry 
in the language of scattering theory.

\sheadC{Orthonormality of the scattering states}

The $(+)$ states form a complete orthonormal basis. 
Also the $(-)$ states form a complete orthonormal basis. 
The orthonormality relation and the transformation 
that relates the two basis sets are   
\beq
\langle E_1,a_1,+|E_2,a_2,+\rangle &=& 2\pi\delta(E_1-E_2) \delta_{a_1,a_2} 
\\ 
\langle E_1,a_1,-|E_2,a_2,-\rangle &=& 2\pi\delta(E_1-E_2) \delta_{a_1,a_2} 
\\ 
\langle E_1,a_1,-|E_2,a_2,+\rangle &=& 2\pi\delta(E_1-E_2) S_{a_1,a_2} 
\eeq
The last equality follows directly 
from the definition of the $S$ matrix.
Without loss of generality we prove 
this lemma for the 3 wire system. 
For example let us explain why it is true 
for ${a_1=1}$ and ${a_2=2}$. 
By inspection of the figure in a previous section
we see that the singular overlaps comes  
only from the first and the second channels. 
Disregarding the "flux" normalization factor 
the singular part of the overlap is 
\beq
\langle E_1,1,-|E_2,2,+\rangle \Big|_{\tbox{singular}} 
\ \ &=& \ \ 
\int_0^{\infty} 
\Big[-\eexp{ikr}\Big]^* \Big[-S_{12}\eexp{+ik_0r}\Big] dr
+
\int_0^{\infty} 
\Big[(S^{-1})_{21} \eexp{-ikr}\Big]^* \Big[\eexp{-ik_0r}\Big] dr
\\ \nonumber
\ \ &=& \ \ 
\int_{0}^{\infty}  S_{12} \eexp{-i(k-k_0)r} dr
+ \int_{-\infty}^{0}  S_{12} \eexp{-i(k-k_0)r} dr
\ \ = \ \ 
2\pi\delta(k-k_0)S_{12}
\eeq
where is the second line we have changed   
the dummy integration variables  
of the second integral from $r$ to $-r$, 
and used $(S^{-1})_{21}^*=S_{12}$. 
If we restore the "flux normalization" factor we get the desired result.
One can wonder what about the non-singular contributions to the overlaps. 
These may give, and indeed give, an overlap that goes like $1/(E \pm E_0)$. 
But on the other hand we know that for $E \neq E_0$ the overlap should be 
exactly zero due to the orthogonality of states with different energy 
(the Hamiltonian is Hermitian). If we check whether all the non-singular 
overlaps cancel, we find that this is not the case. What is wrong?  
The answer is simple. In the above calculation we disregarded a non-singular 
overlap which is contributed by the scattering region. This must 
cancel the non-singular overlap in the outside region because as we 
said, the Hamiltonian is Hermitian.

\sheadC{Getting the $S$ matrix from the $T$ matrix}

In this section we present a relation  
between the $S$ matrix and the $T$ matrix. 
The derivation of this relations 
goes as follows: On the one hand 
we express the overlap of the ingoing and outgoing 
scattering states using the $S$ matrix. On the other 
hand we express it using the $T$ matrix. Namely, 
\beq
\langle \Psi^{E_{1},a_{1},-}|\Psi^{E_{2},a_{2},+}\rangle 
&=& 2\pi\delta(E_{1}-E_{2}) \, S_{a_{1}a_{2}}
\\ \nonumber
\langle \Psi^{E_{1},a_{1},-}|\Psi^{E_{2},a_{2},+}\rangle 
&=& 2\pi\delta(E_{1}-E_{2}) \, (\delta_{a_{1}a_{2}}-iT_{a_{1}a_{2}}) 
\eeq
By comparing the two expressions we deduce the relation between 
the $S$ matrix and the $T$ matrix:   
\beq
S_{a_{1}a_{2}} \ \ = \ \ \delta_{a_{1}a_{2}}-iT_{a_{1}a_{2}}
\eeq
or in abstract notation $S=1-iT$.  
Another way to re-phrase this relation is 
to say that the $S$ matrix can be obtained 
from the matrix elements of an $\hat{S}$ operator:
\beq
\langle \phi^{E_{1},a_{1}} | \hat{S} | \phi^{E_{2},a_{2}}\rangle
\ \ = \ \  2\pi\delta(E_{1}-E_{2}) \, S_{a_{1}a_{2}}
\eeq
where the $\hat{S}$ operator is the evolution operator 
in the interaction picture. Within the framework 
of the time dependent approach to scattering theory the 
latter relation is taken as the definition for the $S$~matrix. 
The two identities that we prove in the previous and in this 
section establish that the time-independent definition 
of the $S$~matrix and the time dependent approach are equivalent.    
The rest of this section is dedicated to the derivation 
of the $T$ matrix relation. 

{\bf First order derivation.-- } 
We recall that the scattering states 
are defined as the solution of $\mathcal{H}\Psi=E\Psi$ 
and they can be expressed as  
\beq
\Psi^{E_2,a_2,+} \ \ &=& \ \ \left(1+G_0^{+}(E_2)V + ...\right) \ \phi^{E_2,a_2}
\\
\Psi^{E_1,a_1,-} \ \ &=& \ \ \left(1+G_0^{-}(E_1)V + ...\right) \ \phi^{E_1,a_1}
\eeq
Hence
\beq
\langle \Psi^{E_1,a_1,-}|\Psi^{E_2,a_2,+} \rangle 
\ \ &=& \ \ 
\langle \phi^{E_1,a_1}|\phi^{E_2,a_2} \rangle 
+ \langle \phi^{E_1,a_1}|VG_0^{+}(E_1)+G_0^{+}(E_2)V|\phi^{E_2,a_2} \rangle + ...
\\
\ \ &=& \ \
\langle \phi^{E_1,a_1}|\phi^{E_2,a_2} \rangle
+ \left[ \frac{1}{E_1-E_2+i0} + \frac{1}{E_2-E_1+i0} \right] \langle \phi^{E_1,a_1}|V|\phi^{E_2,a_2} \rangle + ...
\\
\ \ &=& \ \
2\pi\delta(E_{1}-E_{2}) \delta_{a_{1}a_{2}}
- i2\pi\delta(E_{1}-E_{2})
\langle \phi^{E_1,a_1} \left|V\right|\phi^{E_2,a_2}\rangle + ...
\eeq

{\bf Full derivation.-- }
We recall that the scattering states can be expressed as  
\beq
&& \Psi^{E_2,a_2,+}
\ \ = \ \ (1+G^{+}(E_2)V) \phi^{E_2,a_2} 
\ \ = \ \ \left(1+ \frac{1}{E_2 - \mathcal{H} +i0}  V \right) \phi^{E_2,a_2}
\\ 
&& \Psi^{E_1,a_1,-}
\ \ = \ \ (1+G^{-}(E_1)V) \phi^{E_1,a_1}
\ \ = \ \ \left(1 + \frac1{E_1 - \mathcal{H}_0 - i0}  T(E_1)^{\dag} \right) \phi^{E_1,a_1}
\eeq
where the last equality relays on $GV=G_0T$. 
In the following calculation we use the first identity for the "ket", 
and after that the second identity for the "bra":
\beq
\left\langle \Psi^{E_1,a_1,-}|\Psi^{E_2,a_2,+} \right\rangle
&=& 
\left\langle \Psi^{E_1,a_1,-}
\left|1+\frac{1}{E_2-\mathcal{H}+i0}V\right|
\phi^{E_2,a_2}\right\rangle
\\ \nonumber
&=& 
\left\langle \Psi^{E_1,a_1^-}
\left|1+\frac{1}{E_2-E_1+i0}V\right|
\phi^{E_2,a_2}\right\rangle
\\ \nonumber
&=& 
\left\langle
\phi^{E_1,a_1}
\left|
\left[1+T(E_1)\frac{1}{E_1-\mathcal{H}_0+i0}\right]
\left[1+\frac{V}{E_2-E_1+i0}\right]
\right|
\phi^{E_2,a_2}\right\rangle
\\ \nonumber
&=&
\left\langle\phi^{E_1,a_1}
\left|
\left[1+\frac{T(E_1)}{E_1-E_2+i0}\right]
\right|\phi^{E_2,a_2}\right\rangle
+\left\langle\phi^{E_1,a_1}
\left|
\frac{V+T(E_1)G_0^{+}(E_1)V}{E_2-E_1+i0}
\right|
\phi^{E_2,a_2}\right\rangle
\\ \nonumber
&=&
\left\langle \phi^{E_1,a_1}
\left|1+\frac{T(E_1)}{E_1-E_2+i0}\right|
\phi^{E_2,a_2}\right\rangle
+\left\langle \phi^{E_1,a_1}
\left|\frac{T(E_1)}{E_2-E_1+i0}\right|
\phi^{E_2,a_2}\right\rangle
\\ \nonumber
&=&
2\pi\delta(E_{1}-E_{2}) \delta_{a_{1}a_{2}}
- i2\pi\delta(E_{1}-E_{2})
\left\langle \phi^{E_1,a_1} \left|T(E_1)\right|\phi^{E_2,a_2}\right\rangle
\eeq
where before the last step we have used 
the relation $V+TG_0V=T$.

\sheadC{Subtleties in the notion of cross section}

Assume that we have a scattering state $\Psi$. 
We can write it as a sum of "ingoing" and "outgoing"
waves or as a sum of "incident" and "scattered"  waves. 
This is not the same thing! 
\beq
\Psi &=& \Psi_{ingoing} + \Psi_{outgoing} 
\\ \nonumber
\Psi &=& \Psi_{incident} + \Psi_{scattered}
\eeq
The "incident wave" is a "free wave" that contains 
the "ingoing" wave with its associated "outgoing" component. 
It corresponds to the $\mathcal{H}_0$ Hamiltonian.  
The "scattered wave" is what we have to add in order 
to get a solution to the scattering problem 
with the Hamiltonian $\mathcal{H}$. 
In the case of the usual boundary conditions it 
contains only an "outgoing" component.  

\begin{center}
\putgraph[0.3\hsize]{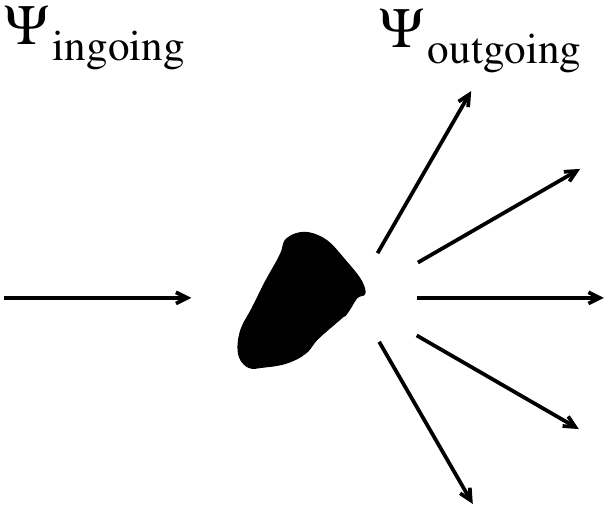} 
\hspace{0.1\hsize}
\putgraph[0.4\hsize]{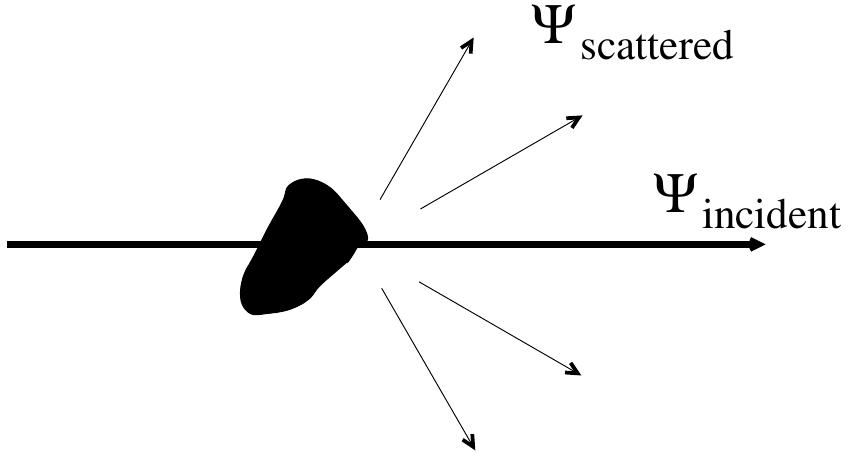}
\end{center}

Below is an illustration of the 1D case 
where we have just two directions of propagation 
(forwards and backwards):

\begin{center}
\putgraph[0.45\hsize]{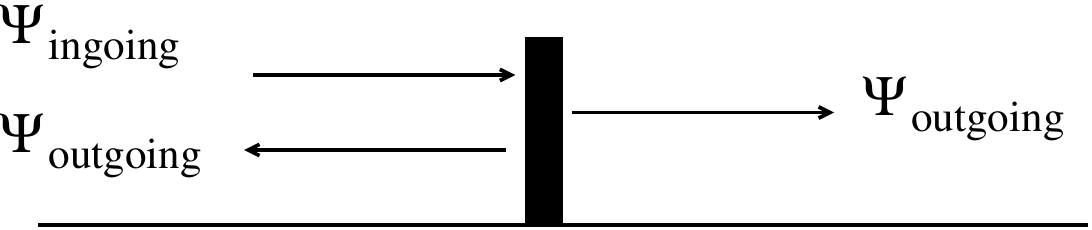} 
\hspace{0.05\hsize}
\putgraph[0.45\hsize]{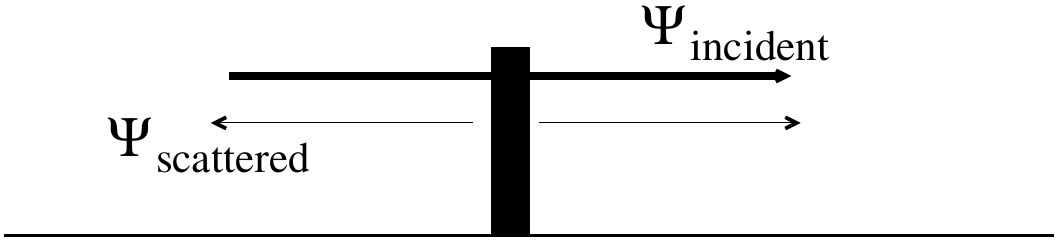}
\end{center}

The $S$ matrix gives the amplitudes  
of the outgoing wave, while the $T = i(\hat{1}-S)$ 
matrix gives the amplitudes of the 
scattered component (up to a phase factor).
Let us consider extreme case in order to clarify 
this terminology. If we have in the above 1D geometry 
a very high barrier, then the outgoing wave on the right 
will have zero amplitude.  This means that 
the scattered wave must have the same 
amplitude as the incident wave but with 
the opposite sign.

In order to define the "cross section" we assume that 
the incident wave is a density normalized ($\rho=1$) 
plane wave. This is like having a current density 
$J=\rho v_E$. If we regard the target as having 
some "area" $\sigma$ then the scattered current is 
\beq
I_{scattered} \,\, =  \,\,  (\rho v_E) \times \sigma 
\eeq
It is clear from this definition that the units 
of the cross section are $[\sigma]=L^{d-1}$. 
In particular in 1D geometry the "differential 
cross section" into channel $a$, assuming an incident 
wave in channel $a_0$ is simply 
\beq
\sigma_a \,\, =  \,\,  |T_{a,a_0}|^2 
\eeq
and by the "optical theorem" the total cross section is 
\beq
\sigma_{\tbox{total}} \,\, =  \,\,  -2\im[T_{a_0,a_0}]
\eeq
The notion of "cross section" is problematic 
conceptually because it implies that it is possible for 
the scattered flux to be larger than 
the incident flux. This is most evident in the 1D 
example that we have discussed above. We see that the 
scattered wave is twice the incident wave,  
whereas in fact the forward scattering cancels the 
outgoing component of the incident wave. 
Later we calculate the cross section of a sphere 
in 3D and get twice the classical cross section ($2\pi a^2$). 
The explanation is the same - there must be 
forward scattering that is equal in magnitude 
(but opposite in sign) to the outgoing component 
of the incident wave in order to create a "shadow region" 
behind the sphere.

\sheadC{The Wigner time delay}

A given element of the $S$ matrix can be written as 
\beq 
S_{ab} \ \ = \ \ \sqrt{g}\eexp{i\theta}
\eeq
where $0<g<1$ is interpreted as either 
transmission or reflection coefficient, 
while $\theta$ is called phase shift.
The cross section is related to~$g$    
while the Wigner time delay that we discuss     
below is related to~$\theta$. 
The Wigner time delay is defined as follows: 
\beq 
\tau_{\tbox{delay}}(E) = \hbar\frac{d\theta}{dE} 
\eeq
Consider for example the time delay in the case 
of a scattering on a ``hard" sphere of radius~$R$. 
We shall see that in such case we have a smooth energy 
dependence $\theta \approx -2kR$ and consequently  
we get the expected result $\tau \approx -2R/v_E$.  
On the other hand we shall consider the case 
of a scattering on a shielded well. 
If the energy is off resonance we get 
the same result $\tau \approx -2R/v_E$. 
But if $E$ is in the vicinity of a resonance, 
then we can get a very large (positive) 
time delay $\tau \sim \hbar/\Gamma_r$, 
where $\Gamma_r$ is the so called ``width" 
of the resonance.  This (positive) time delay 
reflects "trapping" and it is associated 
with an abrupt variation of the cross section 
as discussed in a later section.

Let us explain the reasoning that leads 
to the definition of the Wigner time delay.
For this purpose we consider the propagation of 
a Gaussian wavepacket in a given channel: 
\beq
\Psi(x) \ \ = \ \ \int dk \ \eexp{-\sigma^2(k-k_0)^2} 
\, \sqrt{g}\exp\left[ i\left( k(x-x_0) + \theta - Et \right)\right]
\eeq 
where both $g$ and $\theta$ and $E$ are functions of $k$. 
In order to determine the position $\bar{x}$ of the wavepaket 
we use the stationary phase approximation.  
Disregarding the Gaussian envelope, and regarding $g$ 
as constant at the energy range of interest, 
most of contribution to the $dk$ integral comes from 
the $k$ region where the phase is stationary. This leads to 
\beq 
x-x_0 + \frac{d\theta}{dk} - \frac{dE}{dk}t = 0
\eeq 
The position $\bar{x}$ of the wavepacket is 
determined by the requirement that the above 
equation should have a solution for $k \sim k_0$. 
Thus we get
\beq 
\bar{x} = x_0  + v_{\tbox{group}} \times (t-\tau_{\tbox{delay}})
\eeq 
where $v_{\tbox{group}}=(dE/dk)$ is the group velocity 
for $k=k_0$ and $\tau_{\tbox{delay}}$ is the Wigner time 
delay as defined above. 
How this is related to a scattering problem?
Consider for example scattering on one dimensions 
where we have "left" and "right" channels.
Assume that the wavepacket at $t=0$ 
is an undistorted Gaussian with $\theta=0$. 
The initial wavepacket is centered on the "left" 
around $\bar{x}=x_0$ with momentum $\bar{k}=k_0$. 
We would like to know how  $\bar{x}$ evolves  
with time before and after the scattering. 
Say that we observe the transmitted part of 
the wavepacket that emerges in the "right" channel.
Following the above analysis we conclude that 
the effect of introducing $\theta \ne 0$ 
is the Wigner time delay.

\sheadC{The Friedel phase and the DOS}

The issue that we discuss in this section is how 
to determine the number $\mathcal{N}(E)$  
of energy levels in the scattering region if we know 
how the $S$~matrix depends on the energy~$E$.
This would be the number of no interacting Fermions 
that can be accommodated there at zero temperature.
More precisely what we can extract from $S(E)$ 
is the density of states $\gdos(E)$ in the scattering region. 
To have a clear idea of the physical picture it is best 
to imagine a single lead system, 
where the lead is a waveguide with $\mathcal{M}$ modes, 
and the ``depth" of the scattering region is $L$.
In such system 
\beq 
\gdos(E) 
\ \ \sim \ \ 
\mathcal{M} \times v_{\tbox{E}} \frac{\pi}{L}
\ \ \sim \ \
\mathcal{M}  \times \tau_{\tbox{delay}}
\eeq
where $\tau_{\tbox{delay}} \sim L/v_{\tbox{E}}$ 
is the semiclassical estimate for the time delay.  
We shall see that the so called Friedel-sum-rule 
is merely a refinement of the latter formula.
In complete analogy with the Wigner formula,  
that express $\tau_{\tbox{delay}}$ as 
the $E$~derivative of the scattering phase shift, 
we are going to express $\gdos(E)$ as 
the $E$~derivative of the ``Friedel phase".

The most illuminating way to define the energy levels 
of the scattering region is to impose Dirichlet boundary 
condition at the $r=0$ section. This leads to 
the equation (say for 3~modes):
\beq 
\mathbf{S} 
\left(\amatrix{A_1\cr A_2\cr A_3}\right)
\ \ = \ \ 
\left(\amatrix{A_1\cr A_2\cr A_3}\right)
\eeq
This equation has a non-trivial solution if and on if 
\beq 
\det\Big( \mathbf{S}(E) - \bm{1} \Big) \ \ = \ \ 0
\eeq
Form this condition we can extract the eigen-energies. 
Recall that the $S$ matrix is unitary.  
Therefore its $\mathcal{M}$ complex eigenvalues ${\eexp{i\theta_r}}$ 
are located on a circle. We get zero determinant 
in the above equation each time that one of the 
eigen-phases cross through ${\theta=0}$. 
Thus the mean level spacing is simply $2\pi/\mathcal{M}$ 
divided by the average ``velocity" $d\theta/dE$.  
This leads to 
\beq 
\gdos(E)
\ \ = \ \ 
\frac{1}{2\pi} \left(\frac{d\theta_{\tbox{E}}}{dE}\right)
\ \ = \ \ 
\frac{1}{2\pi i}\trc\left(\frac{dS}{dE}S^{\dagger}\right)
\eeq 
where the Friedel phase is ${\theta_{\tbox{E}}=\sum_r \theta_r}$.
The last equality is easily derived by calculating the 
trace in the basis in which $S$ is diagonal.

A straightforward generalization of the above considerations 
allow to calculate the number of fermions that are emitted 
from the scattering region if a parameter~$X$ is being varied 
very very slowly such that at any moment we have zero temperature 
occupation. The result can be written as ${dN=-GdX}$ where 
\beq 
G
\ \ = \ \ 
\frac{1}{2\pi i}\trc\left(\frac{dS}{dX}S^{\dagger}\right)
\eeq 
The further generalization of this relation to multi-lead geometry 
has been discussed by Brouwer following the work of Buttiker Pretre and Thomas, 
and is known as the scattering approach to quantum pumping.

\sheadC{Phase shifts and resonances}

If we have only one wire (hence one channel) 
the $S$ matrix is ${1\times1}$, and accordingly 
we can write 
\beq
S_{00} \ \ &=& \ \ \exp[i2\delta_0(E)] \\
T_{00} \ \ &=& \ \ -\eexp{i\delta_{0}} 2\sin(\delta_{0})
\eeq     
where $\delta_0(E)$ is known as the phase shift. 
If the scattering potential is short range one 
can use a matching procedure in order to determine 
the phase shift. This will be discussed in the next 
lecture. Another possibility is to use perturbation 
theory via the $T$ matrix formalism:
\beq
T_{00} \ \ = \ \ V_{00} + (VGV)_{00} \ \ = \ \ \mbox{Born expansion}
\eeq     
It is essential to calculate the matrix elements 
using properly flux-normalized free waves:
\beq 
|\phi^{E}\rangle 
\ \ = \ \ 
\frac{1}{\sqrt{v_E}}\eexp{-ik_Er} - \frac{1}{\sqrt{v_E}}\eexp{+ik_Er} 
\ \ = \ \
-i \frac{1}{\sqrt{v_E}} 2\sin(k_Er) 
\eeq
There are two limiting cases of particular interest.  
\begin{itemize}
\setlength{\itemsep}{0mm}
\item 
First order non-resonant scattering by a weak potential $V$  
\item
Resonant scattering which is dominated by a single pole of $G$  
\end{itemize}
In the case of scattering by a weak potential 
we can use the first order Born approximation 
for the $T$ matrix: 
\beq
T_{00} 
\ \ \approx \ \  
V_{00} 
\ \ = \ \ 
\langle \phi^{E} | V |  \phi^{E} \rangle
\ \ = \ \ 
\frac{4}{v_E} \int_{0}^{\infty} V(r)\left(\sin(k_Er)\right)^2 dr 
\eeq
The assumption of weak scattering 
implies $\delta_{0} \ll 1$, leading to the 
first order Born approximation for the phase shift: 
\beq
\delta^{\tbox{Born}}_{0} \approx
-\frac{2}{v_E} \int_{0}^{\infty} V(r)\left(\sin(k_Er)\right)^2 dr 
\eeq
This formula is similar to the WKB phase shift formula. 
It has a straightforward generalization to any $\ell$ 
which we discuss in the context of spherical geometry.   
We note that we have manged above to avoid the standard 
lengthy derivation of this formula, 
which is based on the Wronskian theorem (Messiah p.404).

The other interesting case is resonant scattering where 
\beq
T_{00} \ \  \approx \ \ (VGV)_{00}
\ \ = \ \ 
\frac{ \langle \phi^{E} | V | r \rangle \langle \tilde{r} | V | \phi^{E} \rangle }
{E-E_{r}+i(\Gamma_{r}/2)}
\eeq
From the optical theorem $2\im[T_{00}]=-|T_{00}|^{2}$  
we can deduce that the numerator, if it is real,  
must equal $\Gamma_r$. 
Thus we can write this approximation 
in one of the following equivalent ways:  
\beq
T_{00} \ \ &=& \ \ \frac{\Gamma_{r}}{E-E_{r}+i(\Gamma_{r}/2)}
\\
S_{00} \ \ &=& \ \ \frac{E-E_{r}-i(\Gamma_{r}/2)}{E-E_{r}+i(\Gamma_{r}/2)}
\\ 
\tan(\delta_{\ell}) \ \ &=& \ \ -\frac{\Gamma_r/2}{E-E_r}
\eeq
Note that if the optical 
theorem were not satisfied 
by the approximation 
we would not be able 
to get a meaningful expression 
for the phase shift.   
In order to prove the equivalence 
of the above expressions note 
that $\delta_0$ can be regarded  
as the polar phase of the complex 
number ${z= (E{-}E_{r})-i(\Gamma_{r}/2)}$.

The Wigner time delay is easily obtained by taking 
the derivative of $\delta_0(E)$ with respect to the energy.
This gives a Lorentzian variation of $\tau_{\tbox{delay}}$ 
as a function of $(E{-}E_r)$. The width of 
the Lorentzian is $\Gamma_r$, and the time delay 
at the center of the resonance is of order $\hbar/\Gamma_r$.

\newpage

\sheadB{Scattering in quasi 1D geometry}

In this lecture we consider various scattering 
problems is which the radial functions are $\exp(\pm i k_ar)$, 
where $a$ is the channel index. In order to find a 
scattering solution we have to use a {\em matching} procedure. 
Namely the ingoing and outgoing free wave solutions 
should match an interior solution within 
the scattering region. This requirement imposes 
a boundary condition that relates the wavefunction 
amplitude and its derivative on the boundary.   
The simplest example for this class of systems  
is a network that consists of $\mathcal{M}$ semi 1D wires 
that are connected at one junction. 
The problem of s-scattering ($\ell=0$) in spherical 
geometry is formally the same as  $\mathcal{M}{=}1$ "semi 1D wire" problem. 
We shall discuss the general case (any $\ell$) in a later lecture. 
The 1D scattering problem on a line, 
where we have "left" and "right" leads, 
is formally an $\mathcal{M}{=}2$ "semi 1D wire" system.  
Then it is natural to consider less trivial $\mathcal{M}$~channel 
systems that can be regarded as generalizations of the 1D problem, 
including scattering in waveguides and multi-lead geometries. 
Also {\em inelastic} scattering in 1D can be formally re-interpreted 
as an $\mathcal{M}$~channel "semi 1D wire" problem.

\sheadC{The matching procedure}

In this section we would like to outline 
a procedure for finding an exact result 
for the phase shift is quasi 1D one channel geometry. 
This procedure will allow us to analyze s-scattering 
by ``hard" spheres as well as by ``deep" wells. 
The key assumption is that we have a finite 
range potential which is contained within 
the region ${r<R}$. The boundary of the scattering 
region at ${r=R}$ is fully characterized by the logarithmic 
derivative $\tilde{k}_{0}(E)$. 
The definition of the latter is as follows: 
given the energy $E$, one finds the regular 
solution $\psi(r)$ of the Schrodinger 
equation in the interior (${r \le R}$) region;  
Typically this solution is required to 
satisfy Dirichlet boundary conditions 
at the origin ${r=0}$; Once the regular solution 
is known, the logarithmic derivative 
at the boundary is calculated as follows:
\beq
\tilde{k}_{0} \ \ = \ \ \left[\frac{1}{\psi(r)}\frac{d\psi(r)}{dr}\right]_{r=R}
\eeq
The derivative should be evaluated  
at the {\em outer} side of the boundary. 
For each problem $\tilde{k}_{0}(E)$
should be evaluated from scratch. 
But once it is known we can find both 
the ${E<0}$ bound states 
and the ${E>0}$ scattering states 
of the system. It is useful the consider the 
simplest example: square well of radius~$R$  
that has depth~$|V_0|$. The regular solution 
in the interior region is ${\Psi(r) \propto \sin(\alpha_{\text{in}} r)}$, 
and accordingly the logarithmic derivative 
at the boundary is 
\beq
\tilde{k}_{0}(E) \ = \ \alpha_{\text{in}} \cot(\alpha_{\text{in}} R), 
\hspace{2cm} \text{where} \ \alpha_{\text{in}}  = \sqrt{2\mass(E+|V_0|)}
\eeq

{\bf Finding the bound states:}
In the case of a bound 
state the wavefunction at the outside 
region is ${\psi(r) \propto \exp(-\alpha_E r)}$. 
The matching with the interior solution  
gives the equation 
\beq
\tilde{k}_{0}(E) \ \ = \ \ -\alpha_E
\hspace{2cm} \text{where} \ \alpha_E  = \sqrt{2\mass|E|}
\eeq
This equation determines the eigen-energies  
of the system. In order to have 
bound states $\tilde{k}_{0}$ should become  
negative. This is indeed the case 
with square well as $|V_0|$ being increased. 
It should be opposed with the case of 
a square barrier.

{\bf Finding the scattering states:}
For positive energies we look 
for scattering states.    
The scattering solution    
in the outside region is 
\beq
\Psi(r) 
\ \ = \ \ 
A\eexp{-ik_Er}-B\eexp{ik_Er} 
\ \ = \ \ 
A (\eexp{-ik_Er}- \eexp{i2\delta_0} \eexp{ik_Er}) 
\ \ =  \ \ 
C \sin(k_Er+\delta_0)
\eeq
where $\delta_0$ is the phase shift.
The matching with the interior solution 
gives the equation 
\beq
k_E\cot(k_ER+\delta_0) \ \ = \ \ \tilde{k}_{0}
\eeq
This equation can be written as 
$\tan(\delta_0-\delta_0^{\infty}) = {k_E}/{\tilde{k}_{0}}$, 
where $\delta_0^{\infty} = -k_ER$ is the solution 
for ``hard sphere scattering". Thus the explicit solution 
for the phase shift is 
\beq
\delta_0 \ \ = \ \ \delta_0^{\infty} + \arctan(\frac{k_E}{\tilde{k}_{0}})
\eeq

It is clear that in practical problems  
the boundary ${r=R}$ is an arbitrary radius 
in the "outside" region. It is most natural 
to extrapolate the outside solution into the 
the region ${r<R}$ as if the potential there is zero. 
Then it is possible to define the logarithmic 
derivative $\bar{k}_0$ of the extrapolated 
wavefunction at ${r=0}$. This function contains 
the same information as $\tilde{k}_0$, 
while the subtlety of fixing an arbitrary $R$ 
is being avoided.  From the above analysis we get
\beq
\bar{k}_0(E) 
\ \ = \ \ \left[\frac{1}{\psi(r)}\frac{d\psi(r)}{dr}\right]_{r=0}
\ \ = \ \ k_E\cot(\delta_0(E)) 
\ \ = \ \ \frac{\tilde{k}_{0}+k_E\tan(k_ER)}{k_E-\tilde{k}_{0}\tan(k_ER)}
\eeq 
At low energies, in the limit $k_E\rightarrow 0$, 
this expression takes the simpler form ${\bar{k}_0=((1/\tilde{k}_0)-R)^{-1}}$.

\sheadC{Low energy scattering}

Of particular interest is the long-wavelength limit, 
which means ${k_ER \ll 1}$ where $R$ is the radius 
of the scattering region. Then the regular solution 
for ${r \sim R}$, where the potential is zero,   
has the form ${\psi(r)\propto r-a_{\tbox{s}}}$.
This implies by definition that for low energy~${\delta_0=-k_Ea_{\tbox{s}}}$.  
The parameter~$a_{\tbox{s}}$ is called the scattering length. 
Given $a_s$, it follows from the above definition 
that the log derivative, referenced to the origin, 
is ${\bar{k}_0=-1/a_{\tbox{s}}}$. 
More generally it is customary to expand  
the log-derivative at the origin as follows:  
\beq
\bar{k}_0(E) \ \ \equiv \ \ k_E\cot(\delta_0(E)) 
\ \ = \ \ -\frac{1}{a_{\tbox{s}}} \ + \ \frac{1}{2}r_{\tbox{s}} k_E^2 + ...
\eeq
Thus $a_s$ is merely a parameter in the expression 
for phase shift in the limit of low energies. 
If this phase shift is small it can be  
expressed as ${\delta_0=-k_Ea_{\tbox{s}}}$.
For a sphere of radius~$R$ and potential~$V_0$
one obtains ${a_s = R}$ in the hard sphere limit (${V_0=\infty}$).
As $V_0$ is decreased one obtains ${a_s \approx (2/3)\mass R^3 V_0 }$.
For negative $V_0$ (``spherical well") one obtains ${a_s=R-(1/\alpha)\tan(\alpha R)}$, 
where ${\alpha=\sqrt{2\mass |V_0|}}$.

At this stage we mention a common terminology that might 
look strange at first sight. Referring to a square well 
it is clear that is $|V_0|$ is very small the potential 
has no bound state. We can regard such potential as a broadened 
delta function $V(r) = u \delta(r)$ that has ${u<0}$. 
In such situation we have ${\delta_0>0}$ and $a_s<0$, 
which should be contrasted with hard sphere. 
Such ${u<0}$ scattering potential is called ``attractive".
As $|V_0|$ becomes larger it can accommodate 
a bound state, and it behaves like a ${u>0}$
delta potential with ${\delta_0<0}$ and $a_s>0$. 
Such scattering potential is called ``repulsive".
The bound state energy of the such ``repulsive"  
potential is determined from the matching condition
that has been discussed in the previous section, leading to  
\beq
E_{\tbox{bound}} \ \ = \ \ -\frac{1}{2\mass a_{\tbox{s}}^2}
\eeq
It should be clear that this solution is 
meaningful provided the assumptions ${a_{\tbox{s}} \gg R}$ 
and $|E_{\tbox{bound}}|<\Delta$ are satisfied.
Here $\Delta$ is the range within which $a_s$ 
can be regarded as a constant.
 
From the definition of the scattering length 
it follows that for weak scattering it can be 
expressed as ${\delta_0=-k_Ea_{\tbox{s}}}$.
In a later section we shall see that in 3D 
this means that the total s-scattering cross section is 
\beq
\sigma_{\tbox{total}} 
\ \ = \ \ \frac{4\pi}{k_E^2}\sin^2(\delta_0)  
\ \ \approx \ \  4\pi a_{\tbox{s}}^2 
\eeq
A delta function $V(x) = u\,\delta^3(x)$ in 3D has a zero total 
cross section. Still from the first order Born approximation 
for the $T$ matrix we get as an artifact 
\beq
\sigma_{\tbox{total}} \ \ = \ \ 4\pi\left(\frac{\mass}{2\pi}u\right)^2 
\eeq
It is therefore common to describe an s-scatterer 
that is characterize by a scattering length $a_{\tbox{s}}$, 
as a delta scatterer with ${u=2\pi a_{\tbox{s}}/\mass}$.   
One should be very careful with the latter point of view.

\sheadC{Inelastic scattering by a delta scatterer}

We consider inelastic scattering on a delta scatterer  
in one dimension [\href{http://arxiv.org/abs/0801.1202}{arXiv:0801.1202}]
\beq
\mathcal{H} \ \ = \ \ \frac{p^2}{2\mass} 
+ Q\delta(x) + \mathcal{H}_{\tbox{scatterer}}
\eeq
The scatterer is assumed to have energy levels $n$ 
with eigenvalues $E_n$. It should be clear that 
inelastic scattering of a spinless particle 
by a multi level atom is mathematically equivalent 
to inelastic scattering of a multi level atom 
by some static potential. 
Outside of the scattering region the total energy 
of the system (particle plus scatterer) is 
\beq
\mathcal{E} \ \ = \ \ \epsilon_k + E_{n} 
\eeq
We look for scattering states that satisfy the equation
\beq
\mathcal{H}|\Psi\rangle=\mathcal{E}|\Psi\rangle 
\eeq
The scattering channels are labeled as 
\beq
\bm{n} = (n_0,n) 
\eeq
where $n_0=\mbox{left,right}$. We define
\beq
k_{\bm{n}} =& \sqrt{2m(\mathcal{E}-E_n)}
\ \ \ \ \ & \mbox{for $n\in$ open}
\\ 
\alpha_{\bm{n}} =& \sqrt{-2m(\mathcal{E}-E_n)}
\ \ \ \ \ & \mbox{for $n\in$ closed}
\eeq
later we use the notations
\beq
v_{\bm{n}} &=&  k_{\bm{n}} / \mass \\
u_{\bm{n}} &=&  \alpha_{\bm{n}} / \mass
\eeq
and define diagonal matrices $\bm{v}=\mbox{diag}\{v_n\}$ 
and $\bm{u}=\mbox{diag}\{u_n\}$. 
The channel radial functions are written as 
\beq
R(r) =& A_n \eexp{-ik_nr} + B_n \eexp{+ik_nr}  
\ \ \ \ \ & \mbox{for $n\in$ open} \\
R(r) =& C_n \eexp{-\alpha_n r} 
\ \ \ \ \ & \mbox{for $n\in$ closed}
\eeq
where $r=|x|$. Next we derive expression for the $2N\times 2N$ 
transfer matrix $\bm{T}$ and for the $2N\times 2N$ 
scattering matrix $\bm{S}$, where $N$ is the number 
of open modes. 
The wavefunction can be written as 
\beq
\Psi(r, n_0, Q) = \sum_{n} R_{n_0,n}(r) \chi^{n}(Q)
\eeq 
The matching equations are
\beq
\Psi(0,\mbox{right},Q) = \Psi(0,\mbox{left},Q) \\
\frac{1}{2\mathsf{m}}
\left[\Psi'(0,\mbox{right},Q) + \Psi'(0,\mbox{left},Q)\right] 
= \hat{Q} \Psi(0,Q) 
\eeq
The operator $\hat{Q}$ is represented 
by the matrix $Q_{nm}$ that has the block structure 
\beq
Q_{nm}
= \left(\amatrix{ 
Q_{vv} & Q_{vu} \\  
Q_{uv} & Q_{uu}
} \right)
\eeq
For sake of later use we define
\beq
M_{nm}
= \left(\amatrix{ 
\frac{1}{\sqrt{v}}Q_{vv}\frac{1}{\sqrt{v}} 
& \frac{1}{\sqrt{v}}Q_{vu}\frac{1}{\sqrt{u}} \\  
\frac{1}{\sqrt{u}}Q_{uv}\frac{1}{\sqrt{v}} 
&\frac{1}{\sqrt{u}}Q_{uu}\frac{1}{\sqrt{u}}
} \right)
\eeq
The matching conditions lead to the following set of matrix equations
\beq
A_R + B_R &=& A_L + B_L \\
C_R &=& C_L \\
-i\bm{v} (A_R-B_R + A_L-B_L) &=& 2Q_{vv}(A_L+B_L) + 2Q_{vu}C_L \\
-\bm{u}(C_R + C_L) &=& 2Q_{uv}(A_L+B_L) + 2Q_{uu} C_L
\eeq
from here we get 
\beq
A_R + B_R &=& A_L + B_L \\
A_R-B_R + A_L-B_L &=& i2 (\bm{v})^{-1} \mathcal{Q} (A_L+B_L) 
\eeq
where
\beq
\mathcal{Q} = Q_{vv} 
- Q_{vu}
\frac{1}{(\bm{u} + Q_{uu})} 
Q_{uv}
\eeq
The set of matching conditions
can be expressed using a transfer matrix 
formalism that we discuss in the next 
subsection, where 
\beq
\mathcal{M}
= 
\frac{1}{\sqrt{\bm{v}}} 
\mathcal{Q} 
\frac{1}{\sqrt{\bm{v}}} 
=
M_{vv} 
- M_{vu}
\frac{1}{1 + M_{uu}} 
M_{uv}
\eeq

\sheadC{Finding the $S$ matrix from the transfer matrix}

The set of matching conditions in the 
delta scattering problem can be written as 
\beq
\left( \amatrix{ \tilde{B}_R \\ \tilde{A}_R } \right) 
= \bm{T} \left(\amatrix{ \tilde{A}_L \\ \tilde{B}_L } \right) 
\eeq
where $\tilde{A_n} = \sqrt{v_n} A_n$ 
and $\tilde{B_n} = \sqrt{v_n} B_n$. 
The transfer $2N \times 2N$ matrix can be  
written in block form as follows:
\beq
\bm{T}
=
\left( \amatrix{ 
\bm{T}_{++} & \bm{T}_{+-} \\
\bm{T}_{-+} & \bm{T}_{--}      
} \right)
=
\left( \amatrix{ 
1-i\mathcal{M} & -i\mathcal{M} \\
i\mathcal{M} & 1 + i\mathcal{M}      
} \right)
\eeq
The $\bm{S}$ matrix is defined via 
\beq
\left( \amatrix{ \tilde{B}_L \\ \tilde{B}_R } \right) 
= \bm{S} \left( \amatrix{ \tilde{A}_L \\ \tilde{A}_R } \right) 
\eeq
and can be written in block form as 
\beq
\bm{S}_{\bm{n},\bm{m}} \ \ = \ \  
\left(\amatrix{ \bm{S}_R & \bm{S}_T \\ \bm{S}_T & \bm{S}_R } \right)
\eeq
A straightforward elimination gives 
\beq
\bm{S} \ \ = \ \ 
\left( \begin{array}{ccc} 
-\bm{T}_{--}^{-1}\bm{T}_{-+} 
& \ & 
\bm{T}_{--}^{-1} 
\\ 
\bm{T}_{++} {-} \bm{T}_{-+}\bm{T}_{--}^{-1}\bm{T}_{+-} 
& \ & 
\bm{T}_{+-} \bm{T}_{--}^{-1} 
\end{array}\right) 
\ \ = \ \ 
\left(\begin{array}{ccc}
\left(1+i\mathcal{M}\right)^{-1}-1 & \ & \left(1+i\mathcal{M}\right)^{-1} 
\\
\left(1+i\mathcal{M}\right)^{-1} & \ & \left(1+i\mathcal{M}\right)^{-1}-1
\end{array}\right)
\eeq 
Now we can write expressions for $\bm{S}_R$ and for $\bm{S}_T$ 
using the $\mathcal{M}$ matrix. 
\beq
\bm{S}_T &=& \frac{1}{1+i\mathcal{M}} 
= 1 - i\mathcal{M} - \mathcal{M}^2 +i\mathcal{M}^3 + ... \\
\bm{S}_R&=& \bm{S}_T - \bm{1}
\eeq

\sheadC{Elastic scattering by a delta in a waveguide}

The {\em elastic} scattering of a spinless particle 
by a regularized delta scatterer in a waveguide is mathematically 
the same problem as that of the previous section. We have 
\beq
Q=c\delta(y-y_0)
\eeq 
for which 
\beq
{Q}_{{\mathrm{nm}}} 
\ \ = \ \ c \int 
{\mathrm{ \chi }}_{{n}}\mathrm{ \delta } ( y-y_0 ) 
{\mathrm{ \chi }}_{{m}}\mathrm{dy} 
\ \ = \ \ c \ \chi^n(y_0) \chi^m(y_0)
\eeq
Given the total energy we define 
\beq
M_{nm} \ \ = \ \ 
\frac{1}{\sqrt{|v_n|}} 
Q_{nm} 
\frac{1}{\sqrt{|v_m|}}
\ \ \equiv \ \  
\left(
\amatrix{ M_{vv} & M_{vu} \cr M_{uv} & M_{uu} } 
\right)
\eeq
Regularization means that one impose  
a cutoff on the total number of coupled 
channels, hence $M$ is a finite (truncated) matrix.  
Using the formula for inverting a matrix 
of the type $1-aa^{\dagger}$, 
we first obtain $\mathcal{M}$ and then obtain  
\beq
\bm{S}_R = \frac{iM_{vv}}{1+i \ \trc[M_{vv}]+\trc[M_{uu}]}
\eeq
Let us consider what happens as we change the total 
energy: Each time that a new channels is opened  
the scattering cross section becomes zero.
Similarly, if we remove the regularization we get zero scattering 
for any energy because of the divergent contribution of the closed channels.

\sheadC{Transmission through a tight-binding network}

Consider a network that consists of $N$ sites and described 
by Hamiltonian $\mathcal{H_0}$. 
We can attach a lead to a site of the network, 
or in general we attach several leads to different points.  
We would like to calculate the $S$ matrix in this configuration.  
For this purpose we apply the $T$ matrix formalism, 
combined with the "P+Q" formalism where "P" is the network 
and "Q" are the leads. 

The lead is regarded as a chain of sites ($x=1,2,3,...$)
with hopping amplitude $-c_0$. Accordingly a free "wave" 
with wavenumber $k$ has energy and velocity as follows:
\beq
\epsilon \ &=& \ -2c_0\cos(k) \\
v\ &=& \ \ 2c_0\sin(k), 
\hspace*{2cm} \text{mass} \equiv 1/(2c_0) 
\eeq
The flux-normalized free wave in a disconnected lead is 
\beq
\varphi(x) \ \ = \ \ \frac{2}{\sqrt{v}} \sin(kx),
\hspace*{2cm}
\mbox{Note:} \  \varphi(1)=\frac{\sqrt{v}}{c_0}
\eeq
The Green function due a source at the first site is 
\beq
G^{\text{lead}}(x|1) \ \ = \ \ -\frac{1}{c_0}\eexp{ikx} 
\hspace*{2cm}
\mbox{Note:} \ G(1|1)= \frac{\epsilon-iv}{2c_0^2}
\eeq
The hopping amplitude between the ${x=1}$ site of the lead 
that is labeled as $a$, and a contact point the network 
that is labeled as $n_a$, will be denoted $-c_{\gamma}$. 
Later we shall define the dimensionless coupling parameter ${\gamma=(c_{\gamma}/c_0)^2}$. 
If we have $M$ leads it is convenient to pack the couplings 
in a rectangular ${M\times N}$ contact matrix $Q$ that is composed 
of zeros and ones,  such that $V_{a,n}=-c_{\gamma} Q_{a,n}$.  
The Green function of the network can be calculated using the "P+Q" expression 
\beq
G(n|m) 
\ \ = \ \ \left[ \frac{1}{\epsilon-\mathcal{H}_0-V^{\dag}G^{\text{lead}}V} \right]_{n,m}
\ \ = \ \ \left[ \frac{1}{\epsilon-\mathcal{H}_0-\frac{\gamma}{2}(\epsilon-iv) Q^{\dag}Q} \right]_{n,m}
\eeq
The scattering matrix is expressed as $\bm{S}=\bm{1}-i\bm{T}$, 
where the $T$ matrix elements are 
\beq
\bm{T}_{a,b} =  
\ \ = \ \ \left\langle \varphi^{a} \left| VGV^{\dag} \right| \varphi^{b} \right\rangle 
\ \ = \ \ \left[\frac{\gamma v}{\epsilon-\mathcal{H}_{\text{eff}}+i\frac{\gamma v}{2} Q^{\dag}Q} \right]_{n_a,n_b}
\eeq
where $n_a$ and $n_b$ are the contact sites of lead $a$ and lead $b$, 
and the definition of $\mathcal{H}_{\text{eff}}$ is implied from the 
expression for $G(n|m)$. Note that the $Q^{\dag}Q$ term implies 
that the on-site potential at the contact sites becomes complex.  

Assume that the network represents a 1D wire segment to which leads 
are attached at the endpoint. Ideal contacts corresponds to $\gamma=1$. 
It is instructive to consider the continuum limit of the $T$ matrix 
expression for such a setup. This limit corresponds to small~$k$ 
at the bottom of the band where ${\epsilon\approx-2c_0}$. 
One observes that the effective Hamiltonian becomes 
\beq
\mathcal{H}_{\text{eff}} \ \ = \ \ \mathcal{H}_0 -c_0 Q^{\dag}Q
\eeq
This Hamiltonian corresponds to Neumann boundary conditions. 
The reasoning is as follows: If we place a negative 
delta $-[1/(2\mass a)]\delta(x-a)$ at an infinitesimal distance $a$ 
from the endpoint $x=0$ of the segment, 
it forces the derivative to be zero. In a tight-binding model 
this is like having a potential $-c_0$ at the first site.

\sheadC{Transmission through a cavity}

Consider a cavity to which a lead is attached.
We would like to calculate the $S$ matrix in this configuration.  
One possibility is to take the continuum limit 
of the network model that we have solved in the
previous section. This leads to the so-called Weidenmuller formula:
\beq
\bm{S} 
\ \  = \ \ \bm{1} \ - \ iWGW 
\ \  = \ \ \bm{1} \ - \ i W \frac{1}{E-\mathcal{H}_{\text{eff}}+i(W^{\dag}W/2)} W^{\dag} 
\eeq
where $\mathcal{H}_{\text{eff}}$ is the Hamiltonian 
of the cavity with Neumann boundary conditions on the 
surface of the scattering region. It is convenient
to find a complete set $\varphi^{(n)}(x)$ 
of cavity eigenfunctions. Then the expression for 
the contact matrix takes the form 
\beq
W_{a,n} \ \ = \ \ \sqrt{v_a} \int  \chi^a(s) \varphi^{(n)}(x(s)) \ ds
\eeq
where $\chi^a(s)$ are the channel functions 
and $s$ is the transverse coordinate.
Conventionally the Weidenmuller formula is  
derived from the so called $\bm{R}$ matrix 
formalism, which we outline below.

{\bf The Fisher-Lee relation.-- } 
Before we go on it is appropriate to point out 
the duality between the Weidenmuller formula
and the Fisher-Lee relation. 
The latter expresses the Green function $G(s|s_0)$
using the $S$ matrix as given. 
The surface coordinates $s$ and $s_0$ indicate 
points $x(s)$ and $x(s')$ on the boundary 
of the scattering region.  
 The standard derivation goes as follows [Datta]: 
We place a source at the lead and use the $S$ matrix 
to define the boundary conditions on the surface~$x(s)$  
of the scattering region. We solve for the outgoing 
amplitudes and find that 
\beq
G(s|s_0) \ \ = \ \  i\sum_{ab} 
\frac{1}{\sqrt{v_a}}\chi^a(s) 
\ (\bm{S}-\bm{1})_{ab} \  
\frac{1}{\sqrt{v_b}}\chi^b(s_0)      
\eeq
This relation can be inverted:
\beq
\bm{S}_{ab} 
\ \ = \ \ \delta_{ab} 
\ - \ i \sqrt{v_a v_b} 
\int \chi^a(s) \ G(s|s_0) \ \chi^b(s_0) 
\ ds ds_0
\ \ \equiv \ \ 
\Big[ 1 - iWGW^{\dag} \Big]_{a,b}  
\eeq
The latter expression has the same expected structure 
that is anticipated from $T$ matrix theory, 
and therefore should be equivalent to the Weidenmuller formula. 
Inserting a complete basis of cavity eigenstates for 
the representation of the Green function, 
the integral over $s$ becomes a sum over $n$,
and the definition of $W$ is implied. 
Still this perspective does not provide a practical 
procedure for the choice of the $n$ basis, 
and for the evaluation of $G_{n,n'}$.

{\bf The $R$ matrix formalism.-- } 
We would like to describe a formulation that opens 
the way for a powerful numerical procedure for 
finding the $\bm{S}$ matrix of a cavity-lead system. 
The idea is to reduce the scattering problem 
to a bound state problem by chopping the leads.  
It can be regarded as a generalization 
of the one-dimensional phase shift method where 
the outer solution is matched to an interior solution.
The latter is characterized by its log derivative 
on the boundary. In the same spirit the $\bm{R}$ matrix 
is defined through the relation 
\beq
\Psi(s) \ \ = \ \ \int \bm{R}(s,s') \partial \Psi(s') \ ds'  
\eeq
If we decompose this relation into channels 
we can rewrite it as 
\beq
\Psi_a \ \ = \ \ \sum_b \bm{R}_{ab} \partial \Psi_b 
\eeq
Expressing $\Psi_a$ and $\partial \Psi_a$ 
as the sum and the difference of the ingoing 
and the outgoing amplitudes $A_a$ and $B_a$,  
one finds a simple relation between the $\bm{R}$ matrix 
and the  $\bm{S}$ matrix:
\beq
\bm{R}_{ab} \ \ = \ \ i\frac{1}{\sqrt{k_a k_b}}
\left( \frac{1-\bm{S}}{1+\bm{S}} \right)_{ab} 
\eeq
The inverse relation is 
\beq
\bm{S} \ \ =  \ \
\frac
{1+i\sqrt{\bm{k}}\bm{R}\sqrt{\bm{k}}}
{1-i\sqrt{\bm{k}}\bm{R}\sqrt{\bm{k}}}
\eeq

From the Green theorem it follows that 
\beq
\bm{R}(s,s') \ \ = \ \ - \frac{\hbar^2}{2\mass} G^N(s'|s)
\eeq
where $G^N$ is the Green function of the interior 
with Neumann boundary conditions on the surface of the 
scattering region. If we find a complete set 
of interior eigenfunctions then 
\beq
G^N(s'|s) \ \ = \ \ \sum_n \frac{\varphi^{(n)}(s') \varphi^{(n)}(s) }{E-E_n}
\eeq
and consequently 
\beq
\bm{R}_{ab} \ \ = \ \ - \frac{1}{2} \sum_n 
\left(\frac{W_{an}}{\sqrt{k_a}}\right) 
\frac{1}{E-E_n}
\left(\frac{W_{bn} }{\sqrt{k_b}}\right)
\eeq
The corresponding result for the $\bm{S}$ matrix 
is obtained by expanding ${(1+x)/(1-x)=1+2(...)}$ 
with the identification of ${(...)}$ as the diagrammatic 
expression of the resolvent. 
This leads to the Weidenmuller formula.

\newpage

\sheadB{Scattering in a spherical geometry}

Of special interest is the scattering problem in 3D geometry.
We shall consider in this lecture the definitions  
of the channels and of the $S$ matrix for this geometry.
Later analyze in detail the scattering by a spherically 
symmetric target. In the latter case the potential $V(r)$ 
depends only on the distance from the origin. 
In order to find the scattering states we can perform 
separation of variables,   
where $\ell$ and $m$ are good quantum numbers.   
In the $(\ell,m)$ subspace the equation for 
the radial function $u(r)=rR(r)$ 
reduces to a one dimensional Schr\"{o}dinger equation 
on the $0<r< \infty$ axis. 
To avoid divergence in $R(r)$ 
we have to use the boundary condition $u(0)=0$, 
as if there is an infinite wall at $r=0$.

Most textbooks focus on the Coulomb interaction
${V(r,\theta,\varphi)=-\alpha/r}$
for which the effective radial potential 
is ${ V_{\tbox{eff}}(r) = -\alpha/r + \beta/r^2 }$.
This is an extremely exceptional potential 
because of the following: 

\bitem There is no centrifugal barrier. \\
\bitem Therefore there are no resonances with the continuum. \\
\bitem It has an infinite rather than a finite number of bound states. \\
\bitem The frequencies of the radial and the angular motions are degenerate. \\
\bitem Hence there is no precession of the Kepler ellipse.

We are going to consider as an example 
scattering on a spherical target 
which we call either "sphere" or "well".
The parameters that characterize the sphere 
are its radius $a$, the height of the 
potential floor $V$, and optionally 
the "thickness" the shielding $U$. Namely, 
\beq
V(r,\theta,\varphi) \ \ = \ \ V \Theta(a-r) + U\delta(r-a)
\eeq
Disregarding the shielding the effective radial potential 
is ${ V_{\tbox{eff}}(r) = V\Theta(a-r) + \beta/r^2 }$.
We consider first hard sphere ($V_0=\infty$) and later 
on the scattering on a spherical well ($V_0<0$).
The effective potential for 3 representative  
values of $V$ is illustrated in panels (a)-(b)-(c) 
of the following figure. In panels (d) we illustrate,  
for sake of comparison, the effective potential 
in case of a Coulomb interaction.

\begin{center}
\putgraph[0.7\hsize]{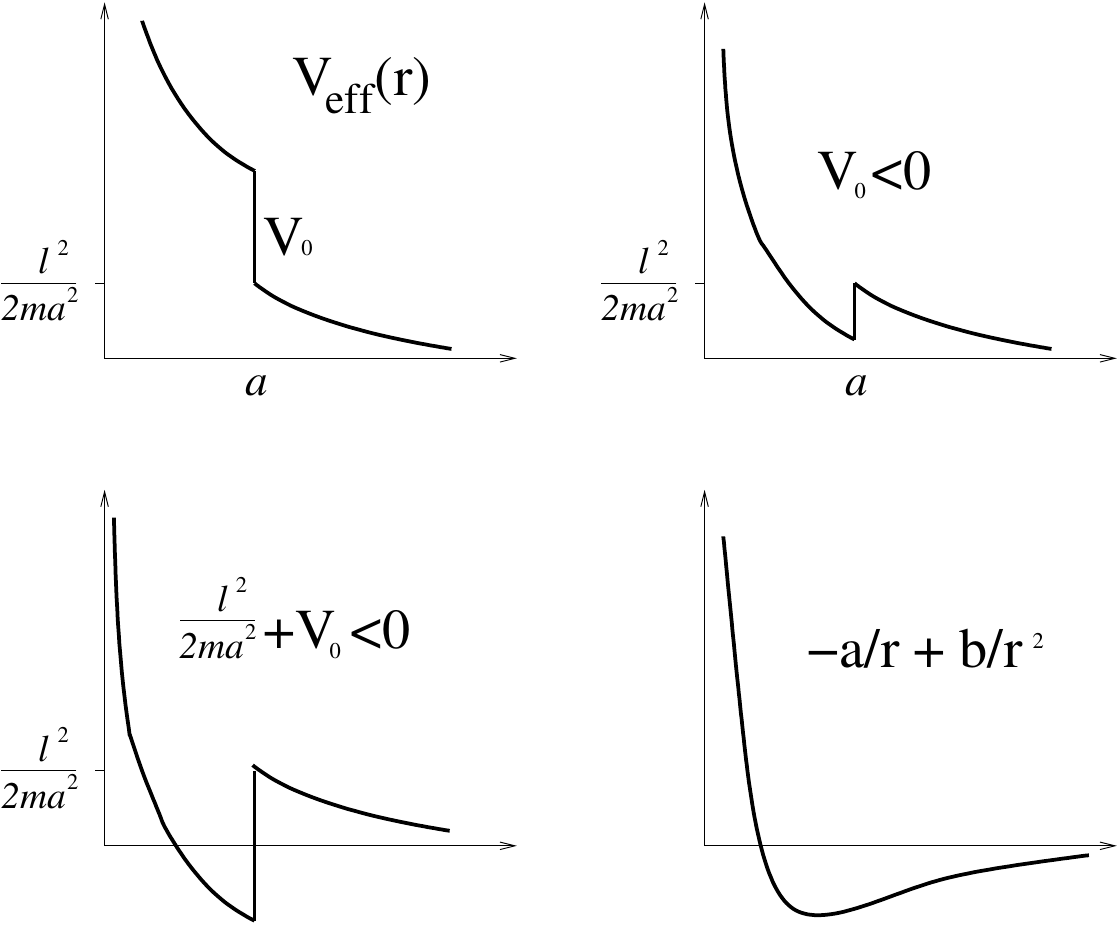}
\end{center}

Classically it is clear that if the impact parameter~$b$
of a particle is larger than the radius~$a$ of the 
scattering region then there is no scattering at all.
The impact parameter is the distance of the particle 
trajectory (extrapolated as a straight line) from 
form the scattering center. Hence its angular momentum 
is $\ell=\mass v_E b$. Thus the semiclassical condition 
for  non-negligible scattering can be written as 
\beq
b<a 
\hspace*{1cm}\Leftrightarrow\hspace*{1cm}
\ell < k_Ea
\hspace*{1cm}\Leftrightarrow\hspace*{1cm} 
\frac{\ell^2}{2\mass a^2} < E
\eeq
The last version is interpreted as the condition 
for reflection from the centrifugal barrier (see Fig). 
In the channels where the semiclassical condition 
is not satisfied we expect negligible scattering. 
(no phase shift). This semiclassical expectation 
assumes that the "forbidden" region is unaccessible 
to the particle, hence it ignores the possibility 
of tunneling and resonances that we shall discuss later on.  
Whenever we neglect the latter possibilities, the scattering 
state is simply a free wave which is described by 
the spherical Bessel function $j_{\ell}(k_E r)$.  
For $k_E r \ll \ell$ this spherical Bessel function 
is exponentially small due to the centrifugal barrier,
and therefore it is hardly affected by the presence 
of the sphere.

\sheadC{The spherical Bessel functions}

Irrespective of whether the scattering 
potential is spherically symmetric 
or not the Schrodinger equation is 
separable outside of the scattering region.
This means that we can expand 
any wavefunction that satisfies 
${\mathcal{H}\Psi =E\Psi}$ 
in the outside region as follows:
\beq
\Psi(x)= \sum_{\ell,m} R_{\ell, m}(r) Y^{\ell m}(\theta,\varphi)
\eeq
The channel index is $a=(\ell,m)$ while $\Omega=(\theta,\varphi)$ 
is analogous to the $s$ of the 2D lead system. 
The  $Y^{\ell m}$ are the channel functions. 
In complete analogy with the case of 1D geometry 
we can define the following set of functions: 
\beq
&& h^{+}_{\ell}(k_Er)   \leftrightarrow \eexp{ik_Er} 
\\ \nonumber
&& h^{-}_{\ell}(k_Er) \leftrightarrow \eexp{-ik_Er} 
\\ \nonumber
&& j_{\ell}(k_Er) \leftrightarrow \sin(k_Er)   
\\ \nonumber
&& n_{\ell}(k_Er) \leftrightarrow \cos(k_Er)
\eeq
Note that the right side equals the 
left side in the special case $\ell=0$, 
provided we divide by $r$. This is because 
the semi-1D radial equation becomes literally 
the 1D Schr\"{o}dinger equation only after the 
substitution $R(r)=u(r)/r$.

In what follows we use Messiah convention p.489. Note that 
other textbooks may use different sign convention.    
The relation between the functions above is defined as follows:
\beq
h_{\ell}^{\pm} \,\,=\,\, n_{\ell}(kr) \pm i j_{\ell}(kr) 
\eeq
We note that the $j_n(r)$ are regular at the origin, 
while the $n_n(r)$ are singular at the origin.
Therefore only the former qualify as global "free waves".
The $l=0$ functions are:
\beq
j_0(kr) &=& \frac{\sin(kr)}{kr} 
\\ \nonumber
n_0(kr) &=& \frac{\cos(kr)}{kr} 
\eeq
The asymptotic behavior for $kr \gg \ell{+}1$ is:
\beq
&& h_{\ell}^{\pm}(kr) \sim (\mp i)^{\ell} \,\, \frac{\eexp{\pm ikr}}{kr} 
\\ \nonumber
&& n_{\ell}(kr) \sim  \frac{\cos(kr-\frac{\pi}{2}\ell)}{kr} 
\\ \nonumber
&& j_{\ell}(kr) \sim  \frac{\sin(kr-\frac{\pi}{2}\ell)}{kr} 
\eeq
The short range $kr \ll \ell{+}1$ behavior is:
\beq
n_{\ell}(kr) &\approx& 
(2l-1)!! 
\left(\frac{1}{kr}\right)^{\ell+1} 
\left[1+\frac{1}{2(2l-1)}(kr)^2+ \dots \right]
\\ \nonumber
j_{\ell}(kr) &\approx& 
\frac{(kr)^{\ell}}{(2l+1)!!}
\left[1-\frac{1}{2(2l+3)}(kr)^2+ \dots \right] 
\eeq

\sheadC{Free spherical waves}

On the energy shell we write the radial wavefunctions as 
\beq
R_{\ell m}(r) = A_{\ell m} R^{E,\ell m,-}(r) - B_{\ell m} R^{E,\ell,m,+}(r)
\eeq
where in complete analogy with the 1D case we define 
\beq
R^{E,\ell m,\pm}(r) = \frac{k_E}{\sqrt{v_E}} \ h_{\ell}^{\pm} (k_E r)
\eeq
The asymptotic behavior of the spherical Hankel functions 
is $(\mp i)^{\ell} \eexp{\pm i k_Er}/(k_Er)$. 
From this follows that the flux of the above radial functions 
is indeed normalized to unity as required.
Also the sign convention that we use for $R^{E,\ell m,\pm}(r)$ 
is appropriate because the free waves are indeed given by 
\beq
|\phi^{E,\ell m}\rangle 
\,\,=\,\,
[R^{lm-}(r)-R^{lm+}(r)] \, Y^{\ell m}(\theta,\varphi) 
\,\,=\,\,
-i\frac{k_E}{\sqrt{v_E}} \, 2j_{\ell}(k_Er) \, Y^{\ell m}(\theta,\varphi) 
\eeq
This spherical free wave solution is analogous to the planar 
free wave ${|\phi^{E,\Omega}\rangle \mapsto \eexp{ik_E\vec{n}_\Omega\cdot\vec{x}}}$.
If we decide (without loss of generality) 
that the planar wave is propagating in the $z$ direction,  
then we can use the following expansion in order 
to express a planar wave as a superposition of spherical waves:  
\beq
\eexp{ik_E z} \,\,=\,\, 
\sum_{\ell}(2\ell+1) \, (i)^{\ell} \, P_{\ell}( \cos(\theta) ) \, j_{\ell}(k_Er)
\eeq
We note that we have only $m=0$ basis functions because there 
is no dependence on the angle $\varphi$.  In different phrasing  
one may say that a plane wave that propagates in the $z$ direction 
has $L_z=0$ angular momentum. Using the identity
\beq
Y^{\ell 0} = \sqrt{\frac{2\ell+1}{4\pi}} P_{\ell}(\cos(\theta))
\eeq
we can write 
\beq
\eexp{ik_E z} \,\,=\,\, 
\sum_{\ell,m=0}\sqrt{(2l+1)\pi} 
\, (i)^{\ell{+}1}  \frac{\sqrt{v_E}}{k_E}   
\, \phi^{E,\ell,m}(r,\theta, \varphi)
\eeq
which makes it easy to identify the ingoing 
and the outgoing components: 
\beq
&& (\eexp{ikz})_{\tbox{ingoing}} 
= \sum_{\ell m} A_{\ell m} Y^{\ell m}(\theta,\varphi) R^{\ell m-}(r) 
\\ \nonumber
&& (\eexp{ikz})_{\tbox{outgoing}} 
= -\sum_{\ell m} B_{\ell m} Y^{\ell m}(\theta,\varphi) R^{\ell m+}(r)
\eeq
where
\beq
B_{\ell m} \,\,=\,\, A_{\ell m} \,\,=\,\,  
\delta_{m,0} 
\,\, \sqrt{(2l+1)\pi} 
\,\, (i)^{\ell{+}1} 
\,\, \frac{\sqrt{v_E}}{k_E}
\eeq
This means that the incident flux in channel $(\ell,0)$ is simply
\beq
i_{\tbox{incident}} = \left[ \frac{\pi}{k_E^2} (2\ell+1) \right] v_E
\eeq
The expression in the square brackets has 
units of area, and has the meaning of cross section.
The actual cross section (see next section) contains  
an additional factor that express how much 
of the incident wave is being scattered. 
The maximum fraction that can be scattered is $400\%$ 
as explained in the previous section.

\sheadC{The scattered wave, phase shifts, cross section}

In the past we were looking for a solution which consists 
of incident plane wave plus scattered component. Namely, 
\beq
\Psi(r) = \eexp{ik_0z} + f(\Omega) \frac{\eexp{ik_E r}}{r}
\eeq
From the decomposition of the incident plane 
wave it is implied that the requested solution is 
\beq
&& \Psi_{\tbox{ingoing}} 
= \sum_{\ell m} A_{\ell m} Y^{\ell m}(\theta,\varphi) R^{\ell m-}(r) 
\\ \nonumber
&& \Psi_{\tbox{outgoing}} 
= -\sum_{\ell m} B_{\ell m} Y^{\ell m}(\theta,\varphi) R^{\ell m+}(r)
\eeq
where
\beq
A_{\ell m} \,\, &=& \,\,  
\delta_{m,0} 
\,\, \sqrt{(2l+1)\pi} 
\,\, (i)^{\ell{+}1} 
\,\, \frac{\sqrt{v_E}}{k_E}
\\ \nonumber
B_{\ell m} \,\, &=& \,\,  
S_{\ell m,\ell' m'} \ A_{\ell' m'} 
\eeq
If we are interested in the scattered wave then 
we have to subtract from the outgoing wave 
the incident component. This means that 
in the expression above we should 
replace $S_{\ell m,\ell' m'}$ by $-iT_{\ell m,\ell' m'}$.

Of major interest is the case where 
the target has spherical symmetry. 
In such case the $S$ matrix is diagonal: 
\beq
S_{\ell m,\ell' m'}(E) &=& 
\delta_{\ell\ell'}\delta_{mm'} 
\, \eexp{2i\delta_{\ell}}   
\\ \nonumber
T_{\ell m,\ell' m'}(E) &=&
- \delta_{\ell\ell'}\delta_{mm'}
\, \eexp{i\delta_{\ell}} \, 2\sin(\delta_{\ell})    
\eeq
Consequently we get 
\beq
\Psi_{\tbox{scattered}} 
\,\,=\,\, 
-\sum_{\ell m} \, T_{\ell\ell} \, 
\sqrt{(2\ell+1)\pi} \, (i)^{\ell} \, 
Y^{\ell 0}(\theta,\varphi) h_{\ell}^{+}(r)
\,\,\sim\,\, 
f(\Omega) \frac{\eexp{ikr}}{r}
\eeq
with 
\beq
f(\Omega)  = -\frac{1}{k_E} \sum_{\ell} \sqrt{(2\ell+1)\pi} 
\, T_{\ell\ell} \, Y^{\ell 0}(\theta,\varphi) 
\eeq
It follows that 
\beq
\sigma_{\tbox{total}} 
\,\,=\,\, 
\int |f(\Omega)|^2 d\Omega 
\,\,=\,\, 
\frac{\pi}{k_E^2} \sum_{\ell} (2\ell+1) |T_{\ell\ell}|^2   
\,\,=\,\,
\frac{4\pi}{k_E^2} \sum_{\ell} (2\ell+1) |\sin(\delta_{\ell})|^2  
\eeq
By inspection of the scattered wave expression 
we see that the scattered flux 
in each of the ${(\ell, m{=}0)}$ channels 
can be written as ${i_{\ell} = \sigma_{\ell}v_E}$,  
where the partial cross section is  
\beq
\sigma_{\ell} 
\ \ = \ \ 
|T_{\ell\ell}|^2 \times \left[ (2\ell+1) \frac{\pi}{k_E^2} \right] 
\ \ = \ \ 
(2\ell+1) |\sin(\delta_{\ell})|^2 \ \frac{4\pi}{k_E^2} 
\eeq
It is important to realize 
that the scattered flux can be 
as large as $4$~times the corresponding incident flux.
The maximum is attained if the scattering 
induces a $\pi/2$ phase shift which inverts 
the sign of the incident wave. In such case  
the scattered wave amplitude should be twice 
the incident wave with an opposite sign.

\sheadC{Finding the phase shift from $T$ matrix theory}

We can get expressions for ${T_{\ell\ell}=V_{\ell\ell}+(VGV)_{\ell\ell}}$ 
using the Born expansion, and hence to get approximations 
for the phase shift $\delta_{\ell}$ 
and for the partial cross section $\sigma_{\ell}$. 
The derivation is the same as in the quasi 1D $\ell{=}0$ case. 
The flux-normalized free waves are 
\beq 
|\phi^{E\ell m}\rangle 
= -i  \frac{k_E}{\sqrt{v_E}} 2j_{\ell}(k_Er) 
Y^{\ell m}(\theta,\varphi)
\eeq
The first order result for the phase shift is  
\beq
\delta^{\tbox{Born}}_{\ell} \approx
-\frac{2}{\hbar v_E}\int^{\infty}_{0}V(r)\left(k_Erj_l(k_Er)\right)^2dr 
\eeq
while in the vicinity of resonances we have  
\beq
\delta_{\ell} = \delta_{\ell}^{\infty} 
- \arctan\left( \frac{\Gamma_r/2}{E-E_r}  \right)
\eeq
where~$\delta_{\ell}^{\infty}$ is 
a slowly varying ``background" phase
that represent the non-resonant contribution.

There are two physical quantities which are 
of special interest and can be deduced 
from the phase shift. One is the time 
delay that we already have discussed in the quasi 
one dimensional context. The other is the cross section: 
\beq 
\sigma_{\ell}(E) = 
(2\ell+1) \frac{4\pi}{k_E^2} |\sin(\delta_{\ell})|^2  
\eeq
For resonant scattering the "line shape" of
the cross section versus energy is 
typically Breit-Wigner and more generally of Fano type.
The former is obtained if we neglect $\delta_{\ell}^{\infty}$, 
leading to 
\beq
\sigma_{\ell} = 
(2\ell+1) \frac{4\pi}{k_E^2}
\frac{(\Gamma_r/2)^2}{(E-E_r)^2+(\Gamma_r/2)^2}  
\eeq
We see that due to a resonance 
the partial cross section $\sigma_{\ell}$  
can attain its maximum value 
(the so called "unitary limit").
If we take $\delta_{\ell}^{\infty}$ into 
account we get a more general result 
which is known as Fano line shape:
\beq
\sigma_{\ell}= (2\ell+1) \frac{4\pi}{k_E^2}
[\sin(\delta_{\ell}^{\infty})]^2
\frac{[\varepsilon+q]^2}{\varepsilon^2+1}  
\eeq
where $\varepsilon=(E-E_r)/(\Gamma/2)$ is the scaled energy, 
and $q=-\cot(\delta_{\ell}^{\infty})$ is the so called 
Fano asymmetry parameter.  
The Breit-Wigner peak is obtained in the limit $q\rightarrow\infty$,  
while a Breit-Wigner dip (also known as anti-resonance) is obtained for $q=0$.

\sheadC{Scattering by a soft sphere}

Assuming that we have a soft sphere ($|V|$ is small), 
we would  like to evaluate the phase shift using the Born approximation. 
Evidently the result makes sense only if the 
phase shift comes out small ($\delta_{\ell} \ll 1$). 
There are two limiting cases. 
If $ka \ll \ell{+}1$  we can use the short 
range approximation of the spherical Bessel 
function to obtain 
\beq
\delta^{\tbox{Born}}_{\ell}\approx
-\frac{2}{(2\ell+3)\left[(2l+1)!!\right]^2}
\frac{\mass V}{\hbar^2}a^2(ka)^{2\ell+1} 
\eeq
On the other hand, if $ka \gg \ell{+}1$ we can use 
the "far field" asymptotic approximation of 
the spherical Bessel function which implies 
$\left[ kr j_{\ell}(kr) \right ]^2 \approx [\sin(kr)]^2 \approx 1/2$. 
Then we obtain
\beq
\delta^{\tbox{Born}}_{\ell} \,\,\approx\,\,
-\frac{1}{\hbar v_E} V a \,\,=\,\,
-\frac{\mass V}{\hbar^2} a^2 (ka)^{-1}  
\eeq

\sheadC{Finding the phase shift by matching} 

We would like to generalize the quasi one-dimensional 
matching scheme to any $\ell$.  
First we have to find the radial function 
for $0<r<a$ and define the logarithmic derivative  
\beq
k_{\ell} = \left(\frac{1}{R(r)}\frac{dR(r)}{dr}\right)_{r=a}
\eeq
Note that we use here $R(r)$ and not $u(r)$ 
and therefore in the $\ell=0$ case we
get ${k_0=\tilde{k}_0-(1/a)}$. The solution in 
the outside region is
\beq
R(r) &=& A h_{\ell}^{-}(k_E r) - B h_{\ell}^{+}(k_E r) 
\\ \nonumber
&=&  A( h_{\ell}^{-}(k_E r) - \eexp{i2\delta_{\ell}} h_{\ell}^{+}(k_E r) ) 
\\ \nonumber
&=& C (\cos(\delta_{\ell})j_{\ell}(k_Er) + \sin(\delta_{\ell})n_{\ell}(k_Er)) 
\eeq
We do not care about the normalization because 
the matching equation involves only logarithmic derivatives:
\beq
k_E\frac{\cos(\delta)j'+\sin(\delta)n'}{\cos(\delta)j+\sin(\delta)n}=k_{\ell}
\eeq
solving this equation for $\tan(\delta_{\ell})$ we get
\beq
\tan(\delta_{\ell})
=-\frac{k_{\ell}j_{\ell}(ka)-k_E j_{\ell}'(ka)}{k_{\ell} n_{\ell} (ka) -k_E n_{\ell}'(ka)}
\eeq
which can also be written as:
\beq
\eexp{i2\delta_{\ell}} = 
\left( \frac{h_{\ell}^{-}}{h_{\ell}^{+}} \right) 
\frac
{ k_{\ell} - ( h_{\ell}'^{-} / h_{\ell}^{-} ) k_E}
{ k_{\ell} - ( h_{\ell}'^{+} / h_{\ell}^{+} ) k_E}
\eeq
In order to go from the first expression to the second 
note that if ${\tan(\delta)=a/b}$ determines 
the phase of a complex number ${z=a+ib}$   
then ${(a+ib)/(a-ib)=\eexp{i2\delta}}$. 
Thus we have expressions for the phase shift given 
the log derivative of the regular solution on the 
boundary of the scattering region.

\sheadC{Scattering by a hard sphere}
 
The phase shifts for a hard sphere ($V\rightarrow\infty$)  
can be found from the phase shift formula of the previous 
section using $k_{\ell}\rightarrow\infty$, leading to 
\beq 
\tan(\delta_{\ell}^\infty)= 
-\frac{j_{\ell}(k_Ea)}{n_{\ell}(k_Ea)} 
\eeq
or equivalently 
\beq 
\eexp{i2\delta_{\ell}^\infty}= \frac{h^-(k_Ea)}{h^+(k_Ea)} 
\eeq
From the first version it is convenient 
to derive the result  
\beq
\delta_{\ell}^{\infty}  \approx 
-\frac{1}{(2\ell+1)!!(2\ell-1)!!} 
(k_Ea)^{2\ell+1}
\ \ \ \ \ \ \ \ \mbox{for $\ell \gg ka$}
\eeq
where we have used the short range 
expansions $j_{\ell}\propto r^{\ell}$ 
and $n_{\ell}\propto 1/r^{\ell+1}$. 
From the second version it is convenient
to derive the result  
\beq 
\delta_{\ell}^{\infty} 
=-\arg(h^+(k_Ea))  
\approx
-(k_Ea-\frac{\pi}{2}{\ell}) 
\ \ \ \ \ \ \ \ \mbox{for $\ell \ll ka$} 
\eeq
where we have used the asymptotic 
expression $h(r) \sim (-\imath)^\ell \ \eexp{ik_Er}/{r}$.

{\bf Small hard sphere:} In the case of a small sphere ($k_E a \ll 1$) 
we have  $1 \gg \delta_0 \gg \delta_1 \gg \delta_2 \dots $ 
and the $\ell=0$ cross section is dominant
\beq 
\delta_0=-(k_Ea) 
\eeq
Hence 
\beq 
\sigma_{\mbox{total}}
=\frac{4\pi}{k^2}\sum_{l=0}^{\infty}{(2l+1)\sin^2{(\delta_\ell)}} 
\approx 
\frac{4\pi}{k^2} 
\sin^{2}{(\delta_0)}
\approx 4 \pi a^2 
\eeq
We got a $\sigma$ that is 4 times bigger than the classical one. 
The scattering is isotropic because only the $\ell=0$ component 
contributes to the scattered wave.

{\bf Large hard sphere:} Now we turn to the case of a large sphere ($k_Ea\gg1$). 
We neglect all $\delta_{\ell}$ for which $\ell > k_Ea$. 
For the non-vanishing phase shifts we get
\beq 
\delta_{\ell} =& -(k_Ea)  & \ \ \ \mbox{for $\ell=0,2,4, \dots $}    
\\ \nonumber
\delta_{\ell} =& -(k_Ea)+\frac{\pi}{2} & \ \ \ \mbox{for $\ell=1,3,5, \dots $}     
\eeq
hence 
\beq 
\sigma_{total} &=&
\frac{4\pi}{k^2}
\left(
\sum_{\ell=0,2 \dots ka}(2l+1)\sin^2(k_Ea)
+\sum_{\ell=1,3 \dots ka}(2l+1)\cos^2(k_Ea)
\right)
\\ \nonumber
& \approx & 
\frac{4\pi}{k^2}
\sum_{\ell=0,1 \dots ka}^{\ell=ka}(2l+1)\frac{1}{2}
\approx 
\frac{2\pi}{k^2}\int_0^{ka}2xdx 
=\frac{2\pi}{k^2}(k_Ea)^2=2\pi a^2 
\eeq
This time the result is 2 times the classical result. 
The factor of 2 is due to forward scattering which partially 
cancels the incident wave so as to create a shadow region 
behind the sphere.

\sheadC{Summary of results for scattering on a sphere} 

We have discussed scattering on a "hard sphere", on 
a "deep well", and finally on a "soft sphere". 
Now we would like to put everything together 
and to draw an $(a,V)$ diagram of all the different regimes.

\begin{center}
\putgraph[0.4\hsize]{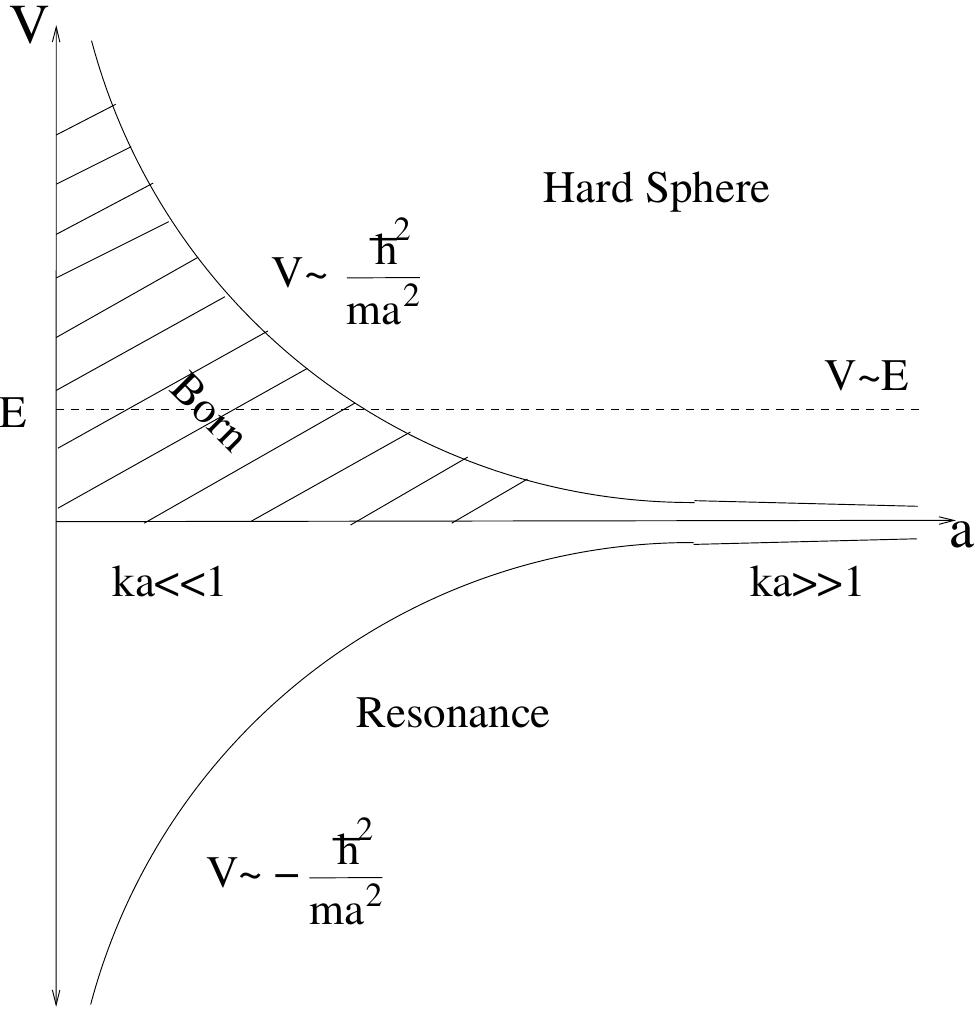} 
\end{center}

Let us discuss the various regimes in this diagram. 
It is natural to begin with a small sphere. 
A small sphere means that the radius is small 
compared with the De-Broglie wavelength ($ka < 1$).  
This means that disregarding resonances the cross 
section is dominated by s-scattering ($\ell=0$). 
On the one hand we have the Born approximation 
and on the other hand the hard sphere result:   
\beq
\delta^{\tbox{Born}}_{0} 
&\approx&
-\frac{2}{3}
\left[\frac{\mass V}{\hbar^2}a^2\right] (ka)
\\ \nonumber
\delta^{\tbox{Hard}}_{0} 
&\approx& 
-(ka)
\eeq
Thus we see that the crossover from 
"soft" to "hard" happens at $V \sim \hbar^2/(\mass a^2)$. 
What about $V<0$? It is clear that the Born approximation
cannot be valid once we encounter 
a resonance (at a resonance the phase shift 
becomes large). So we are safe 
as long as the well is not deep enough 
to contain a quasi bound state. 
This leads to the sufficient condition 
$-\hbar^2/(\mass a^2) < V < 0$. 
For more negative $V$ values we have 
resonances on top of a "hard sphere" behavior.   
Thus we conclude that for $ka<1$ soft sphere means
\beq
|V| \ \ < \ \ {\hbar^2}/{(\mass a^2)}
\eeq
We now consider the case of a large sphere ($ka \gg 1$). 
In the absence of resonances we have 
for small impact parameter  ($\ell \ll ka$) 
either the Born or the hard sphere approximations:
\beq
\delta^{\tbox{Born}}_{\ell} 
& \,\,\approx\,\, &
-\frac{V}{\hbar v_E} a 
\\ \nonumber
\delta^{\tbox{Hard}}_{\ell}  
& \,\, =\,\, & 
\mathcal{O}(1)
\eeq
Also here we have to distinguish 
between two regimes. Namely,  
for $ka \gg 1$ soft sphere means
\beq
|V| \ \ < \ \ {\hbar v_E}/{a}
\eeq
If this condition breaks down we expect to 
have a crossover to the Hard sphere result. 
However, one should be aware that if $V<E$, 
then one cannot trust the hard sphere approximation 
for the low~$\ell$ scattering.

\sheadA{QM in Practice (part I)}

\sheadB{Overview of prototype model systems}

Below we consider some simple one-particle closed systems in 0D/1D/2D/3D. 
Later we further illuminate and analyze some prototype problems. 

{\bf Discrete site systems:} Consider an $N$ site systems. Each site is regarded as a zero dimensional (0D) object, 
sometimes called "dot". In the condense matter context these are known as "tight binding models". 
The simplest is the two site (${N=2}$) system. 
If we want to have non-trivial topology we must consider at least a 3-site (${N=3}$) system. 
Next we can construct $N$-site chains, 
or rings or networks or so called tight-binding arrays. If we regard these models 
as a coarse grained description of motion in free space, then the hopping 
amplitude per unit time between two nearby sites~$i$ and~$j$ 
is determined by the mass of the particle ($\mass$)
and by the hopping distance ($a$):
\beq
K_{ij}  \ \ =  \ \ -\frac{1}{2\mass a^2}
\eeq

{\bf Network systems:} A one dimensional (1D) segment can be regarded 
as the continuum limit of a 1D chain. If the particle in confined between 
two ends, the model is know as "infinite well" or as 1D~box with hard walls.  
If there are periodic boundary conditions, the model is know as "1D ring". 
Several "1D~boxes" can be connected into a network: 
each 1D segment is called "bond", and the junctions are called "vortices".
The simplest is a "delta junction" that connects two wires:  
If we regard the junction as a barrier $u\delta(x)$ the coupling 
between wire wavefunctions is     
\beq
U_{nm} \ \ = \ \ 
-\frac{1}{4\mass^2 u}
\,\,
\Big[\frac{d\psi^{(n)}}{dr}\Big]_{0}
\,\,
\Big[\frac{d\psi^{(m)}}{dr}\Big]_{0}
\eeq
An unconnected endpoint of a 1D segment can be regarded
as "hard wall". If this hard wall can be displaced, it can be regarded 
as a "piston". The perturbation due to a small displacement $dL$ is      
\beq
W_{nm} \ \ = \ \ 
-\frac{dL}{2\mass}
\,\,
\Big[\frac{d\psi^{(n)}}{dr}\Big]_{0}
\,\,
\Big[\frac{d\psi^{(m)}}{dr}\Big]_{0}
\eeq

{\bf 1D systems:} In the more general class of 1D systems 
the confining potential is not stepwise, but changes smoothly. 
The simplest example is "a well with soft walls", 
such as the harmonic oscillator, where the potential is $V(x)=x^2$. 
The prototype example for anharmonic potential, with a dividing separatrix 
in phase space, is the pendulum where ${V(x)=\cos(x)}$. 

{\bf 2D systems:} Genuine 2D systems are chaotic. If the potential floor
is flat, and the walls are "hard", they are called  "billiards". 
These serve as prototype models for "quantum chaos" studies. 
But if the system has (say) rectangular or cylindrical symmetry  
then the problem reduces back to 1D.

{\bf 3D systems:} Genuine 3D systems are rarely considered. 
Typically people consider models with spherical symmetry (e.g. the Hydrogen atom) 
where the problem is reduced to 1D, while the perturbations that do not 
respect the high symmetry are treated using a few-level approximation.
 
{\bf Interacting particles:} The treatment of non-interacting 
system with Bosons or with Fermions reduces to one-particle analysis. 
In order to have new physics we have to take into account interactions. 
The prototype models (Kondo / Hubburd) assume that the interaction 
is of the type $U \propto (\hat{n}-1)\hat{n}$ 
where $\hat{n}=a_1^{\dag}a_1$  is the simplest non-trivial possibility 
in the case of Bosons, while ${\hat{n}=a_1^{\dag}a_1+a_2^{\dag}a_2}$
is the simplest non-trivial possibility in the case of Fermions.
 
{\bf System-bath models:} If we have few distinguished degrees of freedom 
that we call "system", and all the other degrees of freedoms form a 
weakly coupled background that we call "environment", then there 
is a sense in using a system-bath model in order to describe 
the (reduced) dynamics of the system. Typically the environment 
is characterized by its fluctuations, and then modeled as a large collection 
of harmonic oscillators.

\newpage
\sheadB{Discrete site systems}

{\bf Two site model.--} 
The simplest non-trivial network that we can imagine 
is of course a two site system. Without loss of generality   
we can write the Hamiltonian with a real hopping amplitude $c$, namely,      
\beq
\mathcal{H} 
= 
\left( 
\amatrix{
\epsilon/2 & c \cr 
c & -\epsilon/2 }
\right) 
\ \ = \ \ \frac{\epsilon}{2}\sigma_z + c\sigma_x
\ \ = \ \ \vec{\Omega} \cdot \vec{S} 
\eeq
where ${\vec{\Omega}=(2c,0,\epsilon)}$. 
We can define the occupation operator of (say) the first site as 
\beq
\hat{\mathcal{N}}_1 \ \ \equiv \ \  |1\rangle \langle 1|  \ \ \equiv \ \ \frac{1}{2}(1+\sigma_z)
\eeq
The definition of the current that comes out from this site is implied 
by the "rate of change formula", namely  
\beq
\hat{\mathcal{I}}_{1\rightarrow 2} \ \ \equiv \ \  -i[\mathcal{H},\mathcal{N}_1]
\ \ = \ \  -c \sigma_y
\eeq
So we have the continuity equation 
\beq
\frac{d}{dt} \langle \hat{\mathcal{N}}_1  \rangle \ \ = \ \ -  \langle \hat{\mathcal{I}}_{1\rightarrow 2} \rangle,
\ \ \ \ \ \ \ \ \ \ \ \ \ \ \ \ \ \ \ \ 
\langle \hat{\mathcal{N}}_{1} \rangle = |\psi_1|^2,
\ \ \ \ \ \ \ \ \ \ \ \ \ \ \ \ \ \ \ \ 
\langle \hat{\mathcal{I}}_{1\rightarrow 2} \rangle = 2c\,\im[\psi_2^*\psi_1] 
\eeq

{\bf General site model.--} 
The generalization from a two site system to an $N$ site 
system is straightforward. Assuming that the one-particle 
Hamiltonian in the standard (position) basis is represented 
by a matrix $H_{ij}$, we define occupation operators 
and current operators as follow: 
\beq
\hat{\mathcal{N}}_i \ \ &\equiv& \ \  |i\rangle \langle i| 
\\
\hat{\mathcal{I}}_{i\rightarrow j} \ \ &\equiv& \ \ -i \Big[ |j\rangle H_{ji} \langle i| - |i\rangle H_{ij} \langle j| \Big]
\eeq
so as to have the continuity equation 
\beq
\frac{d}{dt} \langle \hat{\mathcal{N}}_i  \rangle 
\ \ = \ \ - \sum_j \langle \hat{\mathcal{I}}_{i\rightarrow j} \rangle
\eeq

{\bf Chain model.-- }
Of particular important is the the standard tight binding model 
where we have a chain of $N$ sites with near neighbor hopping. 
For simplicity we assume periodic boundary conditions, 
which is like saying that we have a closed ring. 
We also use length units such that the sites are 
positioned at ${x=1,2,3...}$. The Hamiltonian, 
with real hopping amplitudes $-c$, can be written as 
\beq
\mathcal{H} \ \ = \ \  -c(\hat{D}+\hat{D}^{-1}) + V(\hat{x}) 
\ \ = \ \ -2c\cos(\hat{p}) + V(\hat{x})
\eeq
where $D$ is the one site displacement operator.
The definition of the velocity operator is implied 
by the "rate of change formula", namely  
\beq
\hat{v} \ \ = \ \ i[\mathcal{H},\hat{x}] 
\ \ = \ \ ic (\hat{D}-\hat{D}^{-1}) \ \ = \ \  2c\sin(p) 
\eeq
Note that for small velocities we have linear 
dispersion relation ${v=(1/\mass)p}$, where the mass 
is ${\mass=1/(2c)}$. If distance between the sites 
were $a$ rather then unity one would conclude 
the identification ${c=1/(2\mass a^2)}$.

{\bf Lattices.-- }
It is interesting to consider a generalizations of the simple chain model, 
where the motion is on a lattice. For example: the case of 1D chain  
with alternating lattice constant ${(a,b,a,b,....)}$. In the latter case 
we get two energy bands rather than one. 
For further information consult textbooks on Solid State Physics.

\newpage
\sheadB{Two level dynamics} 

\sheadC{Bloch sphere picture of the dynamics} 

The most general Hamiltonian for a particle with spin $\frac{1}{2}$ is:
\beq
\mathcal{H} \ \ = \ \ \vec{\Omega}\cdot \vec{S}= \Omega_x S_x + \Omega_y S_y + \Omega_z S_z 
\eeq
where ${\vec{S} = \frac{1}{2} \vec{\sigma}}$.
The evolution operator is:
\beq
U(t) \ \ = \ \ \eexp{-it\hat{\mathcal{H}}} \ \ = \ \ \eexp{-i( \vec{\Omega}t)\cdot\vec{S}} \ \ = \ \ R(\vec{\Phi }(t)) 
\eeq
where $\vec{\Phi}(t) = \vec{\Omega}t$. 
This means that the spin makes precession.
It is best to represent the state of the spin 
using the polarization vector $\vec{M}$. 
Then we can describe the precession  
using a classical-like, so called "Bloch sphere picture". 
The formal derivation of this claim is based on the 
relation between $M(t)$ and $\rho(t)$. 
We can write it either as $M(t)=\trc(\sigma\rho(t))$
or as an inverse relation ${\rho(t)=(1+\vec{M}(t)\cdot\sigma)/2}$. 
In the former case the derivation goes as follows:
\beq
M_i(t)&=&\trc(\sigma_i\rho(t)) 
= \trc(\sigma_i(t)\rho) 
\\ \nonumber
&=& \trc((R^{-1} \sigma_i R) \rho) 
= \trc((R^E_{ij} \sigma_j) \rho) 
=  R^E_{ij}(\Phi(t)) \ M_j(0)
\eeq
where we have used the evolution law ${\rho(t)=U(t)\rho(0)U(t)^{-1}}$
and the fact that $\vec{\sigma}$ is a vector operator.

We notice that the evolution of any system 
whose states form a dim$=$2 Hilbert space 
can always be described using a precession picture.
The purpose of the subsequent exposition is    
(i) to show the power of the precession picture as opposed to diagonalization; 
(ii) To explain the notion of small versus large perturbations. 
With appropriate gauge of the standard basis 
the Hamiltonian can be written as: 
\beq
\mathcal{H} 
\ \ = \ \ 
\left( 
\amatrix{
\epsilon/2 & c \cr 
c & -\epsilon/2 }
\right) 
\ \ = \ \ \frac{\epsilon}{2}\sigma_z + c\sigma_x
\ \ = \ \ \vec{\Omega}\cdot\vec{S}, 
\hspace{2cm}  \vec{\Omega}=(2c,0,\epsilon)
\eeq
In the case of a symmetric system ($\epsilon=0$) 
we can find the eigenstates and then find 
the evolution by expanding the initial state 
at that basis. The frequency of the oscillations 
equals to the energy splitting of the eigen-energies.
But once ($\epsilon \ne 0$) this scheme becomes 
very lengthy and intimidating. It turns out that 
it is much much easier to use the analogy with spin~1/2. 
Then it is clear, just by looking at the 
Hamiltonian that the oscillation frequency is
\beq
\Omega \ \ = \ \ \sqrt{(2c)^2+\epsilon^2} 
\eeq
and hence the eigenenergies are ${E_{\pm}=\pm\Omega/2}$. 
Furthermore, it is clear that the precession axis 
is tilted relative to the $z$~axis with an angle 
\beq
\theta \ \ = \ \ \arctan(2c/\epsilon)
\eeq
Accordingly the eigenstates can be obtained 
from the $\uparrow$ and from the $\downarrow$ states 
by rotating them an angle $\theta$ around the $y$~axis:
\beq
|+\rangle = \left( \begin{array}{cc}  \cos(\theta/2) \\ \sin(\theta/2)  \end{array} \right),  
\ \ \ \ \ \ \ \ \ \ \ \ \ \ 
|-\rangle =  \left( \begin{array}{cc} -\sin(\theta/2) \\ \cos(\theta/2)  \end{array} \right)
\eeq
If we are interested in studying the dynamics there is no need 
to expand the initial state at this basis. It is much easier 
to use write the explicit expression for the evolution operator:
\beq
U(t) \ \ = \ \ R(\vec{\Phi }(t)) \ \ = \ \ \cos(\Omega t/2) - i\sin(\Omega t/2)\Big[ \cos\theta \ \bm{\sigma}_z + \sin\theta \ \bm{\sigma}_x \Big]
\eeq
Let us assume that initially the system is in the "up" state.
We get that the state after time~$t$ is 
\beq
|\psi(t)\rangle \ \ = \ \ U(t)|\uparrow\rangle 
\ \ = \ \ \Big[\cos(\Omega t/2)-i\cos\theta \sin(\Omega t/2)\Big]|\uparrow\rangle -i \Big[\sin\theta \sin(\Omega t/2) \Big]|\downarrow\rangle
\eeq
Let us define ${P(t)}$ as the probability to 
be found after time~$t$ in the same state. 
Initially we have ${P(0)=1}$.
We can easily find the explicit expression for $P(t)$ 
without having to do any diagonalization of the 
Hamiltonian. Rather we exploit the explicit expression above and get
\beq
P(t) \ \ = \ \ 
\Big|\langle\uparrow| \psi(t)\rangle\Big|^2
\ \ = \ \ 
1 - \sin^2(\theta) \sin^2\left(\frac{\Omega t}{2}\right)
\eeq
This result is called the Rabi Formula. 
We see that in order to have nearly complete 
transitions to the other site after half a period, 
we need a very strong coupling ($c \gg \epsilon$).
In the opposite limit  ($c \ll \epsilon$)  
the particle tends to stay in the same site, 
indicating that the eigenstates are barely affected.

We note that the Rabi Formula can be optionally derived  
by considering the precession of the polarization vector $M(t)$.  
This procedure is on the one hand more illuminating, 
but on the other requires better geometrical insight:  
One should realize that the inclination angle of~$M(t)$ 
oscillates between the values $0$ and $2\theta$.  
Therefore $M_z(t)$ oscillates between 
the maximal value $M_z=1$ and the minimal value $M_z=\cos(2\theta)$. 
Using the relation  $P(t)=(1+M_z(t))/2$
one concludes that $P(t)$ oscillates 
with the frequency $\Omega$ between the maximal value~$1$ 
and the minimal value~$(\cos(\theta))^2$ 
leading to the Rabi Formula.

\sheadC{Landau-Zener dynamics} 
\label{sZener}

A prototype adiabatic process is the 
Landau-Zener crossing of two levels. 
The Hamiltonian is 
\beq
\mathcal{H}
\ \ = \ \ 
\frac{1}{2}\left(\amatrix{\alpha t & \kappa \cr \kappa & -\alpha t}\right)
\ \ = \ \ 
\frac{1}{2}\alpha t \sigma_3 + \frac{1}{2}\kappa \sigma_1 
\eeq
Let us assume that the system is prepared at ${t=-\infty}$
in the lower ("-") level (which is the "up" state). 
We want to find what is the probability to find 
the system at ${t=\infty}$ in the upper ("+") level 
(which is again the "up" state!). The exact result is 
know as the Landau-Zener formula:
\beq 
P_{\tbox{LZ}} \ \ = \ \ P_{\infty}(+|-)  \ \ = \ \ P_{\infty}(\uparrow|\uparrow)
\ \ = \ \ \exp\left[-\frac{\pi}{2}\frac{\kappa^2}{\alpha} \right]
\eeq
The straightforward way to derive this formula is to 
sum the Texp terms up to infinite order:
\beq \nonumber
\langle \uparrow | U(t,t_0) | \uparrow \rangle = \!\! &&  
\eexp{-i\frac{\alpha}{4} (t^2-t_0^2)} 
\ \ + \ \ \iint_{t_0<t_1<t_2<t} dt_2dt_1 \ \left(-i\frac{\kappa}{2}\right)^2 \eexp{-i\frac{\alpha}{4} \left((t^2-t_2^2)-(t_2^2-t_1^2)+(t_1^2-t_0^2)\right)}
\\ && +\iint\iint dt_4 dt_3 dt_2 dt_1 \ 
\left(-i\frac{\kappa}{2}\right)^4 \eexp{-i\frac{\alpha}{4}\left( (t^2-t_4^2)-(t_4^2-t_3^2)+(t_3^2-t_2^2)-(t_2^2-t_1^2)+(t_1^2-t_0^2)\right)}
\ \ + \ \ \cdots
\eeq 
where time ordering of the integration variables is required. 
Each term represent a sequence of spin flips that are induced by the perturbation. 
Obviously only even orders appear in the calculation of $P(\uparrow|\uparrow)$. 
Dropping phase factor, rescaling the time variables, and taking 
the integration limits to infinity we get   
\beq 
\langle \uparrow | U(t,t_0) | \uparrow \rangle \ \ = \ \ 
\sum_{n=0}^{\infty}  \left(-i\frac{\kappa}{2}\frac{1}{\sqrt{\alpha}}\right)^{2n} \iint \cdots \iint dt_{2n} \cdots dt_1 \ 
\exp\left[-\frac{i}{2}\left(t_1^2-t_2^2+t_3^2 \cdots -t_{2n}^2 \right)\right]
\eeq
Following Kayanuma [\href{http://journals.jps.jp/doi/abs/10.1143/JPSJ.53.108}{Appendix}]  
we note that the time-ordered integral can be calculated using the 
following trick: the $n$th term can be regarded as describing a train 
of $n$ pulses; New integration variables $\tau_j=t_{2j}-t_{2j-1}$ and $\tilde{t}_j$ are defined;   
The $\tau_j$ is the length of the pulse and the $\tilde{t}_j$  is its reduced time; 
The latter is the value of $t_j$ if the pulses were of zero length.
Consequently one obtains integrals that are symmetric under time permutation, 
and can be factorized into elementary Gaussian integrals, leading to $\pi^n/n!$. 
Thus we get  
\beq
\langle \uparrow | U(t,t_0) | \uparrow \rangle \ \ = \ \ 
\sum_{n=0}^{\infty}  \frac{1}{n!}\left(-\frac{\pi}{4}\frac{\kappa^2}{\alpha}\right)^{n} 
\ \ = \ \ \exp\left[-\frac{\pi}{4}\frac{\kappa^2}{\alpha} \right]
\eeq
For pedagogical purpose we consider below the standard procedures 
for obtaining {\em approximations} for the Landau-Zener formula.            
For a ``fast" (non-adiabatic) process the problem can be 
treated using conventional (fixed basis) perturbation theory.
The first order result requires the calculation of  ${p=|\langle \downarrow | U(t,t_0) | \uparrow \rangle|^2}$.
Performing the time integration one obtains ${p \approx (\pi/2)[\kappa^2/\alpha]}$, 
in consistency with the leading order term in the expansion of $P_{\tbox{LZ}}$. 
Optionally the same result is obtained from the Schrodinger equation 
for the amplitude $c_{\downarrow}(t)$ in the interaction picture: 
\beq
\frac{dc_{\downarrow}(t)}{dt} \ \ = \ \ 
-i\frac{\kappa}{2} \exp\left[-i\frac{1}{2}\alpha t^2\right]
\ c_{\uparrow}(t)
\eeq
Setting $c_{\uparrow}(t)\mapsto 1$ and performing the time integration,  
one obtains the same first order estimate for~$p$.
In order to analyze the Landau-Zener transition in the adiabatic regime   
we write the Schrodinger equation in the adiabatic basis, 
and write the analogous equation for the amplitude $c_{+}(t)$. 
The instantaneous eigenenergies are $E_{\pm} = \pm (1/2)\Omega$, with  
\beq
\Omega(t) = \sqrt{(\alpha t)^{2}+\kappa^{2}}
\eeq
and the associated eigenstates are $|+\rangle$ and $|-\rangle$,  
as defined in the previous section, with 
\beq
\theta(t) \ \ = \ \ \arctan\left({\kappa}/{(\alpha t)}\right).
\eeq
Note that~$\theta{=}\pi$ at ${t{=}-\infty}$ evolves to~$\theta{=}0$ at ${t{=}\infty}$.
In our choice of gauge the adiabatic eigenstates 
have real amplitudes,  
and therefore the Berry ``vector potential" comes out zero.   
At the same time the non-zero element of the 
perturbation matrix $W_{nm}$ is 
\beq
W_{+-} \ \ = \ \ i\frac{\alpha}{E_{+}-E_{-}} 
\left[\frac{1}{2}\sigma_3\right]_{+-} 
\ \ = \ \ -i\frac{\alpha\sin(\theta)}{2\Omega}
\ \ = \ \ -i\frac{\alpha\kappa}{2\Omega^2} 
\ \ = \ \ -\frac{i}{2}\frac{\alpha}{\kappa}\left[\frac{1}{(\alpha t/\kappa)^{2}+1}\right] 
\eeq
Following the standard procedure as in time-dependent 
perturbation theory we substitute 
\beq
a_{\pm}(t) \ \ = \ \ c_{\pm}(t) \ \exp\left[-i \int^t E_{\pm}dt' \right]    
\eeq
Using the rescaled variable $\tau=\alpha t/\kappa$, 
and changing to ${\tau = \sinh(z)}$,   
the expression for the dynamical phase is  
\beq
\Phi(t) 
\ \ = \ \ 
\frac{\kappa^2}{\alpha} \int^{\tau} d\tau \sqrt{\tau^2+1}
\ \ = \ \ 
\frac{\kappa^2}{\alpha}\times \frac{1}{2}\left(z+\frac{1}{2}\sinh(2z)\right)
\eeq
Thus for the amplitude $c_{+}(t)$ we get the equation 
\beq
\frac{dc_{+}(t)}{dt} \ \ = \ \   
-\frac{1}{2}\frac{\alpha}{\kappa}\left[\frac{1}{(\alpha t/\kappa)^{2}+1}\right]  
\, \eexp{i\Phi(t)} \, c_{-}(t)
\eeq
Starting with the zero order solution ($c_{-}(t)\mapsto -1$),  
and performing the time integration,  
one obtains the first order adiabatic estimate for the 
amplitude $c_{+}(t)$, leading to
\beq
P_{\tbox{LZ}}
\ \ \approx \ \ 
\left|\frac{1}{2}\int_{-\infty}^{\infty}\frac{1}{\tau^2+1}\,\eexp{i\Phi(\tau)} d\tau\right|^2
\ \ = \ \ 
\left|\frac{1}{2}\int_{-\infty}^{\infty} \frac{1}{\cosh(z)} \,\eexp{i\Phi(z)} dz\right|^2
\eeq
In order to evaluate this integral one can use contour integration.  
The explicit expression for the phase $\Phi$ as a function of ${z=x+iy}$ is 
\beq
\Phi(z) \ \ = \ \ 
\frac{\kappa^2}{\alpha} \times \frac{1}{2}
\left[\left(x+\frac{1}{2}\sinh(2 x)\cos(2y)\right)
+i\left(y+\frac{1}{2}\cosh(2 x)\sin(2y)\right)\right]
\eeq
The original contour of integration is $y=0$, 
but we would like to deform it into the complex plane 
so as to get rid of the rapid oscillations of the phase factor, 
and have instead a smooth monotonic variation. 
The details can be found in section~VI of [\href{http://arxiv.org/abs/0708.4237}{arXiv}].
The deformed contour is displayed in Fig4 there.
The phase is pure imaginary along the 
curves $C_{-}$ and $C_{+}$. At ${z_0=0+i(\pi/2)}$ 
we have ${\Phi=i(\pi/4)\kappa^2/\alpha}$, 
while $\cosh{z} \approx i(z-z_0)$. 
Consequently we have a pole leading to the same exponent as 
in the Landau-Zener result. The prefactor come out~$(\pi/3)^2$. 
The derivation of the correct prefactor, which is unity,  
requires either exact solution of the differential equation 
using parabolic cylinder functions, or otherwise iterations 
of the above procedure to infinite order.

\sheadC{Adiabatic transfer from level to level} 

A practical problem which is encountered 
in Chemical Physical applications is how to 
manipulate coherently the preparation of a system.
Let us as assume that an atom is prepared 
in level $|E_a\rangle$ and we want to have it  
eventually at level $|E_b\rangle$. 
We have in our disposal a laser source. 
This laser induce AC driving that can couple the two levels. 
The frequency of the laser is $\omega$ and 
the induced coupling is $\Omega$.
The detuning is defined as ${\delta=\omega-(E_b-E_a)}$.
Once the laser is turned ``on" the system 
starts to execute Bloch oscillation.    
The frequency of these oscillation 
is ${\tilde{\Omega}= (\Omega^2+\delta^2)^{1/2}}$.
This is formally like the coherent oscillations      
of a particle in a double well system. 
Accordingly, in order to simplify the following discussion 
we are going to use the terminology of a site-system.    
Using this terminology we say that with a laser 
we control both the energy difference $\delta$ 
and the coupling $\Omega$ between the two sites.

By having exact resonance ($\delta=0$) we can 
create ``complete" Bloch oscillations with 
frequency $\Omega$. This is formally like the 
coherent oscillations of a particle 
in a symmetric double well system. 
In order to have $100\%$ transfer from 
state $|E_a\rangle$ to state $|E_b\rangle$ 
we have to keep $\delta=0$ for a duration 
of exactly half period (${t=\pi/\Omega}$). 
In practice this is impossible to achieve.
So now we have a motivation to find 
a practical method to induce the desired 
transfer.

There are two popular methods  
that allow a robust $100\%$ transfer from 
state $|E_a\rangle$ to state $|E_b\rangle$. 
Both are based on an adiabatic scheme.
The simplest method is to change $\delta$ gradually  
from being negative to being positive. 
This is called ``chirp". Formally this 
process is like making the two levels ``cross" 
each other. This means that a chirp induced transfer 
is just a variation of the Landau-Zener 
transition that we have discussed in the previous section.

There is another so called ``counter intuitive scheme" 
that allows a robust $100\%$ transfer from 
state $|E_a\rangle$ to state $|E_b\rangle$, 
which does not involve a chirp. 
Rather it involves a gradual turn-on 
and then turn-off of two laser sources. 
The first laser source should couple 
the (empty) state $|E_b\rangle$ to a  
third level $|E_c\rangle$. 
The second  laser source should couple 
the (full) state $|E_a\rangle$ to  
the same third level $|E_c\rangle$. 
The second laser is tuned on while 
the first laser it turned off. 
It is argued in the next paragraph that 
this scheme achieves the desired transfer. 
Thus within the framework of this scheme 
it looks as if a transfer 
sequence ${a\mapsto c\mapsto b}$ 
is realized using a counter intuitive 
sequence ${c\mapsto b}$ followed by ${a\mapsto c}$.

The explanation of the ``counter intuitive scheme"  
is in fact very simple. All we have to do is to 
draw the adiabatic energy levels $E_{-}(t)$ 
and $E_0(t)$ and $E_{+}(t)$ as a function of time, 
and then to figure out what is the ``identity" 
of (say) the middle level at each stage. 
Initially only the first laser in ``on" 
and therefore $|E_b\rangle$ and $|E_c\rangle$ 
split into ``even" and ``odd" superpositions. 
This means that initially $E_0(t)$ corresponds  
to the full state $|E_a\rangle$. 
During the very slow switching process 
an adiabatic evolution takes place.  
This means that the system remains in the middle 
level. At the end of the process only the second 
laser is ``on" and therefore, using a similar 
argumentation, we conclude that $E_0(t)$ corresponds  
to the state $|E_b\rangle$. The conclusion is 
that a robust $100\%$ transfer from 
state $|E_a\rangle$ to state $|E_b\rangle$ 
has been achieved.

\newpage
\sheadB{A few site system with Bosons}

\sheadC{A two site system with $N$ Bosons}

We discuss in what follows the Bose-Hubbard Hamiltonian 
for $N$~Bose particles in a two site system.
Defining ${j=N/2}$ we shall see that the Hamiltonian 
can be regarded as describing a spin entity.  
The dimension of the Hilbert space 
is ${\mathcal{N}=N+1=2j+1}$. Later we assume ${N \gg 1}$. 
The Hamiltonian is 
\beq
\mathcal{H} \ = \ \sum_{i=1,2}\left[ \mathcal{E}_i \hat n_i+ \frac{U}{2} \hat n_i(\hat n_i-1)\right]  
- \frac{K}{2}(\hat a_2^{\dag}\hat a_1+ \hat a_1^{\dag}\hat a_2 ) 
\eeq
where $K$ represents the hopping, and $U$ is the interaction, 
and ${\mathcal{E}=\mathcal{E}_2-\mathcal{E}_1}$ is the potential bias.
The total number of particles ${n_1+n_2=N}$ is constant of the motion. 
The Hamiltonian for a given~$N$ is formally 
equivalent to the Hamiltonian of a spin~$j$ entity. Defining   
\beq
J_z \ \ &\equiv& \ \ \frac{1}{2}(\hat{n}_1-\hat{n}_2) \ \ \equiv \ \ \bm{n} \\
J_+ \ \ &\equiv& \ \ \hat{a}_1^{\dag}\hat{a}_2 
\eeq
we can re-write the Hamiltonian as 
\beq
\mathcal{H} \ = \  U\hat J_z^2 - \mathcal{E} \hat J_z - K \hat J_x +\text{const}
\eeq
In the absence of interaction this is 
like having a spin in magnetic field with precession
frequency ${\Omega=(K,0,\mathcal{E})}$. In order to analyze 
the dynamics for finite $U$ it is more convenient to re-write this 
Hamiltonian with canonically conjugate variables. 
Formally we are dealing with two coupled oscillators. 
So for each we can define action-angle variables 
in the following way (no approximations involved):
\beq
\hat a_i \ \ = \ \ 
\left( \begin{array}{cccc} 
0 & 1 & 0 & 0 \\  0 & 0 & 1 & 0 \\
0 & 0 & 0 & 1 \\  0 & 0 & 0 & 0
\end{array}\right)
\left( \begin{array}{cccc} 
0 & 0 & 0 & 0 \\  0 & \sqrt{1} & 0 & 0 \\
0 & 0 & \sqrt{2} & 0 \\  0 & 0 & 0 & \sqrt{3}
\end{array}\right)
\ \ = \ \ 
\left( \begin{array}{cccc} 
0 & \sqrt{1} & 0 & 0 \\  0 & 0 & \sqrt{2} & 0 \\
0 & 0 & 0 & \sqrt{3} \\  0 & 0 & 0 & 0
\end{array}\right)
\ \ \equiv \ \ 
\eexp{i\varphiJ_i}\sqrt{\hat{n}_i}
\eeq
One should be aware that the translation-like 
operator $\eexp{i\varphiJ_i}$ is in fact non-unitary 
because it annihilates the ground state. 
We shall denote its adjoint operator as $\eexp{-i\varphiJ}$, 
so we have ${\hat a_i^{\dag}=\sqrt{\hat n_i}\eexp{-i\varphiJ_i}}$.
Now it is possible to re-write the Hamiltonian 
using the conjugate variables $\hat{n}$
and ${\varphiJ \equiv \varphiJ_1-\varphiJ_2}$. 
Ignoring the issue of commutation the result is 
\beq
\mathcal{H} \ \ = \ \ 
U  \bm{n}^2 \ - \ \mathcal{E} \bm{n} 
\ - \ K
\sqrt{(N/2)^2-\bm{n}^2} \ \cos(\varphiJ) 
\eeq
If we did not ignore the proper ordering of the operators 
the above would be exact. But we are interested in 
treating  ${N\gg1}$ system, where a semiclassical picture 
is applicable, hence small corrections that are associated 
with the ordering are neglected. 
In the ${n \sim N/2}$ region of phase space $\mathcal{H}$ resembles   
the so-called Josephson Hamiltonian, which is essentially    
the Hamiltonian of a pendulum 
\beq
\mathcal{H}_{\tbox{Josephson}} \ \ = \ \ E_C(\bm{n}-n_0)^2 \ - \ E_J\cos(\varphiJ)
\eeq
with $E_c=U$ and $E_J=KN/2$, while $n_0$ is linearly related to $\mathcal{E}$.
The Josephson Hamiltonian is an over-simplification because 
it does not capture correctly the global topology of phase space.
In order to shed better light on the actual Hamiltonian  
it is convenient to define in addition to the $\varphiJ$ 
operator also a $\thetaJ$ operator such that ${J_z \equiv (N/2)\cos(\thetaJ)}$ 
while ${J_x \approx (N/2)\sin(\thetaJ)\cos(\varphiJ)}$. 
It is important to realize that  $\varphiJ$  and $\thetaJ$ do not commute.
Using the ${(\varphiJ,\thetaJ)}$ coordinates, the Hamiltonian takes the form  
\beq
\mathcal{H} \ \ \approx \ \  
\frac{NK}{2}\left[\frac{1}{2} u (\cos\thetaJ)^2 - \varepsilon \cos\thetaJ  - \sin\thetaJ\cos\varphiJ \right]
\eeq
where the scaled parameters are 
\beq
u \equiv \displaystyle{ \frac{NU}{K}},  
\hspace*{3cm}
\varepsilon \equiv \displaystyle{\frac{\mathcal{E}}{K}}
\eeq
We can describe phase space using the canonical 
coordinates ${(\varphiJ,\bm{n})}$, so the 
total area of phase space is~$2\pi N$ 
and Planck cell is ${2\pi\hbar}$ with ${\hbar=1}$. 
Optionally we can use spherical coordinates ${(\varphiJ,\thetaJ)}$, 
so the total area of phase space is~$4\pi$ 
and Planck cell ${4\pi/N}$.
Note that an area element in phase space 
can be written as $d\varphiJ d\cos\thetaJ$.

Within the framework of the semiclassical picture 
a quantum state is described as a distribution 
in phase space, and the eigenstates are associated 
with strips that are stretched along 
contour lines ${H(\varphiJ,\thetaJ)=E}$ of the Hamiltonian. 
The $|n\rangle$ states are the eigenstates of the $K=0$ Hamiltonian. 
In particular the ${|n{=}N\rangle}$ state (all the particles are in the first site) 
is a Gaussian-like wavepacket which is concentrated in the NorthPole. 
By rotation we get from it a family of states ${|\thetaJ,\varphiJ\rangle}$
that are called coherent states. 
If $U$ is large, $\mathcal{E}=0$, and $K$ is very small, 
then the eigenstates are the symmetric and the antisymmetric 
superpositions of the $|\pm n\rangle$ states.  
In particular we have "cat states" of the type:
\beq
| \mbox{CatState} \rangle \ \ = \ \ |n_1{=}N,n_2{=}0\rangle \ \ + \ \ \eexp{i\mbox{phase}} |n_1{=}0,n_2{=}N\rangle 
\eeq
If a coherent state evolves, then the non-linearity (for non zero $U$) 
will stretch it. Sometimes the evolved state might resemble a cat state.
We can characterize the one-body coherence of a state 
by defining the one-body reduced probability matrix as 
\beq
\rho^{(1)}_{ji} 
\ \ = \ \  \frac{1}{N}\langle \hat{a}_i^{\dagger} \hat{a}_j \rangle   
\ \ = \ \  \frac{1}{2}(\bm{1} +  S \cdot \bm{\sigma}) 
\eeq
where  ${S_i=(2/N)\langle J_i \rangle}$, and the so called polarization vector 
is ${S = ( S_x,  S_y, S_z)}$. Note that
\beq
\mbox{AverageOccupation} &=& (N/2)\left[ 1 + S_z \right]
\\
\mbox{OneBodyPurity} &=&  (1/2) \left[ 1 + S_x^2 + S_y^2 + S_z^2 \right]
\eeq
The coherent states have maximum OneBodyPurity. 
This purity can be diminished due to 
non-linear effects or due to interaction 
with environmental degrees of freedom.  
{\bf First example:} 
If we prepare the particles in the ground state 
of the ${U=0}$ Hamiltonian of a symmetric double well, 
then we have a coherent state "polarized" in the $x$ direction.
If we turn on $U$, then this state will be stretched 
and eventually it will spread over the whole equator, 
thus leading to loss of the OneBodyPurity.
{\bf Second example:} 
The interaction with the environment leads 
to the loss of OneBodyPurity due to the loss of the ManyBodyPurity. 
If we have an environmental collisional effect that 
"measures" in which well are the particles ($J_z$ interaction), 
then it leads to dephasing towards the $z$ axis, 
while if the collisional effect is like measuring whether 
the particles are in the barrier region  ($J_z$ interaction) 
then the dephasing is towards the $x$ axis.

\sheadC{An $M$ site system with $N$ Bosons}

A double well system with Bosons has formally the same Hilbert space 
as that of spin~$N/2$ particle. 
The generalization of this statement is straightforward.   
The following model systems have formally the same Hilbert space: \\
 
\bitem The ``$N$ particle" states of a Bosonic system with $M$ sites. \\
\bitem The ``$N$ quanta" states of $M$ Harmonic oscillators. \\
\bitem The states of a $\mbox{dim}(N)$ "spin" of the $SU(M)$ group. \\

We shall explain and formulate mathematically the above statements, 
and later we shall use the formal analogy with "spin" in order to shed 
light on the dynamics of $N$ Bosons in ${M=2}$ site system. 

To regard Bosonic site as an Harmonic "mode"  
is just a matter of language: 
We can regard the "creation" operator $a_i^{\dagger}$
as a "raising" operator, and we can regard the "occupation" 
of the $i$th site $\hat{n}_i= a_i^{\dagger}a_i$ 
as the number of quanta stored in $i$th mode. 
The one particle states of an $M$ site system form 
an $M$ dimensional Hilbert space. The set of unitary 
transformations within this space is the $SU(M)$ group.
Let us call these transformation generalized "rotations". 
If we have $N$ particles in $M$ sites then 
the dimensionality of Hilbert space is ${\mbox{dim}(N)=(N{+}1)!/[(M{-}1)!N{-}M{+}2)!]}$.
For example for $M=2$ system we have ${\mbox{dim}(N)=N{+}1}$ 
basis states ${|n_1,n_2\rangle}$ with ${n_1+n_2=N}$. 
We can "rotate" the whole system using $\mbox{dim}(N)$
matrices. Thus we obtain a $\mbox{dim}(N)$ representation of the $SU(M)$ group. 
By definition these "rotations" can be expressed as a linear combination 
of the $SU(M)$ generators $J_{\mu}$, and they all commute with~$\hat{N}$.
We shall see that the many body Hamiltonian $\mathcal{H}$ may contain ``non linear" 
terms such as ${J_{\mu}^2}$ that correspond to interactions 
between the particles. Accordingly $\mathcal{H}$ of an interacting 
system is not merely a "rotation". 
Still we assume that $\mathcal{H}$ commutes with~$\hat{N}$, 
so we can work within $\mbox{dim}(N)$ subspace.

There is a common {\em mean field approximation} which is used 
in order to analyze the dynamics of many Bosons system.
In the semiclassical framework the dynamics 
in phase space is generated by the Hamilton equations 
of motion for the action-variables $\dot{n}_i$ and $\dot{\varphiJ}_i$.
These are the ``polar coordinates" that describe each of the oscillators.  
Optionally one can define a ``macroscopic wavefunction" 
\beq
\psi_i \ \ \equiv \ \  \sqrt{n_i} \, \eexp{i\varphiJ}
\ \ \ \ \ \ \ \ \ \ \ \ \ \ \ \ \ \ 
\mbox{representing a single point in phase space}
\eeq
The equation for $\dot{\Psi}_i$ is known as the 
discrete version of the non-linear Schrodinger (DNLS) equation:
\beq
i\frac{d\psi_i}{dt} \ \ = \ \ \Big(\mathcal{E}_i + U|\psi_i|^2 \Big) \Psi_i 
- \frac{K}{2}\left(\psi_{i+1}+\psi_{i-1} \right)
\eeq
In the continuum limit this equations is known 
as the Gross-Pitaevskii equation:
\beq
i\frac{d\psi(x)}{dt} \ \ = \ \ \Big[V(x) + g_s|\psi(x)|^2  -\frac{1}{2\mass}\nabla^2 \Big] \psi(x) 
\eeq
The potential $V(x)$ corresponds to the $\mathcal{E}_i$ of the DNLS equation,   
and the mass $\mass$ is associated with the hopping amplitude $K$ using the standard prescription.  
In addition to that we have the interaction parameter $g_s$ that 
corresponds to $U$ multiplied by the volume of a site. It also can be related to 
the scattering length using the relation $g_s = 4\pi a_s/\mass$.

Within the framework of the {\em proper} semiclassical 
treatment the quantum state is described as a distribution 
of points in phase space. 
The {\em proper} semiclassical treatment goes beyond 
the conventional mean field approximation. 
The latter assumes that in any time the state of 
the system looks like a coherent state. 
Such a state corresponds to a Gaussian-like 
distribution is phase space (``minimal wavepacket") 
and accordingly it is characterized 
by  ${\overline{n}_i = \langle n_i \rangle}$ 
and ${\overline{\varphiJ}_i = \langle \varphiJ_i \rangle}$ 
or equivalently by the mean field  
\beq
\overline{\psi}_i = \langle \psi_i \rangle
\ \ \ \ \ \ \ \ \ \ \ \ \ \ \ \ \ \ 
\mbox{representing the center of a wavepaket}
\eeq
To the extend that the mean field assumption can be trusted 
the equation of motion for $\overline{\psi}_i$ is the DNLS.
Indeed if $U=0$ there is no non-linear spreading 
and the mean field description becomes exact. 
Otherwise it is a crude approximation.

\newpage
\sheadB{A few site system with Fermions}

\sheadC{A two-site system with one particle}

The problem of "positioning a particle of spin $1/2$ 
in a specific location" is formally identical to the problem 
of "putting a particle in a two-site system". In both cases 
the system is described by a two-dimensional 
Hilbert space ${\mbox{dim}=2}$. Instead of discussing 
an electron that can be either "up" or "down", 
we shall discuss a particle that can be either 
in site 1 or in site 2. In other words, we identify 
the states as: ${|1\rangle = | \uparrow \rangle}$ 
and ${ |2\rangle = |\downarrow \rangle}$.

\begin{center}
\putgraph{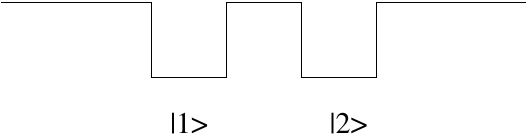}
\end{center}

The standard basis is the position 
basis ${|x=1\rangle, \, |x=2\rangle}$. 
The representation of the operator ${\hat{x}}$ is:
\beq
\hat{x}\rightarrow \left(\amatrix{1&0\cr 0&2}\right) =1+\frac{1}{2}\left(1-\sigma_3\right),
\hspace{2cm}
\sigma_3 = \left(\amatrix{1&0\cr 0&-1}\right)
\eeq
The translation operator can be optionally regarded as a reflection operator:
\beq
\hat{R} \, = \, \hat{D} \, \rightarrow \left(\amatrix{0&1\cr 1&0}\right) = \sigma_1 
\eeq
where ${\sigma_1}$ is the first Pauli matrix.
The $k$ states are defined as the eigenstates 
of the latter operator, hence 
they are the even and the odd superpositions:
\beq
|k =0\rangle = |+\rangle &=& \frac{1}{\sqrt{2}}\left(|1\rangle+|2\rangle\right)
\\ \nonumber
|k =\pi\rangle = |-\rangle &=& \frac{1}{\sqrt{2}}\left(|1\rangle-|2\rangle\right)
\eeq
Note the formal analogy with spin 1/2 system: ${|+\rangle=|\rightarrow\rangle}$ represents spin polarized right, 
while ${|-\rangle=|\leftarrow\rangle}$ represents spin polarized left.

\sheadC{A two site system with two different particles}

In this case the Hilbert space is four dimensional: ${\mbox{dim}=(2\times2) = 4 }$.
If the two particles are different (for example, a proton and a neutron) 
then each state in the Hilbert space is "physical". The standard basis is: 

${|1,1\rangle = |1\rangle \otimes |1\rangle}$ - particle A in site 1, particle B in site 1 \\
${|1,2\rangle = |1\rangle \otimes |2\rangle}$ - particle A in site 1, particle B in site 2 \\
${|2,1\rangle = |2\rangle \otimes |1\rangle}$ - particle A in site 2, particle B in site 1 \\
${|2,2\rangle = |2\rangle \otimes |2\rangle}$ - particle A in site 2, particle B in site 2

The transposition operator ${\hat{T}}$ swaps the location of the particles:
\beq
\hat{T}|i,j\rangle \, = \, |j,i\rangle
\eeq
We must not make confusion between 
the transposition operator and the reflection operators:
\beq
\hat{T} \mapsto 
\left(
\amatrix{1 & 0 & 0 & 0 \cr 0 & 0 & 1 & 0 \cr 0 & 1 & 0 & 0 \cr 0 & 0 & 0 & 1 } 
\right), 
\hspace{2cm} 
\hat{R} \mapsto 
\left(
\amatrix{0 & 0 & 0 & 1 \cr 0 & 0 & 1 & 0 \cr 0 & 1 & 0 & 0 \cr 1 & 0 & 0 & 0 }
\right) 
\eeq
Instead of the basis ${|1,1\rangle, \, |1,2\rangle, \, |2,1\rangle, \, |2,2\rangle}$, 
we may use the basis ${|A\rangle, \, |1,1\rangle, \, |S\rangle, \, |2,2\rangle}$, 
where we have defined:
\beq
|A\rangle=\frac{1}{\sqrt{2}}\left(|1,2\rangle-|2,1\rangle\right),
\hspace{2cm} 
|S\rangle=\frac{1}{\sqrt{2}}\left(|1,2\rangle+|2,1\rangle\right) 
\eeq
The state ${|A\rangle}$ is anti-symmetric under transposition, 
and all the others are symmetric under transposition.

\sheadC{Placing together two identical particles}

The motivation for discussing this system stems from the 
question: is it possible to place two electrons in the 
same location so that one of the spins is up and the other 
is down, or maybe they can be oriented differently. 
For example, one spin left and one right, or one right 
and the other up. We shall continue using the 
terminology of the previous section. We may deduce, 
from general symmetry considerations that the quantum 
state of identical particles must be an eigenstate of 
the transposition operator (otherwise we could 
conclude that the particles are not identical). 
It turns out that we must distinguish between two 
types of identical particles. According to the "spin 
and statistics theorem", particles with half-odd-integer 
spins (fermions) must be in an antisymmetric state. 
Particles with integer spins (bosons) must be in a symmetric state.  

Assume that we have two spin zero particles. 
Such particles are Bosons. There is no problem 
to place two (or more) Bosons at the same site. 
If we want to place two such particles in two sites, 
then the collection of possible states is 
of dimension 3 (the symmetric states), as 
discussed in the previous section.

Electrons have spin 1/2, and therefore they are Fermions.
Note that the problem of placing "two electron in one site" 
is formally analogous to the hypothetical system 
of placing "two spinless electrons in two sites".   
Thus the physical problem is formally related to 
the discussion in the previous section, and we can 
use the same notations. From the requirement of 
having an antisymmetric state it follows that 
if we want to place two electrons at the same location 
then there is only one possible state which is ${|A\rangle}$. 
This state is called the "singlet state". 
We discuss this statement further below.

Let us try to be "wise guys". Maybe there is 
another way to squeeze to electrons into one site? 
Rather than placing one electron with spin "up" and the other 
with spin "down" let us try a superposition of the 
type ${|\rightarrow\leftarrow\rangle-|\leftarrow\rightarrow\rangle}$. 
This state is also antisymmetric under 
transposition, therefore it is as "good" as ${|A\rangle}$.
Let us see what it looks like in the standard basis.
Using the notations of the previous section:   
\beq
\frac{1}{\sqrt{2}}\left(|+-\rangle-|-+\rangle\right) 
&=& \frac{1}{\sqrt{2}}\left(|+\rangle\otimes|-\rangle-|-\rangle\otimes|+\rangle\right) 
\\ \nonumber
&=& \frac{1}{2\sqrt{2}}(\left(|1\rangle+|2\rangle\right)\otimes\left(|1\rangle-|2\rangle\right))
- \frac{1}{2\sqrt{2}}(\left(|1\rangle-|2\rangle\right)\otimes\left(|1\rangle+|2\rangle\right))
\\ \nonumber
&=& - \frac{1}{\sqrt{2}}\left(|1\rangle\otimes|2\rangle-|2\rangle\otimes|1\rangle\right)
= -|A\rangle 
\eeq
So, we see that mathematically it is in fact the same state. 
In other words: the antisymmetric state is a single state 
and it does not matter if we put one electron  
"up" and the other "down", or one electron "right" and the other "left". 
Still let us try another possibility. Let us try to 
put one electron "up" and the other "right". 
Writing ${| \uparrow\rightarrow\rangle }$ in the standard  
basis using the notation of the previous section we get 
\beq
|1\rangle\otimes|+\rangle = \frac{1}{\sqrt{2}}|1\rangle\otimes\left(|1\rangle+|2\rangle\right)= \frac{1}{\sqrt{2}}( |1,1\rangle+|1,2\rangle) 
\eeq
This state is not an eigenstate of the 
transposition operator, it is neither symmetric 
nor anti-symmetric state. Therefore it is not physical.

{\bf Anti-bunching.--} If we place two (spinless) fermions is a two site system, the probability to find them both in the same site is zero. This is the opposite of the bosonic case where we have {\em bunching}. Say that we place one boson in oribial $\ket{+}$ and one in $\ket{-}$. The 2-body state is $\ket{+-}+\ket{-+}$. Going back to the site basis 
we get $\ket{11}-\ket{22}$, with zero probability to find them one in site~1 and one in site~2.

\newpage
\sheadB{Boxes and Networks}

A 1D segment can be regarded as the continuum 
limit of a discrete tight-binding chain. 
If the particle in confined between two ends, 
the model is know as "infinite well" or as "1D box".  
If there are periodic boundary conditions, 
the model is know as "1D ring". 
Several "1D boxes" can be connected into a network: 
each 1D segment is called "bond", 
and the junctions are called "vortices".

The ends of an isolated bond can be regarded as {\em hard~walls}.
It is easily argued that the wavefunction should satisfy Dirichlet 
boundary conditions at the boundary of a hard wall. 
The perturbation which is created by an infinitesimal 
displacement of the wall is finite (see section below).

The coupling between two wavefunctions in a network due 
to the introduction of a connecting {\em junction} can be deduced 
by considering a wire which is divided by 
a high $u\delta(x)$ barrier (see section below).

We characterize a network by specifying the lengths $\{L_a\}$ 
of the bonds, and the the potentials $u$ at each vertex.   
Given this information we would like to find the eigenfunctions 
and the eigenenergies. 
We describe below two procedures for that purpose: 
one is based on a straightforward generalization 
of the standard matching conditions, 
while the other is based on the scattering matrix formalism.

\sheadC{Hard walls} 
\label{sWalls}

Let us assume that we have a particle in a one dimensional box 
of length~$L$, such that ${V(x)=0}$ within the interval ${0<x<L}$. 
We assume Dirichlet boundary conditions at ${x=0}$, 
while at ${x=L_0}$ we assume a potential step, 
such that ${V(x)=V_0}$ for ${x>L_0}$. 
We are going to see that if we take the limit ${V_0 \rightarrow \infty}$, 
then this implies Dirichlet boundary conditions at ${x=L_0}$ too.  

The wavefunction of $n$th eigenstate has to 
satisfy the Dirichlet boundary conditions at ${x=0}$, 
and therefore has to be of the form     
\beq
\psi(x) \ \ = \ \ 
\left\{ \amatrix{ 
A\sin(kx)  & \mbox{for} & 0<x<L \cr
B \eexp{-\alpha x} & \mbox{for} &  x>L} 
\right. 
\eeq
where 
\beq
k &=& \sqrt{2\mass E}
\\ \nonumber
\alpha &=& \sqrt{2\mass(V_0-E)} \approx \sqrt{2\mass V_0} 
\eeq
The normalization factor 
at the limit ${V_0 \rightarrow \infty}$ 
is ${A = ({2}/{L})^{1/2} }$. 
It would be convenient later 
to make the replacement ${A \mapsto (-1)^n A}$, 
in order to have the same sign 
for $\psi(x)$ at ${x \sim L}$.   
The matching condition at ${x=L}$ is:
\beq
\left. \frac{\psi`(x)}{\psi(x)} \right|_{x=L-0} 
\,\, =\,\, 
\left. \frac{\psi`(x)}{\psi(x)} \right|_{x=L+0} 
\,\, = \,\, -\alpha 
\eeq
Consequently the eigenvalue equation is 
\beq
k\cot(kL) \ = \ -\alpha 
\eeq
In the limit ${V_0 \rightarrow \infty}$  
the equation becomes ${\sin(kL) = 0}$ 
which implies that the unperturbed 
eigen-energies are ${E_n=k_n^2/2\mass}$
with ${k_n=(\pi/L)n}$. In this limit we 
see that the matching condition 
at ${x=L}$ simply forces the wavefunction 
to be zero there. Thus we have established 
that Hard walls implies Dirichlet boundary 
conditions.

{\bf Perturbation due to wall displacement.-- }  
We would like to find the perturbation 
due to a small shift $dL$ in the position of the wall. 
We keep $V_0$ finite, but assume that it is very very large. 
We shall take the strict limit $V_0\rightarrow\infty$ 
only at the end of the calculation.
For $L=L_0$ the unperturbed Hamiltonian 
after digonalization is  
\beq 
[\mathcal{H}_0]_{nm} \ \ = \ \ \frac{1}{2\mass}\left( \frac{\pi}{L_0} n \right)^2 \delta_{nm} 
\eeq
If we displace the wall a distance $dL$ the new Hamiltonian 
becomes ${\mathcal{H}=\mathcal{H}_0 + dL W}$. We ask 
what are the matrix elements $W_{nm}$ of this perturbation. 
At first sight it looks as if to displace an "infinite wall" 
constitutes "infinite perturbation" and hence ${W_{nm}=\infty}$.  
But in fact it is not like that. We shall see that 
\beq 
W_{nm} \ \ = \ \  -\frac{\pi^2}{\mass L_0^3}nm, 
\ \ \ \ \ \ \ \ \ \ \ \mbox{[note sign convention]} 
\eeq
It is easily verified that the diagonal terms $W_{nn}dL$
of the perturbation matrix give correctly the first order 
shift $dE_n$ of the energy levels due to the $dL$ displacement.   

{\bf Derivation.-- }  
The Hamiltonian of the unperturbed system, and the Hamiltonian 
after we have displaced the wall a distance ${dL}$ are respectively 
\beq
\mathcal{H}_0 = \frac{p^2}{2\mass}+V(x), 
\hspace{2cm}
\mathcal{H} = \frac{p^2}{2\mass} + \tilde{V}(x)
\eeq
The perturbation is:
\beq
\delta V(x) \ \ = \ \ \tilde{V}(x) - V(x) \ \ \equiv \ \ dL \times W 
\eeq
which is a rectangle of width ${dL}$ and height ${-V_0}$.
It follows that the matrix elements of the perturbation are   
\beq 
W_{nm} 
\ \ = \ \ 
\frac{1}{dL} \int_{L}^{L+dL} \psi^{(n)}(x) [-V_0] \psi^{(m)}(x) dx 
\ \ = \ \ 
-V_0 \psi^{(n)}(L) \psi^{(m)}(L) 
\eeq
The "matching conditions" allow to express $\psi(0)$ 
using its derivative as $-(1/\alpha)\psi'(0)$, 
leading to  
\beq 
W_{nm} 
\ \  = \ \   
-\frac{1}{2\mass} \ 
\left( \frac{d}{dx} \psi^{(n)} (L) \right) 
\left( \frac{d}{dx} \psi^{(m)} (L) \right) 
\eeq
The last expression has only an implicit dependence 
on $V_0$, through the derivatives of the wavefunctions
at ${x=L}$. Obviously the limit ${V_0 \rightarrow \infty}$ 
gives a well defined finite result:
\beq
\left. \frac{d}{dx} \psi^{(n)}(x) \right|_{x=L} \ \ = \ \ \sqrt{\frac{2}{L}} k_n  \ \ = \ \ \left[\frac{2\pi^2}{L^3}\right]^{1/2}n
\eeq
leading to the result for $W_{nm}$ that has been cited 
in the introduction to this derivation.

\sheadC{The two-terminal delta junction}

The simplest non-trivial example for an $S$ matrix, 
is for a system which is composed of two 1D wires 
labeled as ${i=1,2}$, attached by a junction
that is modeled as a delta barrier $u\delta(x)$.   
Using the standard "s-scattering" convention 
the channel wavefunctions are written 
as  ${\Psi(r) \propto A_i\exp(-ikr)-B_i\exp(ikr)}$ 
where ${r=|x|}$. The linear relation between 
the $A$s and the $B$s is determined by the 
matching conditions. The matching conditions 
for the channel wavefunction at the origin ${r=0}$ 
in the case of a delta barrier are further discussed 
in the next subsection, where it is also generalized 
to the case of having a junction with $\mathcal{M}$~wires.   
The $S$ matrix which is implied by these matching 
conditions is  
\beq
\bm{S} \ \ = \ \ -\left(
\amatrix{
r & t \cr
t & r
}\right), 
\ \ \ \ \ \ \ \ \ \ \ \ \ \ \ \ \  
\mbox{[such that $\bm{S}{=}\bm{1}$ for $u{=}\infty$]}
\eeq
with
\beq
v &\equiv& (2E/\mass)^{1/2} \\
t &=& \frac{1}{1+i(u/v)} \\
r &=& -1+t 
\eeq
There is a more elegant way to re-write this expression.
Define the phase 
\beq
\gamma \ \ \equiv \ \ \mbox{phase}(v+iu) \ \ \in \ \ [0,\pi/2]
\eeq
Then it follows that 
\beq
|t|^2 = (\cos(\gamma))^2  \ \ \ \ \ \ \ && \mbox{phase}(t) = -\gamma \\
|r|^2 = (\sin(\gamma))^2  \ \ \ \ \ \ \ && \mbox{phase}(r) = -\gamma -(\pi/2)
\eeq

In practice it is more convenient to use 
in the present context an ad-hoc convention 
of writing a raw-swapped $\tilde{\bm{S}}$ matrix,  
corresponding to the re-definition of the outgoing 
amplitudes as ${\tilde{B}_1=-B_2}$
and ${\tilde{B}_2=-B_1}$, namely, 
\beq
\tilde{\bm{S}} = \left(
\amatrix{
t & r \cr
r & t
}\right)
 \,\,=\,\,
\eexp{-i\gamma} \left(
\amatrix{
\sqrt{g} \eexp{i\phi}  & -i\sqrt{1{-}g}\eexp{-i\alpha} \cr
-i\sqrt{1{-}g}\eexp{i\alpha} & \sqrt{g}\eexp{-i\phi} 
}\right)
\ \ \ \ \ \ \ \ \ \ \ \ \ \ \ \ \  
\mbox{[such that $\tilde{\bm{S}}{=}\bm{1}$ for $u{=}0$]}
\eeq
with $\alpha=\phi=0$ and 
\beq
g \ \ = \ \ |t|^2 \ \ = \ \ \frac{1}{1+(u/v_E)^2} \ \ = \ \ (\cos(\gamma))^2 
\eeq
The advantage of the ad-hoc convention 
is the possibility to regard~${u=0}$ as 
the unperturbed Hamiltonian, and to use 
the $\bm{T}$~matrix formalism to find the $\tilde{\bm{S}}$~matrix.
All the elements of the $\bm{T}$ matrix equal 
to the same number~$T$, and the equality 
${\tilde{\bm{S}}=\bm{1}-i\bm{T}}$ implies 
that ${r=-iT}$ and ${t=1-iT}$.

\sheadC{The multi-terminal delta junction}

The simplest junction is composed of $\mathcal{M}=2$ wires 
that are connected at one point. We can model such a junction 
as a delta barrier $V(x)=u\delta(x)$.  
The matching condition at ${x=0}$  
is implied by the Schrodinger equation, 
and relates the jump in the derivative 
to the value of the wavefuntion at that point.     
\beq
\frac{1}{2\mass} \ \left[ \partial\psi({+}0)-\partial\psi({-}0) \right]
\ \ = \ \ u \ \psi(0)
\eeq
A more elegant way of writing this relation is  
\beq
\sum_{a=1}^{M} \left. \frac{d\psi_a}{dr} \right|_{r{=}0} 
\ \ = \ \ \mathcal{M} \, \mass u \, \psi(0)
\eeq
with $\mathcal{M}=2$. The derivative 
of the radial functions ${\psi_1(r)=\psi(-r)}$
and ${\psi_2(r)=\psi(+r)}$ 
is with respect to the radial coordinate $r=|x|$. 
It is implicit that these wavefuntions should  
have the same value at the meeting point ${r=0}$.
If we have $\mathcal{M}$ wires connected at one 
point we can define a generalized ``delta junction'' 
using the same matching condition. 
Say that we have ${a=1,2,3}$ 
leads. The matching conditions at the junction 
are ${\psi_1(0)=\psi_2(0)=\psi_3(0)\equiv\psi(0)}$ 
while the sum over the radial derivatives should 
equal $\mathcal{M} \mass u \psi(0)$.  
This means that the sum over the outgoing currents is zero.

More generally a junction can be fully characterized by its $S$~matrix.  
So we can ask what is the $S$~matrix of a delta junction.  
It is straightforward to find out that the $S$~matrix which is implied 
by the above matching conditions is 
\beq
\bm{S}_{ab} = \delta_{ab} 
- \frac{2}{\mathcal{M}} \left(\frac{1}{1+i(u/v_E)}\right)
\ \ \ \ \ \ \ \ \ \ \ \ \ \ \ \ \  
\mbox{[such that $\bm{S}{=}\bm{1}$ for $u{=}\infty$]}
\eeq
For $u=0$ we get zero reflection if $\mathcal{M}=2$, 
while if $\mathcal{M}=3$ we get 
\beq
\bm{S} = 
\left(
\amatrix{
+1/3 & -2/3 & -2/3  \cr
-2/3 & +1/3 & -2/3  \cr
-2/3 & -2/3 & +1/3  \cr
}\right)
\eeq
In the limit $\mathcal{M}\rightarrow\infty$ we get total reflection.

Sometimes we want to treat the connecting junction 
as a perturbation. Consider for example 
the simplest possibility of having two 1D~boxes 
connected at ${x=0}$ hence forming a double well structure
which is described by a Hamiltonian ${\cal H}(x,p;u)$ 
in which the barrier is represented by $V(x)=u\delta(x)$.
The eigenstates~$n$ of the unperturbed Hamiltonian ${\cal H}(x,p;\infty)$
are those of the left box together with those of the right box.     
We want to have an explicit expression for the 
perturbation $W_{nm}$ due to the coupling at ${x=0}$. 
Note that $W$ is defined as 
the difference ${{\cal H}(u)-{\cal H}(\infty)}$ 
so it is not the same as  ${V={\cal H}(u)-{\cal H}(0)}$.  
We can obtain the total perturbation $W$ 
from  a sequence of infinitesimal variations 
of the barrier height starting from ${u=\infty}$. 
Namely, 
\beq
{\cal H}(u) \ \ = \ \ 
{\cal H}(\infty)
-\int^{\infty}_{u}
\left(\frac{\partial{\cal H}}{\partial u}\right) \ \mbox{d}u' 
\ \ \equiv \ \ {\cal H}(\infty)+W
\eeq
For any value of $u$ the Hilbert space of the system 
is spanned by a set of (real) eigenfunction labeled by~$n$. 
The matrix elements for an infinitesimal variation 
of the barrier height is 
\beq
\left(\frac{\partial{\cal H}}{\partial u}\right)_{nm}
\ \ = \ \ \psi^{(n)}(0) \,\, \psi^{(m)}(0)
\ \ = \ \ 
\frac{1}{(2\mass u)^2}
\,\,
\Big[\frac{d\psi^{(n)}}{dr}\Big]_{0}
\,\,
\Big[\frac{d\psi^{(m)}}{dr}\Big]_{0}
\eeq
where is the last step we have used the matching conditions 
in order to express the wave function 
by the total radial derivative.
As long as the barrier has small transmission ($g{\ll 1}$), 
the $n$th and the $m$th states 
remain similar to the unperturbed states and accordingly, 
upon integration, we get the result [\href{http://arxiv.org/abs/0807.2572}{arXiv:0807.2572}]
\beq
W_{nm} \ \ = \ \ 
-\frac{1}{4\mass^2 u}
\,\,
\Big[\frac{d\psi^{(n)}}{dr}\Big]_{0}
\,\,
\Big[\frac{d\psi^{(m)}}{dr}\Big]_{0}
\eeq
The $d/dr$ in this expression is the total radial derivative, 
but in fact the contribution comes from one term only, 
because the unperturbed wavefunction of a given eigenstate $\psi^{(n)}$ 
is non-zero only in one box.

\sheadC{Finding the eigenstates of a network}

Consider a network which is composed of~$b$ bonds and $v$~vertexes.
The wavefunction on bond ${a\equiv(i,j)}$  
that connects vertex~$i$ to vertex~$j$ is written as
\beq
\psi_a(x) \ \ = \ \ 
B_a \eexp{ik_a x} + A_a \eexp{-ik_{\tilde{a}} x} 
\eeq
where $0<x<L_a$ is the position of the particle along the bond 
with the origin at vertex~$i$. Note that the wavefunction 
on $\tilde{a}=(j,i)$ is with ${A_{\tilde{a}}=B_a\eexp{ik_aL_a}}$
and ${B_{\tilde{a}}=A_a\eexp{-ik_{\tilde{a}}L_a}}$.
The wavenumber is $k_a = \sqrt{2\mass E} + (\phi_a/L_a)$, 
where $\phi_a$ is determined by the vector potential.  
Note that $\phi_{\tilde{a}}=-\phi_a$.
For simplicity we assume below that there is no magnetic 
field, and accordingly ${k_{\tilde{a}}=k_a}$.

The most economical way to represent the wavefunction is 
by a vector ${\psi \mapsto \{\psi_1,\psi_2,...,\psi_v\}}$ 
that contains the wave amplitudes at the vertexes. 
If we know this vector then we can construct the whole 
wavefunction on bond ${a=(i,j)}$ using the formula
\beq
\psi_a(x) \ \ = \ \ 
\frac{1}{\sin(k_aL_a)}
\Big[ 
\psi_i \sin(k_a(L_a{-}x))+\psi_j \sin(k_a x)
\Big]
\eeq
The in order to find the eigenstates we simply have 
to write the $v$~matching conditions of the vertexes, 
and look for the energies~$E$ for which there is non-trivial 
solution for this homogeneous set of equations.

Optionally the wavefunction can be represented 
by the vector $A = \{A_a\}$ of length $2b$, 
or equivalently by the vector $B = \{ B_a \}$. 
The wavefunction and its gradient in a particular vertex 
are related to the latter as follows:    
\beq
\psi \ &=& \ B + A \\
\partial\psi \ &=& \ ik \times (B-A) 
\eeq

If the junctions are represented by a scattering matrix and 
not by a simple matching condition there is an optional 
procedure for finding the eigenstates. 
The $A$ and the $B$ vectors are related by 
\beq
B \ &=& \ \bm{S} \ A \\
A \ &=& \ \bm{J} \ \eexp{i\bm{k}\bm{L}} \ B 
\eeq
where $\bm{S}$ is a $2b\times 2b$ matrix 
that relates the outgoing to the ingoing fluxes,  
and $\bm{J}$ is a $2b\times 2b$ permutation matrix 
that induces the mapping $a\mapsto\tilde{a}$, 
and $\bm{L} = \mbox{diag}\{ L_a \}$, 
and  $\bm{k} = \mbox{diag}\{ k_a \}$.
The equation for the eigenstates is
\beq
\left(\bm{J}\eexp{i\bm{k}\bm{L}} \bm{S}-1\right) A \ = \ 0
\eeq
We can get from this equation 
a set of eigenvalues $E_n$ with the corresponding 
eigenvectors $A^{(n)}$ and the associated  
amplitudes $B^{(n)}=\bm{S}A^{(n)}$.

The $\bm{S}$ matrix of the network (if appropriately ordered) 
has a block structure, and can be written as 
$\bm{S} = \sum_j \bm{S}^j$, where $S^j$ is the $v_j\times v_j$ block 
that describes the scattering in the $j$th vertex.
$v_j$ is the number of leads that stretch out of that vertex.
A delta function scatterer is regarded as a $v_j=2$ vertex. 
Let us construct a simple example. 
Consider a ring with two delta barriers.
Such ring can be regarded as a network with 
two bonds. The bonds are labeled as   
$12,12',21,21'$, 
where the prime distinguishes the second arm.
The matrices that define the system are 
\beq
\bm{S} = \left(
\amatrix{
r_1 & t_1 & 0 & 0 \cr
t_1 & r_1 & 0 & 0 \cr
0 & 0 & r_2 & t_2 \cr
0 & 0 & t_2 & r_2 \cr
}\right),
\ \ \ \ \ \ \ \ \ \ \ \ \ \ \ \ \ 
\bm{L} = \left(
\amatrix{
L & 0 & 0 & 0 \cr
0 & L' & 0 & 0 \cr
0 & 0 & L & 0 \cr
0 & 0 & 0 & L' \cr
}\right), 
\ \ \ \ \ \ \ \ \ \ \ \ \ \ \ \ \ 
\bm{J} = \left(
\amatrix{
0 & 0 & 1 & 0 \cr
0 & 0 & 0 & 1 \cr
1 & 0 & 0 & 0 \cr
0 & 1 & 0 & 0 \cr
}\right)
\eeq

Finally we note that $\bm{k} = \sqrt{2\mathsf{m}E}+\bm{\phi}/\bm{L}$, 
where the fluxes matrix is $\bm{\phi} = \mbox{diag}\{ \phi_a \}$. 
If there is a single flux line $\phi$ one can 
write $\bm{\phi} = \phi \bm{P}$, 
where $\bm{P}$ can be expressed as a linear 
combination of the channel projectors $\bm{P}_a$. 
For example, if only one wire $a$ encloses the flux line, 
then $\bm{P}=\bm{P}_a-\bm{P}_{\tilde{a}}$ and we get 
\beq
\bm{k} \  = \ \frac{1}{\hbar}\sqrt{2\mathsf{m}E}+\phi\frac{\bm{P}}{\bm{L}}
\eeq
The obvious application of this procedure is for the analysis 
of a multi-mode Aharonov-Bohm ring, where the "bonds" are 
the propagation-modes of the ring, and they all enclose the same flux line.

\sheadA{QM in Practice (part II)}

\sheadB{Approximation methods for finding eigenstates}

\sheadC{The WKB approximation}

In the next sections we discuss the canonical 
version of perturbation theory. The small parameter 
is the strength of the perturbation. Another 
family of methods to find approximate expressions 
for the eigen-functions is based on treating ``$\hbar$" 
as the small parameter. The $\hbar$ in this context 
is not the $\hbar$ of Planck but rather 
its scaled dimensionless version that controls    
quantum-to-classical correspondence. For a particle 
in a box the scaled $\hbar$ is the ratio 
between the De-Broglie wavelength and the linear 
size of the box.  Such  approximation methods  
are known as semi-classical. The most elementary 
example is known as the WKB (Wentzel, Kramers, Brillouin) 
approximation {\bf [Messiah p.231]}.  
It is designed to treat slowly varying 
potentials in 1D where the wavefunction looks 
locally like a plane wave (if $V(x)<E$) or as 
a decaying exponential (in regions where $V(x)>E$). 
The refined version of WKB, which is known 
as  "uniform approximation", allows also 
to do the matching at the turning points 
(where $V(x)\sim E$).  
The generalization of the $d=1$ WKB 
for $d>1$ dimensions integrable systems is known 
as the EBK scheme. There is also a different 
type of generalization for $d>1$ chaotic systems, 
via the Wigner Weyl formalism.

Assuming free wave propagation, hence neglecting 
back reflection, the WKB wavefunction is written as 
\beq 
\Psi(x) \ \  = \ \ \sqrt{\rho(x)} \eexp{iS(x)} 
\eeq
This expression is inserted into the 1D Schr\"{o}dinger 
equation. In leading order in $\hbar$ we get 
a continuity equation for the probability density $\rho(x)$, 
while the local wavenumber should be as expected 
\beq 
\frac{dS(x)}{dx} \ \ = \ \ p(x) \ \ = \ \ \sqrt{2\mass (E-V(x))} 
\eeq
Hence (for a right moving wave) one obtains the WKB approximation
\beq 
\psi(x) \ \ = \ \ \frac{1}{\sqrt{p(x)}}\mbox{EXP}\Big[i\int_{x_0}^{x}p(x')dx'\Big] 
\eeq
where $x_0$ is an arbitrary point. 
For a standing wave the "EXP" can be replaced 
by either "sin" or "cos". 
It should be clear that for a "flat" potential 
floor this expression becomes exact.
Similarly in the "forbidden" region
we have a decaying exponential. Namely,  
the local wavenumber $\pm p(x)$ is replaced 
by ${\pm i\alpha(x)}$, 
where  $\alpha(x) = (2 \mass (V(x)-E))^{1/2}$.

{\bf Scattering.-- } 
If we have a scattering problem in one dimension 
we can use the WKB expression (with $\exp$) in order 
to describe (say) a right moving wave. 
It should be realized that there is no back-reflection 
within the WKB framework, but still we can calculate 
the phase shift for the forward scattering: 
\beq 
\theta_{\tbox{WKB}} & \ \ = \ \ & 
\int_{-\infty}^{\infty}p(x)dx -\int_{-\infty}^{\infty} p_E dx 
\ \ = \ \ 
\int_{-\infty}^{\infty}\left[\sqrt{2\mass(E-V(x))}-\sqrt{2\mass E}\right]dx 
\\ \nonumber
& \ \ = \ \ &
\sqrt{2\mass E}\int_{-\infty}^{\infty} \left[\sqrt{1-\frac{V(x)}{E}}-1\right]dx        
\ \ \approx \ \ 
-\sqrt{\frac{\mass}{2E}}\int_{-\infty}^{\infty}V(x)dx  
\eeq
Hence we get 
\beq 
\theta_{\tbox{WKB}} \ \ = \ \ -\frac{1}{\hbar v_E}\int_{-\infty}^{\infty}V(x)dx 
\eeq
It should be noted that $\theta^{\tbox{WKB}}_{1D}$ 
is the phase shift in a 1D scattering geometry ($-\infty<x<\infty$). 
A similar looking result for semi-1D geometry ($r>0$) 
is known as the Born approximation for the phase shift.

{\bf Bound states.-- } 
If we have a particle in a well, 
then there are two turning points $x_1$ and $x_2$. 
On the outer sides of the well we have WKB decaying 
exponentials, while in the middle we have 
a WKB standing wave. 
As mentioned above the WKB scheme can be extended 
so as to provide matching conditions at the 
two turning points. Both matching conditions 
can be satisfied simultaneously if 
\beq
\int_{x_1}^{x_2} p(x)dx \ \ = \ \  \left(\frac{1}{2}+n\right)\pi\hbar
\eeq
where $n=0,1,2, \dots $ is an integer.
Apart from the $1/2$ this is a straightforward 
generalization of the quantization condition
of the wavenumber of a particle in a 1D box with hard walls 
(${k\times(x_2-x_1)=n\pi}$). 
The $(1/2)\pi$ phase shift arise because 
we assume soft rather than hard walls. 
This $1/2$ becomes exact in the case 
of harmonic oscillator.

The WKB quantization condition has an obvious 
phase space representation, and it coincides     
with the "Born Oppenheimer quantization condition":
\beq
\oint p(x)dx \ \ = \ \ \left(\frac{1}{2}+n\right)2\pi\hbar
\eeq
The integral is taken along the energy   
contour which is formed by the curves $p=\pm p(x)$.  
This expression implies that the number 
of states up to energy $E$ is 
\beq
\mathcal{N}(E) \ \ = \ \ \iint_{H(x,p)<E} \frac{dxdp}{2\pi\hbar} 
\eeq
The $d>1$ generalization of this idea 
is the statement that the number of states up to 
energy $E$ is equal to the phase space volume 
divided by $(2\pi\hbar)^d$. The latter statement is 
known as Weyl law, and best derived using 
the Wigner-Weyl formalism.

\sheadC{The variational scheme} 

The variational scheme is an approximation method that 
is frequently used either as an alternative or in combination 
with perturbation theory. It is an extremely powerful method 
for the purpose of finding the ground-state. 
More generally we can use it to find the lowest energy state 
within a subspace of states. 

The variational scheme is based on the trivial observation 
that the ground state minimize the energy functional 
\beq
F[\psi] \ \ \equiv \ \ \langle \psi | \mathcal{H} | \psi \rangle 
\eeq  
If we consider in the variational scheme 
the most general $\psi$ we simply recover 
the equation $\mathcal{H}\psi=E\psi$ and hence 
gain nothing. But in practice we can substitute 
into $F[]$ a trial function $\psi$ that depends 
on a set of parameters ${X=(X_1,X_2,...)}$. 
Then we minimize the function $F(X)=F[\psi]$ 
with respect to~$X$.   

The simplest example is to find the ground state 
on an harmonic oscillator. If we take the trail function 
as a Gaussian of width $\sigma$, then the minimization 
of the energy functional with respect to $\sigma$ 
will give the exact ground state. If we consider 
an anharmonic oscillator we still can get a very good approximation.  
A less trivial example is to find bonding orbitals 
in a molecules using as a trial function a combination 
of hydrogen-like orbitals.

\newpage \sheadC{Perturbation theory - motivation} 

Let us consider a particle in a two dimensional box. 
On the left a rectangular box, and on the right a chaotic box.

\begin{center}
\putgraph[0.2\hsize]{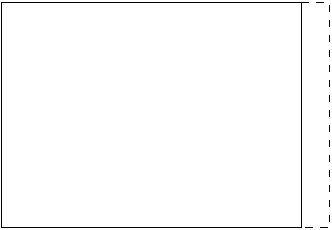} 
\hspace*{0.2\hsize}
\putgraph[0.2\hsize]{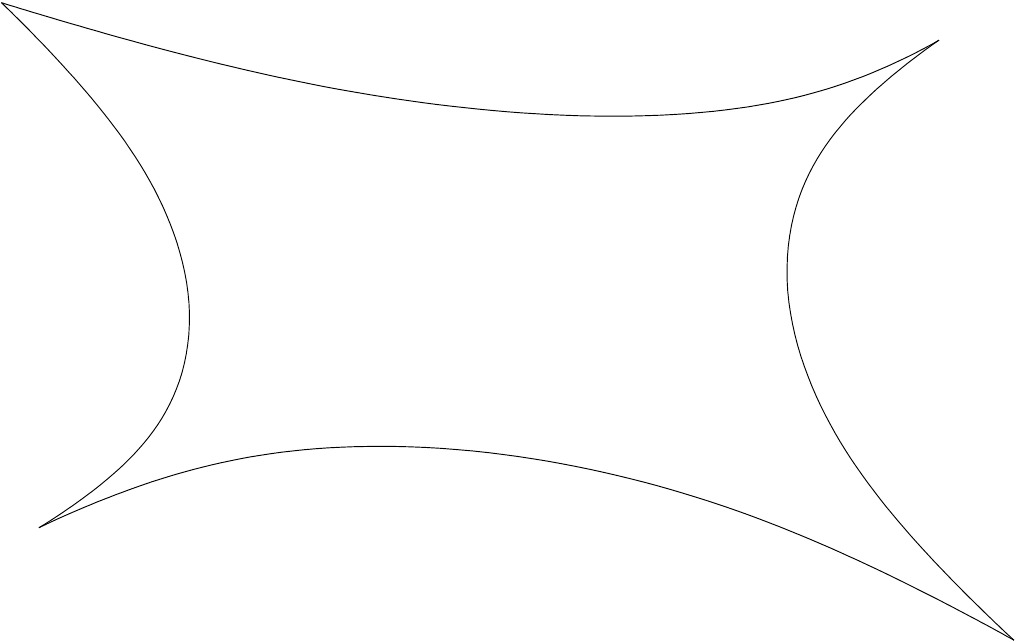}
\end{center}

For a regular box (with straight walls) we found: 
${ E_{n_x,n_y} \propto (n_x/L_x)^2 + (n_y/L_y)^2 }$, 
so if we change $L_x$ we get the energy level scheme  
which is drawn in the left panel of the following figure. 
But if we consider a chaotic box, we shall get 
energy level scheme as on the right panel.  

\begin{center}
\putgraphr{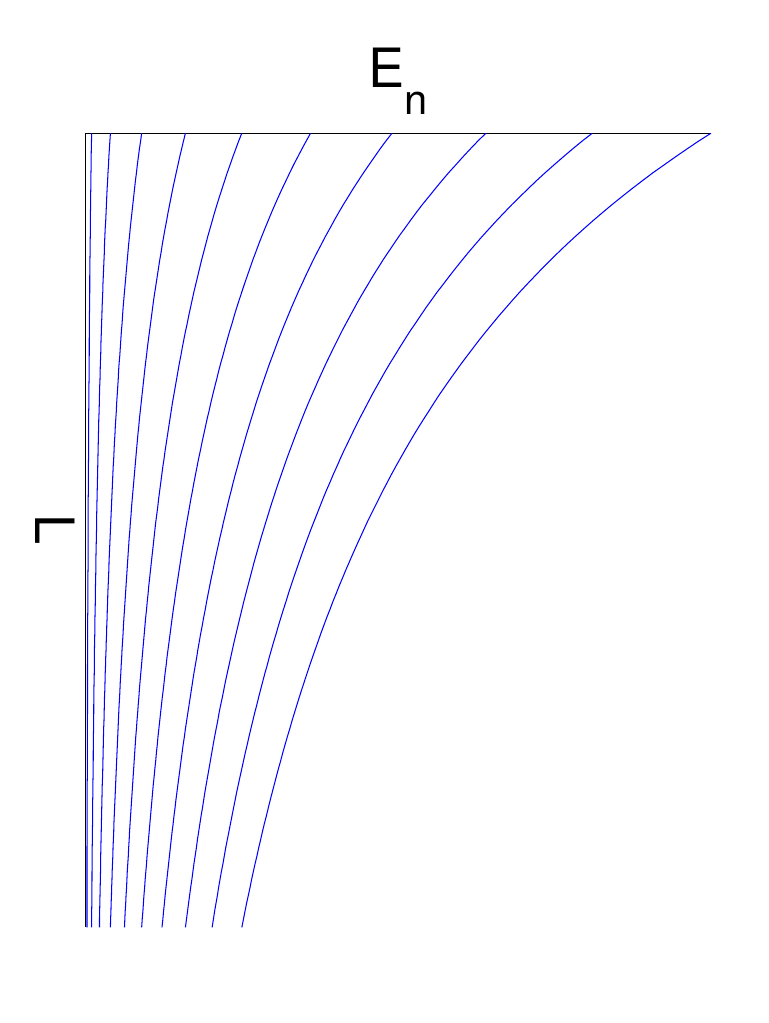} 
\hspace*{0.2\hsize}
\putgraphr{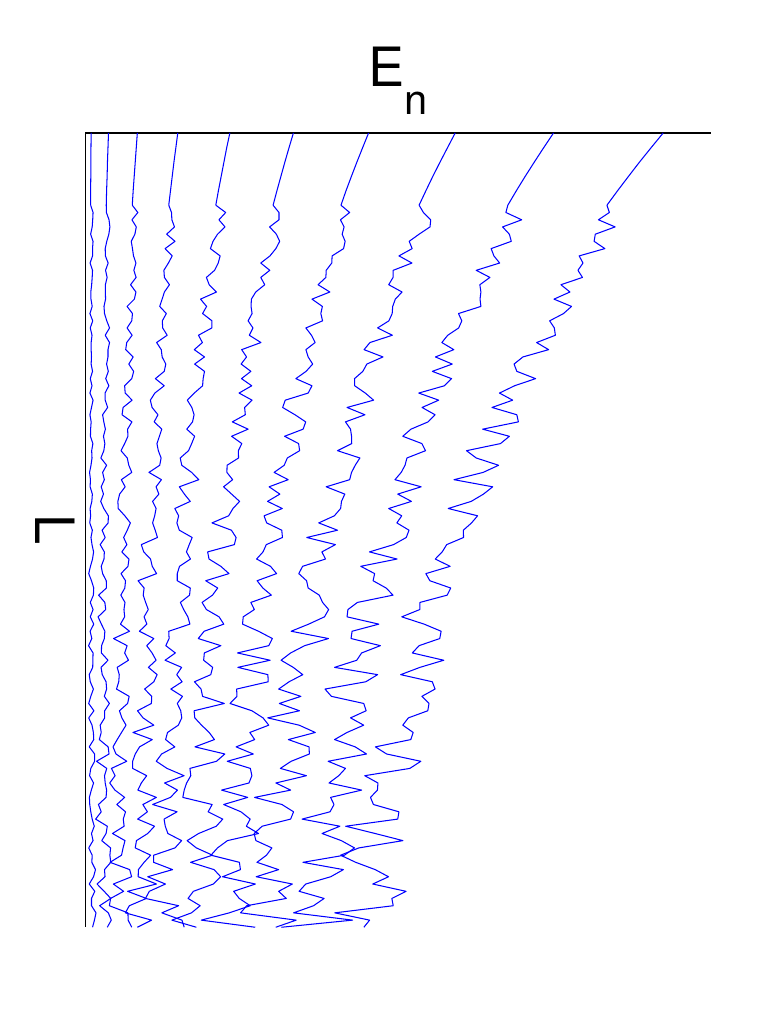} 
\end{center}

The spectrum is a function of a control parameter, 
which in the example above is the position of a wall. 
For generality let us call this parameter~$X$. 
The Hamiltonian is $\mathcal{H}(\hat{Q},\hat{P};X)$. 
Let us assume that we have calculated the levels 
either analytically or numerically for $X=X_0$. 
Next we change the control parameter to the 
value ${X=X_0+\delta X}$, and use the 
notation $\delta X=\lambda$ for the small parameter.   
Possibly, if $\lambda$ is small enough, we can linearize 
the Hamiltonian as follows:
\beq
\mathcal{H} \ \ = \ \  
\mathcal{H}(Q,P;X_0) + \lambda V(Q,P)
\ \ = \ \ \mathcal{H}_0 + \lambda V
\eeq 
With or without this approximation we can try  
to calculate the new energy levels. 
But if we do not want or cannot diagonalize $\mathcal{H}$ 
for the new value of $X$ we can try to 
use a perturbation theory scheme. Obviously 
this scheme will work only for small enough $\lambda$ 
that do not "mix" the levels too much.
There is some "radius of convergence" (in $\lambda$) 
beyond which perturbation theory fails completely.
In many practical cases $X=X_0$ is 
taken as the value for which the Hamiltonian is simple 
and can be handled analytically. In atomic physics 
the control parameter $X$ is usually either 
the prefactor of the spin-orbit term, or an 
electric field or a magnetic field which are turned on.

There is another context in which perturbation theory 
is very useful. Given (say) a chaotic system, 
we would like to predict its response to a small change 
in one of its parameters. For example we may ask  
what is the response of the system to an external 
driving by either an electric or a magnetic field. 
This response is characterized by a quantity called "susceptibility". 
In such case, finding the energies without the perturbation 
is not an easy task (it is actually impossible analytically, 
so if one insists heavy numerics must be used). 
Instead of "solving" for the eigenstates it turns out 
that for any practical purpose it is enough to characterize 
the spectrum and the eigenstates in a statistical 
way. Then we can use perturbation theory in order 
to calculate the "susceptibility" of the system.

\newpage 
\sheadC{Perturbation theory - a mathematical digression}

Let us illustrate how the procedure 
of perturbation theory is applied in order to find the 
roots of a toy equation. Later we shall apply 
the same procedure to find the eigenvalues and the 
eigenstates of a given Hamiltonian.
The toy equation that we consider is 
\beq
x+\lambda x^5=3 
\eeq
We assume that the magnitude of the perturbation (${\lambda}$) is small. 
The Taylor expansion of $x$ with respect to $\lambda$ is:
\beq
x(\lambda) \ \ = \ \ x^{(0)}+x^{(1)}\lambda+x^{(2)}\lambda^2+x^{(3)}\lambda^3+ \dots  
\eeq
The zero-order solution gives us the solution for the case ${\lambda = 0}$:
\beq
x^{(0)} \ \ = \ \ 3 
\eeq
To find the perturbed solution substitute the expansion:
\beq
\Big[x^{(0)}+x^{(1)}\lambda+x^{(2)}\lambda^2+...\Big] 
\ + \ \lambda\Big[x^{(0)}+x^{(1)}\lambda+x^{(2)}\lambda^2+...\Big]^5 \ \ = \ \ 3 
\eeq
This can be re-arranged as follows: 
\beq
\Big[x^{(0)}-3\Big] 
\ + \ \Big[x^{(1)}+(x^{(0)})^5\Big]\lambda
\ + \ \Big[5(x^{(0)})^4x^{(1)}+x^{(2)}\Big]\lambda^2
\ + \ \mathcal{O}(\lambda^3) \ \ = \ \ 0 
\eeq
By comparing coefficients we get a system of equations 
that can be solved iteratively order by order: 
\beq
x^{(0)} &\ \ = \ \ & 3 \\
x^{(1)} & \ \ = \ \ & -(x^{(0)})^5 \ \ = \ \ -3^5  \\
x^{(2)} & \ \ = \ \ & -5(x^{(0)})^4x^{(1)} \ \ = \ \ 5 \times 3^{9} 
\eeq
It is obviously possible to find the corrections 
for higher orders by continuing in the same way.

\newpage
\sheadB{Perturbation theory for the eigenstates}

\sheadC{Degenerate perturbation theory (zero-order)}

Consider a diagonal Hamiltonian matrix, with a small added 
perturbation that spoils the diagonalization:
\beq
\mathcal{H}
= \left(\amatrix{ 
2    & 0.03 & 0   & 0    & 0   & 0.5  & 0 \cr 
0.03 & 2    & 0   & 0    & 0   & 0    & 0 \cr 
0    & 0    & 2   & 0.1  & 0.4 & 0    & 0 \cr 
0    & 0    & 0.1 & 5    & 0   & 0.02 & 0 \cr 
0    & 0    & 0.4 & 0    & 6   & 0    & 0 \cr 
0.5  & 0    & 0   & 0.02 & 0   & 8    & 0.3 \cr 
0    & 0    & 0   & 0    & 0   & 0.3  & 9 } 
\right) 
\eeq
The Hamiltonian can be visualized using an energy level diagram:

\begin{center}
\putgraph[0.4\hsize]{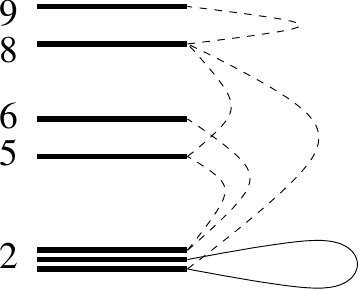}
\end{center}

The eigenvectors without the perturbation are:
\beq
\left(\amatrix{ 1 \cr 0 \cr 0 \cr 0 \cr 0 \cr 0 \cr 0} \right)\,\,\,\,\, 
\left(\amatrix{ 0 \cr 1 \cr 0 \cr 0 \cr 0 \cr 0 \cr 0} \right) \,\,\,\,\, 
\left(\amatrix{ 0 \cr 0 \cr 1 \cr 0 \cr 0 \cr 0 \cr 0} \right) \cdots 
\eeq
The perturbation spoils the diagonalization. The question we 
would like to answer is what are the new eigenvalues and 
eigenstates of the Hamiltonian. We would like to find them "approximately", 
without having to diagonalize the Hamiltonian again. First 
we will take care of the degenerated blocks. The perturbation 
can remove the existing degeneracy. In the above example we 
make the following diagonalization:
\beq
2 \cdot 
\left(
\amatrix{ 
1 & 0 & 0 \cr 
0 & 1 & 0 \cr 
0 & 0 & 1 } 
\right) 
\,\,+ \,\, 
0.03 \cdot 
\left(
\amatrix{ 
0 & 1 & 0 \cr
1 & 0 & 0 \cr 
0 & 0 & 0 } 
\right) 
\ \ \ \rightarrow \ \ \ 
\left(
\amatrix{ 
1.97 & 0 & 0 \cr 
0 & 2.03 & 0 \cr 
0 & 0 & 2 } 
\right)
\eeq
We see that the perturbation 
has removed the degeneracy. 
At this stage our achievement is that there 
are no matrix elements that couple degenerate states. 
This is essential for the next steps:  
we want to ensure that the perturbative calculation 
would not diverge. 

For the next stage we have to transform the Hamiltonian 
to the new basis. See the calculation in the Mathematica 
file "diagonalize.nb". If we diagonalize numerically 
the new matrix we find that the eigenvector that 
corresponds to the eigenvalue $E \approx 5.003$ is 
\beq
|\Psi\rangle  
\,\,\,\,\,\,\,\rightarrow\,\,\,\, 
\left(\amatrix{ 0.0008 \cr 0.03 \cr 0.0008 \cr 1 \cr -0.01 \cr -0.007 \cr 0.0005 } \right) 
\,\,\,\,\,\,\,= \,\,\,\, 
\left(\amatrix{ 0 \cr 0 \cr 0 \cr 1 \cr 0 \cr 0 \cr 0 } \right) 
\,\,\,+ \,\,\, 
\left(\amatrix{ 0.0008 \cr 0.03 \cr 0.0008 \cr 0 \cr -0.01 \cr -0.007 \cr 0.0005 } \right) 
\,\,\,\,\,\,\, \equiv \,\,\,\, 
\Psi_n^{[0]}+ \Psi_n^{[1,2,3, \dots ]} 
\eeq
We note that within the scheme of perturbation theory  
it is convenient to normalize the eigenvectors according 
to the zero order approximation. We also use the convention 
that all the higher order corrections have zero overlap 
with the zero order solution. Else the scheme of 
the solution becomes ill defined.

\sheadC{Perturbation theory to arbitrary order} 

We write the Hamiltonian as ${\mathcal{H}=\mathcal{H}_{0}+ \lambda V}$
where $V$ is the perturbation and ${\lambda}$ is the control parameter.  
Note that $\lambda$ can be "swallowed" in $V$. We keep it during 
the derivation in order to have clear indication for the "order" of the 
terms in the expansion. The Hamiltonian is represented in 
the unperturbed basis as follows: 
\beq
\mathcal{H}
\ \ = \ \ \mathcal{H}_{0}
+ \lambda V 
\ \ = \ \ \sum_{n} | n \rangle \varepsilon_{n} \langle n |
\,\,\, + \,\, 
\lambda \sum_{ n,m} | n \rangle V_{n,m} \langle m | 
\eeq
which means
\beq
\mathcal{H}
\ \ \ \rightarrow \ \ \  
\left(
\amatrix{
\varepsilon_{1} & 0 & 0 & 0 \cr 
0 & \varepsilon_{2} & 0 & 0 \cr 
0 & 0 & \varepsilon_{3} &0 \cr 
0 & 0 & 0 & \dots } 
\right) 
+ 
\lambda
\left(
\amatrix{
V_{1 , 1} & V_{1 , 2} & \dots  & \dots  \cr 
V_{2 , 1} & V_{2 , 2} & \dots  & \dots  \cr 
\dots  & \dots  & \dots  & \dots  \cr 
\dots  & \dots  & \dots  & \dots } 
\right) 
\eeq
In fact we can assume without loss of generality  
that ${V_{n,m} = 0}$ for ${n=m}$, because 
these terms can be swallowed into the diagonal part. 
Most importantly we assume that none of the 
matrix element couples degenerated states. 
Such couplings should be treated in the preliminary 
"zero order" step that has been discussed 
in the previous section.

We would like to introduce a perturbative scheme 
for finding the eigenvalues and the eigenstates of    
the equation 
\beq
(\mathcal{H}_{0}+ \lambda V) | \Psi \rangle \ = \ E | \Psi \rangle
\eeq
The eigenvalues and the eigenvectors are expanded as follows:
\beq
E
&=& E^{[0]} 
+ \lambda E^{[1]} 
+ \lambda^2 E^{[2]} + \cdots 
\\ \nonumber
\Psi_{n} 
&=& \Psi_{n}^{[0]} 
+ \lambda \Psi_{n}^{[1]} 
+ \lambda^2 \Psi_{n}^{[2]} 
\eeq
where it is implicit that the zero order 
solution and the normalization are such that  
\beq
&& E^{[0]} \ \ = \ \ \varepsilon_{n_0}  
\\ \nonumber
&& \Psi_n^{[0]} \ \ = \ \ \delta_{n,n_0}
\\ \nonumber
&& \Psi_n^{[1,2,3, \dots ]} \ \ = \ \  0 \,\,\,\, \mbox{for} \,\,\,\, n=n_0 
\eeq
It might be more illuminating to rewrite the 
expansion of the eigenvector using Dirac notations.
For this purpose we label the unperturbed 
eigenstates as $|\varepsilon_n\rangle$ and the 
perturbed eigenstates as $|E_n\rangle$. 
Then the expansion of the latter is written as  
\beq
{|E_{n_0} \rangle} 
\ \ = \ \ {|E_{n_0}^{[0]} \rangle} 
+ \lambda { |E_{n_0}^{[1]} \rangle}
+ \lambda ^2{ |E_{n_0}^{[2]} \rangle} 
+ \cdots 
\eeq
hence
\beq
\langle \varepsilon_n | E_{n_0} \rangle 
\ \ = \ \ \delta_{n,n_0}
+ \lambda \langle \varepsilon_n | E_{n_0}^{[1]} \rangle 
+ \lambda^2 \langle \varepsilon_n | E_{n_0}^{[2]} \rangle 
+ \cdots 
\eeq
which coincides with the traditional notation.
In the next section we introduce a derivation that 
leads to the following practical results (absorbing~$\lambda$ into the definition of~$V$):
\beq
\Psi_n^{[0]} &=& \delta_{n,n_0}
\\ \nonumber
\Psi_n^{[1]} &=& \frac {V_{n,n_0}}{\varepsilon_{n_0}-\varepsilon_{n}}
\\ \nonumber
E^{[0]} &=& \varepsilon_{n_{0}}
\\ \nonumber
E^{[1]} &=& V_{n_{0},n_{0}}
\\ \nonumber
E^{[2]} &=& \sum_{m (\neq n_{0})} 
\frac{V_{n_{0},m} V_{m,n_{0}}}{\varepsilon_{n_{0}}-\varepsilon_{m}}
\eeq
The calculation can be illustrated graphically 
using a "Feynman diagram".  For the calculation of the 
second order correction to the energy we should 
sum all the paths that begin with the state ${n_{0}}$ 
and also end with the state ${n_{0}}$. 
We see that the influence of the nearer levels 
is much greater than the far ones.
This clarifies why we cared to treat the couplings 
between degenerated levels in the zero order stage 
of the calculation.  
The closer the level the stronger the influence. 
This influence is described as "level repulsion". 
Note that in the absence of first order correction 
the ground state level always shifts down.

\sheadC{Derivation of the results}

The equation we would like to solve is 
\beq
\left(
\amatrix{
\varepsilon_{1} & 0 & 0 & 0 \cr 
0 & \varepsilon_{2} & 0 & 0 \cr 
0 & 0 & \varepsilon_{3} &0 \cr 
0 & 0 & 0 & \dots} 
\right) 
\left(
\amatrix{
\Psi_{1} \cr 
\Psi_{2} \cr 
\dots} 
\right) 
+ \lambda 
\left(
\amatrix{
V_{1 , 1} & V_{1 , 2} & \dots & \dots \cr 
V_{2 , 1} & \dots & \dots & \dots \cr 
\dots & \dots & \dots & \dots \cr 
\dots & \dots & \dots & \dots} 
\right) 
\left(
\amatrix{
\Psi_{1} \cr 
\Psi_{2} \cr 
\dots} 
\right) 
= E \left(\amatrix{\Psi_{1} \cr \Psi_{2} \cr \dots} \right) 
\eeq
Or, in index notation:
\beq
\varepsilon_{n}\Psi_{n} \,+\, 
\lambda \sum_{m} V_{ n , m } \Psi_{m}
\ = \ E \Psi_{n} 
\eeq
This can be rewritten as 
\beq
(E- \varepsilon_{n})\Psi_{n}
\ = \ \lambda \, \sum_{m}V_{ n , m }\Psi_{m} 
\eeq
We substitute the Taylor expansion:
\beq
&& E \ \ = \ \ \sum_{k=0} \lambda^k E^{[k]} 
\ \ \ = \ \  E^{[0]}+\lambda E^{[1]}+\dots 
\\ \nonumber
&& \Psi_{n}
\ \ = \ \ \sum_{k=0} \lambda^k \Psi_{n}^{[k]}
\ \ = \ \
\Psi_{n}^{[0]}
+\lambda \Psi_{n}^{[1]}
+\dots 
\eeq
We recall that ${E^{[0]} = \varepsilon_{n_0}}$, and
\beq
\Psi_{n}^{(0)}
=\delta_{n,n_{0}} 
\,\rightarrow\, 
\left(\amatrix{\dots \cr 0 \cr 0 \cr 1\cr 0 \cr 0 \cr \dots } 
\right),
\ \ \ \ \ \ \ \ \ \ \ \ \ \  
\Psi_{n}^{[k\neq 0]} 
\,\, \rightarrow \,\, 
\left(\amatrix{\dots \cr ? \cr ? \cr 0\cr ? \cr ? \cr \dots} \right) 
\eeq
After substitution of the expansion we use on the left side the identity
\beq
(a_{0} +\lambda a_{1}+ \lambda^2 a_{2}+ \dots)( b_{0} +\lambda b_{1}
+ \lambda^2 b_{2}+ \dots) \ \ = \ \ \sum_k \lambda^k\sum_{k'=0}^k a_{k'} b_{k-k'} 
\eeq
Comparing the coefficients of ${\lambda^k}$ we get 
a system of equations $k=1,2,3...$ 
\beq
\sum_{k'=0}^k E^{[k']} \Psi_{n}^{[k- k']}-\varepsilon_{n}\Psi_{n}^{[k]}
\ = \ \sum_{m} V_{n,m}\Psi_{m}^{[k-1]} 
\eeq
We write the $k$th equation in a more expanded way:
\beq
(E^{[0]}- \varepsilon_{n}) \Psi_{n}^{(k)}+E^{[1]}\Psi_{n}^{[k-1]} 
+E^{[2]}\Psi_{n}^{[k-2]}+\dots+E^{[k]}\Psi_{n}^{[0]}
\ \ = \ \ \sum_{m}V_{n,m}\Psi_{m}^{[k-1]} 
\eeq
If we substitute ${n={n_0}}$ in this equation we get:
\beq
0\,\,+\,\,0\,\,+\dots+E^{[k]} \ \ = \ \ \sum_m V_{n_0,m} \Psi_{m}^{[k-1]}
\eeq
If we substitute ${ n\ne{n_0} }$ in this equation we get:
\beq
(\varepsilon_{n_0}-\varepsilon_{n}) \Psi_{n}^{[k]} 
\ \ = \ \ \sum_{m}V_{n,m} \Psi_{m}^{[k-1]} 
- \sum_{k'=1}^{k-1} E^{[k']} \Psi_{n}^{[k-k']} 
\eeq
Now we see that we can solve the system of equations 
that we got in the following order:
\beq
\Psi^{[0]} 
\ \ \rightarrow \ \ E^{[1]} 
\ \ , \ \ \Psi^{[1]}
\ \ \rightarrow \ \ E^{[2]} 
\ \ , \ \ \Psi^{[2]}
\ \ \rightarrow \ \ E^{[3]} 
\ \ , \ \ \Psi^{[3]}
\ \ \rightarrow \ \ \dots 
\eeq
where:
\beq
E^{[k]} 
&=& 
\sum_m V_{n_0,m} \Psi_{m}^{[k-1]}
\\ \nonumber
\Psi_{n}^{[k]} 
&=& 
\frac{1}{ (\varepsilon_{n_0}-\varepsilon_{n})} 
\left[
\sum_{m}V_{n,m} \Psi_{m}^{[k-1]} 
- \sum_{k'=1}^{k-1} E^{[k']} \Psi_{n}^{[k-k']}
\right] 
\eeq
The practical results that were cited in the previous 
sections are easily obtained from this iteration scheme.

\newpage
\sheadC{Delta scatterer in a box}

Possibly the simplest non-trivial example for the demonstration 
of perturbation theory and its limitations is to find the 
effect of introducing a delta scatterer $V(r)=\delta(r)$ in a box. 
We consider first the exact solution in the 
one dimensional ($d=1$) case, where the delta is placed in 
the centre of segment $|r|<R$, and the $d=3$ case, 
where the delta is placed in the centre of sphere $|r|<R$.
Later we consider the case of a delta in the centre 
of a rectangular box ${[-R,R]^d}$. Below we use units 
such that the mass is unity, and also ${R=1}$.

In the one dimensional case the wavefunction 
has to satisfy Dirichlet boundary conditions 
at the walls, and the matching condition across 
the delta barrier at the origin, leading to 
the eigenvalue equation $k\cot(k)=u$ for $E>0$ 
and $k\coth(k)=u$ for $E<0$ where ${k=\sqrt{2E}}$. 
One observes that for ${u<-1}$ the ground state 
become bounded ($E_0<0$), while all the other  
states are extended (with ${E>0}$). For very 
negative $u$ we get $E_0 \approx -u^2/2$ 
and  $E_{n,-} = (\pi n)^2/2$ and $E_{n,+}\approx E_{n,-}$
that are slightly shifted downwards.   

For $d\ge2$ the effect of introducing a delta scatterer 
is zero. This is complementary to the statement that a delta 
function has zero scattering cross-section. 
In order to have a non-zero effect the scatterer 
should have a finite size. 
To clarify the "zero" effect of a point scatterer 
let us consider the eigenfunctions of a ${d=3}$ sphere 
that has an infinite barrier $V_0$ of radius $a$ in its centre.
The zero angular momentum wavefuntions in the extreme case 
of an infinite barrier are
\beq 
\Psi(r,\theta,\varphi) \ \ = \ \ \const \frac{\sin(k|r-a|)}{r}  
\eeq
The healing length is of order $a$ and in 
the limit $a\rightarrow0$ the ``hole" 
in the wavefunction disappears irrespective 
of whether $V_0a^3$ is kept constant or becomes infinite.

Now we would like to see how/whether the above results 
can be obtained formally via perturbation theory. 
For this purpose it is most convenient 
to consider a delta function in a centre 
of a rectangular box (segment if $d=1$, square if $d=2$). 
Excluding all the states that have node at the centre 
the Hamiltonian matrix takes the following form:
\beq 
H \ \ = \ \ \left(\amatrix{
\varepsilon_1 & 0 & 0 & 0 & ... \cr
0 & \varepsilon_2 & 0 & 0 & ... \cr
0 & 0 & \varepsilon_3 & 0 & ... \cr
0 & 0 & 0 & \varepsilon_4 & ... \cr
... & ... & ... & ... & ... 
}\right)
\ + \ u\left(\amatrix{
1 & 1 & 1 & 1 & ... \cr
1 & 1 & 1 & 1 & ... \cr
1 & 1 & 1 & 1 & ... \cr
1 & 1 & 1 & 1 & ... \cr
... & ... & ... & ... & ... 
}\right)
\eeq
We note that the same Hamiltonian (with ${u<0}$) 
emerges in the analysis of the {\em Copper pair problem},  
which has inspired the BCS theory of superconductivity.  
In turns out (see below) that the important 
information for the analysis is the density 
of states ${\varrho \propto \varepsilon^{\alpha}}$, 
where $\alpha=(d/2){-}1$ with $d=1,2,3$. 
If we introduce a cutoff and consider the 
finite-size Hamiltonian that consists 
of the levels ${\varepsilon_n<\omega_c}$, it is like 
to give the scatterer a finite width ${a \sim (2\mass\omega_c)^{-1/2}}$.

The ground state energy  
using second order perturbation is 
\beq 
E_0 \ \ = \ \ \varepsilon_1 + u + \sum_n \frac{u^2}{E_0-\varepsilon_n} 
\ \ \sim \ \   - \sum_n \frac{u^2}{E_n}  
\eeq
In the one dimensional case this expression is finite 
and give a meaningful result if $u\ll1$. 
But if $\alpha\ge 1$ this sum diverges in the $\omega_c\rightarrow\infty$ limit:
\beq 
E_0  \ \ \sim \ \   -\sum_n \frac{u^2}{E_n}  \ \ \sim \ \  -\infty 
\eeq
This divergence obviously applies to all the levels, 
and to all orders of perturbation theory 
(we consider above the ground state just as an example).  
It turns out from the exact analysis  (below) 
that the this divergence indicates that the perturbation 
in an infinite order (exact) calculation has a zero effect. 
One way to think about this puzzling observation is to 
realize that the geometric sum ${1+x+x^2+x^3+...}$ 
is formally zero (and not infinite) in the limit $x\rightarrow\infty$ 
because it equals ${1/(1-x)}$.

The exact solution of the eigenvalue problem requires 
to solve the matrix equation ${H\Psi=E\Psi}$, namely 
\beq 
\varepsilon_n \Psi_n -u\sum_m\Psi_m  \ = \ E\Psi_n
\eeq
Introduction the notation $C=\sum_m\Psi_m$, 
which later will be determined via normalization, 
we can solve for the wavefunction:  
\beq 
\Psi_n  \ \ = \ \ C\frac{u}{E-\varepsilon_n}
\eeq
From the definition of $C$ it follows 
that the eigenvalues should be the solution 
of the following equation: 
\beq 
\sum \frac{u}{E-\varepsilon_n} \ = \ 1
\eeq
The equation has a nice graphical visualization, 
and one can easily be convinced that in the 
limit $\omega_c\rightarrow\infty$ the perturbed eigenvalues   
approach the unperturbed values.

In view of the interest in the Cooper pair problem 
it is of interest to see what happens if we have 
constant density of states $\varrho$ with 
finite cutoff $\omega_c$ and negative $u$.
The graphical visualization implies that the 
ground state separates from the quasi-continuum and 
forms a bound state of energy $E_0=-\Delta$, 
as in the familiar $d=1$ case. In order 
to estimate $\Delta$ we approximate the sum 
by an integral and get the equation 
\beq 
|u|\varrho \ln\left(\frac{\Delta+\omega_c}{\Delta}\right) \ = \ 1
\eeq
leading to 
\beq 
\Delta \ \ = \ \ \frac{\omega_c}{\exp\left(\frac{1}{|u|\varrho}\right)-1} 
\eeq
In the limit of small $u$ we have the BCS-like 
result $\Delta = \exp(-1/(|u|\varrho))$, while 
in the other extreme of very strong $u$ 
we get $\Delta=Nu$ where $N=\varrho\omega_c$ 
is the number of levels in the quasi-degenerated band. 
The latter result is easily verified by direct 
diagonalization of of the Hamiltonian. Namely,  
if the unperturbed  energies were degenerated 
the Hamiltonian would become diagonal 
in the ``momentum" representation, where 
only the zero momentum state is affected.

\newpage
\sheadB{Perturbation theory / Wigner}

\sheadC{The overlap between the old and the new states} 

We have found that the perturbed eigenstates 
to first-order are given by the expression
\beq
|E_n\rangle 
\ \ \approx \ \ |\varepsilon_n\rangle 
+\sum_m \frac{V_{m,n}}{\varepsilon_{n}-\varepsilon_m} |\varepsilon_m\rangle 
\eeq
So, it is possible to write:
\beq
\langle \varepsilon_m  | E_n\rangle 
\ \ \approx \ \ \frac{V_{mn}}{\varepsilon_{n}-\varepsilon_m} 
\ \ \ \ \ \ \mbox{for $m \ne n$}
\eeq
which implies 
\beq
P(m|n) \ \ \equiv \ \ 
|\langle \varepsilon_m | E_n \rangle|^2 
\ \ \approx \ \ \frac{|V_{mn}|^2}{(E_{m}-E_n)^2} 
\ \ \ \ \ \ \mbox{for $m\ne n$}
\eeq
In the latter expression we have replaced in the denominator
the unperturbed energies by the perturbed energies. 
This is OK in leading order treatment. 
The validity condition is  
\beq
|V| \ \ \ll \ \ \Delta 
\eeq
In other words, the perturbation must be 
much smaller than the mean level spacing. 
We observe that once this condition breaks down
the sum $\sum_m P(m|n)$ becomes much 
larger than one, whereas the exact value 
should have unit normalization. 
This means that if $|V| \gg \Delta$ the above 
first order expression cannot be trusted.

Can we do better? In principle we have to go to 
higher orders of perturbation theory, which might 
be very complicated. But in fact the generic result 
that comes out is quite simple: 
\beq
P(m|n) \ \ \approx \ \
\frac{|V_{m,n}|^2}{(E_m-E_n)^2 + (\Gamma/2)^2 } 
\eeq
This is called "Wigner Lorentzian". 
As we shall see later it is related to an exponential 
decay law that is called "Wigner decay".
The expression for the "width" of this Lorentzian 
is implied by normalization:
\beq
\Gamma \ \ = \ \ \frac{2\pi}{\Delta} |V|^2 
\eeq
The Lorentzian expression is not exact. It is implicit 
that we assume a dense spectrum (high density of states).  
We also assume that all the matrix elements are of the same 
order of magnitude. Such assumption can be 
justified e.g. in the case of a chaotic system. 
In order to show that $\sum_m P(m|n) = \sum_n P(m|n) = 1$ one use the recipe:
\beq
\sum_{n} f(E_n) \ \ \approx \ \ \int \frac{dE}{\Delta} f(E) 
\eeq
where $\Delta$ is the mean level spacing.
In the following we shall discuss further 
the notion Density of States (DOS) and 
Local  Density of States (LDOS) which are 
helpful in further clarifying the significance 
of the Wigner Lorentzian.

\sheadC{The DOS and the LDOS} 

When we have a dense spectrum, we can characterize 
it with a density of states (DOS) function:
\beq
\gdos(E) \ \ = \ \ \sum_n \delta(E-E_n) 
\eeq
We notice that according to this definition:
\beq
\int^{E+dE}_E \gdos(E')dE' 
\ \ = \ \ \mbox{number of states with energy}\,\, E < E_n < E+dE 
\eeq
If the mean level spacing ${\Delta}$ is approximately 
constant within some energy interval then ${\gdos(E)=1/\Delta}$.

The local density of states (LDOS) is a weighted version 
of the DOS. Each level has a weight which is proportional 
to its overlap with a reference state:  
\beq
\rho(E) \ \ = \ \ \sum_{n} |\langle \Psi | n \rangle|^2  \delta(E-E_n) 
\eeq
The index $n$ labels as before the eigenstates 
of the Hamiltonian, while $\Psi$ is the reference state.
In particular $\Psi$ can be an eigenstates $|\varepsilon_0 \rangle$ 
of the unperturbed Hamiltonian. In such case 
the Wigner Lorentzian approximation implies
\beq
\rho(E) \ \ = \ \ \frac{1}{\pi}\frac{(\Gamma/2)}{ (E-\varepsilon_0)^2 + (\Gamma/2)^2} 
\eeq
It should be clear that by definition we have 
\beq
\int_{-\infty}^{\infty} \rho(E) dE \ \ = \ \ \sum_{n} |\langle \Psi | n \rangle|^2  \ \ = \ \ 1 
\eeq

\begin{center}
\putgraphr[0.4\hsize]{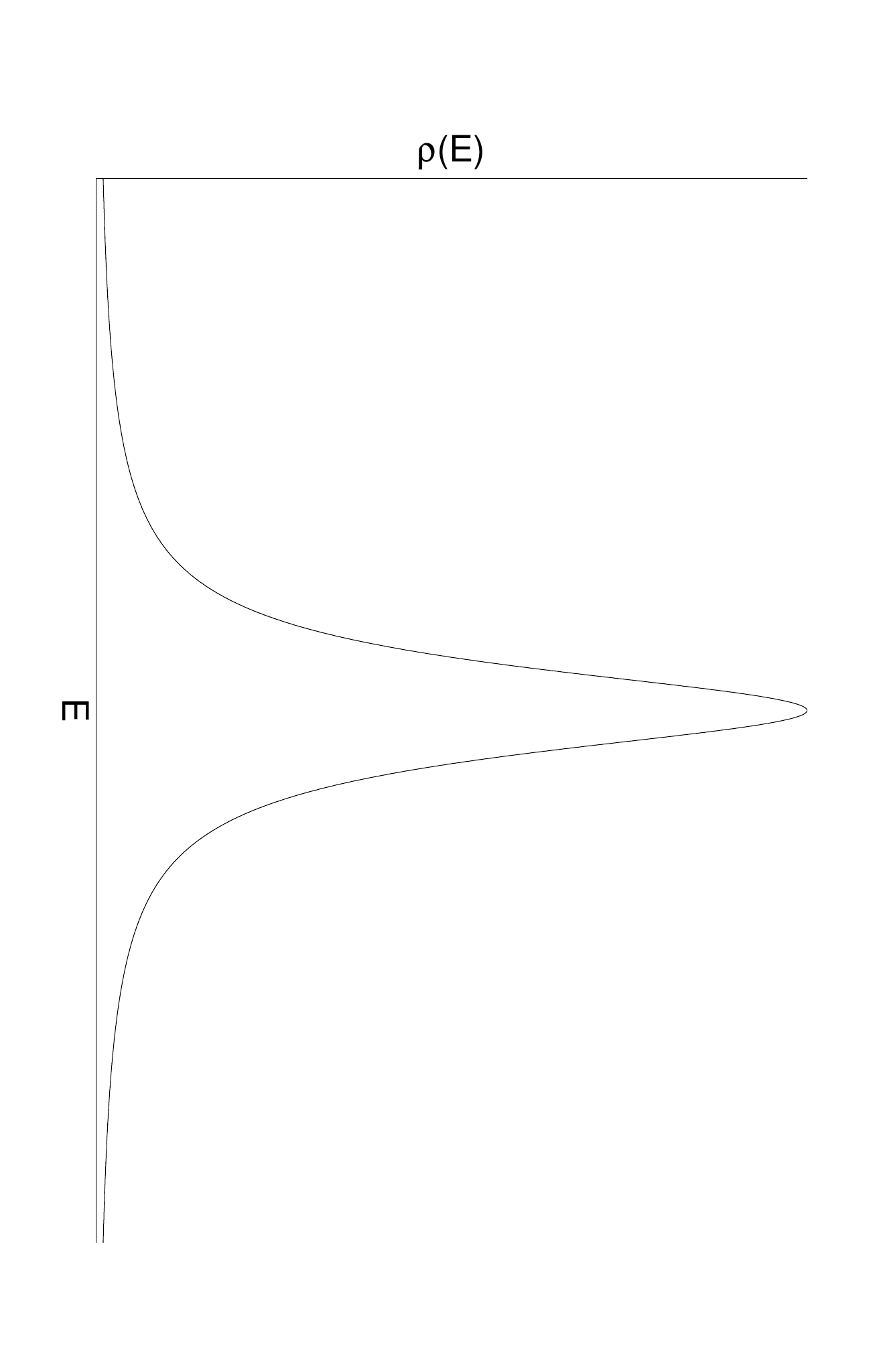} 
\end{center}

We have defined $\rho(E)$ such that it is normalized 
with respect to the measure $dE$. In the next section we 
use it inside a Fourier integral, and use $\omega$ instead
of~$E$. We note that sometimes $\rho(\omega)$ is conveniently re-defined 
such that it is normalized with respect to the measure $d\omega/(2\pi)$. 
We recall the common convention which is used 
for time-frequency Fourier transform in this course:
\beq
F(\omega) 
&=& \int f(t) \eexp{i \omega t}dt
\\ \nonumber
f(t) 
&=& \int \frac{d\omega}{2\pi} F(\omega) \eexp{-i \omega t} 
\eeq

\sheadC{Wigner decay and its connection to the LDOS}

Let us assume that we have a system with many 
energy states. We prepare the system in the 
state ${|\Psi\rangle}$. Now we apply a field 
for a certain amount of time, and then turn it off. 
What is the probability ${P(t)}$ that the system 
will remain in the same state? This probability 
is called the survival probability. By definition:
\beq
P(t) \ \ = \ \ |\langle \Psi(0) |\Psi(t) \rangle|^2 
\eeq
Let ${\mathcal{H}_0}$ be the unperturbed Hamiltonian, 
while ${\mathcal{H}}$ is the perturbed Hamiltonian (while the field is "on"). 
In what follows the index $n$ labels the eigenstates 
of the perturbed Hamiltonian ${\mathcal{H}}$. 
We would like to calculate the survival amplitude:
\beq
\langle \Psi(0) | \Psi(t) \rangle 
\ \ = \ \ \langle \Psi| U(t) | \Psi \rangle 
\ \ = \ \ \sum_n \langle n | \Psi \rangle|^2 \eexp{-iE_nt} 
\eeq
We notice that:
\beq
\langle \Psi(0)|\Psi(t) \rangle 
\ \ = \ \ \mbox{FT}\left[ \sum_n \langle n | \Psi \rangle|^2 2\pi\delta(\omega-E_n) \right] 
\ \ = \ \ \mbox{FT} \left[ 2\pi \rho(\omega) \right] 
\eeq
If we assume that the LDOS is given by Wigner Lorentzian then:
\beq
P(t) 
\ \ = \ \ \Big| \mbox{FT} \left[ 2\pi\rho(E) \right] \Big|^2 
\ \ = \ \ \eexp{-\Gamma t} 
\eeq
The Wigner decay appears when we "break" first-order 
perturbation theory. The perturbation should be strong 
enough to create transitions to other levels. 
Else the system stays essentially at the same 
level all the time (${P(t) \approx 1}$).

\newpage
\sheadB{Decay into a continuum}

\sheadC{Definition of the model} 

In the problem of a particle in a two-site system, we saw that 
the particle oscillates between the two sites. We now turn 
to solve a more complicated problem, where there is one site 
on one side of the barrier, and on the other side there is 
a very large number of energy levels (a "continuum").

\begin{center}
\putgraph[0.4\hsize]{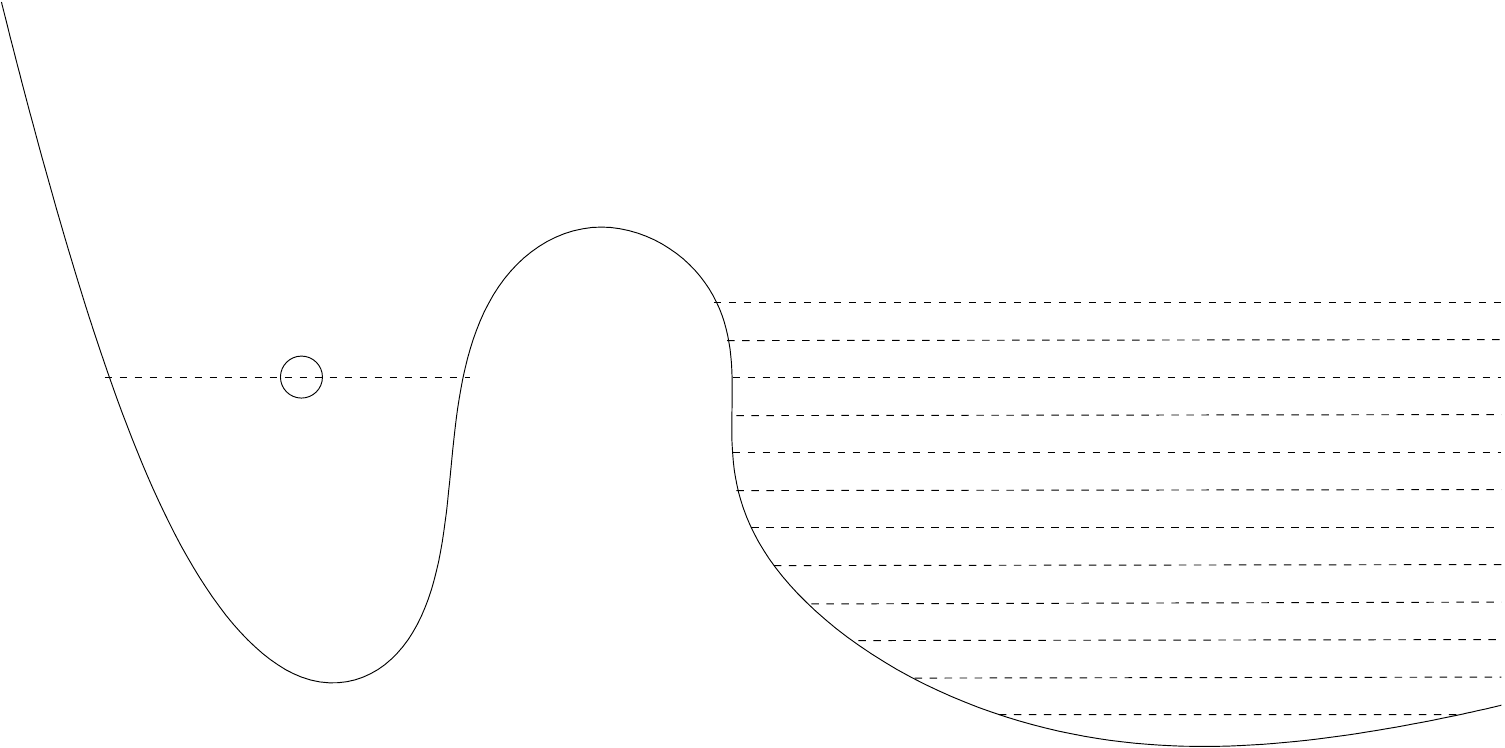}
\end{center}

We shall find that the particle decays into the continuum. 
In the two site problem, the Hamiltonian was:
\beq
\mathcal{H} \ \ = \ \ \left(\amatrix{\epsilon_0 &\sigma \cr \sigma& \epsilon_1}\right) 
\eeq
where ${\sigma}$ is the transition amplitude through the barrier. 
In the new problem the Hamiltonian is:
\beq
\mathcal{H} 
\ \ = \ \ 
\left( 
\amatrix{ 
\epsilon_0 & \sigma_1 & \sigma_2 & \sigma_3 & . & . & . \cr 
\sigma_1 & \epsilon_1 & 0 & 0 & . & . & . \cr 
\sigma_2 & 0 & \epsilon_2 & 0 & . & . & . \cr 
\sigma_3 & 0 & 0 & \epsilon_3 & . & . & . \cr . & . & . & . & . & . & . \cr
. & . & . & . & . & . & . } 
\right) 
\ \ = \ \ \mathcal{H}_0 + V 
\eeq
where the perturbation term ${V}$ includes the coupling elements ${\sigma_k}$. 
Without loss of generality we use gauge such that $\sigma_k$ are real numbers.
We assume that the mean level spacing between the continuum states is ${\Delta}$. 
If the continuum states are of a one-dimensional box with length ${L}$, 
the quantization of the wavenumber is ${{\pi}/{L}}$, 
and from the dispersion relation ${dE = v_E dp}$ we get:
\beq
\Delta \ \ = \ \ v_E \frac{ \pi}{L} 
\eeq
From the Fermi golden rule (FGR) we expect a decay constant 
\beq
\Gamma \ [\mbox{FGR}] \ \ = \ \ 2\pi \frac{1}{\Delta} \sigma^2 
\eeq
Below we shall see that this result is exact. 
We are going to solve both the 
eigenstate equation ${\mathcal{H}\Psi = E \Psi}$, 
and the time dependent Schr\"{o}dinger's equation.
Later we are going to derive the Gamow formula 
\beq
\Gamma \ [\mbox{Gamow}] \ \ = \ \ \mbox{AttemptFrequency} \times \mbox{BarrierTransmission}
\eeq
By comparing with the FGR expression we shall deduce what 
is the coupling $\sigma$ between states that touch  
each other at the barrier. We shall use this result in 
order to get an expression for the Rabi frequency $\Omega$ 
of oscillations in a double well system.

\sheadC{An exact solution - the eigenstates} 

The unperturbed basis is $|0\rangle,|k\rangle$ 
with energies $\epsilon_0$ and $\epsilon_k$. The level spacing 
of the quasi continuum states is $\Delta$.    
The couplings between the discrete state 
and the quasi-continuum levels are $\sigma_k$.  
The set of equations for the eigenstates $|E_n\rangle$ is  
\beq
\epsilon_0 \Psi_0 + \sum_{k'} \sigma_{k'} \Psi_{k'} &=& E\Psi_0 \\
\epsilon_k \Psi_k + \sigma_k \Psi_0 &=& E\Psi_k, \ \ \ \ k=1,2,3,...
\eeq
From the $k$th equation we deduce that 
\beq
\Psi_k  \ \ = \ \ \frac{\sigma_k}{E-\epsilon_k}\Psi_0
\eeq
Hence the expression for the eigenstates is 
\beq
|\Psi\rangle    
\ \ = \ \ 
\sqrt{p} \, |0\rangle 
\  + \ 
\sqrt{p} \sum_{k} \frac{\sigma_k}{E-\epsilon_k} \, |k\rangle 
\eeq
where $p \equiv |\Psi_0|^2$ is determined by normalization.
Substitution of $\Psi_k$ into the 0th equation leads 
to the secular equation for the eigenenergies
\beq
\sum_k \frac{\sigma_k^2}{E-\epsilon_k} \ \ = \ \  E-\epsilon_0
\eeq
This equation can be illustrated graphically.Clearly the roots $E_n$
interlace the the unperturbed values $\epsilon_k$.    
For equal spacing $\Delta$ and equal couplings $\sigma$ 
the secular equation can be written as 
\beq
\cot\left( \pi \frac{E}{\Delta} \right) 
\ = \ \frac{1}{\pi} \frac{\Delta}{\sigma^2} (E-\epsilon_0)
\hspace{2cm} \leadsto \hspace{2cm} E_n = \left(n+\frac{1}{\pi}\varphi_n\right) \Delta
\eeq
where $\varphi$ changes monotonically from $\pi$ to $0$. 
Above and below we use the following identities:
\beq
\sum_{n=-\infty}^{\infty}\frac{1}{x-\pi n} = \cot(x) \equiv S, 
\hspace{2cm} \sum_{n=-\infty}^{\infty}\frac{1}{(x-\pi n)^2} =\frac{1}{\sin^2(x)} = 1 + S^2  
\eeq
Using the second identity and the secular equation 
one obtains a compact expression for the normalization constant 
\beq
p \ \ = \ \ |\Psi_0|^2  
\ \ = \ \ |\langle 0 | E_n \rangle|^2  
\ \ = \ \ \frac{\sigma^2}{(E_n-\epsilon_0)^2 + (\Gamma/2)^2}
\eeq
where 
\beq
\frac{\Gamma}{2} \ \ = \ \ \sqrt{ \sigma^2 + \left(\frac{\pi}{\Delta}\sigma^2 \right)^2 }
\eeq
The plot of $|\Psi_0|^2$ versus $E_n$ is the Wigner Lorentzian. 
It gives the overlap of the eigenstates with the unperturbed discrete state.

\sheadC{An exact solution - the time dependent decay} 

We switch to the interaction picture:
\beq
\Psi_k(t) \ \ = \ \ c_k(t) \eexp{-i\epsilon_k t}, 
\ \ \ \ \ \ \ \ k=0,1,2,3,...
\eeq
We distinguish between ${c_k}$ and ${c_0}$. 
From here on the index runs over the values ${k=1,2,3,...}$, 
and we use the notation ${V_{k,0}=\sigma_k}$ for the couplings.   
We get the system of equations:
\beq
&& i\frac{dc_0}{dt}  \ \ = \ \  \sum_k \eexp{i(\epsilon_0-\epsilon_k)t} \, V_{0,k} \, c_k(t)
\\ \nonumber
&& i\frac{dc_k}{dt}  \ \ = \ \  \eexp{i(\epsilon_k-\epsilon_0)t} \, V_{k,0} \, c_0(t), \ \ \ \ \ \ \ \ k=1,2,3,...
\eeq
From the second equation we get:
\beq
c_k(t)  \ \ = \ \  0 \ -i \, \int_0^t \eexp{i(\epsilon_k-\epsilon_0)t'} \, V_{k,0} \, c_0(t') \, dt' 
\eeq
By substituting into the first equation we get:
\beq
\frac{dc_0}{dt}  \ \ = \ \  -\int_0^t C(t-t') \, c_0(t') \, dt' 
\eeq
where
\beq
C(t-t')  \ \ = \ \  \sum_k |V_{k,0}|^2 \, \eexp{-i(\epsilon_k-\epsilon_0)(t-t')} 
\eeq
The Fourier transform of this function is:
\beq
\tilde{C}(\omega) 
 \ \ = \ \  
\sum_k |V_{k,0}|^2 \, 2\pi \delta(\omega-(\epsilon_k-\epsilon_0)) 
 \ \  \approx  \ \  
\frac{2\pi}{\Delta} \sigma^2 
\eeq
Accordingly 
\beq
C(t-t')  \ \  \approx  \ \  
\left[ \frac{2\pi}{\Delta} \sigma^2 \right] \delta(t-t') 
 \ \ = \ \  \Gamma \delta(t-t') 
\eeq
We notice that the time integration only "catches" half 
of the area of this function. Therefore, the equation for ${c_0}$ is:
\beq
\frac{dc_0}{dt}  \ \ = \ \  -\frac{\Gamma}{2}c_0(t) 
\eeq
This leads us to the solution:
\beq
P(t)  \ \ = \ \  |c_0(t)|^2  \ \ = \ \  \eexp{-\Gamma t} 
\eeq

\newpage
\sheadC{An exact solution - the rezolvent and its pole} 

Using the "P+Q" formalism we can eliminate the
continuum and find an explicit expression for 
the rezolvent of the decaying state: 
\beq
G(z)  
\ \ = \ \ \left[ \frac{1}{z-\mathcal{H}} \right]_{0,0} 
\ \ = \ \ \frac{1}{z \ - \ \epsilon_0 \ - \ g(z)}
\eeq
where
\beq
g(z) \ \ = \ \ \sum_k \frac{|V_{k,0}|^2}{z-\epsilon_k} 
\eeq
The equation for the pole of the rezolvent 
takes the from ${z-\epsilon_0 = g(z)}$. 
This equation has a simple visualization.
In the upper complex-plane perspective (${z>0}$)   
one can regard $g(z)$ as an "electric field" 
that originates from an "image charge" at ${z=-i\infty}$. 
This electric field pushes the zero at $z =\epsilon_0$ into 
the lower plane a distance $\Gamma_0$. 
If the weighted density of $\epsilon_k$ is non-uniform, 
it is like having a tilted "electric field", 
hence there is an additional shift $\Delta_0$ 
in the location of the zero. This correction, that arises 
due to the interaction of the discrete level with the continuum,
is known as the Lamb shift. Accordingly, 
for the pole that appears in the analytic continuation 
of $G(z)$ from the upper sheet we write  
\beq
z_{\text{pole}} \ \ = \ \ \epsilon_0 \ + \ \Delta_0 \ - i \frac{\Gamma_0}{2}  
\eeq
Using leading order perturbation theory the expressions for 
the decay rate $\Gamma_0$ and for the Lamb shift $\Delta_0$ are:
\beq
\Gamma_0 \ &=& \ \sum_k 2\pi \delta(\epsilon_0-\epsilon_k) \ |V_{k,0}|^2 
\\
\Delta_0 \ &=& \ \sum_k \frac{|V_{k,0}|^2}{\epsilon_0-\epsilon_k} 
\eeq
The solution above is appealing, but still it does not illuminate 
the relation to the characteristics of the barrier. 
We therefore turn in the next section to consider a less artificial example.

\sheadC{Related matrix models} 

We have discussed a $2\times 2$ matrix model for Rabi oscillations, 
for which we found 
\beq
\Omega \ \ = \ \ \sqrt { |2V_{R,L}|^2 + (\epsilon_L-\epsilon_R)^2 } 
\eeq
where $V_{\text{R,L}}$ is the coupling between the "left" and the "right" eigenstates.   
Above we have discussed the decay of a discrete "left" state 
into a continuum of "right" states, getting
\beq
\Gamma  \ \ = \ \ 2\pi\varrho_R \, |V_{R,L}|^2
\eeq
where $\varrho_R$ is the density of states. 
It is also possible to consider the scattering problem,  
involving a junction that has two attached leads of length~$L$. 
The transmission can be calculated using the $T$-matrix 
formalism. In leading order the result is 
\beq
g \ \ = \ \  |T_{L,R}|^2 \ \ \approx \ \ (2\pi\varrho_L) (2\pi\varrho_R) |V_{R,L}|^2 
\eeq
In order to obtain this result notice that 
the matrix elements of the $T$ matrix should be 
taken with flux-normalized states ${(2/L)^{1/2}\sin(kx)}$. 
while the matrix elements of $V$ are defined  
with standard normalized states ${ (2/v_E^{1/2}) \sin(kx)}$. 
Note also that within a lead the density 
of states is ${\varrho=L/(\pi v_E)}$, where $v_E$ 
is the velocity at the energy of interest.    
If $g$ is large, one should include higher orders in the $T$ matrix calculation. 
Accordingly the result depends on the details of the junction. 
For a simple point-contact junction the result 
of the geometric summation implies ${g \mapsto g/(1+g)}$, 
as expected from the delta-barrier expression.

It is illuminating to realize that $(2\pi\varrho)^{-1}$ 
can be interpreted as the attempt frequency: 
the number of collisions with the barrier per units time. 
For a lead of length $L$ it equals ${v_E/2L}$. 
By the above analysis one can deduce that the decay constant 
can be calculated using the Gamow formula:
\beq
\Gamma \ \ = \ \ \frac{1}{2\pi\varrho_R} \ g  \ \ = \ \ \frac{v_E}{2a} \ g 
\eeq   
where $a$ is the length of the "left" lead, 
and $v_E/(2a)$ is the attempt frequency. 
In fact this semi-classically expected result 
is quite general as discussed in later sections.

Consider a 1D Hamiltonian with a potential $V(x)$ that 
describes free "left" and "right" regions that are 
separated by a "barrier". 
In practice one one would like to be able to represent 
it by a matrix-type model that involves a "junction":
\beq
\mathcal{H} \ \ = \ \ \frac{p^2}{2\mass} + V(x) \ \ \mapsto \ \ \mathcal{H}_L + \mathcal{H}_J +  \mathcal{H}_R
\eeq
Namely, the unperturbed Hamiltonian is the sum of "left" and "right" segments, 
and the perturbation is the coupling at a "junction" that couples the 
two segments. This approach is most popular in STM applications. 
The question arises how exactly to define the 3 terms in the Hamiltonian. 
Apparently we have to selects a good basis that is composed of "left" and "right" eigenstates, 
and figure out how they are coupled at the junction. 
If the a "point contact" junction is modelled as a delta function ${u\delta(x-x_0)}$
it is most natural to define the  "left" and "right" eigenstates 
as ${\varphi(x) = \sin(k(x-x_0))}$, and one can show that the couplings 
at the energy of interest are given by the formula
\beq
V_{R,L} \ \ = \ \ \frac{1}{4\mass^2 u} (\partial \varphi^R)(\partial \varphi^L)
\eeq
where the derivative $\partial$ is taken at the point ${x=x_0}$. 
Bardeen has found that for a wide barrier it is possible to 
use the following approximation:
\beq
V_{R,L} \ \ = \ \ \frac{1}{2\mass} \left[ \varphi^R (\partial \varphi^L) - (\partial \varphi^R) \varphi^L \right]_{x_0} 
\eeq
where ${x=x_0}$ is an arbitrary point within the barrier region. 
Note that the point-contact junction formula can be regarded 
as a limit of the latter. 
It should be clear that the implied matrix representation of the Hamiltonian 
is somewhat problematic. Effectively we consider a smaller Hilbert space, 
from which all the high lying states and the barrier region are truncated.
This is possibly not very important for transmission or decay rate calculation, 
but might be disastrous for Lamb shift calculation. 
We note that within the matrix model analysis the Lamb shift is     
\beq
\Delta \ \ \approx \ \ \sum_k \frac{|V_{k,0}|^2}{\epsilon_0-\epsilon_k} \ \ =  \ \ \text{prefactor} \times \Gamma
\eeq
where the prefactor depends on the cutoff of the $dk$ integration. 
Thus $\Delta$, unlike $\Gamma$ depends not only on the transmission~$g$ 
of the junction but also on the $k$ dependence and on the global variation 
of the density of states.

\newpage
\sheadC{Decay out of a square well}

A particle of mass $\mass$ in 1D is confined from the left 
by an infinite potential wall and from the right 
by a delta barrier $U(x)=u\delta(x-a)$. 
We assume large $u$ such that the two regions 
are weakly coupled. In this section we shall 
derive an expression for the decay constant $\Gamma$. 
We shall prove that it equals the "attempt frequency" 
multiplied by the transmission of the barrier. 
This is called Gamow formula.

\begin{center}
\putgraph[0.45\hsize]{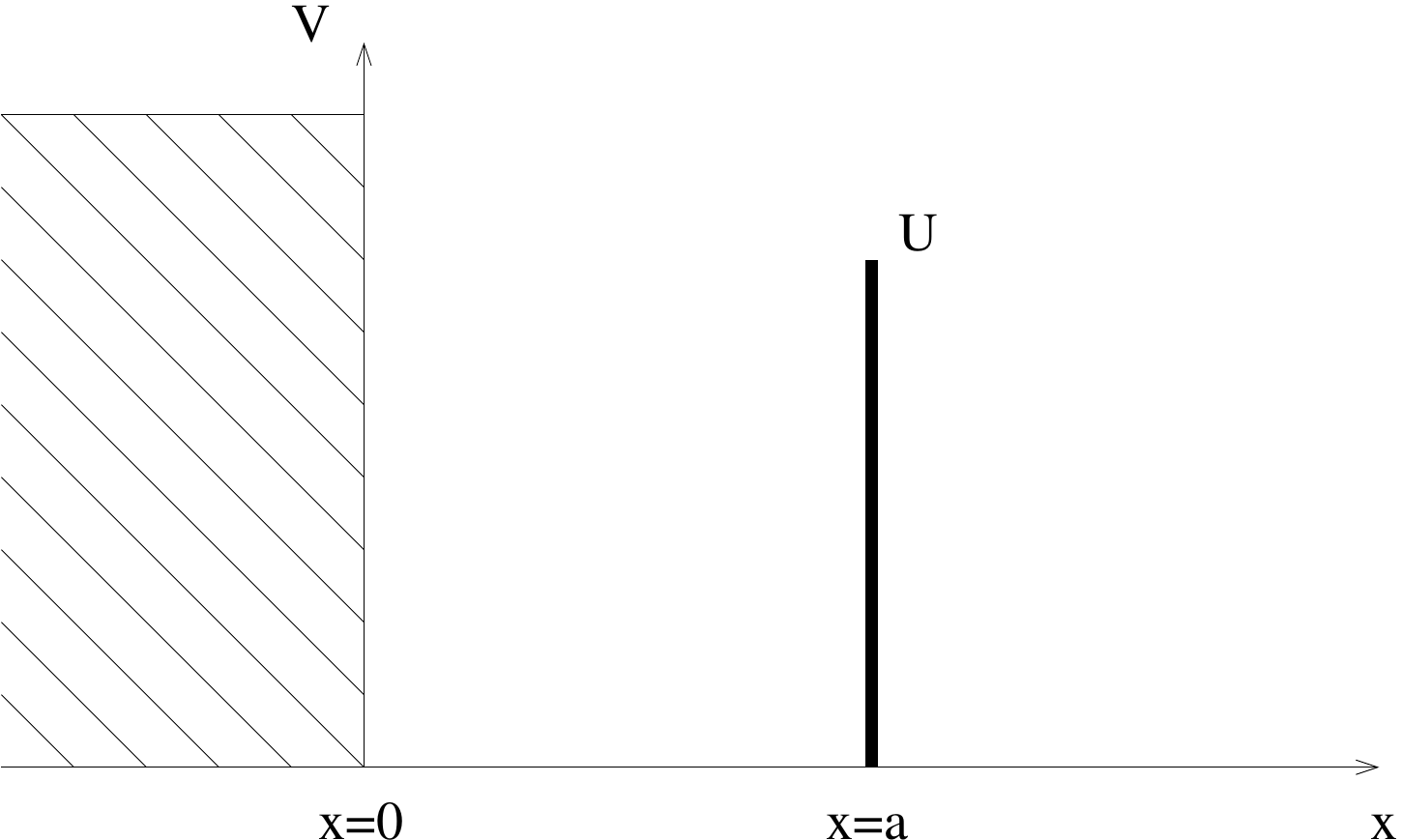}
\end{center}

We look for the complex poles of the rezolvent. 
We therefore look for stationary solutions 
of the equation ${\mathcal{H}\psi = E\psi}$  
that satisfy "outgoing wave" boundary conditions.
The wavenumber of the outgoing wave is written as  
\beq
k \ \ = \ \ k_n - i \gamma_n 
\eeq
which implies complex energies 
\beq
E \ &=& \ \frac{k^{2}}{2\mass} \ \ = \ \ E_n -i \frac{\Gamma_n}{2}, \quad \quad n=0,1,2,3...
\\
E_n \ &=& \ \frac{1}{2\mass}(k^{2}_n - \gamma_n^{2} ) 
\\
\Gamma_n \ &=& \ \frac{2}{\mass}k_n\gamma_n \ \ \equiv \ \  2 v_n \gamma_n
\eeq
Within the well the most general stationary solution 
is ${\psi(x)=A\eexp{ikx} + B\eexp{-ikx}}$. 
Taking into account the boundary condition $\psi(0)=0$ at the hard wall, 
and the outgoing wave boundary condition at infinity we write 
the wavefunction as 
\beq
\psi(x) \ &=& \ C\sin(kx) 
\ \ \ \ \ \ \ \mbox{for $0<x<a$}
\\ \nonumber
\psi(x) \ &=& \ D\eexp{ikx}
\ \ \ \ \ \ \ \ \ \ \ \ \mbox{for $x>a$}
\eeq
The matching conditions across the delta barrier are:
\beq
\psi(a+0)-\psi(a-0) \ &=& \ 0 
\\ 
\psi'(a+0)-\psi'(a-0) \ &=& \ 2\mass u  \ \psi(a)
\eeq
Thus at $x=a$ the logarithmic derivative should have a jump:  
\beq
\left.\frac{\psi'}{\psi}\right|_{+} \ - \ \left. \frac{\psi'}{\psi}\right|_{-} \ \ = \ \ 2\mass u
\eeq
leading to the equation 
\beq
ik - k\cot(ka) \ \ = \ \ 2\mass u
\eeq
We can write the last equation as:
\beq
\tan(ka) \ = \ -\frac{\frac{k}{2\mass u}}{1-i\frac{k}{2\mass u}}
\eeq
The zero order solution in the coupling ($u=\infty$) 
are the energies of an isolated well corresponding to 
\beq
k_n^{(0)} \ \ = \ \ \frac{\pi}{a} n 
\ \ \ \ \ \ \ \ \ \ \ \mbox{[zero order solution]}
\eeq
We assume small $k_n/(2\mass u)$, 
and expand both sides of the equation 
around $k_n$. Namely we set ${k = (k_n + \delta k) -i \gamma}$ 
where $\delta k$ and $\gamma$ are small corrections to the unperturbed 
energy of the isolated state. To leading order the equation takes the form
\beq
a \delta k  - i a \gamma \ \ = \ \ -\frac{k_n}{2\alpha}-i\left(\frac{k_n}{2\alpha}\right)^2
\eeq
Hence we get in leading order 
\beq
k_n \ &=& \   k_n^{(0)} - \frac{1}{a}\left(\frac{k_n^{(0)}}{2\mass u}\right)
\\ 
\gamma_n \ &=& \ \frac{1}{a}\left(\frac{k_n^{(0)}}{2\mass u}\right)^2
\eeq
From here we can calculate both 
the shift and the "width" of the energy.
To write the result in a more attractive way we recall  
that the transmission of the delta barrier at 
the energy $E=E_n$ is 
\beq
g  \ \ = \ \  \frac{1}{1+(u/v_n)^2} 
 \ \ \approx \ \  
\left(\frac{v_n}{u}\right)^2
\eeq
hence 
\beq
\Gamma_n \ \ = \ \ 2v_n\gamma_n 
 \ \  \approx  \ \  \frac{v_n}{2a} \ g
\eeq
This is called Gamow formula. 
It reflects the following semiclassical picture: 
The particle oscillates with 
velocity $v_n$ inside the well, 
hence $v_n/(2a)$ is the number 
of collisions that it has with the barrier 
per unit time. The Gamow formula 
expresses the decay rate as a product of 
this``attempt frequency" with the transmission 
of the barrier. 
It is easy to show that the assumption of 
weak coupling can be written as $g \ll 1$.

\newpage
\sheadC{The Gamow Formula}

We consider a particle in a well of width $a$ that can decay 
to the continuum through a general barrier that has transmission $g$ 
and reflection amplitude 
\beq
\mbox{reflection amplitude} \ \ = \ \ -\sqrt{1-g} \ \eexp{i \theta_0}
\eeq
where both $g$ and the phase shift $\theta_0$ can depend on energy.
We would like to derive the Gamow Formula in this more general setup.
Our starting point as before is the zero order solution 
of an isolated square well ($g=0$) for which the unperturbed eigenstates 
are $\psi(x)=\sin(k_n x)$ with 
\beq
k_n^{(0)} \ \ = \ \ \left[n-\frac{\theta_0}{2\pi}\right]\frac{\pi}{a}
\ \ \ \ \ \ \ \ \ \ \ \mbox{[zero order solution]}
\eeq
But for finite barrier ($g>0$) the poles of the rezolvent 
become complex. 
The equation that determines these poles is obtained 
by matching of the inside solution ${\exp(ikx)-\exp(-ikx)}$ 
with the barrier at ${x=a}$. Namely, 
the  reflected amplitude $B= -\eexp{-ika}$ 
should match the incident amplitude $A=\eexp{ika}$ as follows:  
\beq
B \ \ = \ \  -\sqrt{1-g} \, \eexp{i \theta_0}   \ A
\eeq
This leads to the equation
\beq
\exp\left[-i (2ka + \theta_0) \right] \ \ = \ \ \sqrt{1-g}
\eeq
Assuming that the real part $k_n$ of the solution 
is known we solve for $\gamma_n$ which is assumed 
to be small. In leading order the solution is  
\beq
\gamma_n \ \ = \ \ \frac{1}{4a_n}\ln\left[ \frac{1}{1-g}\right] 
\eeq
In the latter expression we have taken into account
that the phase shift might depend on energy, 
defining the effective width of the well as  
\beq
a_n \ \ = \ \ a \ + \ \frac{1}{2}\frac{d}{dk}\theta_0(k)
\eeq
From here we get
\beq
\Gamma_n \ \ \approx \ \ \frac{v_n}{2 a_n} \ g 
\eeq
This is the Gamow formula, and it is in agreement with 
the semi-classical expectation.
Namely, the interpretation of the prefactor as 
the attempt frequency is consistent with the definition 
of the Wigner delay time: the period 
of the oscillations within the well is 
\beq
\text{TimePeriod} 
\ \ = \ \  \frac{2a}{v_n} + \frac{d}{dE}\theta_0(E) 
\ \ = \ \ \frac{2a_n}{v_n}
\eeq

\newpage \sheadC{From Gamow to the double well problem}

Assume a double well which is divided by the same delta 
function as in the Gamow decay problem. Let us use the solution 
of the Gamow decay problem in order to deduce 
the oscillation frequency in the double well.

\begin{center}
\putgraph[0.45\hsize]{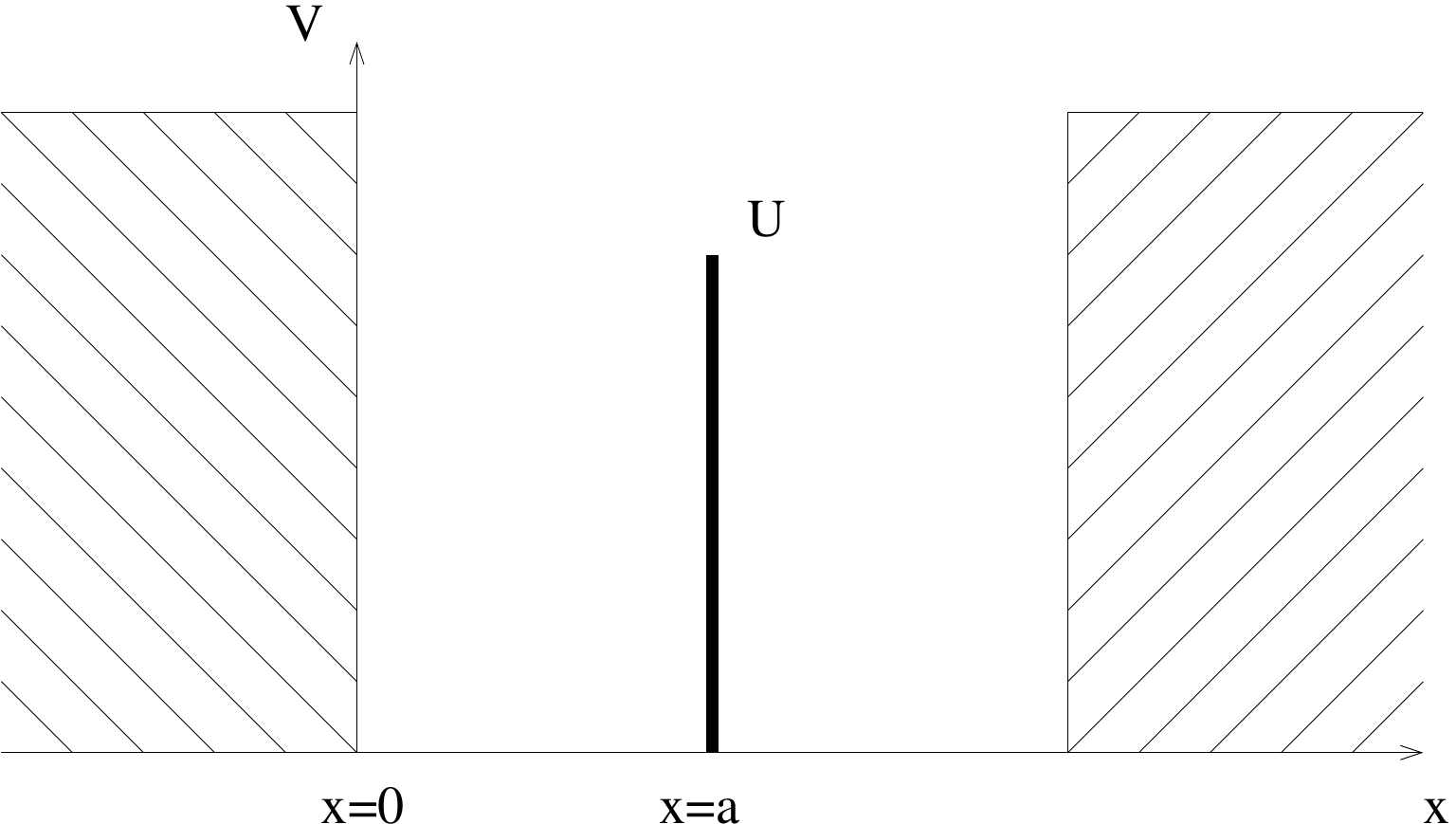}
\end{center}

Re-considering the Gamow problem we assume 
that the region outside of the well is very large, 
namely it has some length $L$ much larger than $a$. 
The $k$ states form a quasi-continuum 
with mean level spacing $\Delta_L$.
By Fermi golden rule the decay rate is  
\beq
\Gamma \ \ = \ \ \frac{2\pi}{\Delta_L} \, |V_{nk}|^2 
\ \ = \ \ \frac{2L}{v_E} \, |V_{nk}|^2
\eeq
where $V_{nk}$ is the probability amplitude 
per unit time to make a transitions 
form level $n$ inside the well to 
any of the $k$ states outside of the well.
This expression should be compared with 
the Gamow formula, which we write as  
\beq
\Gamma \ \ = \ \ \frac{v_E}{2a} \, g 
\eeq
where $g$ is the transmission of the barrier, 
The Gamow formula should agree with  
the Fermi golden rule. Hence we deduce 
that the over-the-barrier coupling is 
\beq
|V_{nk}|^2 \ \ = \ \ \left(\frac{v_E}{2L}\right) \, \left(\frac{v_E}{2a}\right) \, g
\eeq
Once can verify that this is consistent 
with the formula for the coupling between 
two wavefunctions at the point of 
a delta junction [see Section~34]:
\beq
V_{nk}  \ \ = \ \  -\frac{1}{4\mass^2 u} [\partial \psi^{(n)}] [\partial \psi^{(k)}]
\eeq
where $\partial \psi$ is the radial derivative 
at the point of the junction. This formula works 
also if both functions are on the same 
side of the barrier. 

Now we can come back to the double well problem. 
For simplicity assume a symmetric double well. 
In the two level approximation $n$ and $k$ are 
``left" and ``right" states with the same 
unperturbed energy. Due to the coupling 
we have coherent Bloch oscillation whose frequency is   
\beq
\Omega \ \ = \ \ 2|V_{nk}| 
\ \ = \ \ 
\frac{v_E}{a} \, \sqrt{g}
\eeq

\newpage
\sheadB{Scattering resonances}

\sheadC{Fabry Perrot interference / transmission resonance}

The Fabry-Perrot problem is to find the transmission of a double barrier 
given the transmission of each barrier and their "optical distance" $\phi$  
(see definition below). We could assume that the barriers are represented 
by delta functions $u\delta(x\pm(a/2))$. The conventional way of solving 
this problem is to match together the solutions in the three segments.
This procedure is quite lengthy and better to do it with Mathematica.    
The optional (short) way of solving this problem is to "sum over paths", 
similar to the way one solves the interference problem in the two slit geometry.  

\begin{center}
\putgraph[0.4\hsize]{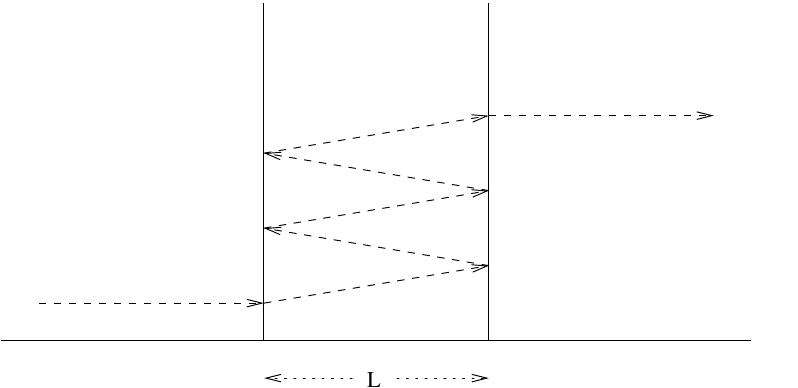} 
\end{center}

We take as given the transmission coefficient ${T=|t|^2}$, 
and the reflection coefficient ${ R=|r|^2=1-T }$. 
If the distance between the two barriers is ${a}$,  
then a wave accumulates a phase ${ka}$ 
when going from the one barrier to the other. 
The transmission of both barriers together is:
\beq
\mbox{transmission}= | t \times \eexp{ika} \times (1+(r\eexp{ika})^2 + (r\eexp{ika})^4 +  \dots  ) \times t |^2 
\eeq
Every round trip between the barriers includes two reflections, 
so the wave accumulates a phase factor ${(\eexp{i\phi})^2}$, where
\beq
\phi \ \ = \ \ ka + \text{phase}(r)
\eeq
We have a geometrical series, and its sum is:
\beq
\mbox{transmission}= \left|t \times \frac{\eexp{ika}} { 1 - \left( |r| \eexp{i\phi} \right)^2 } \times t \right|^2 
\eeq
After some algebra we find the Fabry Perrot expression:
\beq
\mbox{transmission}= \frac{1}{1+4[R/T^2](\sin(\phi))^2} 
\eeq
We notice that this is a very "dramatic" result. 
If we have two barriers that are almost absolutely opaque ${R\sim1}$, 
then as expected for most energies we get a very low transmission 
of the order of magnitude of ${T^2}$. But there are energies 
for which ${\phi=\pi \times \mbox{integer}}$ and then we find 
that the total transmission is $100\%$!
In the following figure we compare the two slit 
interference pattern (left) to the Fabry Perrot result (right):

\begin{center}
\putgraph[0.3\hsize]{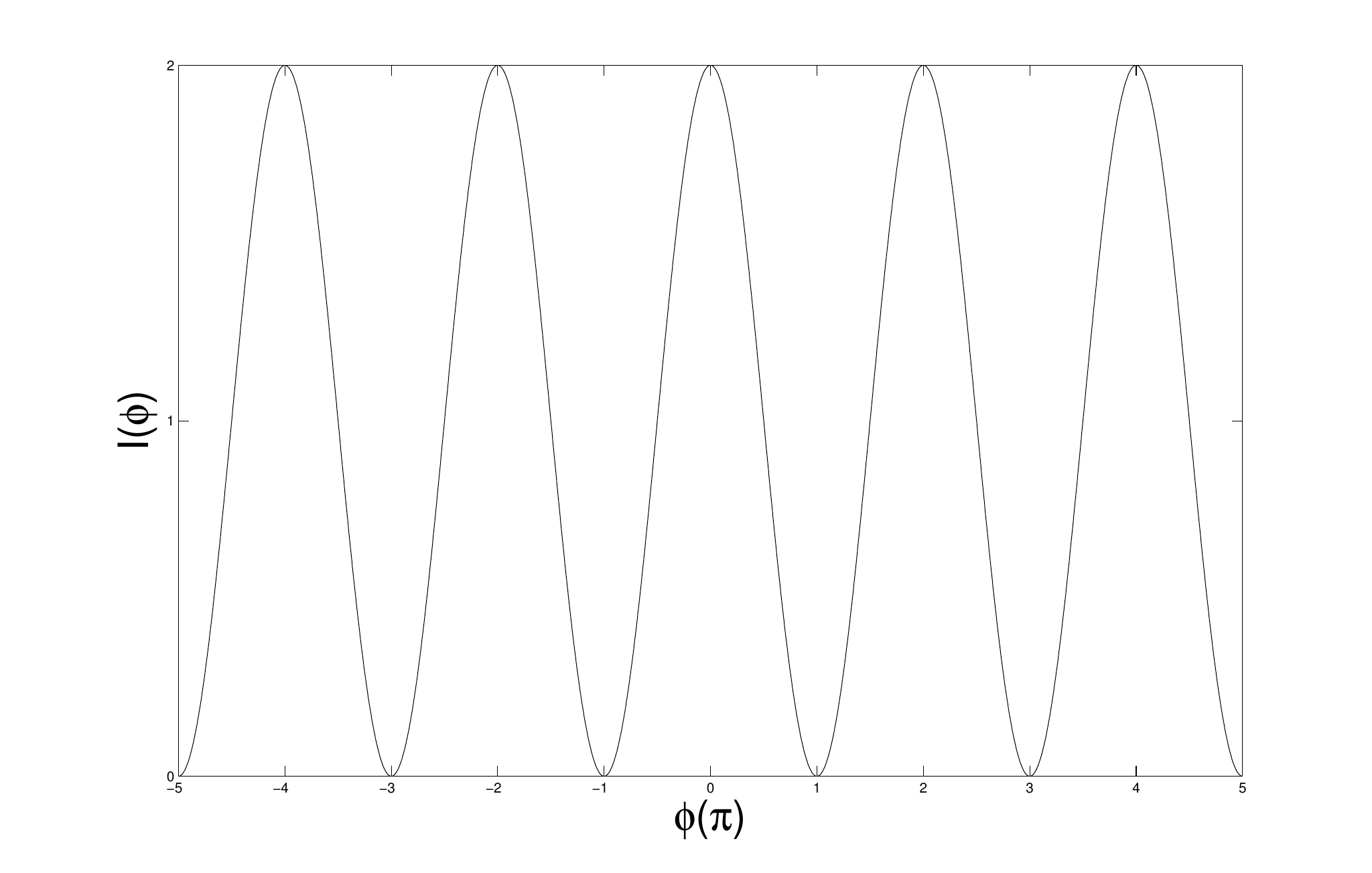}
\ \ \ \ 
\putgraphr[0.3\hsize]{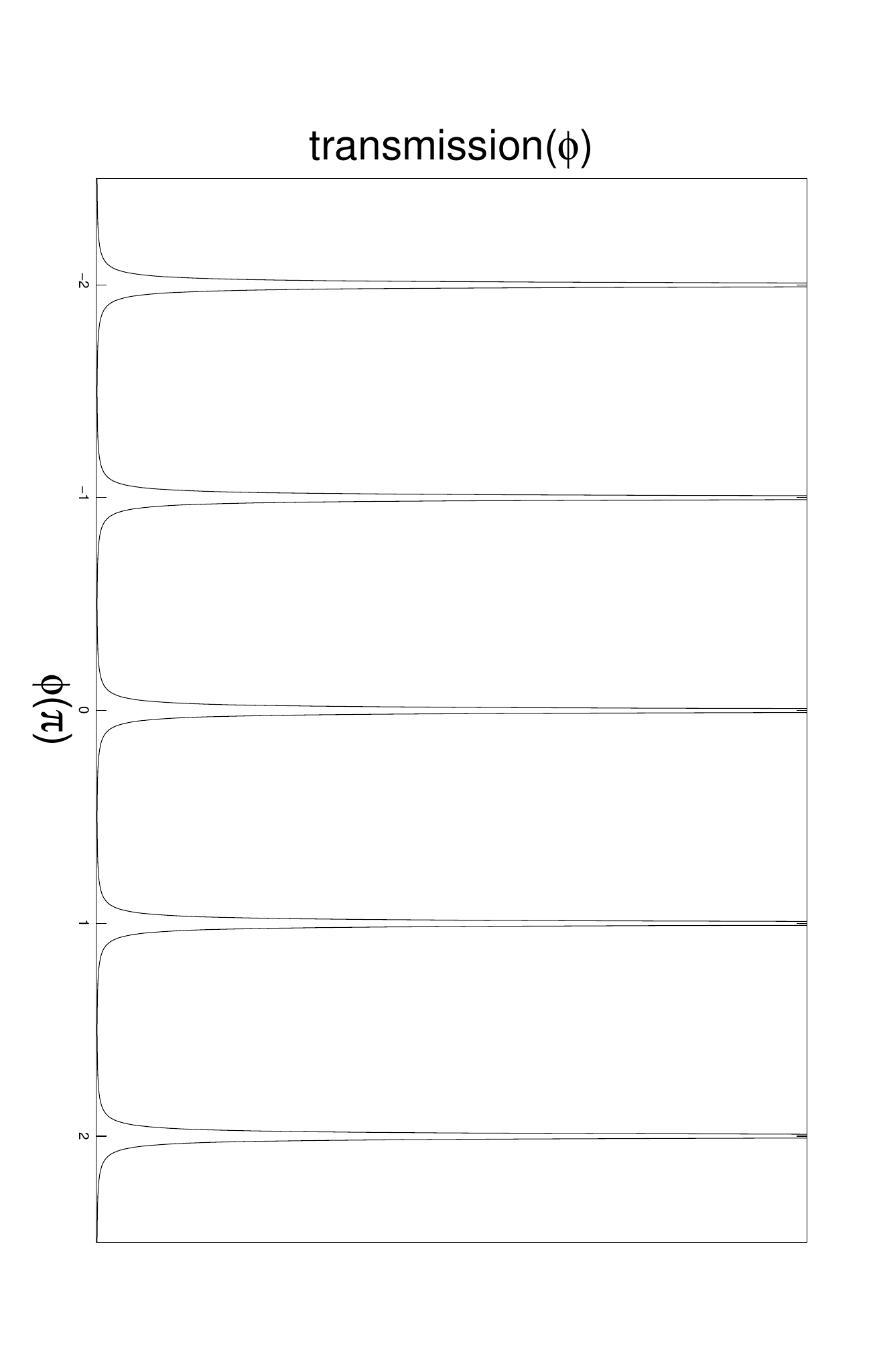} 
\end{center}

\sheadC{Scattering on a single level system}

The most elementary example for a resonance is when 
the scattering region contains only one level. 
This can be regarded as a simplified version of 
the Fabry-Perrot double barrier problem, 
or it can be regarded as simplified version of 
scattering on a shielded well which we discuss 
later on.

We assume, as in the Wigner decay problem that we had been analyzed 
in a previous lecture, that the scattering region contains 
a single level of energy $E_0$. Due to the coupling 
with the continuum states of the lead, 
this level acquire withs $\Gamma_0$. This width would 
determine the Wigner decay rate if initially the particle 
were prepared in this level. But we would like to 
explore a different scenario in which the particle 
is scattered from the outside. Then we shall see 
that $\Gamma_0$ determines the Wigner delay time of the scattering.

In order to perform the analysis using the a formal 
scattering approach we need a better characterization 
of the lead states $|k\rangle$. For simplicity of presentation  
it is best to imagine the leas as a 1D segment of length~$L$  
(later we take~$L$ to be infinite). The density 
of states in the energy range of interest is $L/(\pi v_E)$. 
The coupling between the $E_0$ state and any of the 
volume-normalized $k$ states is ${V_{0,k}=w/\sqrt{L}}$. 
The coupling parameter $w$ can be calculated 
if the form the barrier or its transmission are known
(see appropriate "QM is practice~I" sections).    
Consequently we get the the FGR width of the level $E_0$ 
is  ${\Gamma_0=(2/v_E)w^2}$. Using a formal language 
it means that the resolvent of the $E_0$ subsystem is 
\beq
G(E) \ \ = \ \ \frac{1}{E-E_0+i(\Gamma_0/2)}
\eeq 
In order to find the $S$ matrix, the element(s)  
of the $T$ matrix should be calculated in a 
properly normalized basis. The relation between the 
flux-normalized states and the volume-normalized states is: 
\beq
|\phi^E\rangle \ \ = \ \ \sqrt{\frac{2L}{v_E}} |k\rangle 
\ \ \longmapsto \ \  \frac{1}{\sqrt{v_E}}\eexp{-ikx} - \frac{1}{\sqrt{v_E}}\eexp{+ikx}
\eeq 
Consequently we get for the $1\times1$ scattering matrix the expected result 
\beq
S(E) \ \ = \ \ 1-iT \ \ = \ \ 1-iV_{k_E,0}G(E)V_{0,k_E} \ \ = \ \ 1-i\frac{\Gamma_0}{E-E_0+i(\Gamma_0/2)}
\eeq 
which implies a phase shift
\beq
\delta_{0} \ \ = \ \ \arctan\left[-\frac{\Gamma_0/2}{E-E_0}\right] 
\eeq 
and accordingly the time delay is 
\beq
\tau_0 \ \ = \ \ \frac{\Gamma_0}{(E-E_0)^2+(\Gamma_0/2)^2}
\eeq 
Note that at resonance the time delay is of order $1/\Gamma_0$.

The procedure is easily generalized in order to handle 
several leads, say two leads as in the double barrier problem.
Now we have to use an index ${a=1,2}$ in order to distingush 
the left and right channels. The width of the $E_0$ levels 
is ${\Gamma_0=(2/v_E)[w_1^2+w_2^2]}$. 
The freewaves solution is the leads are labled as $|\phi^{E,a}\rangle$, 
and the $S$ matrix comes out  
\beq
S_{ab}(E) \ \ = \ \ \delta_{a,b}-iT_{ab} \ \ = \ \ 
\delta_{a,b}-i\frac{\Gamma_0/2}{E-E_0+i(\Gamma_0/2)}
\left[\frac{2w_aw_b}{w_1^2+w_2^2}\right]
\eeq 
We see that the maximum transmission is for scattering with ${E=E_0}$,  
namely ${|2w_aw_b/(w_1^2+w_2^2)|^2}$, which becomes $100\%$ transmission  
if ${w_1=w_2}$ as expected from the Fabry-Perrot analysis.

\sheadC{Scattering resonance of a shielded 1D well}

A shielded 1D well is defined by 
\beq
V(r) \ \ = \ \ V \Theta(R-r) + U \delta(r-R) 
\eeq
where $V$ is the potential floor inside 
the scattering region ${0<r<R}$, 
and $U$ is a shielding potential 
barrier at the boundary ${r=R}$ 
of the scattering region. 
We add the shield in order to have 
distinct narrow resonances.   
The interior wave function is 
\beq
\psi(r) \ \ = \ \ \mbox{SIN}(\alpha r)
\eeq
where ${\alpha = \sqrt{2\mass |E-V|}}$, 
and SIN is either $\sin$ or $\sinh$ 
depending on whether $E$ is larger or 
smaller than $V$. 
Taking the logarithmic derivative at ${r=R}$ 
and realizing that the effect of the 
shield $U$ is simply to boost the result we get: 
\beq
\tilde{k}_0(E; V,U) 
\ \ \equiv \ \ \left[\frac{1}{\psi(r)}\frac{d\psi(r)}{dr}\right]_{r=R+0}
\ \ = \ \ \alpha \mbox{CTG}(\alpha R) + 2\mass U
\eeq
where CTG is either $\cot$ or $\coth$ 
depending on the energy $E$. 
It should be realized that $\tilde{k}_0$ 
depends on the energy as well as on $V$ 
and on $U$. For some of the discussions 
below, and also for some experimental application 
it is convenient to regard~$E$ 
as fixed (for example it can be the Fermi energy 
of an electron in a metal), 
while~$V$ is assumed to be controlled 
(say by some gate voltage). 
The dependence of $\tilde{k}_0$     
on~$V$ is illustrated in the following figure.  
At $V=E$ there is a smooth crossover 
from the "cot" region to the "coth" region.
If $V$ is very large then $\tilde{k}_0=\infty$. 
This means that the wavefunction has to 
satisfy Dirichlet (zero) boundary conditions at ${r=R}$. 
This case is called ``hard sphere scattering": 
the particle cannot penetrate the scattering 
region. If $U$ is very large then still for most 
values of $V$ we have ``hard sphere scattering".
But in the latter case there are narrow strips 
where $\tilde{k}_0$ has a wild variation, 
and it can become very small or negative.
This happens whenever the CTG term 
becomes negative and large enough in absolute
value to compensate the positive shield term.   
We refer to these strips as ``resonances". 
The locations ${V \sim V_r}$ of the resonances 
is determined by the 
equation $\tan(\alpha R) \sim 0$. 
We realize that this would be the condition 
for having a bound state inside the scattering 
region if the shield were of infinite height.

\begin{center}
\putgraph[0.4\hsize]{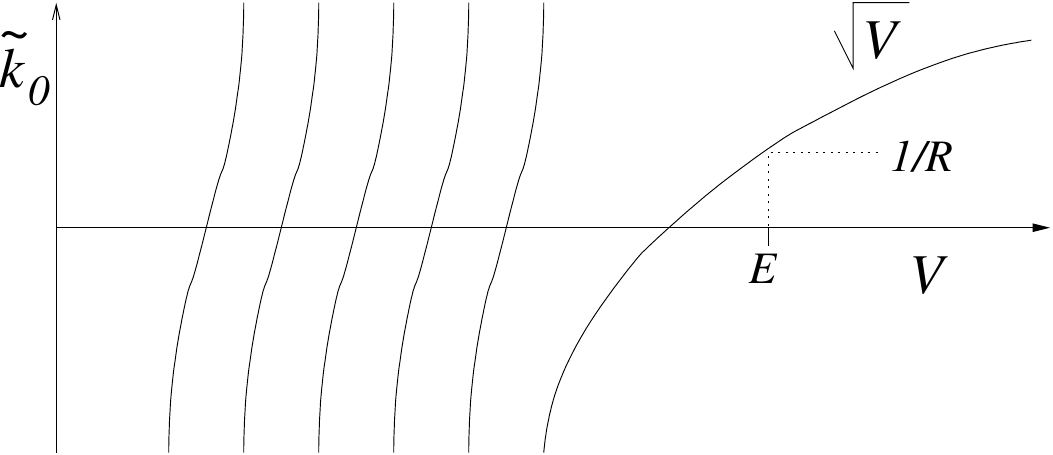}
\end{center}

In order to find the phase shift we use the standard 
matching procedure. In the vicinity of a resonance 
we can linearize the logarithmic derivative 
with respect to the control parameter~$V$ 
as ${\tilde{k}_0(V) \approx (V-V_r)/v_r}$
or if~$V$ is fixed, then with respect to the energy~$E$ 
as ${\tilde{k}_0(E) \approx -(E-E_r)/v_r}$, 
where the definitions of either $V_r$ or $E_r$ and $v_r$ are implied 
by the linearization procedure.
The two procedures are equivalent 
but for clarity we prefer the former.      
Thus in the vicinity of the resonance 
we get for the the phase shift
\beq
\delta_{0} \ \ = \ \ 
\delta_{0}^{\infty} + \arctan\left[ \frac{k_E}{\tilde{k}_0(V)}  \right]
\ \ = \ \ 
-k_E R + \arctan\left[\frac{\Gamma_r/2}{V-V_r}\right]
\eeq
where $\Gamma_r = 2 v_r k_E$.
The approximation above assumes 
well separated resonances.   
The distance between 
the locations $V_r$ of the resonances   
is simply the distance between the 
metastable states of the well. 
Let us call this level spacing $\Delta_0$. 
The condition for having a narrow resonance 
is $\Gamma_r < \Delta_0$. 
By inspection of the plot it should be clear 
that shielding (large $U$) shifts 
the plot upwards, and consequently $v_r$ 
and hence $\Gamma_r$ become smaller. 
Thus by making $U$ large enough we can 
ensure the validity of the above approximation.

In order to get the Wigner time delay we  
regard~$V$ as fixed, and plot the variation 
of $k_0(E)$ as a function of~$E$. 
This plot looks locally the same 
as the~$\tilde{k}_0(V)$ plot, with ${E \leftrightarrow -V}$. 
Then we can obtain $\delta_{0}$ as a function of~$E$, 
which is illustrated in the figure below. 
The phase shift is defined modulo $\pi$, 
but in the figure it is convenient 
not to take the modulo so as to have a continuous plot. 
At low energies the s-scattering   
phase shift is ${\delta_0(E)=-k_ER}$ 
and the time delay is $\tau \approx -2R/v_E$.
As the energy is raised there is an extra $\pi$~shift 
each time that~$E$ goes through a resonance. 
In order to ensure narrow resonance one should  
assume that the well is shielded by a large barrier. 
At the center of a resonance the time delay 
is of order $1/\Gamma_r$.

\begin{center}
\putgraph[0.36\hsize]{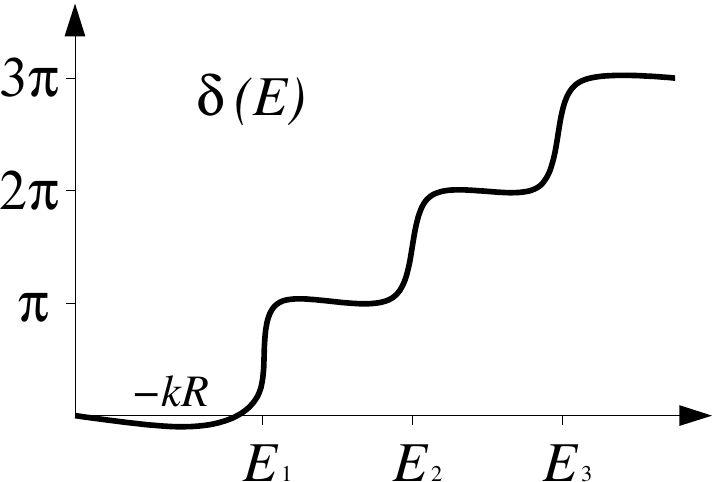}
\end{center}

\sheadC{Scattering resonance of a spherical shielded well}

The solution of the ${\ell>1}$ version of 
the shielded well scattering problem goes 
along the same lines as in the ${\ell=0}$ 
case that has been discussed above. 
The only modification is a change of convention: 
we work below with the radial functions $R(r)=u(r)/r$ 
and not with $u(r)$. Accordingly for the logarithmic 
derivative on the boundary of the scattering region  
we use the notation $k_{\ell}$ instead of $\tilde{k}_{\ell}$. 
Once we know the $\tilde{k}_{\ell}$ the phase 
shift can be calculated using the formula 
\beq
\eexp{i2\delta_{\ell}} =  
\left( \frac{h_{\ell}^{-}}{h_{\ell}^{+}} \right)  
\frac 
{ k_{\ell}(V) - ( h_{\ell}'^{-} / h_{\ell}^{-} ) k_E} 
{ k_{\ell}(V) - ( h_{\ell}'^{+} / h_{\ell}^{+} ) k_E} 
\eeq
In what follows we fix the energy $E$ of the scattered 
particle, and discuss the behavior of the phase shift 
and the cross section as a function of $V$. 
In physical applications $V$ can be interpreted 
as some "gate voltage". Note that in most 
textbooks it is customary to fix~$V$ and to change $E$.
We prefer to change~$V$ because then the expansions 
are better controlled. In any case our strategy  
gives literally equivalent results. 
Following Messiah p.391 and using the notations  
\beq
\left( \frac{h_{\ell}^{-}}{h_{\ell}^{+}} \right)  
\equiv \eexp{i2\delta_{\ell}^\infty}, 
\ \ \ \ \ \ \ \ \ \ \ \ \ 
k_E\left(\frac{h_{\ell}'^{+}}{h_{\ell}^{+}}\right)_{r=a}
\equiv\epsilon+i\gamma 
\eeq
we write 
\beq
\eexp{i2\delta_{\ell}} \ \ = \ \   
\left(\eexp{i2\delta_{\ell}^\infty} \right)  
\frac{ k_{\ell}(V) - \epsilon+i\gamma}{k_{\ell}(V) - \epsilon-i\gamma} 
\eeq
which gives 
\beq 
\delta_{\ell} \ \ = \ \ 
\delta_{\ell}^{\infty} +
\arctan\left[\frac{\gamma}{k_{\ell}(V)-\epsilon}\right] 
\eeq
We can plot the right hand side of the last equation 
as a function of $V$.  If the shielding is large 
we get typically $\delta_{\ell} \approx \delta_{\ell}^\infty$ 
as for a hard sphere. But if $V$ is smaller than~$E$
we can find narrow resonances as in the ${\ell=0}$ 
quasi 1D problem. The analysis of these resonances is 
carried out exactly in the same way. Note that for $\ell>0$ 
we might have distinct resonances even without 
shielding thanks to the centrifugal barrier.

\sheadA{QM in Practice (part III)}

\sheadB{The Aharonov-Bohm effect}

\sheadC{The Aharonov-Bohm geometry}

In the quantum theory it is natural to describe the electromagnetic 
field using the potentials ${V,A}$ and regard ${\mathcal{E},\mathcal{B}}$ 
as associated observables. Below we discuss the case 
in which ${\mathcal{E}= \mathcal{B}=0}$ in the region where 
the particle is moving. According to the classical theory 
one expects that the motion of the particle would not 
be affected by the field, since the Lorentz force is zero. 
However, we shall see that according to the quantum theory 
the particle is affected due to the non-zero circulation of~${A}$. 
This is a topological effect that we are going to clarify. 
Specifically we consider ring that is penetrated by a magnetic 
flux ${\Phi}$ through its center. This is the so-called Aharonov-Bohm 
geometry. To have a flux through the ring means that:
\beq
\oint \vec{A} \cdot d\vec{l} \ \ = \ \ \iint \mathcal{B} \cdot d\vec{s} \ \ = \ \ \Phi 
\eeq
The simplest gauge choice for the vector potential is  
\beq
A \ \ = \ \ \frac{\Phi}{L}  \ \ \ \ \  \ \mbox{[tangential]}
\eeq
where $L$ is the length of the ring. 
Below we treat the ring as a 1D segment ${0<x<L}$ 
with periodic boundary conditions. 
The Hamiltonian is 
\beq
\mathcal{H} \ \ = \ \ \frac {1}{2\mass}\left(\hat{p}-\frac{e\Phi}{cL}\right)^2 
\eeq
The eigenstates of $\mathcal{H}$ 
are the momentum states ${|k_n \rangle}$ where:
\beq
k_n \ = \ \frac {2\pi }{L} n, 
\ \ \ \ \ \ \ \ \ n=0,\pm1,\pm2,...
\eeq
The eigenvalues are (written with $\hbar$ for historical reasons):
\beq
E_n \ \ = \ \ \frac{1}{2\mass}\left(\frac{2\pi\hbar}{L}n-\frac{e\Phi}{cL}\right)^2
\ \ = \ \ \frac{1}{2\mass}\left(\frac {2\pi\hbar}{L}\right)^2 \left(n-\frac{e\Phi}{2\pi\hbar c}\right)^2 
\eeq
The unit ${{2\pi \hbar c}/{e}}$ is called "fluxon". It is the basic unit of flux in nature. 
We see that the energy spectrum is influenced by the presence of the magnetic flux. 
On the other hand, if we draw a plot of the energies as a function of the flux 
we see that the energy spectrum repeats itself every time the change in the flux is 
an integer multiple of a fluxon. (To guide the eye we draw the ground state energy with thick line). 

The fact that the electron is located in an area where there is no Lorentz 
force ${\mathcal{E}= \mathcal{B}=0}$, but is still influenced by 
the vector potential is called the Aharonov-Bohm Effect. 
This is an example of a topological effect.

\begin{center}
\putgraph[0.2\hsize]{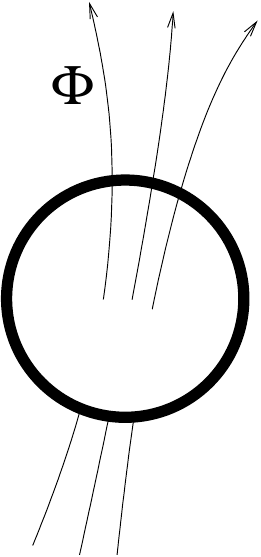} 
\hspace*{0.2\hsize}
\putgraph[0.4\hsize]{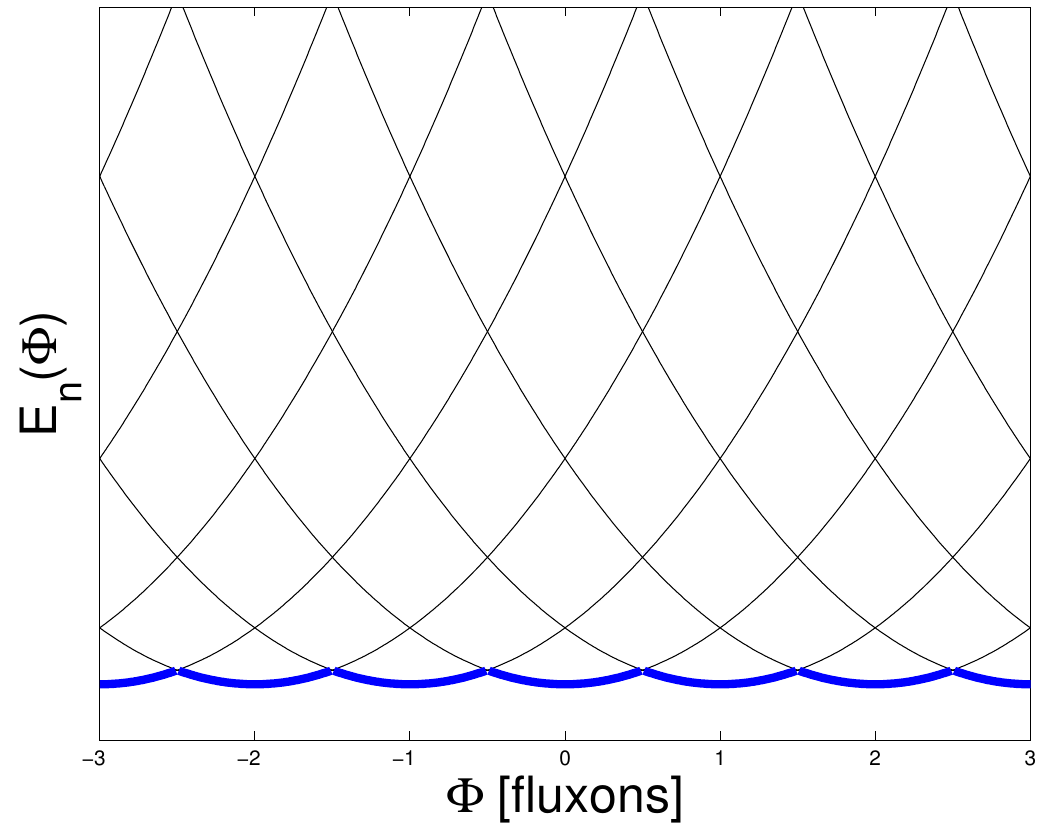}
\end{center}

\sheadC{The energy levels of a ring with a scatterer}

Consider an Aharonov-Bohm ring with (say) a delta scatterer:
\beq
\mathcal{H} \ \ = \ \ 
\frac{1}{ 2\mass} 
\left( p - \frac{e\Phi}{L} \right)^2 
+ u \delta(x)
\eeq
We would like to find the eigenenergies of the ring.
The standard approach is to write the general solution 
in the empty segment, and then to impose the matching 
condition over the delta scatterer. An elegant procedure 
for solution is based on the scattering formalism. 
In order to characterize the scattering within the system, 
the ring is cut at some arbitrary point and the $S$ matrix of the 
open segment is specified. It is more convenient to  
use the row-swapped matrix, such that the transmission 
amplitudes are along the diagonal:
\beq
\tilde{\mathbf{S}} 
\ \ = \ \ 
\eexp{i\gamma}
\left( \amatrix{\sqrt{g}\eexp{i\phi} &  -i\sqrt{1-g}\eexp{-i\alpha} \cr
-i\sqrt{1-g}\eexp{i\alpha} & \sqrt{g}\eexp{-i\phi}} \right)
\eeq
where the transmission is 
\beq
g(E)=
\left[ 1 
+ \left( \frac{u}{v_E}\right)^2 
\right]^{-1},
\hspace{2cm} v_E = \mbox{velocity at energy $E$} 
\eeq
We include ``legs" to the delta scatterer, 
hence the total transmission phase is  
\beq
\gamma(E) \ \ = \ \ k_E L -\arctan\left( \frac{u}{v_E} \right) 
\eeq
More precisely, with added flux the transmission 
phases are $\gamma\pm\phi$ where $\phi=e\Phi/\hbar$. 
The reflection phases are ${\gamma-(\pi/2) \pm \alpha}$, 
where ${\alpha=0}$ if we cut the ring symmetrically, 
such that the two legs have the same length.  

The periodic boundary conditions imply the "matching condition" 
\beq
\left( \amatrix{ A \cr B } \right) \ \ = \ \  
\tilde{\mathbf{S}}
\left( \amatrix{ A \cr B } \right)
\eeq
This equation has a non-trivial solution if and only if   
\beq
\det(\tilde{\mathbf{S}}(E) -\bm{1}) \ \ = \ \ 0
\eeq
For the calculation is is useful to note that 
\beq
\det(\tilde{\mathbf{S}}-\bm{1}) &=& \det(\tilde{\mathbf{S}})-\trc{(\tilde{\mathbf{S}})}+1 \\
\det(\tilde{\mathbf{S}}) &=& (\eexp{i\gamma})^2  \\
\trc{(\tilde{\mathbf{S}})} &=& 2\sqrt{g}\eexp{i\gamma}\cos{\phi}
\eeq
Hence we get an equation for the eigen-energies:
\beq
\cos(\gamma(E)) \ \ = \ \ \sqrt{g(E)} \cos(\phi)
\eeq
In order to find the eigen-energies we plot both sides as 
a function of $E$. The left hand side oscillates
between $-1$ and $+1$,  while the right hand side 
is slowly varying monotonically. It is easily verified 
that the expected results are obtained for clean ring ($g=1$) 
and for infinite well ($g=0$).

\sheadC{Perturbation theory for a ring + scatterer}

Let us consider a particle with mass $\mass$ 
on a 1D ring. A flux ${\Phi}$ goes through the ring. 
In addition, there is a scatterer that is described 
by a delta function. The Hamiltonian that 
describes the system is:
\beq
\mathcal{H} \ \ = \ \ \frac{1}{2\mass}\left(p-\frac{\Phi}{L}\right)^2 +u\delta(x) 
\eeq
For $\Phi=u=0$  the symmetry group of this Hamiltonian is ${O(2)}$. 
This means symmetry with respect to rotations and reflections. 
Note that in one-dimension ring $=$ circle $=$ torus, 
hence rotations and displacements are the same. 
Only in higher dimensions they are different (torus ${\ne}$ sphere). 

Degeneracies are an indication for symmetries of the 
Hamiltonian. If the eigenstate has a lower symmetry 
than the Hamiltonian, a degeneracy appears. Rotations 
and reflections do not commute, that is why we have degeneracies. 
When we add flux or a scatterer, the degeneracies open up. 
Adding flux breaks the reflection symmetry, 
and adding a scatterer breaks the rotation symmetry. 
Accordingly, depending on the perturbation, 
it would be wise to use one of the following two bases:

{\bf The first basis:} 
The first basis complies with the rotation (=translations) symmetry:
\beq
|n=0\rangle &=& \frac{1}{\sqrt{L}}
\\ \nonumber
| n, \mbox{anticlockwise} \rangle &=& \frac{1}{\sqrt{L}} \eexp{+ik_n x}, \ \ \ \ \ \ \ \ \ \ n=1,2,...
\\ \nonumber
|n,\mbox{clockwise} \rangle &=& \frac{1}{\sqrt{L}} \eexp{-ik_n x}, \ \ \ \ \ \ \ \ \ \ n=1,2,...
\eeq
The degenerate states are different under reflection. 
Only the ground state ${|n=0\rangle}$ is symmetric
under both reflections and rotations, and therefore it 
does not have to be degenerate. 
It is very easy to calculate the perturbation matrix 
elements in this basis:
\beq
\langle n |\delta(x)|m \rangle 
\ \ = \ \  \int\Psi^n(x)\delta(x)\Psi^m(x)dx = \Psi^n(0)\Psi^m(0) 
\ \ = \ \ \frac{1}{L} 
\eeq
so we get:
\beq
V_{nm} \ \ = \ \ \frac{u}{L} \, 
\left(
\amatrix{
1 & 1 & 1 & 1 & \dots \cr 
1 &1& 1 &1 & \dots \cr
1 & 1 & 1 & 1 & \dots \cr 
1 & 1 & 1 & 1 & \dots \cr 
\dots & \dots & \dots & \dots & \dots } 
\right)  
\eeq

{\bf The second basis:} 
The second basis complies with the reflection symmetry:
\beq
|n=0\rangle &=& \frac{1}{\sqrt{L}}
\\ \nonumber
| n, + \rangle &=& \sqrt{\frac{2}{L}} \cos(k_n x), \ \ \ \ \ \ \ \ \ \ n=1,2,...
\\ \nonumber
| n, - \rangle &=& \sqrt{\frac{2}{L}} \sin(k_n x), \ \ \ \ \ \ \ \ \ \ n=1,2,...
\eeq
The degeneracy is between the even states 
and the odd states that are displaced by 
half a wavelength with respect to each other. 
If the perturbation is not the flux but 
rather the scatterer, then it is better 
to work with the second basis, which complies 
with the potential's symmetry. The odd states 
are not influenced by the delta function, 
and they are also not "coupled" to the even states. The reason is that:
\beq
\langle m |\delta(x)|n \rangle &=& \int\Psi^m(x)\delta(x)\Psi^n(x)dx=0, 
\ \ \ \ \ \ \ \ \ \ \mbox{if $n$ or $m$ are "sin"}
\eeq
Consequently the subspace of odd states 
is not influenced by the perturbation,
i.e.  ${V_{nm}^{(-)}= \bm{0}}$, 
and we only need to diagonalize 
the block that belongs to the even states. 
It is very easy to write the perturbation 
matrix for this block:
\beq
V_{nm}^{(+)} \ \ = \ \ \frac{u}{L} \ \  
\left( 
\amatrix{
1 & \sqrt{2} & \sqrt{2} & \sqrt{2} & \dots \cr 
\sqrt{2} & 2 & 2 & 2 & \dots \cr 
\sqrt{2} & 2 &2 &2 & \dots \cr 
\sqrt{2} & 2 & 2 & 2 & \dots \cr 
\dots & \dots & \dots & \dots & \dots } 
\right) 
\eeq

{\bf Energy levels:} 
Without a scatterer the eigenenergies are:
\beq
E_n(\Phi, u{=}0)
\ \ = \ \ \frac{1}{2\mass}
\left(\frac{2\pi}{L}\times \mbox{integer}-\frac{\Phi}{L}\right)^2,
\ \ \ \ \ \ \ \ \ \ \mbox{integer}=0,\pm1,\pm2,...
\eeq
On the other hand, in the limit ${u\rightarrow\infty}$ 
the system does not "feel" the flux, and the ring becomes 
a one-dimensional box. The eigenenergies in this limit are:
\beq
E_n(\Phi, u{=}\infty) 
\ \ = \ \ \frac{1}{2\mass}\left(\frac{\pi}{L} \times \mbox{integer}\right)^2,
\ \ \ \ \ \ \ \ \ \ \mbox{integer}=1,2,...
\eeq
As one increases ${u}$ the number of the energy 
levels does not change, they just move. See figure below. 
We would like to use perturbation theory in order 
to find corrections to the above expressions. 
We consider how we do perturbation theory with respect 
to the  ${u{=}0}$ Hamiltonian. It is also possible 
to carry out perturbation theory with respect 
to the  ${u{=}\infty}$ Hamiltonian (for that one 
should use the formula for the interaction at a junction).

\begin{center}
\putgraph[0.4\hsize]{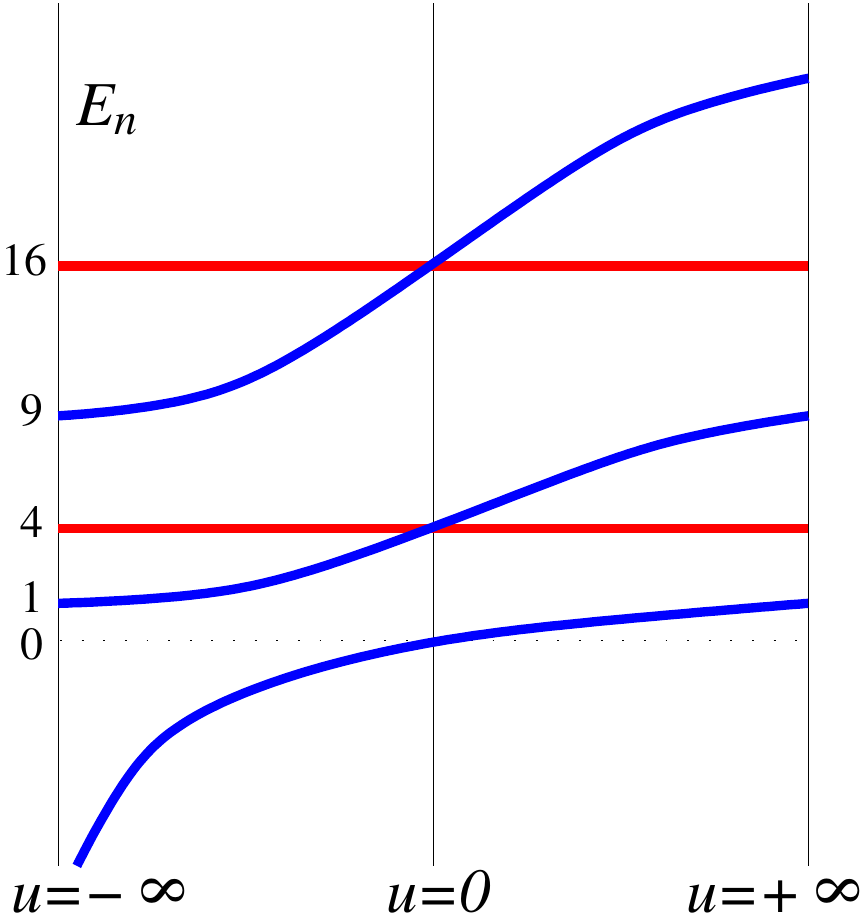} 
\end{center}

{\bf The corrections to the energy:} 
Let us evaluate the first-order correction to the energy 
of the eigenstates in the absence of an external flux.
\beq
E_{n{=}0} \ \ &=& \ \ E_{n{=}0}^{[0]} + \frac{u}{L}
\\ \nonumber
E_{n{=}2,4,\dots} \ \ &=& \ \ E_n^{[0]} + \frac{2u}{L} 
\eeq
The correction to the ground state 
energy, up to the second order is:
\beq
E_{n{=}0} \ \ = \ \ 0 + \frac{u}{L} + \left(\frac{u}{L}\right)^2 \, 
\sum_{k=1}^{\infty}
\frac{(\sqrt{2})^2} {0-\frac{1}{2\mass}
\left(\frac{2\pi}{L}k\right)^2} 
\ \ = \ \ 
\frac{u}{L}\left (1-\frac{1}{6}u \mass L \right) 
\eeq
where we have used the identity:
\beq
\sum_{k=1}^{\infty}\frac{1}{k^2} \ = \ \frac{\pi^2}{6} 
\eeq

{\bf Optional calculation:} 
We will now assume that we did not notice the 
symmetry of the problem, and we chose to work with 
the first basis. Using perturbation theory 
on the ground state energy is simple in this basis:
\beq
E_{n{=}0} \ \ = \ \ 0 + \frac{u}{L} + \left(\frac{u}{L}\right)^2 \, 
2\sum_{k=1}^{\infty}\frac{(1)^2} {0-\frac{1}{2\mass}\left(\frac{2\pi}{L}k\right)^2} 
\ \ = \ \ 
\frac{u}{L}\left (1-\frac{1}{6}u \mass L \right) 
\eeq
But using perturbation theory on the rest of 
the states is difficult because there are degeneracies. 
The first thing we must do is "degenerate perturbation theory". 
The diagonalization of each degenerate energy level is:
\beq
\left(\amatrix{1 & 1 \cr 1 & 1 } \right) \ \  \rightarrow \ \ \left(\amatrix{2 & 0 \cr 0 & 0 } \right) 
\eeq
Now we should transform to a new basis, where the 
degeneracy is removed. This is exactly the 
basis that we chose to work with due to symmetry 
considerations. The moral lesson is: understanding the 
symmetries in the system can save us work in the 
calculations of perturbation theory.

\sheadC{The AB effect in a closed geometry}

The eigen-energies of a particle in a closed ring 
are periodic functions of the flux.
In particular in the absence of scattering   
\beq
E_n \ \ = \ \ \frac {1}{2\mass} 
\left( \frac {2 \pi\hbar}{L} \right)^2 
\left( n- \frac {e\Phi}{2\pi\hbar c}\right)^2 
\ \ = \ \ \frac{1}{2}\mass v_n^2 
\eeq
That is in contrast with classical mechanics, 
where the energy can have any positive value:
\beq
E_{\mbox{classical}} \ \ = \ \ \frac {1}{2} \mass v^2 
\eeq
According to classical mechanics the lowest energy 
of a particle in a magnetic field is zero, with velocity zero. 
This is not true in the quantum case. 
It follows that an added magnetic flux has 
an detectable effect on the system. 
The effect can be described in one of the following ways:

\bitem The spectrum of the system changes 
(it can be measured using spectroscopy) \\
\bitem For flux that is not an integer or half integer number 
there are persistent currents in the system. \\
\bitem The system has either a diamagnetic 
or a paramagnetic response (according to the occupancy).

We already have discussed the spectrum of the system.
So the next thing is to derive an expression for the 
current in the ring. The current operator is 
\beq
\hat{I}\ \ \equiv \ \  -\frac{\partial \mathcal{H}}{\partial \Phi} 
\ \ = \ \  \frac{e}{L} \left[\frac{1}{\mass}\left(\hat{p}-\frac{e\Phi}{L}\right)\right]
\ \ = \ \  \frac{e}{L}\hat{v} 
\eeq
It follows that the current which is created 
by an electron that occupies the $n$th level is:
\beq
I_n \ \ = \ \
\left\langle  n \left| \left(-\frac{\partial \mathcal{H}}{\partial \Phi}\right) \right| n \right\rangle 
\ \ = \ \  -\frac{d E_n}{d \Phi} 
\eeq
The proof of the second equality is one line of algebra.
If follows that by looking at the plot of the energies ${E_n(\Phi)}$ 
as a function of the flux, one can determine (according to the slope) 
what is the current that flows in each occupied energy level.
If the flux is neither integer nor half integer, 
all the states "carry current" so that in equilibrium 
the net current is not zero. This phenomenon is 
called "persistent currents". The equilibrium current 
in such case cannot relax to zero, 
even if the temperature of the system is zero.

There is a statement in classical statistical mechanics that 
the equilibrium state of a system is not affected 
by magnetic fields. The magnetic response of any system 
is a quantum mechanical effect that has to do with the quantization 
of the energy levels (Landau magnetism) or with the spins (Pauly magnetism). 
Definitions: 

\bitem Diamagnetic System - in a magnetic field, the system energy increases. \\
\bitem Paramagnetic System - in a magnetic field, the system energy decreases. 

The Aharonov Bohm geometry provides the simplest example 
for magnetic response. If we place one electron in a ring, 
and add a weak magnetic flux, the system energy increases. 
Accordingly we say that the response is "diamagnetic". 
The electron cannot "get rid" of its kinetic energy, 
because of the quantization of the momentum.

\newpage
\sheadC{Dirac Monopoles} 
\label{sMonopoles}

Yet another consequence of the "Aharonov Bohm" effect 
is the quantization of the magnetic charge. 
Dirac has claimed that if magnetic monopoles exist, 
then there must be an elementary magnetic charge. 
The formal argument can be phrased as follows:
If a magnetic monopole exists, it creates a vector potential 
field in space (${A(x)}$). The effect of the field of 
the monopole on an electron close by is given by the 
line integral ${\oint \vec{A}\cdot d\vec{r}}$. 
We can evaluate the integral by calculating the 
magnetic flux ${\Phi}$ through a Stokes surface. 
The result should not depend on the choice of the surface, 
otherwise the phase is not be well defined. 
In particular we can choose Stokes surfaces that 
pass above and below the monopole,  
and deduce that the phase difference ${\phi = {e\Phi}/{\hbar c}}$ 
should be zero modulo ${2\pi}$.
Hence the flux $\Phi$ should be an integer multiple of ${{2\pi\hbar c}/{e}}$. 
Using "Gauss law" we conclude that the monopole 
must have a magnetic charge that is quantized 
in units of ${{\hbar c}/{2e}}$. 

Dirac's original reasoning was somewhat more constructive.  
Let us assume that a magnetic monopole exist.
The magnetic field that would be created by this 
monopole would be like that of a tip of a solenoid. 
But we have to {\em exclude} the region in space 
where we have the magnetic flux that goes through the solenoid. 
If we want this "flux line" to be unobservable 
then it should be quantized in units of  ${{2\pi\hbar c}/{e}}$. 
This shows that Dirac "heard" about the Aharonov Bohm effect, 
but more importantly this implies that the "tip" would have 
a charge which equals an integer multiple of ${{\hbar c}/{2e}}$.

\sheadC{The AB effect: path integral formulation}

We can optionally illustrate the Aharonov-Bohm Effect 
by considering an open geometry. In an open geometry the energy 
is not quantized:  it is determined by scattering arrangement.
If the energy potential floor is taken as a reference - 
the energy $E$ can be any positive value.  
We are looking for stationary states that solve the Schr\"{o}dinger equation 
for a given energy.  These states are called "scattering states". 
Below we discuss the Aharonov-Bohm effect in a "two slit" geometry, 
and later refer to a "ring" geometry (with leads). 

First we notice the following rule: 
if we have a planar wave ${\psi(x) \propto \eexp{ikx}}$, 
and if the amplitude at the point ${x=x_1}$ is ${\psi (x_1)=C}$, 
then at another point ${x=x_2 }$ the 
amplitude is ${\psi (x_2)= C \eexp{ik(x_2-x_1)}}$. 

Now we generalize this rule for the case in which 
there is a vector potential $A$. 
For simplicity, we assume that the motion 
is in one-dimension. The eigenstates of the Hamiltonian are 
the momentum states. 
If the energy of the particle is~$E$ then the wavefunctions   
that solve the Schr\"{o}dinger's equation 
are ${\psi (x) \propto \eexp{ik_{\pm} x}}$, where  
\beq
k_{\pm} \ \ = \ \  \pm \sqrt {2\mass E } + A \ \  \equiv \ \  \pm k_E + A 
\eeq
Below we refer to the advancing wave:  
if at point ${x=x_1}$ the amplitude is ${\psi (x_1)=C}$, 
then at another point ${x=x_2}$ the amplitude 
is ${\psi (x_2)= C \eexp{ ik_E(x_2-x_1) + A \cdot (x_2-x_1) }}$. 
It is possible to generalize the idea to three 
dimensions: if a wave advances along a certain path 
from point $x_1$ to point $x_2$, then the accumulated phase is:
\beq
\phi \ \ = \ \  k_E L + \int_{x_1}^{x_2} A \cdot dx, 
\ \ \ \ \ \ \ \ \mbox{$L$ = length of the path}  
\eeq
If there are two different paths that connect 
the points ${x_1}$ and ${x_2}$, then the phase difference is:
\beq
\Delta \phi  \ \ = \ \ k_E \Delta L + \int_{L_2} A \cdot dx - \int_{L_1} A \cdot dx  
\ \ = \ \ k_E \Delta L + \oint A \cdot dx 
\ \ = \ \ k_E \Delta L + \frac{e}{\hbar c}\Phi 
\eeq
where in the last term we "bring back" the standard physical units. 
The approach which was presented above for calculating 
the probability of the particle to go from one point 
to another is called "path integrals". This approach 
was developed by Feynman, and it leads to what is 
called "path integral formalism" - an optional approach 
to do calculations in quantum mechanics. The conventional method 
is to solve the Schr\"{o}dinger's equation with the 
appropriate boundary conditions.

\sheadC{The AB effect in a two slits geometry} 

We can use the path integral point of view 
in order to analyze the interference 
in the two slit experiment. A particle  
that passes through two slits, splits into two partial waves 
that unite at the detector. 
Each of these partial waves passes a different optical path.
Hence the probability of reaching the detector, 
and consequently the measured intensity of the beam is 
\beq
\mbox{Intensity} 
\ \ = \ \ \Big|1\times \eexp{ ikr_1} + 1\times \eexp{ikr_2}\Big|^2 
\ \ \propto \ \ 1+\cos(k(r_2-r_1)) 
\eeq
Marking the length difference as $\Delta L$, 
and the associated phase difference as ${\Delta\phi}$, 
we rewrite this expression as:
\beq
\mbox{Intensity} \ \ \propto \ \ 1 + \cos(\Delta\phi), 
\eeq
Changing the location of the detector results in a change in the phase difference ${\phi}$. 
The "intensity", or more precisely the probability that the particle will reach the detector, 
as a function of the phase difference ${\phi}$, is called "interference pattern". 
If we place a solenoid between the slits, then the formula for the phase difference becomes:
\beq
\Delta\phi \ \ = \ \ k \Delta L + \frac{e}{\hbar c}\Phi 
\eeq
If we draw a plot of the "intensity" as a function 
of the flux we get the same "interference pattern".

\begin{center}
\putgraph[0.5\hsize]{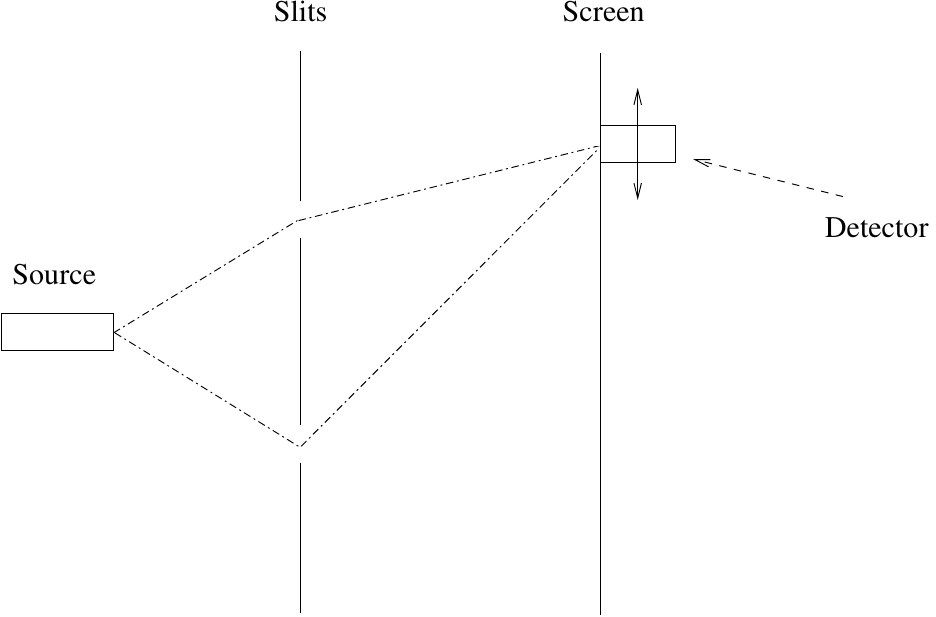} 
\end{center}

If we want to find the transmission of an Aharonov-Bohm 
device (a ring with two leads) then we must sum all the 
paths going from the first lead to the second lead. 
If we only take into account the two shortest 
paths (the particle can pass through one arm or the other arm), 
then we get a result that is formally identical to the result 
for the two slit geometry. In reality we must take into account 
all the possible paths. That is a very long calculation, 
leading to a Fabry-Perrot type result (see previous lecture).
In the latter case the transmission is an even function of~$\Phi$, 
even if the arms do not have the same length.
Having the same transmission for $\pm\Phi$ in the case  
of a closed device is implied by time reversal symmetry. 
 
\newpage
\sheadB{Motion in uniform magnetic field (Landau, Hall)}

\sheadC{The two-dimensional ring geometry}

Let us consider a three-dimensional box 
with periodic boundary conditions 
in the $x$ direction, and zero boundary 
conditions on the other sides.
In the absence of magnetic field we assume that 
the Hamiltonian is:
\beq
\mathcal{H}
\ \ = \ \ \frac{1}{2\mass} \hat{p}_x^2 
+ \left[ \frac{1}{2\mass}\hat{p}_y^2 + V(y) \right]
+ \left[ \frac{1}{2\mass} \hat{p}_z^2  + V_{\tbox{box}}(z)\right]
\eeq
The eigenstates of a particle in such box are labeled as 
\beq
&& | k_x , n_y , n_z \rangle
\\ \nonumber
&& k_x = \frac {2\pi}{L} \times \mbox{integer}
\\ \nonumber
&& n_y ,n_z =1,2,3\dots 
\eeq
The eigenenergies are:
\beq
E_{k_x,n_y,n_z} \ \ = \ \ \frac {k_x^2}{2\mass} 
+ \varepsilon_{n_y} 
+ \frac {1}{2\mass}\left(\frac{ \pi}{L_z}n_z\right)^2 
\eeq
We assume $L_z$ to be very small 
compared to the other dimensions.
We shall discuss what happens when 
the system is prepared in low energies 
such that only $n_z=1$ states are relevant. 
So we can ignore the $z$ axis.

\sheadC{Classical Motion in a uniform magnetic field}

Consider the motion of an electron in 
a two-dimensional ring. We assume that 
the vertical dimension is "narrow", 
so that we can safely ignore it, as was explained 
in the previous section. For convenience  
we "spread out" the ring susch that it forms a rectangle 
with periodic boundary conditions over ${0<x<L_x}$, 
and an arbitrary confining potential ${V(y)}$ 
in the perpendicular direction. 
Additionally we assume that there is a uniform 
magnetic field ${\mathcal{B}}$ along the $z$ axis. 
Therefore the electron is affected by 
a Lorentz force ${F = - (e/c) \mathcal{B} \times v}$. 
If there is no electrical potential, the electron 
performs a circular motion with the cyclotron frequency:
\beq
\omega_B \ \ = \ \ \frac{e\mathcal{B}}{\mass c} 
\eeq
If the electron has a kinetic energy ${E}$, its velocity is:
\beq
v_E \ \ = \ \ \sqrt{ \frac{2E}{\mass}} 
\eeq
Consequently it moves along a circle of radius
\beq
r_E \ \ = \ \ \frac{v_E}{\omega_B} \ \ = \ \ \frac{\mass c}{e\mathcal{B}}v_E 
\eeq
If we take into account a non-zero electric field
\beq
\mathcal{E}_y \ \  = \ \ -\frac{dV}{dy} 
\eeq
we get a motion along a cycloid with the 
drift velocity (see derivation below):
\beq
v_{\mbox{drift}} \ \ = \ \ c\frac{\mathcal{E}_y}{\mathcal{B}} 
\eeq

Let us remind ourselves why the motion is along 
a cycloid. The Lorentz force in the laboratory reference 
frame is (from now on we absorb the $(e/c)$
of the electron into the definition of the field):
\beq
F \ = \ \mathcal{E} - \mathcal{B} \times v 
\eeq
If we transform to a reference frame 
that is moving at a velocity ${v_0}$ we get:
\beq
F \ \ = \ \ \mathcal{E} - \mathcal{B} \times ( v' + v_0 ) 
\ \ = \ \ (\mathcal{E} + v_0 \times \mathcal{B}) - \mathcal{B} \times v' 
\eeq
Therefore, the non-relativistic transformation 
of the electromagnetic field is:
\beq
\mathcal{E}' &=& \mathcal{E} + v_0 \times \mathcal{B} 
\\ \nonumber
\mathcal{B}' &=& \mathcal{B} 
\eeq
If there is a field in the $y$ direction in the laboratory reference frame, 
we can transform to a new reference frame where the field is zero. 
From the transformation above we conclude that in order 
to have a zero electrical field, the velocity of the "new" 
frame of reference should be: 
\beq
v_0 \ \ = \ \ \frac{\mathcal{E}}{\mathcal{B}}c \ \ \ \ \ \ \ \ \ \ \text{[restoring CGS units for clarity]} 
\eeq
In the new reference frame the particle 
moves along a circle. Therefore, in the laboratory 
reference frame it moves along a cycloid.

\begin{center}
\putgraphv[0.25\hsize]{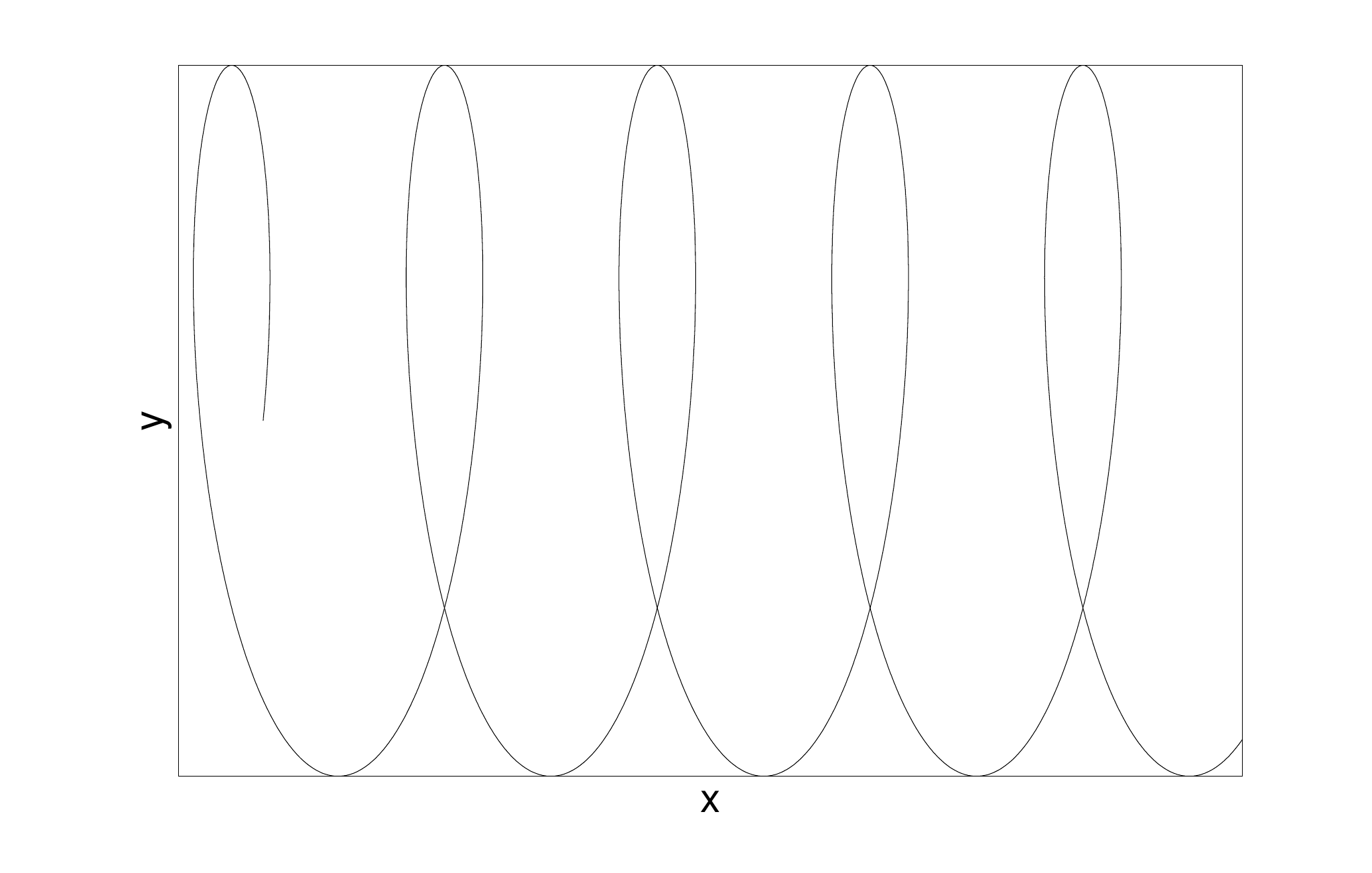} 
\putgraphv[0.25\hsize]{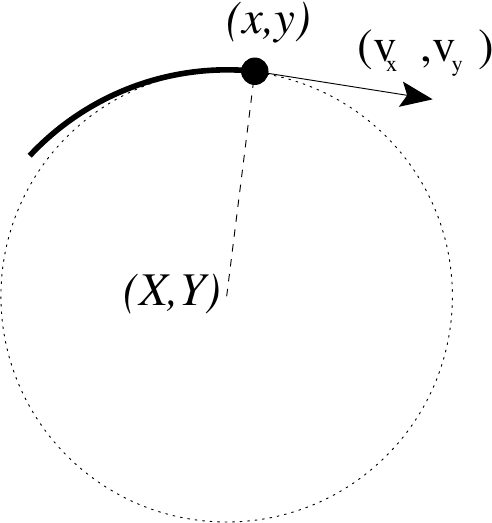} 
\end{center}

Conventionally the classical state of the particle 
is described by the coordinates ${r=(x,y)}$
and ${v=(v_x,v_y)}$. 
But from the above discussion it follows that a simpler 
description of the trajectory is obtained if we follow 
the motion of the moving circle. 
The center of the circle ${R=(X,Y)}$ 
is moving along a straight line, 
and its velocity is ${v_{\mbox{drift}}}$. 
The transformation that relates $R$ 
to ${r}$ and ${v}$ is 
\beq
\vec{R} \ \ = \ \ \vec{r} - \vec{e}_z \times \frac{1}{\omega_B}\vec{v}
\eeq
where $\vec{e}_z$ is a unit vector in the $z$ direction.
The second term in the right hand side is 
a vector of length $r_E$ in the radial direction 
(perpendicular to the velocity).
Thus instead of describing the motion with the canonical 
coordinates ${(x,y,v_x,v_y)}$, we can use 
the new coordinated ${(X,Y,v_x,v_y)}$.

\sheadC{The Hall Effect} 

If we have particles spread out in a uniform 
density ${\rho}$ per unit area, then  
the current density per unit length is:
\beq
J_x \ \ = \ \ e \rho v_{\mbox{drift}} 
\ \ = \ \ \rho \frac{ec}{\mathcal{B}}\mathcal{E}_y 
\ \ = \ \ - \rho \frac{ec}{\mathcal{B}} \frac{dV}{dy} 
\eeq
where we keep CGS units and $V$ is the electrical potential (measured in Volts). 
The total current is:
\beq
I_x \ \ = \ \ \int_{y_1}^{y_2} J_x dy 
\ \ = \ \ - \rho \frac{ec}{\mathcal{B}} ( V(y_2) - V(y_1) ) 
\ \ = \ \ - \rho \frac{c}{\mathcal{B}} ( \mu_2 - \mu_1 )
\eeq
Here $\mu=eV$ is the chemical potential.
Accordingly the Hall conductance is:
\beq
G_{\mbox{Hall}} \ \ = \ \ - \rho \frac{ec}{\mathcal{B}} 
\eeq
In the quantum analysis we shall see that the electrons 
occupy "Landau levels". The density of electrons 
in each Landau Level is ${{e\mathcal{B}}/{2\pi\hbar c}}$. 
From this it follows that the Hall conductance is quantized 
in units of ${e^2/{2\pi\hbar}}$, which is the universal 
unit of conductance in quantum mechanics. 
The experimental observation is illustrated in the 
figure below [taken from the web]. The right panels shows 
that for very large field fractional values are observed.
The explanation of this fraction quantum Hall effect (FQHE) 
requires to taken into account the interactions between the electrons.

\includegraphics[height=5cm]{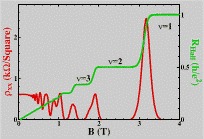}
\ \ \ \ \ \ \ \ \ \ 
\includegraphics[height=5cm]{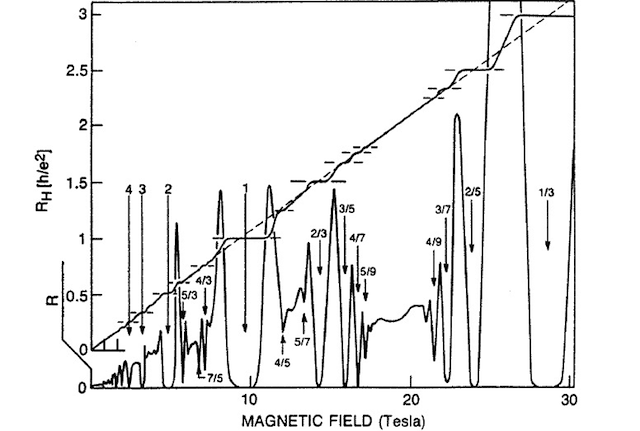} \\

We note that both Ohm law and Hall law should be written as:
\beq
I \ \ = \ \ G \times \frac{1}{e}( \mu_2 - \mu_1 ) 
\eeq
and not as:
\beq
I \ \ = \ \ G \times ( V_2 - V_1 ) 
\eeq
where ${\mu}$ is the electrochemical potential. 
If the electrical force is the only cause for the current, 
then the electrochemical potential is simply the electrical potential 
(multiplied by the charge of the electron).  
At zero absolute temperature ${\mu}$ can be identified with the Fermi energy. 
In metals in equilibrium, according to classical mechanics, 
there are no currents inside the metal. 
This means that the electrochemical potential must be uniform. 
This does not mean that the electrical potential is uniform! 
For example: when there is a difference in concentrations of 
the electrons (e.g. different metals) 
then there should be a "contact potential" 
to balance the concentration gradient, so as to  
have a uniform electrochemical potential. 
Another example: in order to balance the gravitation 
force in equilibrium, there must be an electrical force
such that the total potential is uniform. 
In general, the electrical field in a metal 
in equilibrium cannot be zero.

\sheadC{Electron in Hall geometry: Landau levels}

In this section we show that there is 
an elegant formal way of treating the problem 
of an electron in Hall geometry using 
a canonical transformation. 
This method of solution is valid both in classical 
mechanics and in quantum mechanics (all one has to 
do is to replace the Poisson brackets with commutators). 
In the next lecture we solve the quantum problem 
again, using the conventional method of "separation of variables". 
Here and later we use the Landau gauge: 
\beq
\vec{A} & \ \ = \ \ & (-\mathcal{B}y,0,0)
\\ \nonumber
\mathcal{B} & \ \ = \ \ & \nabla \times \vec{A} \ \ = \ \ (0,0,\mathcal{B}) 
\eeq
Recall that we absorb the charge of the electron 
in the definition of the fields. 
Consequently the Hamiltonian is 
\beq
\mathcal{H} 
\ \ = \ \ \frac{1}{2\mass}( \hat{p}_x+\mathcal{B}y)^2 
+ \frac{1}{2\mass}( \hat{p}_y)^2 + V(y) 
\eeq
We define a new set of operators:
\beq
&& v_x \ \ = \ \ \frac{1}{\mass}( p_x + \mathcal{B}y )
\\ \nonumber
&& v_y \ \ = \ \ \frac{1}{\mass} p_y
\\ \nonumber
&& X \ \ = \ \ x + \frac{1}{\omega_B} v_y = x + \frac{1}{\mathcal{B}}p_y
\\ \nonumber
&& Y \ \ = \ \ y - \frac{1}{\omega_B} v_x = - \frac{1}{\mathcal{B}} p_x 
\eeq
Recall that from a geometrical perspective, ${(X,Y)}$ represent 
the center of the circle along which the particle is moving. 
Note also that the operators ${X,Y}$ 
commute with ${v_x,v_y}$. On the other hand:
\beq
&& [X,Y] \ \ = \ \ -i\mathcal{B}^{-1} 
\\ \nonumber
&& [v_x, v_y] \ \ = \ \ i\frac{1}{\mass^2} \mathcal{B} 
\eeq
Consequently we can define a new set of canonical coordinates:
\beq
Q_1 \ = \ \frac{\mass}{\mathcal{B}} v_x,
\hspace*{1cm}
P_1 \ = \ \mass v_y,
\hspace*{1cm}
Q_2 \ = \ Y,
\hspace*{1cm}
P_2 \ = \ \mathcal{B} X 
\eeq
The Hamiltonian takes the form
\beq
\mathcal{H}(Q_1,P_1,Q_2,P_2) 
\ \ = \ \ \frac{1}{2\mass}P_1^2 
+ \frac{1}{2} \mass \omega_B^2 Q_1^2 
+ V(Q_1 + Q_2) 
\eeq
We see, as expected, that ${Q_2 = Y}$ is a constant of motion. 
Let us further assume that the potential $V(y)$ is zero within 
a large box of area $L_xL_y$. Then also ${P_2 \propto X}$ 
is a constant of motion. The coordinates ${(X,Y)}$ are 
conjugate with $\hbar_B=B^{-1}$ and therefore we deduce that 
there are $g_{\text{Landau}}$ states that have the same energy, where 
\beq
g_{\text{Landau}} 
\ \ = \ \ \frac{L_x L_y}{2\pi \hbar_B}
\ \ = \ \ \frac{L_x L_y}{2\pi}\mathcal{B} 
\eeq
Later we shall explain that it leads to the $e^2/(2\pi\hbar)$ 
quantization of the Hall conductance.  
Beside the ${(X,Y)}$ degree of freedom we have of course 
the kinetic ${(v_x,v_y)}$ degree of freedom. 
Form the transformed Hamiltonian in the new coordinates  
we clearly see that the kinetic energy is quantized 
in units of ${\omega_B}$.  
Therefore, it is natural to label the eigenstates 
as ${|\ell , \nu \rangle}$, where: ${\nu = 0,1,2,3 \dots }$, 
is the kinetic energy index, and $\ell$ labels 
the possible values of the constant of motion~$Y$.

\sheadC{Electron in Hall geometry: The Landau states} 

We go back to the Hamiltonian that describes the motion of a particle 
in the Hall bar geometry (see beginning of previous section).  
Recall that we have periodic boundary conditions 
in the $x$ axis, and an arbitrary confining potential $V(y)$ 
in the $y$ direction. We would like to find the eigenstates and 
the eigenvalues of the Hamiltonian. 
The key observation is that in the Landau gauge the 
momentum operator $p_x$ is a constant of motion.
It is more physical to re-phrase this statement 
in a gauge independent way. Namely, the constant 
of motion is in fact   
\beq
Y \ \ = \ \ \hat{y}-\frac{1}{\omega_B}\hat{v}_x
\ \ = \ \ -\frac{1}{\mathcal{B}}\hat{p}_x 
\eeq
which represents the $y$ location of the classical cycloid. 
In fact the eigenstates that we are going to find 
are the quantum mechanical analogue of the classical cycloids. 
The eigenvalues of ${\hat{p}_x }$ 
are ${ {2 \pi}/{L_x} \times \mbox{\small integer} }$. 
Equivalently, we may write that the eigenvalues 
of ${\hat{Y} }$ are:
\beq
Y_{\ell} \ \ = \ \ \frac{2 \pi}{\mathcal{B}L_x}\ell,  
\ \ \ \ \ \ \ \ \ \ \ \ 
\mbox{[$\ell =$ integer]} 
\eeq
That means that the $y$ distance between the eigenstates  
is quantized. According to the "separation of variables theorem"  
the Hamiltonian matrix is a block diagonal matrix 
in the basis in which the ${\hat{Y}}$ matrix is diagonal. 
It is natural to choose the basis ${|\ell, y \rangle}$ 
which is determined by the operators ${\hat{p}_x, \ \ \hat{y}}$.
\beq
\langle \ell, y | \mathcal{H} | \ell', y' \rangle \ \ 
= \ \ \delta_{\ell,\ell'} \, \mathcal{H}^{\ell}_{yy'} 
\eeq
It is convenient to write the Hamiltonian 
of the block ${\ell}$ in abstract notation (without indexes):
\beq
\mathcal{H}^{\ell} 
\ \ = \ \ \frac{1}{2\mass}\hat{p}^2_y 
+ \frac{\mathcal{B}^2}{2\mass} ( \hat{y}-Y_{\ell} )^2 + V( \hat{y} ) 
\eeq
Or, in another notation:
\beq
\mathcal{H}^{\ell} 
\ \ = \ \ \frac{1}{2\mass}\hat{p}^2_y + V_{\ell} ( \hat{y} ) 
\eeq
where the effective potential is:
\beq
V_{\ell} (y) 
\ \ = \ \ V(y) + \frac{1}{2}\mass \omega_\mathcal{B}^2 ( y - Y_{\ell} )^2 
\eeq
For a given ${\ell}$, we find the 
eigenvalues ${| \ell, \nu \rangle}$ of 
the one-dimensional Hamiltonian ${\mathcal{H}^{\ell}}$. 
The running index ${\nu=0,1,2,3, \dots }$ 
indicates the quantized values of the kinetic energy.

For a constant electric field we notice that this 
is the Schr\"{o}dinger equation of a displaced 
harmonic oscillator. More generally, the harmonic 
approximation for the effective potential is 
valid if the potential ${V(y)}$ is wide compared 
to the quadratic potential which is contributed 
by the magnetic field. In other words, we assume that 
the magnetic field is strong. We write the wave functions as:
\beq
|\ell,\nu \rangle 
\ \ \ \rightarrow \ \ \ 
\frac{1}{\sqrt{L_x}} \  
\eexp{-i (\mathcal{B}Y_{\ell}) x } \  
\varphi^{(\nu)}(y-Y_{\ell}) 
\eeq
We notice that ${\mathcal{B}Y_{\ell}}$ are 
the eigenvalues of the momentum operator. 
If there is no electrical field then 
the harmonic approximation is exact, 
and then ${\varphi^{(\nu)}}$ are the 
eigenfunctions of a harmonic oscillator. 
In the general case, we must "correct" them
(in case of a constant electric field 
they are simply shifted). 
If we use the harmonic approximation 
then the energies are:
\beq
E_{\ell,\nu} \ \ \approx \ \ V( Y_{\ell} ) \ + \ \left( \frac{1}{2} + \nu \right) \omega_B 
\eeq

\begin{center}
\putgraph[0.6\hsize]{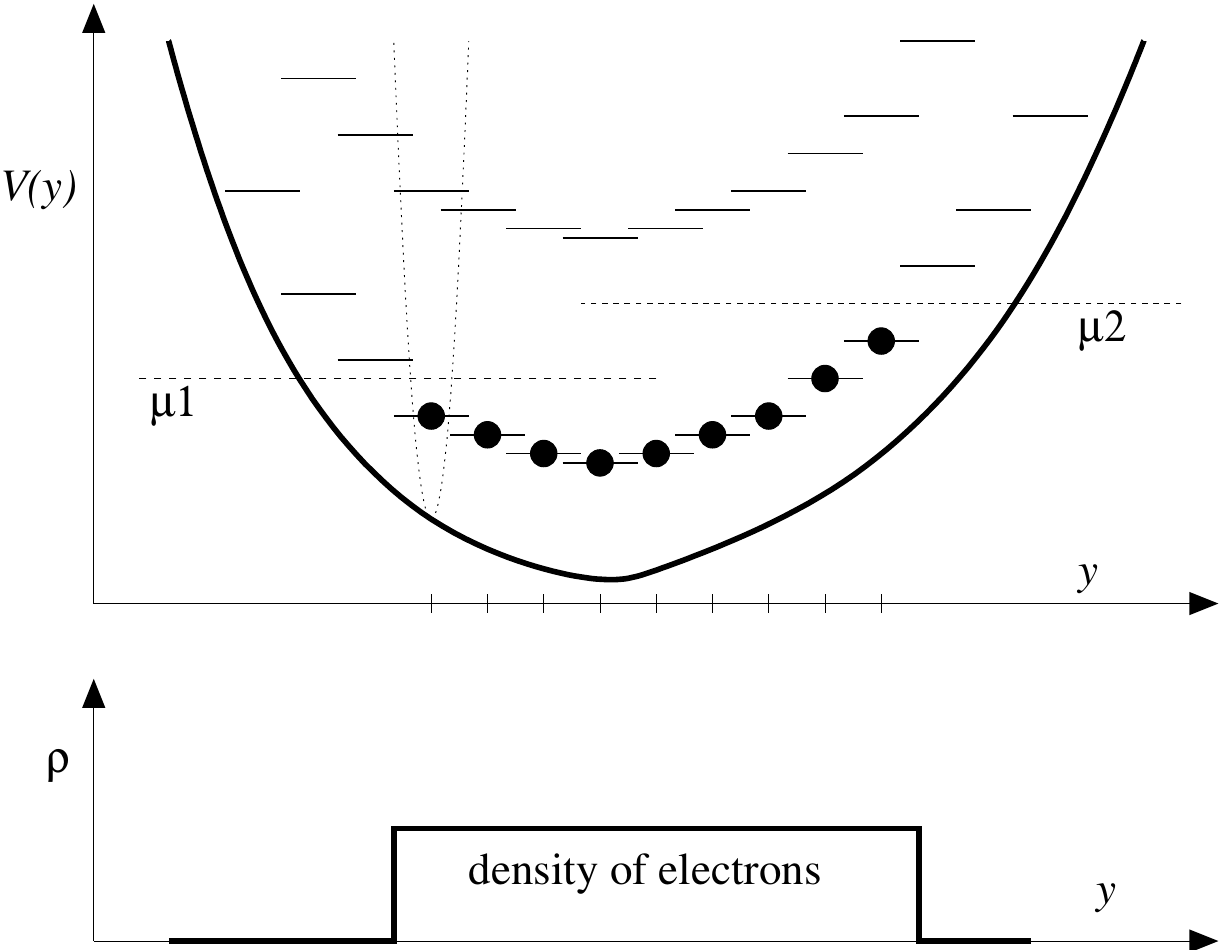} 
\end{center}

Plotting $E_{\ell,\nu}$ against $Y_{\ell}$ 
we get a picture of "energy levels" that 
are called "Landau levels" (or more precisely they 
should be called "energy bands"). 
The first Landau level is the collection of states with ${\nu=0}$. 
We notice that the physical significance of the term 
with ${\nu}$ is kinetic energy. The Landau levels 
are "parallel" to the bottom of the potential ${V(y)}$. 
If there is a region of width ${L_y}$ where the electric potential 
is flat (no electric field), then the eigenstates 
in that region (for a given ${\nu}$) would be degenerate 
in the energy (they would have the same energy). 
Because of the quantization of $Y_{\ell}$  
the number of particles that can occupy a Hall-bar of 
width ${ L_y }$ in each Landau level 
is $g_{\text{Landau}}=L_y/\Delta Y$, 
leading to the same result that we deduced earlier.   
In different phrasing, and restoring CGS units, 
the density of electrons in each Landau level is 
\beq
\rho_{\text{Landau}} \ \ = \ \ \frac{g_{\text{Landau}}}{L_xL_y}
\ \ = \ \ \frac{e}{2\pi\hbar c}\mathcal{B} 
\eeq
This leads to the $e^2/(2\pi\hbar)$ quantization of the Hall conductance.

\sheadC{Hall geometry with AB flux}

Here we discuss a trivial generalization 
of the above solution that helps later (next section) 
to calculate the Hall current. 
Let us assume that we add a magnetic flux ${\Phi}$ through the ring, 
as in the case of Aharonov-Bohm geometry. In this case, 
the vector potential is:
\beq
\vec{A} \ \ = \ \ \left(\frac{\Phi}{L_x}-\mathcal{B}y ,0,0\right) 
\eeq
We can separate the variables in the same way, and get:
\beq
E_{\ell,\nu} \ \ \approx \ \ V\left( Y_{\ell}
+ \frac{1}{\mathcal{B}L_x} \Phi \right ) 
\ + \ \left[ \frac{1}{2} + \nu \right] \omega_B 
\eeq

\sheadC{The Quantum Hall current}

We would like to calculate the current for an electron 
that occupies a Landau state ${| \ell, \nu \rangle}$:
\beq
&& \hat{J}_{x}(x,y) 
\ \ = \ \ 
\frac{1}{2}(v_x\delta(\hat{x}-x)\delta (\hat{y}-y)+h.c.) 
\\ \nonumber
&& \hat{J}_{x}(y) 
\ \ = \ \ 
\frac{1}{L_x} \int{\hat{J}_x(x,y) dx} 
\ \ = \ \ 
\frac{1}{L_x} \hat{v}_x \delta (\hat{y}-y) 
\\ \nonumber
&& \hat{I}_{x} 
\ \ = \ \ 
- \frac{\partial \mathcal{H}}{ \partial \Phi } 
\ \ = \,\,
\frac{1}{L_x}v_x
\ \ = \ \ 
\int \hat{J}_x(y) dy 
\eeq
Recall that ${v_x = (\mathcal{B}/\mass) (\hat{y}-\hat{Y})}$,  
and that the Landau states are eigenstates of ${\hat{Y}}$. 
Therefore, the current density of an occupied state is given by:
\beq
J_{x}^{\nu \ell}(y) 
\ \ = \ \ \langle {\ell\nu} | \hat{J}_{x}(y) | \ell \nu \rangle
\ \ = \ \ \frac{\mathcal{B} }{L_x\mass} 
\left\langle (\hat{y}-\hat{Y}) \ 
\delta (\hat{y}-y) \right \rangle 
\ \ = \ \ \frac{\mathcal{B} }{\mass L_x } (y-Y_{\ell}) \ 
\left| \varphi^{\nu}(y-Y_{\ell}) \right|^2 
\eeq
If we are in the region (${Y_{\ell}<y}$) we observe 
current that flows to the right (in the direction of the positive $x$ axis), 
and the opposite if we are on the other side. This is consistent 
with the classical picture of a particle moving clockwise in a circle. 
If there is no electric field, the wave function is symmetric 
around ${Y_{\ell}}$, so we get zero net current. 
If there is a non-zero electric field, it shifts the wave 
function and then we get a net current that is not zero. 
The current is given by:
\beq
I^{\ell\nu}_x \ \ = \ \ \int J^{\ell\nu}_x(y) dy 
\ \ = \ \ -\frac{\partial{E_{\ell\nu}}}{ \partial{\Phi} }
\ \ = \ \ - \frac{ 1 }{\mathcal{B}L_x} \left. \frac{dV(y)}{dy} \right|_{y=Y_{\ell}} 
\eeq
For a Fermi occupation of the Landau level we get:
\beq
I_{x} & \ \ = \ \ & \sum_{\ell} I_{x}^{\ell\nu}
\ \ = \ \ \int_{ y_1}^{y_2} \frac{dy } { 2\pi / (BL_x) } \left( - \frac{1}{\mathcal{B}L_x} \frac{ dV(y) } { dy } \right ) 
\\ \nonumber
& \ \ = \ \ & - \frac{1}{ 2\pi } (V(y_2)-V(y_1)) 
\ \ = \ \ - \frac{e}{2\pi\hbar} (\mu_2-\mu_1) 
\eeq
In the last equation we have restored the standard physical units. 
We see that if there is a chemical potential difference we get 
a current ${I_x}$. Writing $\mu=eV$, the Hall coefficient 
is ${e^2/(2\pi\hbar)}$ times the number of full Landau levels.

\sheadC{Hall effect and adiabtic transport}

The calculation of the Hall conductance is possibly the simplest 
non-trivial example for adiabatic non-dissipative response. 
The standard geometry is the 2D "hall bar" of dimension ${L_x\times L_y}$. 
We have considered what happens if the electrons are confined in 
the transverse direction by a potential $V(y)$. 
Adopting the Landauer approach it is assumed that the edges are 
connected to leads that maintain a chemical 
potential difference. Consequently there is a net current in the~$x$
direction. From the "Landau level" picture it is clear that the 
Hall conductance $G_{xy}$ is quantized in units $e^2/(2\pi\hbar)$.
The problem with this approach is that the more complicated
case of disordered $V(x,y)$ is difficult for handling. We therefore 
turn to a formal linear response approach [see section~22]. 
From now on we use  units such that ${e=\hbar=1}$.

We still consider a Hall bar ${L_x\times L_y}$, but now 
we impose periodic boundary condition such that ${\psi(L_x,y)= \eexp{i\phi_x}\psi(0,y)}$ 
and ${\psi(x,L_y)= \eexp{i\phi_y}\psi(x,0)}$. 
Accordingly the Hamiltonian depends on the parameters ${(\phi_x,\phi_y,\Phi_B)}$, 
where $\Phi_B$ is the uniform magnetic flux through 
the Hall bar in the $z$~direction. The currents ${I_x=(e/L_x)v_x}$ 
and $I_y=(e/L_y)v_y$ are conjugate to $\phi_x$ and $\phi_y$.
We consider the linear response relation ${I_y=-G_{yx}\dot{\phi_x}}$.
This relation can be written as ${dQ_y=-G_{yx}d\phi_x}$. 
The Hall conductance quantization means that a $2\pi$ variation 
of $\phi_x$ leads to one particle transported in the $y$ direction.
The physical picture is very clear in the standard $V(y)$ geometry: 
the net effect is to displace all the filled Landau level 
"one step" in the $y$ direction.   

We now proceed with a formal analysis to show that the Hall conductance
is quantized for general $V(x,y)$ potential. We can define 
a "vector potential" $A_n$ on the $(\phi_x,\phi_y)$ manifold. 
If we performed an adiabatic cycle the Berry phase would be 
a line integral over $A_n$. By Stokes theorem this can be 
converted into a  $d\phi_x d\phi_y$ integral over~$B_n$.  
However there are two complementary domains over which the 
surface integral can be done. Consistency requires that the 
result for the Berry phase would come out the same modulo~$2\pi$.
It follows that 
\beq
\frac{1}{2\pi}\int_0^{2\pi}\int_0^{2\pi} B_n d\phi_x d\phi_y 
\ \ = \ \ \mbox{integer} \ \ \mbox{[Chern number]} 
\eeq
This means that the $\phi$ averaged $B_n$ is quantized in units 
of $1/(2\pi)$. If we fill several levels the Hall conductance 
is the sum ${\sum_n B_n}$ over the occupied levels, namely
\beq
G_{yx} \ \ = \ \ \sum_{n \in \mbox{band}}\sum_m 
\frac{2\im[I^y_{nm}I^x_{mn}]}{(E_m-E_n)^2}
\eeq
If we have a quasi-continuum it is allowed to 
average this expression over $(\phi_x,\phi_y)$.
Then we deduce that the Hall conductance of 
a filled band is quantized. The result is of physical 
relevance if non-adiabatic transitions 
over the gap can be neglected.

\sheadC{The Hofstadter butterfly}

Let us consider the same Hall geometry but assume that the motion of 
the electron is bounded to a periodic lattice, such that the $x$ and $y$ 
spacings between sites are $a$ and $b$ respectively. In the absence 
of magnetic field the Hamiltonian is ${H=-2\cos(ap_x)-2\cos(bp_y)}$ 
where the units of times are chosen such that the hopping coefficient is unity.
The eigenenergies occupy a single band ${E \in [-4,+4]}$.
We now turn on a magnetic field. Evidently if the flux $\Phi$ through 
a unit-cell is increases by $2\pi$ the spectrum remains the same, 
accordingly it is enough to consider ${0< \Phi \le 2\pi}$. 
It is convenient to write the flux per unit cell as $\Phi = 2\pi \times (p/q)$.
Given $q$ we can define a super-cell that consists of $q$ unit cells, 
such that the Hamiltonian is periodic with respect to it. It follows from Bloch 
theorem that the spectrum consists of $q$ bands. Plotting the bands 
as a function of $\Phi$ we get the Hofstadter butterfly [Hofstadter, PRB 1976]. 
See figure [taken from the homepage of Daniel Osadchy]: the horizontal
axis is the flux and the vertical axis is the energy.   
Note that for ${q=2}$ the two bands are touching with zero gap at ${E=0}$, 
and therefore cannot be resolved. One can see how for intermediate values of $\Phi$ 
the Landau bands become distinct as expected from the standard analysis.    
  
\includegraphics[height=9cm]{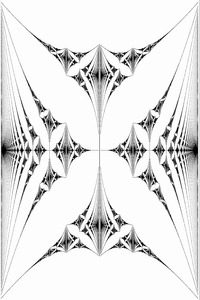}

\sheadC{The fractional Hall effect}

Instead of using rectangular geometry let us assume that we 
have a circular geometry. Using the symmetric gauge one observes 
that ${L_z=m}$ is a good quantum number. Hence the eigenstates 
can be written as $\psi(x,y) = R(r) \eexp{im\varphi}$, 
where $m$ is integer, and $R(r)$ is a radial function.   
Given $m$ the lowest energy is found to be $E=(1/2)\omega_B$, 
and for the radial function one obtains ${R^{(m)}(r) \propto r^m \exp(-(1/4)[r/r_B]^2)}$,  
where $r_B = (eB)^{-1/2}$ is identified as the cyclotron radius in the lower Landau level.
This radial function is picked at ${r \approx (2m)^{1/2} r_B}$. 
If the electrons are confined in a disc of radius $R$, 
then $m$ is allowed to range from zero to $(1/2)[R/r_B]^2$, 
which is the expected degeneracy of the first Landau level.
Using the notation ${z=x+iy}$ and working with length units
such that ${r_B=1}$, the degenerate set of orbitals in the 
lowest Landau level can be written compactly 
as ${\psi^{(m)}(z) \propto z^m \exp(-(1/4)z^2)}$.

If the potential looks like a shallow bowl, then $N$ non-interacting 
Fermions would occupy at zero temperature the lowest orbitals ${m=1,2,...,N}$.    
The many body wavefunction of $N$ fermions is a Slater determinant  
\beq
\Psi(z_1,z_2,...,z_N)
\ \ = \ \ f(z_1,z_2,...,z_N) \ \exp\left[- \frac{1}{4}\sum_i z_i^2 \right]  
\ \ \ = \ \ \ \left( \prod_{\langle ij \rangle} (z_i-z_j)^q \right) \ \exp\left[- \frac{1}{4}\sum_i z_i^2 \right]  
\eeq
with $q=1$. Let us take into account that the electrons repel each other.
In such a case it might be advantageous to have a more dilute population 
of the $m$ orbitals. We first note that if we make a Slater determinant 
out of states that have total angular momentum ${M=m_1+m_2+...+m_N}$ the 
result would be with a polynomial $f$ that all his terms are of degree~$M$. 
Obviously $M$ is a good quantum number, and there are many ways to form 
an $M$ states. So a general $M$ state is possibly a superposition 
of all possible Slater determinants. Laughlin [PRL 1983] has made an 
educated guess that odd values of $q$, that are consistent with the 
anti-symmetry requirement,  would minimize the cost of the repulsion. 
His guess turns out to be both a good approximation and an exact result 
for a delta repulsion. In such states the occupation extends up to 
the orbital ${m=Nq}$. It corresponds to filling fraction ${\nu=1/q}$.
Indeed Hall plateaus have been observed for such values. This is known 
as the fractional Hall effect.

\sheadC{The spin Hall effect}

Consider an electron that is constraint to move in a 2D sample, 
experiencing an in-plane electric field $\mathcal{E}$ 
that is exerted by the confining potential $V(x,y)$. 
There is no magnetic field, yet there is always a spin-orbit interaction:
\beq
H_{\mbox{SO}} 
\ \ = \ \ C \ \bm{\mathcal{E}} \times \bm{p} \cdot \bm{\sigma}
\ \ = \ \ C \ \sigma_z \ (\bm{n} \times \bm{\mathcal{E}}) \cdot \bm{p}
\eeq  
where $C$ is a constant, and $n$ is a unit vector in the $z$ direction. 
Note that in this geometry only the $z$ component of the spin is involved, 
hence $\sigma_z=\pm 1$ is a good quantum number. One observes that
the "down" electrons experience a vector potential ${\bm{A} = C \bm{n} \times \bm{\mathcal{E}}}$, 
that looks like a rotated version of $\mathcal{E}$. 
This has formally the same effect as that of a perpendicular magnetic field. 
The "up" electrons experience the same field with an opposite sign.    

\includegraphics[height=4cm]{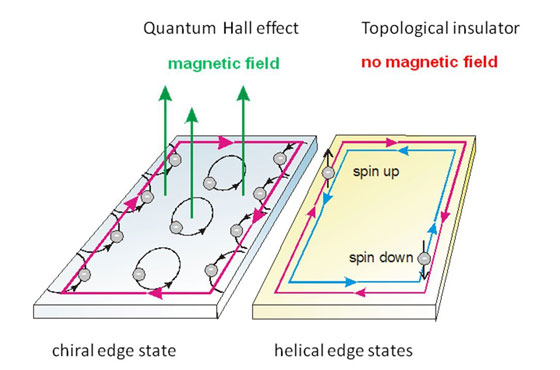}

Considering an electron that has an "up" or "down" spin 
one observes that its direction of motion 
along the edge of a potential wall is clockwise or anticlockwise
respectively. The spin is "locked" to the direction of the motion.    
This implies that it is very difficult to back-scattered such an electron: 
If there is a bump in the potential along the wall it can 
reverse the direction, but not the spin, hence there is a zero 
matrix element for back-scattering from $k\uparrow$ to $-k\downarrow$.
In order to have a non-zero matrix element one has 
to add magnetic field or magnetic impurities, 
breaking the time reversal symmetry that "protected" 
the undisturbed motion of the electron.

\newpage
\sheadB{Motion in a central potential}

\sheadC{The Hamiltonian} 

Consider the motion of a particle under 
the influence of a spherically symmetric 
potential that depends only on  
the distance from the origin. We can write 
the Hamiltonian in ``spherical coordinates" 
as a sum of radial term and an additional 
azimuthal term that involves the 
generators $\vec{L}= \vec{r} \times \vec{p}$. 
Namely,    
\beq
\mathcal{H} 
\ \ = \ \ \frac{1}{2\mass} p^2+eV(r) 
\ \ = \ \ \frac{1}{2\mass}\left(p_r^2 + \frac{1}{r^2} L^2\right) + eV(r) 
\eeq
In order to go form the first to the second expression 
we have used the vector style version of ${1=\cos^2+\sin^2}$, 
which is ${ A^2 B^2 = (A \cdot B)^2 +  (A \times B)^2}$, 
with $A=r$ and $B=p$. One should be careful about operator ordering. 
Optionally we can regard this identity as the differential 
representation of the Laplacian in spherical coordinates, with      
\beq
p_r^2 \ \ \rightarrow \ \ -\frac{1}{r}\frac{\partial^2}{\partial r^2} r 
\eeq
The Hamiltonian commutes with rotations:
\beq
[\mathcal{H},\hat{R}] \ \ = \ \ 0
\eeq
And in particular:
\beq
[\mathcal{H},L^2] \ \ = \ \ 0
\\ \nonumber
[\mathcal{H},L_z] \ \ = \ \ 0
\eeq
According to the separation of variables theorem 
the Hamiltonian becomes block diagonal in 
the basis which is determined by $L^2$ and $L_z$.   
The states that have definite $\ell$ 
and $m$ quantum numbers are of the 
form ${\psi(x,y,z) = R(r)Y^{\ell m}(\theta,\varphi)}$, 
so there is some freedom in the choice 
of this basis. The natural choice 
is ${ |r, \ell, m \rangle }$
\beq
\langle r, \theta, \varphi | r_0,\ell_0,m_0 \rangle 
\ \ = \ \ Y^{\ell_0,m_0}(\theta,\varphi) \frac{1}{r} \delta(r-r_0) 
\eeq
These are states that "live" on spherical shells. 
Any wavefunction can be written as a linear 
combination of the states of this basis. Note that 
the normalization is correct (the volume element 
in spherical coordinates includes ${r^2}$). 
The Hamiltonian becomes 
\beq
&& \langle r,\ell,m | \mathcal{H} | r', \ell', m' \rangle 
\ \ = \ \ \delta_{\ell,\ell' }\delta_{m,m'} \mathcal{H}^{(\ell,m)}_{r,r'}
\\ 
&& \mathcal{H}^{(\ell,m)}
\ \ = \ \ \frac{1}{2\mass}\hat{p}^2 
+ \left( \frac{\ell(\ell+1)}{2\mass r^2}+V(r)\right ) 
\ \ = \ \ \frac{1}{2\mass}p^2 + V^{(\ell)}(r) 
\eeq
where ${p \rightarrow -i(d/dr)}$.
The wavefunction in the basis which has been 
defined above are written as  
\beq
|\psi\rangle 
\ \ = \ \ 
\sum_{r,\ell,m} u_{\ell m}(r) \ | r, \ell, m \rangle
\ \ \longmapsto \ \ 
\sum_{\ell m} \frac{u_{\ell m}(r)}{r} Y^{\ell,m}(\theta,\varphi)
\eeq 
In the derivation above we have made a "shortcut". 
In the approach which is popular in textbooks the basis 
is not properly normalized, 
and the wave function is written as 
${\psi(x,y,z) = \psi(r,\theta,\varphi) = R(r) Y^{\ell,m}(\theta,\varphi)}$, 
without taking the correct normalization measure 
into account. Only in a later stage they 
define ${u(r) = rR(r)}$.  
Eventually they get the same result. 
By using the right normalization of the basis 
we have saved an algebraic stage.

By separation of variables the Hamiltonian has 
been reduced to a semi-one-dimensional Schr\"{o}dinger operator 
acting on the wave function ${u(r)}$.   
By "semi-one-dimensional" we mean that ${0<r<\infty}$.  
In order to get a wave function ${\psi(x,y,z)}$ 
that is continuous at the origin, we must require  
the radial function ${R(r)}$ to be finite, 
or alternatively the function ${u(r)}$ has to be 
zero at ${r=0}$.

\sheadC{Eigenstates of a particle on a spherical surface}

The simplest central potential that we can consider 
is such that confine the particle to move within 
a spherical shell of radius $R$. Such potential 
can be modeled as ${V(r)=-\lambda\delta(r-R)}$. 
For ${\ell=0}$ we know that a narrow deep well has 
only one bound state. We fix the energy of this state 
as the reference. The centrifugal potential 
for ${\ell>0}$ simply lifts the potential 
floor upwards. Hence the eigen-energies are 
\beq
E_{\ell m} \ \  = \ \   \frac{1}{2\mass R^2} \ell(\ell+1) 
\eeq
We remind ourselves of the considerations leading 
to the degeneracies. The ground state ${Y^{00}}$, 
has the same symmetry as that of the Hamiltonian: 
both invariant under rotations. 
This state has no degeneracy. On the other hand, 
the state ${Y^{10}}$ has a lower symmetry and by rotating 
it we get~3 orthogonal states with the same energy. 
The degeneracy is a "compensation" for the low symmetry 
of the eigenstates: the symmetry of the energy level 
as a whole (i.e. as a subspace) is maintained.

The number of states ${N(E)}$ up to energy~$E$, 
that satisfy ${E_{\ell m}<E}$, is easily found. 
The density of states turns out to be constant:
\beq
\frac{dN}{dE} \ \   \approx \ \   \frac{\mass}{2\pi\hbar^2}\mathsf{A}, 
\ \ \ \ \ \ \ \ \ \ \  \mathsf{A}=4\pi R^2
\eeq
It can be proved that this formula is valid also for 
other surfaces: to leading order only the surface area $\mathsf{A}$ 
is important.  The most trivial example is obviously a square for which 
\beq
E_{n,m} 
\ \  = \ \   
\frac{\pi^2 }{2\mass L^2} (n^2+m^2) 
\eeq
The difference between the $E_{\ell m}$ spectrum 
of a particle on a sphere, and the $E_{n,m}$ spectrum 
of a particle on a square, is in the way that 
the eigenvalues are spaced, not in their average density.
The degeneracies of the spectrum are determined 
by the symmetry group of the surface on which the motion 
is bounded. If this surface has no special symmetries 
the spectrum is expected to be lacking systematic degeneracies.

\sheadC{The Hydrogen Atom}

The effective potential ${V^{\ell}(r)}$ that appears 
in the semi-one-dimensional problem includes the 
original potential plus a centrifugal potential 
(for ${\ell \ne 0}$). Typically, the centrifugal 
potential ${ +{1}/{r^2} }$ leads to the appearance 
of a potential barrier. 
Consequently there are "resonance states" 
in the range ${E>0}$, that can "leak" out through 
the centrifugal barrier (by tunnelling) 
into the outside continuum.
But in the case of the Hydrogen atom the attractive 
potential ${-{1}/{r}}$ wins over the centrifugal potential, 
and there is no such barrier. 
Moreover, unlike typical short range potentials, 
there are infinite number of bound states in the ${E<0}$ range. 
Another special property of the Hydrogen 
atom is the high degree of symmetry: the Hamiltonian commutes 
with the Runge-Lentz operators. This is manifested in the 
degeneracy of energy levels, which is much greater 
than expected from SO(3).

For sake of later reference we write the potential as:
\beq
V(r) \ \ = \ \ -\frac{\alpha c}{r}, \hspace{2cm} \alpha=\frac{e^2}{\hbar c} \approx  \frac{1}{137}
\eeq
Below we use as usual ${\hbar=1}$ units. 
Solving the radial equation (for each ${\ell}$ separately) one obtains:
\beq
E_{\ell ,m, \nu} \ \ = \ \ [\mass c^2] \ - \frac{\alpha^2 \mass c^2}{2(\ell+\nu)^2} 
\eeq
where ${ \nu = 1,2,3, \dots  }$.
The rest-mass of the atom has been added here in square brackets 
as a reminder that from a relativistic point of view we are dealing 
here with an expansion with respect to the fine structure constant~$\alpha$.
In the non-relativistic treatment this term is omitted. 
The energy levels are illustrated in the diagram below.
It is customary in text books to use the quantum number ${n=\ell+\nu}$ 
in order to label the levels. One should bare in mind that the~$n$  
quantum number has significance only for the~$1/r$ potential 
due to its high symmetry: it has no meaning if we have say ${1/r^5}$ potential.

\begin{center}
\putgraph[0.45\hsize]{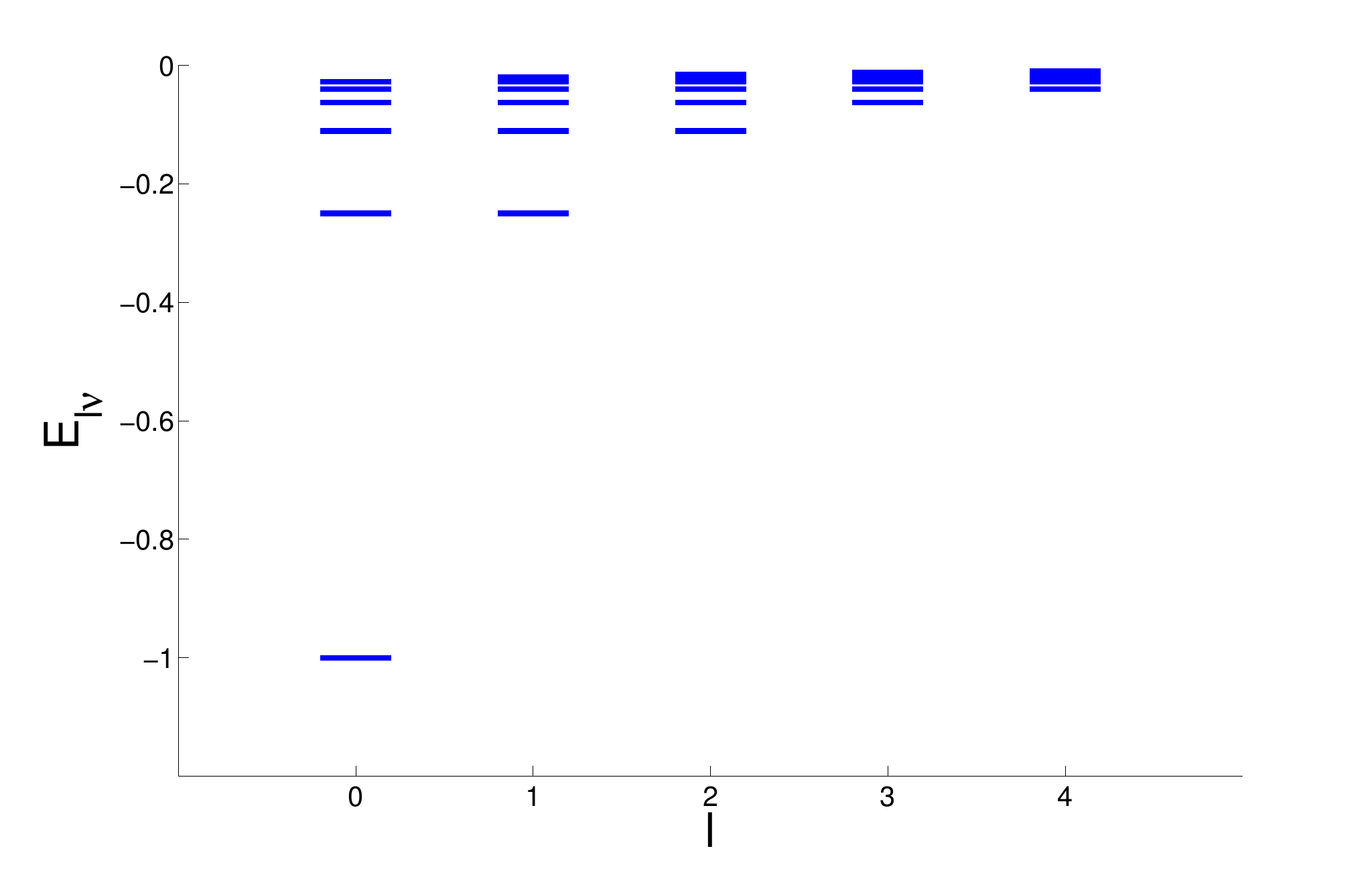} 
\end{center}

{\bf Degeneracies.-- }
We would like to remind the reader 
why there are degeneracies in the energy spectrum 
(this issue has been discussed in general in the section 
about symmetries and their implications).  
If the Hamiltonian has a single constant 
of motion, there will usually not be a degeneracy. 
According to the separation of variables theorem 
it is possible to transform to a basis in which 
the Hamiltonian has a block structure, 
and then diagonalize each block separately. 
There is no reason for a conspiracy amongst the 
blocks. Therefore there is no reason for a degeneracy. 
If we still get a degeneracy it is 
called an accidental degeneracy.  

But, if the Hamiltonian commutes with 
a "non-commutative group" then there 
are necessarily degeneracies that are 
determined by the dimensions of the irreducible 
representations. In the case of a central potential, 
the symmetry group is the "rotation group". 
In the special case of the potential ${-1/r}$ 
a larger symmetry group exists.

A degeneracy is a compensation for having  
eigenstates with lower symmetry compared with the Hamiltonian. 
The degree of degeneracy is determined 
by the dimensions of the irreducible representations. 
These statements were discussed 
in the Fundamentals~II section.
Let us paraphrase the argument in the present context: 
Assume that we have found an eigenstate of the Hamiltonian. 
If it has full spherical symmetry then there is no reason for degeneracy. 
These are the states with ${\ell=0}$. 
But, if we have found a state with a lower symmetry 
(the states with ${ \ell \ne 0 }$) then we can rotate 
it and get another state with the same energy level. 
Therefore, in this case there must be a degeneracy.

Instead of "rotating" an eigenstate, it is simpler (technically) 
to find other states with the same energy by using ladder 
operators. This already gives an explanation why the degree 
of the degeneracy is determined by the dimension of the irreducible 
representations. Let us paraphrase the standard argument 
for this statements in the present context: 
The Hamiltonian ${\mathcal{H}}$ commutes with all the rotations, 
and therefore it also commutes with all their generators and 
also with ${L^2}$. We choose a basis ${|n ,\ell, \mu \rangle}$ 
in which both ${\mathcal{H}}$ and ${L^2}$ are diagonal. 
The index of the energy levels $n$ is determined by ${\mathcal{H}}$, 
while the index $\ell$ is determined by ${L^2}$. 
The index ${\mu}$ differentiates states with the same 
energy and the same ${\ell}$. 
According to the "separation of variables theorem" every rotation 
matrix will have a "block structure" in this basis: each level 
that is determined by the quantum numbers ${(n,\ell)}$ 
is an invariant subspace under rotations. 
In other words, ${\mathcal{H}}$ together with ${L^2}$ induce 
a decomposition of the group representation. 
Now we can apply the standard procedure in order to conclude 
that the dimension of the (sub) representation 
which is characterized by the quantum number ${\ell}$, is ${2\ell+1}$. 
In other words, we have ``discovered" that the degree of the 
degeneracy must be ${2\ell+1}$ (or a multiple of this number). 

\ \\

{\bf Corrections.-- }
The realistic expression for the energy levels of Hydrogen atom contains 
a term that originates from the relativistic Dirac equation, 
plus a Lamb-Shift due to the interaction with the fluctuations 
of the electromagnetic vacuum, plus additional "hyperfine" correction 
due to the interaction with the nucleus:
\beq
E_{\ell,j,m,\nu} = mc^2\left(1+\left[\dfrac{\alpha}{\nu+\ell-(j+s)+\sqrt{(j+s)^2-\alpha^2}}\right]^2\right)^{-1/2} + \text{LambShift} + \text{HyperFine}
\eeq
Disregarding the hyperfine interaction, 
the good quantum numbers are ${\ell}$, and ${j}$, and ${s=1/2}$.
Defining ${n=\ell+\nu}$ and using ${(n,\ell,j,m)}$ as good quantum numbers,    
one observes that the eigen-energies are independent of $\ell$.
It is customary to expand the Dirac expression 
with respect to the the fine-structure constant ${\alpha\equiv e^2/(\hbar c)}$.
The leading ${\alpha^2}$ term is the same as that of Bohr.
The next ${\alpha^4}$ term can be regarded as the sum 
of relativistic kinetic ${p^4}$ correction,  
a Darwin term ${ \delta^3(r)}$ that affects $s$-orbitals,   
and a spin-orbit term ${ (1/r^3) L{\cdot}S}$. 
On top we have the ${\alpha^5}$ Lamb-shift correction, 
that can be approximated as an additional ${ \delta^3(r)}$ term.
The Darwin and the Lamb-shift terms can be interpreted 
as arising from the smearing of the ${1/r}$ interaction, 
either due to zitterbewegung or due to the fluctuations 
that are induced by the electromagnetic field, respectively.
The Lamb shift splits the degeneracy of 
the ${j=1/2}$ levels "2s" and "2p".  
The Lamb shift physics is also responsible for the 
anomalous value of the spin-orbit coupling (${g \approx 2 + (\alpha/\pi)}$).

The spin-related fine-structure will be discussed in the next lectures. 
It will be regarded as arising from "corrections" to the non-relativistic Schroedinger Hamiltonian. 
In particular we shall see how the quantum number $j$ arises due 
to the addition of the angular momentum of the spin, and the presence 
of the spin-orbit interaction.

\newpage
\sheadB{The Hamiltonian of a spin 1/2 particle}

\sheadC{The Hamiltonian of a spinless particle} 

The Hamiltonian of a spinless particle can be written as:
\beq
\mathcal{H} \ \ = \ \ \frac{1}{2\mass}( \vec{p}-e\vec{A}(r))^2+eV(r) 
\ \ = \ \ \frac{p^2}{2\mass}
- \frac{e}{2\mass}(\vec{A}\cdot \vec{p}
+ \vec{p}\cdot \vec{A})+ \frac{e^2}{2\mass}A^2+eV(r) 
\eeq
where $c{=}1$. We assume that the field is 
uniform ${\mathcal{B}=(0,0,\mathcal{B}_0)}$. 
In the previous lectures we saw that 
this field can be derived from ${\vec{A}= (-\mathcal{B}_0y,0,0)}$, 
but this time we use a different gauge, called "symmetrical gauge":
\beq
\vec{A}
\ \ = \ \ \left(-\frac{1}{2}\mathcal{B}_0y, \frac{1}{2}\mathcal{B}_0x ,0\right)
\ \ = \ \ \frac {1}{2} \mathcal{B}\times \vec{r} 
\eeq
Triple vector multiplication is associative, 
and therefore we have the following identity:
\beq
\vec{A}\cdot\vec{p} 
\ \ = \ \ \frac{1}{2}( \mathcal{B}\times \vec{r}) \cdot\vec{p} 
\ \ = \ \ \frac{1}{2} \mathcal{B}\cdot (\vec{r}\times\vec{p}) 
\ \ = \ \ \frac{1}{2} \mathcal{B}\cdot \vec{L} 
\eeq
Substitution into the Hamiltonian gives:
\beq
\mathcal{H} 
\ \ = \ \ \frac{p^2}{2\mass}+eV(r) 
-\frac{e}{2\mass}\mathcal{B}\cdot\vec{L} 
+ \frac{e^2}{8\mass}(r^2 \mathcal{B}^2-(\vec{r} \cdot \mathcal{B})^2 ) 
\eeq
Specifically for a homogeneous field in the $z$ axis we get
\beq
\mathcal{H} 
\ \ = \ \ \frac{p^2}{2\mass}+eV(r) 
-\frac{e}{2\mass}\mathcal{B}_0 L_z 
+\frac{e^2}{8\mass}\mathcal{B}_0^2(x^2+y^2) 
\eeq
The two last terms are called the "Zeeman term" and the "diamagnetic term". 
\beq
\mathcal{H}_{\mbox{Zeeman,orbital motion}} 
\ \ = \ \ -\frac{e}{2\mass c} \mathcal{B} \cdot \vec{L} 
\hspace{3cm} \text{[here $c$ is restored]}
\eeq

\sheadC{The additional Zeeman term for the spin} 

Spectroscopic measurements on atoms have shown, 
in the presence of a magnetic field,  
a double (even) Zeeman splitting of the levels, 
and not just the expected "orbital" splitting 
(which is always odd).
From this Zeeman has concluded that the electron has 
another degree of freedom which is called "spin $1/2$".  
The Hamiltonian should include an additional term:
\beq
\mathcal{H}_{\mbox{Zeeman,spin}} 
\ \ =  \ \ -g\frac{e}{2\mass c} \mathcal{B} \cdot \vec{S}
\eeq
The spectroscopic measurements of the splitting 
make it possible to determine the gyromagnetic 
coefficient to a high precision. The same 
measurements were conducted also for protons, 
neutrons (a neutral particle!) and other particles:

Electron: ${g_e = 2.0023 }$  \ \ \ \ [with $e=-|e|$ reflecting its negative charge] \\
Proton: \ \ ${g_p= 5.5854 }$  \ \ \ \ [with $e=+|e|$ reflecting its positive charge]  \\
Neutron: ${g_n= 3.8271 }$ \ \ \ \ [with $e=-|e|$ as if it were charged negatively] 

The implication of the Zeeman term in the Hamiltonian 
is that the wavefunction of the electron 
precesses with the frequency
\beq
\Omega \ \ = \ \ -\frac{e}{2\mass c}\mathcal{B} 
\eeq
while the spin of the electron precesses 
with a twice larger frequency
\beq
\Omega \ \ = \ \ -g_e\frac{e}{2\mass c}\mathcal{B}, 
\ \ \ \ \ \ \ \ \ \ \ \Big[g_e \approx 2\Big]
\eeq

\sheadC{The spin orbit term}

The added Zeeman term describes the interaction of the spin with the 
magnetic field. In fact, the "spin" degree of freedom 
(and the existence of anti-particles) is inevitable because 
of relativistic considerations of invariance under 
the Lorentz transformation. These considerations 
lead to Dirac's Hamiltonian. There are further "corrections" 
to the non-relativistic Hamiltonian that are required in order 
to make it "closer" to Dirac's Hamiltonian. 
The most important of these corrections is the "spin orbit interaction":
\beq
\mathcal{H}_{spin-orbit} 
\ \ = \ \ -(g{-}1)\frac{e}{2(\mass c)^2} \ \vec{\mathcal{E}} \times \vec{p} \ \cdot \ \vec{S} 
\eeq
In other words, the spin interacts with the 
electric field. This interaction depends on its  
velocity. This is why the interaction is called 
spin-orbit interaction. If there is also 
a magnetic field then we have the additional 
interaction which is described by the Zeeman term.

We can interpret the "spin-orbit" interaction 
in the following way: even if there is 
no magnetic field in the "laboratory" reference 
frame, still there is a magnetic field 
in the reference frame of the particle, 
which is a moving reference frame.  
This follows from Lorentz transformation:
\beq
\tilde{\mathcal{B}}
\ \ = \ \ \mathcal{B} - (1/c)\vec{\mathbf{v}}_{\mbox{frame}} \times \mathcal{E}
\eeq
It looks by this argument that the spin-orbit term should be 
proportional to $g$, while in fact it is proportional to ${(g{-}1)}$.
The extra contribution is called ``Thomas precession" and has 
a purely kinematical reason [discussed in the book of Jackson]. 
The physical picture is as follows: the spin-orbit term originates 
from the component of the electric field that is perpendicular
to the velocity; This leads to a rotated motion; 
In the rotating rest-frame of the particle one observes
precession due to the Zeeman interaction; Going back to 
the laboratory frame the Lorentz transformation implies 
that the spin experiences an extra magnetic-like field, 
analogous to Coriolis. This is because the laboratory frame 
is rotating with respect to the rest frame of the particle.

We summarize this section by writing the common non-relativistic 
approximation to the Hamiltonian of a particle with spin $1/2$. 
\beq
\mathcal{H} \ \ = \ \ \frac{1}{2\mass}\left( \vec{p}-\frac{e}{c}\vec{A}(r) \right)^2
\ +eV(r) 
\ -g\frac{e}{2(\mass c)} \mathcal{B} \cdot \vec{S} 
\ -(g{-}1)\frac{e}{2(\mass c)^2} (\mathcal{E}\times \vec{p}) \cdot \vec{S} 
\eeq
In the case of spherically symmetric potential $V(r)$ 
the electric field is 
\beq
\mathcal{E} \ \ = \ \ - \frac{V'(r)}{r}\vec{r}
\eeq
Consequently the Hamiltonian takes the form (here again $c{=}1$):
\beq
\mathcal{H} 
\ = \  
\frac{1}{2\mass}\left( p_r^2 + \frac{1}{r^2} L^2 \right) 
+ eV(r) + \frac{e^2}{8\mass}\mathcal{B}^2 r_{\perp}^2
-\frac{e}{2\mass} \mathcal{B} \cdot \vec{L}
-g\frac{e}{2\mass} \mathcal{B} \cdot \vec{S}
+ (g{-}1)\frac{e}{2\mass^2}\frac{V'(r)}{r} \vec{L} \cdot \vec{S} 
\eeq

\newpage
\sheadC{The Dirac Hamiltonian}

In the absence of an external electromagnetic field the 
Hamiltonian of a free particle should be a function of 
the momentum operator alone $\mathcal{H}=h(\hat{p})$ 
where ${\hat{p}=(\hat{x},\hat{y},\hat{z})}$. 
Thus $p$ is a good quantum number. The reduced Hamiltonian 
within a $p$ subspace is $\mathcal{H}^{(p)}=h(p)$. 
If the particle is spineless $h(p)$ is a number 
and the dispersion relation is $\epsilon=h(p)$. 
But if the particle has an inner degree of freedom (spin) 
then $h(p)$ is a matrix. 
In the case of Pauli Hamiltonian ${h(p)=(p^2/(2\mass))\hat{1}}$ 
is a ${2 \times 2}$ matrix. We could imagine a more complicated 
possibility of the type ${h(p)= \sigma \cdot p + ...}$. 
In such case $p$ is a good quantum number, 
but the spin degree of freedom is no longer degenerated:   
Given~$p$, the spin should be polarized either 
in the direction of the motion (right handed polarization) 
or in the opposite direction (left handed polarization).
This quantum number is also known as helicity.
The helicity might be a good quantum number, 
but it is a "Lorentz invariant" feature only for 
a massless particle (like a photon) that travels 
in the speed of light, else one can always transform 
to a reference frame where $p=0$ 
and the helicity become ill defined.

Dirac has speculated that in order to have a Lorentz 
invariant Schrodinger equation (${d\psi/dt=...}$) for the evolution,  
the matrix $h(p)$ has to be linear (rather than quadratic) 
in $p$. Namely $h(p)=\alpha \cdot p + \const\beta$. 
The dispersion relation should be consistent with 
${\epsilon^2=\mass^2+p^2}$ which implies ${h(p)^2=(\mass^2+p^2)\hat{1}}$.
It turns out that the only way to satisfy the latter 
requirement is to assume that $\alpha$ and $\beta$ 
are $4\times4$ matrices:
\beq
\alpha_j = \left[ \amatrix{\bm{0} & \bm{\sigma}_j \cr \bm{\sigma}_j & \bm{0}} \right]
\hspace*{2cm} 
\beta = \left[ \amatrix{\bm{1} & \bm{0} \cr \bm{0} & -\bm{1}} \right]
\eeq
Hence the Dirac Hamiltonian is 
\beq
\mathcal{H} \ \ = \ \ \alpha \cdot p + \beta \mass 
\eeq
It turns out that the Dirac equation, 
which is the Schrodinger equation with Dirac's Hamiltonian,   
is indeed invariant under Lorentz. 
Given $p$ there are 4~distinct eigenstates 
which we label as ${|p,\lambda\rangle}$.
The $4$~eigenstates  are determined via the diagonalization 
of $h(p)$. Two of them have the dispersion 
${\epsilon = +\sqrt{p^2+\mass^2}}$ 
and the other two have the dispersion 
${\epsilon = -\sqrt{p^2+\mass^2}}$. 
It also turns out that the helicity (in a give reference frame) is a good quantum number.  
The helicity operator is $\Sigma\cdot p$ where
\beq
\Sigma_j = \left[ \amatrix{\bm{\sigma}_j & \bm{0} \cr \bm{0} & \bm{\sigma}_j} \right]
\eeq
This operator commutes with Dirac Hamiltonian.   
Thus the electron can be right or left handed  
with either positive or negative mass. 
Dirac's interpretation for this result 
was that the "vacuum" state of the universe 
is like that of an intrinsic semiconductor 
with gap $2\mass c^2$. 
Namely, instead of talking about electrons 
with negative mass we can talk about 
holes (positrons) with positive mass.  
The transition of an electron from an 
occupied negative energy state to an empty     
positive energy state is re-interpreted 
as the creation of an electron positron pair.
The reasoning of Dirac has lead to the conclusion 
that particles like the electron must have 
a spin as well as antiparticle states.

\newpage
\sheadB{Implications of having "spin"}

\sheadC{The Stern-Gerlach effect} 

We first discuss what effect the Zeeman term 
has on the dynamics of a "free" particle. 
We shall see that because of this term, 
there is a force acting on the particle 
if the magnetic field is non-homogeneous. 
For simplicity of presentation we assume 
that the magnetic field is mainly in the Z~direction,  
and neglect its other components.
Defining ${r=(x,y,z)}$ the Hamiltonian takes the form
\beq
\mathcal{H} \ \ = \ \ \frac { \vec{p}^2}{2\mass} -g \frac{e}{2\mass} \mathcal{B}_z(r) S_z 
\eeq
We see that ${S_z}$ is a constant of motion. 
If particle is prepared with spin "up" it experiences   
an effective potential:
\beq
V_{\mbox{eff}} \ \ = \ \ -\frac{1}{2} g \frac{e}{2\mass} \mathcal{B}_z(r) 
\eeq
A a particle with spin "down" experiences an inverted potential  
(with the opposite sign). That means that the direction 
of the force depends on the direction of the spin. 
We can come to the same conclusion by looking at the 
equations of motion. The velocity of the particle is 
\beq
\frac{d}{dt} \langle r \rangle 
\ \ = \ \ \langle i [ \mathcal{H},r ] \rangle 
\ \ = \ \ \left\langle \frac{1}{\mass} \left ( \vec{p}- A(r) \right ) \right\rangle 
\eeq
This still holds with no change. But what about the 
acceleration? We see that there is a new term:
\beq
\frac{d}{dt} \langle v \rangle 
\ \ = \ \ \langle i [ \mathcal{H},v ] \rangle 
\ \ = \ \ \frac{1}{\mass} \left\langle \,\mbox{Lorentz force}
\ + \ 
g \frac{e}{2\mass} (\nabla \mathcal{B}_z ) S_z \, \right\rangle 
\eeq
The observation that in inhomogeneous magnetic field  
the force on the particle depends on the spin orientation 
is used in order to measure the spin using a Stern-Gerlach apparatus.

\sheadC{The reduced Hamiltonian in a central potential}

We would like to consider the problem of electron in a central potential, 
say in the Hydrogen atom, taking into account the spin-orbit interaction.
This add an $L\cdot S$ term to the Hamiltonian. 
We first would like to clarify what are the surviving constants of motion.
The system still has symmetry to rotations, 
and therefore the full Hamiltonian as well as $L\cdot S$ 
commutes with ${J=L+S}$. The $J_i$ generate rotations 
of the wavefunction and the spin as one "package".  
The Hamiltonian does not commute with the generators 
${L_x,L_y,L_z}$ separately. For example ${[L \cdot S,L_x] \neq 0}$.
Still the Hamiltonian commutes with~$L^2$. 
In particular ${[L \cdot S,L^2]=0}$. This is clear 
because $L^2$ is a Casimir operators (commutes with all the generators).    
Note also that  
\beq
\vec{L} \cdot \vec{S} \ \ = \ \ \frac{1}{2} \left (J^2 - L^2 - S^2 \right ) 
\eeq
Form the above we deduce that $\ell$ is still a good quantum number, 
and therefore it makes sense to work with the orbital states  
\beq
|\ell m \nu \rangle \ \ \rightarrow \ \ R^{\ell\nu}(r) \, Y^{\ell m}(\theta,\varphi) 
\eeq
The Hamiltonian in the $\ell$ subspace is 
\beq
\mathcal{H}^{(\ell)} 
\ \ = \ \ \mathcal{H}_0^{(\ell)} 
- \frac{e}{2\mass} \mathcal{B} L_z 
- g\frac{e}{2\mass}\mathcal{B} S_z 
+ f(r) L\cdot S 
\eeq
So far no approximations were involved. 
If we further assume that the last terms in the Hamiltonian 
is a weak perturbation that does not "mix" energy levels, 
then we can make an approximation that reduces further Hamiltonian 
into the subspace of states that have the same energy:
\beq
\mathcal{H}^{(\ell\nu)} \ \ = \ \ -h L_z - g h S_z + v L \cdot S +\mbox{const} 
\eeq
where the first term with ${h=e\mathcal{B}/(2\mass)}$ 
is responsible for the orbital Zeeman splitting,  
and the second term with ${gh}$ is responsible 
to the spin-related Zeeman splitting. 
Note that $g=2.0023$. We also use the notation  
\beq
v \ \ = \ \ \langle \ell, \nu | f(r) | \ell, \nu \rangle 
\eeq

If the spin-orbit interaction did not exist, the dynamics 
of the spin would become independent of the dynamics 
of the wave function. But even with the spin-orbit 
interaction, the situation is not so bad. ${L \cdot S}$ couples 
only states with the same ${\ell}$. 
Furthermore the ${\ell=0}$ for which ${L \cdot S=0}$
are not affected by the spin-orbit interactions.

\sheadC{The Zeeman Hamiltonian}

From now we focus on the ${\ell=1}$ subspace in the 
second energy level of the Hydrogen atom. 
The Hamiltonian matrix is $6 \times 6$. 
The reduced Hamiltonian can be written 
in the standard basis ${ |\ell=1, \nu=1 , m_{\ell}, m_s \rangle}$. 
It is easy to write the matrices for the Zeeman terms:
\beq
&& L_z 
\ \ \rightarrow \ \  
\left( 
\amatrix{ 
1 & 0 & 0 \cr 
0 & 0 & 0 \cr 
0 & 0 & -1 } 
\right) 
\otimes 
\left( 
\amatrix{ 
1 & 0 \cr 
0 & 1 } 
\right) 
\ \ = \ \ \left( 
\amatrix{ 
1 & 0 & 0 & 0 & 0 & 0 \cr 
0 & 1 & 0 & 0 & 0 & 0 \cr 
0 & 0 & 0 & 0 & 0 & 0 \cr 
0 & 0 & 0 & 0 & 0 & 0 \cr 
0 & 0 & 0 & 0 & -1 & 0 \cr 
0 & 0 & 0 & 0 & 0 & -1 } 
\right) 
\\ \nonumber
&& S_z 
\ \ \rightarrow \ \ 
\left( 
\amatrix{ 
1 & 0 & 0 \cr 
0 & 1 & 0 \cr 
0 & 0 & 1 } 
\right) 
\otimes 
\frac{1}{2} 
\left( 
\amatrix{ 
1 & 0 \cr 
0 & -1 } 
\right) 
\ \ = \ \  
\frac{1}{2} 
\left( \amatrix{ 
1 & 0 & 0 & 0 & 0 & 0 \cr 
0 & -1 & 0 & 0 & 0 & 0 \cr 
0 & 0 & 1 & 0 & 0 & 0 \cr 
0 & 0 & 0 & -1 & 0 & 0 \cr 
0 & 0 & 0 & 0 & 1 & 0 \cr 
0 & 0 & 0 & 0 & 0 & -1 } 
\right) 
\eeq
But the spin-orbit term is not diagonal. 
The calculation of this term is more demanding:
\beq
\vec{L} \cdot \vec{S} 
\ \ = \ \ \frac{1}{2} \left(J^2 - L^2 - S^2 \right) 
\ \ = \ \ \frac{1}{2} \left(J^2 -\frac{11}{4}\right)
\eeq
The calculation of the matrix that represent $J^2$ 
in the standard basis is lengthy, 
though some shortcuts are possible (see lecture 
regarding "addition of angular momentum").
Doing the calculation, the result for the total Hamiltonian is  
\beq
\mathcal{H} 
\ \ \rightarrow \ \
-h
\left(
\begin{array}{cccccc}
1 & 0 & 0 & 0 & 0 & 0 \\
0 & 1 & 0 & 0 & 0 & 0 \\
0 & 0 & 0 & 0 & 0 & 0 \\
0 & 0 & 0 & 0 & 0 & 0 \\
0 & 0 & 0 & 0 & -1 & 0 \\
0 & 0 & 0 & 0 & 0 & -1 
\end{array}
\right)
-gh
\left(
\begin{array}{cccccc}
\frac{1}{2} & 0 & 0 & 0 & 0 & 0 \\
0 & -\frac{1}{2} & 0 & 0 & 0 & 0 \\
0 & 0 & \frac{1}{2} & 0 & 0 & 0 \\
0 & 0 & 0 & -\frac{1}{2} & 0 & 0 \\
0 & 0 & 0 & 0 & \frac{1}{2} & 0 \\
0 & 0 & 0 & 0 & 0 & -\frac{1}{2} 
\end{array}
\right)
+v
\left(
\begin{array}{cccccc}
\frac{1}{2} & 0 & 0 & 0 & 0 & 0 \\
0 & -\frac{1}{2} & \frac {1}{\sqrt{2} } & 0 & 0 & 0 \\
0 & \frac {1}{\sqrt{2} } & 0 & 0 & 0 & 0 \\
0 & 0 & 0 & 0 & \frac {1}{\sqrt{2} }& 0 \\
0 & 0 & 0 & \frac {1}{\sqrt{2} } & -\frac{1}{2} & 0 \\
0 & 0 & 0 & 0 & 0 & \frac{1}{2} 
\end{array}
\right)
\eeq
At this stage we can diagonalize the Hamiltonian, 
and find exact results for the eigen-energies.  
However, we shall see in the next section that 
it is possible to find some shortcuts that save 
much of the technical burden.

\sheadC{The Zeeman energies}

We write again the Hamiltonian that we want to diagonalize:
\beq
\mathcal{H} \ \ = \ \  \frac{v}{2} \left(J^2 -\frac{11}{4}\right) \ - h L_{z}  \ - g h S_{z}  
\eeq
There is a relatively simple way 
to figure out the representation of ${J^2}$ 
using the "addition theorem". 
Namely, after diagonalization it should become:
\beq
J^2 \ \ \rightarrow \ \ \left( 
\amatrix{ 
(15/4) & 0 & 0 & 0 & 0 & 0 \cr 
0 & (15/4) & 0 & 0 & 0 & 0 \cr 
0 & 0 & (15/4) & 0 & 0 & 0 \cr 
0 & 0 & 0 & (15/4) & 0 & 0 \cr 
0 & 0 & 0 & 0 & (3/4) & 0 \cr 
0 & 0 & 0 & 0 & 0 & (3/4) } 
\right) 
\eeq
It follows that the exact eigenenergies of the Hamiltonian 
in the absence of a magnetic field are:
\beq
&& E_{j=\frac{3}{2}} \ \ = \ \ v/2, \hspace{2cm} \mbox{[degeneracy $=4$]} 
\\ \nonumber
&& E_{j=\frac{1}{2}} \ \ = \ \ -v, \hspace{2cm} \mbox{[degeneracy $=2$]} 
\eeq
On the other hand, in a strong magnetic field 
the spin-orbit term is negligible, and we get: 
\beq
E_{m_{\ell},m_s} \ \ \approx \ \ -(m_{\ell} + g m_s)h 
\eeq
In fact there are two levels that are exact 
eigensates of the Hamiltonian for any $h$. These are:
\beq 
E_{j=\frac{3}{2},m_j=\pm \frac{3}{2}} \ \ = \ \  \frac{v}2 \mp \left(1 + \frac{g}{2}\right) h 
\eeq
The spectrum of $\mathcal{H}$ can be found for a range of $h$ values. 
See the Mathematica file \mbox{zeeman.nb}. 
The results (in units such that $v=1$) are illustrated 
in the following figure:

\begin{center}
\putgraph[0.7\hsize]{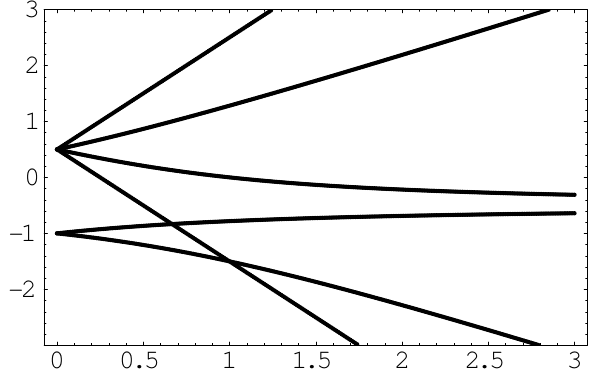}
\end{center}

\sheadC{Calculation of the Zeeman splitting}

For a weak magnetic field it is better to write 
the Hamiltonian in the  $|j,m_{j}\rangle$ basis, 
so as to have $\vec{L}\cdot \vec{S}$ diagonal, while  
the Zeeman terms are treated as a perturbation.
The calculation, using Mathematica, leads to 
\beq
\nonumber
\mathcal{H} \rightarrow 
\frac{v}{2}
\left(
\begin{array}{cccccc}
1 & 0 &  0 & 0  & 0  &  0 \\
0 & 1 &  0 & 0  & 0  &  0 \\
0 & 0 &  1 & 0  & 0  &  0 \\
0 & 0 &  0 & 1  & 0  &  0 \\
0 & 0 &  0 & 0  & -2 &  0 \\
0 & 0 &  0 & 0  & 0  & -2
\end{array}
\right)
-\frac{h}{3}
\left(
\begin{array}{cccccc}
3 & 0 &  0 & 0  & 0 &  0 \\
0 & 1 &  0 & 0  & -\sqrt{2} &  0 \\
0 & 0 & -1 & 0  & 0 &  -\sqrt{2} \\
0 & 0 &  0 & -3 & 0 &  0 \\
0 & -\sqrt{2} &  0 & 0  & 2 &  0 \\
0 & 0 &  -\sqrt{2} & 0  & 0 & -2 
\end{array}
\right)
-g\frac{h}{6}
\left(
\begin{array}{cccccc}
3 & 0 &  0 & 0  & 0  &  0 \\
0 & 1 &  0 & 0  & 2\sqrt{2}  &  0 \\
0 & 0 & -1 & 0  & 0  & 2\sqrt{2}  \\
0 & 0 &  0 & -3 & 0  &  0 \\
0 & 2\sqrt{2} &  0 & 0  & -1 &  0 \\
0 & 0 &  2\sqrt{2} & 0  & 0  &  1 
\end{array}
\right)
\eeq
Optionally, if we want to avoid the above numerical task,   
we can determine approximately the splitting of the $j$ 
multiplets using degenerate perturbation theory. 
In order to do so we only need to find the $j$ sub-matrices of the 
the Hamiltonian. We already know that they 
should obey the Wigner Eckart theorem. 
By inspection of the Hamiltonian we see 
that this is indeed the case. 
We have $g_L=\frac{2}{3}$ and $g_S=\frac{1}{3}$ for $j=\frac{3}{2}$, 
while $g_L=\frac{4}{3}$ and $g_S=-\frac{1}{3}$ for $j=\frac{1}{2}$.
Hence we can write 
\beq 
E_{j,m_j} \ \ = \ \ E_{j} - (g_M m_j) h 
\eeq
where $g_M = g_L + g g_S$  is associated 
with the vector operator $\vec{M} = \vec{L} + g \vec{S}$. 
In order to calculate $g_L$ and $g_S$  
we do not need to calculate the above $6\times 6$ matrices. 
We can simply can use the formulas 
\beq
g_{L} \ &=& \ \frac {\langle \vec{J}\cdot \vec{L}\rangle}{j(j+1)}
\ \ = \ \ \frac{j(j+1)+\ell(\ell+1)-s(s+1)}{2j(j+1)} \\ 
g_{S} \ &=& \ \frac {\langle \vec{J}\cdot \vec{S}\rangle}{j(j+1)}
\ \ = \ \ \frac{j(j+1)+s(s+1)-\ell(\ell+1)}{2j(j+1)}
\eeq

\sheadA{Special Topics}

\sheadB{Quantization of the EM Field}

\sheadC{The Classical Equations of Motion}

The equations of motion for a system which 
is composed of non-relativistic classical particles 
and EM fields are:
\beq
\mass_i \frac{d^{2}\mathbf{x}_i}{dt^{2}}
&=&e_i\mathcal{E}-e_i\mathcal{B}\times\dot{\mathbf{x}_i} 
\\ \nonumber
\nabla\cdot\mathcal{E}
&=&4\pi\rho 
\\ \nonumber
\nabla\times\mathcal{E}
&=&-\frac{\partial\mathcal{B}}{\partial t} 
\\ \nonumber
\nabla\cdot\mathcal{B}
&=&0 
\\ \nonumber
\nabla\times\mathcal{B}
&=&\frac{\partial\mathcal{E}}{\partial t} + 4\pi\vec{J}
\eeq
where
\beq
\rho(x) &=& \sum_i e_i\delta(x-\mathbf{x}_i) 
\\ \nonumber
J(x) &=& \sum_i e_i\dot{\mathbf{x}}_i\delta(x-\mathbf{x}_i)    
\eeq
We also note that there is a continuity equation which 
is implied  by the above definition and also 
can be regarded as a consistency requirement for the Maxwell equation: 
\beq
\frac{\partial\rho}{\partial t} = - \nabla \cdot J 
\eeq
It is of course simpler to derive the 
EM field from a potential $(V,A)$ as follows:
\beq
\mathcal{B} &=& \nabla \times A  
\\ \nonumber
\mathcal{E} &=& - \frac{ \partial A}{\partial t} - \nabla V 
\eeq
Then we can write an equivalent system of equations of motion
\beq
\mass_i \frac{d^{2}\mathbf{x}_i}{dt^{2}}
&=&e_i\mathcal{E}-e_i\mathcal{B}\times\dot{\mathbf{x}_i} 
\\ \nonumber
\frac{\partial^{2} A}{\partial t^{2}} 
&=& \nabla^{2}\vec{A} - \nabla \left(\nabla\cdot\vec{A} + \frac{\partial}{\partial t} V  \right) + 4 \pi J 
\\ \nonumber
\nabla^2 V &=&  -\frac{\partial}{\partial t} \left( \nabla\cdot\vec{A} \right) - 4\pi \rho 
\eeq

\sheadC{The Coulomb Gauge}

In order to further simplify the equations 
we would like to use a convenient gauge 
which is called the "Coulomb gauge". 
To fix a gauge is essentially like 
choosing a reference for the energy.   
Once we fix the gauge in a given reference 
frame ("laboratory") the formalism 
is no longer manifestly Lorentz invariant. 
Still the treatment is exact.

Any vector field can be written as a sum of 
two components, one that has 
zero divergence and another that has zero curl. 
For the case of the electric 
field $\mathcal{E}$, the sum can be written as: 
\beq
\mathcal{E}=\mathcal{E}_{\shortparallel}+\mathcal{E}_{\perp}
\eeq
where $\nabla\cdot\mathcal{E}_{\perp}=0$ 
and  $\nabla\times\mathcal{E}_{\shortparallel}=0$. 
The field $\mathcal{E}_{\perp}$ is called 
the transverse or solenoidal or "radiation" component, 
while the field $\mathcal{E}_{\shortparallel}$ 
is the longitudinal or irrotational or "Coulomb" component. 
The same treatment can be done for the magnetic 
field $\mathcal{B}$ with the observation 
that from Maxwell's equations $\nabla\cdot\mathcal{B}=0$ 
yields $\mathcal{B}_{\parallel}=0$. That means 
that there is no magnetic charge and hence 
the magnetic field $\mathcal{B}$ is entirely 
transverse. So now we have 
\beq
\mbox{EM field} = 
(\mathcal{E}_{\shortparallel}, \mathcal{E}_{\perp}, \mathcal{B}) 
= \mbox{Coulomb field} + \mbox{radiation field}
\eeq
Without loss of generality we can derive 
the radiation field from a transverse
vector potential. Thus we have: 
\beq
&& \mathcal{E}_{\shortparallel} = - \nabla V 
\\ \nonumber
&& \mathcal{E}_{\perp} = - \frac{\partial A}{\partial t} 
\\ \nonumber
&& \mathcal{B} = \nabla\times A 
\eeq
This is called the Coulomb gauge, 
which we use from now on.
We can solve the Maxwell equation  
for $V$ in this gauge, leading to the potential 
\beq
V(x) = \sum_{j} \frac{e_j} {|x-\mathbf{x}_j|}
\eeq
Now we can insert the solution 
into the equations of motion of the particles. 
The new system of equations is
\beq
\mass_i \frac{d^{2}\mathbf{x}_i}{dt^{2}}
&=&
\left[\sum_j \frac{e_ie_j \ \vec{n}_{ij}} {|\mathbf{x}_i-\mathbf{x}_j|^2}\right]
+ e_i \mathcal{E}_{\perp} - e_i\mathcal{B}\times\dot{\mathbf{x}_i}
\\ \nonumber
\frac{\partial^{2} A}{\partial t^{2}} 
&=& 
\nabla^{2}\vec{A} + 4 \pi J_{\perp} 
\eeq
It looks as if $J_{\shortparallel}$ is missing 
in the last equation. In fact it can be easily shown 
that it cancells with the $\nabla V$ term due 
to the continuity equation that relates $\partial_t\rho$
to $\nabla\cdot J$,  and hence $\partial_t \nabla V$
to $J_{\shortparallel}$ respectively.

\sheadC{Hamiltonian for the Particles}

We already know from the course in classical mechanics 
that the Hamiltonian, from which the equations of motion   
of {\em one} particles in EM field are derived, is    
\beq
\mathcal{H}^{(i)} \ \ = \ \  \frac{1}{2\mass_i} 
(\mathbf{p}_i - e_i A(\mathbf{x}_i))^{2} 
+ e_iV(\mathbf{x}_i)   
\eeq
This Hamiltonian assumes the presence of~$A$  
while~$V$ is potential which is created by all the other 
particles. Once we consider the many body system, 
the potential~$V$ is replaced by a mutual Coulomb interaction term,  
from which the forces on any of the particles are derived:
\beq
\mathcal{H} = \sum_i \frac{1}{2\mass_i} (\mathbf{p}_i-e_iA(\mathbf{x}_i))^{2} 
+  \frac{1}{2} \sum_{i,j} 
\frac{e_i e_j}{|\mathbf{x}_{i}-\mathbf{x}_{j}|}   
\eeq
By itself the {\em direct Coulomb interaction} between 
the particles seems to contradict relativity.  
This is due to the non-invariant way that we 
had separated the electric field into components.    
In fact our treatment is exact: as a whole the Hamiltonian 
that we got is Lorentz invariant, 
as far as the EM field is concerned. 

The factor ${1/2}$ in the Coulomb interaction is there in order 
to compensate for the double counting of the interactions.
What about the diagonal terms $i=j$? We can keep them 
if we want because they add just a finite constant 
term to the Hamiltonian. The "self interaction" infinite 
term can be regularized by assuming that each particle 
has a very small radius. To drop this constant from 
the Hamiltonian means that "infinite distance" between 
the particles is taken as the reference state for the energy.  
However, it is more convenient to keep this infinite 
constant in the Hamiltonian, because then we can write:
\beq
\mathcal{H} = \sum_i \frac{1}{2\mass_i} (\mathbf{p}_i-e_iA(\mathbf{x}_i))^{2} 
+  \frac{1}{8\pi} \int \mathcal{E}_{\shortparallel}^{2} d^3x
\eeq
In order to get the latter expression we have used 
the following identity:
\beq
\frac{1}{2}\int \frac{\rho(x)\rho(x')}{|x-x'|} d^3x d^3x' 
\ \ = \ \ 
\frac{1}{8\pi} \int \mathcal{E}_{\shortparallel}^{2} d^3x 
\eeq
The derivation of this identity is based on Gauss law, 
integration by parts, and using the fact that the Laplacian 
of $1/|x-x'|$ is a delta function.  
Again we emphasize that the integral diverges 
unless we regularize the physical size 
of the particles, or else we have to subtract 
an infinite constant 
that represents the "self interaction".

\sheadC{Hamiltonian for the Radiation Field}

So now we need another term in the Hamiltonian 
from which the second equation of motion is derived  
\beq
\frac{\partial^{2} A}{\partial t^{2}} 
-\nabla^{2}\vec{A}
= 4\pi \vec{J}_{\perp}
\eeq
In order to decouple the above equations into "normal modes" 
we write $\vec{A}$ in a Fourier series: 
\beq
\vec{A}(x) 
= \frac{1}{\sqrt{\mbox{volume}}} \sum_{k} \vec{A}_{k} \ \eexp{ikx} 
= \frac{1}{\sqrt{\mbox{volume}}} \sum_{k,\alpha} A_{k,\alpha}
\ \bm{\varepsilon}^{k,\alpha} \eexp{ikx} 
\eeq
where $\bm{\varepsilon}^{k,\alpha}$ for a given $k$ is a set 
of two orthogonal unit vectors. If $A$ were a general vector field 
rather than a transverse field, then we would have to include 
a third unit vector. Note that $\nabla \cdot A=0$  
is like  $ k \cdot \vec{A}_k =0$. Now we can rewrite 
the equation of motion as 
\beq
\ddot{A}_{k,\alpha} + \omega_k^2 A_{k,\alpha} = 4\pi J_{k,\alpha}
\eeq
where $\omega_k=|k|$.
The disadvantage of this Fourier expansion is that 
it does not reflect that $A(x)$ is a real field. 
In fact the $A_{k,\alpha}$ should satisfy 
$A_{-k,\alpha}=(A_{k,\alpha})^*$. In order to 
have proper "normal mode" coordinates we have 
to replace each pair of complex 
$A_{k,\alpha}$ and $A_{-k,\alpha}$ by 
a pair of real coordinates  
$A'_{k,\alpha}$ and $A''_{k,\alpha}$. Namely
\beq
A_{k,\alpha}  = \frac{1}{\sqrt{2}} [A'_{k,\alpha} + i A''_{k,\alpha}] 
\eeq
We also use a similar decomposition for $J_{k,\alpha}$. 
We choose the $1/\sqrt{2}$ normalization so as to have 
the following identity:
\beq
\int J(x) \cdot A(x) \ dx \ \ = \ \ 
\sum_{k,\alpha} J_{k,\alpha}^{*} A_{k,\alpha} = 
\sum_{[k],\alpha} 
(J'_{k,\alpha} A'_{k,\alpha} + J''_{k,\alpha} A''_{k,\alpha})
\ \ \equiv \ \ \sum_r J_r Q_r
\eeq
In the sum over degrees of freedom $r$ 
we must remember to avoid double counting. 
The vectors $k$ and $-k$ represent the 
same direction which we denote as $[k]$. 
The variable $A'_{-k,\alpha}$ 
is the same variable as $A'_{k,\alpha}$, 
and the variable $A''_{-k,\alpha}$ 
is the same variable as $-A''_{k,\alpha}$. 
We denote this set of coordinates $Q_r$, 
and the conjugate momenta as $P_r$.  
We see that each of the normal coordinates 
$Q_r$ has a "mass" that equals $1/(4\pi)$ [CGS!!!]. 
Therefore the conjugate "momenta" are
$P_r=[1/(4\pi)]\dot{Q}_r$, which up to a factor 
are just the Fourier components of the electric field. 
Now we can write the Hamiltonian as 
\beq
\mathcal{H}_{rad} = 
\sum_{r} \left[
\frac{1}{2\cdot\mbox{mass}} P_r^2 
+ \frac{1}{2}\mbox{mass}\cdot \omega_r^2 \ {Q_r^2} 
- J_r Q_r \right]
\eeq
where $r$ is the sum over all the degrees of freedom:  
two independent modes for each direction and polarization.
By straightforward algebra the sum can be written as  
\beq
\mathcal{H}_{rad} = 
\frac{1}{8\pi} \int (\mathcal{E}_{\perp}^{2}+\mathcal{B}^{2})d^3x
- \int \vec{J}\cdot A d^3x
\eeq
More generally, if we want to write the total Hamiltonian 
for the particle and the EM field we have to 
ensure that ${-\partial\mathcal{H}/\partial A(x) = J(x)}$. 
It is not difficult to figure out the the 
following Hamiltonian is doing the job: 
\beq
\mathcal{H} =  \mathcal{H}_{\tbox{particles}} 
+ \mathcal{H}_{\tbox{interaction}} 
+ \mathcal{H}_{\tbox{radiation}} =
\sum_i \frac{1}{2\mass_i} (\mathbf{p}_i-e_iA(\mathbf{x}_i))^{2} 
+ \frac{1}{8\pi} \int (\mathcal{E}^{2}+\mathcal{B}^{2}) d^3x
\eeq
The term that corresponds to $\vec{J}\cdot A$ is  
present in the first term of the Hamiltonian.
As expected this terms has a dual role: 
on the one hand it gives the Lorentz force 
on the particle, while on the other hand it 
provides the source term that drives the EM field.   
I should be emphasized that the way we write 
the Hamiltonian is somewhat misleading:  
The Coulomb potential term  
(which involves $\mathcal{E}_{\shortparallel}^{2}$) 
is combined with the "kinetic" term of the radiation 
field (which involves $\mathcal{E}_{\perp}^{2}$).

\sheadC{Quantization of the EM Field} 
 
Now that we know the "normal coordinates" of 
the EM field the quantization is trivial. 
For each "oscillator" of "mass" $1/(4\pi)$ 
we can define $a$ and $a^{\dag}$ operators
such that $Q_r=(2\pi/\omega)^{1/2}(a_r+a_r^{\dag})$.  
Since we have two distinct variables for each direction, 
we use the notations $(b, b^{\dag})$ and $(c, c^{\dag})$ respectively:
\beq
Q_{[k]'\alpha}
&= A_{k\alpha}'=A_{-k\alpha}' 
&= \sqrt{\frac{2\pi}{\omega_k}} \left(b_{[k]\alpha}+b_{[k]\alpha}^{\dag}\right)
\\ 
Q_{[k]''\alpha}
&= A_{k\alpha}''=-A_{-k\alpha}'' 
&= \sqrt{\frac{2\pi}{\omega_k}} \left(c_{[k]\alpha}+c_{[k]\alpha}^{\dag}\right)
\eeq
In order to make the final expressions look more elegant 
we use the following canonical transformation: 
\beq
a_{+} &=& \frac{1}{\sqrt{2}} (b+ic) 
\\ \nonumber
a_{-} &=& \frac{1}{\sqrt{2}} (b-ic) 
\eeq
It can be easily verified by calculating the commutators 
that the transformation from $(b,c)$ to $(a_{+},a_{-})$ 
is canonical. Also note that 
$b^{\dag}b+c^{\dag}c=a_{+}^{\dag}a_{+}+a_{-}^{\dag}a_{-}$. 
Since the oscillators (normal modes) are uncoupled 
the total Hamiltonian is a simple sum over all the modes: 
\beq
\mathcal{H} = \sum_{[k],\alpha} (\omega_kb_{k,\alpha}^{\dag}b_{k,\alpha}
+ \omega_kc_{k,\alpha}^{\dag} c_{k,\alpha})
= \sum_{k,\alpha} \omega_k a_{k,\alpha}^{\dag}a_{k,\alpha}
\eeq
For completeness we also write the expression for the field operators:
\beq
A_{k,\alpha} = \frac{1}{\sqrt{2}} (A'+iA'')
= \frac{1}{\sqrt{2}}\left[ 
\sqrt{\frac{2\pi}{\omega_k}}(b+b^{\dag}) 
+ i\sqrt{\frac{2\pi}{\omega_k}}(c+c^{\dag}) \right]
= \sqrt{\frac{2\pi}{\omega_k}}(a_{k,\alpha} + a_{-k,\alpha}^{\dag})
\eeq
and hence
\beq
\vec{A}(x) 
= \frac{1}{\sqrt{\mbox{volume}}} \sum_{k,\alpha} 
\sqrt{\frac{2\pi}{\omega_k}}(a_{k,\alpha} + a_{-k,\alpha}^{\dag})
\ \bm{\varepsilon}^{k,\alpha} \eexp{ikx} 
\eeq

The eigenstates of the EM field are
\beq
|n_{1},n_{2},n_{3}, \dots ,n_{k,\alpha}, \dots \rangle 
\eeq
We refer to the ground state as the vacuum state: 
\beq
\left\vert \mbox{vacuum} \right\rangle =
\left\vert 0,0,0,0, \dots  \right\rangle  
\eeq
Next we define the one photon state as follows: 
\beq
\left\vert \mbox{one photon state} \right\rangle =
\hat{a}_{k\alpha}^{\dag}
\left\vert \mbox{vacuum} \right\rangle  
\eeq
and we can also define two photon states (disregarding normalization): 
\beq
\left\vert \mbox{two photon state} \right\rangle =
\hat{a}_{k_2\alpha_2}^{\dag}\hat{a}_{k_1\alpha_1}^{\dag} 
\left\vert  \mbox{vacuum} \right\rangle 
\eeq
In particular we can have two photons in the same mode:
\beq
\left\vert \mbox{two photon state} \right\rangle =
(\hat{a}_{k_1\alpha_1}^{\dag})^2 
\left\vert  \mbox{vacuum} \right\rangle 
\eeq
and in general we can have $N$ photon states or any 
superposition of such states. 

An important application of the above formalism 
is for the calculation of spontaneous emission.
Let us assume that the atom has an excited 
level $E_B$ and a ground state $E_A$. The atom 
is prepared in the excited state, and the 
electromagnetic field is assume to be initially 
in a vacuum state.  According to Fermi Golden Rule   
the system decays into final states  
with one photon $\omega_k=(E_B-E_A)$. 
Regarding the atom as a point-like object 
the interaction term is 
\beq
\mathcal{H}_{\tbox{interaction}} 
\ \ \approx  \ \ 
-\frac{e}{c} A(0) \cdot \hat{v}
\eeq
where $\hat{v}=\hat{p}/\mass$ is the velocity 
operator. It is useful to realize  
that ${\hat{v}_{AB} = i(E_B-E_A) \hat{x}_{AB}}$. 
The vector $\vec{D}=\hat{x}_{AB}$ is known 
as the dipole matrix element. It follows 
that matrix element for the decay is   
\beq
|\langle   
n_{k\alpha}=1,A
|\mathcal{H}_{\tbox{interaction}}|
\mbox{vacuum},B
\rangle|^2
=
\frac{1}{\mbox{volume}} 
\left(\frac{e}{c}\right)^2
2\pi\omega_k |\bm{\varepsilon}^{k,\alpha} \cdot D |^2 
\eeq
In order to calculate the decay rate we have 
to multiply this expression by the density of 
the final states, to integrate over all the 
possible directions of $k$, and to sum 
over the two possible polarizations $\alpha$.

\newpage
\sheadB{Quantization of a many body system}

\sheadC{Second Quantization}

If we regard the electromagnetic field as a collection of 
oscillators then we call $a^{\dag}$ and $a$ raising and lowering  
operators.  This is "first quantization" language. 
But we can also call  $a^{\dag}$ and $a$ creation and destruction 
operators. Then it is "second quantization" language.
So for the electromagnetic field the distinction between 
"first quantization" and "second quantization" is merely 
a linguistic issue. Rather than talking about "exciting" an oscillator 
we talk about "occupying" a mode.

For particles the distinction between "first quantization" 
and "second quantization" is not merely a linguistic issue. 
The quantization of {\em one} particle is called "first quantization". 
If we treat several distinct particles (say a proton and an electron) 
using the same formalism then it is still "first quantization".  

If we have many (identical) electrons then a problem 
arises. The Hamiltonian commutes with "transpositions" 
of particles, and therefore its eigenstates can be categorized 
by their symmetry under permutations. In particular there 
are two special subspaces: those of states that are symmetric 
for {\em any} transposition, and those that are antisymmetric 
for {\em any} transposition. It turns out that in nature 
there is a "super-selection" rule that allows only one 
of these two symmetries, depending on the type of particle. 
Accordingly we distinguish between Fermions and Bosons.
All other sub-spaces are excluded as "non-physical".

We would like to argue that the "first quantization" approach, 
is simply the wrong language to describe a system of identical particles. 
We shall show that if we use the "correct" language, 
then the distinction between Fermions and Bosons comes out 
in a natural way. Moreover, there is no need for the super-selection rule!

The key observation is related to the definition of Hilbert space. 
If the particles are distinct it makes sense to ask 
"where is each particle". But if the particles are identical 
this question is meaningless. The correct question is 
"how many particles are in each site". The space of all possible 
occupations is called "Fock space". Using mathematical language 
we can say that in "first quantization", Hilbert space is the 
external product of "one-particle spaces". In contrast to that, 
Fock space is the external product of "one site spaces".

When we say "site" we mean any "point" in space. Obviously we 
are going to demand at a later stage "invariance" 
of the formalism with respect to the choice of one-particle basis. 
The formalism should look the same if we talk about occupation of 
"position states" or if we talk about occupation 
of "momentum states". Depending on the context we 
talk about occupation of {\em "sites"} or of {\em "orbitals"} 
or of {\em "modes"} or we can simply use the term {\em "one particle states"}.

Given a set of orbitals $|r\rangle$ the Fock space is spanned 
by the basis ${ \{ |...,n_r,...,n_s,...\rangle \} }$. 
We can define a subspace of all $N$~particles states  
\beq    
\mbox{span}_N\{ |...,n_r,...,n_s,...\rangle  \}
\eeq
that includes all the superpositions of 
basis states with $\sum_r n_r=N$ particles.
On the other hand, if we use the first quantization approach,  
we can define Hilbert subspaces that contains 
only totally symmetric or totally anti-symmetric states:
\beq    
\mbox{span}_S\{ |r_1,r_2,...,r_N\rangle  \} 
\\ \nonumber
\mbox{span}_A\{ |r_1,r_2,...,r_N\rangle  \}
\eeq
The mathematical claim is that there is a one-to-one 
correspondence between Fock $\mbox{span}_N$ states    
and Hilbert $\mbox{span}_S$ or $\mbox{span}_A$ states 
for Bosons and Fermions respectively. 
The identification is expressed as follows:
\beq 
|\Psi\rangle = |...,n_r,...,n_s,...\rangle 
\ \ \Longleftrightarrow  \ \
\frac{1}{N!}\sqrt{C^n_N}\sum_P \xi^P \ P |r_1,r_2,...,r_N\rangle 
\eeq
where ${r_1,...,r_N}$ label the occupied orbitals, 
$P$ is an arbitrary permutation operator, 
$\xi$ is $+1$ for Bosons and $-1$ for Fermions, 
and ${C^n_N=N!/(n_r!n_s!...)}$.    
We note that in the case of Fermions the 
formula above can be written as a Slater determinant.
In order to adhere to the common notations 
we use the standard representation:  
\beq
\left\langle x_{1},...,x_{N}| \Psi \right\rangle 
= 
\frac{1}{\sqrt{N!}} 
\begin{vmatrix} 
\left\langle x_{1}|r_{1}\right\rangle  
& \cdots & \left\langle x_{1} |r_{N}\right\rangle \\ 
\vdots &  & \vdots\\ 
\left\langle x_{N}|r_{1}\right\rangle  
& \cdots & \left\langle x_{N} |r_{N}\right\rangle 
\end{vmatrix} 
= 
\frac{1}{\sqrt{N!}} 
\begin{vmatrix} 
\varphi^{(1)}(x_{1})  
& \cdots & \varphi^{(N)}(x_{1}) \\ 
\vdots &  & \vdots\\ 
\varphi^{(1)}(x_{N})  
& \cdots &  \varphi^{(N)}(x_{N})  
\end{vmatrix} 
\eeq
In particular for occupation of $N=2$ particles 
in orbitals $r$ and $s$ we get  
\beq
\Psi(x_1,x_2) = \frac{1}{\sqrt{2}}
\Big(\varphi^r(x_1)\varphi^s(x_2) -\varphi^s(x_1)\varphi^r(x_2)\Big)   
\eeq
In the following section we discuss only the Fock space formalism.
Nowadays the first quantization Hilbert space approach is used 
mainly for the analysis of two particle systems. For larger number 
of particles the Fock formalism is much more convenient, and all the 
issue of "symmetrization" is avoided.

\sheadC{Raising and Lowering Operators} 

First we would like to discuss the mathematics 
of a single "site". The basis states $|n\rangle$ 
can be regarded as the eigenstates of a number operator:
\beq
&& \hat{n}|n\rangle = n |n\rangle 
\\ \nonumber
&& \hat{n}\longrightarrow 
\begin{pmatrix} 
0 &  &  & 0\\ 
& 1 &  & \\ 
&  & 2 & \\ 
0 &  &  & \ddots 
\end{pmatrix} 
\eeq
In general a lowering operator 
has the property  
\beq
\hat{a}\left\vert n\right\rangle 
=f(n)  \left\vert n-1\right\rangle 
\eeq
and its matrix representation is: 
\beq
\hat{a}\longrightarrow 
\begin{pmatrix} 
0 & \ast &  & \\ 
& \ddots & \ddots & \\ 
&  & \ddots & \ast\\ 
0 &  &  & 0 
\end{pmatrix} 
\eeq
The adjoint is a raising operator:
\beq
\hat{a}^{\dagger}\left\vert n\right\rangle 
= f(n+1) \left\vert n+1 \right\rangle 
\eeq
and its matrix representation is:
\beq
\hat{a}^{\dag}
\longrightarrow 
\begin{pmatrix} 
0 &  &  & 0\\ 
\ast & \ddots &  & \\ 
& \ddots & \ddots & \\ 
&  & \ast & 0 
\end{pmatrix} 
\eeq
By appropriate gauge we can assume without loss 
of generality that $f(n)$ is real and non-negative.
so we can write 
\beq
f(n)=\sqrt{g(n)}
\eeq
From the definition of $\hat{a}$ it follows that 
\beq
\hat{a}^{\dag}\hat{a} |n\rangle = g(n) |n\rangle   
\eeq
and therefore
\beq
\hat{a}^{\dag}\hat{a} = g\left(\hat{n}\right)  
\eeq
There are 3 cases of interest
\begin{itemize}
\item
The raising/lowering is unbounded ($-\infty<n<\infty$)
\item
The raising/lowering is bounded from one side (say $0 \le n < \infty$)
\item
The raising/lowering is bounded from both sides (say $0 \le n < \mathcal{N}$)
\end{itemize}
The simplest choice for $g(n)$ in the first case is 
\beq
g(n)=1 
\eeq
In such a case $\hat{a}$ becomes the translation operator,  
and the spectrum of $n$ stretches from $-\infty$ to $\infty$. 
The simplest choice for $g(n)$ in the second case is 
\beq
g(n)=n
\eeq
this leads to the same algebra as in the case of an harmonic 
oscillator.  The simplest choice for $g(n)$ in the third case is 
\beq
g(n)=(\mathcal{N}-n)n
\eeq
Here it turns out that the algebra  
is the same as that for angular momentum. 
To see that it is indeed like that define 
\beq
m \,\,=\,\, n - \frac{\mathcal{N}-1}{2} \,\,=\,\, -s, \dots ,+s
\eeq
where $s=(\mathcal{N}-1)/2$.
Then it is possible to write 
\beq
g(m)  = s(s+1)-m(m+1)  
\eeq
In the next sections we are going to discuss 
the ``Bosonic" case $\mathcal{N}=\infty$ with $g(n)=n$, 
and the ``Fermionic" case $\mathcal{N}=2$ with $g(n)=n(1-n)$. 
Later we are going to argue that these are the 
only two possibilities that are relevant to the 
description of many body occupation. 

\begin{center}
\putgraph[0.55\hsize]{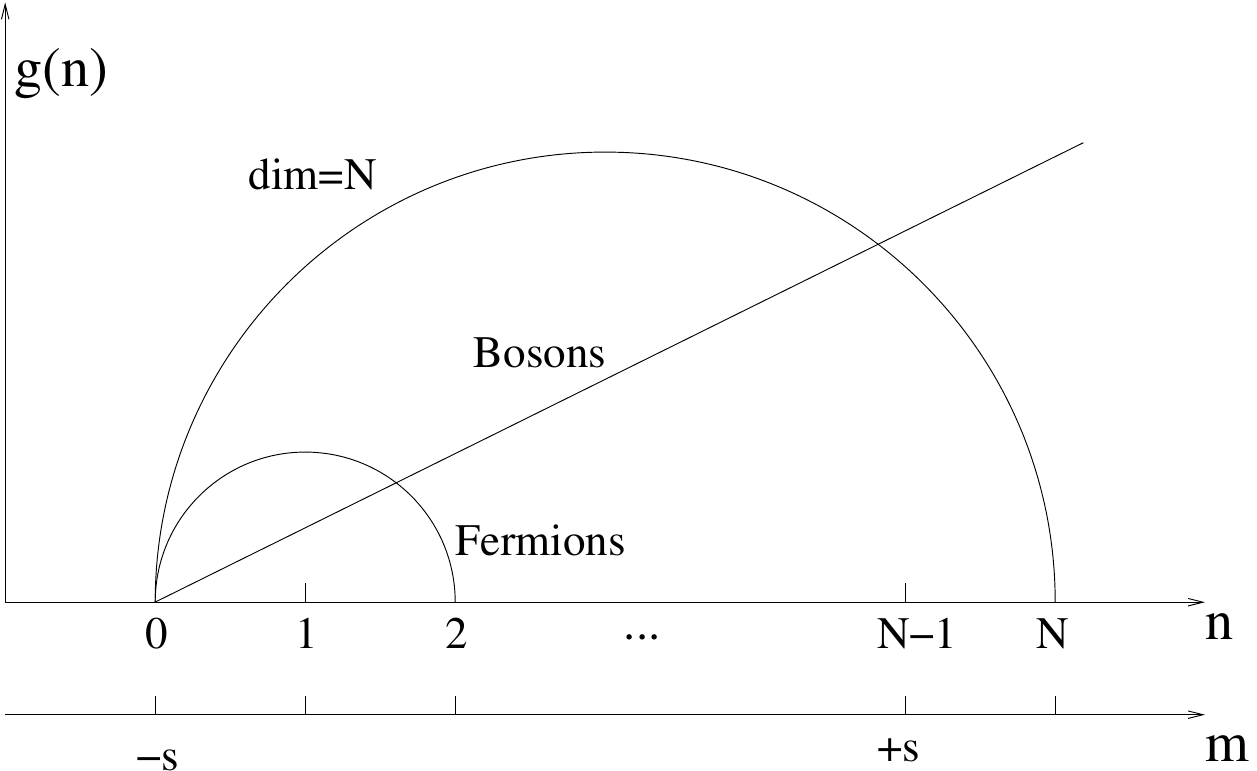}
\end{center}

It is worthwhile to note that the algebra of ``angular momentum" 
can be formally obtained from the Bosonic algebra 
using a trick due to Schwinger.  
Let us define two Bosonic operators $a_1$ and $a_2$, and  
\beq
c^{\dag} \ \ = \ \ a_2^{\dag}a_1
\eeq
The $c^{\dag}$ operator moves a particle from site~1 
to site~2. Consider how $c$ and $c^{\dag}$  
operate within the subspace of $(N-1)$ particle states. 
It is clear that $c$ and $c^{\dag}$ act 
like lowering/raising operators with respect 
to $\hat{m} = (a_2^{\dag}a_2 - a_1^{\dag}a_1)/2$.
Obviously the lowering/raising operation in 
bounded from both ends. In fact it is easy 
to verify that $c$ and $c^{\dag}$ have the same 
algebra as that of ``angular momentum".

\sheadC{Algebraic characterization of field operators} 

In this section we establish some 
mathematical observations that we 
need for a later reasoning regarding 
the classification of field operators   
as describing Bosons or Fermions.
By field operators we mean either 
creation or destruction operators, 
to which we refer below 
as raising or lowering (ladder) operators.    
We can characterize a lowering operator as follows:  
\beq
\hat{n}\Big(  \hat{a}\left\vert n\right\rangle \Big)  
=(n-1) \Big(  \hat{a}\left\vert n\right\rangle \Big) 
\ \ \ \ \ \ \mbox{for any $n$} 
\eeq
which is equivalent to  
\beq
\hat{n}\hat{a} = \hat{a}(\hat{n}-1)  
\eeq
A raising operator is similarly 
characterized by ${\hat{n}\hat{a}^{\dag} = \hat{a}^{\dag}(\hat{n}+1)}$.

It is possible to make a more 
interesting statement. Given that 
\beq
[a,a^{\dag}] \ \ = \ \ aa^{\dag}-a^{\dag}a \ \ = \ \  1
\eeq
we deduce that $a$ and $a^{\dag}$ 
are lowering and raising operators 
with respect to $\hat{n}=a^{\dag}a$.
The prove of this statement 
follows directly form the observation 
of the previous paragraph.
Furthermore, from 
\beq
|| a|n\rangle || &=& \langle n| a^{\dag} a |n\rangle  \ \ = \ \ n
\\ 
|| a^{\dag} |n\rangle || &=& \langle n| a a^{\dag} |n\rangle  \ \ = \ \ 1 + n
\eeq
Since the norm is a non-negative value it follows 
that $n=0$ and hence also all the positive 
integer values $n=1,2,...$ form its spectrum. 
Thus in such case $a$ and $a^{\dag}$  
describe a system of Bosons.

Let us now figure our the nature of 
an operator that satisfies the analogous 
anti-commutation relation:
\beq
[a,a^{\dag}]_{+} \ \ = \ \ aa^{\dag}+a^{\dag}a \ \ = \ \  1
\eeq
Again we define $\hat{n}=a^{\dag}a$
and observe that $a$ and $a^{\dag}$ are characterized 
by ${\hat{n}\hat{a} = \hat{a}(1-\hat{n})}$
and ${\hat{n}\hat{a}^{\dag} = \hat{a}^{\dag}(1-\hat{n})}$. 
Hence we deduce that both $a$ and $a^{\dag}$ 
simply make transposition of two $n$ 
states ${|\epsilon\rangle}$ and ${|1{-}\epsilon\rangle}$. 
Furthermore 
\beq
|| a|n\rangle || &=& \langle n| a^{\dag} a |n\rangle  \ \ = \ \ n
\\ 
|| a^{\dag} |n\rangle || &=& \langle n| a a^{\dag} |n\rangle  \ \ = \ \ 1 - n
\eeq
Since the norm is a non-negative value 
it follows that ${0\le\epsilon\le1}$. 
Thus we deduce that the irreducible 
representation of $a$ is 
\beq
\hat{a}= 
\begin{pmatrix} 
0 & \sqrt{\epsilon}\\ 
\sqrt{1{-}\epsilon} & 0 
\end{pmatrix} 
\eeq
One can easily verify the the desired 
anti-commutation is indeed satisfied.
We can always re-define $\hat{n}$ 
such that ${n{=}0}$ would correspond 
to one eigenvalue and ${n{=}1}$ 
would correspond to the second one. 
Hence it is clear that $a$ and $a^{\dag}$  
describe a system of Fermions.

\sheadC{Creation Operators for "Bosons"} 

For a "Bosonic" site we define 
\beq
\hat{a}\left\vert n\right\rangle =\sqrt{n}\left\vert n-1\right\rangle 
\eeq
hence
\beq
\hat{a}^{\dag}\left\vert n\right\rangle =\sqrt{n+1}\left\vert 
n+1\right\rangle 
\eeq
and
\beq
\left[ \hat{a},\hat{a}^{\dag}\right] 
=\hat{a}\hat{a}^{\dag}-\hat{a}^{\dag}\hat{a}=1
\eeq
If we have many sites then we define
\beq
\hat{a}_{r}^{\dag}
=1\otimes1\otimes \dots \otimes\hat{a}^{\dag}\otimes  \dots \otimes1
\eeq
which means 
\beq
\hat{a}_{r}^{\dag}\left\vert n_{1},n_{2}, \dots ,n_{r},\dots \right\rangle 
=\sqrt{n_{r}+1} \, \left\vert n_{1},n_{2}, \dots ,n_{r}+1,\dots \right\rangle 
\eeq
and hence
\beq
\left[  \hat{a}_{r},\hat{a}_{s}\right] =0 
\eeq
and
\beq
\left[  \hat{a}_{r},\hat{a}_{s}^{\dag} \right] = \delta_{r,s} 
\eeq

We have defined our set of creation operators using 
a particular one-particle basis. What will happen 
if we switch to a different basis?  
Say from the position basis to the momentum basis? 
In the new basis we would like to have
the same type of "occupation rules", namely, 
\beq
\left[ \hat{a}_{\alpha},\hat{a}_{\beta}^{\dag}\right]  =\delta_{\alpha\beta}
\eeq
Let's see that indeed this is the case. 
The unitary transformation matrix 
from the original $\left\vert r\right\rangle$ basis 
to the new $\left\vert \alpha\right\rangle$ basis is
\beq
T_{r,\alpha}=\left\langle r|\alpha\right\rangle 
\eeq
Then we have the relation 
\beq
\left\vert \alpha\right\rangle =\underset{r}{\sum}\left\vert r\right\rangle 
\left\langle r|\alpha\right\rangle =\underset{r}{\sum}T_{r,\alpha}\left\vert 
r\right\rangle 
\eeq
and therefore
\beq
\hat{a}_{\alpha}^{\dag}=\underset{r}{\sum}T_{r,\alpha}\hat{a}_{r}^{\dag}
\eeq
Taking the adjoint we also have 
\beq
\hat{a}_{\alpha} = \underset{r}{\sum}T_{r,\alpha}^* \hat{a}_{r}
\eeq
Now we find that 
\beq
\left[  \hat{a}_{\alpha},\hat{a}_{\beta}^{\dag}\right]  =\underset{r,s}{\sum 
}\left[  T_{r\alpha}^{\ast}\hat{a}_{r},T_{s\beta}\hat{a}_{s}^{\dag}\right] 
=T_{r\alpha}^{\ast}T_{s\beta}\delta_{rs}=\left(  T^{\dag}\right)  _{\alpha 
r}T_{r\beta}=\left(  T^{\dag}T\right)  _{\alpha\beta}=\delta_{\alpha\beta}
\eeq
This result shows that $\hat{a}_{\alpha}$ and $\hat{a}_{\beta}^{\dag}$ 
are indeed destruction and creation operators of the same "type" 
as $\hat{a}_r$ and $\hat{a}_r^{\dag}$. Can we have the same 
type of invariance for other types of occupation? We shall see 
that the only other possibility that allows "invariant" description 
is $\mathcal{N}=2$.

\sheadC{Creation Operators for "Fermions"} 
 
In analogy with the case of a "Boson site" we define a "Fermion site" using 
\beq
\hat{a}\left\vert n\right\rangle =\sqrt{n}\left\vert n-1\right\rangle 
\eeq
and
\beq
\hat{a}^{\dag}\left\vert n\right\rangle =\sqrt{n+1}\left\vert 
n+1\right\rangle 
\ \ \ \ \ \ \ \ \mbox{with mod(2) plus operation}
\eeq
The representation of the operators is, using Pauli matrices:
\beq
&& \hat{n}=  
\begin{pmatrix} 
1 & 0\\ 
0 & 0 
\end{pmatrix} 
=\frac{1}{2}\left(\hat{1}+\sigma_{3}\right) 
\\ \nonumber
&& \hat{a}= 
\begin{pmatrix} 
0 & 0\\ 
1 & 0 
\end{pmatrix} 
=\frac{1}{2}\left(  \sigma_{1}-i\sigma_{2}\right) 
\\ \nonumber
&& \hat{a}^{\dag}=  
\begin{pmatrix} 
0 & 1\\ 
0 & 0 
\end{pmatrix} 
=\frac{1}{2}\left(\sigma_{1}+i\sigma_{2}\right) 
\\ \nonumber
&& \hat{a}^{\dag}\hat{a} = \hat{n}
\\ \nonumber
&& \hat{a}\hat{a}^{\dag} = \hat{1} - \hat{n}
\\ \nonumber
&& \left[ \hat{a},\hat{a}^{\dag}\right]_{+}
=\hat{a}\hat{a}^{\dag}+\hat{a}^{\dag}\hat{a} = 1 
\eeq
while 
\beq
\left[\hat{a},\hat{a}^{\dag}\right] = 1-2\hat{n}
\eeq

Now we would like to proceed with the many-site system  
as in the case of "Bosonic sites". But the problem is that 
the algebra 
\beq
\left[\hat{a}_r,\hat{a}_s^{\dag}\right] 
= \delta_{r,s} (1-2\hat{a}_r^{\dag}\hat{a}_r)
\eeq
is manifestly not invariant under a change 
of one-particle basis. The only hope is to 
have 
\beq
\left[\hat{a}_r,\hat{a}_s^{\dag}\right]_{+} = \delta_{r,s} 
\eeq
which means that $a_r$ and $a_s$ for $r\ne s$ 
should anti-commute rather than commute. 
Can we define the operators $a_r$ in such 
a way? It turns out that there is such a possibility:
\beq
\hat{a}_{r}^{\dag}
\left\vert n_{1},n_{2}, \dots ,n_{r}, \dots \right\rangle 
= (-1)^{ \sum_{s(>r)} n_s }\sqrt{1+n_r}\,  
\left\vert n_{1},n_{2}, \dots ,n_{r}+1, \dots \right\rangle
\eeq
For example, it is easily verified that we have:
\beq
a_2^{\dag} a_1^{\dag} |0,0,0, \dots \rangle =
-a_1^{\dag} a_2^{\dag} |0,0,0, \dots \rangle =
|1,1,0, \dots \rangle 
\eeq
With the above convention if we create 
particles in the "natural order" then the 
sign comes out plus, while for any  
"violation" of the natural order we get 
a minus factor.

\sheadC{One Body Additive Operators} 
 
Let us assume that we have an additive quantity 
$V$ which is not the same for different 
one-particle states. One example is the (total) kinetic energy,  
another example is the (total) potential energy. 
It is natural to define the many body operator 
that corresponds to such a property in the basis 
where the one-body operator is diagonal. 
In the case of potential energy it is 
the position basis:   
\beq
V= \sum_{\alpha} V_{\alpha,\alpha} \hat{n}_{\alpha} 
= \sum_{\alpha} \hat{a}_{\alpha}^{\dag} V_{\alpha,\alpha} \hat{a}_{\alpha}     
\eeq
This counts the amount of particles in each $\alpha$ 
and multiplies the result with the value of $V$ 
at this site.   
If we go to a different one-particle basis 
then we should use the transformation
\beq
&& \hat{a}_{\alpha} = \sum_{k} T_{k,\alpha}^* \hat{a}_{k} 
\\ \nonumber
&& \hat{a}_{\alpha}^{\dag} = \sum_{k'} T_{k',\alpha} \hat{a}_{k'}^{\dag} 
\eeq
leading to 
\beq
V=\sum_{k,k`}\hat{a}_{k'}^{\dag} V_{k',k} \hat{a}_{k} 
\eeq

Given the above result we can calculate the matrix elements 
from a transition between two different occupations:
\beq
\left\vert\left\langle 
n_{1}-1,n_{2}+1|V|n_{1},n_{2}
\right\rangle\right\vert^{2}=
\left(n_{2}+1\right) n_{1} \left\vert V_{2,1}\right\vert^{2}
\eeq
What we get is quite amazing: in the case of Bosons 
we get an amplification of the transition 
if the second level is already occupied.  
In the case of Fermions we get "blocking" if the 
second level is already occupied. Obviously 
this goes beyond classical reasoning. 
The latter would give merely $n_1$ as a prefactor.

\sheadC{Two Body ``Additive" Operators} 

It is straightforward to make a generalization to the 
case of two body ``additive" operators. Such operators 
may represent the two-body interaction between the particles.
For example we can take the Coulomb interaction, 
which is diagonal in the position basis. 
Thus we have
\beq
U=
\frac{1}{2}\sum_{\alpha\neq\beta} 
U_{\alpha\beta,\alpha\beta}  
\hat{n}_{\alpha}\hat{n}_{\beta}
+\frac{1}{2}\sum_{\alpha} 
\,\,U_{\alpha\alpha,\alpha\alpha}
\,\,\hat{n}_{\alpha}\left(\hat{n}_{\alpha}-1\right)  
\eeq
Using the relation 
\beq
\hat{a}_{\alpha}^{\dag}\hat{a}_{\beta}^{\dag}
\hat{a}_{\beta}\hat{a}_{\alpha}
=\left\{ 
\amatrix{ 
\hat{n}_{\alpha} \hat{n}_{\beta}  
& \,\,\,\mbox{for $\alpha\ne\beta$}  \cr 
\hat{n}_{\alpha} (\hat{n}_{\alpha}-1)  
& \,\,\,\mbox{for $\alpha = \beta$} }
\right.
\eeq
We get the simple expression
\beq
U=\frac{1}{2}\sum_{\alpha,\beta}
\hat{a}_{\alpha}^{\dag}\hat{a}_{\beta}^{\dag}
\,\,U_{\alpha\beta,\alpha\beta} 
\,\,\hat{a}_{\beta}\hat{a}_{\alpha}
\eeq
and for a general one-particle basis
\beq
U=\frac{1}{2}\sum_{k'l',kl}
\hat{a}_{k'}^{\dag}\hat{a}_{l'}^{\dag}
\,\,U_{k'l',kl}
\,\,\hat{a}_{l}\hat{a}_{k}
\eeq
We call such operator ``additive" (with quotations) 
because in fact they are not really additive. 
An example for a genuine two body additive operator 
is $[A,B]$, where $A$ and $B$ are one body operators. 
This observation is very important in the theory 
of linear response (Kubo).

\sheadC{Matrix elements with $N$ particle states}

Consider an $N$ particle state of a Fermionic system,  
which is characterized by a definite occupation of $k$~orbitals:
\beq
\left\vert R_{N}\right\rangle = 
\hat{a}_{N}^{\dag} \dots \hat{a}_{2}^{\dag}\hat{a}_{1}^{\dag}
\left\vert 0\right\rangle 
\eeq
For the expectation value of a one body operator we get
\beq
\left\langle R_N |V| R_N \right\rangle = 
\sum_{k\in R} \left\langle k | V | k \right\rangle 
\eeq
because only the terms with $k=k'$ do not vanish.
If we have two $N$ particle states with definite 
occupations, then the matrix element of~$V$ would 
be in general zero unless they differ by a single 
electronic transition, say from an orbital $k_0$ 
to another orbital $k_0'$. In the latter case 
we get the result ${V_{k_0',k_0}}$ 
as if the other electrons are not involved.

For the two body operator we get for the expectation value  
a more interesting result that goes beyond the naive expectation.
The only non-vanishing terms in the sandwich calculation 
are those with either $k'=k$ and $l'=l$ or with $k'=l$ and $l'=k$.  
All the other possibilities give zero. Consequently    
\beq
\left\langle R_N |U| R_N \right\rangle
=\frac{1}{2}
\sum_{k,l \in R} 
\underset{direct}{\left\langle kl|U| kl \right\rangle }
-\underset{exchange}{\left\langle lk|U|kl \right\rangle }
\eeq
A common application of this formula is in 
the context of multi-electron atoms and molecules, 
where~$U$ is the Coulomb interaction. 
The direct term has an obvious electrostatic 
interpretation, while the exchange term reflects 
the implications of the Fermi statistics.      
In such application the exchange term is non-vanishing 
whenever two orbitals have a non-zero spatial overlap. 
Electrons that occupy well separated orbitals 
have only a direct electrostatic interaction.

\sheadC{Introduction to the Kondo problem}

One can wonder whether the Fermi energy, due to the Pauli  
exclusion principle, is like a lower cutoff that ``regularize" 
the scattering cross section of of electrons in a metal. 
We explain below that this is not the case unless the scattering 
involves a spin flip. The latter is known as the Kondo effect.    
The scattering is described by 
\beq
V = \sum_{k',k} a^{\dag}_{k'} V_{k',k} a_{k}
\eeq
hence:
\beq
T^{[2]} = 
\left\langle k_{2} \left| 
V \frac{1}{E-\mathcal{H}+i0} V 
\right|k_{1} \right\rangle
\ \ = \ \ 
\sum_{k_b',k_b}
\sum_{k_a',k_a}
\left\langle k_{2} \left| 
a^{\dag}_{k_b'} V_{k_b',k_b} a_{k_b}
\frac{1}{E-\mathcal{H}+i0}
a^{\dag}_{k_a'} V_{k_a',k_a} a_{k_a}
\right|k_{1}\right\rangle
\eeq
where both the initial and the final 
states are zero temperature Fermi sea 
with one additional electron above 
the Fermi energy. The initial and final 
states have the same energy:  
\beq
E \ \ = \ \  E_0 + \epsilon_{k_1} \ \ = \ \ E_0 + \epsilon_{k_2}
\eeq
where $E_0$ is the total energy of 
the zero temperature Fermi sea.
The key observation is that 
all the intermediate states are 
with definite occupation. 
Therefore we can pull out the resolvent:
\beq
T^{[2]}
=  \sum_{k_b',k_b,k_a',k_a}
\frac{V_{k_b',k_b} V_{k_a',k_a}}{E-E_{k_a,k_a'}}
\left\langle k_{2} \left| 
a^{\dag}_{k_b'}  a_{k_b}
a^{\dag}_{k_a'}  a_{k_a}
\right|k_{1}\right\rangle
\eeq
where 
\beq
E_{k_a,k_a'}\ \ = \ \ E_0 + \epsilon_{k_1} - \epsilon_{k_a} + \epsilon_{k_a'}
\eeq

\begin{center}
\putgraph[0.6\hsize]{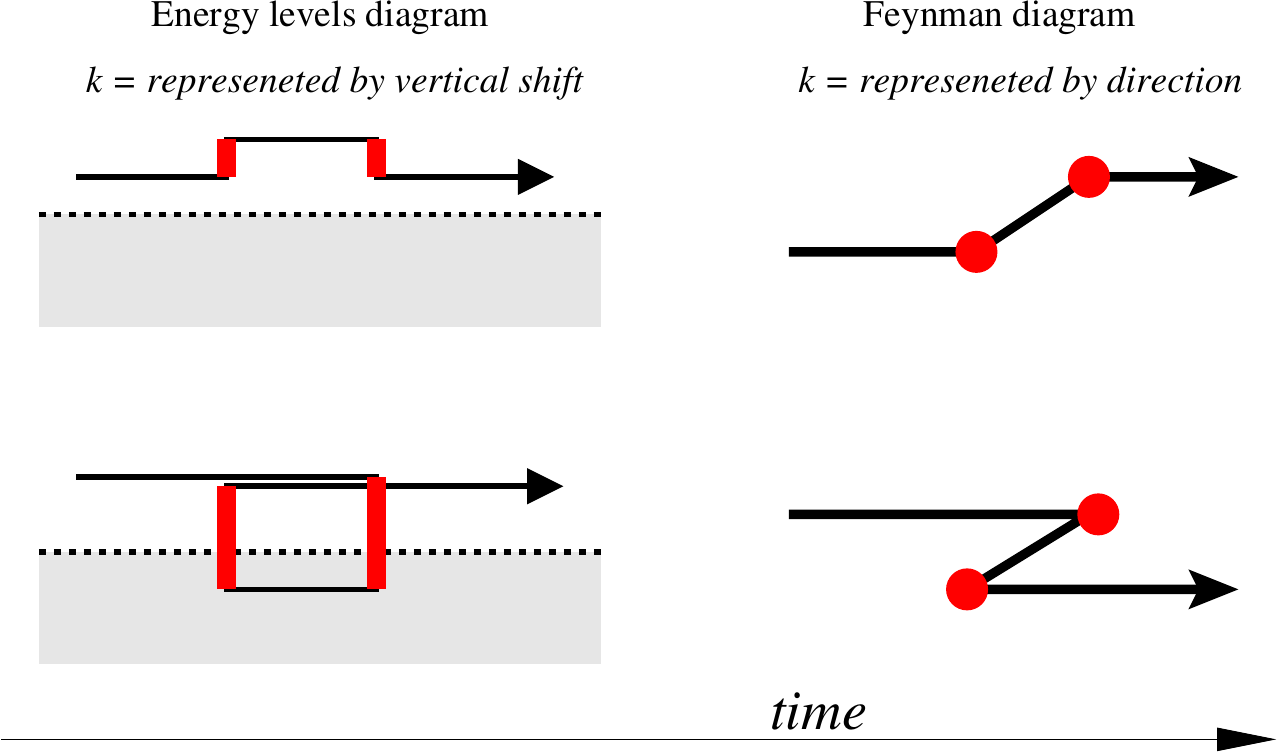}
\end{center}

As in the calculation of ``exchange" 
we have two non-zero contribution to the sum.
These are illustrated in the figure above:  
Either  $(k_b',k_b,k_a',k_a)$ equals 
$(k_2,k',k',k_1)$ with $k'$ above the Fermi 
energy, or $(k',k_1,k_2,k')$ with $k'$ below 
the Fermi energy. Accordingly $E-E_{k_a,k_a'}$ 
equals either ${(\epsilon_{k_1}-\epsilon_{k'})}$ 
or  ${-(\epsilon_{k_1}-\epsilon_{k'})}$. Hence we get
\beq
T^{[2]}
=  \sum_{k'}
\left[
\frac{ V_{k_2,k'} V_{k',k_1} }{+(\epsilon_{k_1}-\epsilon_{k'})+i0}
\left\langle k_{2} \left| 
a^{\dag}_{k_2}  a_{k'}
a^{\dag}_{k'}  a_{k_1}
\right|k_{1}\right\rangle
+
\frac{ V_{k',k_1} V_{k_2,k'} }{-(\epsilon_{k_1}-\epsilon_{k'})+i0}
\left\langle k_{2} \left| 
a^{\dag}_{k'}  a_{k_1}
a^{\dag}_{k_2}  a_{k'}
\right|k_{1}\right\rangle
\right]
\eeq
Next we use 
\beq 
\left\langle k_{2} \left| 
a^{\dag}_{k_2}  a_{k'}
a^{\dag}_{k'}  a_{k_1}
\right|k_{1}\right\rangle 
=
\left\langle k_{2} \left| 
a^{\dag}_{k_2} (1-n_{k'}) a_{k_1}
\right|k_{1}\right\rangle 
= 
+1 \times  
\left\langle k_{2} \left| 
a^{\dag}_{k_2}a_{k_1}
\right|k_{1}\right\rangle 
\eeq
which holds if $k'$ is above the Fermi energy 
(otherwise it is zero).  And 
\beq 
\left\langle k_{2} \left| 
a^{\dag}_{k'}  a_{k_1}
a^{\dag}_{k_2}  a_{k'}
\right|k_{1}\right\rangle
= 
\left\langle k_{2} \left| 
a_{k_1} ( n_{k'} ) a^{\dag}_{k_2} 
\right|k_{1}\right\rangle
= 
-1 \times 
\left\langle k_{2} \left| 
a^{\dag}_{k_2}a_{k_1}
\right|k_{1}\right\rangle 
\eeq
which holds if $k'$ is below the Fermi energy 
(otherwise it is zero). Note that without 
loss of generality we can assume gauge 
such that ${\langle k_{2}| a^{\dag}_{k_2}a_{k_1}|k_{1}\rangle =1}$.
Coming back to the transition matrix we   
get a result which is not divergent at the Fermi energy: 
\beq
T^{[2]} = 
\sum_{k'\in \tbox{above}}
\frac{ V_{k_2,k'} V_{k',k_1} }{\epsilon_{k_1}-\epsilon_{k'}+i0}
+ 
\sum_{k'\in \tbox{below}}
\frac{ V_{k_2,k'} V_{k',k_1} }{\epsilon_{k_1}-\epsilon_{k'}-i0}
\eeq
If we are above the Fermi energy, 
then it is as if the Fermi energy does not exist at all. 
But if the scattering involves a spin flip, 
as in the Kondo problem, 
the divergence for $\epsilon$ close to the Fermi energy 
is not avoided. Say that we want to calculate 
the scattering amplitude   
\beq
\left\langle k_{2} \uparrow,\Downarrow\left|T\right|k_{1},\uparrow,\Downarrow\right\rangle
\eeq
where the double arrow stands for the spin of 
a magnetic impurity. It is clear that the only 
sequences that contribute are those that 
take place {\em above} the Fermi energy. 
The other set of sequences, that involve 
the creation of an electron-hole pair do not exist: 
Since we assume that the magnetic impurity 
is initially "down", it is not possible to 
generate a pair such that the electron spin is "up".

\sheadC{Green functions for many body systems}

The Green function in the one particle formalism 
is defined via the resolvent as the Fourier transform 
of the propagator. In the many body   
formalism the role of the propagator is taken 
by the time ordered correlation of field operators. 
In both cases the properly defined Green function  
can be used in order to analyze scattering problems 
in essentially the same manner. 
It is simplest to illustrate this observation 
using the example of the previous section. 
The Green function in the many body context 
is defined as 
\beq
G_{k_2,k_1}(\epsilon) \ \ = \ \ 
-i \mbox{FT}\Big[ \big\langle \Psi \big|\mathcal{T} 
a_{k_2}(t_2)a_{k_1}^{\dag}(t_1) \big|\Psi \big\rangle \Big]   
\eeq
If $\Psi$ is the vacuum state this coincides 
with the one particle definition of the Green function: 
\beq
G_{k_2,k_1}(\epsilon) \ \ = \ \ 
-i \mbox{FT}\Big[\Theta(t_2{-}t_1) \big\langle k_2|U(t_2-t_1)|k_1\rangle\Big] 
\eeq
But if $\Psi$ is (say) a non-empty zero temperature Fermi sea 
then also for $t_2<t_1$ we get a non-zero contribution 
due to the possibility to annihilate an 
electron in an occupied orbital. Thus we get 
\beq
G_{k_2,k_1}(\epsilon) = 
\sum_{\epsilon_k > \epsilon_{\tbox{F}}}
\frac{ \delta_{k_1,k} \delta_{k_2,k} }{\epsilon-\epsilon_{k}+i0}
+ 
\sum_{\epsilon_k < \epsilon_{\tbox{F}}}
\frac{ \delta_{k_1,k} \delta_{k_2,k} }{\epsilon-\epsilon_{k}-i0}
\eeq
One should observe that the many-body definition 
is designed so as to reproduce the correct $T$ matrix 
as found in the previous section. 
The definition above allows us to adopt an optional 
point of view of the scattering process: 
{\em a one particle point of view instead 
of a many body point of view!}  
In the many body point of view an electron-hole 
pair can be created, and later the hole 
is likely to be annihilated with the injected electron.  
In the one particle point of view 
the injected electron can be ``reflected" 
to move backwards in time and then is likely 
to be scattered back to the forward time direction. 
The idea here is to regard antiparticles as 
particles that travel backwards in time.  
This idea is best familiar in the context of the Dirac equation.

\newpage
\sheadB{Wigner function and Wigner-Weyl formalism}

\sheadC{The classical description of a state}

Classical states are described by a probability function.
Given a random variable $\hat{x}$, we define $\rho(x) = \mathrm{Prob}(\hat{x}=x)$ 
as the probability of the event ${ \hat{x} = x }$,  
where the possible values of $x$ are called the spectrum 
of the random variable $\hat{x}$. 
For a random variable continuous spectrum we 
define the  probability density function 
via ${\rho(x)dx =\mathrm{Prob}( x < \hat{x}< x+dx)}$. 
The expectation value of a random variable $\hat{A}=A(\hat{x})$ 
is defined as  
\beq
\langle\hat{A}\rangle = \sum{ \rho(x) \ A(x) }
\eeq
and for a continuous variable as. 
For simplicity we use, from now on, a notation 
as if the random variables have a discrete spectrum, 
with the understanding that in the case of a continuous 
spectrum we should replace the $\sum$ by an integral with 
the appropriate {\em measure} (e.g. $dx$ or $dp/(2\pi\hbar)$).  
Unless essential for the presentation we set ${\hbar=1}$.

Let us consider two random variables ${ \hat{x}, \hat{p} }$.
One can ask what is the joint probability distribution 
for these two variables.  This is a valid question only in classical
mechanics.  In quantum mechanics one usually cannot ask this question
since not all variables can be measured in the same measurement.  
In order to find the joint probability function of these two variables
in quantum mechanics one needs to build a more sophisticated method.
The solution is to regard the expectation value as the fundamental
outcome and to define the probability function as an expectation value.
In the case of one variable such as $\hat{x}$ or $\hat{p}$, we define
probability functions as
\beq
\rho(X) &\equiv& \langle \ \delta(\hat{x}-X) \ \rangle 
\\ \nonumber
\rho(P) &\equiv& \langle \ 2\pi\delta(\hat{p}-P) \ \rangle
\eeq
Now we can also define the joint probability function as
\beq
\rho(X,P) = \langle 2\pi\delta(\hat{p}-P) \ \delta(\hat{x}-X) \ \rangle
\eeq
This probability function is normalized so that
\beq
\int \rho(X,P)\frac{dX dP}{2\pi} 
\ \ = \ \ \left\langle\int{\delta(\hat{p}-P)dP}\int{\delta(\hat{x}-X)dX}\right\rangle
\ \ = \ \ 1
\eeq
In the next section we shall define essentially the same object 
in the framework of quantum mechanics.

\sheadC{Wigner function}

Wigner function is a real normalized function which is defined as
\beq
\rho_{\tbox{W}}(X,P)
=\Big\langle \ \Big[ 
\ 2\pi\delta(\hat{p}-P) \ \delta(\hat{x}-X) \ 
\Big]_{\tbox{sym}} \ \Big\rangle
\eeq
In what follows we define what we mean 
by symmetrization (``sym''), and we 
relate $\rho_{\tbox{W}}(X,P)$ to the conventional 
probability matrix $\rho(x',x'')$.  
We recall that the latter is defined as
\beq
\rho(x',x'')
=\Big\langle P^{x'x''}\Big\rangle
=\Big\langle \ |x''\rangle\langle x'| \ \Big\rangle
\eeq
The ``Wigner function formula'' that we are going to prove is 
\beq
\rho_{\tbox{W}}(X,P) = \int  \rho\left( X+\frac{1}{2}r, X-\frac{1}{2}r \right) \eexp{-iPr} dr
\eeq
Thus to go from the probability matrix to the Wigner function 
is merely a Fourier transform, and can be loosely regarded 
as a change from ``position representation'' to ``phase space representation''.

Moreover we can use the same transformation to switch 
the representation of an observable $\hat{A}$ 
from $A(x',x'')$ to $A(X,P)$ . Then we shall prove 
the ``Wigner-Weyl formula"
\beq
\trc (A \rho) = \int  A(X,P) \rho_{\tbox{W}}(X,P) \frac{dX dP}{2\pi}
\eeq
This formula implies that expectation values of an observable  
can be calculated using a semi-classical calculation. 
This extension of the Wigner function formalism is known  
as the Wigner-Weyl formalism.

\sheadC{Mathematical derivations}

Fourier transform reminder:
\beq
F(k) &=& \int f(x) \eexp{-ikx}dx 
\\ \nonumber
f(x) &=& \int \frac{dk}{2\pi} F(k) \eexp{ikx}
\eeq
The inner product is invariant under change of representation
\beq
\int{f^*(x)g(x)dx} \ = \ \int{\frac{dk}{2\pi}F^*(k)G(k)}
\eeq
For the matrix representation of an operator $A$ we use the notation
\beq
A(x',x'') 
=  \langle x' | A | x'' \rangle 
\eeq 
It is convenient to replace the row and column indexes 
by diagonal and off-diagonal coordinates:
\beq
X&=&\frac{1}{2}\left(x'+x''\right) = \mbox{the diagonal coordinate} 
\\ \nonumber
r&=&x'-x'' = \mbox{the off diagonal coordinate} 
\eeq
such that $x'=X+(r/2)$ and  $x''=X-(r/2)$,  
and to use the alternate notation
\beq
A(X,r) 
=  \left\langle X+\frac{1}{2}r \Big| A \Big| X-\frac{1}{2}r \right\rangle 
\eeq 
Using this notation, the transformation to 
phase space representations can be written as 
\beq
A(X,P) = \int  A(X,r) \eexp{-iPr} dr
\eeq
Consequently the inner product of two matrices 
can be written as a phase space integral 
\beq
\trc(A^{\dag} B)
= \int A^*(x',x'') B(x',x'') dx'dx''
= \int A^*(X,r) B(X,r) dXdr
= \int A^*(X,P) B(X,P) \frac{dXdP}{2\pi}
\eeq
This we call the "Wigner-Weyl formula". 
Note that if $A$ is hermitian then $A(X,-r)=A^*(X,+r)$, 
and consequently $A(X,P)$ is a real function.

For the derivation of the ``Wigner function formula'' 
we have to explain how the operators ${\delta(\hat{p}-P)\delta(\hat{x}-X)}$
are symmetrized, and that that they form a complete set 
of observables that are linearly related to the standard 
set ${|x'\rangle\langle x''|}$. First we recall that  
a projector can be written as a delta function. 
This is quite transparent in discrete notation, 
namely  ${|n_0\rangle\langle n_0| = \delta_{\hat{n},n_0}}$.
The left hand side is a projector  
whose only non-zero matrix element is $n=m=n_0$. 
The right hand side is a function of $f(\hat{n})$ 
such that $f(n)=1$ for $n=n_0$ and zero otherwise.  
Therefore the right hand side is also diagonal in $n$ 
with the same representation. In continuum notation 
\beq
|X\rangle\langle X| \ \ = \ \ \delta(\hat{x}-X) \ \ = \ \  \int\frac{dp}{2\pi} \eexp{ip(\hat{x}-X)}
\eeq
With the same spirit we {\em define} the symmetrized operator 
\beq
\Big[ 
\ 2\pi\delta(\hat{p}-P)
\ \delta(\hat{x}-X) 
\ \Big]_{\tbox{sym}} 
\ \ \equiv \ \ 
\int \frac{dr dp}{2\pi} 
\ \eexp{ir(\hat{p}-P) + ip(\hat{x}-X)} 
\ \ = \ \ 
\int dr \eexp{-irP} \int\frac{dp}{2\pi} \eexp{ir\hat{p} + ip(\hat{x}-X)}
\eeq
The inner integral can be carried out:
\beq
\int\frac{dp}{2\pi} \eexp{ir\hat{p} + ip(\hat{x}-X)}
\ \ = \ \ 
\int\frac{dp}{2\pi} \eexp{i(r/2)\hat{p}}\eexp{ip(\hat{x}-X)}\eexp{i(r/2)\hat{p}}
\ \ = \ \ 
\eexp{i(r/2)\hat{p}} \delta(\hat{x}-X) \eexp{i(r/2)\hat{p}}
\eeq
where we used the 
identity ${\eexp{\hat{A}+\hat{B}}=\eexp{\hat{A}} \eexp{\hat{B}}\eexp{\frac{1}{2}[\hat{A},\hat{B}]}}$,
and the commutation $[\hat{x},\hat{p}]=i$, 
in order to split the exponents, as in the 
equality ${\eexp{\hat{x}+\hat{p}} = \eexp{\hat{p}/2} \eexp{\hat{x}} \eexp{\hat{p}/2}}$. 
Replacing the $\delta(\hat{x}-X)$ by $|X\rangle\langle X|$, 
and operating on the ket and on the bra with the displacement 
operators $\eexp{\pm i(r/2)\hat{p}}$ we find that   
\beq
\Big[ 
\ 2\pi\delta(\hat{p}-P)
\ \delta(\hat{x}-X) 
\ \Big]_{\tbox{sym}} 
\ \ = \ \
\int dr \eexp{-irP}
\ \ \Big| X{-}(r/2) \Big\rangle\Big\langle X{+}(r/2) \Big| 
\eeq
Taking the trace of both sides with $\rho$ we deduce 
that $\rho_{\tbox{W}}(X,P)$ is related to $\rho(X,r)$ 
via a Fourier transform in the $r$~variable.

\sheadC{Applications of the Wigner Weyl formalism}

In analogy with the definition of the Wigner function $\rho_{\tbox{W}}(X,P)$ 
which is associated with $\hat{\rho}$ we can define a Wigner-Weyl 
representation $A_{\tbox{WW}}(X,P)$  of any hermitian operator $\hat{A}$.  
The phase space function $A_{\tbox{WW}}(X,P)$ is obtained from 
the standard representation ${A(x',x'')}$ 
using the same ``Wigner function formula" recipe.  
Let us consider the simplest examples.    
First consider how the recipe works for the operator $\hat{x}$:   
\beq
\langle \ x'|\hat{x} |x'' \rangle
\ \ = \ \ 
x'\delta(x'-x'')
\ \ = \ \ X\delta(r)
\ \ \stackrel{\tbox{WW}}{\longmapsto} \ \ X
\eeq
Similarly $\hat{p}\stackrel{WW}{\rightarrow}P$.
Further examples are: 
\beq
f(\hat{x})
& \ \ \stackrel{\tbox{WW}}{\longmapsto} \ \  
& f(X)
\\ 
g(\hat{p})
& \ \ \stackrel{\tbox{WW}}{\longmapsto} \ \  
& g(P)
\\ 
\hat{x}\hat{p} 
& \ \ \stackrel{\tbox{WW}}{\longmapsto} \ \  
& XP +\frac{1}{2}i 
\\ 
 \hat{p}\hat{x} 
& \ \ \stackrel{\tbox{WW}}{\longmapsto} \ \ 
& XP -\frac{1}{2}i 
\\ 
\frac{1}{2}(\hat{x}\hat{p}+\hat{p}\hat{x}) 
& \ \ \stackrel{\tbox{WW}}{\longmapsto} \ \ 
& XP 
\eeq
In general for appropriate ordering we get that $f(\hat{x},\hat{p})$ 
is represented by $f(X,P)$. But in practical applications $f(\hat{x},\hat{p})$  
will not have the desired ordering, and therefore this recipe 
should be considered as a leading term in a semiclassical $\hbar$ expansion. 

There are two major applications of the Wigner Weyl formula. 
The first one is the calculation of the partition function.
\beq
\hspace{2 mm} \mathcal{Z}(\beta)
=\sum_{r}  \eexp{-\beta E_{r}}
= \sum_{r}  \left\langle r | \eexp{-\beta H} |r \right\rangle
=\trc(\eexp{-\beta H})
= \int\frac{dXdP}{2\pi}(\eexp{-\beta H(X,P)}) + \mathcal{O}(\hbar)
\eeq
The second one is the calculation of the number of eigenstates 
up to a given energy~$E$
\beq
\mathcal{N}(E)
&=& \sum_{E_r\leq E}1
=\sum_{r}\Theta(E-E_{r})
=\trc[\Theta(E-\hat{H})]
\\ \nonumber
& \approx & \int\frac{dXdP}
{2\pi}\Theta(E-H(X,P)) 
=\int_{{H(X,P)\leq E}}\frac{dXdP}{2\pi}
\eeq
Below we discuss some further applications
that shed light on the dynamics of 
wavepackets, interference, and on the nature 
of quantum mechanical states. We note that 
the time evolution of Wigner function is 
similar but not identical to the evolution 
of a classical distribution unless the 
Hamiltonian is a quadratic function 
of~$\hat{x}$ and~$\hat{p}$.

\sheadC{Wigner function for a Gaussian wavepacket}

A Gaussian wavepacket in the position representation is written as  
\beq
\Psi(x) = \frac{1}{\sqrt{\sqrt{2 \pi} \sigma }}  \eexp{-\frac{(x-x_0)^2}{4\sigma^2}} \eexp{ip_0 x}
\eeq
The probability density matrix is 
\beq
\rho(X,r)
&=&\Psi\left(X{+}\frac{1}{2}r\right) \Psi^*\left(X{-}\frac{1}{2}r\right) 
\\ \nonumber
&=& \frac{1}{\sqrt{2 \pi} \sigma }  
\eexp{-\frac{((X-x_0)+\frac{1}{2}r)^2}{4\sigma^2} 
- \frac{((X-x_0)-\frac{1}{2}r)^2}{4\sigma^2} +ip_0 r}
= \frac{1}{\sqrt{2 \pi} \sigma } 
\eexp{-\frac{(X-x_0)^2}{2 \sigma^2} - \frac{r^2}{8\sigma^2} +ip_0 r}
\eeq
Transforming to the Wigner representation
\beq
\rho_{\tbox{W}}(X,P) = \int \frac{1}{ \sqrt{2\pi} \sigma } 
\eexp{ -\frac{(X-x_0)^2}{2 \sigma^2} - \frac{r^2}{8\sigma^2} -i (P-p_0) r}  dr 
= \frac{1}{\sigma_x \sigma_p}  
\eexp{ -\frac{(X-x_0)^2}{2 \sigma_x^2} - \frac{(P-p_0)^2}{2\sigma_p^2}}
\eeq
where $\sigma_x = \sigma$ and $\sigma_p = \frac{1}{2\sigma}$. 
It follows that $\sigma_x \sigma_p = 1/2$. 
Let us no go backwards.  Assume that have a Gaussian in phase
space, which is characterized by some $\sigma_x$ and $\sigma_p$. 
Does it represent a legitimate quantum mechanical state?
The normalization condition ${\trc(\rho)=1 }$ is automatically satisfied. 
We also easily find that 
\beq
\trc(\rho^2) 
=\int{\frac{1}{\sigma_x^2 \sigma_p^2} 
\eexp{-\frac{(X-x_0)^2}{\sigma_x^2} 
- \frac{(P-p_0)^2}{\sigma_p^2}} 
\frac{dX dP}{2\pi}} 
= \frac{1}{2 \sigma_x \sigma_p }
\eeq
We know that ${\trc(\rho^2)=1 }$ implies pure state. 
If ${\trc(\rho^2)<1 }$ it follows that we have a mixed state, 
whereas $\trc(\rho^2) > 1 $ is not physical.
It is important to remember that not any $\rho(X,P)$ 
corresponds to a legitimate quantum mechanical state. 
There are classical states that do not have 
quantum mechanical analog (e.g. point like preparation). 
Also the reverse is true: not any quantum state 
has a classical analogue. The latter is implied 
by the possibility to have negative regions in 
phase space. These is discussed in the next example.

\newpage
\sheadC{The Winger function of a bounded particle}

Wigner function may have some modulation on 
a fine scale due to an interference effect. 
The simplest and most illuminating example 
is the Wigner function of the $n$th eigenstate 
of a particle in a one dimensional box ($0<x<L$).
The eigen-wavefunction that correspond 
to wavenumber ${k=(\pi/L)\times \mbox{\small integer}}$ 
can be written as the sum of a right moving 
and a left moving wave 
${\psi(x)= (1/\sqrt{2})(\psi_1(x)+\psi_2(x))}$
within $0<x<L$, and $\psi(x)=0$ otherwise.
The corresponding Wigner function
is zero outside of the box. Inside the box
it can be written as
\beq
\rho_{\tbox{W}}(X,P) \ \ = \ \ 
\frac{1}{2}\rho_1(X,P)+\frac{1}{2}\rho_2(X,P)+\rho_{12}(X,P)
\eeq
where $\rho_{12}$ is the interference component.
The semiclassical components are concentrated
at $P=\pm k$, while the interference component
is concentrated at $P=0$.
The calculation of $\rho_1(X,P)$ in the interval $0<x<L/2$ 
is determined solely by the presence of the hard wall at $x=0$.
The relevant component of the wavefunction is
\beq
\psi_1(x) = \frac{1}{\sqrt{L}}\Theta(x) \eexp{ikx}
\eeq
and hence
\beq
\rho_1(X,P) &=&
\int_{-\infty}^{\infty}
\psi_1(X+(r/2))\psi_1^*(X-(r/2)) \eexp{-iPr} dr
\ \ = \ \ 
\frac{1}{L}
\int_{-\infty}^{\infty}
\Theta(X+(r/2))\Theta(X-(r/2))\eexp{-i(P-k)r} dr
\nonumber \\ \ &=&
\frac{1}{L}
\int_{-2X}^{2X}\eexp{-i(P-k)r} dr
\ \ = \ \ 
\frac{4X}{L} \sinc(2X(P-k))
\eeq
This shows that as we approach the sharp feature
the non-classical nature of Wigner function
is enhanced, and the classical (delta) approximation
becomes worse. The other components of Wigner function
are similarly calculated,
and for the interference component we get
\beq
\rho_{12}(X,P) =-2\cos(2kX) \times \frac{4X}{L} \sinc(2XP)
\eeq
It is easily verified that integration of $\rho_{\tbox{W}}(X,P)$
over $P$ gives $\rho(x)=1+1-2\cos(2kX) = 2(\sin(kX))^2$.

In many other cases the energy surface in phase space 
is ``soft'' (no hard walls) and then one can derive a 
uniform semiclassical approximation [Berry, Balazs]:  
\beq
\rho_{\tbox{W}}(X,P) = \frac{2\pi}{\Delta_{sc}(X,P)}\mathrm{Ai}
\left(\frac{{\cal H}(X,P)-E}{\Delta_{sc}(X,P)}\right)
\eeq
where for ${\cal H}=p^2/(2\mass)+V(x)$ 
\beq
\Delta_{sc} = \frac{1}{2}\left[
\hbar^2 \left(\frac{1}{\mass}|\nabla V(X)|^2
+\frac{1}{\mass^2} (P\cdot \nabla)^2 V(X)\right)
\right]^{1/3}
\eeq
What can we get out of this expression? 
We see that $\rho_{\tbox{W}}(X,P)$ decays 
exponentially as we go outside of the 
energy surface. Inside the energy surface 
we have oscillations due to interference.

The interference regions of the Wigner function 
might be very significant. A nice example is given by Zurek. 
Let us assume that we have a superposition 
of $N\gg1$ non-overlapping Gaussian.
we can write the Wigner function as
$\rho = (1/N)\sum \rho_j + \rho_{\tbox{intrfr}}$.
We have $\trc(\rho)=1$ and also $\trc(\rho^2)=1$.
This implies that $\trc(\rho_{\tbox{intrfr}})=0$,
while $\trc(\rho_{\tbox{intrfr}}^2) \sim 1$.
The latter conclusion stems from the observation that
the classical contribution is $N\times(1/N)^2 \ll 1$. 
Thus the interference regions of the Wigner function 
dominate the calculation.

\newpage
\sheadC{The Winger picture of a two slit experiment}

The textbook example of a two slit 
experiment will be analyzed below. 
The standard geometry is described in 
the upper panel of the following figure.
The propagation of the wavepacket is in the $y$ direction.
The wavepacket is scattered by the slits 
in the $x$ direction.  
The distance between the slits is $d$. 
The interference pattern is resolved 
on the screen. In the lower panel the phase-space 
picture of the dynamics is displayed. 
Wigner function of the emerging wavepacket is projected 
onto the $(x,p_x)$ plane. 

\begin{center}
\putgraph[0.4\hsize]{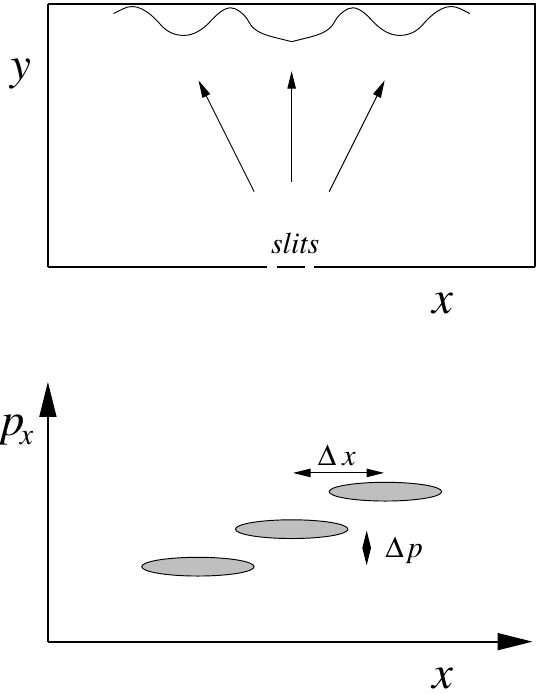}
\end{center}

The wavepacket that emerges from the 
two slits is assumed to be a superposition 
\beq
\Psi(x) \approx \frac{1}{\sqrt{2}} (\varphi_1(x)+\varphi_2(x)) 
\eeq
The approximation is related to the normalization 
which assumes that the slits are well separated.
Hence we can regard ${ \varphi_1(x) = \varphi_0(x+(d/2)) }$  
and ${ \varphi_2(x) = \varphi_0(x-(d/2)) }$
as Gaussian wavepackets with a vanishingly small overlap. 
The probability matrix of the superposition is 
\beq
\rho(x',x'')
= \Psi(x')\Psi^*(x'') 
= (\varphi_1(x')+\varphi_2(x'))(\varphi_1^*(x'')+\varphi_2^*(x''))
=\frac{1}{2}\rho_1+\frac{1}{2}\rho_2+\rho_{\tbox{interference}}
\eeq
All the integrals that are encountered in the calculation are 
of the Wigner function are of the type 
\beq
\int 
\varphi_0\left((X-X_0)+\frac{1}{2}(r-r_0)\right) 
\ \varphi_0\left((X-X_0)-\frac{1}{2}(r-r_0)\right) 
\ \eexp{-iPr} dr 
\ \ \equiv \ \ 
\rho_0(X-X_0,P) \ \eexp{-iPr_0}  
\eeq 
where $X_0=\pm d/2$  and $r_0=0$ for the 
classical part of the Wigner function, 
while $X_0=0$ and  $r_0=\pm d/2$ 
for the interference part.  
Hence we get the result   
\beq
\rho_{\tbox{W}}(X,P)
\ \ = \ \ \frac{1}{2}\rho_0\left(X+\frac{d}{2},P\right)
\ + \ \frac{1}{2}\rho_0\left(X-\frac{d}{2},P\right)
\ + \ \cos(Pd) \ \rho_0(X,P)
\eeq
Note that the momentum distribution can be obtained by integrating over~$X$
\beq
\rho(P) \ \ = \ \ (1+\cos(Pd)) \rho_0(P) \ \ = \ \ 2\cos^2(\frac{Pd}{2})\rho_0(P)
\eeq
In order to analyze the dynamics it is suggestive to write $\rho(X,P)$ 
schematically as a sum of partial-wavepackets, each characterized   
by a different transverse momentum:
\beq
\rho_{\tbox{W}}(X,P) \ \ = \ \ \sum_{n=-\infty}^{\infty} \rho_{n}(X,P)
\eeq
By definition the partial-wavepacket $\rho_{n}$  
equals $\rho$ for $|P-n\times(2\pi\hbar/d)| < \pi\hbar/d$ 
and equals zero otherwise. Each partial wavepacket 
represents the possibility that the particle, being 
scattered by the slits,  
had acquired a transverse momentum which is 
an integer multiple of 
\beq
\Delta p \ \ = \ \ \frac{2\pi\hbar}{d}
\eeq
The corresponding angular separation is 
\beq
\Delta\theta \ \ = \ \ \frac{\Delta p}{P} \ \ = \ \ \frac{\lambda_B}{d}
\eeq
as expected. The associated spatial separation 
if we have a distance~$y$ from the slits to the screen 
is ${\Delta x = \Delta\theta y}$. 
It is important to distinguish between the 
``preparation'' zone $y<d$, and the far-field 
(Franhaufer) zone $y \gg d^2/\lambda_B$.  
In the latter zone we have ${\Delta x \gg d}$ 
or equivalently ${\Delta x\Delta p \gg 2\pi\hbar}$.

\sheadC{Thermal states}

A stationary state $\partial \rho / \partial t$ has to satisfy $[\mathcal{H},\rho]=0$. 
This means that $\rho$ is diagonal in the energy representation. It can be further 
argued that in typical circumstances the thermalized mixture is of the canonical type.
Namely    
\beq 
\hat{\rho} 
= \sum |r\rangle p_r \langle r| 
= \frac{1}{\mathcal{Z}}\sum |r\rangle  \eexp{-\beta E_r}  \langle r| 
= \frac{1}{\mathcal{Z}} \eexp{-\beta \hat{H}} 
\eeq
Let us consider some typical examples. 
The first example is spin 1/2 in a magnetic field. 
In this case the energies are 
$E_\uparrow = \epsilon/2$ and $E_\downarrow = \epsilon/2$.
Therefore $\rho$ takes the following form:
\beq
\rho = \frac{1}{2 \cosh( \frac{1}{2} \beta \epsilon )} 
\left( \begin{array}{cc}
\eexp{\beta \frac{\epsilon}{2}} & 0 \\ 
0 & \eexp{- \beta \frac{\epsilon}{2}} \end{array} \right)
\eeq
Optionally one can represent the state of the spin   
by the polarization vector 
\beq
\vec{M} \ \ = \ \ \Big(0,0,\tanh (\frac{1}{2} \beta \epsilon) \Big) 
\eeq

The next example is a free particle.
The Hamiltonian is $H = \hat{p}^2/2\mass$. 
The partition function is  
\beq
\mathcal{Z} 
\ \ = \ \ \int \frac{dk}{(2\pi)/L} \eexp{-\beta \frac{k^2}{2m}} 
\ \ = \ \ \int \frac{dXdP}{2 \pi} \eexp{-\beta \frac{P^2}{2m}} 
\ \ = \ \ L \left( \frac{m}{2 \pi \beta} \right)^{\frac{1}{2}} 
\eeq
The probability matrix  $\rho$ is diagonal in the $p$ 
representation, and its Wigner function representation 
is identical with the classical expression. 
The standard $\hat{x}$ representation can be calculated either 
directly or via inverse Fourier transform of the Wigner function:
\beq
\rho (x',x'') & = & \ \left( \frac{1}{L} \right)  \eexp{-\frac{m}{2 \beta }[x'-x'']^2}
\eeq
Optionally it can be regarded as a special case of 
the harmonic oscillator case. In the case of an 
harmonic oscillator the calculation is 
less trivial because the Hamiltonian is not diagonal neither 
in the $x$ nor in the $p$ representation. 
The eigenstates of the Hamiltonian are $H |n\rangle = E_n |n\rangle$ 
with $E_n = \left( \frac{1}{2} + n \right) \omega$. 
The probability matrix $\rho_{nn'}$ is 
\beq
\rho_{nn'} \ \ = \ \ \frac{1}{\mathcal{Z}} \delta_{nn'} 
\eexp{-\beta \omega \left( \frac{1}{2} + n \right) } \eeq
where the partition function is 
\beq
\mathcal{Z} \ \ = \ \ \sum_{n=0}^{\infty} \eexp{-\beta E_n} 
\ \ = \ \ \left( 2 \sinh \left( \frac{1}{2} \beta \omega \right) \right)^{-1}
\eeq
In the $\hat{x}$ representation
\beq
\rho (x',x'') 
\ \ = \ \ \sum_n \langle x' | n \rangle p_n \langle n | x'' \rangle  
\ \ = \ \ \sum_n p_n \varphi^ n (x')\varphi^ n (x'') 
\eeq
The last sum can be evaluated by using properties 
of Hermite polynomials, but this is very complicated. 
A much simpler strategy is to use of the Feynman path integral 
method. The calculation is done as for the 
propagator $\langle x' | \exp(-i\mathcal{H} t) | x'' \rangle$
with the time $t$ replaced by $-i\beta$.  
The result is  
\beq
\rho (x',x'')  \ \ \propto \ \   
\eexp{-\frac{m \omega}{2 \sinh (\beta \omega)} 
\left[ \cosh(\beta \omega) \left( ({x''}^2 + {x'}^2)  - 2 x' x'' \right)\right]}
\eeq
which leads to the Wigner function 
\beq
\rho_{\tbox{W}}(X,P)  \ \ \propto  \ \ \eexp{-\beta 
\left( \frac{\tanh \left( \frac{1}{2} \beta \omega \right)}{\frac{1}{2} \beta \omega} \right) 
\left[ \frac{P^2}{2m} + \frac{1}{2} m \omega^2 X^2 \right]}
\eeq
It is easily verified that in the zero temperature 
limit we get a minimal wavepacket that represent the 
pure ground state of the oscillator, while in high 
temperatures we get the classical result which represents 
a mixed thermal state.

\newpage
\sheadB{Quantum states, operations and measurements}

\sheadC{The reduced probability matrix}

In this section we consider the possibility 
of having a system that has interacted 
with its surrounding. So we have ``system $\otimes$ environment'' 
or ``system $\otimes$ measurement device'' or simply 
a system which is a part of a larger thing which 
we can call ``universe''. The question that we would 
like to ask is as follows: Assuming that we know 
what is the sate of the ``universe'', what is the way 
to calculate the state of the ``system''?    
 
The mathematical formulation of the problem is as follows. 
The pure states of the "system" span $N_{\tbox{sys}}$ dimensional 
Hilbert space, while the states of the "environment" 
span $N_{\tbox{env}}$ dimensional Hilbert space. So the state 
of the "universe" is described by  $N\times N$ probability 
matrix $\rho_{i\alpha,j\beta}$, where $N=N_{\tbox{sys}} N_{\tbox{env}}$. 
This means that if we have operator $A$ which is represented 
by the matrix $A_{i\alpha,j\beta}$, then it expectation value is 
\beq
\langle A \rangle 
\ \ = \ \ 
\trc (A \rho ) 
\ \ = \ \ 
\sum_{i,j, \alpha , \beta} A_{i \alpha ,j \beta} \ \rho_{j \beta ,i \alpha}
\eeq
The probability matrix of the "system" is defined in the usual 
way. Namely, the matrix element  $\rho^{\tbox{sys}}_{j,i}$ is defined 
as the expectation value of ${P^{ji}=|i\rangle\langle j| \otimes \bm{1}}$. Hence
\beq
\rho^{\tbox{sys}}_{j,i} 
= 
\langle P^{ji} \rangle
= 
\trc ( P^{ji} \rho ) 
= 
\sum_{k,\alpha,l,\beta} 
P^{ji}_{k\alpha,l\beta} 
\rho_{l\beta,k\alpha} 
= 
\sum_{k,\alpha,l,\beta} 
\delta_{k,i}\delta_{l,j} \delta_{\alpha,\beta} 
\ \rho_{l\beta,k\alpha} 
= 
\sum_{\alpha} \rho_{j\alpha,i\alpha}
\eeq    
The common terminology is to say that $\rho^{\tbox{sys}}$ 
is the {\em reduced} probability matrix, which is obtained  
by {\em tracing out} the environmental degrees of freedom.
Just to show mathematical consistency we note that for a general 
system operator of the type  ${ A = A^{\tbox{sys}} \otimes \mathbf{1}^{\tbox{env}} }$ 
we get as expected
\beq
\langle A \rangle 
\ \ = \ \ 
\trc(A \rho)
\ \ = \ \ 
\sum_{i,\alpha,j,\beta} 
A_{i\alpha,j\beta}
\rho_{j\beta,i\alpha} 
\ \ = \ \ 
\sum_{i,j} 
A^{\tbox{sys}}_{i,j} 
\rho^{\tbox{sys}}_{j,i} 
\ \ = \ \ 
\trc (A^{\tbox{sys}} \rho^{\tbox{sys}})
\eeq

\sheadC{Entangled superposition}

Of particular interest is the case where the universe  
is in a pure state $\Psi$. Choosing for the system $\otimes$ environemnt  
an arbitrary basis ${|i\alpha\rangle = |i\rangle \otimes |\alpha\rangle }$,  
we can expand the wavefunction as 
\beq
|\Psi\rangle  \ \ = \ \ \sum_{i,\alpha} \Psi_{i\alpha} |i\alpha\rangle
\eeq
By summing over $\alpha$ we can write
\beq
|\Psi\rangle \ \ = \ \ \sum_{i} \sqrt{p_i} |i\rangle \otimes | \chi^{(i)} \rangle
\eeq
where $| \chi^{(i)} \rangle \propto \sum_{\alpha} \Psi_{i\alpha} |\alpha\rangle$ 
is called the "relative state" of the environment with respect 
to the $i$th state of the system, while $p_i$ is the associated normalization factor.
Note that the definition of the relative state implies 
that ${\Psi_{i\alpha} = \sqrt{p_j} \chi^{(i)}_{\alpha}}$. 
Using these notations it follows that the reduced probability matrix of the system is 
\beq
\rho^{\tbox{sys}}_{j,i} 
\ \ = \ \  
\sum_{\alpha} \Psi_{j\alpha}\Psi_{i\alpha}^* 
\ \ = \ \ 
\sqrt{p_ip_j} \ \langle \chi^{(i)} | \chi^{(j)} \rangle
\eeq

The prototype example for a system-environment entangled state 
is described by the superposition  
\beq
|\Psi\rangle \ \ = \ \ 
\sqrt{p_1}\ |1\rangle \otimes |\chi^{(1)}\rangle 
+  \sqrt{p_2}\ |2\rangle \otimes |\chi^{(2)}\rangle
\eeq
where $|1\rangle$ and $|2\rangle$ are orthonormal states of the system. 
The singlet state of two spin 1/2 particles is possibly the simplest 
example for an entangled superposition of this type. 
Later on we shall see that such entangled superposition may come out as 
a result of an interaction between the system and the environment. 
Namely, depending on the state of the system the environment, 
or the measurement apparatus, ends up in a different state $\chi$.
Accordingly we do not assume that $\chi^{(1)}$ and $\chi^{(2)}$ are orthogonal, 
though we normalize each of them and pull out the normalization 
factors as $p_1$ and $p_2$. The reduced probability matrix of the system is 
\beq
\rho^{\tbox{sys}}_{j,i} 
\ \ = \ \ 
\left(\amatrix{ 
p_1 & \lambda^* \sqrt{p_1p_2} \cr
\lambda \sqrt{p_1p_2} & p_2
}\right)
\eeq
where ${\lambda=\langle \chi^{(1)} | \chi^{(2)} \rangle}$. 
At the same time the environment is in a mixture 
of non-orthogonal states:  
\beq
\rho^{\tbox{env}}
\ \ = \ \ 
p_1 \ |\chi^{(1)}\rangle\langle \chi^{(1)}| 
+ p_2 \ |\chi^{(2)}\rangle\langle \chi^{(2)}|  
\eeq
The purity of the state of the system in the above 
example is determined by $|\lambda|$, and can be 
characterized by ${\trc(\rho^2)=1-2p_1p_2(1{-}|\lambda|^2)}$. 
The value ${\trc(\rho^2)=1}$ indicates a pure state, 
while ${\trc(\rho^2)=1/N}$ with ${N=2}$ 
characterizes a 50\%-50\% mixture.
Optionally the purity can be characterized by 
the Von Neumann entropy as discussed in a later section:  
This gives $S[\rho]=0$ for a pure state and ${S[\rho]=\log(N)}$ 
with ${N=2}$ for a 50\%-50\% mixture.

\sheadC{Schmidt decomposition}

If the "universe" is in a pure state 
we cannot write its $\rho$ as a {\em mixture} 
of product states, but we can write its $\Psi$ 
as an entangled superposition of product states. 
\beq
|\Psi\rangle \ \ = \ \ \sum_{i} \sqrt{p_i} |i\rangle \otimes | B_i \rangle
\eeq
where the $| B_i \rangle$ is the "relative state" of subsystem $B$ 
with respect to the $i$th state of subsystem $A$, 
while $p_i$ is the associated normalization factor. 
The states $| B_i \rangle$ are in general not orthogonal.
The natural question that arise is whether we can find a decomposition 
such that the  $| B_i \rangle$ are orthonormal. The answer is positive: 
Such decomposition exists and it is unique. It is called Schmidt 
decomposition, and it is based on singular value decomposition (SVD).
Let us regard $\Psi_{i\alpha}=W_{i,\alpha}$ as an $N_A \times N_B$ matrix.   
From linear algebra it is known that any matrix can be written 
in a unique way as a product:
\beq
W_{(N_A \times N_B)} = U_{(N_A \times N_A)}^{A} D_{(N_A \times N_B)} U_{ (N_B\times N_B)}^{B} 
\eeq
where $U^{A}$ and $U^{B}$ are the so called left and right unitary matrices, 
while $D$ is a diagonal matrix with so called (positive) singular values. 
Thus we can re-write the above matrix multiplication as 
\beq
\Psi_{i\alpha} 
\ \ = \ \ 
\sum_r U^{A}_{i,r} \ \sqrt{p_r} \ U^{B}_{r,\alpha} 
\ \ \equiv \ \ 
\sum_r  \sqrt{p_r} \ u^{A_r}_{i} \ u^{B_r}_{\alpha} 
\eeq
Substitution of this expression leads to the result
\beq
|\Psi\rangle 
\ \ = \ \ 
\sum_{i,\alpha} \Psi_{i\alpha} | i \alpha \rangle
\ \ = \ \ 
\sum_{r} \sqrt{p_r} |A_r\rangle \otimes | B_r \rangle
\eeq
where $| A_r \rangle$ and $| B_r \rangle$ are implied by the unitary 
transformations. We note that the normalization of $\Psi$ implies $\sum p_r=1$. 
Furthermore the probability matrix is $\rho^{A+B}_{i\alpha,j\beta}=W_{i,\alpha}W_{j,\beta}^*$, 
and therefore the calculation of the reduced probability matrix can be written as:
\beq
\rho^{A} =&  WW^{\dagger}   &= (U^{A}) D^2 (U^{A})^{\dag} 
\\ \nonumber
\rho^{B} =& (W^{T})(W^{T})^{\dagger} &= [(U^{B})^{\dag} D^2 (U^{B})]^*
\eeq
This means that the matrices $\rho^A$ and $\rho^B$ 
have the same non-zero eigenvalues~$\{p_r\}$, or in other words 
it means that the degree of purity of the two subsystems is the same.

\sheadC{The violation of the Bell inequality}

We can characterize the entangled state using
a correlation function as in the EPR thought experiment. 
The correlation function ${C(\theta)=\langle \hat{C}\rangle}$  
is the expectation value of a so-called ``witness operator". 
If we perform the EPR experiment with two spin 1/2 particles 
(see the Fundamentals~II section), then the witness
operator is $\hat{C}=\sigma_z \otimes \sigma_{\theta}$, 
and the correlation function comes out ${C(\theta)=-\cos(\theta)}$, 
which violates Bell inequality.

Let us see how the result for ${C(\theta)}$ is derived. 
For pedagogical purpose we present 3 versions of the 
derivation. One possibility  is to perform a straightforward 
calculation using explicit {\em standard-basis} representation:        
\beq
|\Psi\rangle  \mapsto \frac{1}{\sqrt{2}}\left(\amatrix{0 \cr 1 \cr -1 \cr 0}\right),
\ \ \ \ \ \ \ \ \ \ \     
\hat{C} = \sigma_z \otimes \sigma_{\theta} 
\mapsto \left(\amatrix{ \sigma_{\theta} & 0 \cr 0 & -\sigma_{\theta} }\right),
\ \ \ \ \ \ \ \ \ \ \  
\langle \hat{C}\rangle = \frac{1}{2}\Big(\langle\sigma_{\theta}\rangle_{\downarrow}-\langle\sigma_{\theta}\rangle_{\uparrow}\Big), 
\eeq
leading to the desired result. The second possibility
is to use the ``appropriate" basis for the $C$ measurement:
\beq
\text{MeasurementBasis} \ \ = \ \ 
\Big\{ 
|z \theta\rangle, \ \ 
|z \bar{\theta}\rangle, \ \ 
|\bar{z} \theta\rangle, \ \ 
|\bar{z} \bar{\theta}\rangle
\Big\} 
\eeq
where $\bar{z}$ and $\bar{\theta}$ label polarization 
in the $-z$ and $-\theta$ directions respectively.
The singlet state in this basis is 
\beq 
|\psi \rangle 
\ \ = \ \ \frac{1}{\sqrt{2}} \left( |\theta\bar{\theta}\rangle-|\bar{\theta}\theta\rangle \right) 
\ \ = \ \ 
\frac{1}{\sqrt{2}}\cos\left(\frac{\theta}{2}\right) \left( |z \bar{\theta}\rangle - |\bar{z}\theta\rangle \right) 
+\frac{1}{\sqrt{2}}\sin\left(\frac{\theta}{2}\right) \left( |z \theta\rangle + |\bar{z}\bar{\theta}\rangle \right)
\eeq
Therefore the probabilities to get $C{=}1$ and $C{=}-1$ 
are $|\sin(\theta/2)|^2$ and $|\cos(\theta/2)|^2$ respectively, 
leading to the desired result for the average value $\langle \hat{C}\rangle$.

Still there is a third version of this derivation, which is more physically 
illuminating. The idea is to relate correlation functions to {\em conditional} calculation 
of expectation values. Let ${A=|a_0\rangle\langle a_0|}$ be a projector 
on state~$a_0$ of the first subsystem, and let $B$ some 
observable which is associated with the second subsystem. 
We can write the state of the whole system as an entangled superposition
\beq
\Psi \ \ = \ \ \sum_a \sqrt{p_a}\ |a \rangle \otimes |\chi^{(a)}\rangle
\eeq
Then it is clear that ${\langle A\otimes B \rangle = p_{a_0} \langle \chi^{(a_0)} |B| \chi^{(a_0)} \rangle}$. 
More generally if ${A=\sum_a |a\rangle a \langle a|}$ is any operator then 
\beq
\langle A\otimes B \rangle \ \ = \ \ \sum_a p_{a} \ a \ \langle \chi^{(a)} |B| \chi^{(a)} \rangle
\eeq
Using this formula 
with $A=\sigma_z=|\uparrow\rangle\langle \uparrow|-|\downarrow\rangle\langle \downarrow|$,  
and $B=\sigma_{\theta}$ we have ${p_{\uparrow}=p_{\downarrow}=1/2}$, 
and we get the same result for $\langle \hat{C}\rangle$ as in the previous derivation.

\sheadC{Quantum entanglement}

Let us consider a system consisting of two sub-systems, 
"A" and "B", with no correlation between them. 
Then, the state of the system can be factorized: 
\beq
\rho^{A+B} = \rho^{A}\rho^{B} 
\eeq
But in reality the state of the two sub-systems 
can be correlated. In classical statistical 
mechanics  $\rho^{A}$ and $\rho^{B}$ are 
probability functions, while $\rho^{A+B}$ is the joint 
probability function. In the classical state 
we can always write 
\beq
\rho^{A+B}(x,y) 
\ \ = \ \ 
\sum_{x',y'} \rho^{A+B}(x',y') \ \delta_{x,x'} \ \delta_{y,y'} 
\ \ \equiv \ \ 
\sum_r p_r \ \rho^{A_r}(x) \ \rho^{B_r}(y) 
\eeq
where $x$ and $y$ label classical definite states 
of subsystems $A$ and $B$ respectively, 
and $r=(x',y')$ is an index that distinguish pure classical 
states of $A \otimes B$. 
The probabilities $p_r=\rho^{A+B}(x',y')$ satisfy $\sum p_r=1$. 
The distribution $\rho^{A_r}$ represents a pure classical state of subsystem $A$, 
and $\rho^{B_r}$ represents a pure classical state of subsystem $B$.
Thus any classical state of $A \otimes B$ can be expressed 
as a {\em mixture} of product states. 

By definition a quantum state is {\em not entangled} 
if it is a product state or a mixture of product states.
Using explicit matrix representation it means that 
it is possible to write 
\beq
\rho_{i\alpha,j\beta}^{A+B} \ \ = \ \ 
\sum_{r} p_{r} \ \rho_{i,j}^{(A_r)} \ \rho_{\alpha,\beta}^{(B_r)} 
\eeq
It follows that an entangled state, unlike a non-entangled state, cannot have a classical interpretation. 
This means that it cannot be described by a classical joint probability function. 
The latter phrasing highlights the relation between entanglement and 
the failure of the hidden-variable-hypothesis of the EPR experiment. 

The question how to detect an entangled state is still open.
Clearly the violation of the Bell inequality indicates entanglement.
Adopting the GHZ-Mermin perspective (see ``Optional tests of realism" section)   
this idea is presented as follows: Assume you have two sub-systems $(A,B)$.
You want to characterize statistically the outcome of 
possible measurements using a {\em joint} probability function ${f(a_1,a_2,a_3,...;b_1,b_2,b_3,...)}$. 
You measure the correlations ${C_{ij} = \langle a_i b_j \rangle}$. 
Each $C$ imposes a restriction on the hypothetical~$f$s that could describe the state.
If the state is non-classical ({\em entangled})
the logical conjunction of all these restrictions gives NULL.

\sheadC{Purity and the von-Neumann entropy}

The purity of a state can be characterized 
by the von-Neumann entropy:  
\beq
S[\rho] = -\trc(\rho\log\rho)  = -\sum_{r} p_{r} \log{p_{r}} 
\eeq
In the case of a pure state we have $S[\rho]=0$, 
while in the case of a uniform mixture of $N$ states we have $S[\rho]=\log(N)$.  
From the above it should be clear that while the "universe" 
might have zero entropy, it is likely that a subsystem  
would have a non-zero entropy. For example if the universe 
is a zero entropy singlet, then the state of each spin 
is unpolarized with $\log(2)$ entropy. 

We would like to emphasize that the von-Neumann entropy $S[\rho]$ 
should not be confused with the Boltzmann entropy $S[\rho|A]$. 
The definition of the latter requires to introduce 
a partitioning~$A$ of phase space into cells.  
In the quantum case this ``partitioning"  is realized 
by introducing a complete set of projectors (a basis).  
The $p_r$ in the case of the Boltzmann entropy 
are probabilities in a given basis and not eigenvalues.
In the case of an isolated system out of equilibrium 
the Von Neumann entropy is a constant of the motion, 
while the appropriately defined Boltzmann entropy increases with time. 
In the case of a canonical thermal equilibrium  
the Von Neumann entropy $S[\rho]$ turns out to be equal 
to the thermodynamic entropy~$\mathcal{S}$. 
The latter is defined via the equation $dQ=Td\mathcal{S}$, 
where $T=1/\beta$ is an integration factor 
which is called the absolute temperature.

If the von-Neumann entropy were defined for a classical 
distributions ${\rho=\{p_r\}}$, it would have all the 
classical ``information theory" properties of the Shanon entropy.
In particular if we have two subsystems~$A$ and~$B$ one 
would expect 
\beq
S[\rho^{\tbox{A}}], \ S[\rho^{\tbox{B}}] 
\ \ \le \ \ 
S[\rho^{\tbox{AB}}]
\ \ \le \ \ 
S[\rho^{\tbox{A}}]+S[\rho^{\tbox{B}}]
\eeq 
Defining $N=\exp(S)$, the inequality ${N_{\tbox{AB}}<N_{\tbox{A}}N_{\tbox{B}}}$ 
has a simple graphical interpertation.

The above ineqiality is satisfied also in the  
quantum case provided the subsystems are not entangled. 
We can use this mathematical observation in order 
to argue that the zero entropy singlet state 
is an entangled state: It cannot be written as a product 
of pure states, neither it cannot be a mixture of product states. 

The case where $\rho^{A+B}$ is 
a zero entropy pure state 
deserves further attention.
As in the special case of a singlet, 
we can argue that if the state 
cannot be written as a product, then 
it must be an entangled state. 
Moreover for the Schmidt decomposition 
procedure it follows 
that the entropies of the subsystems  
satisfy $S[\rho^{\tbox{A}}]=S[\rho^{\tbox{B}}]$. 
This looks counter intuitive at first 
sight because subsystem $A$ might be 
a tiny device which is coupled  
to a huge environment $B$. We emphasize  
that the assumption here is that the 
"universe" $A \otimes B$ is prepared 
in a zero order pure state. 
Some further thinking leads to the conjecture 
that the entropy of the subsystems in 
such circumstance is proportional to the 
area of surface that divides the two regions 
in space. Some researchers speculate that 
this observation is of relevance to the discussion 
of black holes entropy.

\sheadC{Quantum operations}

The quantum evolution of an isolated system is described 
by a unitary operator, hence ${\tilde{\rho} = U \rho U^{\dag}}$.
We would like to consider a more general case.
The system is prepared in some well-controlled initial state  $\rho$, 
while the environment is assumed to be in some mixture 
state ${\sigma = \sum_{\alpha} |\alpha\rangle p_{\alpha} \langle\alpha|}$.
The state of the universe is ${R = \rho \otimes \sigma}$. 
The evolution of the universe is represented by $U(n\alpha|n'\alpha')$.     
Hence the evolution of the reduced probability matrix 
can be written as a linear operation, so called "quantum operation", namely 
\beq 
\tilde{\rho}_{n,m} \ = \ 
\sum_{n',m'} \mathcal{K}(n,m|n',m') \, \rho_{n',m'}, 
\ \ \ \ \ \ \ \ \ 
\mathcal{K}(n,m|n',m') \equiv \sum_{\alpha\alpha'}p_{\alpha'}U(n,\alpha|n'\alpha') U(m,\alpha|m',\alpha')^*
\eeq
With slight change of notations this can be re-written 
in a way that is called "Kraus representation" 
\beq
\tilde{\rho} \ \ = \ \  \sum_r [K^r] \, \rho \, [K^r]^{\dagger},
\ \ \ \ \ \ \ \ \ \ \ \ \ \ 
\sum_r [K^r]^{\dagger} \, [K^r] \ = \ \bf{1}
\eeq
where the sum rule reflects trace preservation (conservation of probability). 
Below we are using the following terminology: 
The quantum operation is induced by a linear $\mathcal{K}$-map, 
that is represented by a $\mathcal{K}$-kernel ${\mathcal{K}(n,m|n',m')}$,
with an associated $\mathcal{K}$-matrix ${\mathcal{K}_{nn',mm'} \equiv \mathcal{K}(n,m|n',m')}$. 
Note the order of indices.

Not any matrix $\rho$ can be regarded as representing a quantum state. 
Disregarding normalization ${\trc(\rho)=1}$ one requires that ${\trc(\rho P)>0}$ for any projector, 
or equivalently ${\BraKet{\psi}{\rho}{\psi}>0}$ for any $\psi$.
This is called "positivity". Upon diagonalization the eigenvalues $p_r$ have to be non-negative. 
It follows that one can define a matrix $V$ such that ${\rho_{n,m} = \sum_r V_{n,r} V_{m,r}^*}$, 
or in abstract notations ${\rho = VV^{\dag}}$.

It can be easily verify that the linear kernel $\mathcal{K}$ preserves the positivity of $\rho$. 
This positivity is essential for the probabilistic interpretation of~$\rho$. 
In fact $\mathcal{K}$ is ``completely positive". 
This means that if we consider any ``positive" matrix $R$ that describes the universe, 
possibly an entangled state, the result of the operation ${\mathcal{K} \otimes \bm{1}}$ 
would be ``positive" too. Explicitly it means that for any $R$ we get  
a positive matrix ${\tilde{R}_{n\alpha,m\beta} = \sum_{n'm'} \mathcal{K}_{nn',mm'} R_{n'\alpha,m'\beta}}$.

It is now possible to turn things around, and claim 
that any trace-preserving completely-positive linear mapping
of Hermitian matrices has a "Kraus representation". 
Given $\mathcal{K}$, it follows from the Hermiticity 
requirement that ${\mathcal{K}_{nn',mm'}}$ is Hermitian matrix. 
Therefore it has real eigenvalues $\lambda_r$, 
with transformation matrix $T(nn'|r)$ such that   
\beq
\mathcal{K}(n,m|n',m') \ \ \equiv \ \  \mathcal{K}_{nn',mm'}
\ \ = \ \ \sum_r  T(nn'|r) \, \lambda_r \, T(r|mm')
\ \ = \ \ \sum_r  \lambda_r [K^r_{nn'}] [K^r_{mm'}]^*
\eeq
where $\trc([K^r]^{\dag}[K^s]])=\delta_{r,s}$, 
due to the orthonormality of the transformation matrix. 
From the "complete positivity" it follows that all the $\lambda_r$ 
are positive (see proof below) hence we can absorb $\lambda_r^{1/2}$
into the definition of the $K^r$, and get the Kraus representation. 
We note that in this derivation the Kraus representation comes  
out orthogonal. One can switch to an optional Kraus representation 
using a linear transformation ${\tilde{K}^r = \sum_{s} u_{rs} K^s}$  
were $u_{rs}$ is any unitary matrix. After such transformation 
the $\tilde{K}^r$ are no longer orthogonal.

The statement that complete positivity implies ${\lambda_r>0}$,  
is paraphrasing of Choi's theorem. The theorem states 
that if the $\mathcal{K}$-map is completely positive
then the $\mathcal{K}$-matrix is positive, 
hence we get the Kraus representation as described above.
The proof is misleadingly simple: write the condition of 
complete positivity for the spacial case ${R = |\Psi\rangle \langle \Psi| }$, 
with the entangled state ${|\Psi\rangle = \sum_n |n,\alpha_n \rangle }$, 
where the relative states ${\alpha_n}$ are orthogonal. 
Then it follows that ${\tilde{R}_{n\alpha,m\beta} = \mathcal{K}_{n\alpha,m\beta}}$ 
has to be positive.      
  
Extension of the above reasoning is used in order to 
derive a master equation that describes the 
evolution of~$\rho$. The key non-trivial assumption 
is that the environment can be regarded as 
effectively factorized from the system at any moment.
Then we get the Lindblad equation:
\beq 
\frac{d\rho}{dt} \ = \  -i[\mathcal{H}_0,\rho] 
\ + \sum_r [W^r] \rho [W^r]^{\dagger} 
\ - \frac{1}{2}\left[\Gamma\rho + \rho \Gamma \right],
\ \ \ \ \ \ \ \ \ \ \ \ \ \ 
\Gamma=\sum_r [W^r]^{\dagger}[W^r]
\eeq
where the Lindblad operators $W^r$ parallel the Kraus operators $K^r$, 
and $\Gamma$ is implied by conservation of probability.
The Lindblad equation is the most general form of 
a Markovian master equation for the probability matrix.
We emphasize that in general the Markovian assumption 
does not hold, hence Lindblad is not as satisfactory as the Kraus description.

\sheadC{Measurements, the notion of collapse}

In elementary textbooks the quantum measurement process 
is described as inducing ``collapse" of the wavefunction. 
Assume that the system is prepared in state 
${\rho_{\tbox{initial}}=|\psi\rangle \langle \psi|}$ and that one 
measures ${\hat{P}=|\varphi\rangle \langle \varphi|}$. If the result 
of the measurement is $\hat{P}=1$ then it is said 
that the system has collapsed into the 
state  ${\rho_{\tbox{final}}=|\varphi\rangle \langle \varphi|}$. 
The probability for this ``collapse" is given by the projection 
formula ${\mbox{Prob}(\varphi | \psi) = |\langle \varphi | \psi \rangle|^2}$.

If one regard $\rho(x,x')$ or $\psi(x)$ as representing {\em physical reality},  
rather than a probability matrix or a probability amplitude, 
then one immediately gets into puzzles. Recalling the EPR experiment 
this world imply that once the state of one spin is measured at Earth, 
then immediately the state of the other spin (at the Moon) 
would change from unpolarized to polarized.  This would suggest that 
some spooky type of ``interaction" over distance has occurred. 

In fact we shall see that the quantum theory of measurement 
does not involve any assumption of spooky ``collapse" mechanism. 
Once we recall that the notion of quantum state has 
a statistical interpretation the mystery fades away. 
In fact we explain (see below) that {\em there is ``collapse" also in classical physics}!
To avoid potential miss-understanding it should be clear 
that I do not claim that the classical ``collapse" 
which is described below is an explanation of the 
the quantum collapse. The explanation of quantum collapse  
using a quantum measurement (probabilistic) point of view will 
be presented in a later section. The only claim of this section 
is that in probability theory a correlation is frequently mistaken 
to be a causal relation: ``smokers are less likely to have Alzheimer" not because cigarettes 
help to their health, but simply because their life span is smaller. 
Similarly quantum collapse is frequently mistaken to 
be a spooky interaction between well separated systems.

Consider the thought experiment 
which is known as the ``Monty Hall Paradox".
There is a car behind one of three doors. 
The car is like a classical "particle", 
and each door is like a "site". 
The initial classical state is such that 
the car has equal probability to be behind 
any of the three doors. You are asked to make a guess.
Let us say that you peak door \#1. Now the organizer 
opens door \#2 and you see that there is no car behind it. 
This is like a measurement. Now the organizer allows 
you to change your mind. The naive reasoning is that 
now the car has equal probability to be behind either 
of the two remaining doors. So you may claim that 
it does not matter. But it turns out that this simple 
answer is very very wrong! The car is no longer 
in a state of equal probabilities: Now the probability 
to find it behind door \#3 has increased. A standard   
calculation reveals that the probability to find 
it behind door \#3 is twice large compared with 
the probability to find it behind door \#2.
So we have here an example for a classical collapse. 

If the reader is not familiar with this well known "paradox", 
the following may help to understand why we have this 
collapse (I thank my colleague Eitan Bachmat for providing 
this explanation).  Imagine that there are billion doors. 
You peak door \#1. The organizer opens all the other doors 
except door \#234123. So now you know that the 
car is either behind door \#1 or behind door \#234123. 
You want the car. What are you going to do?    
It is quite obvious that the car is almost definitely 
behind door \#234123. It is also clear the that 
the collapse of the car into site \#234123 does not 
imply any physical change in the position of the car.

\sheadC{Quantum measurements}

What do we mean by quantum measurement? In order to 
clarify this notion let us consider a system that is prepared 
in a superposition of states $a$. Additionally we have 
a detector that is prepared  independently in a state $q{=}0$. 
In the present context the detector is called "von-Neumann pointer". 
The initial state of the system and the detector is  
\beq
|\Psi\rangle \ \ = \ \ \left[ \sum_a \psi_a |a\rangle \right] \otimes | q=0 \rangle 
\eeq
As a result of an interaction we assume that 
the pointer is displaced. Its displacement 
is proportional to~$a$. Accordingly the  
detector correlates with the system as follows:
\beq
\hat{U}_{\tbox{measurement}}\Psi \ \ = \ \ \sum\psi_a |a\rangle  \otimes  | q=a \rangle 
\eeq
We call such type of unitary evolution an {\em ideal projective measurement}. 
If the system is in a definite $a$ state, then it is not 
affected by the detector. Rather, we gain information on 
the state of the system. One can think of $q$ as representing 
a memory device in which the information is stored.
This memory device can be of course the brain of a human 
observer. From the point of view of the observer, 
the result at the end of the measurement process 
is to have a definite $a$.  This is interpreted 
as a ``collapse" of the state of the system. Some people 
wrongly think that ``collapse" is something that goes 
beyond unitary evolution. But in fact this term just 
makes over-dramatization of the above unitary process.

The concept of measurement in quantum mechanics involves 
psychological difficulties which are best 
illustrated by considering the ``Schroedinger cat" experiment.  
This thought experiment involves a radioactive nucleus, 
a cat, and a human being.  
The half life time of the nucleus is an hour. 
If the radioactive nucleus decays it triggers 
a poison which kills the cat. 
The radioactive nucleus and the cat are inside 
an isolated box. At some stage the human observer 
may open the box to see what happens with the cat...   
Let us translate the story into a mathematical language.  
A time $t=0$ the state of the universe (nucleus$\otimes$cat$\otimes$observer) is 
\beq
\Psi \ \ = \ \ 
|\uparrow=\mbox{radioactive}\rangle \ \otimes 
\ |{q=1=\mbox{alive}}\rangle 
\ \otimes |{Q=0=\mbox{ignorant}}\rangle
\eeq
where $q$ is the state of the cat, and $Q$ is the state 
of the memory bit inside the human observer.    
If we wait a very long time the nucleus would 
definitely decay, and as a result we will have a 
definitely dead cat:
\beq
U_{\tbox{waiting}} \Psi 
\ \ = \ \
|\downarrow=\mbox{decayed}\rangle \ \otimes 
\ |{q=-1=\mbox{dead}}\rangle 
\ \otimes \ |{Q=0=\mbox{ignorant}}\rangle
\eeq
If the observer opens the box he/she would see a dead cat:
\beq
U_{\tbox{seeing}} U_{\tbox{waiting}} \Psi 
\ \ = \ \ 
|\downarrow=\mbox{decayed}\rangle \ \otimes 
\ |{q=-1=\mbox{dead}}\rangle 
\ \otimes \ |{Q=-1=\mbox{shocked}}\rangle
\eeq
But if we wait only one hour then 
\beq
U_{\tbox{waiting}} \Psi \ = 
\ \frac{1}{\sqrt{2}} \Big[ 
|\uparrow\rangle  \otimes |q=+1\rangle  + |\downarrow\rangle  \otimes |q=-1\rangle  
\Big]  
\ \otimes \ |{Q=0=\mbox{ignorant}}\rangle
\eeq
which means that from the point of view 
of the observer the system (nucleus+cat) 
is in a superposition. The cat at this 
stage is neither definitely alive 
nor definitely dead. But now the 
observer open the box and we have:
\beq
U_{\tbox{seeing}} U_{\tbox{waiting}} \Psi \ = 
\ \frac{1}{\sqrt{2}} \Big[ 
|\uparrow\rangle  \otimes |q=+1\rangle \otimes |Q=+1 \rangle
\ + \ 
|\downarrow\rangle  \otimes |q=-1\rangle \otimes |Q=-1\rangle
\Big]  
\eeq
We see that now, from the point of view of the observer,  
the cat is in a definite(!) state. This is regarded by 
the observer as ``collapse" of the superposition.
We have of course two possibilities: one possibility is that  
the observer sees a definitely dead cat, while the other 
possibility is that the observer sees a definitely alive cat. 
The two possibilities "exist" in parallel, which leads to 
the "many worlds" interpretation.  Equivalently one may 
say that only one of the two possible scenarios is realized 
from the point of view of the observer,   
which leads to the "relative state" concept of Everett.
Whatever terminology we use, "collapse" or "many worlds" 
or "relative state", the bottom line is that we have 
here merely a unitary evolution.

\sheadC{Measurements and the macroscopic reality}

The main message of the Schroedinger's cat thought experiment 
is as follows: if one believes that a microscopic object (atom) 
can be prepared in a superposition state, then also a macroscopic 
system (atom+cat) can be prepared in a superposition state.
Accordingly the quantum mechanical reasoning should be applicable  
also in the macroscopic reality.

In fact there are more sophisticated schemes that allow to perform  
so called "quantum teleportation" of state from object to object.
However, one can prove easily the ``no cloning" theorem:  
a quantum state cannot be copied to other objects.
Such duplication would violate unitarity. 
The proof goes as follows: 
Assume that there were a transformation $U$ that maps (say) 
a two spin state ${|\theta\rangle \otimes |0\rangle}$  
to ${|\theta\rangle \otimes |\theta\rangle}$. 
The inner product ${\langle \theta' | \theta \rangle}$ 
would be mapped to ${\langle \theta' | \theta \rangle^2}$.
This would not preserve norm, hence a unitary cloning 
transformation is impossible. 

Many textbooks emphasize that in order to say that we have 
a measurement, the outcome should be macroscopic.  
As far as the thought experiment of the previous section 
is concerned this could be achieved easily: 
we simply allow the system to interact with one pointer, 
then with a second pointer, then with a third pointer, etc. 
We emphasize again that during an {\em ideal} measurement  
the pointer is not affecting the system, 
but only correlates with it. In other words: 
the measured observable $\hat{A}$ is a constant of the motion.
 
A more interesting example for a measurement with a macroscopic outcome is as follows: 
Consider a ferromagnet that is prepared at temperature ${T>T_c}$.
The ferromagnet is attached to the system and cooled down 
below~$T_c$. The influence of the system polarization (spin up/down) 
on the detector is microscopically small. But because of the 
symmetry breaking, the ferromagnet (huge number of coupled spins) will become 
magnetized in the respective (up/down) direction.  One may say that this 
is a generic model for a {\em macroscopic pointer}.

\sheadC{Measurements, formal treatment}

In this section we describe mathematically how an ideal projective measurement 
affects the state of the system. First of all let us write 
how the $U$ of a measurement process looks like. The formal expression is  
\beq
\hat{U}_{\tbox{measurement}} \ \ = \ \ \sum_a \hat{P}^{(a)}\otimes \hat{D}^{(a)}
\eeq
where $\hat{P}^{(a)}=|a\rangle\langle a|$ is the projection 
operator on the state $|a\rangle$, and $\hat{D}^{(a)}$ is a 
translation operator. Assuming that the measurement device 
is prepared in a state of ignorance $|q=0\rangle$, 
the effect of $\hat{D}^{(a)}$ is to get $|q=a\rangle$. Hence 
\beq
\hat{U}\Psi
\ \ &=& \ \ 
\left[\sum_a \hat{P}^{(a)} \otimes \hat{D}^{(a)}\right] 
\left(\sum_{a'} \psi_{a'} |a'\rangle  \otimes  | q=0 \rangle \right)
\nonumber \\ 
\ \ &=& \ \ 
\sum_a \psi_a |a\rangle \otimes \hat{D}^{(a)}|q=0\rangle
\ \ = \ \ 
\sum_a \psi_a |a\rangle \otimes |q=a\rangle
\eeq
A more appropriate way to describe the state 
of the system is using the probability matrix. 
Let us describe the above measurement process using this 
language. After "reset" the state 
of the measurement apparatus is ${\sigma^{(0)}=|q{=}0\rangle\langle q{=}0|}$.  
The system is initially in an arbitrary state $\rho$. 
The measurement process correlates that state of the 
measurement apparatus with the state of the system as follows:
\beq
\hat{U}\rho\otimes \sigma^{(0)} \hat{U}^\dagger
\ \ = \ \ 
\sum_{a,b} \hat{P}^{(a)} \rho \hat{P}^{(b)} 
\otimes [\hat{D}^{(a)}] \sigma^{(0)} [\hat{D}^{(b)}]^{\dagger}
\ \ = \ \ 
\sum_{a,b} \hat{P}^{(a)} \rho \hat{P}^{(b)} 
\otimes |q{=}a\rangle\langle q{=}b|
\eeq
Tracing out the measurement apparatus we get
\beq
\rho^{\tbox{final}}  
\ \ = \ \ 
\sum_a \hat{P}^{(a)} \rho \hat{P}^{(a)} 
\ \ \equiv \ \ \sum_a p_a \rho^{(a)}
\eeq
where $p_a$ is the trace of the projected probability 
matrix $\hat{P}^{(a)} \rho \hat{P}^{(a)}$, while $\rho^{(a)}$ 
is its normalized version.
We see that the effect of the measurement is to turn the 
superposition into a mixture of $a$ states, 
unlike unitary evolution for which 
\beq
\rho^{\tbox{final}}  
\ \ = \ \ 
U \  \rho \ U^{\dag}
\eeq
So indeed a measurement process looks like a non-unitary process: 
it turns a pure superposition into a mixture. 
A simple example is in order. Let us assume that the 
system is a spin 1/2 particle. The spin is prepared 
in a pure polarization state $\rho=|\psi \rangle \langle \psi|$ 
which is represented by the matrix 
\beq
\rho_{ab} \ \ = \ \ \psi_a \psi_b^{*} 
\ \ = \ \  \left( \begin{array}{cc}
|\psi_1|^2 & \psi_1 \psi_2^{*} \\
\psi_2 \psi_1^{*} & |\psi_2|^2 
\end{array}\right)
\eeq
where $1$ and $2$ are (say) the "up" and "down" states.
Using a Stern-Gerlach apparatus we can measure 
the polarization of the spin in the up/down direction.
This means that the measurement apparatus projects 
the state of the spin using     
\beq
P^{(1)} = \left( \begin{array}{cc}
1 & 0 \\
0 & 0 \end{array} \right)
\hspace{15mm} \mbox{and} \hspace{15mm}
P^{(2)} = \left( \begin{array}{cc}
0 & 0 \\
0 & 1 \end{array} \right)
\eeq
leading after the measurement to the state 
\beq
\rho^{\tbox{final}} 
\ \ = \ \ 
P^{(1)} \rho P^{(1)} + P^{(2)} \rho P^{(2)} 
\ \ = \ \ 
\left( \begin{array}{cc}
|\psi_1|^2 & 0 \\
0 & |\psi_2|^2 \end{array} \right)
\eeq
Thus the measurement process has eliminated the off-diagonal terms
in $\rho$ and hence turned a pure state into a mixture.  
It is important to remember that this non-unitary non-coherent evolution 
arise because we look only on the state of the system. 
On a universal scale the evolution is in fact unitary.

\sheadC{Weak measurement with post-selection}

In the previous section we have assumed that the measurement operation 
takes zero time, and results in the translation of a pointer $q$.
Such measurement is generated by an interaction term  
\beq
\mathcal{H}_{\tbox{measurement}} \ \ = \ \ -\lambda g(t) \ \mathcal{A} \ \hat{x}
\eeq
In this expression $\mathcal{A}$ is the system observable 
that we want to measure. It has a spectrum of values $\{a_i\}$. 
This observable is coupled  to a von-Nuemann {\em pointer} 
whose canonical coordinates are ${(\hat{x},\hat{q})}$.
The coupling constant is $\lambda$, and its temporal 
variation is described by a short time normalized function $g(t)$. 
If $\mathcal{A}=a$ this interaction shifts
the pointer ${q \mapsto q + \lambda a}$. 
Note that $x$ unlike $q$ is a constant of the motion.
Note also that $q$ is a dynamical variable, 
hence it has some uncertainty. For practical 
purpose it is useful to assume that the pointer 
has been prepared as minimal wave-packet that 
is initially centered at ${x=q=0}$.  

It is easily shown that for general preparation 
the average shift of the pointer 
is ${\lambda \langle \mathcal{A} \rangle}$. 
We would like to know how this result is modified 
if a {\em post-selection} is performed. 
Aharonov has suggested this scheme in order to treat 
the past and the future on equal footing. 
Namely: one assumes that the system is prepared in a state $|\Psi\rangle$, 
that is regarded as pre-selection; 
and keeps records of~$q$ only for events 
in which the final state of the system is post-selected as $|\Phi\rangle$.   
Below we shall prove that the average shift of the pointer 
is described by the complex number 
\beq
\langle\mathcal{A}\rangle_{\Phi,\Psi}^{\tbox{weak}} 
\ \ = \ \  
\frac{ \langle \Phi |\mathcal{A}| \Psi \rangle}
{\langle \Phi | \Psi \rangle}
\eeq
It is important to realize that this ``weak value" is not bounded:
it can exceed the spectral range of the observable.

The evolution of the system with the pointer is described by 
\beq
U_{\tbox{measurement}} \ \ = \ \ \eexp{i\lambda \mathcal{A} \hat{x}}  
\ \ \approx \ \ 1+i\lambda \mathcal{A} \hat{x}
\eeq
where the latter approximation holds for what we regard here as ``weak measurement".
The $x$ position of the von-Nuemann pointer is 
a constant of the motion. Consequently the representation 
of the evolution operator is  
\beq
\langle \Phi, x|U| \Psi, x_0 \rangle \ \ = \ \ U[x]_{\Phi,\Psi} \ \delta(x-x_0)
\eeq
where $U[x]$ is a system operator that depends 
on the constant parameter~$x$.    
If the system is prepared in state $P^{\Psi} = |\Psi\rangle\langle\Psi|$, 
and the pointer is prepared in state $\rho^{(0)}$, 
then after the interaction we get 
\beq
\mbox{final state of the universe} \ \ = \ \  U \ \Big[ \ P^{\Psi} \ \rho^{(0)} \ \Big] \ U^{\dag}
\eeq
The reduced state of the pointer after post selection is 
\beq
\rho(x',x'') 
\ \ = \ \ \trc\Big[ \ P^{\Phi} P^{x',x''} \  U P^{\Psi} \rho^{(0)} U^{\dag} \  \Big]
\ \ = \ \ \tilde{K}(x',x'') \ \rho^{(0)}(x',x'') 
\eeq
where $P^{x',x''}=|x'' \rangle\langle x'|$.
Note that this reduced state is not normalized: 
the trace is the probability to find the system 
in the post-selected state $\Phi$.
In the last equality we have introduced the notation    
\beq
\tilde{K}(x',x'') \ \ = \ \ \langle \Phi | U[x'] | \Psi \rangle \ \langle \Psi | U[x'']^{\dag} | \Phi \rangle
\eeq
Defining $X$ and $r$ as the average and the difference 
of $x''$ and $x'$ respectively, we can write the evolution 
of the pointer in this representation as 
\beq
\rho(r,X) \ = \ \tilde{K}(r,X) \ \rho^{(0)}(r,X),
\hspace{2cm}
\tilde{K}(r,X) \ \equiv \ \Big\langle \Psi \Big| U\left[X-\frac{r}{2}\right]^{\dag} \Big| \Phi \Big\rangle  
\ \Big\langle \Phi \Big| U\left[X+\frac{r}{2}\right] \Big| \Psi \Big\rangle  
\eeq
Optionally we can transform to the Wigner 
representation, and then the multiplication 
becomes a convolution: 
\beq
\rho(q,X) \ = \  \int K(q{-}q';X) \ \rho^{(0)}(q',X) \ dq',
\hspace{2cm}
K(q{-}q';X) \ \equiv \ \int  \tilde{K}(r,X) \ \eexp{-i(q{-}q')r} dr
\eeq
If we summed over $\Phi$ we would get the standard result 
that apply to a measurement without post-selection, namely, 
\beq
K(q{-}q';X) 
\ \ = \ \ \sum_a |\langle a | \Psi \rangle|^2 \ \delta(q-q'-a) 
\ \ \sim \ \ \delta\Big(q-q'-\lambda \langle \mathcal{A} \rangle \Big)
\eeq
The last rough equality applies since our interest 
is focused on the {\em average} pointer displacement.    
With the same spirit we would like to obtain 
a simple result in the case of post-selection. 
For this purpose we assume that the measurement is weak, leading to 
\beq
\tilde{K}(r,X) \ \ = \ \  
|\langle \Phi|\Psi \rangle|^2 
+ i \lambda \re\left[ \langle \Psi|\Phi \rangle \langle \Phi| \mathcal{A} |\Psi \rangle   \right] \, r  
- \lambda \im\left[  \langle \Psi|\Phi \rangle \langle \Phi| \mathcal{A} |\Psi \rangle \right] \, X
\eeq
Bringing the terms back up to the exponent, 
and transforming to the Wigner representation we get
\beq
K(q{-}q';X) \ \ = \ \ |\langle \Phi|\Psi \rangle|^2 
\ \eexp{-\lambda \im\langle \mathcal{A} \rangle^{\tbox{weak}} X}
\ \delta\Big(q-q'-\lambda \re\langle \mathcal{A} \rangle^{\tbox{weak}} \Big)
\eeq
we see that the real and the imaginary parts if the ``weak value" 
determine the shift of the pointer in phase space. 
Starting with a minimal Gaussian of width $\sigma$, 
its center is shifted as follows:
\beq
\mbox{q-shift} \ \ &=& \ \ \lambda \ \re\left[\langle \mathcal{A} \rangle_{\Phi,\Psi}^{\tbox{weak}}\right]
\\
\mbox{x-shift} \  \ &=& \ \ \frac{\lambda}{2\sigma^2} \ \im\left[\langle \mathcal{A} \rangle_{\Phi,\Psi}^{\tbox{weak}}\right]
\eeq
We emphasize again that the ``weak value" manifest itself
only if the coupling is small enough to allow linear approximation 
for the shift of the pointer. In contrast to that, 
without post-selection  the average $q$-shift is ${\lambda \langle \mathcal{A} \rangle}$ 
irrespective of the value of $\lambda$.

\sheadC{Weak continuous measurements}

A more interesting variation on the theme of weak measurements  
arises due to the possibility to perform a continuous measurement.
This issue has been originally discussed by Levitov in connection 
with theme of ``full counting statistics" (FCS). 
Namely, let us assume that we have an interest in the current $\mathcal{I}$ that flows 
through a section of a wire. We formally define the counting 
operator as follows:  
\beq
\mathcal{Q} \ \ = \ \ \int_{-\infty}^{\infty} \mathcal{I}(t) \ dt
\eeq
The time window is defined by a rectangular function $g(t)$  that 
equals unity during the measurements. The interaction 
with the von-Nuemann {\em pointer} is  
\beq
\mathcal{H}_{\tbox{measurement}} \ \ = \ \ -\lambda g(t) \ \mathcal{I} \ \hat{x}
\eeq
Naively we expect a shift ${q \mapsto q + \lambda \langle \mathcal{Q} \rangle}$, 
and more generally we might wonder what is the probability distribution of~$Q$.
It turns out that these questions are ill-defined. 
The complication arises here because $\mathcal{Q}$ is an integral 
over a time dependent operator that does not have to commute 
with itself in different times.

The only proper way to describe the statistics of $Q$ 
is to figure out what is the outcome of the measurement 
as reflected in the final state of the von-Nuemann pointer.
The analysis is the same as in the previous section, 
and the result can be written as 
\beq
\rho(q,X) \ \ = \ \  \int K(q{-}q';X) \ \rho^{(0)}(q',X) \ dq' 
\eeq
where  
\beq
K(Q;x) \ \ = \ \ \frac{1}{2\pi} \int 
\left\langle \psi \Big| U[x-(r/2)]^{\dag} U[x+(r/2)] \Big| \psi \right\rangle \eexp{-iQr} dr
\eeq
In the expression above it was convenient 
to absorb the coupling $\lambda$ into the definition 
of $\mathcal{I}$. The derivation and the 
system operator $U[x]$ are presented below.   
The FCS kernel $K(Q;x)$ is commonly calculated for $x=0$. 
It is a quasi-probability distribution (it might have negative values).  
The $k$th quasi-moment of $Q$ can be obtained by taking the $k$th 
derivative of the bra-ket expression in the integrand 
with respect to~$r$, then setting $r=0$.

We follow here the formulation of Nazarov. 
Note that his original derivation has been based on 
an over-complicated path integral approach.
Here we present as much simpler version. 
The states of system can be expanded in some arbitrary basis $|n\rangle$,  
and accordingly for the system with the detector 
we can use the basis ${|n,x\rangle}$. 
The $x$ position of the von-Nuemann pointer is 
a constant of the motion. Consequently the representation 
of the evolution operator is  
\beq
U(n,x|n_0,x_0) \ \ = \ \ U[x]_{n,n_0} \ \delta(x-x_0)
\eeq
where $U[x]$ is a system operator that depends 
on the constant parameter~$x$.    
We formally write the explicit expression for $U[x]$ both in the Schrodinger picture
and also in the interaction picture using time ordered exponentiation:  
\beq
U[x] 
\ \ = \ \ \mathcal{T}\exp\left[-i\int_0^t (\mathcal{H} - x\mathcal{I})dt' \right]   
\ \ = \ \ U[0] \,\mathcal{T}\exp\left[i x \int_0^t \mathcal{I}(t')dt' \right]  
\eeq
The time evolution of the detector is described by its 
reduced probability matrix 
\beq
\rho(x',x'') 
\ \ = \ \ \trc\Big[ \ P^{x',x''} \  U \rho^{\psi} \rho^{(0)} U^{\dag} \  \Big]
\ \ = \ \ \tilde{K}(x',x'') \ \rho^{(0)}(x',x'') 
\eeq
where ${\rho^{\psi} = |\psi\rangle\langle\psi|}$ is the 
initial state of the system, and ${\rho^{(0)}}$ 
is the initial preparation of the detector, and 
\beq
\tilde{K}(x',x'') 
\ \ = \ \ \sum_n \langle n | U[x'] | \psi \rangle \ \langle \psi | U[x'']^{\dag} | n \rangle
\ \ = \ \ \langle \psi | U[x'']^{\dag} \ U[x'] | \psi \rangle
\eeq
Transforming to the Wigner function representation as in the previous section 
we get the desired result.

\sheadC{Interferometry}

Interferometry refers to a family of techniques whose purpose 
is to deduce the ``relative phase" in a superposition.
The working hypothesis is that there is some preferred ``standard" basis 
that allows measurements. In order to clarify this concept 
let us consider the simplest example, which is a ``two slit" experiment.
Here the relative phase of being in either of the two slits 
has the meaning of transverse momentum. Different momenta have different 
velocities. Hence the interferometry here is straightforward: 
one simply places a screen far away, such that transverse momentum 
transforms into transverse distance on the screen. This way one can 
use position measurement in order to deduce momentum. 
Essentially the same idea is used in ``time of flight" measurments 
of Bose-Einstein condensates: The cloud is released and expands, 
meaning that its momentum distribution translates into position distribution.  
The latter can be captured by a camera. Another example for interferometry
concerns the measurement of the relative phase of Bose condensed 
particles in a double well superposition. Here the trick is to induce 
a Rabi type ``rotation" in phase space, ending up in a population imbalance
that reflects the relative phase of the preparation.

\newpage
\sheadB{Theory of quantum computation}

\sheadC{Motivating Quantum Computation}

Present day secure communication is based on the RSA two key encryption 
method. The RSA method is based on the following observation: 
Let $N$ be the product of two unknown big prime numbers $p$ and $q$.
Say that we want to find are what are its prime factors. 
The simple minded way would be to try to divide $N$ 
by $2$, by $3$, by $5$, by $7$, and so on. 
This requires a huge number ($\sim N$) of operations. 
It is assumed that $N$ is so large such that in practice the 
simple minded approached is doomed.  
Nowadays we have the technology to build classical computers 
that can handle the factoring of numbers as large as $N \sim 2^{300}$  
in a reasonable time. But there is no chance to factor 
larger numbers, say of the order $N \sim 2^{308}$.
Such large numbers are used by Banks for the encryption of important 
transactions. In the following  sections we shall see that 
factoring of a large number $N$ would become 
possible once we have a quantum computer.

{\bf Computational complexity:}
A given a number $N$ can be stored in an $n$-bit 
register. The size of the register should be 
$n \sim \log(N)$, rounded upwards such that $N \le 2^n$. 
As explained above in order to factor a number 
which is stored in an $n$ bit register by a classical computer 
we need an exponentially large number ($\sim N$) 
of operations. Obviously we can do some of the operations 
in parallel, but then we need an exponentially 
large hardware. Our mission is to find an efficient 
way to factor an $n$-bit number that do not 
require exponentially large resources. It turns
out that a quantum computer can do the job 
with hardware/time resources that scale like 
power of $n$, rather than exponential in $n$.      
This is done by finding a number $N_2$ that has a common 
divisor with $N$. Then it is possible to use 
Euclid's algorithm in order to find this common 
divisor, which is either $p$ or $q$.

{\bf Euclid's algorithm:}
There is no efficient algorithm to factor a large 
number ${N \sim 2^n}$. The classical computational 
complexity of this task is exponentially large in $n$. 
But if we have two numbers $N_1=N$ and $N_2$ we can quite 
easily and efficiently find their greater common divisor ${\mbox{GCD}(N_1,N_2)}$ 
using Euclid's algorithm.  
Without loss of generality we assume that ${N_1>N_2}$.  
The two numbers are said to be co-prime if ${\mbox{GCD}(N_1,N_2)=1}$.      
Euclid's algorithm is based on the observation that 
we can divide $N_1$ by $N_2$ and take 
the reminder ${N_3=\mbox{mod}(N_1,N_2)}$ 
which is smaller than $N_2$. 
Then we have ${\mbox{GCD}(N_1,N_2)=\mbox{GCD}(N_2,N_3)}$. 
We iterate this procedure generating a sequence 
${N_1>N_2>N_3>N_4>\cdots}$ until the reminder is zero.  
The last non-trivial number in this sequence 
is the greater common divisor.

{\bf The RSA encryption method:}
The RSA method is based on the following mathematical observation. 
Given two prime numbers $p$ and $q$ define $N=pq$. 
Define also $a$ and $b$ such that ${ab=1 \mod[(p-1)(q-1)]}$. 
Then we have the relations
\beq
B =&  A^a & \ \mod[N] \\
A =&  B^b & \ \mod[N] 
\eeq  
This mathematical observation can be exploited as follows. 
Define
\beq
\mbox{public key} &=& (N,a) \\
\mbox{private key} &=& (N,b) 
\eeq  
If anyone want to encrypt a message $A$, 
one can use for this purpose the public key. 
The coded message $B$ cannot be decoded 
unless one knows the private key. 
This is based on the assumption that 
the prime factors $p$ and $q$ and hence $b$ 
are not known.

\sheadC{The factoring algorithm}

According to Fermat's  theorem, if $N$ is prime, 
then $M^N=M \mod(N)$ for any number $M (<N)$.
If the "seed" $M$ is not divisible by $N$, 
this can be re-phrased as saying that the period  
of the function $f(x)=M^x \mod(N)$ is $r=N{-}1$.
That means ${f(x+r)=f(x)}$. 
To be more precise the primitive period can be smaller 
than~$N$ (not any "seed" is "generator"). 
More generally if $N$ is not prime, 
and the seed $M$ has no common divisor with it, 
then the primitive period of $f(x)$ 
is called ``the order".

The quantum computer is a black box that allows 
to find the period $r$ of a function $f(x)$. 
How this is done will be discussed in the 
next section. Below we explain how this ability 
allows us to factor a given large number $N$. 
\begin{itemize}
\item[{\bf (1)}] 
We have to store $N$ inside an $n$-bit register. 
\item[{\bf (2)}] 
We pick a large number $M$, so called seed, which is smaller than $N$. 
We assume that $M$ is co-prime to $N$. This assumption can be 
easily checked using Euclid'd algorithm. If by chance the chosen $M$ is not 
co-prime to $N$ then we are lucky and we can factor $N$ without quantum computer. 
So we assume that we are not lucky, and $M$ is co-prime to $N$.
\item[{\bf (3)}] 
We build a processor that can calculate the function $f(x)=M^x \mod(N)$. 
This function has a period~$r$ which is smaller than $N$. 
\item[{\bf (4)}] 
Using a quantum computer we find one of the Fourier components 
of $f(x)$ and hence its period~$r$. This means that $M^r = 1 \mod(N)$.
\item[{\bf (5)}] 
If $r$ is not even we have to run the quantum computer a second 
time with a different $M$. Likewise if ${a^{r/2}=-1}$. 
There is a mathematical theorem that guarantees 
that with probability of order one we should be able 
to find $M$ with which we can continue to the next step. 
\item[{\bf (6)}] 
We define $Q=M^{r/2} \mod(N)$.  We have $Q^2=1 \mod(N)$, 
and therefore ${(Q-1)(Q+1)=0 \mod(N)}$. 
Consequently both $(Q-1)$ and $(Q+1)$ must have either $p$ or $q$  
as common divisors with $N$. 
\item[{\bf (6)}] 
Using Euclid's algorithm we find the GCD of $N$ and $\tilde{N}=(Q-1)$, 
hence getting either $p$ or $q$.        
\end{itemize}

The bottom line is that given $N$ and $M$ an an input,  
we would like to find the period $r$ of the functions 
\beq
f(x) = M^x \mod(N) 
\eeq
Why do we need a quantum computer to find the period?   
Recall that the period is expected to be of order $N$.
Therefore the $x$ register should be $n_c$ bits, 
where $n_c$ is larger or equal to $n$. 
Then we have to make order of $2^{n_c}$ operations 
for the purpose of evaluating $f(x)$ so as to find out 
its period. It is assumed that $n$ is large enough such that this 
procedure is not practical. We can of course try to do 
parallel computation of $f(x)$. But for that we need 
hardware which is larger by factor of $2^n$. It is assumed that 
to have such computational facility is equally not practical.
We say that factoring a large number has an exponentially complexity.

The idea of quantum processing is that the calculation of $f(x)$ 
can be done ``in parallel" without having to duplicate 
the hardware. The miracle happens due to the superposition principle. 
A single quantum register of size $n_c$ can be prepared at $t=0$ with 
all the possible input $x$ values in superposition. 
The calculation of $f(x)$ is done in parallel on the 
prepared state. The period of $f(x)$ in found via a Fourier analysis. 
In order to get good resolution $n_c$ should be larger 
than $n$ so as to have $2^{n_c}\gg 2^n$. 
Neither the memory, nor the number of operation, 
are required to be exponentially large.

\sheadC{The quantum computation architecture}

We shall regard the computer as a machine that has registers 
for memory and gates for performing operations. The complexity 
of the computation is determined by the total 
number of memory bits which has to be used times 
the number of operations. In other words the complexity 
of the calculation is the product ${ \mbox{memory}\times\mbox{time} }$. 
As discussed in the previous 
section classical parallel computation do not  
reduce the complexity of a calculation.  
Both classical and quantum computers 
work according to the following scheme:
\beq
|\mbox{output}\rangle \ \ = \ \ U[\mbox{input}] \ |0\rangle
\eeq
This means that initially we set all the bits or all the qubits 
in a zero state (a reset operation), and then we operate 
on the registers using gates, taking the input 
into account. It is assumed that there is a well define 
set of elementary gates. An elementary gate (such as "AND")  
operates on few bits (2 bits in the case of AND) 
at a time. The size of the hardware is determined by how 
many elementary gates are required in order to realize 
the whole calculation.

The quantum analog of the digital bits ("0" or "1") 
are the qubits, which can be regarded as spin~1/2 elements.
These are "analog" entities because the "spin" can point 
in any direction (not just "up" or "down").   
The set of states such that each spin is aligned 
either "up" or "down" is called the \textit{computational basis}. 
Any other state of a register can be written a superposition: 
\beq
|\Psi \rangle \ \ = \ \ \sum_{x_0,x_1,x_2,...} \psi(x_0,x_1,x_2...) \ |x_0,x_1,x_2,...\rangle
\eeq
The architecture of the quantum computer which is 
requited in order to find the period $r$ of the function $f(x)$ 
is illustrated in the figure below. We have two registers:
\beq
x &=& (x_0,x_1,x_2,...,x_{n_c-1}) \\
y &=& (y_0,y_0,y_2,...,y_{n-1})
\eeq
The registers $x$ and $y$ can hold binary numbers 
in the range ${x<N_c}$ and ${y<\bar{N}}$ respectively, 
where ${N_c=2^{n_c}}$ and ${\bar{N}=2^{n} > N}$.
The $y$ register is used by the CPU for processing $\mbox{mod}(N)$ 
operations and therefore it requires a minimal number of $n$~bits. 
The $x$ register has $n_c$ bits and it is used to store 
the inputs for the function $f(x)$.
In order to find the period of $f(x)$ the 
size $n_c$ of the latter register should be 
significantly larger compared with $n$.
Note that that ${n_c=n+10}$ implies that 
the $x$~range becomes roughly ${\times 1000}$ larger 
than the expected period. 
Large $n_c$ is required if we want 
to determine the period with large accuracy.

\begin{center}
\putgraph[0.6\hsize]{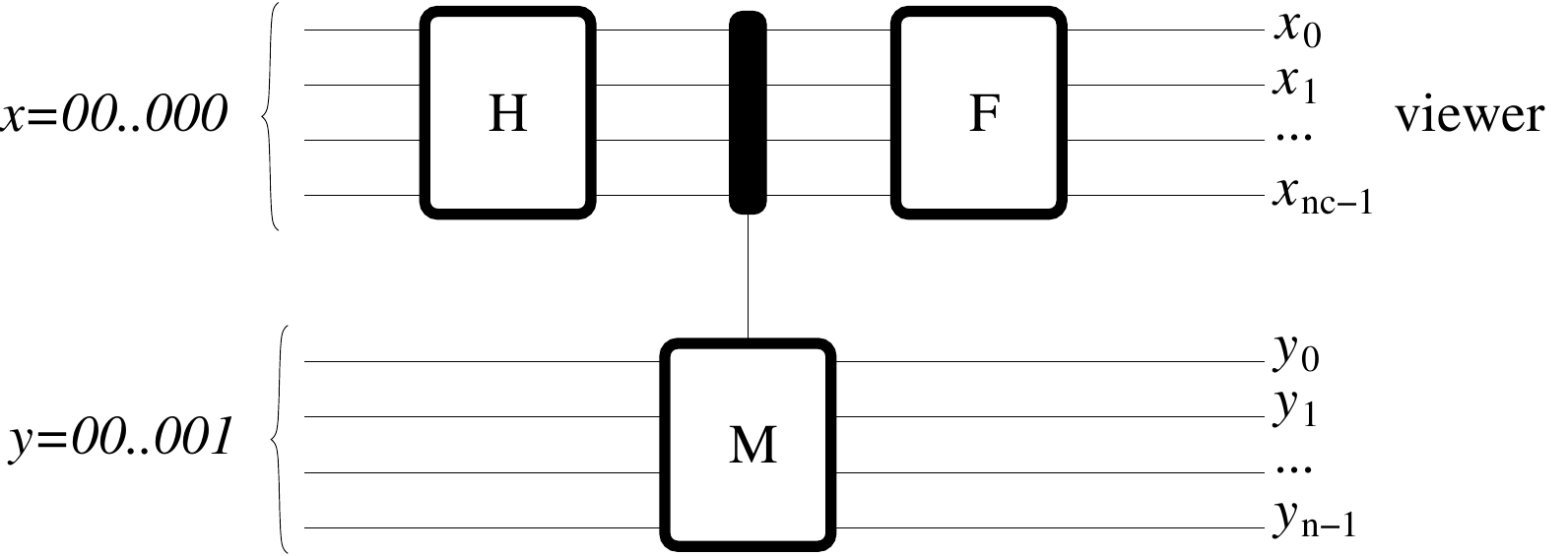}
\end{center}

We are now ready to describe the quantum computation.
In later sections we shall give more details, 
and in particular we shall see that the realization 
of the various unitary operations which are 
listed below does not require an exponentially 
large hardware.   
The preliminary stage is to make a "reset" 
of the registers, so as to have 
\beq
|\Psi \rangle \ \ = \ \ |x ; y\rangle = | 0,0,0,...,0,0 ; 1,0,0,...,0,0 \rangle
\eeq
Note that it is convenient to start with $y=1$ rather than $y=0$. 
Then come a sequence of unitary operations 
\beq
U \ \ = \ \ U_F U_M U_H
\eeq
where
\beq
U_H \ \ &=& \ \ U_{\tbox{Hadamard}} \otimes \bm{1} \\
U_M \ \ &=& \ \ \sum_x |x\rangle\langle x| \otimes  U_M^{(x)} \\
U_F \ \ &=& \ \ U_{\tbox{Fourier}} \otimes \bm{1}
\eeq
The first stage is a unitary operation $U_H$ that 
sets the $x$ register in a democratic state.
It can realized by operating on $\Psi$ with 
the Hadamard gate. Disregarding normalization we get   
\beq
|\Psi \rangle \ \ = \ \ \sum_x |x \rangle \otimes   | y{=}1 \rangle 
\eeq
The second stage is an $x$ controlled operation $U_M$.
This stage is formally like a quantum measurement: 
The $x$ register is "measured" by the $y$ register.  
The result is 
\beq
|\Psi \rangle \ \ = \ \ \sum_x |x \rangle \otimes  | y{=}f(x) \rangle 
\eeq
Now the $y$ register is entangled with the $x$ register. 
The fourth stage is to perform Fourier transform 
on the $x$ register:
\beq
|\Psi \rangle \ \ = \ \ \sum_x 
\left[ \sum_{x'}  \eexp{i\frac{2\pi}{N_c}xx'} |x' \rangle \right] 
\otimes   | f(x) \rangle 
\eeq
We replace the dummy integer index $x'$ by $k=(2\pi/N_c)x'$ 
and re-write the result as  
\beq
|\Psi \rangle \ \ = \ \ \sum_k   |k \rangle  \otimes   
\left[ \sum_x  \eexp{ikx} |f(x) \rangle \right] 
\ \ \equiv \ \ 
\sum_k p_k |k\rangle \otimes |\chi^{(k)}\rangle
\eeq
The final stage is to measure the $x$ register. 
The probability to get $k$ as the result is 
\beq  
p_k \ \ = \ \ 
\left|\left|
\sum_x  \eexp{ikx} |f(x) \rangle
\right|\right|^2 
\eeq
The only non-zero probabilities are associated 
with $k=\mbox{integer} \times (2\pi/r)$.
Thus we are likely to find one of these $k$ values, 
from which we can deduce the period $r$. 
Ideally the error is associated with the 
finite length of the $x$ register. By making 
the $x$ register larger the visibility of the Fourier 
components becomes better.

\sheadC{Single qubit quantum gates}

The simplest gates are one quibit gates.
They can be regarded as spin rotations. 
Let us list some of them:
\beq
T &=& \left(
\begin{array}
[c]{cc}1 & 0\\
0 & \eexp{i\pi/4}\end{array}
\right)  
\\ \nonumber
S &=& T^{2}=\left(
\begin{array}
[c]{cc}1 & 0\\
0 & i
\end{array}
\right)  
\\ \nonumber
Z &=& S^{2}=\left(
\begin{array}
[c]{cc}
1 & 0\\
0 & -1
\end{array}
\right) =  \sigma_{z} = i\eexp{-i\pi S_{z}}
\\ \nonumber
X &=& \left(
\begin{array}
[c]{cc} 
0 & 1\\
1 & 0
\end{array}
\right)  = \sigma_{x} = i\eexp{-i\pi S_{x}} = \mbox{NOT gate}
\\ \nonumber
Y &=& \left(
\begin{array}
[c]{cc}0 & -i\\
i & 0
\end{array}
\right)  = \sigma_{y} =  i\eexp{-i\pi S_{y}} = iR^2
\\ \nonumber
H &=& 
\frac{1}{\sqrt{2}}
\left(
\begin{array}
[c]{cc}1 & 1\\
1 & -1
\end{array} 
\right) 
= \frac{1}{\sqrt{2}}(\sigma_x+\sigma_z)
= i\eexp{-i\pi S_{n}} 
= \mbox{Hadamard gate}
\\ \nonumber
R &=& 
\frac{1}{\sqrt{2}}
\left(
\begin{array}
[c]{cc}1 & -1\\
1 & 1
\end{array} 
\right) 
= \frac{1}{\sqrt{2}}(1-i\sigma_y)
= \eexp{-i(\pi/2) S_{y}} 
= \mbox{90deg Rotation}
\eeq
We have $R^4=-1$ which is a $2\pi$ rotation in $SU(2)$.
We have $X^{2}=Y^{2}=Z^{2}=H^{2}=1$ which implies 
that these are $\pi$ rotations in $U(2)$.  
We emphasize that though the operation of $H$ on "up" 
or "down" states looks like $\pi/2$ rotation, 
it is in fact a $\pi$ rotation around a $45^o$ inclined axis:   
\beq
H =
\frac{1}{\sqrt{2}}\left(
\begin{array}
[c]{cc}1 & 1\\
1 & -1
\end{array}
\right)  =\left(  \frac{1}{\sqrt{2}},0,\frac{1}{\sqrt{2}}\right)
\cdot\overset{\rightarrow}{\sigma}=\overset{\rightarrow}{n}\cdot
\overset{\rightarrow}{\sigma}
\eeq

\sheadC{The CNOT quantum gate}

More interesting are elementary gates that operate 
on two qubits. In classical computers it is 
popular to use "AND" and "OR" gates.
In fact with "AND" and the one-bit operation "NOT" 
one can build any other gate. The "AND" cannot be 
literally realized as quantum gate because it does not 
correspond to unitary operation. But we can build 
the same logical circuits with  "CNOT" (controlled NOT) 
and the other one-qubit gates. The definition of CNOT 
is as follows: 
\beq
U_{\tbox{CNOT}}=\left(
\begin{array}
[c]{cc}\begin{array}
[c]{cc}1 & 0\\
0 & 1
\end{array}
& 0\\
0 &
\begin{array}
[c]{cc}0 & 1\\
1 & 0
\end{array}
\end{array}
\right)    
\eeq
The control bit is not affected 
while the $y$ bit undergoes NOT 
provided the $x$ bit is turned on. 
The gate is schematically illustrated in 
the following figure:

\begin{center}
\putgraph[0.3\hsize]{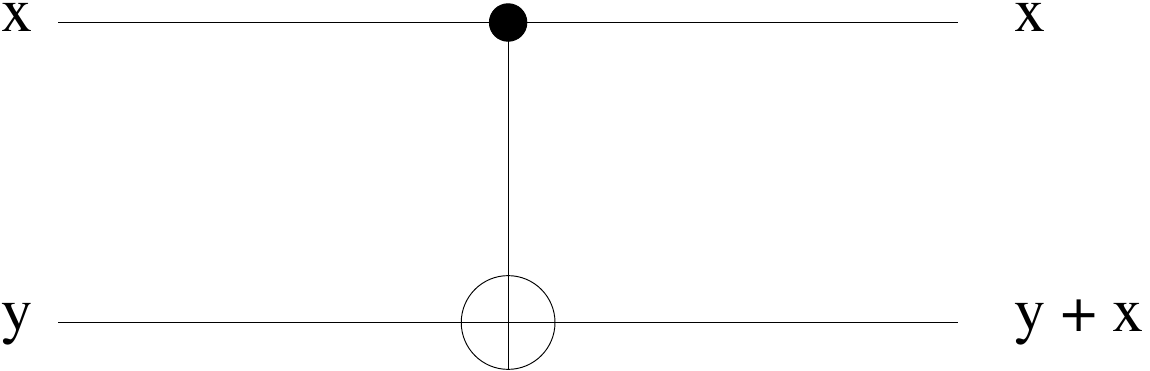}
\end{center}

One may wonder how a CNOT gate can be realized in practice. 
We first recall that any single qubit operation can be regarded 
as a rotation ${\phi}$ around some axis ${n}$, namely, 
\beq
R_n(\phi) \ \ = \ \ \exp\left[-i \frac{\phi}{2} \sigma_n \right] 
\eeq
If we add a control qubits ${\sigma^c}$ we can exploit standard spin-spin interaction 
in order realize the following gate operation:
\beq
U_z(\varphi) \ \ = \ \ \exp\left[i \frac{\varphi}{2} \sigma_z^c \otimes \sigma_z \right] \ \ = \ \  |0\rangle\langle 0| \otimes R_z(\varphi) +  |1\rangle\langle 1| \otimes R_z(-\varphi)   
\eeq
From that we can construct a simple controlled phase operation:
\beq
R_z(\pi/2) U_z(\pi/2) \ \ = \ \ |0\rangle\langle 0| \otimes 1 +  |1\rangle\langle 1| \otimes R_{z}(\pi) 
\eeq  
and then to construct a CNOT-like gate 
\beq
R_y(-\pi/2) R_z(\pi/2) U_z(\pi/2) R_y(\pi/2)  \ \ = \ \ |0\rangle\langle 0| \otimes 1 +  |1\rangle\langle 1| \otimes R_{x}(\pi) 
\eeq
This is not the standard CNOT. The standard CNOT requires 
an additional controlled phase ${\pi/2}$ operation to get  
\beq
\text{CNOT}  \ \ = \ \ |0\rangle\langle 0| \otimes 1 +  |1\rangle\langle 1| \otimes \sigma_{x} 
\eeq

\sheadC{The SWAP and the Toffoli gates}

Having the ability to realize single-qubit operations, and the CNOT gate, 
we can construct all other possible gates and circuits. 
It is amusing to see how SWAP gate can be realized by combining 3 CNOT gates:
\beq
U_{\tbox{SWAP}}=\left(
\begin{array}
[c]{ccc}1 &  & \\
&
\begin{array}
[c]{cc}0 & 1\\
1 & 0
\end{array}
& \\
&  & 1
\end{array}
\right) 
\eeq
which is illustrated is the following diagram: 

\begin{center}
\putgraph[0.4\hsize]{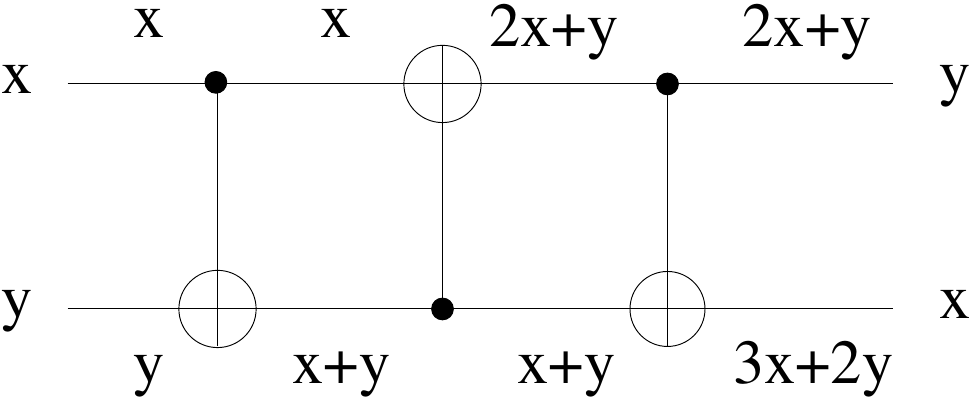}
\end{center}

The generalization of CNOT to the case where 
we have two qubit control register is known 
as Toffoli. The NOT operation is performed 
only if both control bits are turned on: 

\begin{center}
\putgraph[0.8\hsize]{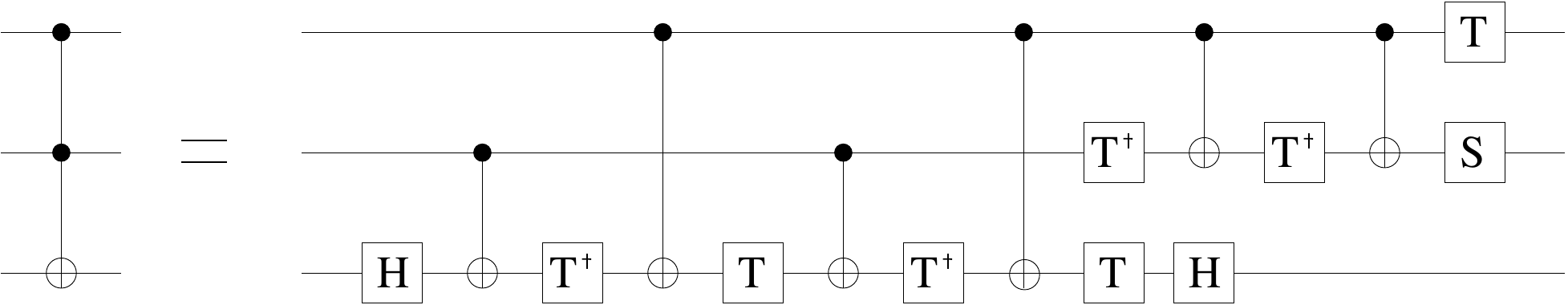}
\end{center}

The realization of the Toffoli gate 
opens the way to the quantum realization 
of an AND operation. 
Namely, by setting $y=0$ at the input,  
the output would be $y = x_1 \wedge x_2$.  
For generalization of the Toffoli gate 
to the case where the $x$ register 
is of arbitrary size see p.184 of Nielsen and Chuang.

\sheadC{The Hadamard Transform}

In the following we discuss the Hadamard and the Fourier transforms. 
These are unitary operations that are defined on the multi-qubit $x$ register. 
A given basis state $|x_0,x_1,x_3,...\rangle$ can be regarded 
as the binary representation of an integer number:  
\beq
x=\sum_{r=0}^{n_c-1} x_r 2^r
\eeq
We distinguish between the algebraic multiplication 
for which we use the notation $x x'$,  
and the scalar product for which we use the notation $x \cdot x'$, 
\beq
x \cdot x'  &=& \sum_r x_r x'_r 
\\ \nonumber
x x' &=& \sum_{r,s}  x_r x'_s 2^{r+s} 
\eeq

So far we have defined the single qubit Hadamard gate. 
If we have an multi-qubit register it is natural to define 
\beq
U_{\tbox{Hadamard}} \ \ = \ \ H \otimes H \otimes H \otimes \cdots
\eeq   
The operation of a single-qubit Hadamard gate can be written as 
\beq
\left\vert x_1 \right\rangle 
\ \ \overset{H}{\longrightarrow} \ \ 
\frac{1}{\sqrt{2}}
\left( \left\vert 0\right\rangle +(-1)^{x_1}
\left\vert1\right\rangle \right) 
\ \ = \ \
\sqrt{1}{2} \sum_{k_1=0,1}  (-1)^{k_1x_1} |k_1\rangle
\eeq
If we have a multi-qubit register we simply have to perform 
(in parallel) an elementary  Hadamard transform on each qubit:
\beq
\left\vert x_0, x_1, ..., x_r, ... \right\rangle 
\overset{H}{\longrightarrow} 
&& 
\prod_r 
\frac{1}{\sqrt{2}}
\left(  \left\vert 0\right\rangle +(-1)^{x_r}
\left\vert 1\right\rangle \right)  
\ \ = \ \ 
\frac{1}{\sqrt{2^{n_c}}} 
\prod_r 
\left[\sum_{k_r=0,1}
(-1)^{k_r x_r}  \left\vert k_r \right\rangle\right] 
\\ \nonumber
&& =
\frac{1}{\sqrt{N_c}} \sum_{k_0,k_1,...} (-1)^{k_0x_0+k_1x_1+...} 
\left\vert k_0, k_1, ..., k_r, ... \right\rangle
\ \ = \ \  
\frac{1}{\sqrt{N_c}} \sum_k (-1)^{k\cdot x} \left\vert k \right\rangle
\eeq
The Hadmard transform is useful in order to prepare 
a "democratic" superposition state as follows:
\beq
|0,0,...,0\rangle  
\ \ \overset{H}{\longrightarrow} \ \ 
\frac {1} {\sqrt{2}}(|0\rangle+|1\rangle)
\otimes 
\frac {1} {\sqrt{2}}(|0\rangle+|1\rangle)
\otimes ... \otimes 
\frac {1} {\sqrt{2}}(|0\rangle+|1\rangle)
\ \ \mapsto \ \ 
\frac {1} {\sqrt{N_c}}
\left( \begin{array}{c} 1 \\ 1 \\ . \\ . \\ . \\ 1  \end{array} \right)
\eeq
To operate with a unitary operator on this state 
is like making parallel computation 
on all the possible $x$ basis states.

\sheadC{The quantum Fourier transform}

The definitions of the Hadamard transform 
and the quantum Fourier transform are very similar in style:
\beq
U_{\tbox{Hadamard}}|x\rangle 
\ &=& \ \frac{1}{\sqrt{N_c}} \sum_k (-1)^{k\cdot x}|k\rangle
\\
U_{\tbox{Fourier}}|x\rangle 
\ &=& \ \frac{1}{\sqrt{N_c}} \sum_k \eexp{-i\frac{2\pi}{N_c}kx}|k\rangle
\eeq
Let us write the definition of the quantum Fourier transform
using different style so as to see that it is indeed a Fourier 
transform operation in the conventional sense. 
First we notice that its matrix representation is 
\beq
\langle x' |U_{\tbox{Fourier}}| x \rangle 
\ \ = \ \ \frac{1}{\sqrt{N_c}}  \eexp{-i \frac {2\pi}{N_c} x' x}
\eeq
If we operate with it 
on the state  $|\psi\rangle = \sum_x \psi_x |x\rangle$
we get $|\varphi \rangle = \sum_x \varphi_x |x\rangle$, 
where the column vector $\varphi_x$ is obtained from $\psi_x$ by 
a multiplication with the matrix that represents $U_{\tbox{Fourier}}$. 
Changing the name of the dummy index form $x$ to $k$ we get     
the relation
\beq
\varphi_k \ \ = \ \ \frac{1}{\sqrt{N_c}} \sum_{x=0}^{N_c-1} \eexp{-i \frac {2\pi}{N_c} kx} \psi_x
\eeq
This is indeed the conventional definition of 
\beq
\left( 
\begin{array}{c} 
\psi_0 \\ \psi_1 \\ \psi_2 \\ . \\ . \\ . \\ \psi_{N_c-1} 
\end{array} \right)
\ \ \ \overset{FT}{\longrightarrow} \ \ \ 
\left( 
\begin{array}{c} 
\varphi_0 \\ \varphi_1 \\ \varphi_2 \\ . \\ . \\ . \\ \varphi_{N_c-1} 
\end{array} \right)
\eeq
The number of memory bits which are required 
to store these vectors in a classical register 
is of order $N \sim 2^n$.  
The number of operations which is involved in 
the calculation of a Fourier transform  
seems to be of order $N^2$. 
In fact there is an efficient 
``Fast Fourier Transform" (FFT) algorithm that 
reduces the number of required operations to $N\log N = n2^n$. 
But this is still an exponentially large number in~$n$. 
In contrast to that a quantum computer can store  
these vectors in $n$ qubit register. Furthermore,  
the "quantum" FT algorithm can perform the calculation  
with only $n^2 \log n \log \log n$ operations. 
We shall not review here how the Quantum Fourier transform 
is realized. This can be found in the textbooks. 
As far as this presentation is concerned 
the Fourier transform can be regarded 
as a complicated variation of the Hadamard transform.

\sheadC{Note on analog or optical computation}

A few words are in order here regarding quantum 
computation versus classical analog computation.  
In an analog computer every analog "bit" can have 
a voltage within some range, so ideally each analog bit can 
store infinite amount of information. This is of course 
not the case in practice because the noise in the circuit   
defines some effective finite resolution.  Consequently 
the performance is not better compared with a digital computers.
In this context the analog resolution is a determining factor 
in the definition of the memory size.   
Closely related is optical computation. This can be regarded as  
a special type of analog computation. Optical Fourier Transform  
of a "mask" can be obtained on a "screen" 
that is placed in the focal plane of a lens.  
The FT is done in one shot. However, also here we have 
the same issue: Each pixel of the mask and each pixel  
of the screen is a hardware element. Therefore we still need 
an exponentially large hardware just to store the vectors. 
At best the complexity of FT with optical computer is of order $2^n$.

\sheadC{The $U_M$ operation}

The CNOT/Toffoli architecture can be generalized 
so as to realize any operation of the type $y = f(x_1,x_2,...)$, 
as an $x$-controlled operation, where $y$ is a single qubit.   
More generally we have 
\beq
x &=& \left(  x_{0},x_{1},x_{2}...x_{n_{c}-1}\right) 
\\ 
y &=& \left(  y_{0},y_{1},y_{2},...y_{n-1}\right)
\eeq
and we would like to realize a unitary controlled operation 
\beq
U 
\ \ = \ \ 
\sum_x |x\rangle\langle x| \otimes U^{(x)} 
\ \ \equiv \ \ 
P^{(0)} \otimes U^{(0)} +  P^{(1)} \otimes U^{(1)} + P^{(2)} \otimes U^{(2)} + ...
\eeq
This is formally like a measurement of the $x$ register by the $y$ register.
Note that $x$ is a constant of motion, and that $U$ has a block diagonal form:
\beq
\langle x^{\prime},y^{\prime} | U | x,y\rangle
\ \ = \ \ 
\delta_{x^{\prime},x}U_{y^{\prime},y}^{(x)}
\ \ = \ \ 
\begin{pmatrix}
U^{(0)} &  &  & \\
& U^{(1)} & & \\
&  & U^{(2)} &  \\
& & & ... 
\end{pmatrix}
\eeq
Of particular interest is the realization of a unitray 
operation that maps ${y=1}$ to ${y=f(x)}$. Let us look on 
\beq
U_M^{(x)} \Big| y \Big\rangle \ \ = \ \ \Big| M^x y \mod(N) \Big\rangle
\eeq
If $M$ is co-prime to $N$, then $U$ is merely  
a permutation matrix, and therefore it is unitary.   
The way to realize this operation is implied by the formula 
\beq
M^{x} \ \ = \ \ M^{\sum_s x_{s}2^{s}}
\ \ = \ \ \prod_s \left(M^{2^s}\right)^{x_{s}}
\ \ = \ \ \prod_{s=0}^{n_c-1} M_{s}^{x_{s}}
\eeq
which requires $n_c$ stages of processing.
The circuit is illustrated in the figure below.
In the $s$ stage we have to perform a controlled 
multiplication of $y$ by ${M_s\equiv M^{2^s} \mod(N)}$.

\begin{center}
\putgraph[0.6\hsize]{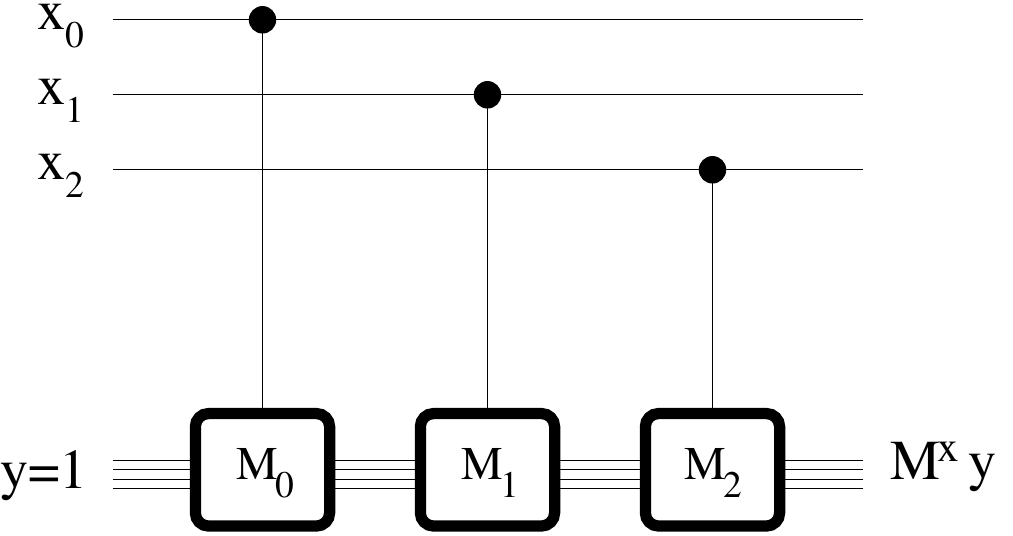}
\end{center}

\newpage
\sheadB{The foundation of statistical mechanics}

\sheadC{The canonical formalism}

Consider some system, for example 
particles that are confined in a box. 
The Hamiltonian is 
\beq
{\cal H} \ \ = \ \ {\cal H}(\bm{r},\bm{p} ; X)
\eeq
where $X$ is some control parameter, 
for example the length of the box.
The energy of the system is defined as  
\beq
E \ \ \equiv \ \ \langle {\cal H} \rangle 
\ \ = \ \ \trc({\cal H}\rho) 
\ \ = \ \ \sum_r p_r E_r
\eeq
where is the last equality we have assume 
that we are dealing with a stationary state.
Similarly the expression for the 
generalized force $y$ is   
\beq
y \ \ \equiv \ \ 
\left\langle 
-\frac{\partial \mathcal{H}}{\partial X}
\right\rangle 
\ \ = \ \ 
\sum_r p_r \left(-\frac{E_r}{dX}\right)
\eeq

It is argued that the weak interaction 
with an environment that has a huge 
density of state $\gdos_{\tbox{env}}$ 
leads after relaxation to a canonical state 
which is determined by the parameter 
${\beta= d\log(\gdos_{\tbox{env}}(E))/dE}$ 
that characterizes the environment. 
The argument is based on the assumption 
that the universe (system+environment) 
is a closed system with some total energy $E_{\tbox{total}}$.
After ergodization the system get into 
a stationary-like state. The probability $p_r$ 
to find the system in state $E_r$ is proportional 
to ${\gdos_{\tbox{env}}(E_{\tbox{total}}{-}E_r) 
\approx \gdos_{\tbox{env}}(E_{\tbox{total}})\eexp{-\beta E_r}}$. 
Accordingly 
\beq
p_{r}=\frac{1}{Z}\eexp{-\beta E_r}
\eeq
where the so-called partition function 
provides the normalization
\beq
Z\left(\beta ; X\right)=\sum_{r}\eexp{-\beta E_{r}}
\eeq
One observes that the energy can be obtained from 
\beq
E \ \ = \ \ -\frac{\partial \ln Z}{\partial \beta}
\eeq
while the generalized force is 
\beq
y \ \ = \ \ 
\frac{1}{\beta} \frac{\partial \ln Z}{\partial X}
\eeq

If we slightly change $X$ and $\beta$ and assume that 
the state of the system remains canonical then    
\beq
dE \ \ = \ \ \sum_{r}dp_{r}E_{r} + \sum p_{r}dE_{r}
\ \ \equiv \ \  TdS - ydX
\eeq
where the absolute temperature is {\em defined} 
as the integration factor of the first term  
\beq
T \ \ = \ \ \mbox{integration factor} \ \ = \ \ \frac{1}{\beta} 
\eeq
and the implied {\em definition} of the thermodynamic entropy is  
\beq
S \ \ = \ \ -\sum p_{r}\ln p_{r} 
\eeq
Note that the thermodynamic entropy is an extensive 
quantity in the thermodynamic limit.
At this state it is convenient to define 
the Helmholtz generating function
\beq
F(T,X) 
\ \ \equiv \ \ 
-\frac{1}{\beta}\ln Z(\beta;X)
\eeq
which allows to write the state equation in an elegant way:
\beq
S &=& -\frac{\partial F}{\partial T}
\\ 
y &=& -\frac{\partial F}{\partial X}
\eeq
and 
\beq
E &=& F+TS
\eeq

\sheadC{Work}

In the definition of work the system and the environment 
are regarded as one driven {\em closed} unit. 
On the basis of the ``rate of change formula" we have  
the following exact expression: 
\beq
\mathcal{W} \ \ = \ \ 
- \int \langle {\cal F} \ \rangle_t \ dX
\eeq
where ${\cal F}=-d{\cal H}/dX$. 
Note that $\langle {\cal F} \rangle_t$ 
is calculated for the time dependent 
(evolving) state of the system.  
From linear response theory 
of driven closed systems 
we know that in leading order  
\beq
\langle {\cal F} \rangle_t \ \approx \ \langle {\cal F} \rangle_X \ - \eta \dot{X} 
\eeq
The first terms is the conservative force, 
which is a function of $X$ alone. 
The subscript implies that the expectation 
value is taken with respect to the instantaneous 
adiabatic state. The second term is the leading 
correction to the adiabatic approximation. 
It is the ``friction" force which is proportional 
to the rate of the driving. The net conservative work 
is zero for a closed cycle while the ``friction" leads 
to irreversible dissipation of energy 
with a rate $\eta \dot{X}^2$.

The above reasoning implies that for a quasi static process 
we can calculate the work as the sum of two contributions:
${{\cal W} = -W  + {\cal W}_{\tbox{irreversible}}}$. 
The conservative work is defined as 
\beq
W \ \ = \ \ \int_{A}^{B} y(X) dX  
\eeq
The rate of irreversible work is 
\beq 
\dot{{\cal W}}_{\tbox{irreversible}} \ = \ \eta \dot{X}^2
\eeq
where $\eta$ is the ``friction" coefficient, 
which can be calculated using linear response theory.

\sheadC{Heat}

In order to understand which type of statements 
can be extracted form the canonical formalism 
we have to discuss carefully the physics 
of work and heat. We distinguish between the system 
and the environment and write the Hamiltonian 
in the form
\beq
\mathcal{H}_{\tbox{total}}
=\mathcal{H}(\bm{r},\bm{p};X(t))+\mathcal{H}_{\tbox{int}}+\mathcal{H}_{\tbox{env}}
\eeq
It is implicit that the interaction term is extremely 
small so it can be ignored in the calculation 
of the total energy. We define 
\beq
\mathcal{W} = \mbox{work}  
&\equiv& 
\Big(
\langle \mathcal{H}_{\tbox{total}} \rangle_{B}
-\langle \mathcal{H}_{\tbox{total}}\rangle_{A}
\Big)
\\
\mathcal{Q} = \mbox{heat}
&\equiv &
-\Big(\langle  
\mathcal{H}_{\tbox{env}}\rangle_{B}
-\langle  \mathcal{H}_{\tbox{env}}\rangle_{A}
\Big)
\\
E_{\tbox{final}}-E_{\tbox{initial}}  
& \equiv & 
\langle \mathcal{H}\rangle_{B}
-\langle \mathcal{H}\rangle_{A}
\ \ = \ \ \mathcal{Q} \ + \ \mathcal{W}
\eeq
If for a general process we know the work ${\cal W}$  
and the change in the energy of the system, 
then we can deduce what was the heat flow $\mathcal{Q}$. 
If we compare $dE = TdS - ydX$
with the expression $dE = \dbar \mathcal{Q} + \dbar \mathcal{W}$
we deduce that ${TdS=\dbar\mathcal{Q}+\dbar{\cal W}_{\tbox{irreversible}}}$.
This implies that the change in the entropy of the system 
is ${dS=(\dbar\mathcal{Q} + \dbar{\cal W}_{\tbox{irreversible}})/T}$.

\sheadC{The second law of thermodynamics}

The discussion of irreversible processes has been differed to
{\em Lecture Notes in Statistical Mechanics and Mesoscopic}, \href{http://arxiv.org/abs/1107.0568}{arXiv:1107.0568}

\sheadC{Fluctuations}

The partition function and hence the thermodynamic 
equations of state give information only on the 
spectrum $\{E_n\}$ of the system. In the classical 
context this observation implies that a magnetic field    
has no influence on the equilibrium state of a system 
because the spectrum remains ${E=\mass v^2/2}$ 
with $0<|v|<\infty$. In order to probe the dynamics 
we have to look on the 
fluctuations ${S(t)=\langle \mathcal{F}(t)\mathcal{F}(0)\rangle}$, 
where $\mathcal{F}$ is some observable. 
The Fourier transform of $S(t)$ describes 
the power spectrum of the fluctuations:
\beq
\tilde{S}(\omega) \ = \ \int_{-\infty}^{\infty} S(t) \eexp{i\omega \tau} d\tau 
\ = \ \sum_n p_n \sum_m |{\cal F}_{mn}|^2 
\ 2\pi \delta\left(\omega-(E_m{-}E_n)\right)
\eeq
This is the same object that appears in the 
Fermi-Golden-rule for the rate of transitions 
due to a perturbation term $V = -f(t) {\cal F}$. 
In the above formula $\omega>0$ corresponds 
to absorption of energy (upward transitions), 
while $\omega<0$ corresponds to emission (downward transitions).
It is a straightforward algebra to show that 
for a canonical preparations with ${p_n\propto \exp(-E_n/T)}$ 
there is a detailed balance relation:
\beq
\tilde{S}(\omega) 
= \tilde{S}(-\omega) 
\ \exp\left(\frac{\hbar\omega}{T}\right)
\eeq
This implies that if we couple to the system 
another test system (e.g. a two level ``thermometer")  
it would be driven by the fluctuations 
into a canonical state with the same temperature.

The connection with Fermi-Golden-rule is better formalized 
within the framework of the so called fluctuation-dissipation 
relation. Assume that the system is driven by varying a parameter $X$, 
and define ${\cal F}$ as the associated generalized force. 
The Kubo formula (see {\em Dynamics and driven systems} lectures) 
relates the response kernel to $S(t)$. 
In particular the dissipation coefficient is:
\beq
\eta(\omega) 
\ \ = \ \ 
\frac{\tilde{S}(\omega)-\tilde{S}(-\omega)}{2\omega}  
\eeq
If the system is in a canonical state it 
follow that the zero frequency response 
is $\eta_0= \tilde{S}(0)/(2T)$. 
If we further assume ``Ohmic response", 
which means having constant ${\eta(\omega)=\eta_0}$ 
up to some cutoff frequency, then the above 
relation can be inverted:
\beq
\tilde{S}_{\tbox{ohmic}}(\omega)
\ \ = \ \  
\eta_0 \frac{2\omega}{1-\eexp{-\omega/T}}
\eeq
The best known application of this relation is known 
as the Nyquist theorem. If a ring is driven 
by an electro-motive force $-\dot{\Phi}$,   
then the rate of heating is ${\dot{\mathcal{W}}=G\dot{\Phi}^2}$, 
which is know as Joule law. The generalized 
force which is associated with $\Phi$ is the 
current $\mathcal{I}$, and $G$ is known as the conductance.
Note that Joule law is equivalent 
to Ohm law $\langle\mathcal{I}\rangle = -G\dot{\Phi}$.    
It follows from the fluctuation-dissipation relation 
that the fluctuations of the current at equilibrium 
for ${\omega \ll T}$ are described by ${\tilde{S}(\omega)=2GT}$. 
It should be clear that in non-equilibrium we might have 
extra fluctuations, which in this example are known 
as shot noise.

\sheadC{The modeling of the environment}

It is common to model the environment as a huge 
collection of harmonic oscillators, and to say that 
the system is subject to the fluctuations 
of a field variable ${\cal F}$ which is a linear 
combination of the bath coordinates: 
\beq
{\cal F} 
\ \ = \ \ 
\sum_{\alpha} c_{\alpha}Q_{\alpha} 
\ \ = \ \ 
\sum_{\alpha} c_{\alpha} 
\left(\frac{1}{2 \mathsf{m}_{\alpha} \omega_{\alpha}} \right)^{1/2} 
(a_{\alpha} + a_{\alpha}^{\dag})
\eeq
For preparation of the bath 
in state $\bm{n}=\{n_{\alpha}\}$
we get 
\beq
\tilde{S}(\omega)
\ \ = \ \ 
\sum_{\alpha}\sum_{\pm} 
c_{\alpha}^2
\ |\langle n_{\alpha}{\pm}1 | Q_{\alpha} | n_{\alpha} \rangle|^2 
\ 2\pi\delta(\omega\mp\omega_{\alpha})
\eeq
Using
\beq
\langle n_{\alpha}{+}1 | Q_{\alpha} | n_{\alpha} \rangle &=& 
\left(\frac{1}{2 \mathsf{m}_{\alpha} \omega_{\alpha}} \right)^{1/2}
\ \sqrt{1 + n_{\alpha}}
\\
\langle n_{\alpha}{-}1 | Q_{\alpha} | n_{\alpha} \rangle &=& 
\left( \frac{1}{2 \mathsf{m}_{\alpha} \omega_{\alpha}} \right)^{1/2}
\ \sqrt{n_{\alpha}}
\eeq
we get
\beq
\tilde{S}(\omega) 
\ \ = \ \
\sum_{\alpha} 
\frac{1}{2\mathsf{m}_{\alpha} \omega_{\alpha}}
2\pi c_{\alpha}^2  \Big[ 
(1 {+} n_{\alpha}) \delta(\omega-\omega_{\alpha})
+ n_{\alpha} \delta(\omega+\omega_{\alpha})
\Big]
\eeq
For a canonical preparation of the bath 
we have $\langle n_{\alpha} \rangle = f(\omega_{\alpha}) \equiv 1/(\eexp{\omega/T}-1)$. 
It follows that  
\beq
\tilde{S}(\omega) 
\ \ = \ \ 
2J(|\omega|) \times 
\left\{\amatrix{   
(1 + f(\omega)) \cr
f(-\omega) 
}\right. 
\ \ = \ \ 
2J(\omega)\frac{1}{1-\eexp{-\beta\omega}} 
\eeq
where we used ${f(-\omega)=-(1+f(\omega))}$, 
and defined the spectral density of the bath as 
\beq
J(\omega) \ = \ \frac{\pi}{2} \sum_{\alpha}
\frac{c_{\alpha}^2}{\mathsf{m}_{\alpha} \omega_{\alpha}}
\delta(\omega-\omega_{\alpha}) 
\eeq
with anti-symmetric continuation. 
For an Ohmic bath $J(\omega)=\eta\omega$,  
with some cutoff frequency $\omega_c$.

\end{document}